\documentclass[11pt]{article}
\pdfoutput=1
\usepackage{jheppub}
\usepackage{tabu}
\usepackage[vcentermath]{youngtab}
\usepackage[usenames,dvipsnames,table]{xcolor}
\usepackage{graphicx,amsmath,amssymb,amsthm,multirow,array,bm,bbm,esint}
\usepackage[mathscr]{eucal}
\usepackage[bbgreekl]{mathbbol}
\usepackage{epsf,amsfonts}
\usepackage{slashed}
\usepackage[numbers,sort&compress]{natbib}
\usepackage{minibox}

\usepackage{rotating}
\usepackage{pdflscape}
\usepackage{array}
\newcolumntype{H}{>{\setbox0=\hbox\bgroup}c<{\egroup}@{}}

\usepackage{tikz}
\usetikzlibrary{shapes.geometric, arrows}
\tikzstyle{thread} = [rectangle, minimum width=.3\textwidth, minimum height=.6cm, text centered, text width=.3\textwidth, draw=black, fill=blue!10]
\tikzstyle{threadSmall} = [rectangle, minimum width=.15\textwidth, minimum height=.6cm, text centered, text width=.15\textwidth, draw=black, fill=green!5]
\tikzstyle{details} = [rectangle, minimum width=.23\textwidth, minimum height=.6cm, text centered, text width=.23\textwidth, draw=black, fill=green!5]
\tikzstyle{arrow} = [thick,->,>=stealth']
\tikzstyle{arrowT} = [ultra thick,->,>=stealth']
\tikzstyle{arrow2} = [thick,dashed,->,>=stealth']

\usepackage{rotfloat}
\usepackage{afterpage}
\usepackage{ccaption}
  \captionnamefont{\bfseries}
  \captiontitlefont{\small\sffamily}
  \captiondelim{: }
  \hangcaption

\newtheorem{theorem}{Theorem}

 \def\shadeB{\cellcolor{blue!5}}
\def\shadeR{\cellcolor{red!5}}

\definecolor{rust}{rgb}{0.8,0.2,0.2}
\def\MR#1{{\color{rust}{ #1}}}

\usepackage{tocloft}
\newcommand{\beq}{\begin{equation}}
\newcommand{\eeq}{\end{equation}}
\newcommand{\bea}{\begin{eqnarray}}
\newcommand{\eea}{\end{eqnarray}}

\newcommand{\prn}[1]{\left ( #1 \right )}
\newcommand{\brk}[1]{\left [ #1 \right ]}
\newcommand{\bigbr}[1]{\Bigl\{ #1 \Bigr\} }

\newcommand{\half}{\frac{1}{2}}
\newcommand{\quarter}{\frac{1}{4}}

\newcommand{\form}[1]{\bm{#1}}

\newcommand{\ic}{\form{i}}
\newcommand{\hodge}{{}^\star}
\newcommand{\lieD}{\pounds}

\newcommand{\fu}{\form{u}}
\newcommand{\fa}{\form{a}}
\newcommand{\fomega}{\form{\omega}}

\newcommand{\fA}{\form{A}}
\newcommand{\Ah}{\hat{A}}
\newcommand{\fAh}{\form{\hat{A}}}

\newcommand{\fF}{\form{F}}

\newcommand{\fFh}{\form{\hat{F}}}

\newcommand{\fB}{\form{B}}

\newcommand{\fBh}{\form{\hat{B}}}

\newcommand{\El}{E}
\newcommand{\fE}{\form{\El}}

\newcommand{\fEh}{\form{\hat{\El}}}

\newcommand{\fGamma}{\form{\Gamma}}
\newcommand{\Gammah}{\hat{\Gamma}}
\newcommand{\fGammah}{\form{\hat{\Gamma}}}

\newcommand{\fR}{\form{R}}

\newcommand{\fRh}{\form{\hat{R}}}

\newcommand{\fBR}{\form{B}_R}

\newcommand{\fER}{\form{\El}_R}

\newcommand{\Kbeta}{{\bm{\beta}}}
\newcommand{\LambdaB}{\Lambda_{\bm{\beta}}}
\newcommand{\Bfields}{{\mathscr B}}
\newcommand{\hcur}{{\mathscr C}_{\cal H}}
\newcommand{\hfields}{{\bm \Psi}}

\def\vs{v_s^2}
\def\vssq{v_s^4}
\newcommand{\acc}{{\mathfrak a}}
\newcommand{\cv}{{\mathfrak v}}
\newcommand{\mue}{{\tilde{\mu}}}

\newcommand{\KEq}{K}
\newcommand{\LambdaEq}{\Lambda_K}
\newcommand{\Eqfields}{{\mathscr K}}
\newcommand{\diffEq}{{\delta_{_\Eqfields}}}

\newcommand{\Sp}{\Sigma}

\newcommand{\PSymplPot}[1]{\slashed{\delta}_{#1}{\varTheta}_{_{\text{PS}}}}

\newcommand{\Lag}{{\mathcal L}}

\newcommand{\Komar}{\mathcal{K}}

\newcommand{\N}{\mathrm{N}}

\newcommand{\diffCons}{{\bbdelta}}
\newcommand{\Mref}{\mathbbm M}
\newcommand{\Kref}{\mathbb \bbbeta}
\newcommand{\Lref}{\Lambda_\Kref}
\newcommand{\Breffields}{\mathbbm B}
\newcommand{\gref}{\mathbb g}
\newcommand{\Aref}{\mathbbm A}
\newcommand{\Tref}{\mathbbm T}
\newcommand{\Jref}{\mathbbm J}
\newcommand{\aheatref}{\mathbbm h}
\newcommand{\achargeref}{\mathbbm n}
\newcommand{\Fref}{\mathbbm F}
\newcommand{\Chref}{\mathbb\Gamma}

\newcommand{\Dref}{\mathbbm D}
\newcommand{\akkref}{\mathbb a}
\newcommand{\pkkref}{\mathbb p}
\newcommand{\Akkref}{\mathbbm a}
\newcommand{\ukkref}{\mathbb u}
\newcommand{\uref}{\mathbbm u}
\newcommand{\Lagref}{\mathbbm L}
\newcommand{\hreffields}{\mathbb  \Psi}
\newcommand{\xref}{\mathbbm x}
\newcommand{\dRLgref}{\mathbb {\bar g}}
\newcommand{\dRLAref}{\mathbb {\bar A}}

\newcommand{\dRLhfref}{{\bar \hreffields}}
\newcommand{\aRLhfref}{{\breve \hreffields}}

\newcommand{\Sigmaref}{\mathbb \Sigma}
\newcommand{\SPref}{\Sigmaref_\fP}
\newcommand{\qPref}{\mathbbm{q}_\fP}
\newcommand{\JPref}{\Jref_\fP}
\newcommand{\SpHref}{\Sigmaref_{H}}
\newcommand{\JHref}{\Jref_{H}}
\newcommand{\JHhref}{\hat{\Jref}_{H}}
\newcommand{\SpHhref}{\hat{\Sigmaref}_{H}}
\newcommand{\diffBref}{\delta_{_{\Breffields}}}

\newcommand{\JTref}{\mathbbm J_\smallT}
\newcommand{\achargeTref}{{\mathbbm n}_\smallT}
\newcommand{\muref}{{\bbmu}}
\newcommand{\Omegaref}{{\mathbb \Omega}}
\newcommand{\sref}{{\mathbbm s}}

\newcommand{\xiref}{\bbxi}
\newcommand{\Lambdaref}{{\mathbb \Lambda}}

\newcommand{\fP}{{\form{\mathcal{P}}}}
\newcommand{\fPh}{{\widehat{\fP}}}
\newcommand{\ICS}{{\form{I}}_{CS}}

\newcommand{\VP}{{\form{V}}_{\fP}}
\newcommand{\WCS}{{\form{W}}_{CS}}

\newcommand{\JP}{J_\fP}
\newcommand{\fJP}{\form{J}_\fP}
\newcommand{\SP}{\Sp_\fP}
\newcommand{\fSP}{\form{\Sp}_\fP}
\newcommand{\qP}{q_{_\fP}}
\newcommand{\fqP}{\form{q}_{_\fP}}

\newcommand{\JH}{\mathrm{J}_{H}}
\newcommand{\JHh}{\hat{\mathrm{J}}_{H}}
\newcommand{\fJH}{\form{\mathrm{J}}_{H}}
\newcommand{\fJHh}{\form{\hat{\mathrm{J}}}_{H}}

\newcommand{\SpH}{\mathrm{\Sp}_{H}}
\newcommand{\SpHh}{\hat{\mathrm{\Sp}}_{H}}
\newcommand{\fSpH}{\form{\mathrm{\Sp}}_{H}}
\newcommand{\fSpHh}{\form{\hat{\mathrm{\Sp}}}_{H}}

\newcommand{\THall}{\mathrm{T}_{H}}

\newcommand{\JBZ}{J_{BZ}}

\newcommand{\SpBZ}{\Sp_{BZ}}

\newcommand{\TBZ}{T_{BZ}}

\newcommand{\TPV}{(T^{\alpha\beta})_{\PV}}
\newcommand{\JPV}{(J^\alpha)_{\PV}}
\newcommand{\fJPV}{\form{J}_{_{\PV}}}
\newcommand{\JpSPV}{(J'_{S})_{\PV}}
\newcommand{\fJpSPV}{\form{J}'_{_{S,\PV}}}
\newcommand{\GpPV}{(\mathcal{G}'^{\,\sigma})_{\PV}}
\newcommand{\fGpPV}{\form{\mathcal{G}}'_{_{\PV}}}
\newcommand{\SPV}{\Sp_{_{\PV}}}
\newcommand{\fSPV}{\form{\Sp}_{_{\PV}}}
\newcommand{\qPV}{q_{_{_{\PV}}}}
\newcommand{\fqPV}{\form{q}_{_{_{\PV}}}}

\newcommand{\fFT}{{\sf \fF^{ \,\!{\scriptscriptstyle{(T)}}}}\!}
\newcommand{\fBT}{{\sf \fB^{ \,\!{\scriptscriptstyle{(T)}}}}\!}

\newcommand{\PVj}{\tiny{(\PV,j)}}
\newcommand{\VPPV}{{\form{V}}_{\fP_{\PV}}}

\newcommand{\skR}{\text{\tiny R}}
\newcommand{\skL}{\text{\tiny L}}
\newcommand{\skRl}{\text{R}}
\newcommand{\skLl}{\text{L}}
\newcommand{\hodgeB}{{}^{\star_{2n+1}}}
\newcommand{\gratio}{\mathfrak{s}}

\newcommand{\bulkM}{{\mathscr M}}
\newcommand{\bulkint}{\fint_{\bulkM}}
\newcommand{\bulkintref}{\fint_{\Mref_{d+1}}}
\newcommand{\fatQ}{{\mathfrak{Q}}}
\newcommand{\fatP}{{\mathfrak{P}}}
\newcommand{\Jbulk}{J_{_{(d+1)}}}
\newcommand{\Tbulk}{T_{_{(d+1)}}}
\newcommand{\aheatbulk}{\aheat^{_{(d+1)}}}
\newcommand{\achargebulk}{\acharge_{_{(d+1)}}}

\newcommand{\Jbulkref}{\Jref_{_{(d+1)}}}
\newcommand{\Tbulkref}{\Tref_{_{(d+1)}}}

\newcommand{\mb}{{\scriptscriptstyle{M}}}
\newcommand{\pb}{{\scriptscriptstyle{P}}}
\newcommand{\nb}{{\scriptscriptstyle{N}}}
\newcommand{\sbb}{{\scriptscriptstyle{S}}}
\newcommand{\rb}{{\scriptscriptstyle{T}}}
\newcommand{\qb}{{\scriptscriptstyle{Q}}}

\newcommand{\smallTbrk}{{\sf \!{\scriptscriptstyle{(T)}}}}
\newcommand{\smallT}{{\sf \!{\scriptscriptstyle{T}}}}
\newcommand{\smallL}{{\scriptscriptstyle{\text{L}}}}

\newcommand{\UT}{U(1)_{\scriptstyle{\sf T}}}
\newcommand{\AT}{{\sf A^{ \!{\scriptscriptstyle{(T)}}}}\!}
\newcommand{\LagT}{\Lag_{ \!{\scriptscriptstyle{\,{\sf T}}}}}
\newcommand{\muT}{{\bm \mu}^{\,\smallTbrk}}
\newcommand{\lT}{{\bm \lambda}^\smallTbrk}

\newcommand{\hfieldsT}{\hfields_{\scriptstyle{\sf T}}}
\newcommand{\hreffieldsT}{\mathbb  \Psi_{\scriptstyle{\sf T}}}
\newcommand{\txi}{\bar \xi}
\newcommand{\tLam}{\bar \Lambda}
\newcommand{\tgdiff}{g'}
\newcommand{\tAdiff}{A'}
\newcommand{\tFdiff}{F'}
\newcommand{\tDdiff}{D'}
\newcommand{\haux}{\Psi_{\sf T}^{\varnothing}}

\newcommand{\bgdiff}{g^\skL}
\newcommand{\bAdiff}{A^\skL}
\newcommand{\bFdiff}{F^\skL}
\newcommand{\bDdiff}{D^\skL}

\newcommand{\dRLg}{(g^\skR-\bgdiff)}
\newcommand{\dRLA}{(A^\skR-\bAdiff)}
\newcommand{\LambdaT}{{\Lambda^{\sf \!{\scriptscriptstyle{(T)}}}}\!\,}
\newcommand{\LambdaTb}{{\bar{\Lambda}^{\sf \!{\scriptscriptstyle{(T)}}}}\!\,}
\newcommand{\LambdaBT}{{\LambdaB^{\sf \!{\scriptscriptstyle{(T)}}}}\!\,}
\newcommand{\diffLB}{\delta_{\LambdaTb  \Bfields}}
\newcommand{\FT}{{\sf F^{ \,\!{\scriptscriptstyle{(T)}}}}\!}
\newcommand{\TL}{T_\smallL}
\newcommand{\JL}{J_\smallL}
\newcommand{\TLc}{T_{\smallL^c}}
\newcommand{\JLc}{J_{\smallL^c}}
\newcommand{\TLLc}{T_{\smallL + \smallL^c}}
\newcommand{\JLLc}{J_{\smallL + \smallL^c}}
\newcommand{\achargeT}{\acharge_{\,\smallT}}
\newcommand{\JT}{J_{\,\smallT}}

\newcommand{\NT}{\N_{\,\smallT}}
\newcommand{\JST}{(J_S)_{\,\smallT}}
\newcommand{\ArefT}{{\mathbbm A}^\smallTbrk}
\newcommand{\LrefT}{\Lambda_\Kref^\smallTbrk}
\newcommand{\tgdiffref}{\gref'}
\newcommand{\tAdiffref}{\Aref'}
\newcommand{\TrefLc}{\Tref_{\smallL^c}}
\newcommand{\JrefLc}{\Jref_{\smallL^c}}
\newcommand{\TrefLLc}{\Tref_{\smallL + \smallL^c}}
\newcommand{\JrefLLc}{\Jref_{\smallL + \smallL^c}}

\newcommand{\tildeg}{\tilde{g}}
\newcommand{\tildeA}{\tilde{A}}
\newcommand{\GT}{{\cal G}_{\,\smallT}}

\newcommand{\PS}{{\rm H}_S}
\newcommand{\PV}{{\rm H}_V}
\newcommand{\PF}{{\rm H}_F}
\newcommand{\LS}{{\overline{\rm H}}_S}
\newcommand{\GV}{{\overline{\rm H}}_V}
\newcommand{\Hs}{{\rm H}}
\newcommand{\Ds}{{\rm D}_s}
\newcommand{\Dv}{{\rm D}_v}
\newcommand{\Diss}{\Delta}
\newcommand{\LT}{{\rm L}_{\,\smallT}}

\newcommand{\diffB}{\delta_{_{\Bfields}}}
\newcommand{\Xfields}{{\mathscr X}}
\newcommand{\diffF}{{\delta_{_\Xfields}}}
\newcommand{\aheat}{{\mathfrak h}}
\newcommand{\acharge}{{\mathfrak n}}

\newcommand{\BerryG}{{\cal N}}
\newcommand{\BerryGA}{{\cal X}}
\newcommand{\BerryA}{{\cal S}}

\newcommand{\DVisc}{\Upsilon}
\newcommand{\cdg}{{\bm \eta}}
\newcommand{\cdgA}{{\bm \kappa}}
\newcommand{\cdA}{{\bm \sigma}}

\newcommand{\wzD}{\mathscr D}
\newcommand{\Dp}{{}^{(\mathfrak{p})}\wzD}
\newcommand{\fGammap}{{}^{(\mathfrak{p})}\fGamma}
\newcommand{\fRp}{{}^{(\mathfrak{p})}\fR}
\newcommand{\JWZ}{{\sf J}_{_{{\text{Euler}}}}}
\newcommand{\JChern}{{\sf J}_{_{{\text{Chern}}}}}

\newcommand{\Wey}{{\scriptscriptstyle\mathcal{W}}}
\newcommand{\LambdaW}{\Lambda_{\Wey}}

\newcommand{\diffFW}{{\delta^\Wey_{_\Xfields}}}

\newcommand{\AWeyl}{\mathcal{W}}
\newcommand{\FWeyl}{{}^\Wey F}
\newcommand{\DWeyl}{\mathscr{D}^{\Wey}}
\newcommand{\RWeyl}{{}^\Wey R}
\newcommand{\GamWeyl}{{}^\Wey \Gamma}
\newcommand{\SWeyl}{{}^\Wey S}

\newcommand{\cnd}{{\mathsf c}}
\newcommand{\Snd}{{\bf S}}
\newcommand{\Lnd}{\Lambda_{\bf S}}
\def\epform#1{\, {}^\varepsilon#1}
\newcommand{\Sndref}{\mathbbm S} 
\newcommand{\Lndref}{\Lambda_\Sndref}
\def\epformref#1{\, {}^{\mathbb e}#1}
\def\pndref{\bar \pkkref}

\title{Adiabatic hydrodynamics:\\ The eightfold way to dissipation}

\author[a]{Felix M. Haehl}
\author[b]{\!, R.\ Loganayagam}
\author[a]{\!, Mukund Rangamani}
\affiliation[\,a]{Centre for Particle Theory \& Department of Mathematical Sciences,\\
Durham University, South Road, Durham DH1 3LE, UK.}
\affiliation[\,b]{Institute for Advanced Study, Einstein Drive, Princeton, NJ 08540, USA.}
\emailAdd{f.m.haehl@gmail.com}
\emailAdd{nayagam@gmail.com}
\emailAdd{mukund.rangamani@durham.ac.uk}

\abstract{
Hydrodynamics is the low-energy effective field theory of any interacting quantum theory, capturing the long-wavelength fluctuations of an equilibrium Gibbs density matrix. Conventionally, one views the effective dynamics in terms of the conserved currents, which should be expressed via the constitutive relations in terms of the fluid velocity and the intensive parameters such as the temperature, chemical potential, etc.. However, not all constitutive relations are acceptable; one has to ensure that the second law of thermodynamics is satisfied on all physical configurations. In this paper,  we provide a complete solution to hydrodynamic transport at all orders in the gradient expansion compatible with the second law constraint.  

The key new ingredient we introduce is the notion of adiabaticity, which allows us to take hydrodynamics off-shell. Adiabatic fluids are such that off-shell dynamics of the fluid compensates for entropy production. The space of adiabatic fluids is quite rich, and admits a decomposition into seven distinct classes. Together with the dissipative class this establishes the eightfold way of hydrodynamic transport. Furthermore, recent results guarantee that dissipative terms beyond leading order in the gradient expansion are agnostic of the second law. While this completes a transport taxonomy, we go on to argue for a new symmetry principle, an Abelian gauge invariance that guarantees adiabaticity in hydrodynamics. We suggest that this symmetry is the macroscopic manifestation of the microscopic KMS invariance. We demonstrate its utility by explicitly constructing effective actions for adiabatic transport. The theory of adiabatic fluids, we speculate, provides a useful starting point for a new framework to describe non-equilibrium dynamics, wherein dissipative effects arise by Higgsing the Abelian symmetry.
}

\preprint{DCPT-15/01}

\begin{document}

\maketitle


\clearpage

\part{An Invitation to Adiabatic Hydrodynamics}
\label{part:intro}
\hspace{1cm}
\section{Introduction}
\label{sec:intro}

Hydrodynamics, as is well known, is the universal long-wavelength effective description of near-equilibrium dynamics of interacting quantum systems. Given its wide range of applicability and the simplicity of its dynamical content, it behooves us to understand the derivation of classical hydrodynamic equations from first principles. While many attempts have been made to distill the essentials of the theory and derive the low energy dynamics following rules of effective field theory, 
it is perhaps fair to say that to date a completely autonomous theory of hydrodynamics remains in absentia.

The traditional approach to hydrodynamics involves identifying the conserved currents such as energy-momentum and charge currents. Firstly, one invokes an appropriate Gibbsian ensemble to describe equilibrium thermodynamics. The Gibbs free 
energy, as a function of temperature $T$ and chemical potentials $\mu_i$, determines the equilibrium data: pressures, internal energies, charge densities, etc., which constitute the components of the currents in the inertial frame chosen by the equilibrium configuration denoted by a unit timelike vector $u^\mu$. This explicitly constructs the ideal fluid currents which satisfy apposite conservation equations.  One then allows arbitrary long-wavelength (infra-red) fluctuations of the intensive variables $(T,\mu_i)$ and the local thermal frame $u^\mu$. The fluctuations of the Gibbs density matrix in a current algebra language translate into higher derivative operators correcting  the ideal fluid constitutive relations. 
As in usual effective field theory one allows these operators (respecting requisite symmetries) with arbitrary coefficients. In the case at hand we should admit local functions of the intensive thermodynamic parameters
$(T,\mu_i)$; these are the transport coefficients of hydrodynamics.  Note that the dynamics is still enforced by the conservation of the currents.

Thus far the construction of the low energy hydrodynamic theory seems analogous to any other effective field theory albeit in a current algebra language. The main novelty of hydrodynamics is that it has a constraint: one expects that the hydrodynamic evolution locally respects the second law of thermodynamics. More abstractly, in addition to the conserved currents which capture the dynamical information, one posits the existence of an entropy current, which is constrained to have non-negative divergence. This approach to hydrodynamics,
which we dub as the current algebra approach, is the canonical method to determining both the constitutive relations and the constraints on the transport coefficients. This viewpoint has been well appreciated for many decades now, cf., \cite{landau} for a clear discussion.

Since the second law of thermodynamics is stated as an inequality one usually finds that it imposes sign-definiteness constraints on transport data. For instance, one learns  at the first non-trivial order in the gradient expansion that viscosities and conductivities are positive definite to ensure entropy production. What is perhaps less familiar is the recent discovery that one also encounters explicit constraints fixing some transport coefficients in terms of others \cite{Bhattacharyya:2012nq}. An example of this is the Gibbs-Duhem and Euler relations obeyed at zeroth order in derivative expansion.
 Said differently, by a careful analysis one can show that the allowed class of operators respecting the second law is smaller than one might a-priori have imagined. In all known examples studied so far, the transport coefficients thus constrained  can be obtained from analyzing general hydrostatic equilibria -- they are determined in terms of the so called {\em hydrostatic}  or {\em thermodynamic response} parameters. In particular, these constraints can be understood in terms of subjecting the fluid to arbitrary stationary sources (background metric, gauge fields) and obtaining the desired relations by writing down the generating function for the current correlators, or equivalently the equilibrium partition function \cite{Banerjee:2012iz,Jensen:2012jh}.

The class of hydrostatic transport coefficients is quite rich. Not only does it comprise of novel constraints on higher order hydrodynamic data, but it also importantly  includes the class of anomalous transport coefficients which provide an interesting insight into the underlying quantum dynamics in thermodynamic systems. While the presence of anomalous transport coefficients was first clearly encountered in fluid/gravity computations of \cite{Erdmenger:2008rm,Banerjee:2008th}, they were soon understood as being necessary from the canonical current algebra framework of fluid dynamics in \cite{Son:2009tf}. More recently, starting from the work of \cite{Loganayagam:2011mu}, it has been appreciated  that the anomalous contribution to transport belongs to the hydrostatic class. We now have clear picture of how to derive the constraints on them using the equilibrium partition function \cite{Banerjee:2012iz,Jain:2012rh,Banerjee:2012cr,Jensen:2012kj,Jensen:2013kka,Jensen:2013rga} (for related work on anomaly induced transport we refer the reader to \cite{Torabian:2009qk,Kharzeev:2009pj,Jensen:2010em,Amado:2011zx,Landsteiner:2011cp,Kharzeev:2011ds,Dubovsky:2011sk,Loganayagam:2012pz,Neiman:2010zi,Jensen:2012jy,Nair:2011mk,Valle:2012em,Landsteiner:2012kd,Banerjee:2013qha,Megias:2013joa,Banerjee:2013fqa,Haehl:2013hoa,Banerjee:2014cya,Jimenez-Alba:2014pea,Yee:2014dxa,Monteiro:2014wsa,Banerjee:2014ita}). We will have occasion to describe these results in due course, but for now we simply record the fact that the equilibrium partition function provides a powerful way to study the constraints on hydrodynamic transport.

Recently, building on the hydrostatic analysis, Sayantani Bhattacharyya \cite{Bhattacharyya:2013lha,Bhattacharyya:2014bha} derived a remarkable theorem about hydrodynamic transport. She proved that:
\begin{itemize}
\item  All the dangerous terms in hydrodynamics which could potentially lead to entropy destruction, are constrained to vanish from the hydrostatic analysis and there are no further equality constraints beyond hydrostatics.
\item Of the entropy producing dissipative terms, only the ones at leading order in the gradient expansion are constrained to be sign-definite. 
\end{itemize}
Essentially the upshot of this analysis is the following: once one understands the leading deviations from a perfect fluid and is able to analyze hydrostatic configurations, one has completed the task of constructing the hydrodynamic effective field theory (at least as a current algebra). 

While this result captures the essence of the second law constraints, it still leaves unanswered questions about the structure of hydrodynamics. One may think of the situation in the following vein: a-priori in the current algebra approach, constructing tensor valued operators which correct the ideal fluid conserved currents is a question in representation theory. Given the intensive parameters $\{T,\mu_i\}$ and the hydrodynamic velocity field $u^\mu$ (which is a unit timelike vector), we simply have to build tensors with suitable symmetries to appear in the energy-momentum tensor $T^{\mu\nu}$ and charge currents $J^\mu_i$. The number of such tensors can be inferred from a straightforward counting exercise at any desired order. Complications start to arise when we impose the constraints of the second law since this poses a non-linear constraint, potentially mixing terms across derivative orders.  Indeed the proof of the statements quoted above in  \cite{Bhattacharyya:2013lha} relies on a careful unpacking of such mixing (see \cite{Bhattacharyya:2012nq,Bhattacharyya:2014bha}). It would be ideal if we could understand the second law constraints in a more straightforward fashion. Ideally, one would like to have a complete classification of hydrodynamic transport, both dissipative and non-dissipative, which respects the democratic ordering of the gradient expansion.\footnote{ In many circumstances one also demands that the Onsager relations \cite{Onsager:1931fk,Onsager:1931uq} are upheld, by invoking the microscopic time-reversal symmetry. We will for the most part be agnostic about these relations, and relegate comments regarding them to \S\ref{sec:discuss}.} 

This compels us to further our understanding of hydrodynamics, a task we will undertake in this work. Our primary result will indeed be a complete classification of hydrodynamic transport at all orders in the gradient expansion.  In the process we will also be able to identify the origins of various curious results that have been uncovered in recent analysis of hydrodynamic transport, both from studies in kinetic theory and in holography. The analysis of transport in holographic fluids, which began with the pioneering work of \cite{Policastro:2001yc} and was extended by the fluid/gravity correspondence to the non-linear level \cite{Bhattacharyya:2008jc}, provides a rather fertile laboratory for testing non-linear hydrodynamics.  

To explain more precisely the rationale behind our analysis, we start with the following observation. In order to ascertain the structure of hydrodynamics, it would be ideal if we could formulate the effective field theory, not in terms of the currents as described above, but rather directly in terms of a Wilsonian effective action. The question then becomes: what are the natural variables for such an action and how does one incorporate the fact that the theory is intrinsically dissipative?

As a toy problem, one can focus on the structure of hydrodynamic effective actions in the absence of dissipation. At  the very least this attempt can help us  learn about the constraints resulting from demanding an off-shell (by definition), off-equilibrium, effective action for hydrodynamics. Happily, such a formalism exists. It was invented in the distant past to formulate the dynamics of ideal fluids coupled to gravitational degrees of freedom \cite{Taub:1954zz,Carter:1973fk}. In recent years this effective action formalism has been revived starting from the work of \cite{Dubovsky:2005xd,Dubovsky:2011sj}. These works formulate the story in terms of the Goldstone degrees of freedom associated with individual fluid elements. A systematic exploration for neutral fluids was undertaken in \cite{Bhattacharya:2012zx} wherein a detailed comparison with the conventional current algebra approach to hydrodynamics was made.\footnote{ This approach has also been used to study parity-odd transport in $3$ dimensions \cite{Nicolis:2011ey,Haehl:2013kra,Geracie:2014iva}.} In particular, it was noticed that demanding the presence of an effective action appears to pose stronger constraints than what would be encountered by the existence of an entropy current with desired properties.\footnote{ For non-dissipative fluids one imposes a strong constraint on the entropy current: we demand it to be divergence free off-shell. So the theory actually has an additional conserved current due to entropy non-production. } Said differently, there were relations amongst transport coefficients (the functions of intensive parameters multiplying higher order tensor structures in the currents) which remained inexplicable. Curiously, some of these relations are also manifested in the class of holographic fluids, which prompted us to examine the situation further.

A crucial check of this effective action formalism was provided in \cite{Haehl:2013hoa} where it was shown that one can recover the anomalous transport data for non-abelian flavour anomalies in arbitrary even dimensions (see \cite{Dubovsky:2011sk} for abelian anomalies in $2$ dimensions). This analysis highlighted two important facts about the effective action approach: (i) the symmetries of the theory (ii) the necessity of  doubling the degrees of freedom, a la Schwinger-Keldysh along with non-trivial cross terms (a.k.a. influence functionals of Feynman-Vernon \cite{Feynman:1963fq}) in the effective action. We will postpone a discussion of the doubling to later, but the symmetries are worth examining at present.

The effective action  for $d$-dimensional hydrodynamics  is constructed in terms of $d-1$ scalar fields $\phi^{_I}$ which capture energy-momentum transport along with a set of fields $\cnd$ which transform in a bifundamental representation of the flavour symmetry. One allows arbitrary {\em volume preserving field reparametrizations} of the fields, $\phi^{_I} \mapsto f^{_I}(\phi)$ and $\cnd\mapsto \,\cnd\, g(\phi) $. This implies that the effective action is invariant under a generalized volume-preserving diffeomorphism group; this  guarantees entropy non-production. More specifically the conservation of configuration space volume $d\phi^1\wedge \cdots \wedge d\phi^{d-1}$ is interpreted as the statement of entropy current conservation.   That is one identifies $\star\form{J}_S = \star(s\, \fu)= d\phi^1\wedge \cdots \wedge d\phi^{d-1}$ as the 1-form entropy current, with $s$ being the entropy density and ${\bf u}$ the fluid velocity. Note that the formalism forces the entropy current to take its ideal fluid form at all orders in the hydrodynamic gradient expansion. The resulting constitutive relation may then said to be presented in the {\em entropy frame}.\footnote{ Frame choices in hydrodynamic current algebra are a reflection of field redefinitions. For example, there being no a-priori canonical choice for the velocity field in a relativistic fluid one can choose to define it conveniently. The above choice is just as natural as the often made choice of Landau frame, wherein the velocity field is taken to be the unit-timelike eigenvector of the energy-momentum current.}

While one can tie in the entropy conservation with the presence of the enlarged symmetry, one would as such like to understand the rationale behind its existence and check the consistency of employing it to define a conserved entropy current. Moreover, if we were to extend the effective action approach to physically relevant dissipative fluids, we need to understand how to allow for entropy production. In fact, empirically the  failure point of the formalism appears to be at an even simpler situation. An attempt to extend the considerations of anomalous transport  to mixed flavour-Lorentz anomalies suggests that modifications to the entropy current from the form $\form{J}_S = s\, \fu$ are imperative.\footnote{ We do not have a complete proof of this statement, but the ease with which we are able to recover all the results in the formalism described below suggests to us that this is the correct intuition.} This being impossible in the framework described above one is led to look elsewhere.

Let us therefore step back and ask the following: what is the canonical choice of variables for a hydrodynamic effective action?
A natural set of hydrodynamic variables in terms of which an effective action ought to be written, one would guess, are simply the intensive thermodynamic variables that characterize the Gibbs ensemble $T,\mu_i$ etc., and the fluid velocity $u^\mu$. In the non-dissipative effective action temperature is viewed as a function of its thermodynamic conjugate entropy density, while the velocity and chemical potential (and entropy itself) are indirectly defined in terms
of the fields $\phi^{_I}$ and $\cnd$.  As such the on-shell action computes not quite the thermodynamic Gibbs potential, but rather, its Legendre transform with respect to the entropy density. If one were interested in allowing deformation of the entropy current,  working with Gibbs potential is more natural.\footnote{ Hence  the common use of Landau-Ginzburg free-energy funtionals to describe condensed matter systems.} However, in the hydrodynamic gradient expansion computing the Legendre transformation is non-trivial.\footnote{ This has the insalubrious effect of making comparisons between the effective actions and the equilibrium partition functions rather complex and involved. In special cases such as the anomalous effective action one can carry out the Legendre transformation trivially owing to independence of such terms from the entropy density, 
cf., \cite{Haehl:2013hoa}.}  Inspired by the structural aspects of the formal Legendre transformation we establish now a new formalism that naturally incorporates the hydrodynamic variables as the basic fields and provides a framework to describe what we call {\it adiabatic fluids}, which are a generalization of the non-dissipative fluids discussed hitherto.

To motivate the study of adiabatic fluids, let us ask the following question: ``what is the most convenient way to implement the second law of thermodynamics, which a-priori is stated as an inequality, in practice?'' As we discussed before the conventional current-algebraic approach is to work on-shell by classifying independent tensors, but this is limiting from the point of view of constructing an action principle. A useful trick for implementing inequality constraints is in fact to go off-shell using a suitable set of Lagrange multipliers, which sometimes is referred to as the Liu procedure \cite{Liu:1972nr}. The basic idea can be explained as follows: suppose we want to constrain the solutions of a set of linear equations with an inequality constraint. We add to the inequality of interest a suitable linear combination of the dynamical equations with Lagrange multipliers. While the new quantity  defined thus, also satisfies the same inequality, it has the distinct advantage that we are no longer on-shell. Said differently, incorporating the  dynamical equations of motion into the inequalities, we can
uplift the constraints off-shell and analyze them without having to solve for the dynamically independent set of data.

Specifically, we take the on-shell statement of non-negative entropy production $\nabla_\mu J_S^\mu \geq0$
and upgrade it to an off-shell statement which reads:
\begin{equation}\label{eq:aintro}
\begin{split}
\nabla_\mu J_S^\mu &+ \Kbeta_\mu\prn{\nabla_\nu T^{\mu\nu}-J_\nu \cdot F^{\mu\nu}-\THall^{\mu\perp}}\\
&+ (\LambdaB+\Kbeta^\lambda A_\lambda) \cdot \prn{D_\nu J^\nu-\JH^\perp}  \equiv \Delta \geq 0 \,.
\end{split}
\end{equation}
Here $\Kbeta_\mu$ is the Lagrange multiplier for the energy-momentum conservation equation involving the energy-momentum tensor $T^{\mu\nu}$ and the charge current $J^\mu$, and $(\LambdaB+\Kbeta^\lambda A_\lambda)$ is the Lagrange multiplier for the charge conservation equation. We have denoted the background metric and flavour gauge fields by $\{g_{\mu\nu},A_\mu\}$ respectively and $F_{\mu\nu}$ denotes the field-strength associated with $A_\mu$. Note that we have written the conservation equations for a general situation including contributions from sources and anomalies (which are captured by the Hall currents $J_H^\perp$ and $T^{\mu\perp}_H$).\footnote{ If the underlying quantum system of interest is anomaly-free then we can set the Hall currents to zero; we refer to the corresponding version of \eqref{eq:aintro} (with $\Diss =0$) as the {\em non-anomalous adiabaticity equation} for definiteness.}  The notation will become clear when we set-up the problem in greater detail in due course. 

It is convenient to take these Lagrange multiplier fields to be the basic hydrodynamic fields. At zeroth order in derivative expansion, thermodynamics demands that $\Kbeta^\mu = u^\mu/T$ be the velocity field rescaled by the temperature and $\LambdaB+\Kbeta^\lambda A_\lambda = \mu/T$ be the chemical potential measured in thermal units. We will use a part of the field redefinition freedom to assume that 
these simple relations hold to arbitrary orders in derivative expansion. In fact, these variables naturally encompass all of the hydrodynamic degrees of freedom; by rescaling the normalized velocity field by the temperature we have ensured that $\Kbeta^\mu$ is an unconstrained vector field. We refer to these fields as the {\em thermal vector} and the {\em thermal twist} respectively; they are the physical degrees of freedom in local equilibrium.

The off-shell rewriting of the second law of thermodynamics in \eqref{eq:aintro} turns out to provide sufficient control to classify all hydrodynamic transport. A-priori, we can distinguish between two kinds of transport coefficients: on the one hand those that contribute to off-shell entropy production, i.e., lead to positive definite $\Delta \neq 0$ -- these are the dissipative (Class D) parts of transport. On the other hand we have those which comprise the marginal case of no entropy production, i.e., those where the production of entropy is compensated for by the flow of energy-momentum and charge. The latter form the boundary of the domain of physically admissible constitutive relations and have $\Delta =0$. Understanding this marginal case turns out to be the crucial step that allows us to complete our classification. This therefore motivates for us the study of 
{\em adiabatic hydrodynamics}, defined as the constitutive relations which solve \eqref{eq:aintro} with $\Delta =0$, which we will refer to as the {\em adiabaticity equation}.\footnote{  The adiabaticity equation we study in some detail below was first introduced in \cite{Loganayagam:2011mu} to aid the analysis of anomaly induced hydrodynamic transport using the standard current algebra approach to hydrodynamics.} The class of adiabatic constitutive relations subsumes (but is not identical to) the pre-existing discussions of non-dissipative fluids.

\begin{figure}[ht!]
\begin{center}
\includegraphics[width=4in]{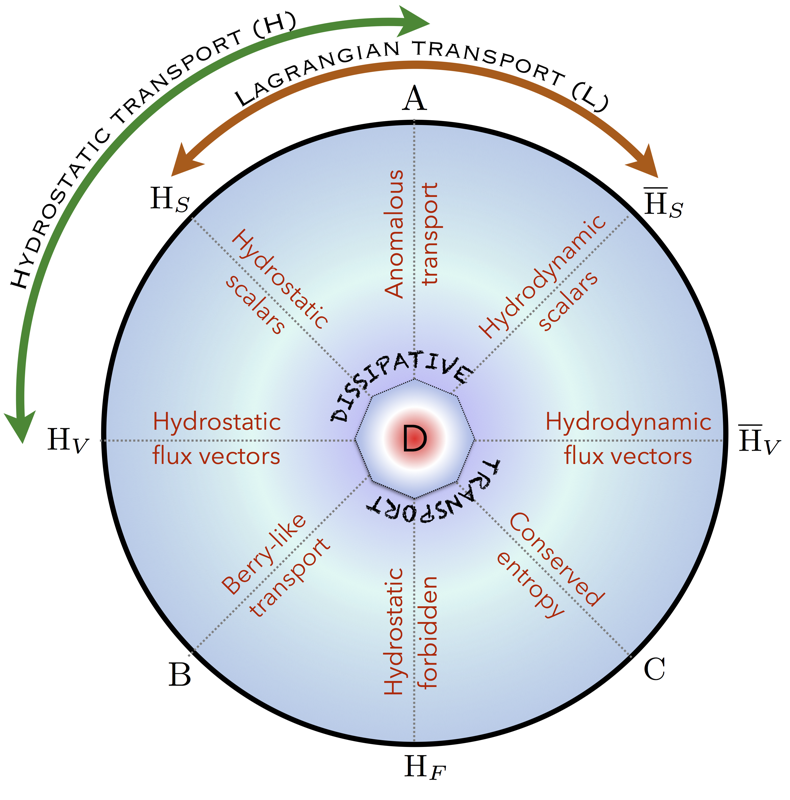}
\caption{The eightfold way of hydrodynamic transport. }
\label{fig:eightfold}
\end{center}
\end{figure}

A large part of our work will be devoted to identifying and classifying all constitutive relations $\{T^{\mu\nu},J^\mu,J^\mu_S\}$ that solve the adiabaticity equation \eqref{eq:aintro} with $\Delta =0$. At a broad brush level there are two types of adiabatic transport:  hydrostatic (Class $\Hs$) and non-hydrostatic or hydrodynamic (Class ${\overline \Hs}$). The former can be inferred from the dynamics of the fluid in hydrostatic equilibrium described earlier by subjecting it to time-independent spatially varying external sources. The latter are more diverse; some can be obtained using a simple Lagrangian formalism involving the hydrodynamic fields $\{\Kbeta^\mu,\LambdaB\}$, but there are others which evade such a simple description. We have found it convenient to isolate the solutions of the adiabaticity equation into seven classes based on their origins. Together with the dissipative Class D,  we are led to the eightfold way of hydrodynamic transport, as illustrated in Fig.~\ref{fig:eightfold}.\footnote{ We give a preview of this classification in some detail in \S\ref{sec:aclassify}.}

We emphasize that we classify (off-shell inequivalent) constitutive relations i.e., give combinations of currents that satisfy the adiabaticity equation.  However, it is convenient for purposes of taxonomy to refer to some more primitive object that generates such constitutive relations. The classification turns out to be canonically motivated by the choice of the grand canonical free energy current (obtained by Legendre transforming the entropy current) which is a spacetime vector. This vector can be either longitudinal (aligned to the thermal vector) or transverse. Since longitudinal vectors are characterized by their magnitude, we can refer to it as the scalar component. Hence all the classes in  Fig.~\ref{fig:eightfold} refer to either scalar or vector structures.  For example, $\PS$ and $\LS$ refer to hydrostatic and hydrodynamic terms that can be encoded in a scalar Lagrangian. Similarly, $\PV$ and $\GV$ are classes of transport that transform as transverse vectors. Whether at the end of the day such terms show up as tensors, vectors or scalars in the constitutive relations, is largely a matter of convention and frame choice.  

We also provide evidence that certain well studied hydrodynamic systems respect the adiabatic classification. In particular, strongly coupled conformal plasmas which can be described holographically via the fluid/gravity correspondence \cite{Bhattacharyya:2008jc,Hubeny:2011hd}, as well as existing results in kinetic theory \cite{York:2008rr} manifestly exhibit the eightfold  path. Furthermore, as noted in our short companion paper \cite{Haehl:2014zda}, second order transport for a neutral holographic fluid, is encapsulated in a simple effective action (built out of the sources and the thermal vector and twist)!  We take this as striking evidence of physical fluid systems being cognizant of the adiabaticity equation and the sevenfold constitutive relations which solve it.

To demonstrate that our classification is exhaustive, we argue that all solutions to the adiabaticity equation can be obtained from a master effective action. This eightfold effective action clearly is a functional of the hydrodynamic fields $ \{\Kbeta^\mu, \LambdaB\}$ and the background sources $\{g_{\mu\nu}, A_\mu\}$. Rather surprisingly, a complete picture emerges only upon including a second set of background sources, $\{ \tildeg_{\mu\nu}, \tildeA_\mu\}$, which morally speaking appear to be a proxy for the 
  Schwinger-Keldysh partners of the basic sources. Furthermore, this doubling of sources comes with an interesting new Abelian gauge  symmetry with an associated gauge field $\AT_\mu$!\footnote{ It is tempting for various reasons to christen this $\UT$ symmetry as {\em KMS-gauge invariance}. We will however refrain from doing so since we won't fully justify the connections with the Schwinger-Keldysh framework in the present work and leave a full exposition for the future \cite{Haehl:2014kq}. } 

 In the thermofield construction one has sources for the left (L) and right (R) degrees of freedom; these are specific linear combinations of the sources $\{g_{\mu\nu}, A_\mu\}$ and $\{ \tildeg_{\mu\nu}, \tildeA_\mu\}$. The necessity to double of the degrees of freedom, whilst curious for adiabatic transport, has already been encountered previously in attempts to construct effective actions for anomalous hydrodynamic transport, which forms a special case, in \cite{Haehl:2013hoa}. What is really intriguing is the  gauge field $\AT_\mu$ and its associated gauge invariance $\UT$, which along with the diffeomorphism and gauge invariance forms the symmetries of the effective action.\footnote{ A clue to the existence of such a structure is provided by the analysis of hydrostatic partition functions satisfying the Euclidean consistency condition in the presence of gravitational anomalies \cite{Jensen:2013rga}.}   The latter act canonically on the fields above, but the $\UT$ gauge symmetry acts non-trivially. All fields carry $\UT$ charges, with the gauge transformation acting as a diffeomorphism or flavour gauge transformation in the direction of $\Kbeta^\mu, \LambdaB$. In addition, $\tildeg_{\mu\nu}$ and $\tildeA$ further undergo transformations depending on the physical fields $\{\Kbeta^\mu, \LambdaB,g_{\mu\nu}, A_\mu\}$. The Bianchi identity associated with $\UT$ gauge invariance immediately leads to the adiabaticity equation. In fact, armed with this enlarged space of fields one can immediately write down a diffeomorphism, flavour gauge and $\UT$ invariant effective action which captures all of the adiabatic transport. 

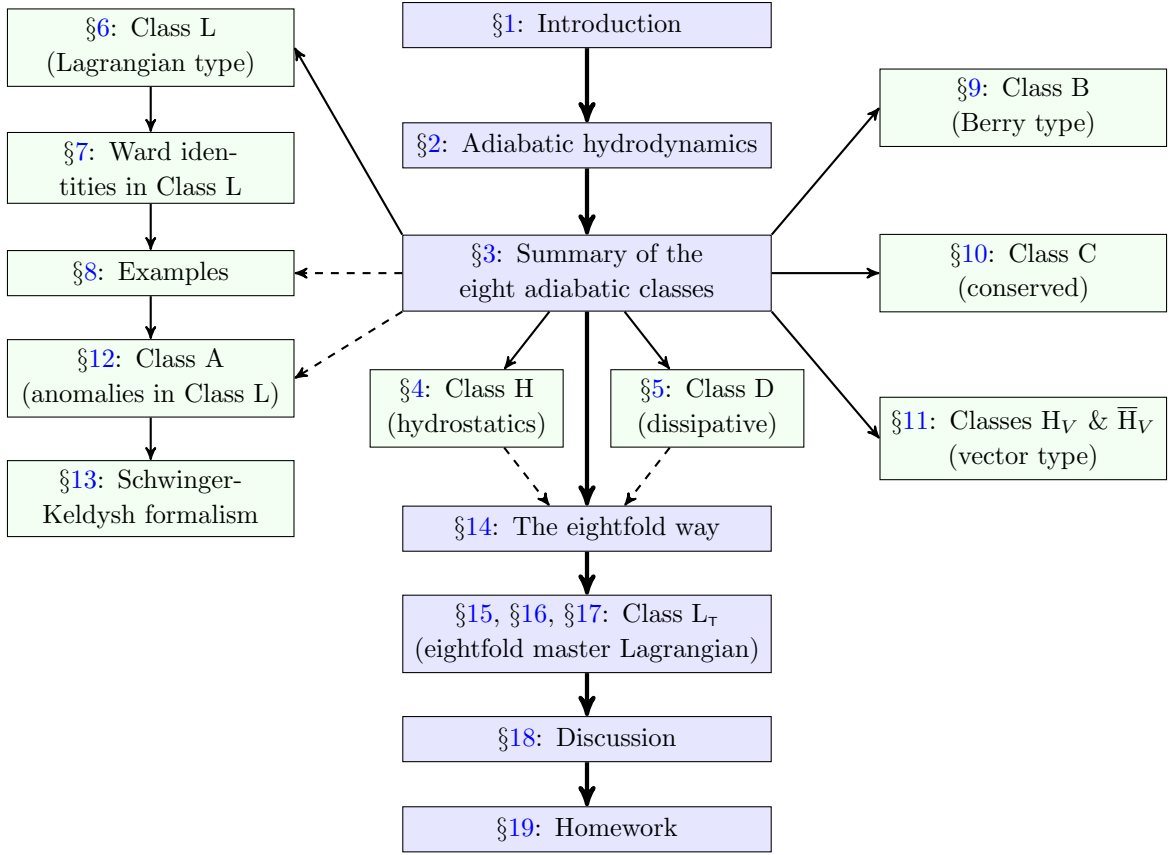
\begin{figure}
\centering
\small{
\begin{tikzpicture}[node distance=1.2cm]
\node (intro) [thread] {\S\ref{sec:intro}: Introduction};
\node (adiabaticity) [thread, below of = intro, yshift = -.4cm] {\S\ref{sec:adiabat}: Adiabatic hydrodynamics};
\node (summary) [thread, below of = adiabaticity, yshift = -.5cm] {\S\ref{sec:aclassify}: Summary of the\\eight adiabatic classes};

\node (classL) [details, left of = intro, yshift = -.3cm, xshift=-4.6cm] {\S\ref{sec:classL}: Class L\\(Lagrangian type)};
\node (Lward) [details, below of = classL, yshift = -.4cm] {\S\ref{sec:Leoms}: Ward identities in Class L};
\node (applications) [details, below of = Lward, yshift = -.2cm] {\S\ref{sec:examples}: Examples};
\node (classA) [details, below of = applications, yshift = -.2cm] {\S\ref{sec:anomalies}: Class A\\(anomalies in Class L)};
\node (SK) [details, below of = classA, yshift = -.4cm] {\S\ref{sec:skdouble}: Schwinger-Keldysh formalism};

\node (classC) [details, right of = summary, xshift = 4.6cm] {\S\ref{sec:consJ}: Class C\\(conserved)};
\node (classB) [details, above of = classC, yshift = 1cm] {\S\ref{sec:classB}: Class B\\(Berry type)};
\node (classV) [details, below of = classC, yshift = -1cm] {\S\ref{sec:Cvector}: Classes $\PV$ \& $\GV$ \\(vector type)};

\node (classH) [threadSmall, below of = summary, xshift = -1.6cm,yshift=-.6cm] {\S\ref{sec:hydrostatics}: Class H\\(hydrostatics)};
\node (classD) [threadSmall, below of = summary, xshift = 1.6cm,yshift=-.6cm] {\S\ref{sec:classD}: Class D\\(dissipative)};
\node (8fold) [thread, below of = summary, yshift = -2.2cm] {\S\ref{sec:8fold}: The eightfold way};
\node (LT) [thread, below of = 8fold, yshift = -.2cm] {\S\ref{sec:classLT}, \S\ref{sec:LTvariational}, \S\ref{sec:eightfoldLT}: Class $\LT$\\(eightfold master Lagrangian)};

\node (discussion) [thread, below of = LT, yshift = -.2cm] {\S\ref{sec:discuss}: Discussion};
\node (HW) [thread, below of = discussion, yshift = 0cm] {\S\ref{sec:homework}: Homework};

\draw [arrowT] (intro) -- (adiabaticity);
\draw [arrowT] (adiabaticity) -- (summary);
\draw [arrow] (summary.north west) -- (classL.east);
\draw [arrow2] (summary.west) -- (applications.east);
\draw [arrow2] (summary.south west) -- (classA.east);
\draw [arrow] (classA) -- (SK);
\draw [arrow] (summary.north east) -- (classB.west);
\draw [arrow] (summary.east) -- (classC.west);
\draw [arrow] (summary.south east) -- (classV.west);
\draw [arrow] (Lward.south) -- (applications.north);
\draw [arrow] (applications) -- (classA);
\draw [arrow] (classL) -- (Lward);
\draw [arrow] ([xshift=-.5cm]summary.south) -- ([xshift=.5cm]classH.north);
\draw [arrow2] ([xshift=.5cm]classH.south) -- ([xshift=-.5cm]8fold.north);
\draw [arrow] ([xshift=.5cm]summary.south) -- ([xshift=-.5cm]classD.north);
\draw [arrow2] ([xshift=-.5cm]classD.south) -- ([xshift=.5cm]8fold.north);
\draw [arrowT] (summary.south) -- (8fold.north);
\draw [arrowT] (8fold) -- (LT);
\draw [arrowT] (LT) -- (discussion);
\draw [arrowT] (discussion) -- (HW);
\end{tikzpicture}
}
\caption{Flowchart giving the structure of this paper. Light blue sections in the middle column form the main thread of our analysis. The sections in light green contain detailed constructions of the various classes and could be skipped on a first read. The sections in the left column are concerned with various Lagrangian descriptions. The classes in the right column are those that do not fit into a simple Lagrangian framework (without $\UT$).}
\label{fig:flowchart}
\end{figure}

We have chosen to structure the paper into six parts owing to the complexities of the results we uncover. To help orient readers through the maze of results we also provide a flowchart in Fig.~\ref{fig:flowchart} to  indicate the inter-relations between various sections. 

\paragraph{Part \ref{part:intro}:} We begin in \S\ref{sec:adiabat} with a definition of the basic statement of the adiabatic fluids and explain some of the general properties of such fluids directly from the study of the adiabaticity equation. In \S\ref{sec:aclassify}  we present a quick overview of the diverse classes of adiabatic constitutive relations; this provides a short synopsis of 
Part \ref{part:adiabatic}. We then turn in \S\ref{sec:hydrostatics} to connecting our construction with the hydrostatic analysis of \cite{Banerjee:2012iz,Jensen:2012jh}. We go on to argue that the adiabaticity equation we introduce can be thought of as an off-equilibrium off-shell extension of the hydrostatic constitutive relations. In \S\ref{sec:classD} we discuss dissipative constitutive relations, reviewing the results of \cite{Bhattacharyya:2013lha,Bhattacharyya:2014bha} in a language adapted to our analysis. This allows us to give an alternate proof of the theorem, classifying dissipative transport coefficients into those constrained by the second law, and those that are agnostic to entropy production.  

 \paragraph{Part \ref{part:adiabatic}:} This is the central part of the paper where we study adiabatic constitutive relations. For the most part we focus on the non-anomalous adiabaticity equation as inclusion of anomalies ends up providing a specific particular solution.  We first show in \S\ref{sec:classL}  how to construct Lagrangian solutions to the adiabaticity equation in the absence of anomalies. Any diffeomorphism and gauge invariant Lagrangian has a set of Bianchi identities which together with the canonical definition of the entropy current leads to the adiabaticity equation. Having established that the background symmetries guarantee adiabaticity we turn to describing how to obtain hydrodynamic Ward identities in \S\ref{sec:Leoms} in terms of a constrained variational principle. To illustrate the efficacy of our formalism, we quickly review some basic examples discussed in the context of non-dissipative fluids in \S\ref{sec:examples}. 

 In \S\ref{sec:classB}-\S\ref{sec:Cvector} we turn to the set of adiabatic constitutive relations which do not belong to Class L, i.e., they don't admit a Lagrangian description. Our first example is the  Berry-like class of adiabatic transport \S\ref{sec:classB} -- these are non-hydrostatic, non-dissipative parts of transport which include well studied examples of transport such as Hall viscosity and Hall conductivity, but also others which have been encountered in holographic fluids. In \S\ref{sec:consJ} we describe the class of conserved entropy currents which are agnostic to physical current transport. We argue in particular that such constitutive relations will be encountered in the presence of topological ground state degeneracy.  In \S\ref{sec:Cvector} constitutive relations determined by transverse vector currents are described; there are some new ingredients here for such transport appears to have never been studied in the literature (outside hydrostatics).

In \S\ref{sec:anomalies} we turn to the problem of finding Lagrangian solutions to the anomalous hydrodynamic transport. We also demonstrate that a complete story for obtaining anomalous Ward identities requires working in a Schwinger-Keldysh doubled theory.
 We review the thermofield doubled Schwinger-Keldysh construction for hydrodynamics in \S\ref{sec:skdouble}, taking the opportunity to  highlight certain obvious tension with adiabaticity and the role of influence functionals.
 We also explain why the terms we introduce in the anomalies both in the current and in our previous discussion 
 \cite{Haehl:2013hoa} are necessary and sensible.

\paragraph{Part \ref{part:8classes}:} We discuss how we can use  adiabatic hydrodynamics to provide a complete classification of hydrodynamic transport in \S\ref{sec:8fold}. Not only do we provide an algorithm for the eightfold classification, amplifying on the results presented in \cite{Haehl:2014zda}, but we also finally prove that the classification is exhaustive. To exemplify our construction, we provide evidence that various hydrodynamic systems are cognizant of the eightfold classification (\S\ref{sec:holofluids}) and give a concise summary of a variety of fluid systems (up to second order in gradients) in \S\ref{sec:8classvar}. We also outline the basic construction of an effective action which encompasses all of the adiabatic constitutive relations in \S\ref{sec:classLT} and \S\ref{sec:LTvariational}, though we leave a fuller exposition of the construction to a later publication \cite{Haehl:2014kq}. In \S\ref{sec:eightfoldLT} we show how the eightfold way of adiabatic transport is captured by this effective action.

\paragraph{Part \ref{part:discuss}:} 
We end the main thread of the paper with a discussion in \S\ref{sec:discuss} and highlight some open questions which we think can be addressed with existing technology in \S\ref{sec:homework}.

\paragraph{Part \ref{part:igeneral}:} There are several extensions to our general analysis of adiabatic constitutive relations which are interesting to explore.  We have chosen to present some of these outside the main line of development so as to keep the flow of the paper more straightforward.  In  Appendix \ref{sec:adcons} we provide a translation from the covariant to the consistent form of anomalous adiabaticity equation.  In  Appendix \ref{sec:ndf} we argue that the Lagrangians we introduce are canonically related via  a Legendre transformation to those described earlier in the literature on non-dissipative fluids. Our presentation clarifies the origins of various symmetries encountered in the previous discussion.  Appendix \ref{sec:WenZee} provides a construction of topological currents which play a role in constructing Class C adiabatic constitutive relations. Finally in  Appendix \ref{sec:aweyl} we give a complete characterization of Weyl invariant adiabatic fluids relevant in the study of holographic fluids,  generalizing the discussion of \cite{Loganayagam:2008is}.

\paragraph{Part \ref{part:technical}:} In these appendices  we provide some technical details which are omitted from the main text. Appendix \ref{sec:varform} provides useful technical details for implementing the variational calculus
in Class L theories. Appendix \ref{sec:neutral2d} gives a detailed discussion of second order neutral fluid hydrodynamics. 
In Appendix \ref{sec:sayantani} we compare our construction with that of \cite{Bhattacharyya:2013lha,Bhattacharyya:2014bha}. Appendix \ref{sec:AnomBianchi} contains details of the derivation of Bianchi identities in the presence of anomalies. Appendix 
\ref{sec:appendixT} verifies that the symmetries of the eightfold Lagrangian are consistent and provides some details for deriving the various Bianchi identities quoted in the text.  

Appendix \ref{sec:Notation} provides a quick reference of the various symbols we introduced through the course of or discussion.

\section{Adiabatic hydrodynamics}
\label{sec:adiabat}

We would like to construct a hydrodynamic effective field theory which not only incorporates the fundamental symmetries present in the underlying microscopic quantum system, but also and in addition is cognizant of the basic constraint of such effective theories, viz., the second law of thermodynamics. As described in \S\ref{sec:intro} the complications of the inequality constraints imposed by the second law can be tackled by taking the constraints off-shell using a suitable combination of the dynamical equations of motion \cite{Liu:1972nr}. While this procedure allows exploration of the off-shell constraints, it is actually more efficacious to  first undertake a full classification
in the marginal situation where the second law is imposed as a statement of entropy conservation (as opposed to entropy production). This split is guided by the fact that hydrodynamic transport can a-priori be categorized as being either adiabatic\footnote{ We use the word adiabatic in a precise technical sense  defined below. Reference \cite{Bhattacharyya:2013lha} uses this word synonymously with hydrostatics, which we prefer not to do. There is more to adiabaticity than equilbirum, as we shall  unearth in the course of our discussion.} or dissipative.

Aided by this intuition we therefore propose to study a class of hydrodynamic theories which we call adiabatic fluids. These are fluids where entropy production is compensated for off-shell by the dynamics of the theory. Having an off-shell formalism allows much insight into how one might construct hydrodynamic effective actions. In fact it will turn out that much of the constraints of the second law can be gleaned from an analysis of adiabatic transport; explicitly dissipative terms will turn out to be quite tractable in the sequel.

\subsection{The adiabaticity equation}
\label{sec:amotive}

Consider a fluid characterized by normalized velocity field  $u^\mu$
(with $u^\mu u_\mu=-1$), temperature $T$ and chemical potential $\mu$ moving in
a background geometry ${\cal M}$ with metric $g_{\mu\nu}$ and a background flavour gauge
field $A_\mu$ which generically will be taken to be non-abelian.\footnote{ Generalizations to arbitrary number of flavour symmetries is straightforward.} We will work in $d$ spacetime dimensions and will assume that the hydrodynamic fields $\{u^\mu, T,\mu\}$ as well as the background sources $\{g_{\mu\nu}, A_\mu\}$ are slowly varying on this spacetime manifold throughout our discussion.

While we could choose to work with the hydrodynamic fields defined above it is in fact convenient to repackage them into an unnormalized vector field and a scalar field. By a simple redefinition we therefore introduce the {\em hydrodynamic fields} (denoted collectively by $\Bfields$)
\begin{equation}
\Bfields  \equiv \{ \Kbeta, \LambdaB\} \,,\qquad
\Kbeta^\mu \equiv \frac{u^\mu}{T}\ ,\qquad
\LambdaB \equiv   \frac{\mu}{T}-\frac{u^\sigma}{T} A_\sigma \,.
\label{eq:hydrofields}
\end{equation}
The fields $\{\Kbeta^\mu,\LambdaB\}$ which we refer to as the {\em thermal vector} and {\em thermal twist}, encode the same hydrodynamic data as the fields $\{u^\mu,T,\mu\}$. We can explicitly invert the above relations to get
\begin{equation}
\begin{split}
u^\mu = \frac{\Kbeta^\mu}{\sqrt{-g_{\alpha\lambda}\Kbeta^\alpha \Kbeta^\lambda}} \ ,\quad
T= \frac{1}{\sqrt{-g_{\alpha\lambda}\Kbeta^\alpha \Kbeta^\lambda}}\ ,\quad
\mu= \frac{\LambdaB +\Kbeta^\sigma A_\sigma}{\sqrt{-g_{\alpha\lambda}\Kbeta^\alpha \Kbeta^\lambda}}
\,.
\end{split}
\label{eq:Tumuinvert}
\end{equation}
Thus for the rest of the discussion, the dynamical content of hydrodynamics is captured by the $d+1$ degrees of freedom in  the vector field $\Kbeta^\mu$ and scalar field $\LambdaB$.

A general hydrodynamic system as reviewed in \S\ref{sec:intro} is characterized by a set of currents: we have the energy-momentum tensor $T^{\mu\nu}$ and a charge current $J^\mu$ which should be considered dynamical. In addition we have an entropy current $J_S^\mu$ which enforces the constraint of the second law. It is also useful to include the free energy current $\mathcal{G}^\mu$, which is a particular linear combination of the above, which we will encounter shortly, cf., \eqref{eq:GDef}. To simplify notation, we will collect the various currents we have introduced into a single set by introducing a collection of tensor fields $\hcur$ 
\begin{equation}
\hcur \equiv\{ T^{\mu\nu}, \;J^\mu, \;J_S^\mu\} \,,
\label{eq:hydrocurrents}
\end{equation}
where instead of $J_S^\mu$ we often equivalently consider the Gibbs free energy current ${\cal G}^\mu$ to be defined in due course.

These currents  should all  be thought of as given by local covariant functionals of  the background and hydrodynamical fields which we also collectively denote as
$\hfields$
\begin{equation}
\hfields \equiv \{g_{\mu\nu},A_\mu,\Kbeta^\mu,\LambdaB\} \,.
\label{eq:hfields}
\end{equation}
Then we can write for our currents $\hcur  = \hcur\brk{\hfields}$ or more explicitly, for the fundamental currents we have
\begin{equation}
\begin{split}
T^{\mu\nu} &= T^{\mu\nu}\brk{\hfields} = T^{\mu\nu} \brk{g_{\alpha\beta},A_\alpha,\Kbeta^\alpha,\LambdaB } \\
J^\mu &= J^\mu \brk{\hfields} = J^\mu \brk{g_{\alpha\beta},A_\alpha,\Kbeta^\alpha,\LambdaB } \\
J_S^\mu &=J_S^\mu\brk{\hfields} = J_S^\mu \brk{g_{\alpha\beta},A_\alpha,\Kbeta^\alpha,\LambdaB } \,.
\end{split}
\end{equation}
These relations are termed \emph{constitutive relations}.

The dynamical information of hydrodynamics comprises simply of the statement of conservation modulo source terms (which do work on the system) and anomalies. In general we can write the conservation equations for a microscopic quantum theory with flavour and Lorentz anomalies in the presence of background sources as:
\begin{equation}
\nabla_\nu T^{\mu\nu} =  J_\nu \cdot F^{\mu\nu}+\THall^{\mu\perp}\, \qquad
D_\nu J^\nu = \JH^\perp  \,.
\label{eq:hydroCons}
\end{equation}
Here, $F_{\mu\nu}$ and $D_\mu$ denote the field-strength and gauge-covariant derivative associated with $A_\mu$ while $\{\THall^{\mu\perp},\JH^\perp\}$
are the covariant Lorentz and flavour anomalies respectively.\footnote{ If $\fP[\fF,\fR]$ is the
anomaly polynomial, then the covariant anomalies are determined using the following equations:
\begin{equation}
\JH^\perp\  \hodge {\bf 1} \equiv  \frac{\partial \fP}{\partial \fF}\ ,\quad
\SpH^{\perp\nu}{}_\mu\  \hodge {\bf 1} \equiv  2\frac{\partial \fP}{\partial \fR^\mu{}_\nu}\ ,\quad
\THall^{\mu\perp} \equiv \half \nabla_\nu \SpH^{\perp\mu\nu}\ .
\end{equation}
Here $\SpH^{\perp\mu\nu}$ is the torque on the system due to Lorentz anomaly. We adopt
a bold-face notation for differential forms. In general our notation follows that of
\cite{Jensen:2013kka,Jensen:2013rga,Haehl:2013hoa} where the reader will find further details on the conventions used herein. We will be more explicit when we solve the anomalous adiabaticity equation in \S\ref{sec:anomalies}. Some further useful details  are collected in Appendices \ref{sec:adcons} 
and \ref{sec:Notation}. }
The center-dot ``$\cdot$'' is reserved for gauge index contraction which we will never write explicitly. 
The gauge-covariant derivative acts on tensors $X^{\mu\cdots\nu}{}_{\rho\cdots\sigma}$ in a familiar fashion, viz., 
\begin{equation}
\begin{split}
D_\alpha X^{\mu\cdots\nu}{}_{\rho\cdots\sigma} &= \nabla_\alpha X^{\mu\cdots\nu}{}_{\rho\cdots\sigma} + [A_\alpha, \, X^{\mu\cdots\nu}{}_{\rho\cdots\sigma}] 
\end{split}
\label{eq:CovDer}
\end{equation}
with 
\begin{equation} \label{eq:CovDer2}
\begin{split}
\nabla_\alpha X^{\mu\cdots\nu}{}_{\rho\cdots\sigma} = \partial_\alpha X^{\mu\cdots\nu}{}_{\rho\cdots\sigma} 
&+ \Gamma^\mu{}_{\lambda\alpha} X^{\lambda\cdots\nu}{}_{\rho\cdots\sigma} + \ldots + \Gamma^\nu{}_{\lambda\alpha}  X^{\mu\cdots\lambda}{}_{\rho\cdots\sigma}  \\
 &  - \Gamma^\lambda{}_{\rho\alpha} X^{\mu\cdots\nu}{}_{\lambda\cdots\sigma} - \ldots - \Gamma^\lambda{}_{\sigma\alpha} X^{\mu\cdots\nu}{}_{\rho\cdots\lambda} \,.
\end{split}
\end{equation}
Here $[\ ,\ ]$ represents the appropriate adjoint action of the flavour algebra. The equations \eqref{eq:hydroCons}, which we term as the {\em hydrodynamic Ward identities}, together with $\nabla_\mu J^\mu_S \geq 0$ capturing the essence of the second law, complete the specification of the hydrodynamic effective field theory in the current algebra language.

The task of a hydrodynamicist is to provide these constitutive relations, order by order in gradients of the fields $\hfields$, subject to symmetry and second law requirements, cf., \cite{landau} for the classic treatment. We will refer the reader to the vast literature on hydrodynamic constitutive relations which have been computed (in certain cases up to the second order in the gradient expansion); see \cite{Rangamani:2009xk,Hubeny:2011hd} for a partial  summary of certain results in the past few years.\footnote{ These computations are typically done by fixing a fluid frame (e.g., in the Landau frame one demands that the non-ideal parts of $T^{\mu\nu}$ and $J^\mu$ are transverse to velocity). We will a-priori make no such assumptions though at various stages of our analysis we will present results by making  certain frame choices.}

While most analyses of the second law constraints are done by classifying first on-shell independent data, as explained in \S\ref{sec:intro} it is useful to work off-shell. To this end we want to extend the statement of the second law, viz.,
\begin{equation}
\exists \; J^\mu_S[\hfields] : \;\; \nabla_\mu J^\mu_S \geq 0 \,,
\label{eq:slaw}
\end{equation}
to a more amenable one which is agnostic of dynamics. The simplest way to proceed is to use the fact that linear combinations of the equations of motion can be added to \eqref{eq:slaw} without affecting the inequality \cite{Liu:1972nr}. All we need is appropriate Lagrange multipliers to ensure that the vectorial energy conservation and the scalar charge conservation equations can be combined with the gradient of the entropy current. The canonical choice is simply to take the Lagrange multipliers to be the hydrodynamic fields
$\Bfields = \{\Kbeta^\mu, \LambdaB\}$ themselves. One way to motivate this choice is to exploit the field redefinition freedom inherent in fluid dynamics, to align the Lagrange multiplier fields to the velocity (rescaled by the temperature) and chemical potential.

This then leads us to the following statement of the second law of thermodynamics:
 \begin{equation}\label{eq:AdiabaticityD}
 \begin{split}
 \nabla_\mu J_S^\mu &+ \Kbeta_\mu\prn{\nabla_\nu T^{\mu\nu}-J_\nu \cdot F^{\mu\nu}-\THall^{\mu\perp}}\\
 &+ (\LambdaB+\Kbeta^\lambda A_\lambda) \cdot \prn{D_\nu J^\nu-\JH^\perp} =\Diss \geq 0\,.
 \end{split}
 \end{equation}
We have introduced $\Diss$ as the placeholder for the entropy production resulting from the  dissipative constitutive relations.

Often when confronted with solving constraints given as inequalities, it is simplest to examine the boundary of the acceptable domain. In the present case this amounts to switching off dissipation by setting 
$\Diss =0$. The part of the constitutive relation which does not contribute to $\Diss$ will be termed adiabatic.

This canonical split allows us to motivate the \emph{adiabaticity equation}. By definition it captures the marginal situation where dissipation is turned off, i.e., $\Diss = 0$:
 \begin{equation}\label{eq:Adiabaticity}
 \begin{split}
 \nabla_\mu J_S^\mu &+ \Kbeta_\mu\prn{\nabla_\nu T^{\mu\nu}-J_\nu \cdot F^{\mu\nu}-\THall^{\mu\perp}}\\
 &+ (\LambdaB+\Kbeta^\lambda A_\lambda) \cdot \prn{D_\nu J^\nu-\JH^\perp} = 0\,.
 \end{split}
 \end{equation}
The constitutive relations which satisfy the adiabaticity equation are called adiabatic constitutive relations.\footnote{ We provide a  translation of the adiabaticity equation in terms of the consistent currents which are sometimes more natural when working with effective actions in Appendix~\ref{sec:adcons}.}  Note that this relation is being imposed off-shell on the hydrodynamical system of interest, a fact that will be of crucial import in our discussion.
For most of this paper we will be concerned with the adiabatic case. However, we will, at some early stage of the discussion (cf.,  \S\ref{sec:classD}), describe the dissipative part of hydrodynamics building 
on the results of  \cite{Bhattacharyya:2013lha,Bhattacharyya:2014bha} using the lessons learned from our adiabatic analysis. 

It is worthwhile recording here a version of the adiabaticity equation that holds when we consider non-anomalous fluids. Since the quantum anomaly manifests itself through the Hall current terms  $T^{\mu\perp}_H$ and $J^\perp_H$ setting them to zero allows us to capture the desired equation for non-anomalous adiabatic fluids, viz.,
 \begin{equation}\label{eq:naadiabatic}
 \begin{split}
 \nabla_\mu J_S^\mu &+ \Kbeta_\mu\prn{\nabla_\nu T^{\mu\nu}-J_\nu \cdot F^{\mu\nu}}
 + (\LambdaB+\Kbeta^\lambda A_\lambda) \cdot D_\nu J^\nu=0 \,.
 \end{split}
 \end{equation}

In the initial part of our discussion we will find it convenient to work with the non-anomalous case first, and then build up to include the presence of anomalies. There is  in fact a useful  perspective that helps to segregate the anomalous contribution from the rest. Apart from anomalies appearing via the Hall currents, the adiabaticity equation is linear in the constitutive relations and relates terms of the same derivative order in the constitutive relations. This means that we can treat anomalous terms in \eqref{eq:Adiabaticity} as ``inhomogeneous source terms". Thus they can be removed by picking a suitable particular solution of adiabaticity equation. As a result we will for the most part assume that such anomalous terms have been appropriately dealt with and focus on the non-anomalous adiabaticity equation by setting them to zero, i.e.,  work with the homogeneous equation \eqref{eq:naadiabatic}. In \S\ref{sec:anomalies} we will describe how the particular solutions to incorporate anomalous effects can be obtained.

It is important to appreciate the following fact: when we refer here and in the sequel to finding solutions to 
\eqref{eq:Adiabaticity} we mean that we would like to find a set of hydrodynamic currents 
$\hcur\brk{\hfields}$ which satisfy this equation off-shell. Thus we would like to determine families of constitutive relations  parameterized by the transport coefficients that are adiabatic.  
As in any structural analysis of hydrodynamic transport we will not be interested in fixing values of transport coefficients. That can only be accomplished once we have an understanding of the microscopic quantum system whose hydrodynamic description we seek.\footnote{ In the interest of full disclosure, we should add  that certain constitutive relations which are forbidden by demanding existence of hydrostatic equilibrium can be viewed as fixing certain transport coefficients to being functions of others (which are the only physical ones).}  With this understanding we will continue to speak of solving the adiabaticity equation, hopefully without causing any confusion.

\subsection{Physical interpretation of adiabatic fluids}
\label{sec:phyad}

Let us physically understand the nature of the fluid systems that satisfy
\eqref{eq:Adiabaticity}, by qualifying the adjective `adiabatic'. 
Suppose  we restrict ourselves to fluid configurations $\{\Kbeta^\mu,\LambdaB \}$ which satisfy the hydrodynamic equations of motion \eqref{eq:hydroCons}. Let us re-characterize them for the present discussion as
\begin{equation}\label{eq:HydroEq}
\begin{split}
\nabla_\nu T^{\mu\nu} &\simeq J_\nu \cdot F^{\mu\nu}+\THall^{\mu\perp}\\
D_\nu J^\nu &\simeq \JH^\perp
\end{split}
\end{equation}
with the symbol $\simeq$ referring to the fact that these equations hold only in this restricted sense (i.e., on-shell).
We can then assign a conserved entropy current to this restricted set of
fluid configurations, i.e., $\nabla_\mu J_S^\mu \simeq 0$. Thus, the constitutive
relations which solve adiabaticity equation describe entropy-conserving (i.e., adiabatic)
transport once hydrodynamic equations are imposed. In this sense the adiabatic fluids are on-shell equivalent to the non-dissipative fluids as defined in \cite{Bhattacharya:2012zx}.
One way to interpret the adiabaticity equation is  to take the view that we have taken entropy conservation off-shell using the hydrodynamic fields as Lagrange multipliers, along the lines espoused in \cite{Liu:1972nr}.

Alternately, the adiabaticity equation is actually a stronger assertion than just entropy
conservation. Say, instead of taking hydrodynamics on-shell via \eqref{eq:HydroEq},
we impose
\begin{equation}\label{eq:HydroEqForced}
\begin{split}
\nabla_\nu T^{\mu\nu} &\simeq J_\nu \cdot F^{\mu\nu}+\THall^{\mu\perp}+f^\mu_{ext}\\
D_\nu J^\nu &\simeq \JH^\perp+ Q_{ext}
\end{split}
\end{equation}
where $f^\mu_{ext}$ is the force per unit volume due to an external system and
$Q_{ext}$ is the charge injected per unit time per unit volume by the external system.
Let us assume that this injection of energy-momentum and charge happens adiabatically
and the entropy injected into the fluid is $\nabla_\mu J_S^\mu\simeq S_{ext}$. The
adiabaticity equation is the statement that all these cannot be together
true for arbitrary $\{f^\mu_{ext},Q_{ext},S_{ext}\}$. In fact this transfer
can be adiabatic if and only if $TS_{ext}+u_\nu f^\nu_{ext}+\mu\cdot Q_{ext} \simeq 0$, i.e.,
if and only if external system satisfies adiabaticity equation. Thus, any two
systems which satisfy adiabaticity equation can be combined to a bigger
system which satisfies adiabaticity equation, in a way reminiscent of the classical
discussions on thermodynamics by Carnot and others.

Thus the adiabaticity hypothesis brings in a sense of linearity into hydrodynamics, much like the superposition principle of quantum mechanics. This allows us to focus the discussion on isolated systems, with the potential downside that we do not have access to the dissipative part of hydrodynamics.

The main motivation for considering adiabatic hydrodynamics is the observation that
non-dissipative parts of many actual hydrodynamic theories coincide with
what one finds in adiabatic hydrodynamics. We note that not all solutions of adiabaticity equation
 might arise in a given microscopic QFT. For example one might want to impose additional constraints
(like Euclidean consistency \cite{Jensen:2012kj,Jensen:2013rga}) and identify on-shell equivalent or fluid frame-equivalent expressions  to eliminate potentially unphysical solutions. Thus, we generally expect the solutions of adiabaticity equation to furnish a super-set of physically admissible non-dissipative constitutive relations up to field redefinitions.

\subsection{Ideal fluids are adiabatic}
\label{sec:ideal}

Having presented the basic equation of interest, we now turn to asking how one might characterize the solutions to the adiabaticity equation. After all we are interested in using these as the first step in understanding more realistic fluid systems (including dissipation). To this end we need to show that we have a non-empty solution set to \eqref{eq:Adiabaticity}.

It  is  natural to study the non-anomalous adiabatic constitutive relations order by order in derivative expansion. Let us illustrate how this works at zeroth order in derivative expansion. The most general constitutive relation with zero derivatives of the hydrodynamic data is\footnote{ We have reverted to $\{u^\mu,T,\mu\}$ so as to write the constitutive relations in their familiar form.}
\begin{equation}
\begin{split}
J_S^\mu =s \, u^\mu\ ,\quad
T^{\mu\nu} = \epsilon\, u^\mu u^\nu + p\,  P^{\mu\nu} \ ,\quad
J^\mu = q \, u^\mu .
\end{split}
\end{equation}
where the entropy density $s$, energy density $\epsilon$, pressure $p$ and charge density $q$ are scalar functions of $T$ and $\mu$. The tensor $P_{\mu\nu} = g_{\mu\nu} + u_\mu\,u_\nu$ is the projector transverse to the velocity. 

The adiabaticity condition \eqref{eq:naadiabatic} can then be written quite simply as
\begin{equation}
u^\alpha \left(T \, \nabla_\alpha  s + \mu \cdot \nabla_\alpha q - \nabla_\alpha
\epsilon\right)
+  \left(T  \, s + \mu \cdot  q -  \epsilon - p\right) \Theta  =0 \,,
\end{equation}
where $\Theta\equiv \nabla_\mu u^\mu$ is the fluid expansion. If we insist that this hold for an arbitrary fluid configuration, then the combination in each of the parentheses should individually vanish. This then implies that the fluid should satisfy the first law
$$ \delta\epsilon = T \,\delta s + \mu \cdot \delta q \,,$$
and the Euler relation
$$ \epsilon+p = T \,s +\mu \cdot q .$$
Thus, we recover standard constitutive relations describing thermodynamics from the formalism of adiabatic hydrodynamics.

We will soon see that the family of adiabatic fluids is far richer as evidenced by our eightfold classification illustrated in 
Fig.~\ref{fig:eightfold}. We will shortly provide a short synopsis of the different classes in \S\ref{sec:aclassify}. The reader impatient to see some more examples is invited to consult \S\ref{sec:examples} where we study neutral fluids and parity-odd charged fluids at higher orders.

\subsection{The adiabatic free energy current}
\label{sec:afree}

We have phrased our discussion of the adiabaticity equation in terms of the entropy current.
However, since we are describing via the hydrodynamic expansion the fluctuations in the Gibbsian  density matrix, it makes more sense to ask about the behaviour of the free energy current itself. This involves using the standard definition of the  grand canonical  free energy current. In terms of the other hydrodynamic currents introduced hitherto:\footnote{ While the physical free energy current is $\mathcal{G}^\sigma$, it is often convenient to write expression for $\N^\sigma$ (which is the free energy rescaled by 
$-T^{-1}$). This quantity naturally appears as a Noether charge in our effective action constructions. As a result we will use both quantities interchangeably for much of our discussion. }
\begin{equation}
\begin{split}
\mathcal{G}^\sigma  &= -T\, \N^\sigma\,,
\\
&\equiv - T\brk{J_S^\sigma+\Kbeta_\nu T^{\nu\sigma}+(\LambdaB+\Kbeta^\nu A_\nu)\cdot J^\sigma} \,.
\label{eq:GDef}
\end{split}
\end{equation}

Assuming we know the free energy current we can solve for the entropy current by inverting the above relation
\begin{equation}
\begin{split}
J_S^\sigma &= - \brk{\Kbeta_\nu T^{\nu\sigma}+(\LambdaB+\Kbeta^\nu A_\nu)\cdot J^\sigma+\frac{\mathcal{G}^\sigma}{T}}
\\
& \equiv (J_S^\sigma)_{can} - \frac{\mathcal{G}^\sigma}{T} \,.
\end{split}
\end{equation}
This expression is useful in that it segregates the various contributions to the entropy current. The terms $ - \brk{\Kbeta_\nu T^{\nu\sigma}+(\LambdaB+\Kbeta^\nu A_\nu)\cdot J^\sigma}$ are usually interpreted as the canonical part of the entropy current $(J_S^\mu)_{can}$. On the other hand the vector  $- \mathcal{G}^\sigma/T$ is called the non-canonical part of the
entropy current. Thus, passing to grand canonical ensemble can be thought
of as focusing our attention on the part of entropy flow which is not
simply related to energy and charge flow. We can think of free energy
(up to a factor of $T$) as just the name given to this part of entropy.

While in the present discussion the grand canonical free energy current appears as a convenient book keeping device for the non-canonical part of the entropy current, it will soon transpire when we consider hydrostatics that it has a natural interpretation in terms of
a partition function.

The notion of the free energy current is quite useful in the context of anomalous hydrodynamics.  While the presence of a quantum  anomaly does not necessarily introduce entropy into the fluid,\footnote{ The anomalous contribution to the entropy current can typically be chosen to vanish for flavour anomalies. The story for Lorentz anomalies is a bit more involved and is discussed in \S\ref{sec:anomalies}.} the charge and energy-momentum injection is inevitably accompanied by a
free energy injection. The free energy per unit time per unit volume injected by
anomalies  is
\begin{equation}
\begin{split}
\mathcal{G}_{_H}^\perp &\equiv  - T\brk{\Kbeta_\nu \THall^{\nu\perp}+(\LambdaB+\Kbeta^\nu A_\nu)\cdot \JH^\perp}\\
&= -\brk{u_\nu\THall^{\nu\perp}+ \mu \cdot \JH^\perp} \,.
\end{split}
\end{equation}
Using this definition, we can now write the grand canonical version of the adiabaticity equation
\eqref{eq:Adiabaticity} as (we include $\Delta$ for completeness)
\begin{equation}\label{eq:AdiabaticityG}
\begin{split}
-\brk{\nabla_\sigma\prn{\frac{\mathcal{G}^\sigma}{T}}-\frac{\mathcal{G}_{_H}^\perp}{T}}&=
\half  T^{\mu\nu}\diffB  g_{\mu\nu} + J^\mu \cdot \diffB  A_\mu + \Delta
\\
&= T^{\mu\nu}\nabla_\mu\prn{\frac{u_\nu}{T}}
+  J^\sigma \cdot \brk{D_\sigma\prn{\frac{\mu}{T}}-\frac{E_\sigma}{T}} + \Delta \,.
\end{split}
\end{equation}
Here $E^\mu = F^{\mu\nu} \,u_\nu$ is the electric field and   $\diffB $ represents the Lie derivatives using the diffeomorphism and flavour transformations generated by $\{\Kbeta^\mu,\LambdaB\}$:
\begin{equation}
\begin{split}
\diffB g_{\mu\nu} &\equiv \lieD_\Kbeta g_{\mu\nu} = \nabla_\mu \Kbeta_\nu
+ \nabla_\nu \Kbeta_\mu  \,, \\
\diffB  A_\mu &\equiv \lieD_\Kbeta A_\mu +\partial_\mu \LambdaB  + [A_\mu,\LambdaB]
=D_\mu(\LambdaB+\Kbeta^\nu A_\nu)+\Kbeta^\nu F_{\nu\mu} \,.
\end{split}
\label{eq:delBdef}
\end{equation}
In this expression, we used  $\lieD_\Kbeta$ to denotes the Lie derivative along the vector field $\Kbeta^\mu$.

It is useful to record the expression for the Lie derivative in terms of the more familiar hydrodynamic decomposition. A quick evaluation leads to
\begin{equation}
\begin{split}
\diffB g_{\mu\nu} &= 2\, \nabla_{(\mu} \Kbeta_{\nu)} 
 = \frac{2}{T}\, \brk{ \sigma_{\mu\nu}  + P_{\mu\nu}\, \frac{\Theta}{d-1} - \left(\acc_{(\mu} + \nabla_{(\mu} \log T \right) u_{\nu)} }
  \\
 \diffB A_\mu &
 = D_\mu(\LambdaB+\Kbeta^\nu A_\nu)+\Kbeta^\nu F_{\nu\mu}
 = u^\alpha\, D_\alpha\left(\frac{\mu}{T} \right)\, u_\mu - \frac{1}{T}\,\cv_\mu \,.
 \end{split}
\label{eq:diffbga}
\end{equation}
We use the standard decomposition of the gradient of the velocity field into the transverse traceless 
shear-strain rate $\sigma_{\mu\nu}$, the anti-symmetric vorticity $\omega_{\mu\nu}$, the vectorial acceleration $\acc_\nu$ and scalar expansion $\Theta$ respectively, viz.,
\begin{align}
\nabla_\mu u_\nu = \sigma_{(\mu\nu)} + \omega_{[\mu\nu]} - u_\mu\, \acc_\nu + P_{\mu\nu}\,
\frac{\Theta}{d-1} \,,
\label{eq:uder}
\end{align}
and the flavour fields decompose as
\begin{align}
\cv^\mu = E^\mu - T\, P^{\mu\nu}\, \nabla_\nu\left(\frac{\mu}{T}\right) \,, \qquad E^\mu = F^{\mu\nu}\, u_\nu \,.
\label{eq:cvdef}
\end{align}

An alternate form of \eqref{eq:AdiabaticityG} can be given by using the fluid
acceleration  $\acc^\alpha \equiv u^\mu\nabla_\mu u^\alpha$ to eliminate the thermal gradients:
\begin{equation}
\begin{split}
-\brk{(\nabla_\sigma  + \acc_\sigma) \,\mathcal{G}^\sigma-\mathcal{G}_{_H}^\perp}
&=
J_S^\sigma(\nabla_\sigma  + \acc_\sigma)T
+ T^{\mu\nu}(\nabla_\nu+\acc_\nu)u_\mu
+  J^\sigma \cdot \brk{D_\sigma \mu +\acc_\sigma \mu-E_\sigma} +T \Delta \,.
\end{split}
\end{equation}
This form of the equation is quite useful in making comparisons with traditional hydrodynamic analysis.\footnote{ Recall that in the current algebra approach one typically chooses to eliminate thermal gradients in favour of velocity derivatives. Some useful formulae and commonly used notation are collected in the tables of Appendix \ref{sec:Notation}.}

\section{Classification of adiabatic transport}
\label{sec:aclassify}
We now  provide a telegraphic summary of the different solutions (i.e., constitutive relations) to the adiabaticity equation \eqref{eq:Adiabaticity} we will encounter in the course of our discussion. 
\paragraph{$\bullet$ Class H (hydrostatic constitutive relations) \S\ref{sec:hydrostatics}:} Consider placing a fluid on a stationary background wherein there exists a timelike Killing vector and a Killing gauge transformation, $\{K^\mu,\Lambda_K\}$, that leave the background sources $g_{\mu\nu}$ and $A_\mu$ invariant. In \S\ref{sec:hydrostatics} we explain our reasoning for taking it as an axiom that the obvious hydrostatic fluid configuration given by $\{\Kbeta^\mu,\LambdaB\} = \{K^\mu,\Lambda_K\}$ immediately gives solutions to the adiabaticity equation. We dub this premise as the {\it hydrostatic principle}. This gives us the hydrostatic (i.e., time-independent) configurations of a generic fluid dynamical system. These configurations lead to hydrostatic constitutive relations which capture the thermodynamic response parameters that are encoded in terms of an equilibrium partition function.

We emphasize here that we want to classify adiabatic constitutive relations, i.e., currents 
$\hcur\brk{\hfields}$ solving \eqref{eq:Adiabaticity}. In some cases the  solutions can be efficiently encoded in a generating function that may be parametrized in terms of other tensor structures. The physical currents should thus be treated as functionals of these auxiliary (solution generating) tensor structures. With this clarification we record here that the hydrostatic class can be further sub-divided into two:
 \begin{itemize}
 \item {\bf Class $\PS$:} Hydrostatic partition functions  transforming as thermal scalars,\footnote{ The transformation property here refers to behaviour under Kaluza-Klein gauge transformations where we treat the Euclidean thermal circle relevant in hydrostatic as the compact direction we reduce on.} which comprise the class of non-anomalous equilibrium generating functions discussed in \cite{Banerjee:2012iz,Jensen:2012jh}.
 \item {\bf Class $\PV$:} Hydrostatic partition functions transforming as thermal vectors such as those relevant for understanding transcendental anomaly induced transport which generalizes the parity-odd part of Cardy formula to higher dimensions \cite{Jensen:2013rga}. Further details are provided in \S\ref{sec:Cvector}.
 \end{itemize}

\paragraph{$\bullet$ Class $\PF$ (hydrostatic forbidden) \S\ref{sec:hydrostatics}:}  At a given order in derivatives, Classes $\PS$ and $\PV$ are exhausted after classifying all independent scalars and vectors that survive the hydrostatic limit. Together they parameterize the (Euclidean) equilibrium partition function. However, this counting generically does not match the classification of all possible tensor structures in the equilibrium constitutive relations at this order. The mismatch is due to a number of hydrostatic relations $\PF$ which arise as constraints from the existence of an equilibrium configuration.

As the simplest example, consider a neutral fluid at zeroth order in derivative expansion. A-priori there are two possible tensor structures in the stress tensor constitutive relations:  
$u^\mu \, u^\nu$ and $P^{\mu\nu}$. The pressure which multiplies $P^{\mu\nu}$ parameterizes hydrostatic partition function. Moreover, the existence of hydrostatic equilibrium at this level is equivalent to the Euler relation, $\epsilon + p = T\,s$, which enforces that the  coefficient of $u^\mu\, u^\nu$ (energy density) is not independent, but rather is determined in terms of pressure (this is the hydrostatic version of the analysis in \S\ref{sec:ideal}). This statement can be viewed as saying that there is one relation between two a-priori independent pieces of transport data, a feature that characterizes
Class $\PF$ more generally. 

 \paragraph{$\bullet$ Class L (Lagrangian solutions) \S\ref{sec:classL}, \S\ref{sec:Leoms}:}
If we are on-shell the content of \eqref{eq:Adiabaticity} with $\Delta = 0$ is simply that of entropy conservation, but we have now allowed ourselves to take this off-shell. By doing so we have gained the distinct advantage of being able to ask the following question: ``What are the Lagrangian theories that respect the adiabaticity equation?'' This turns out to be surprisingly easy to answer. Consider any local diffeomorphic and gauge invariant Lagrangian density $\Lag\left[g_{\mu\nu}, A_\mu, \Kbeta^\mu, \LambdaB\right]$ viewed as a functional of the background metric $g^{\mu\nu}$, background gauge potential $A_\mu$ and the hydrodynamic variables $\{\Kbeta^\mu, \LambdaB \}$.  $\Kbeta^\mu$ is a vector field under diffeomorphisms and $ \LambdaB$ transforms as appropriate for a gauge parameter. We then have a set of Bianchi identities arising from these background symmetries.  These identities
 then simply imply the adiabaticity equation with the entropy current $J_S^\mu = s\, u^\mu$ where  $s$ is taken to be the Euler-Lagrange derivative of the Lagrangian with respect to the temperature, keeping fixed the chemical potentials and the background sources. In other words the adiabaticity equation follows quite trivially in a wide class of Lagrangian hydrodynamic theories.\footnote{ For the moment, we are ignoring situations with anomaly induced transport. We will later explain how to find Lagrangian solutions to anomalous hydrodynamics, which is indeed possible, albeit with some interesting technical complications.}

Given a Lagrangian theory of hydrodynamics, we not only want to show that the adiabaticity equation is satisfied, but also obtain the correct dynamical equations, which as we have emphasized, are simply conservation equations. Unfortunately, the simple unconstrained variational principle with respect to hydrodynamic variables $\{\Kbeta^\mu, \LambdaB \}$ does not result in  the desired energy-momentum and charge conservation Ward identities. So the non-trivial part of the construction involves demonstrating the existence of a novel constrained variational principle that obeys the desired dynamics. In fact, such a principle is easy to state: fixing $\{g_{\mu\nu},A_\mu\}$, and extremizing  $\int_{\cal M} \sqrt{-g}\, \Lag\brk{\hfields}$ under constrained variations that stay on Lie orbits of $\{\Kbeta^\mu,\LambdaB\}$ gives on-shell hydrodynamic configurations.

There is an equivalent but more convenient way of parameterizing such variations. Start by fixing a {\it reference hydrodynamic configuration} $\{\Kref^a, \Lref\}$ on a (fiducial) reference manifold $\Mref$ which is diffeomorphic to the physical spacetime ${\cal M}$. Then consider a diffeomorphism field $\varphi^a(x)$ and a gauge transformation $c(x)$ which relate ${\cal M}$ and $\Mref$. Said differently, the true physical configuration $\{\Kbeta^\mu,\LambdaB\}$ on ${\cal M}$ is given by the pullback of the reference configuration $\{\Kref^a, \Lref\}$ along $\varphi^a(x)$, $c(x)$:
\begin{equation}
\begin{split}
\Kbeta^\mu(x) &= \frac{\partial x^\mu}{\partial \varphi^a} \; \Kref^a[\varphi(x)]  \\
\LambdaB(x) &= c(x)\ \Lref[\varphi(x)]\ c^{-1}(x) + \Kbeta^\sigma(x) \, \partial_\sigma c(x)\ c^{-1}(x)
\end{split}
\end{equation}
The fields $\varphi^a(x)$ and $c(x)$ are just a convenient parametrization of a given Lie orbit of $\{\Kbeta^\mu, \LambdaB\}$. So instead of doing a constrained variation of $\{\Kbeta^\mu,\LambdaB\}$ along Lie orbits, we can equivalently do an unconstrained variation of the fields $\{\varphi^a(x), c(x)\}$ holding fixed the hydrodynamic reference configuration. We demonstrate that this prescription leads to the desired Ward identities. 

In addition, as suggested by choice of notation for the fields, we have a wonderful bonus -- we can derive and explain the symmetries of the non-dissipative effective action in a simple and effective manner, see Appendix \ref{sec:ndf}. This analysis will be tantamount to saying that all transport described by the non-dissipative action formalism (Class ND) is contained within the Class L constitutive relations. Furthermore, we will demonstrate that Class $\PS \subset$ Class L by showing that the equilibrium limit of the Lagrangians we construct, reproduces the (scalar part of) hydrostatic partition function. However, the scalar Lagrangian density also has hydrodynamic scalar densities (Class $\LS$) (which vanish in equilibrium); strictly speaking L = $\PS \cup \LS$ with the hydrostatic scalars forming the coset $\PS = \text{L}/\LS$.

The simplest non-trivial example of non-dissipative fluids is a neutral fluid at second order in gradient expansion. This system has been studied from an effective action perspective in \cite{Bhattacharya:2012zx}. We revisit this example in \S\ref{sec:examples} (see also Appendix \ref{sec:neutral2d}) where we describe how we can exploit our new Lagrangian formalism to understand features of the transport.  The neutral fluid Lagrangian is parameterized by five scalar functions which determine 15 transport coefficients. This implies 10 linear differential relation amongst transport coefficients. 5 of these are hydrostatic and were obtained earlier in \cite{Banerjee:2012iz} -- they make up Class $\PF$. The remaining 5 relations are adiabatic combinations of transport that do not lead to any dissipation. Curiously as reported in  \cite{Haehl:2014zda}  all second order transport for strongly coupled Weyl invariant holographic fluids is obtained from Class L.

\paragraph{$\bullet$ Class B (Berry-like transport) \S\ref{sec:classB}:} These comprise of the class of solutions to the adiabaticity equation which are non-hydrostatic but explicitly non-dissipative. They turn out to satisfy the adiabaticity equation trivially and as a result such transport coefficients are completely unconstrained in the current algebra approach.  The nomenclature owes its origin to the fact that these can be viewed as Berry curvature induced transport. Heuristically these arise when we  adiabatically traverse a closed loop in the fluid configuration space \cite{Read:2008rn}.
To wit, consider the following constitutive relations:\footnote{ To achieve completeness, one needs to consider slightly more general structures, whose discussion we defer to the main text, see Eq.\ \eqref{eq:TJClassVMatB}.}
\begin{equation}\label{eq:introClassB}
\begin{split}
(T^{\mu\nu})_\text{B}
&\equiv -\quarter \prn{ \BerryG^{(\mu\nu)(\alpha\beta)}-\BerryG^{(\alpha\beta)(\mu\nu)} }\diffB  g_{\alpha\beta} + \BerryGA^{(\mu \nu) \alpha} \cdot \diffB  A_\alpha
\\
(J^\alpha)_\text{B}
&\equiv  - \half \BerryGA^{(\mu \nu) \alpha} \diffB  g_{\mu\nu}- \BerryA^{[\alpha\beta]} \cdot \diffB  A_\beta \\
(J_S^\alpha)_\text{B}
&\equiv  - \frac{u_\sigma}{T} (T^{\alpha\sigma})_B - \frac{\mu}{T} \cdot (J^\alpha)_B
\end{split}
\end{equation}
where $\{\BerryG^{\mu\nu\alpha\beta},\BerryGA^{\mu \nu \alpha}, \BerryA^{\alpha\beta} \}$ are arbitrary tensors and $\diffB$ denotes the Lie derivative associated to the diffeomorphism and gauge transformation generated by $\{\Kbeta^\mu,\LambdaB\}$.
A prime example for structures of the type (\ref{eq:introClassB}) are the parity-odd shear tensor in $3$ dimensions which contributes to Hall viscosity. There also turn out to be some novel combinations in parity-even neutral fluids; the (shear)$\times$(vorticity) term which is present in conformal fluids is of this type (as is a combination seen in the analysis of \cite{Bhattacharya:2012zx}).

\paragraph{$\bullet$ Class C (conserved entropy) \S\ref{sec:consJ}:} This is the simplest class of adiabatic transport, comprising of identically conserved entropy current, with vanishing physical currents. Since we want non-trivial entropy current, we classify such terms up to exact gradients, i.e., cohomologically. Wen-Zee currents \cite{Wen:1992ej} and their generalizations discussed by \cite{Golkar:2014wwa,Golkar:2014paa} which we explore (and extend) in \S\ref{sec:WenZee} provide examples of such terms.
One can view this as capturing the transport of topological states of the hydrodynamical system. These are  finite since they arise as solutions to a cohomology problem.

\paragraph{$\bullet$ Class $\GV$ (Hydrodynamic  vectors) \S\ref{sec:Cvector}:} 
Just as Class $\PV$ contains hydrostatic vectors that give rise to transverse free energies, Class $\GV$ does the same for hydrodynamic vectors which are vanishing in equilibrium.\footnote{ Given that these constitutive relations are primarily determined by the free energy current, it might be more natural to call this the Gibbsian vector class.} In particular, consider the following constitutive relations:
\begin{equation}
\begin{split}
(T^{\mu\nu})_{\GV} &\equiv
	  \half \brk{D_\lambda{\mathfrak C}_{\BerryG}^{\lambda (\mu\nu)(\alpha\beta)}
	  \, \diffB  g_{\alpha\beta} + 2\ {\mathfrak C}_{\BerryG}^{\lambda (\mu\nu)(\alpha\beta)}
	  \, D_\lambda \diffB  g_{\alpha\beta}} \\
&\qquad 	+ D_\lambda{\mathfrak C}_{\BerryGA}^{\lambda(\mu \nu) \alpha} \cdot \diffB  A_\alpha
	   + 2\ {\mathfrak C}_{\BerryGA}^{\lambda(\mu \nu) \alpha} \cdot
  	  \, D_\lambda \diffB  A_\alpha \\
(J^\alpha)_{\GV} &\equiv
	 \half \brk{ D_\lambda{\mathfrak C}_{\BerryGA}^{\lambda(\mu \nu) \alpha}  \diffB  g_{\mu\nu}
	   + 2\ {\mathfrak C}_{\BerryGA}^{\lambda(\mu \nu) \alpha}
	  \, D_\lambda \diffB  g_{\mu\nu} } \\
&\qquad 	+ D_\lambda{\mathfrak C}_{\BerryA}^{\lambda(\alpha\beta)} \cdot \diffB  A_\beta
	   + 2\ {\mathfrak C}_{\BerryA}^{\lambda (\alpha\beta)} \cdot
  	  \, D_\lambda \diffB  A_\beta
\end{split}
\end{equation}
where ${\mathfrak C}_{\BerryG}^{\lambda (\mu\nu)(\alpha\beta)}={\mathfrak C}_{\BerryG}^{\lambda (\alpha\beta)(\mu\nu)}$ and $\{{\mathfrak C}_\BerryG^{\lambda\mu\nu\alpha\beta},{\mathfrak C}_\BerryGA^{\lambda\mu\nu\alpha},{\mathfrak C}_\BerryA^{\lambda\alpha\beta}\}$ are transverse in their first index and 
are otherwise arbitrary local functionals of $\hfields$.  This trivially solves the adiabaticity equation with a particular choice of transverse free energy current. The simplest examples of such transport can be found at second order charged fluid dynamics. The constitutive relations are messy, cf., \eqref{eq:HbarVex1} and \eqref{eq:HbarVex2}, but the free energy current is simpler and given by the transverse vectors $\sigma^{\mu\nu} \,\cv_\nu$ and $\Theta\, \cv^\mu$.

\paragraph{$\bullet$ Class A (anomalous transport terms) \S\ref{sec:anomalies}, \S\ref{sec:skdouble}:} Anomalous transport is hydrostatic 
\cite{Jain:2012rh,Banerjee:2012cr,Jensen:2012jy,Jensen:2012kj,Jensen:2013kka,Jensen:2013rga}. It is however instructive to 
extract anomalous transport from an off-shell, off-equilibrium approach, as it provides non-trivial checks on our formalism. By generalizing the arguments of \cite{Haehl:2013hoa} for flavour anomalies we construct an off-shell effective action for mixed flavour and gravitational anomalies in Class $\PS$. This procedure however does not capture the $\PV$ terms that are necessary for ensuring Euclidean consistency (the transcendental terms of \cite{Jensen:2012kj}).

The construction of an action $S_{anom}$, which can be added to any non-anomalous action from Class L to account for the presence of mixed anomalies, proceeds in two steps.
Firstly, we need to use the anomaly inflow mechanism which extends the physical spacetime ${\cal M}$ to a $(d+1)$-dimensional bulk spacetime ${\bulkM}_{d+1}$ with $\partial{\bulkM}_{d+1} = \mathcal{M}$. The gauge and gravitational connections of the latter are such that their boundary inflow compensates for the gauge non-invariance of the physical theory on ${\cal M}$. In \S\ref{sec:anomalies} we consider the following bulk action:
\begin{equation}
\begin{split}
S_{anom} =
\int_{\bulkM_{d+1}} \VP[\fA, \fGamma,\fAh,\fGammah] \,,
\end{split}
\label{eq:sanom1}
\end{equation}
where $\VP$ is the transgression form that interpolates between the gauge and gravitational connections $\fA, \fGamma^\mu{}_\nu$ and their ``shadows" $\fAh = \fA + \mu \, \fu$ and 
$\fGammah^\mu{}_\nu = \fGamma^\mu{}_\nu + \Omega^\mu{}_\nu \, \fu$. Here $\Omega^\mu{}_\nu$ is the out-of-equilibrium spin chemical potential extending the definition of \cite{Jensen:2013kka}. The transgression form $\VP$ is itself given in terms of the anomaly polynomial.

As expected, the stress tensor and charge current derived from $S_{anom}$ are non-equilibrium generalizations of the anomalous currents found in
\cite{Loganayagam:2011mu,Jensen:2013kka}. In the presence of gravitational anomalies the entropy current is modified by an anomalous contribution (which vanishes in equilibrium) from its canonical form $J^\mu_S = s\, u^\mu$.\footnote{ The modification of the entropy current was one of the main reasons for us to upgrade from the non-dissipative effective action formalism used to analyze flavour anomalies in  \cite{Dubovsky:2011sk,Haehl:2013hoa}.} 

Secondly, an anomalous effective action should also lead to the correct hydrodynamic Ward identities; this however fails if we just consider \eqref{eq:sanom1}. As discussed by us in the context of flavour anomalies 
\cite{Haehl:2013hoa}, getting the Ward identities to work requires  a Schwinger-Keldysh doubled theory. We describe the general framework of such doubled constructions in \S\ref{sec:skdouble}  and show how to obtain the correct anomalous constitutive relations and Ward identities using
  two copies (R and L) of the transgression form \eqref{eq:sanom1} together with a  third transgression form that acts as an influence functional connecting the two copies,
\begin{equation}
\begin{split}
  S_{tot,anom} = \int_{\Mref_{d+1}} \left(\VP[\Aref_\skR, \Chref_\skR,{\hat \Aref}_\skR, {\hat \Chref}_\skR] - 
  	\VP[\Aref_\skL, \Chref_\skL,{\hat \Aref}_\skL,{\hat \Chref}_\skL] +
   \VP[ {\hat \Aref}_\skR, {\hat \Chref}{}_\skR; {\hat \Aref}_\skL, {\hat \Chref}{}_\skL]
       \right) .
\end{split}
\end{equation}

A rather pleasant consequence of viewing hydrodynamic fields $\{\varphi^a, c\}$ as maps from a fiducial reference spacetime to a physical configuration is that one has a clear picture of the doubled gauge symmetries (especially diffeomorphisms) encountered in the Schwinger-Keldysh constructions. While there are indeed two sets of gravitational sources, they are both obtained by pushing-forward data from a single reference manifold $\Mref_{d+1}$.

\paragraph{$\star$ The Eightfold Lagrangian (Class $\LT$) \S\ref{sec:classLT}:} The classification of adiabatic constitutive relations described above can be  motivated on physical grounds. Furthermore, one can show that the 7 classes above (Classes $\PS$, $\PV$, $\LS$, $\GV$, B, C, A) exhaust all adiabatic transport. One can in fact obtain all of these solutions from a single master action. This effective action is parameterized by the hydrodynamic fields $\{\Kbeta^\mu, \LambdaB\}$, the  background sources $\{g_{\mu\nu}, A_\mu\}$, their Schwinger-Keldysh like doubles $\{\tildeg_{\mu\nu}, \tildeA_\mu\}$, and a new Abelian gauge field $\AT_\mu$. While the doubling of the sources, whilst  perhaps unanticipated, could be reconciled with the fact that non-equilibrium dynamics is incomplete without such a construction, the new gauge principle hints at some hidden underlying structure. We describe why this is necessary and point out some of its physical implications below. For now we shall simply record that a set of hydrodynamic currents $\hcur\brk{\hfields}$ satisfying the adiabaticity equation can be derived rather straightforwardly from  the effective action
\begin{align}
S_{\,\smallT} = \int\, d^dx\, \sqrt{-g}\, \left( \frac{1}{2} T^{\mu\nu}\, \tilde g_{\mu\nu} + J^\mu\, \tildeA_\mu - \frac{1}{T}\,
{\mathcal G}^\mu \, \AT_\mu\right)\,.
 \label{}
\end{align}

Before proceeding to a detailed discussion of the specific classes of solutions to the adiabaticity equation, let us remark on how this construction is related to the earlier work on non-dissipative fluids and how the knowledge of adiabatic transport enables us to complete the classification of hydrodynamic transport.

\paragraph{$\circ$ Class ND (non-dissipative fluids) (Appendix \ref{sec:ndf}):} 
This class encompasses non-dissipative fluids as described in the formalism of \cite{Dubovsky:2011sj} with the fundamental degrees of freedom expressed in terms of fluid element Goldstone fields as reviewed in
\S\ref{sec:intro}. We explain in Appendix \ref{sec:ndf} how this family of fluids can be embedded into Class L. We show this by starting with the Lagrangian written in terms of the
hydrodynamic fields and implementing a Legendre transformation with respect to temperature. The fundamental variable is then the entropy current itself in the new parameterization. By passing to the reference manifold and examining the redundancies in the map from the fiducial spacetime $\Mref$ to the physical spacetime ${\cal M}$ implemented by the fields $\{\phi^{_I}, \cnd\}$ mentioned in \S\ref{sec:intro}, we unearth the origins of the volume-preserving diffeomorphism and chemical shift symmetries postulated in the effective action constructions of \cite{Dubovsky:2011sj,Bhattacharya:2012zx}.

\paragraph{$\bullet$ Class D (dissipative transport) \S\ref{sec:classD}:} Having classified all terms that solve the adiabaticity equation (\ref{eq:Adiabaticity}) with $\Delta = 0$, we are left with some terms that are genuinely dissipative, i.e., terms for which no choice of entropy current exists that would remove them. These dissipative terms can be further subdivided: on the one hand, we have transport coefficients whose sign is constrained by the requirement of $\nabla_\mu J^\mu_S \geq 0$ for arbitrary fluid flows (Class $\Dv$). Such terms only show up at the leading order in derivative expansion. On the other hand, there are sub-dissipative terms which are completely unconstrained (Class $\Ds$). It transpires, using a result established in \cite{Bhattacharyya:2013lha} that $\Dv$ terms appear only at the first order in gradient expansion. We describe this argument in our language, providing at the same time a simple characterization of dissipative constitutive relations.

\section{Class H: Hydrostatics from adiabaticity}
\label{sec:hydrostatics}

We have defined adiabatic fluids to be the set of hydrodynamic constitutive relations that satisfy \eqref{eq:Adiabaticity}. While in \S\ref{sec:adiabat} we have argued that this set comprises of the obvious example of ideal fluids, we would like to ascertain and classify other solutions to the adiabaticity equation. We will proceed to establish the existence of various classes of solutions to \eqref{eq:Adiabaticity} in the reminder. To keep the logical flow of the arguments simple we will start with  statements that hold in great generality and subsequently specialize to more special cases.

Our first case of interest is what we called Class H in \S\ref{sec:aclassify}. We specialize to time-independent configurations in hydrodynamics (i.e., we limit ourselves to hydrostatics). In order to ascertain non-trivial constraints on fluids from this hydrostatic restriction we need to turn on external sources, e.g., background metric and gauge fields, which themselves are time-independent to begin with.
Therefore let us assume that there exists a Killing vector and Killing
gauge transformation collectively denoted by $\Eqfields \equiv\{\KEq^\mu,\LambdaEq\}$ such that
$\diffEq g_{\mu\nu} = 0$ and $\diffEq A_\mu =0$.  We will further assume that
$\KEq^\mu$ is timelike everywhere on the manifold the fluid propagates on.\footnote{ In particular, we demand by virtue of $\KEq$ being globally timelike on ${\cal M}$ that the background the fluid propagates on is free of ergosurfaces. This is necessary in order for the fluid configuration to have a stationary solution aligned with the Killing field.} To wit, a stationary background source configuration is encoded as
\begin{equation}
\Eqfields \equiv \{\KEq^\mu,\LambdaEq\} \,,\quad g_{\mu\nu} \,\KEq^\mu\, \KEq^\nu \leq 0
\;\longrightarrow\; \diffEq g_{\mu\nu} = \diffEq A_\mu =0 \,.
\label{eq:hstatic}
\end{equation}

There is a natural hydrostatic configuration associated with this
background given by $\{\Kbeta^\mu,\LambdaB\}=\{\KEq^\mu,\LambdaEq\}$. This
configuration is time-independent since $\diffEq \Kbeta^\mu =\diffEq\KEq^\mu=0$
and $\diffEq \LambdaB =\diffEq \LambdaEq=0$. It therefore follows that for any functional
${\cal Z}\brk{\hfields}$ of the fluid dynamical variables we have
\begin{equation}
\diffB  {\cal Z}\brk{g_{\alpha\beta},A_\alpha,\Kbeta^\alpha,\LambdaB}
= \diffEq {\cal Z} \brk{g_{\alpha\beta},A_\alpha,\KEq^\alpha,\LambdaEq}= 0 \,.
\label{eq:hsdelb}
\end{equation}

\paragraph{The Hydrostatic Principle:} We now formulate an important non-trivial statement
about the configurations we have just described. A-priori by aligning the fluid velocity and the gauge parameter to the background Killing structure only results in an off-shell configuration of the system. The \emph{hydrostatic principle} asserts that these off-shell hydrostatic configurations constructed above are also \emph{automatically on-shell},
i.e., they automatically solve the hydrodynamic equations
\eqref{eq:hydroCons}. We will see that this holds true for
all adiabatic constitutive relations that we will consider in the sequel.

Let us now define the hydrostatic limit of various currents we have defined in the
previous sections by just substituting $\{\Kbeta^\mu,\LambdaB\}=\{\KEq^\mu,\LambdaEq\}$.
The hydrodstatic currents are then simply obtained as
\begin{equation}
\left(\hcur\right)_\text{Hydrostatic} = \hcur \big|_{\{\Kbeta^\mu,\LambdaB\}=\{\KEq^\mu,\LambdaEq\}} \,.
\end{equation}
The utility of these currents is that they allow us to write down an expression for the hydrostatic partition function, the generating function for correlators  of the currents $\hcur$.

\subsection{Hydrostatic partition functions}
\label{sec:hpfns}

Consider a fluid on a background manifold ${\cal M}$ with metric $g_{\mu\nu}$ and gauge field $A_\mu$. When the background sources satisfy \eqref{eq:hstatic}  we can construct a Wick-rotated manifold over which a partition function can be defined. We begin by identifying every point $p \in {\cal M}$,  with the point $p'$ in its
future separated from it by a unit affine distance along the vector $\KEq^\mu$.
More precisely, we identify the points $p$  and $p'$ if there is a curve $x^\mu(\tau)$ such that
\begin{equation}
\begin{split}
x^\mu(\tau=0) = x^\mu(p)\ ,\qquad
x^\mu(\tau=1) = x^\mu(p') \ ,\qquad
\frac{dx^\mu}{d\tau} = \KEq^\mu \,.
\end{split}
\end{equation}
We will  also assume that $\{g_{\mu\nu},A_\mu\}$ are sufficiently slowly varying (spatially) so that there are no caustics within a unit affine distance. This identification then
converts the original spacetime ${\cal M}$ into a fibre bundle with a timelike circle fibred
over a spacelike base space $\Sigma_{\cal M}$.\footnote{ In order to do this without any ambiguity, one needs to prescribe how the flavour fibres at $p$ and $p'$ should be identified -- we will identify these flavour fibres with a flavour transformation given by $\LambdaEq$, i.e., we take the local gauge choice  at $p$ and $p'$ to be related by the gauge transformation generated by
$\LambdaEq$. This then gives fibre bundles charged under the flavour group over the spatial base space.}
For definiteness, we will also choose an embedding of the
base space into the original spacetime as a spatial hypersurface (this is equivalent
to fixing a gauge for  the Kaluza-Klein (KK) gauge field which arises when we reduce along the timelike circle). For consistency, we will require that our final results should not depend on
this (arbitrary) choice of embedding.

Since $\KEq^\mu$ is a Killing vector field, we can Wick rotate the background ${\cal M} \mapsto {\cal M}_E$ by a suitable analytic continuation of its  orbits. Likewise we
also Wick-rotate all the hydrostatic currents to obtain suitable Euclidean currents\footnote{ In this section we will use the subscript $E$ uniformly to denote the Wick rotated fields of interest.}
\begin{equation}
\left(\hcur\right)_\text{Hydrostatic}  \mapsto \left(\hcur\right)_\text{E} \,.
\end{equation}

With these manipulations we are now in a position to define the grand canonical partition function $e^{iW_E}$ following \cite{Banerjee:2012iz,Jensen:2012jh}. It is given as the path integral over the manifold ${\cal M}_E$ constructed above with thermal boundary condition along the $\tau$-circle.
This path integral is dominated  by the hydrostatic saddle point we have just described so it is trivial to write down the final answer. Let us distinguish situations with and without anomalies since the final answer for the generating function depends on the microscopics of the theory.

\paragraph{(i). Non-anomalous hydrostatic partition functions:} In the absence of anomalies, $W_E$ is just the Wick rotation of the total hydrostatic free energy, viz.,
\begin{equation}
\begin{split}
W_E &= -\int_{\Sigma_E}   \left(\frac{\mathcal{G}^\mu_E}{T}\right)  d^{d-1}S_\mu\
\end{split}
\label{eq:wenadef}
\end{equation}
where $d^{d-1}S_\sigma$ is the area form on the base space $\Sigma_E$  which is defined using  the choice of base space embedding  we described above (thus  $
\Sigma_E =\Sigma_{\cal M}$).

One can easily check that this answer for $W_E$ is embedding independent.  Setting $\{\Kbeta^\mu,\LambdaB\}=\{\KEq^\mu,\LambdaEq\}$ in \eqref{eq:AdiabaticityG} and using \eqref{eq:hstatic} we get
\begin{equation}
\nabla_\mu\prn{ \frac{\mathcal{G}^\mu}{T}}_\text{Hydrostatic}=0 \,.
\end{equation}
This means that its Wick-rotated counterpart $\frac{\mathcal{G}_E^\mu}{T}$
is also divergenceless (i.e., it is conserved). This then  implies that
$W_E$ is embedding independent.  As advertised earlier, knowledge of the Euclidean Gibbs current is sufficient information to recover the generating function of current correlators. 

\paragraph{(ii). Anomalous hydrostatic partition functions:}  The above argument can be extended to situations where we have quantum anomalies with some minor modifications.
As evident from \eqref{eq:AdiabaticityG}, the covariant free energy current is no longer conserved in the adiabatic limit because of the free energy injection due to anomalies. This issue can be solved however if we choose to work with consistent  free energy current instead.\footnote{ The distinction between the covariant and the consistent currents is that the latter is naturally obtained from a (anomalous) quantum effective action by varying with respect to sources and is so named because it satisfies the Wess-Zumino consistency conditions. We provide a quick review of the translation between the covariant and consistent currents in Appendix \ref{sec:adcons}. A detailed account of the issues can also be found in \cite{Banerjee:2012iz,Banerjee:2012cr} and in standard references 
such as \cite{Bertlmann:1996aa,Harvey:2005it}.}

For the present discussion we simply assume that the consistent current is obtained from the covariant one by a well-defined prescription. Once this is done, the adiabaticity equation can be written directly in terms of the consistent currents, see  \eqref{eq:AdiabConsG}.
Given the consistent currents, using \eqref{eq:hstatic} in \eqref{eq:AdiabConsG} we get
\begin{equation}
\nabla_\mu\prn{ \frac{\mathcal{G}_{cons}^\mu}{T}}_\text{Hydrostatic}=0 \,, 
\end{equation}
provided we  choose to work in hydrostatic gauge defined via
$$ {\text{Hydrostatic Gauge:}} \qquad \partial_\nu \KEq^\mu = 0 , \qquad    \LambdaEq  = 0\,. $$
As in the preceding discussion this then suffices to define a generating functional, since we can consider the integral of $\frac{\mathcal{G}_{cons,E}^\mu}{T}$ over the  base space $\Sigma_E$ in analogy with \eqref{eq:wenadef}. Thus modulo a restricted gauge choice, the consistent free energy current leads to  a Euclidean partition function that is well-defined, and independent of the choice of base-space embedding, even in the presence of anomalies.

It is convenient to perform an inverse Wick-rotation of the Euclidean partition function $W_E$
so that we may use the metric with Lorentzian signature. This amounts
to
\begin{equation}
\begin{split}
W_\text{Hydrostatic} &= -\brk{\int_{\Sigma_E}
\prn{ \frac{\mathcal{G}_{cons}^\mu}{T}}\ d^{d-1}S_\mu\ }_\text{Hydrostatic} \\
\end{split}
\label{eq:Whydro}
\end{equation}
As we will see shortly, this hydrostatic partition function is a very
powerful way to characterize a large subset of adiabatic constitutive
relations.

\paragraph{A classification of hydrostatic partition functions:} The basic lesson from the above discussion is that the choice of hydrostatic partition functions is characterized by the (consistent) free energy current $\mathcal{G}^\mu_{cons}$. Being a vector in spacetime, it can naturally be decomposed into a longitudinal part along $\Kbeta^\mu$ and  transverse part, i.e., ${\mathcal G}^\mu = {\mathfrak S}\, \Kbeta^\mu + {\mathfrak V}^\mu$ with ${\mathfrak V}_\mu\, \Kbeta^\mu =0$. Accounting for the presence of anomalies as well, this prompts a further subdivision of the hydrostatic partition functions into three sub-classes. 
\begin{enumerate}
\item Class $\PS$: These are hydrostatic partition functions that lift to spacetime scalars. In this case we can extend the definition \eqref{eq:Whydro} to write the final answer as a complete spacetime integral (as opposed to a spatial integral) over ${\cal M} = \Sigma_E \times I_\KEq$ where $I_\KEq$ is an interval of unit affine length along the Killing direction
$\KEq$.
\begin{equation}
W_\text{Hydrostatic} = \int_{\Sigma_E} P_S\brk{\hfields_\Eqfields}\  \KEq^\sigma \ d^{d-1}S_\sigma= 
\int_{\Sigma_E \times I_\KEq} \, d^d x \, \sqrt{-g}\; P_S\brk{\hfields_\Eqfields} \,,
\label{eq:wps}
\end{equation}
where $\hfields_\Eqfields = \{ g_{\mu\nu}, A_\mu, \KEq^\mu, \LambdaEq\}$.
This is the canonical form in which we expect to see generating functions for correlators of currents.
\item Class $\PV$: These are terms in the hydrostatic partition function
which lift to transverse spacetime vectors as suggested by \eqref{eq:Whydro}. More explicitly, they are  allowed contributions of the form
\begin{equation}
W_\text{Hydrostatic} = \int_{\Sigma_E} \big(P_V^{\ \sigma} \brk{\hfields_\Eqfields} \big)_\text{Hydrostatic}  \ d^{d-1}S_\sigma\,,
\label{eq:wpv}
\end{equation}
with
\begin{equation}
\big(\Kbeta_\sigma P_V^{\ \sigma} \brk{\hfields_\Eqfields} \big)_\text{Hydrostatic} =\KEq_\sigma \big(P_V^{\ \sigma}\brk{\hfields_\Eqfields} 
\big)_\text{Hydrostatic}=0 \,.
\end{equation}
Note that in order for this to be independent of embedding, we will also require that the vector field $P_V^{\ \sigma}$ be conserved, $\big(\nabla_\sigma P_V^{\ \sigma}\big)_\text{Hydrostatic}=0$. We will discuss these constitutive relations further in \S\ref{sec:JLYtranscend}.

\item Class A: These are non-gauge invariant, non-diffeomorphism invariant
terms that are added to the hydrostatic partition function to reproduce flavour and gravitational anomalies.  We will explain how to obtain the anomalous constitutive relations  in an off-shell Lagrangian formalism extending the  analysis of \cite{Haehl:2013hoa} in \S\ref{sec:anomalies}.
\end{enumerate}

One of the main outcomes of demanding the existence of equilibrium on arbitrary time-independent backgrounds is that it serves to delineate a set of constraints on the transport due to the second law. Certain terms if present in the constitutive relations will give rise to sign-indefinite contributions to $\nabla_\mu J_S^\mu$ -- these have to be forbidden if we want to ensure that the divergence of the entropy current is positive definite. This leads us to an important class of terms which are the hydrostatic forbidden terms ($\PF$). To wit, 
\paragraph{Class $\PF$:} Transport coefficients in Classes $\PS$ and $\PV$ can be enumerated by listing all off-shell independent hydrostatic scalars and vectors at a given order in derivative expansion. The existence of equilibrium imposes that all other hydrostatic tensor structures that can appear in the constitutive relations be constrained to occur as linear combinations of the unconstrained ones. This consistency of the equilibrium partition function gives a number of hydrostatic relations that we call as $\PF$ (for hydrostatic forbidden).\footnote{ Empirically, terms in Class $\PF$ appear to account for about a third of the total number of transport coefficients  at a given order in the gradient expansion (beyond leading order).
See Tables \ref{tab:CountingEven}  and \ref{tab:CountingOdd} for a summary of the counting in a variety of examples.}

We will have more to say about these various categories in the course of our discussion.

\subsection{Currents from the hydrostatic partition function}
\label{sec:hscurrents}

Given a hydrostatic partition function the constitutive relations can be derived from a straightforward variational principle. 
One can effectively think of the partition function as a generating function, from which one derives the currents by 
varying with respect to the sources \cite{Banerjee:2012iz}.  One can also obtain the entropy current using the boundary 
terms of the variational calculus. We review how this can be done as described in the recent discussion of \cite{Bhattacharyya:2013lha,Bhattacharyya:2014bha}. 
We will recast this in a language that will be at once familiar, and at the same time set the stage for further discussions in \S\ref{sec:classL}.

First let us see how one can recover the hydrostatic energy momentum tensor and currents from \eqref{eq:Whydro}. Consider the variation of the
partition function under a small change of sources $\{g_{\mu\nu},A_\mu\}$. It is useful to think of this 
variation as arising  from a very slow time-dependence of the sources.  Indeed any deviation from equilibrium can be measured by the temporal changes -- we will make extensive use of the 
fact that $\diffB g_{\mu\nu}$ and $\diffB A_\mu$ will capture the linear time dependence away from equilibrium in what follows.

We want to calculate the change in $W_\text{Hydrostatic}$ between two time slices separated by an infinitesimal displacement $ \Kbeta^\alpha\ \delta t$. This can be done by using Gauss law:
\begin{equation}
\begin{split}
\delta W_\text{Hydrostatic} &= -\delta\brk{\int_{\Sigma_E}
\prn{ \frac{\mathcal{G}_{cons}^\mu}{T}}\ d^{d-1}S_\mu\ }_\text{Hydrostatic} \\
&= \delta t \brk{\int_{\Sigma_E}
\brk{ -\nabla_\sigma \prn{ \frac{\mathcal{G}_{cons}^\sigma}{T} } }\ \Kbeta^\alpha d^{d-1}S_\alpha\ }_\text{Hydrostatic}\\
&\qquad \qquad +\;\delta t \brk{\int_{\partial\Sigma_E}
\prn{ \frac{\mathcal{G}_{cons}^j}{T}}\ d^{d-2}S_j\ }_\text{Hydrostatic} \\
\end{split}
\label{eq:WhsvarPre}
\end{equation}
The bulk piece can then be simplified using  the adiabaticity equation for the consistent free energy current \eqref{eq:AdiabConsG}.
Restricting to the hydrostatic gauge we obtain with a single variation\footnote{ If we are dealing with non-anomalous systems, the subscript {\em cons} may be freely omitted.}
\begin{equation}
\begin{split}
\delta W_\text{Hydrostatic} 
&=  \brk{\int_{\Sigma_E}
\prn{ \half  \,T_{cons}^{\mu\nu}\;\delta  g_{\mu\nu} + J_{cons}^\mu \cdot \delta  A_\mu }\ \Kbeta^\alpha d^{d-1}S_\alpha\ }_\text{Hydrostatic}\\
&\qquad \qquad +\;\delta t \brk{\int_{\partial\Sigma_E}
\prn{ \frac{\mathcal{G}_{cons}^j}{T}}\ d^{d-2}S_j\ }_\text{Hydrostatic} \\
\end{split}
\label{eq:Whsvar}
\end{equation}
where $\delta  g_{\mu\nu} = \delta t\ \diffB g_{\mu\nu}$ and  $\delta   A_\mu = \delta t\ \diffB  A_\mu$. Further, we have chosen our time slices such that there is no linear time dependence
in $\{\Kbeta^\alpha,\LambdaB\}$, i.e.,  $\delta \Kbeta^\alpha = \delta t\ \diffB \Kbeta^\alpha = 0$ and  $\delta   \LambdaB = \delta t\ \diffB  \LambdaB = 0$.

Since any variation can be mimicked by a slow time-dependence, we conclude that, in general 
\begin{equation}\label{eq:WhsvarPost}
\begin{split}
\delta W_\text{Hydrostatic} 
&=  \brk{\int_{\Sigma_E}
\prn{ \half  \,T_{cons}^{\mu\nu}\;\delta  g_{\mu\nu} + J_{cons}^\mu \cdot \delta  A_\mu }\ \Kbeta^\alpha d^{d-1}S_\alpha\ }_\text{Hydrostatic}\\
&\qquad \qquad + \brk{\int_{\partial\Sigma_E}
 (\PSymplPot{})^j \ d^{d-2}S_j\ }_\text{Hydrostatic} \\
\end{split}
\end{equation}
where $(\PSymplPot{})^j$ is a boundary term linear in variations of fields, arising out of integration by parts. 

For the particular kind of slow time dependence under consideration, we can write $(\PSymplPot{})^j = \delta t (\PSymplPot{\Bfields})^j$ where $\PSymplPot{\Bfields}$
is obtained by changing all the variations $\delta(\ldots)$ into Lie-derivative $\diffB(\ldots)$. A comparison of \eqref{eq:WhsvarPost}
against \eqref{eq:Whsvar} then yields
\begin{equation}
\begin{split}
(\PSymplPot{\Bfields})^j = \frac{\mathcal{G}_{cons}^j}{T}
\end{split}
\end{equation}
Thus, when we vary the sources in the hydrostatic partition function, we get a bulk variation which
allows us to figure out the consistent currents and a boundary variation which gives
us the information about the spatial component of free energy current. Since the temporal
component of free energy current (i.e., free energy density) is already captured by the 
partition function before variation, we can then reconstruct the entire free energy current. 
By using the free energy current  thus obtained as the non-canonical part of the entropy current,
we can finally compute the  entropy current associated with the partition function.

A clear algorithmic procedure for doing this which is inspired by our Class L discussion, can be phrased as follows (cf., also  Appendix \ref{sec:sayantani}):
\begin{enumerate}
\item  From $W_\text{Hydrostatic}$ determine $\mathcal{G}_{cons}^0$. By varying it, determine the currents $\{T^{\mu\nu}, J^\mu \}$ and the boundary term gives $\mathcal{G}_{cons}^j$. When covariantized, the latter is just  the pre-symplectic potential $(\PSymplPot{})^\mu$ which arises as the surface term when varying the hydrostatic partition function. 
\item Having obtained the hydrostatic currents, one then takes them off-shell by giving them linear time dependence. To do so, one adds in  non-hydrostatic terms in $\{T^{\mu\nu}, J^\mu \}$ by unlinking $\Bfields$ from $\Eqfields$. Effectively what this amounts to is that the linear variation of the background fields, in the direction of $\{\Kbeta^\mu, \LambdaB\}$,  i.e.,  $\diffB$ defined in \eqref{eq:delBdef} plays the role of time derivative.
\item One similarly upgrades the boundary term from $(\PSymplPot{})^\mu$ to  $(\PSymplPot{\Bfields})^\mu$ to obtain the linear time dependence in the spatial component of
 free energy current.
\item In Class $\PS$ the non-canonical part of the entropy current is simply obtained by combining the temporal and the spatial components of the free energy current:
\begin{equation}
(J_S^\mu)_{non-can} =\Kbeta^\mu\, P_S\brk{\hfields} -(\PSymplPot{\Bfields})^\mu \,.
\end{equation}	
We note in passing that we can add a total derivative term $\nabla_\nu \Komar^{[\mu\nu]}$ (Komar terms) to the above whilst still retaining a conserved entropy current, cf., \S\ref{sec:Lhstatic}.
\item In Class $\PV$ the vector in the partition function can be just covariantised to give the non-canonical part of the entropy current \cite{Banerjee:2012cr}.
\item  By adding this non-canonical part to the canonical part of entropy current 
$(J^\mu_S)_{can} = - \Kbeta_\nu \,T^{\mu\nu} - \, \frac{\mu}{T}\, J^\mu$ , we thus reproduce the prescription for computing the 
hydrostatic entropy current given in \cite{Bhattacharyya:2014bha}.
\end{enumerate}

With this, we have thus given a direct, covariant and off-shell rederivation  of the entire theory of hydrostatic partition function and its associated entropy current developed in references \cite{Banerjee:2012iz,Bhattacharyya:2014bha}.  Before moving away  from hydrostatics, we remark on  two crucial ideas we have used above to considerably simplify the existing derivations. First is  the relation  between consistent free energy current  and hydrostatic partition function first proposed in \cite{Banerjee:2012cr}. The second idea is  the off-shell adiabaticity  equation \eqref{eq:Adiabaticity}
 (first introduced in \cite{Loganayagam:2011mu}) which  will continue to play a crucial role in what follows.

To summarize, given a hydrostatic partition function which is a functional of the background (time-independent) sources we can recover on-shell currents $\hcur$ which
satisfy the adiabaticity equations \eqref{eq:Adiabaticity}. This provides us with our first class of examples of non-trivial adiabatic fluids, though soon we will be enlarging our repertoire.

\section{Class D: Dissipative terms }
\label{sec:classD}

In our discussion thus far we have motivated focusing on the solutions to the adiabaticity equation \eqref{eq:Adiabaticity} which switches off dissipation, viz.,  $\Diss =0$. While we will explore a rather intricate structure of adiabatic constitutive relations in Part \ref{part:adiabatic}, one would imagine that this would be a small part hydrodynamic transport. After all, most of the phenomena we intuitively grasp in hydrodynamics have to do with dissipation. Strangely enough, this turns out not to be true. Nevertheless,  any complete classification of transport has to tackle the constraints on such dissipative terms as well. We will now argue that this is relatively straightforward and, using a key result proved in \cite{Bhattacharyya:2013lha,Bhattacharyya:2014bha}, make a case for a complete transport classification, once the adiabatic story is complete.

\subsection{Constraints on dissipative transport}
\label{sec:Dcons}

The second law of thermodynamics requires $\Diss \geq 0$, so let us recall what is known about dissipative transport. The simplest dissipative terms are the shear and bulk viscosities and charge conductivity, usually denoted as $\{\eta, \zeta, \sigma\}$ respectively and enter at first order in gradients.  Recall that at leading order in the gradient expansion, the currents are (see \cite{Rangamani:2009xk}):
\begin{align}
T^{\mu\nu} &= \epsilon\, u^\mu\, u^\nu +  p \, P^{\mu\nu} -2\,\eta\, \sigma^{\mu\nu} - \zeta\, \Theta\, P^{\mu\nu}\,,
\nonumber \\
J^\mu & = q\, u^\mu + \sigma_{_{\text{Ohm}}}\, \cv^\mu \,, \qquad  J_S^\mu = s\, u^\mu - \frac{\mu}{T} \,  \sigma_{_{\text{Ohm}}}\, \cv^\mu\,.
\end{align}	
which leads to entropy generation 
\begin{equation}
\Diss = \frac{1}{T} \, \left(2\,\eta\, \sigma_{\mu\nu} \, \sigma^{\mu\nu}  + \zeta\, \Theta^2 +   \sigma_{_{\text{Ohm}}} \, \cv^2 \right) \,.
\label{eq:visent}
\end{equation}	
This is consistent with the second law and positive definite for 
\begin{equation}
\eta \geq 0 \,, \qquad \zeta \geq 0 \,,\qquad \sigma_{_{\text{Ohm}}} \geq 0\,.
\end{equation}	
The fact that there are no adiabatic terms at first order, as well as absence of (non-canonical) corrections to the entropy current, are well known. These statements can be easily derived from the standard current algebra perspective. However, one may intuitively expect the higher order story to be much more complex, making it unclear how to proceed. This intuition turns out to be wrong.

For the moment assume that the adiabatic part of the constitutive relations has been dealt with.
First, let us use the positivity of viscosities and conductivities to define a sub-class $\Dv \subset$ D which contains {\em genuinely dissipative/viscous} transport. These are terms that are constrained to be sign-definite by the second law.  Terms which are not of this form will be called {\em sub-dissipative} ($\Ds \subset$ D); these will be allowed to take on any value without causing trouble for the second law. Much of the material that follows in this subsection was first explained in \cite{Bhattacharyya:2013lha,Bhattacharyya:2014bha}. We are mostly going to paraphrase the results first, before unveiling a more abstract proof inspired by our analysis of adiabaticity in Part \ref{part:adiabatic}.

Let us understand why the split ${\rm D} = \Dv \cup \Ds$ exists.  In the hydrodynamic gradient expansion the divergence of the entropy current captured by $\Diss$ itself admits a gradient expansion (starting at second order as ideal fluids are non-dissipative):
\begin{equation}
\Diss = \Diss_2 + \Diss_3 + \cdots
\end{equation}	
where $\Diss_2$ is the quantity given in the r.h.s. of \eqref{eq:visent} (or generalizations thereof). The higher order $\Diss_k$ arise from gradient corrections to both the entropy current and the constitutive relations. When we compute $\Diss_k$ we can assemble this into a linear combination of scalar operators with exactly $k$ derivatives ($\partial^k$) . There is a basis of such operators; in the current algebraic approach one usually works with a basis of on-shell independent scalars at a given derivative order, but this is not necessary (as we shall see).  The statement about the splitting of Class D is equivalent to the statement that the scalar operators admit such a decomposition. 

Scalars of interest at ${\cal O}(\partial^k)$ are composite operators and can either be (i) `descendants' of  operators constrained at lower orders or (ii) simple `product-composites' of lower order operators. The descendant operators are obtained by acting with derivatives on lower order tensor structures, while product-composites are simply obtained by contraction. Examples of the former are operators such as $\{ \Theta\, u^\mu \nabla_\mu \Theta, \sigma^{\mu\nu}\, u^\alpha \,\nabla_\alpha \sigma_{\mu\nu}\}$, while $\{ \Theta^3, \sigma^{\alpha \beta}\sigma_{\beta\gamma} \sigma^{\gamma}{}_\alpha, \Theta\, \sigma_{\mu\nu} \, \sigma^{\mu\nu}\}$ exemplify the latter at ${\cal O}(\partial^3)$.\footnote{ It is useful at this stage to infer from our hydrostatic discussion that the operator $\diffB$ serves the role of capturing time derivatives about equilibrium configuration. All the aforementioned operators can be written directly in terms of $\diffB g_{\mu\nu}$ (likewise occurrences of  $\cv_\mu$ can be expressed in terms of $\diffB A_\mu$).}

The product-composites are simple; since they are invariants built out of terms that are already constrained, their coefficients can be arbitrary whilst still respecting the second law (in the gradient expansion). 
For example, taking the viscosities and conductivities to be positive, we ensure $\Diss_2 \geq 0$ and thence the contribution to $\Diss_3$ from such product-composite form, is simply sub-dominant and poses no obstruction to the second law. To wit, 
\begin{align}
&2\, \eta\, \sigma_{\mu\nu} \, \sigma^{\mu\nu}  + \zeta\, \Theta^2 + \gamma_1\, \Theta^3 + \gamma_2\, \sigma^{\alpha \beta}\sigma_{\beta\gamma} \sigma^{\gamma}{}_\alpha + \gamma_3\, \Theta\, \sigma_{\mu\nu} \, \sigma^{\mu\nu}
\geq 0 
\nonumber \\
& \qquad \Longrightarrow \qquad \eta, \zeta \geq 0 \,, \text{and}\; \{\gamma_1, \gamma_2,\gamma_3 \}  \;  
\text{unconstrained}.
\end{align}	
The descendants are a-priori trickier to handle; at any given derivative order they give rise to new scalar invariants which have not been encountered at lower orders. Their contribution to $\Diss$ cannot be subsumed into lower order terms. One way to argue for their importance is to note that one can find fluid configurations where the lower order gradients are locally made to vanish, making the descendants important in some domain. Since we want the second law to hold in all possible scenarios, one must therefore control the descendants. The rather non-trivial fact is that these are also easy to handle beyond the leading order. 

Let us understand this a bit more carefully following the impressively clear and complete analysis of dissipative transport of \cite{Bhattacharyya:2014bha} (see \cite{Bhattacharyya:2012nq,Bhattacharyya:2013lha} for earlier results). As explained there, scalar operators contributing to $\Diss$ are of three types:\footnote{ In \cite{Bhattacharyya:2014bha} a fourth type was introduced called $\Delta_{non-diss}$ -- these will be accounted for in our adiabatic story  as they end up having net zero contribution to entropy production. Note that we will not include them explicitly in our counting of Class C constitutive relations as these terms are exact differentials and thus trivial in cohomology.}   
\begin{itemize}
\item Terms that contribute to $\Diss_2$ at leading order which need to be controlled to ensure $\Diss \geq 0$. They belong to $\Dv$ and impose constraints on transport (such terms were called $\Delta_{2nd-order}$ in \cite{Bhattacharyya:2014bha}). Note that we can write them effectively in terms of $(\diffB g )^2$ and $(\diffB A)^2$ for they appear only at quadratic order.
\item Descendant terms at any given order which are composite scalars built from a $(k-1)^{\rm st}$ order independent operator and a first order operator, i.e., of the form $\diffB g\, D{\cal O}_{k-2}$ where ${\cal O}_{k-2}$ could be a composite-product. Such terms were denoted as 
$\Delta_{diss-imp}$ in \cite{Bhattacharyya:2014bha}.
\item Composite-product terms which simply take the form $(\diffB g)^k$ and $(\diffB A)^k$. Terms of this type were called $\Delta_{diss-product}$ in \cite{Bhattacharyya:2014bha}. 
\end{itemize}

Given this decomposition, we have schematically 
\begin{align}
\Diss  &= \alpha_{2g}\, (\diffB g)^2 +  \alpha_{2A}\, (\diffB A)^2 + 
\sum_{k=3}^\infty\, \left[\kappa_{kg} \; \diffB g\; D{\cal O}_{k-2} + \gamma_{kg}\, (\diffB g)^k + \cdots \right]
\nonumber \\
&\sim \alpha_{2g}\, \left[ \diffB g + \sum_{k=3}^\infty\, \frac{\kappa_{kg}}{2\, \alpha_{2g}} \;  
D{\cal O}_{k-2} \right]^2 + 
\sum_{k=3}^\infty\, \gamma_{kg}\, (\diffB g)^k + \cdots 
\label{eq:Dscheme}
\end{align}
where we have only written out the composite higher order terms explicitly for the metric variation and elided over writing the gauge variations. In the second line we have indicated a merger of the descendant and leading order terms into a quadratic form which plays a role in the argument below. With this parameterization 
\begin{itemize}
\item $\alpha_{2g}, \alpha_{2A} \geq 0$ for the second law to hold. 
\item $\gamma_k$ are unconstrained since they multiply terms which are parametrically smaller than the leading order contributions. These are clearly in the sub-dissipative Class $\Ds$.
\item By completing squares, one can take care of the descendant terms as well (they are effectively in Class $\Ds$ despite appearances). Indeed, as written, positivity of the second line of \eqref{eq:Dscheme} is ensured once we demand $\alpha_{2g}, \alpha_{2A}$ etc., to be positive definite. One further needs to ensure that the cross-terms obtained in the process can be assembled into positive-definite quadratic form.
\end{itemize}

The remarkable statement of \cite{Bhattacharyya:2014bha} is that this can always be done, recursively order by order in the gradient expansion! More specifically, the aforementioned reference used the specific example of (parity-even) charged fluid at third order in gradients to illustrate the general picture proposed in \cite{Bhattacharyya:2013lha}. But the general structure emerging from that analysis makes it clear that the construction can be extended to higher orders. The argument in \cite{Bhattacharyya:2014bha} can be worded as follows: given a hydrostatic entropy current, there exists a hydrodynamic correction $J^\mu_{extra}$ which serves to absorb the contributions from the descendant contribution $\Delta_{diss-imp}$ to convert them into a positive-definite quadratic form. Inspired by our construction of solutions to the adiabaticity equation we give a more abstract discussion below in \S\ref{sec:Ddiffops} using a class of tensor-valued differential operators that subsume the above statements efficiently.

The upshot of this discussion is that Class D = $\Dv \cup \Ds$ with $\Dv$ contributions which are sensitive to the second law making their appearance only at the leading order in the gradient expansion. So while there are many potential contributions to Class D, most of them are agnostic to the constraints and thus can be treated democratically.

\subsection{Differential operators for dissipation}
\label{sec:Ddiffops}

We will now give a succinct summary of the results of \cite{Bhattacharyya:2014bha, Bhattacharyya:2013lha}, using a compact notation that is inspired by our construction of adiabatic constitutive relations. The impatient reader may choose to skip this section at first reading. 

We begin our discussion by constructing a set of tensor structures that provide Class D constitutive relations. We will see in the sequel that these will be quite useful in demarcating the second law constraints quite effectively for they provide us with a simple way to assemble both the product-composite and descendant operators into a positive definite form.

Let us begin with the non-anomalous grand canonical adiabaticity equation \eqref{eq:AdiabaticityG}
\begin{equation}
\begin{split}
-\nabla_\sigma\prn{\frac{\mathcal{G}^\sigma}{T}}&=
\half  T^{\mu\nu}\diffB  g_{\mu\nu} + J^\mu \cdot \diffB  A_\mu + \Diss
\end{split}
\label{eq:AGdiss}
\end{equation}
where we have included $\Diss$ denoting entropy production and dissipation. Second law of thermodynamics demands
that $\Diss \geq 0$.  We will now arrange our conserved currents in an appropriate form where this condition
can be imposed readily.

Consider  a differential operator $\DVisc$ which is tensor valued, i.e., it is a map from the space of tensors to themselves involving some derivations. For the moment we will be rather abstract about the precise form of this operator, but we have here in mind operators constructed from $\hfields$ and quantities such as $\diffB g_{\mu\nu}$ and $\diffB A_\mu$. Since in the course of manipulations we will have to reshuffle derivatives around, we also denote the corresponding adjoint differential operator obtained via  integration by parts  as $\DVisc^\dag$. 

Let us examine the following constitutive relations, which we suggestively refer to as Class D constitutive relations:
\begin{equation}\label{eq:TJClassV}
\begin{split}
(T^{\mu\nu})_{\text{D}} &\equiv \; 
	  -\half \; \brk{\DVisc_{\cdg_g}^\dag\ \cdg\  \DVisc_{\cdg_g}
	  + \DVisc_{\cdA_g}^\dag\ \cdA\  \DVisc_{\cdA_g}}^{(\mu\nu)(\alpha\beta)}
	  \, \diffB  g_{\alpha\beta} 
\\&	  \qquad \qquad 
	   -\; \brk{\DVisc_{\cdg_g}^\dag\ \cdg\  \DVisc_{\cdg_A}
	  + \DVisc_{\cdA_g}^\dag\ \cdA\  \DVisc_{\cdA_A}}^{(\mu \nu) \alpha} \cdot 
	  \diffB  A_\alpha
\\
(J^\alpha)_{\text{D}} &\equiv \; 
	-\half \; \brk{\DVisc_{\cdg_A}^\dag\ \cdg\  \DVisc_{\cdg_g}
	  + \DVisc_{\cdA_A}^\dag\ \cdA\  \DVisc_{\cdA_g}}^{\alpha(\mu \nu)}
	  \diffB  g_{\mu\nu} 	  
\\ & 	\qquad \qquad	  
	  -\; \brk{\DVisc_{\cdg_A}^\dag\ \cdg\  \DVisc_{\cdg_A}
	  + \DVisc_{\cdA_A}^\dag\ \cdA\  \DVisc_{\cdA_A}}^{\alpha\beta} \cdot \,
	  \diffB  A_\beta \,.
\end{split}
\end{equation}
By construction these currents vanish in hydrostatics, so they potentially contribute to $\Diss$. Furthermore, by exploiting the structure of \eqref{eq:TJClassV} we have ensured that the r.h.s.\ of \eqref{eq:AGdiss} is at least quadratic in $\diffB g$ and $\diffB A$ as expected.

We need to now specify the representation content of the operators $\DVisc$ that appear above. We denote the vector representation as $\text{Vect}$ and the two-index symmetric tensor representation by $\text{Sym}_2$. Clearly, 
$A_\mu \in  \text{Vect}$ and $g_{\mu\nu} \in \text{Sym}_2$, where we shall use the isomorphism provided by metric on ${\cal M}$ to raise and lower indices when necessary.  With this understanding
\begin{itemize}
\item  $\{\DVisc_{\cdg_g} ,  \DVisc_{\cdA_g}\}$ are tensor-valued differential operators which take two-indexed symmetric
tensor fields ($\text{Sym}_2$) to diffeomorphism representations denoted by $\text{Tens}_\cdg$ and $\text{Tens}_\cdA$ respectively, where the latter are those which are direct sums of tensor representations, viz.,
\begin{align}
\DVisc_{\cdg_g} : \text{Sym}_2 \rightarrow \text{Tens}_\cdg \ ,\qquad
\DVisc_{\cdA_g} : \text{Sym}_2 \rightarrow \text{Tens}_\cdA   
\label{eq:dvisc1}
\end{align}
\item Similarly, $\{\DVisc_{\cdg_A} ,  \DVisc_{\cdA_A}\}$ are tensor-valued differential operators which take vector fields
 ($\text{Vect})$ to  diffeomorphism representations  $\text{Tens}_\cdg$ and $\text{Tens}_\cdA$ respectively.
\begin{align}
\DVisc_{\cdg_A} : \text{Vect} \rightarrow \text{Tens}_\cdg \ ,\qquad
\DVisc_{\cdA_A} : \text{Vect} \rightarrow \text{Tens}_\cdA   
\label{eq:vdisc2}
\end{align}
\item  The adjoints of these differential operators act in the opposite direction, mapping direct sums of tensor representations back to  $\text{Sym}_2$ and $\text{Vect}$ representations respectively. 
 \begin{equation}
\begin{split}
 \DVisc_{\cdg_g}^\dag &:  \text{Tens}_\cdg  \rightarrow \text{Sym}_2 \ ,\qquad
\DVisc_{\cdA_g}^\dag : \text{Tens}_\cdA  \rightarrow \text{Sym}_2  \ ,\\
 \DVisc_{\cdg_A}^\dag &:  \text{Tens}_\cdg \rightarrow  \text{Vect}\ ,\qquad
\DVisc_{\cdA_A}^\dag : \text{Tens}_\cdA  \rightarrow \text{Vect}
\end{split}
\end{equation}
\item Finally, the symbols  $\{\cdg,  \cdA \}$ denote arbitrary tensor fields in the product representations
\[ \{ \text{Tens}_\cdg \otimes \text{Tens}_\cdg, 
\text{Tens}_\cdA \otimes \text{Tens}_\cdA \} \] 
which can be thought of as intertwiners to ensure that the net operator acting on either $\text{Sym}_2$ or $\text{Vect}$ is in the representations indicated by the index structure in \eqref{eq:TJClassV}.
\end{itemize}

It is useful to note that  we can write the Class D constitutive relations also in a matrix form
\begin{equation}\label{eq:TJClassVMat}
\begin{split}
\prn{ \begin{array}{c} T^{\mu\nu} \\ J^\alpha \end{array} }_{\text{D}}
&=
- \prn{ \begin{array}{cc} \DVisc_{\cdg_g}^\dag & \DVisc_{\cdA_g}^\dag \\ \DVisc_{\cdg_A}^\dag & \DVisc_{\cdA_A}^\dag \end{array} }
\prn{ \begin{array}{rr} \cdg\ & 0 \\  0\ & \cdA \end{array} }
\prn{ \begin{array}{cc} \DVisc_{\cdg_g} & \DVisc_{\cdg_A} \\ \DVisc_{\cdA_g} & \DVisc_{\cdA_A} \end{array} }
\prn{ \begin{array}{c} \half \diffB  g  \\ \diffB  A \end{array} }
\end{split}
\end{equation}
which better exhibits the algebraic structure of this class of constitutive relations.

Substituting the above expression back into the right hand side of grand canonical adiabaticity equation and
removing  the adjoint operators via integration by parts, we obtain
\begin{equation}\label{eq:dissIntPart}
\begin{split}
 \half  (T^{\mu\nu})_{\text{D}}\,\diffB  g_{\mu\nu} &+ (J^\alpha)_{\text{D}} \cdot \diffB  A_\alpha
\\
&= -\half \diffB  g_{\mu\nu} \brk{\DVisc_{\cdg_g}^\dag\ \cdg\  \DVisc_{\cdg_g}
	  + \DVisc_{\cdA_g}^\dag\ \cdA\  \DVisc_{\cdA_g}}^{(\mu\nu)(\alpha\beta)}
	  \, \half \diffB  g_{\alpha\beta}  
	  \\
&\qquad 	
	  -\half \diffB  g_{\mu\nu} \brk{\DVisc_{\cdg_g}^\dag\ \cdg\  \DVisc_{\cdg_A}
	  + \DVisc_{\cdA_g}^\dag\ \cdA\  \DVisc_{\cdA_A}}^{(\mu \nu) \alpha} \cdot \diffB  A_\alpha
\\
&\qquad
	-\half \diffB  A_\alpha \cdot \brk{\DVisc_{\cdg_A}^\dag\ \cdg\  \DVisc_{\cdg_g}
	  + \DVisc_{\cdA_A}^\dag\ \cdA\  \DVisc_{\cdA_g}}^{\alpha(\mu \nu)}
	  \diffB  g_{\mu\nu} \\
&\qquad 	
	  - \diffB  A_\alpha \cdot \brk{\DVisc_{\cdg_A}^\dag\ \cdg\  \DVisc_{\cdg_A}
	  + \DVisc_{\cdA_A}^\dag\ \cdA\  \DVisc_{\cdA_A}}^{\alpha\beta} \cdot \,\diffB  A_\beta \\
&= -\brk{ \half \DVisc_{\cdg_g} \diffB  g +\DVisc_{\cdg_A} \diffB  A } \cdg \brk{ \half \DVisc_{\cdg_g} \diffB  g +\DVisc_{\cdg_A} \diffB  A } \\
&\qquad -\brk{ \half \DVisc_{\cdA_g} \diffB  g +\DVisc_{\cdA_A} \diffB  A } \cdA \brk{ \half \DVisc_{\cdA_g} \diffB  g +\DVisc_{\cdA_A} \diffB  A }
 + \nabla_\alpha (\N^\alpha)_{\text{D}}
\end{split}
\end{equation}
where we have kept the total derivative term from integration by parts.

Let us  take  $(\mathcal{G}^\alpha)_{\text{D}} = - T (\N^\alpha)_{\text{D}} $, which in the micro-canonical ensemble  is equivalent to taking the entropy current to be
\begin{equation}\label{eq:JSClassV}
\begin{split}
(J_S^\alpha)_{\text{D}} &\equiv  - \frac{u_\beta}{T}    (T^{\alpha\beta})_{\text{D}} - \frac{\mu}{T} \cdot    (J^\alpha)_{\text{D}} +  (\N^\alpha)_{\text{D}} \,.
\end{split}
\end{equation}

With this choice we have a simple expression for the entropy production within this class of constitutive relations (using \eqref{eq:AGdiss}):
\begin{equation}
\begin{split}
\Diss &= \brk{ \half \DVisc_{\cdg_g} \diffB  g +\DVisc_{\cdg_A} \diffB  A } \cdg \brk{ \half \DVisc_{\cdg_g} \diffB  g +\DVisc_{\cdg_A} \diffB  A } \\
&\qquad +\brk{ \half \DVisc_{\cdA_g} \diffB  g +\DVisc_{\cdA_A} \diffB  A } \cdA \brk{ \half \DVisc_{\cdA_g} \diffB  g +\DVisc_{\cdA_A} \diffB  A } \\
\end{split}
\label{eq:dissUp}
\end{equation}
which is completely parameterized by the differential operators $\DVisc$ and the intertwiners between different representations  $\{\cdg, \cdA\}$. We want to ensure that $\Diss \geq 0$ for the second law, which is easy to insist for \eqref{eq:dissUp} has the structure of a quadratic form with the intertwiners playing the role of the metric. Then demanding that  $\{\cdg,\cdA\}$ transform in appropriately symmetric representations to provide a positive definite quadratic form, viz.,
\begin{equation}\label{eq:etasymm}
\begin{split}
\cdg \in \text{Sym}_+\left( \text{Tens}_\cdg \otimes \text{Tens}_\cdg \right),  \qquad
\cdA  \in \text{Sym}_+\left( \text{Tens}_\cdA \otimes \text{Tens}_\cdA \right)
\end{split}
\end{equation}
with the subscript $+$ denoting that the eigenvalues are non-negative definite. This gives us a solution to $\Diss \geq 0$.

Should we consider intertwiners not transforming in the symmetric tensor product representation we will find that they would correspond to adiabatic or hydrostatic forbidden constitutive relations. For instance taking the anti-symmetric representation will lead to Class B adiabatic constitutive relations as we shall discuss in \S\ref{sec:classB}.

To summarize the above construction, by suitably picking tensor structures we are in a position to engineer constitutive relations that are guaranteed to satisfy the second law of thermodynamics.

\subsection{Examples: Low order Class D differential operators}
\label{sec:classDexamples}
 
 Let us now examine the results of our earlier discussion in \S\ref{sec:Dcons} in light of this renewed understanding. The easiest way to proceed is to start with the leading order in gradients. Since 
\eqref{eq:dissUp} already has two factors of $\diffB$ on the r.h.s., a contribution to $\Diss_2$ necessarily requires $\DVisc$ to be a tensor operator involving no derivatives. 
For instance, picking
\begin{align}
\DVisc_{\cdg_g} = \DVisc_{\cdA_A}  =\text{Id}\,,\qquad 
\DVisc_{\cdg_A} = \DVisc_{\cdA_g}  = 0 \,,  \label{eq:TrivUpsilon}
\end{align}
and  $\cdg \in \text{Sym}_2 \otimes \text{Sym}_2$ and 
$\cdA \in \text{Vect} \otimes \text{Vect}$ to be also zero derivative tensors built from 
the background metric $g_{\mu\nu}$ and the hydrodynamic field $\Kbeta^\mu$ we can recover \eqref{eq:visent}.  To be specific, \eqref{eq:visent} is reproduced by taking \eqref{eq:TrivUpsilon} combined with
\begin{equation}\label{eq:Intertwine0}
\begin{split}
 \cdg^{\mu\nu\rho\sigma}_{_{(0)}} = T\, \zeta \, P^{\mu\nu}P^{\rho\sigma} + 2\,T\,\eta \,P^{\rho<\mu} P^{\nu>\sigma}\,,\qquad
 \cdA^{\alpha\beta}_{_{(0)}} = T\, \sigma_{_{\text{Ohm}}} \, P^{\alpha\beta} \,,
\end{split}
\end{equation}
where the subscript $(0)$ means that these intertwiners are the contributions at lowest order in a derivative expansion (the corresponding constitutive relations are first order, of course) and $<\cdot>$ denotes the symmetric transverse traceless projection.\footnote{ Explicitly, the projector is given by $X^{<\alpha\beta>} = \left( \frac{1}{2} \left(P^\alpha_\mu P^\beta_\nu + P^\beta_\mu P^\alpha_\nu\right) - \frac{1}{d-1} P^{\alpha\beta} P_{\mu\nu}\right)X^{\mu\nu}$.}

At higher order in derivative expansion an interesting subset of Class D constitutive relations can be obtained by restricting to $\Upsilon$-operators which do not act as derivatives. For instance, in the case of second order constitutive relations, we can consider arbitrary first order transverse tensors $\BerryGA^{(\mu\nu)\alpha}$ and the following correction to \eqref{eq:TrivUpsilon}:
\begin{equation}\label{eq:1stUpsilon}
\begin{split}
 \DVisc_{\cdg_g} &= \DVisc_{\cdA_A}  =\text{Id}\,,\qquad 
 \DVisc_{\cdA_g}  = 0 \,,   \\
 (\DVisc_{\cdg_A})_{\rho\sigma}{}^\alpha &= -\frac{1}{T\zeta}\, \frac{1}{(d-1)^2} \, P_{\rho\sigma} \, {\BerryGA}_\lambda{}^{\lambda\alpha} - \frac{1}{2\,T\eta} \, {\BerryGA}_{<\rho\sigma>}{}^\alpha \,,
 \end{split}
\end{equation}
together with the following first order correction to the intertwiners \eqref{eq:Intertwine0}:
\begin{equation}\label{eq:classDberrySimple}
\begin{split}
 \cdg^{\mu\nu\rho\sigma}_{_{(1)}} = \half \left(\BerryG^{(\mu\nu)(\rho\sigma)} +\BerryG^{(\rho\sigma)(\mu\nu)}\right) \,,\qquad
 \cdA^{\alpha\beta}_{_{(1)}} = \BerryA^{(\alpha\beta)} \,,
\end{split}
\end{equation}
where $\{\BerryG^{\mu\nu\alpha\beta},\BerryA^{\alpha\beta}\}$ are arbitrary tensors involving one derivative. From \eqref{eq:1stUpsilon} and intertwiners $\{(\cdg_{_{(0)}}+\cdg_{_{(1)}})^{\mu\nu\rho\sigma}\,,\, (\cdA_{_{(0)}}+\cdA_{_{(1)}})^{\alpha\beta}\}$ we get a large set of Class D constitutive relations at second order:
\begin{align}\label{eq:TJDissSimple}
(T^{\mu\nu})_\text{D} &\equiv \prn{-2\, \eta\,\sigma^{\mu\nu} - \zeta\,\Theta\,P^{\mu\nu} }
	 -\quarter \prn{ \BerryG^{(\mu\nu)(\alpha\beta)}+\BerryG^{(\alpha\beta)(\mu\nu)} }
	 \,  \diffB  g_{\alpha\beta} + \BerryGA^{(\mu \nu) \alpha} \cdot \diffB  A_\alpha + {\cal O}(\partial^3)\,,
\notag \\
(J^\alpha)_\text{D} &\equiv \prn{ \sigma_{_{\text{Ohm}} }\, \cv^\alpha  }+
	\half \BerryGA^{(\mu \nu) \alpha}  \diffB  g_{\mu\nu}
	- \BerryA^{(\alpha\beta)} \cdot\diffB  A_\beta + {\cal O}(\partial^3)\,,
\notag \\
({\cal G}^\sigma)_\text{D} & = 0 \,,
\end{align}
As we will see in \S\ref{sec:counting}, classifying first order transverse tensors $\{\BerryG^{\mu\nu\alpha\beta},\BerryGA^{\mu\nu\alpha},\BerryA^{\alpha\beta}\}$ captures all Class D constitutive relations of the parity-even charged fluids at second order via \eqref{eq:TJDissSimple}. 

At any given order in derivatives we can construct further constitutive relations by considering non-trivial $\Upsilon$-tensors in such a way that the entropy production \eqref{eq:dissUp} is manifestly a quadratic form. For example, a simple subset of Class D constitutive relations at any order is obtained by the parameterization \eqref{eq:TJDissSimple} with arbitrary tensor structures $\{\BerryG^{\mu\nu\rho\sigma},\BerryA^{\alpha\beta}\}$ for the entropy production of these terms automatically assembles itself in a quadratic form. Note that the contribution of the object $\BerryGA^{\mu\nu\alpha}$ is more complicated and it only takes the form of \eqref{eq:TJDissSimple} at second order. At higher orders its mixing with other terms in \eqref{eq:TJDissSimple} must be taken into account. With non-trivial choice of $\DVisc$ we can easily obtain the descendant terms while composite-products are obtained by making judicious choices for the intertwiners 
$\{\cdg,\cdA\}$  themselves. We discuss more examples where Class D transport shows up in \S\ref{sec:counting}.

Before we conclude our discussion of dissipative terms, 
it is worth highlighting the following point. In most of the preceding analyses of hydrodynamics (such as the discussion of  second order Weyl invariant neutral fluids in 
\cite{Bhattacharyya:2008xc,Bhattacharyya:2008mz}, generic neutral fluids in  \cite{Bhattacharyya:2012nq}, or charged fluids in  \cite{Bhattacharyya:2014bha})  the modus operandi has always been to write down the constitutive relations for the conserved currents at a given order in the gradient expansion. One then ascertains the form of the entropy current which ensures that  the second law of thermodynamics is upheld. Usually the entropy currents are determined modulo certain ambiguities (see below). However, to ensure that the entropy production is sign-definite $\Diss \geq 0$, say by completing the contributions into a positive definite quadratic form, one needs to understand contributions to the conserved currents at higher orders as well. This is most clearly illustrated in the neutral fluid analysis of \cite{Bhattacharyya:2012nq}.  One of the advantages of working with the operators $\DVisc$   and the intertwiners $\{\cdg, \cdA\}$ is that we can in one fell swoop ascertain the combination of transport that ensures $\Diss \geq 0$. This makes the analysis at higher orders much more straightforward.

We note that when we shift the entropy current by
\begin{equation}
\begin{split}
J_S^\mu \mapsto J_S^\mu + \frac{1}{\sqrt{-g}}\lieD_V\brk{\sqrt{-g} J_S^\mu} =  J_S^\mu + V^\mu D_\sigma J_S^\sigma + D_\nu \prn{ V^\nu J_S^\mu  - J_S^\nu V^\mu} 
\end{split}
\end{equation}
the amount of entropy produced $\Diss$ shifts by a Lie derivative $\Diss \mapsto \Diss+ \frac{1}{\sqrt{-g}}
\lieD_V \brk{\sqrt{-g} \, \Diss}$ which preserves the condition $\Diss \geq 0$. Thus such a shift moves us within the space of admissible Class D constitutive relations. In the holographic context this is reflected in the pullback ambiguity in the construction of the entropy current   \cite{Bhattacharyya:2008xc}.

The explicit construction of Class D constitutive relations in \S\ref{sec:Ddiffops} establishes the main statement we made at the end of \S\ref{sec:Dcons}; all of the product-composite and descendant scalar operators can be assembled into a form where the only operative restriction from the second law of thermodynamics applies at $\Diss_2$ order. Thus as claimed in \cite{Bhattacharyya:2013lha,Bhattacharyya:2014bha} the only inequality constraints of interest arising from the second law operate at the leading order in the gradient expansion. All higher order terms can be subsumed via the derivative operators into the sub-dissipative Class $\Ds$. 

The reader interested solely in the constraints arising from the second law can stop here. As described in \cite{Bhattacharyya:2013lha} once we figured out the terms in Classes $\PF$ and $\Dv$, one is done. There is no more information from the second law. However, those intrepid souls willing to brave the winding pathways of the eightfold way are encouraged to continue onto Part \ref{part:adiabatic}. 

\newpage
\part{The Classification of Adiabatic Constitutive Relations}
\label{part:adiabatic}
\hspace{1cm}

\section{Class L: Lagrangian solutions to adiabaticity equation}
\label{sec:classL}

Following the analysis of \S\ref{sec:hydrostatics} we have seen that the set of adiabatic fluids is much larger than just the ideal fluid family.  In principle, we could continue on from our analysis at zero derivative level as in \S\ref{sec:ideal}, and solve the adiabaticity equation at  higher derivative orders, to find new families of adiabatic constitutive relations.
In practice, however, the number of terms  proliferates very fast and the analysis becomes complicated. Hence, one therefore would like to seek more practical ways of solving adiabaticity equation or writing down adiabatic constitutive relations. The most elegant solution would be to mimic our discussion in hydrostatics 
(Class H) and construct the generating function for the adiabatic constitutive relations, consistent with our desire of being off-shell and off-equilibrium. 

In this section, we will describe a method to generate a large class of adiabatic constitutive
relations in the absence of anomalies. Though this does not give all possible solutions, at any  given derivative  order, many solutions seem to fall into this class. We will call this class of adiabatic  constitutive relations as Class L (where L stands for Lagrangian-derivable) as one can find a local Lagrangian or Landau-Ginzburg free energy functional which succinctly encodes the constitutive relations. As presaged this will be quite close to the  Euclidean partition function in hydrostatics. In particular, observe that in Class $\PS$, the generating functional was given by the longitudinal part of the free energy current. That is, the natural decomposition of any covariant free energy current,
\begin{align} \label{eq:Gdecomp}
\mathcal{G}^\mu = {\mathfrak S} \,\Kbeta^\mu + {\mathfrak V} ^\mu \,,\qquad {\mathfrak V}^\mu \,\Kbeta_\mu = 0 \,,
\end{align}
gives a natural scalar object ${\mathfrak S}$ which in hydrostatics took the role of the partition function scalar density $P_S\brk{\hfields_\Eqfields}$ in \eqref{eq:wps}. If we consider the full set of scalar invariants (up to field redefinitions), including ones that vanish in equilibrium, then we can  write down an off-shell Lagrangian density $\Lag \sim f_{\mathfrak S}\; {\mathfrak S}$ which parameterizes Class L. The non-hydrostatic part of this construction comprises of those scalars which  identically vanish in equilibrium. All in all such  Lagrangian densities will completely parameterize the longitudinal part of the free energy current in  
\eqref{eq:Gdecomp}. 

Let us now carry out  this construction in detail. 
Constitutive relations in Class L are parametrized by a Lagrangian density
$\Lag\brk{g_{\mu\nu},A_\mu,\Kbeta^\mu,\LambdaB}$ which we will assume to be a
local scalar functional of its arguments, i.e., under gauge transformations and diffeomorphisms
$\Lag$ transforms like a scalar field. Intuitively, $\Lag$ can be thought
of as some sort of a generalized pressure functional for the adiabatic fluid.\footnote{ We will later see that upon restricting to hydrostatic configurations, $\Lag$ reduces to the hydrostatic partition function $W_\text{Hydrostatic}$ which suggests this intuition.}  We may write
\begin{equation}
S_\text{hydro} = \int d^dx \, \sqrt{-g}\ \Lag\brk{\hfields} \,.
\end{equation}

Consider now a variation of this Lagrangian functional which, after sufficient number
of integration by parts, can be brought to the form
\begin{equation}\label{eq:LagVar}
\begin{split}
\frac{1}{\sqrt{-g}}&\delta\prn{\sqrt{-g}\ \Lag} -\nabla_\mu (\PSymplPot{})^\mu\\
&= \half \; T^{\mu\nu}\;\delta g_{\mu\nu} + J^\mu \cdot \delta A_\mu +
T \,\aheat_\sigma \;\delta \Kbeta^\sigma
+ T\, \acharge \cdot \prn{\delta\LambdaB+ A_\sigma \,\delta \Kbeta^\sigma}\,.
\end{split}
\end{equation}
Here $(\PSymplPot{})^\mu$ denotes the surface terms generated due to
integration by parts and is related to the pre-symplectic potential. The symbol $\slashed{\delta}$ denotes that it is
linear in variations of fields. The reader may simply take \eqref{eq:LagVar} as the defining statement of the variational principle.

So far $\aheat_\sigma$ and $\acharge$ which multiply variations of the hydrodynamic fields (and are thus conjugate to them)
are simply defined by the above variational principle; they will have a role to play in the sequel. 
We will refer to them as the {\em adiabatic heat current} and {\em adiabatic charge density} respectively.

The variation of the Lagrangian makes it easy to obtain the currents $\hcur$. For instance we read off $\{T^{\mu\nu},J^\mu\}$ from the above variation and take $J_S^\mu = s \,u^\mu$ with
\begin{equation}
\begin{split}
s  \equiv
\prn{\frac{1}{\sqrt{-g}}\; \frac{\delta}{\delta T}\; \int \sqrt{-g}\; \Lag\brk{\hfields} \; }\bigg|_{\{u^\sigma, \,\mu,
\,g_{\alpha\beta},\,A_\alpha\} = \text{fixed}}
\end{split}
\label{eq:sdef}
\end{equation}
Here $\frac{\delta}{\delta T}$ is the variational (i.e., Euler-Lagrange) derivative. The free energy current can be obtained using  \eqref{eq:GDef}. It is convenient to rewrite this expression in terms of the adiabatic currents $\{\aheat_\sigma,\acharge\}$ for simplification of  future computations. Note that
\begin{equation}\label{KLambdaK_uTmu}
\begin{split}
T \,\aheat_\sigma \, \delta \Kbeta^\sigma
+ T \, \acharge \cdot \prn{\delta\LambdaB+ A_\sigma \delta \Kbeta^\sigma}
& =  (\aheat_\sigma+\acharge \cdot A_\sigma) \, \delta u^\sigma
+\acharge  \cdot \delta\prn{\mu-u^\sigma A_\sigma } \\
& \qquad  -
\brk{\aheat_\sigma \,\Kbeta^\sigma  + \acharge  \cdot  (\LambdaB+A_\sigma \Kbeta^\sigma ) }\ \delta T
\end{split}
\end{equation}
which in turn implies that
\begin{equation}\label{eq:sVZeta}
\begin{split}
s &=- \brk{\aheat_\sigma \, \Kbeta^\sigma  + \acharge  \cdot  (\LambdaB+A_\sigma \Kbeta^\sigma ) }
= - \frac{1}{T}\brk{\aheat_\sigma u^\sigma  + \acharge  \cdot  \mu } \\
& \qquad \quad\Longrightarrow\quad
T\, s+\mu  \cdot \acharge  +u^\sigma \,\aheat_\sigma  =0\,.
\end{split}
\end{equation}
In the above and in  what follows, we will often want to convert general variations of hydrodynamic fields $\{u^\sigma, T, \mu\} $ in terms of variations of $\{\Kbeta^\mu,\LambdaB\}$ and vice versa. This can readily be done by  using the defining equation \eqref{eq:hydrofields} and explicit expressions can be found in \eqref{eq:varrules} for convenience. 
 In sum from $\Lag$  we have access to both the physical and adiabatic currents respectively, with the latter determining the entropy current.

\subsection{Bianchi identities in Class L}
\label{sec:LBianchi}

The invariance of $S_\text{hydro}$ under gauge/diffeomorphisms implies certain
identities obeyed by $\{T_{\mu\nu},J_\mu,\aheat_\sigma,\acharge\}$.\footnote{ The material in this  subsection was worked out in collaboration with Kristan Jensen.}  Interpreting these identities in a particular manner will prove conducive to showing that having a Lagrangian
$\Lag\brk{\hfields}  =  \Lag\left[g_{\mu\nu}, A_\mu, \Kbeta^\mu, \LambdaB\right]$ leads to currents which solve the adiabaticity equation.

To see this, consider the diffeomorphism and gauge variations induced by a pair of arbitrary vector field and scalar collectively denoted as $\Xfields \equiv \{\xi^\mu,\Lambda\}$ on the basic hydrodynamic fields:
\begin{equation}
\begin{split}
\diffF  g_{\mu\nu} &\equiv \lieD_\xi g_{\mu\nu} = \nabla_\mu \xi_\nu + \nabla_\nu \xi_\mu \,,\\
\diffF  A_\mu &\equiv \lieD_\xi A_\mu + [A_\mu,\Lambda]+\partial_\mu \Lambda
=D_\mu\brk{\Lambda+\xi^\nu A_\nu}+\xi^\nu F_{\nu\mu} \,, \\
\diffF  \Kbeta^\mu &\equiv \lieD_\xi \Kbeta^\mu
=\xi^\nu \nabla_\nu \Kbeta^\mu  - \Kbeta^\nu \nabla_\nu \xi^\mu \,,\\
\diffF  \LambdaB + A_\nu \diffF  \Kbeta^\nu
&\equiv \xi^\mu D_\mu\brk{\LambdaB+\Kbeta^\nu A_\nu}-\Kbeta^\mu D_\mu\brk{\Lambda+\xi^\nu A_\nu}  \\
&\qquad\qquad -\xi^\mu \Kbeta^\nu F_{\mu\nu} + [\LambdaB+\Kbeta^\nu A_\nu,\Lambda+\xi^\lambda A_\lambda] \,,
\end{split}
\label{eq:GenericDiffeo}
\end{equation}
where the symbol $\lieD_\xi$ denotes the Lie derivative along the vector field $\xi^\mu$.

Plugging this  variation into the expression appearing on the r.h.s. of \eqref{eq:LagVar}  followed by a  straightforward integration by parts gives
\begin{equation}\label{eq:NoetherHydroPre}
\begin{split}
\half & T^{\mu\nu}\diffF g_{\mu\nu} + J^\mu \cdot \diffF A_\mu +T \, \aheat_\mu  \diffF \Kbeta^\mu
+ T\,\acharge \cdot \prn{\diffF\LambdaB+ A_\mu \diffF \Kbeta^\mu}\\
&= \nabla_\mu \N^\mu[\Xfields]
+ \xi_\mu \brk{-\nabla_\nu T^{\mu\nu}+ J_\nu \cdot F^{\mu\nu}
+\frac{g^{\mu\nu}}{\sqrt{-g}}\diffB \prn{\sqrt{-g}\ T\, \aheat_\nu}
+  g^{\mu\nu} T\, \acharge \cdot \diffB  A_\nu }\\
&\qquad \qquad \qquad+ (\Lambda+\xi^\lambda A_\lambda) \cdot \brk{-D_\nu J^\nu
+\frac{1}{\sqrt{-g}}\diffB \prn{\sqrt{-g}\ T\,\acharge} \ }\,,
\end{split}
\end{equation}
with $\N^\mu[\Xfields]$  defined as
\begin{equation}\label{eq:Nchi}
\begin{split}
\N^\mu[\Xfields] &\equiv \xi_\nu T^{\mu\nu}+ (\Lambda+\xi^\lambda A_\lambda) \cdot J^\mu
- T\Kbeta^\mu \brk{\xi^\nu \aheat_\nu + (\Lambda+\xi^\lambda A_\lambda) \cdot \acharge }\,.
\end{split}
\end{equation}
We have denoted the diffeomorphism and gauge variation induced by
$\{\Kbeta^\mu,\LambdaB\}$ as $\diffB $ which is defined in \eqref{eq:delBdef}.  The Noether current $\N^\mu\brk{\Bfields}$ will play a role in constructing the non-canonical part of the entropy current or equivalently the free energy current.\footnote{ In the process of deriving various variational expressions we find the following identities quite useful to collect terms:
\begin{align}
\frac{1}{\sqrt{-g}}\,\diffB\prn{\sqrt{-g}\, S} = \nabla_\alpha\prn{\Kbeta^\alpha\, S} \,, \qquad
\frac{1}{\sqrt{-g}}\,\diffB\prn{\sqrt{-g}\, V_\sigma} = \lieD_{\Kbeta}\, V_\sigma + V_\sigma\, \nabla_\alpha \Kbeta^\alpha 
\label{eq:liediden}
\end{align}
where $S$ and $V_\alpha$ are arbitrary scalar and one-form fields, respectively. We have made use of the latter in derving 
\eqref{eq:NoetherHydroPre}  and will use the former in, e.g., \eqref{eq:LEntropyEq}.}

We can now plug the expression \eqref{eq:NoetherHydroPre} into \eqref{eq:LagVar} and integrate over the background geometry to obtain a statement for the total variation
\begin{align}
\diffF \int \sqrt{-g}\ \Lag\brk{\hfields} &
= \int \sqrt{-g}\  \xi_\mu \brk{-\nabla_\nu T^{\mu\nu}+ J_\nu \cdot F^{\mu\nu}
+\frac{g^{\mu\nu}}{\sqrt{-g}}\diffB \prn{\sqrt{-g}\ T\, \aheat_\nu}
+  g^{\mu\nu} T\, \acharge \cdot \diffB  A_\nu } \nonumber \\
&\qquad \quad+ \int \sqrt{-g}\ (\Lambda+\xi^\lambda A_\lambda) \cdot \brk{-D_\nu J^\nu
+\frac{1}{\sqrt{-g}}\diffB \prn{\sqrt{-g}\ T\,\acharge} \ }
\nonumber \\
&\qquad\qquad\quad+ \text{Boundary terms} \,.
\end{align}

Since $\Lag$ is a scalar under the background diffeomorphism and gauge transformation,  the integral on the l.h.s.\ has to vanish, $\diffF S_\text{hydro}=0$,  up to boundary terms.
This immediately implies for arbitrary $\{\xi^\mu, \Lambda\}$ one has the diffeomorphism and gauge Bianchi identities:
\begin{equation}\label{eq:LHydroEq}
\begin{split}
\nabla_\nu T^{\mu\nu}&= J_\nu \cdot F^{\mu\nu}
+\frac{g^{\mu\nu}}{\sqrt{-g}}\diffB \prn{\sqrt{-g}\ T\, \aheat_\nu}
+  g^{\mu\nu} \,T\,\acharge \cdot \diffB  A_\nu \\
D_\sigma J^\sigma &= \frac{1}{\sqrt{-g}}\diffB \prn{\sqrt{-g}\ T\,\acharge}
\end{split}
\end{equation}
These are the Bianchi identities we are after and per se they hold off-shell.  If we think of $\{T^{\mu\nu},J^\mu,\aheat_\sigma,\acharge\}$  as tensor-valued functionals of $\hfields$, obtained from the variational principle \eqref{eq:LagVar}, then these equations hold  identically for the currents for all configurations.

We can supplement \eqref{eq:LHydroEq} with another identity which follows from our definition of the entropy current \eqref{eq:sdef}
\begin{equation}\label{eq:LEntropyEq}
\begin{split}
\nabla_\sigma J_S^\sigma &= \nabla_\sigma (T\,s \,\Kbeta^\sigma)= \frac{1}{\sqrt{-g}}\diffB \prn{\sqrt{-g}\ Ts}\,,
\end{split}
\end{equation}
which is again valid off-shell.

We can now easily check that \eqref{eq:LHydroEq} and \eqref{eq:LEntropyEq} together imply the adiabaticity equation \eqref{eq:naadiabatic} in the absence of anomalies, for
\begin{align}
\nabla_\mu J_S^\mu &+ \Kbeta_\mu\prn{\nabla_\nu T^{\mu\nu}-J_\nu \cdot F^{\mu\nu}}
+ (\LambdaB+\Kbeta^\lambda A_\lambda) \cdot D_\nu J^\nu
\nonumber \\
&= \frac{1}{\sqrt{-g}} \, \Big[
\diffB \prn{\sqrt{-g}\ Ts}  + \Kbeta^\sigma \,\diffB \prn{\sqrt{-g}\ T\, \aheat_\sigma} + \sqrt{-g}\, T\, \Kbeta^\sigma \,\acharge \, \cdot \diffB A_\sigma
 + \frac{\mu}{T} \cdot \diffB \prn{\sqrt{-g} \,T\, \acharge} \Big] 
\nonumber\\
& =
\frac{1}{\sqrt{-g}} \;
\diffB \prn{\sqrt{-g}\ \brk{T\,s + u^\sigma\,\aheat_\sigma+ \mu \cdot \acharge } }
\nonumber \\
&=0\,.
\label{eq:Laddemo}
\end{align}
We have used the basic definitions \eqref{eq:hydrofields} and  the relation \eqref{eq:sVZeta} derived earlier. We should emphasize that by virtue of the Bianchi identities \eqref{eq:LHydroEq} holding off-shell we have demonstrated that the Lagrangian system defined by
$\Lag\brk{\hfields}$ satisfies the non-anomalous adiabaticity equation \eqref{eq:naadiabatic} off-shell. We will postpone a more detailed discussion of the anomalous situation until
\S\ref{sec:anomalies}; suffice it to say for now that there is a Lagrangian construction that gives a particular solution to \eqref{eq:Adiabaticity}.

Sometimes it is convenient to write the combinations that occur above in a
conventional hydrodynamic expansion. Upon explicit evaluation one finds
\begin{equation}
\begin{split}
\frac{1}{\sqrt{-g}}&\diffB \prn{\sqrt{-g}\ T\, \aheat_\sigma} + T\, \acharge\cdot\diffB  A_\sigma \\
&=   \nabla_\lambda ( \aheat_\sigma\, u^\lambda) +  \aheat_\lambda \, (\nabla_\sigma+\acc_\sigma) u^\lambda
+s (\nabla_\sigma+\acc_\sigma) T -\acharge\cdot
\brk{E_\sigma-D_\sigma\mu-\acc_\sigma \mu}
\end{split}
\end{equation}
and
\begin{equation}
\begin{split}
\frac{1}{\sqrt{-g}}\diffB \prn{\sqrt{-g}\ T\,\acharge} = D_\sigma (\acharge\, u^\sigma)+[\acharge,\mu]
\end{split}
\end{equation}
In the above expressions we encounter the fluid acceleration vector $\acc_\sigma$
and the rest frame electric field $E_\sigma = F_{\sigma\lambda}u^\lambda$ introduced earlier.

\subsection{Noether current in Class L}
\label{sec:NoetherL}

Having seen that Lagrangian systems of hydrodynamics as formulated above satisfy adiabaticity equation off-shell,
we now proceed to extract some more basic lessons. Most of these follow from the basic variational principle and are
encoded in the Noether current for the Class L constitutive relations which is related to
the free energy current of the system.

We proceed by first deriving the Noether theorem for our Lagrangian system. By substituting
\eqref{eq:LHydroEq} into \eqref{eq:NoetherHydroPre}, we get
\begin{equation}\label{eq:NoetherHydro}
\begin{split}
\nabla_\mu \N^\mu[\Xfields] &= \half  T^{\mu\nu}\,\diffF g_{\mu\nu} + J^\mu \cdot \diffF A_\mu
+ T\,\aheat_\mu  \,\diffF \Kbeta^\mu
+ T\,\acharge \cdot \prn{\diffF\LambdaB+ A_\mu\, \diffF \Kbeta^\mu}
\end{split}
\end{equation}
with $\N^\mu[\Xfields]$ as given in \eqref{eq:Nchi}.   The primary content of Noether theorem is that a current $\N^\mu[\Xfields] $ satisfying the above equation exists.

It is easy to see that every Noether current satisfying \eqref{eq:NoetherHydro} gives
a free energy current satisfying the adiabaticity equation
\eqref{eq:AdiabaticityG} with $\mathcal{G}_{_H}^\perp =0$ (for non-anomalous fluids).
In particular, we see that we solve \eqref{eq:AdiabaticityG} by
identifying $\{\xi^\mu,\Lambda\}=\{\Kbeta^\mu,\LambdaB\}$
(but we will still keep $\{g_{\mu\nu},A_\mu\}$ general) and take
\begin{equation}
\begin{split}
\mathcal{G}^\sigma &=-T\,\N^\sigma[\Bfields] \\
& = 
-\; T \prn{
\Kbeta_\nu \,T^{\sigma\nu}+ (\LambdaB+\Kbeta^\lambda A_\lambda) \cdot J^\sigma
- T\,\Kbeta^\sigma \brk{\Kbeta^\nu \,\aheat_\nu + (\LambdaB+\Kbeta^\lambda A_\lambda)
\cdot \acharge }
}  \\
\mathcal{G}_{_H}^\perp &= 0 \,.
\end{split}
\end{equation}
Thus we see that the free energy current coincides (up to a factor of $T$)
with the Noether current (or the non-canonical part of the entropy current)
$\N^\sigma[\Bfields]$, cf., \eqref{eq:GDef}.

The corresponding entropy current is also easily constructed: we remind
the reader that the non-canonical part of the entropy current is
$-\mathcal{G}^\sigma/T =\N^\sigma[\Bfields]$ so that the total entropy current
is given by
\begin{equation}\label{eq:JSCanonNonCanon}
\begin{split}
J_S^\sigma &= \N^\sigma[\Bfields]- \Kbeta_\lambda \,T^{\sigma\lambda}-
(\LambdaB+\Kbeta^\lambda A_\lambda) \cdot J^\sigma \\
&=\N^\sigma[\Bfields]- \frac{u_\lambda}{T} \,T^{\sigma\lambda}-  \frac{\mu}{T}\cdot J^\sigma \\
&= - T\,\Kbeta^\sigma \brk{\Kbeta^\nu \,\aheat_\nu + (\LambdaB+\Kbeta^\lambda A_\lambda)
\cdot \acharge } \,.
\end{split}
\end{equation}
Thus, the choice of free energy/entropy currents is in one to one correspondence
with the choice of the Noether current, consistent with our identification in \eqref{eq:sVZeta}.

Let us now try to get an alternate expression for $\N^\mu[\Xfields]$ which will
be useful later on. We have from \eqref{eq:LagVar} and \eqref{eq:NoetherHydro} the simple identity
\begin{equation}\label{eq:NoetherHydroCanon}
\begin{split}
\nabla_\mu \N^\mu[\Xfields]
&= \frac{1}{\sqrt{-g}}\diffF\prn{\sqrt{-g}\ \Lag} -\nabla_\mu (\PSymplPot{\Xfields})^\mu\\
&= \nabla_\mu \brk{\xi^\mu \Lag-(\PSymplPot{\Xfields})^\mu} \,,
\end{split}
\end{equation}
where we have assumed that $\Lag$ transforms as a scalar. This shows that the vector
$\xi^\mu \Lag-(\PSymplPot{\Xfields})^\mu$ (which is often called the canonical Noether current)
has the same divergence as $\N^\mu[\Xfields]$. Assuming there are no cohomological obstructions, we can then write
\begin{equation}
\begin{split}
\N^\mu[\Xfields] &=\xi^\mu \Lag-(\PSymplPot{\Xfields})^\mu
+\nabla_\nu \Komar^{\mu\nu}[\Xfields]\,,
\end{split}
\label{eq:KomarDef}
\end{equation}
where $\Komar^{\mu\nu}[\Xfields]=- \Komar^{\nu\mu}[\Xfields]$ is called the Komar charge
of the system. We will call this decomposition of $\N^\mu[\Xfields]$ as Komar decomposition.
This gives an alternate expression for free energy current as
\begin{equation}
\begin{split}
\mathcal{G}^\sigma &=-T\, \N^\sigma[\Bfields] = -T \bigg({\Kbeta^\sigma \Lag-(\PSymplPot{\Bfields})^\sigma +\nabla_\nu \Komar^{\sigma\nu}[\Bfields]}\bigg) \,,
\end{split}
\label{eq:freeGL}
\end{equation}
and
\begin{equation}
\begin{split}
J_S^\mu
&=s\,u^\mu
\\
&=-\Kbeta_\nu T^{\mu\nu}- (\LambdaB+\Kbeta^\lambda A_\lambda) \cdot J^\mu
+\N^\mu[\Bfields] \\
&=-\Kbeta_\nu \,T^{\mu\nu}- (\LambdaB+\Kbeta^\lambda A_\lambda) \cdot J^\mu
+\Kbeta^\mu \Lag-(\PSymplPot{\Bfields})^\mu +\nabla_\nu \Komar^{\mu\nu}[\Bfields] \,.
\end{split}
\label{eq:JSK}
\end{equation}
Note that the pre-symplectic potential $(\PSymplPot{})^\mu$ appears in these expressions only through
$(\PSymplPot{\Bfields})^\mu$. Since $\diffB  \Kbeta^\mu =0 $ and $\diffB  \LambdaB =0 $, this means to get
the free energy current or the entropy current, we need not actually get the contributions to  $(\PSymplPot{\Bfields})^\mu$
from varying $\{\Kbeta^\mu,\LambdaB\}$. The contribution
to the  pre-symplectic potential can be obtained by just varying $\{g_{\mu\nu},A_\mu\}$
and then see what we obtain when we integrate by parts.

\subsection{Hydrostatic partition function for Class L}
\label{sec:Lhstatic}

Our discussion of the Class L solutions to the adiabaticity equation has so far been unconstrained, in that we have only assumed that the currents can be derived from a Lagrangian $\Lag\brk{\hfields}$. We now relate this story to the analysis of \S\ref{sec:hydrostatics} where we also derived the currents from a generating function.  In order to ascertain the connection we now specialize to  hydrostatics. Since we have an explicit expression
for the free energy current in Class L,  we can invoke the arguments that led to \eqref{eq:wenadef} to come up with a hydrostatic partition function for theories in Class L.

We will now argue that Class L provides an off-shell generalization for hydrostatics.  Note however, that Class L can at best incorporate Class $\PS$ as we are required to write the integral of a spacetime scalar density for the effective action $S_\text{hydro}$.

First we constrain the sources to support a Killing field configuration
$\Eqfields \equiv \{\KEq^\mu,\LambdaEq\} $ which we will momentarily identify with
$\Bfields =\{\Kbeta^\mu, \LambdaB\}$. Further using  \eqref{eq:freeGL}  we can write on the base space $\Sigma_E$
\begin{equation}
\begin{split}
W_\text{Hydrostatic}
&=-\brk{\int_{\Sigma_E}  \frac{\mathcal{G}^\sigma}{T}\ d^{d-1}S_\sigma\ }_\text{Hydrostatic}
=\brk{\int_{\Sigma_E}   \N^\sigma[\Bfields]\ d^{d-1}S_\sigma\ }_\text{Hydrostatic}  \\
&= \brk{\int_{\Sigma_E}
\bigg(\Kbeta^\sigma \Lag-(\PSymplPot{\Bfields})^\sigma
+\nabla_\nu \Komar^{\sigma\nu}[\Bfields]
\bigg)
\ d^{d-1}S_\sigma\ }_\text{Hydrostatic} \\
&=  \int_{\Sigma_E}
\Lag_\text{Hydrostatic}\ \KEq^\sigma \;d^{d-1}S_\sigma\ + \brk{\int_{\partial\Sigma_E}
\half \,\Komar^{\sigma\nu}[\Bfields]\ d^{d-2}S_{\sigma\nu}\ }_\text{Hydrostatic}
\end{split}
\end{equation}
where we have used the fact that $\diffB$ annihilates functions in hydrostatics \eqref{eq:hsdelb} to drop  the $(\PSymplPot{\Bfields})^\sigma$ contribution. Here, $\Lag_\text{Hydrostatic}$ denotes  $\Lag\brk{\hfields}$ with $\{\Kbeta^\mu,\LambdaB\}$ replaced by $\{\KEq^\mu,\LambdaEq\}$. We finally obtain
\begin{equation}
W_\text{Hydrostatic}
=  \int_{\Sigma_E \times I_\KEq}
d^dx\sqrt{-g}\ \Lag_\text{Hydrostatic}\   + \text{Boundary contributions} \,,
\label{eq:Lpfn}
\end{equation}
where the integral is performed over the manifold $\Sigma_E\; \times I_\KEq$
where $I_K$ is an interval  of unit affine length along $\KEq^\mu$.
So in the end we get the simple result  that the hydrostatic partition function is just the integral over the Lagrangian after taking the hydrostatic limit of $\Lag\brk{\hfields}$.

Lest the reader be misled into thinking that we recover the complete set of hydrostatic partition functions (Class H) from the Class L family of adiabatic fluids we hasten to add an important caveat alluded to earlier. It should be clear from \eqref{eq:Lpfn} that we obtain from $\Lag\brk{\hfields}$ only those hydrostatic partition functions that can be written as spacetime scalars, since we have an integral over the entire manifold ${\cal M}= \Sigma_E \times I_\KEq$. This is what we called $\PS$ in our discussion in \S\ref{sec:hydrostatics}.  From the categorization explained there, there are two other classes of terms in the partition function which do not obviously arise from Class L Lagrangians: the Class $\PV$ terms involving integrals over transverse vectors and the Class A terms which play a role in anomalous hydrodynamic transport.\footnote{ Since we have focused on Lagrangian solutions to non-anomalous adiabaticity equation \eqref{eq:naadiabatic} it is not surprising that we have not yet encountered Class A. We will encounter Class A terms when we turn to a detailed discussion of anomalies in \S\ref{sec:anomalies}.}  Thus, apart from these terms  (which seem to be a finite set of terms in any spacetime dimension) we recover most of the  theory of partition functions and the adiabatic constitutive relations that they lead to. 

Given the connections to hydrostatics, it is useful therefore to decompose Class L into explicit contributions from hydrostatics $\PS$ and genuine  {\em hydrodynamic scalars} 
$\LS$. To wit, one has $\text{L} = \PS \cup \LS$. The hydrodynamic scalars, which necessarily involve one $\diffB$ insertion (by definition),  are identically vanishing in equilibrium -- they require us to turn on time dependence to contribute. Moreover, as a result the values of the hydrostatic scalars can be freely changed by contributions proportional to the Class $\LS$ terms. Hence, hydrostatic scalars $\PS \subset \text{L}$ take values in a coset: $\PS = \text{L}/\LS$.

Finally, we can make a precise connection between the Noether current construction outlined in \S\ref{sec:NoetherL} and the entropy current constrained by hydrostatics. As explained in \S\ref{sec:hydrostatics} the hydrostatic entropy current has been studied in some detail in \cite{Bhattacharyya:2013lha,Bhattacharyya:2014bha}. As we now understand, varying \eqref{eq:Lpfn} with respect to the metric and gauge field (we do not vary $\{\Kbeta^\mu,\LambdaB\}$  since they are fixed to $\{\KEq^\mu,\LambdaEq\}$ in the hydrostatic limit) we obtain
\begin{equation}
\begin{split}
\delta W_\text{Hydrostatic}
&=  \int
d^dx\sqrt{-g}\ \brk{\half T^{\mu\nu} \,\delta g_{\mu\nu}
+J^\sigma\cdot \delta A_\sigma }_\text{Hydrostatic}  \\
&\qquad \qquad + \text{Boundary contributions}
\end{split}
\end{equation}
which agrees with the rule given in \cite{Banerjee:2012iz}. Further, if we just
keep the first order deviations from hydrostatics in the equation for the
non-canonical part of the entropy current $\N^\sigma[\Bfields]
=\Kbeta^\sigma \Lag-
(\PSymplPot{\Bfields})^\sigma +\nabla_\nu \Komar^{\sigma\nu}[\Bfields]$, we get
the prescription given in \cite{Bhattacharyya:2014bha}. 
This was explained  in detail in \S\ref{sec:hscurrents} and we elaborate on the connections
further in  Appendix \ref{sec:sayantani}. 

\section{Hydrodynamic Ward identities in Class L}
\label{sec:Leoms}

Up to this point our discussion of Class L has been quite abstract. We have only exploited the diffeomorphism and gauge symmetry to extract the Bianchi identities \eqref{eq:LHydroEq}, which in turn led to the adiabaticity equation. 
As such we have in fact been treating the hydrodynamic fields $\{\Kbeta^\mu,\LambdaB\}$  effectively as  non-dynamical fields, thus working off-shell as far as the hydrodynamic fields are concerned.  The only
exception is the hydrostatic limit studied in \S\ref{sec:Lhstatic}, where we went
on-shell by simply setting $\{\Kbeta^\mu,\LambdaB\}=\{\KEq^\mu,\LambdaEq\}$
and invoking the hydrostatic principle.

This is clearly unsatisfactory; the utility of a Lagrangian is that it not only allows us to construct an off-shell action, but that it also comes equipped with a variational principle that captures the on-shell dynamics by giving us the equations of motion. We will now proceed to address this lacunae and give a variational procedure to go on-shell. Our goal is to simply to give the hydrodynamic fields $\{\Kbeta^\mu,\LambdaB\}$ appropriate dynamics which enforces the
conservation equations in \eqref{eq:hydroCons} (with $T^{\mu\perp}_H = J_H^\perp =0$ in the absence of anomalies).

\subsection{A constrained variational  principle for hydrodynamics}
\label{sec:Lvarprin}

Let us go back to the derivation of the Bianchi type identities in \S\ref{sec:LBianchi}.
Inspection of the Bianchi identities \eqref{eq:LHydroEq} which are obeyed by all Class L constitutive relations suggests that on-shell equations of non-anomalous hydrodynamics \eqref{eq:hydroCons} would be satisfied
(with anomaly terms set to zero) if the fields $\{\Kbeta^\mu,\LambdaB\}$ obeyed the following equations:
\begin{equation}\label{eq:LHydroOnShell}
\begin{split}
\frac{1}{\sqrt{-g}}\diffB \prn{\sqrt{-g}\ T\,\aheat_\mu}
+   T\,\acharge \cdot \diffB  A_\mu &\simeq 0 \,,\\
\frac{1}{\sqrt{-g}}\diffB \prn{\sqrt{-g}\ T\,\acharge} &\simeq 0\,.
\end{split}
\end{equation}

These equations have to arise for consistency of our formalism as the dynamical equations of motion obtained by varying the fields $\{\Kbeta^\mu,\LambdaB\}$. It is clear a-priori that this is not going to happen naturally; the basic variational equation \eqref{eq:LagVar} if interpreted na\"ively would lead to $\aheat_\sigma + \acharge \cdot A_\sigma =0$ and
$ \acharge =0$ (assuming $T\neq0$), which is certainly not what we would like to have.  The key point we have to understand is the following: given that the dynamical degrees of freedom comprise of a vector $\Kbeta^\mu$ and a scalar $\LambdaB$, we have to decide what variations of these fields to admit as being physical. Our argument above shows that an unconstrained variation of these fields is inconsistent with the dynamics we seek, so the question is whether a suitable constrained variational principle exists.

We would like to claim now that such a constrained variation of $\{\Kbeta^\mu,\LambdaB\}$ exists and it naturally leads to the correct hydrodynamic Ward identities upon using the Bianchi identities \eqref{eq:LHydroEq}.  To see how the desired equations can be obtained from a variational principle, consider the following:
Fix the  metric and gauge field and extremize  $S_\text{hydro}\brk{\hfields}$  among a family of
$\Bfields= \{\Kbeta^\mu,\LambdaB\}$ which are related to each other via Lie transport.  We will denote this class of variations by $\diffCons$ to distinguish it from the variation we have considered hitherto without the Lie transport constraint.

To wit, given an arbitrary
$\Xfields=\{\xi^\mu,\Lambda\}$  we define this constrained variation as:
\begin{align}
\diffCons: \quad \diffCons \Kbeta^\mu =\diffF \Kbeta^\mu \,,\qquad
\diffCons  \LambdaB=\diffF\LambdaB\,,\qquad
\diffCons g_{\mu\nu} = \diffCons  A_\mu =0 \,.
\label{eq:consLvar}
\end{align}
Our claim is that the on-shell hydrodynamic configurations are precisely those  $\{\Kbeta^\mu,\LambdaB\}$ which satisfy $\diffCons S_\text{hydro}\brk{\hfields}=0$ up to boundary contributions. 

To show this, we use  the definition of $\diffCons$, \eqref{eq:consLvar}, in
\eqref{eq:LagVar} to write
\begin{equation}
\begin{split}
\frac{1}{\sqrt{-g}}&\diffCons\prn{\sqrt{-g}\ \Lag} -\nabla_\mu ({\slashed{\diffCons}}\varTheta_{_{\text{PS}}} )^\mu =  T\, \aheat_\sigma \,\diffF \Kbeta^\sigma
+ T\,\acharge \cdot \prn{\diffF\LambdaB+ A_\sigma \diffF \Kbeta^\sigma} \\
&=  -T \,\aheat_\sigma\, \diffB  \xi^\sigma
-T\,\acharge \cdot \prn{\diffB \Lambda+ A_\sigma \diffB  \xi^\sigma} \,,
\end{split}
\end{equation}
where we have used $\diffF \Kbeta = \lieD_\xi \Kbeta = - \lieD_\Kbeta \xi = -\diffB \xi $.
Integrating the above equation by parts to move the derivatives over from the variational parameters $\{\xi^\mu,\Lambda\} $ results in
\begin{equation}\label{eq:LagConstrainedVar}
\begin{split}
\frac{1}{\sqrt{-g}}\diffCons\prn{\sqrt{-g}\ \Lag\brk{\hfields}} & =
\bigg(\frac{1}{\sqrt{-g}}\diffB  \prn{\sqrt{-g}\ T \, \aheat_\sigma}
+ T\, \acharge \cdot \diffB  A_\sigma \bigg)\, \xi^\sigma
\\
& \qquad + \frac{1}{\sqrt{-g}}\diffB  \brk{\sqrt{-g}\ T\,\acharge} \cdot \prn{\Lambda+
A_\sigma \, \xi^\sigma} \\
& \qquad +\nabla_\mu \bigg\{
({\slashed{\diffCons}}\varTheta_{_{\text{PS}}} )^\mu-T\,\Kbeta^\mu\brk{ \aheat_\sigma \, \xi^\sigma
+ \acharge \cdot \prn{\Lambda+ A_\sigma \, \xi^\sigma}} \bigg\} \,.
\end{split}
\end{equation}
Insisting that this be zero for arbitrary $\Xfields=\{\xi^\mu,\Lambda\}$ then directly leads to the
required equations  \eqref{eq:LHydroOnShell}.

Thus as advertised Lagrangian theories of hydrodynamic fields equipped with a suitable variational principle, give rise to the correct hydrodynamic equations of motion and simultaneously provide an off-shell solution to the adiabaticity equation.

\subsection{Reference fields and conservation equations}
\label{sec:reffields}

In \S\ref{sec:Lvarprin} we gave a constrained variational principle which enabled us to derive the hydrodynamic equations from the Lagrangian $\Lag\brk{\hfields}$. While prescriptive, it is cumbersome to work in the space of constrained variations to derive dynamics. It is somewhat more satisfactory to shift to a description where these constraints are  automatically implemented by an action, rather than being imposed by hand.

To do this, we need to decompose the variations of $\{\Kbeta^\mu,\LambdaB\}$
into those allowed by the constraint,  and those in the orthogonal space of variations (which are forbidden by the constraint). The former lie in the Lie orbit of an admissible configuration. We can exploit  this characterization in decomposing the degrees of freedom into the truly dynamical ones and the ones held rigid under the variation. To ascertain the physical space of variations, we pick  a reference configuration $\{\Kref^\mu,\Lref\}$ in each Lie orbit and then express the
actual $\{\Kbeta^\mu,\LambdaB\}$ by Lie dragging this reference configuration by
a gauge transformation and diffeomorphism. We thus seek to decompose the hydrodynamic
fields into
\begin{itemize}
\item[(i).] A heavy component which is the reference configuration that one does not
vary when extremizing (denoted by the blackboard bold font characters).
\item[(ii).] A light component which is given by the Lie drag modes that one varies
when extremizing.
\end{itemize}

We begin by systematically first establishing a reference configuration. It is
convenient to imagine that these reference configurations live on some other
spacetime $\Mref$ which is gauge equivalent and diffeomorphic to the original spacetime. We will use the (first half of the) lowercase Latin alphabet to denote the spacetime indices on $\Mref$ to distinguish them from lowercase Greek indices used for the original spacetime ${\cal M}$.

\begin{figure}
\setlength{\unitlength}{0.1\columnwidth}
\centerline{\includegraphics[width=.7\textwidth]{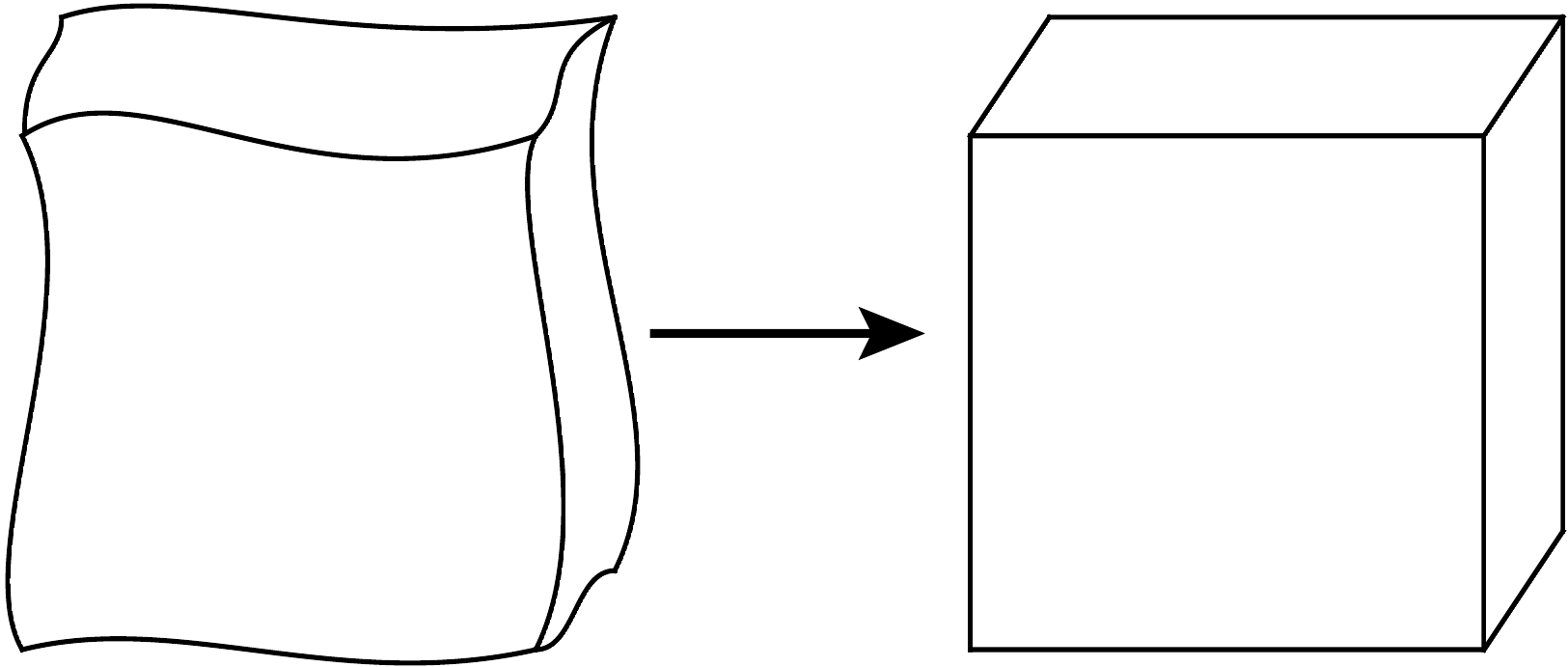}}
\begin{picture}(0.3,0.4)(0,0)
\put(6.9,0.2){\makebox(0,0){$\Mref$}}
\put(2.7,0.2){\makebox(0,0){${\cal M}$}}
\put(2.7,2.4){\makebox(0,0){$x^\mu$}}
\put(2.7,1.7){\makebox(0,0){$\{g_{\mu\nu}, A_\mu\}$}}
\put(2.7,1){\makebox(0,0){$\Bfields \equiv\{\Kbeta^\mu,\LambdaB\}$}}
\put(5,2.2){\makebox(0,0){$\{\varphi^a,c\}$}}
\put(6.9,2.4){\makebox(0,0){$\xref^a = \varphi^a(x)$}}
\put(6.9,1.7){\makebox(0,0){$\{\gref_{ab}, \Aref_a\}$}}
\put(6.9,1){\makebox(0,0){$\Breffields\equiv\{\Kref^a,\Lref\}$}}
\end{picture}
\caption{Illustration of the connection between the physical and reference fields for Class L adiabatic fluids. The fields on the physical spacetime manifold ${\cal M}$ are related to those
on the reference manifold $\Mref$ by a pull-back using the dynamical fields $\{\varphi^a,c\}$. The constrained variation on ${\cal M}$ which gives the correct equations of motion corresponds to varying $\{\varphi^a,c\}$ while holding $\{\Kref^a,\Lref\}$ fixed.
}
\label{fig:ManifoldSketch}
\end{figure}
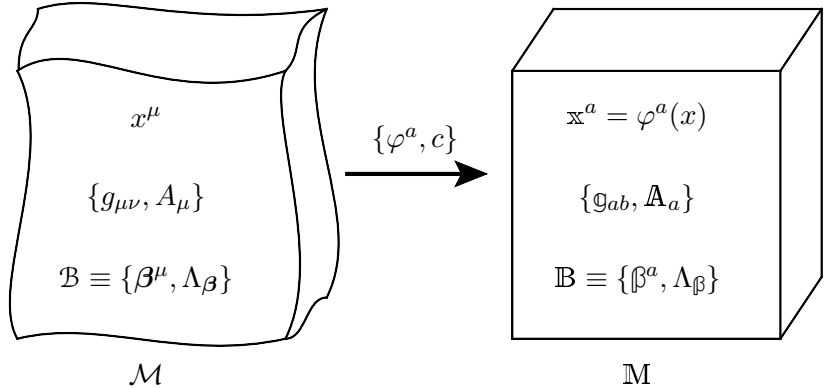

Let $x^\mu$ be coordinates on ${\cal M}$ and $\xref^a$ be coordinates on $\Mref$. $\Breffields \equiv \{\Kref^a,\Lref\}$ be the reference hydrodynamic fields living on  $\Mref$.
The actual $\{\Kbeta^\mu,\LambdaB\}$ are obtained by introducing a diffeomorphism field  $\varphi^a(x)$ and a gauge transformation field  $c(x)$ from physical spacetime ${\cal M}$
to $\Mref$ and then using them to  pull-back
$\{\Kref^a,\Lref\}$. In order to do this, let us introduce the matrix
$\partial_\mu \varphi^a\equiv \frac{\partial\varphi^a}{\partial x^\mu}$ and its inverse
$e^\mu_a\equiv \frac{\partial x^\mu}{\partial\varphi^a}$ that can be used
to pull-back  tensor indices. For definiteness, let us think of these matrices
as functions of $x$, viz., living on the actual spacetime ${\cal M}$. They satisfy
\begin{equation}
\begin{split}
e^\mu_a \, \partial_\nu \varphi^a  &= \delta^\mu_\nu\ ,\qquad
e^\mu_a \, \partial_\mu \varphi^b = \delta^a_b \,.
\end{split}
\end{equation}
With this definition the pull-back of the reference configuration is given by
\begin{equation}\label{eq:KLambdaKPullBack}
\begin{split}
\Kbeta^\mu &= e^\mu_a(x) \; \Kref^a[\varphi(x)]  \\
\LambdaB &= c(x)\ \Lref[\varphi(x)]\ c^{-1}(x) + \Kbeta^\sigma(x) \, \partial_\sigma c(x)\ c^{-1}(x) \,.
\end{split}
\end{equation}
Note that $\LambdaB$ transforms with the correct inhomogeneous piece so that
$\LambdaB+A_\sigma \,\Kbeta^\sigma$ transforms covariantly. More precisely, consider
a flavour transformation $A_\sigma \mapsto g^{-1} A_\sigma\, g+g^{-1}\,\partial_\sigma g$
and $\prn{\LambdaB+A_\sigma \,\Kbeta^\sigma}\mapsto g^{-1}\prn{\LambdaB+A_\sigma \Kbeta^\sigma}g$.
It follows from the above expressions that this corresponds to a left transformation of $c$ given by
$c \mapsto g^{-1} c$ with $\Lref$ kept fixed.

The decomposition given in \eqref{eq:KLambdaKPullBack} means that changing $\{\varphi^a,c\}$ takes
$\{\Kbeta^\mu,\LambdaB\}$  along a Lie orbit whereas changing the functional form of
$\{\Kref^a,\Lref\}$ takes $\{\Kbeta^\mu,\LambdaB\}$
from one Lie orbit to another Lie orbit. So, in order to get the hydrodynamic equations,
we should extremize $S_\text{hydro}\brk{\hfields}$ with respect to variations of the
$\{\varphi^a,c\}$ fields keeping  the functional form of $\{\Kref^a,\Lref\}$ fixed.
See Fig.~\ref{fig:ManifoldSketch} for an illustration of the situation.

It is easy to intuit how this has to work in principle.  We simply have to consider the variation of the relation  \eqref{eq:KLambdaKPullBack} between the physical fields and the reference parameterization introduced above. We would then plug this into \eqref{eq:LagVar} to ascertain the variations of the Lagrangian in the physically admissible directions.
To see how this actually works in practice, let us begin by recording out the variation of $\Kbeta^\mu$ and $\LambdaB$ in terms of the reference and physical fields. Relegating the details of the actual computation to Appendix \ref{sec:varform} we quote here the final result of import:
\begin{equation}\label{eq:deltaKLambdaK}
\begin{split}
\delta \Kbeta^\mu
&=   e^\mu_a \; \delta \Kref^a -\diffB \prn{e^\mu_a \,\delta\varphi^a} \\
\delta\LambdaB
&=
c\ \delta\Lref \ c^{-1}+ e_a^\sigma\; \delta \Kref^a   \; (\partial_\sigma c)c^{-1} +\diffB  \brk{\delta c\ c^{-1}- e^\sigma_a \delta \varphi^a \ \partial_\sigma c\ c^{-1}} ,
\end{split}
\end{equation}
where the variations of $\{\varphi^a,c\}$ appear as variations along Lie orbit as
we wanted.

We can now substitute  the above equation into \eqref{eq:LagVar}. Once again we have to do  an integration by parts to isolate the terms proportional to the dynamical degrees of freedom. After a bit of algebra we get
\begin{equation}\label{eq:phiLagVar}
\begin{split}
\frac{1}{\sqrt{-g}}&\delta\prn{\sqrt{-g}\ \Lag} -\nabla_\mu
\bigbr{ (\PSymplPot{})^\mu
+ u^\mu \acharge \cdot \delta c\,c^{-1}
- u^\mu \,e^\sigma_a \,\delta\varphi^a \brk{\aheat_\sigma+\acharge
\cdot(\partial_\sigma c\ c^{-1} + A_\sigma) }
}\\
&= \half\, T^{\mu\nu}\,\delta g_{\mu\nu} + J^\mu \cdot \delta A_\mu
+e^\sigma_a\,\brk{\aheat_\sigma+\acharge \cdot(\partial_\sigma c\ c^{-1} + A_\sigma) }T\;
\delta \Kref^a
+c^{-1}\,\acharge \,c\cdot T\;\delta \Lref\\
&\quad -
\frac{1}{\sqrt{-g}}\diffB \prn{\sqrt{-g}\ T\,\acharge}
\cdot \bigbr{\delta c\ c^{-1}-e^\sigma_a \,\delta\varphi^a(\;
\partial_\sigma c\ c^{-1} + A_\sigma )}\\
&\quad +e^\sigma_a\,
\delta\varphi^a \brk{  \frac{1}{\sqrt{-g}}\diffB (\sqrt{-g}\ T\, \aheat_\sigma )
+ T\, \acharge \cdot \diffB  A_\sigma
} .
\end{split}
\end{equation}
The dynamical equations of motion can be now read off from the terms proportional to
$\delta \varphi^a$ and $\delta c$ respectively. Inspection of \eqref{eq:phiLagVar} make it clear immediately that isolating these terms we end up with the equations of motion \eqref{eq:LHydroOnShell}.

Thus, extremizing $S_\text{hydro}\brk{\hfields}$ with respect to $\{\varphi^a,c\}$ variations gives the correct energy-momentum and charge conservation equations as required.  As such we have transplanted a constrained variational problem into one where the variation is unconstrained for the physical fields $\{\varphi^a,c\} $
but the price one pays is to introduce a set of reference fields which are held rigid in the process.

\subsection{Gauge redundancy of reference fields}
\label{sec:refgauge}

In \S\ref{sec:reffields} we decomposed the hydrodynamic fields $\{\Kbeta^\mu,\LambdaB\}$
into fixed reference fields $\{\Kref^a,\Lref\}$ and dynamical fields $\{\varphi^a, c\}$ parameterizing the Lie orbits of this reference configuration. Let us now scrutinize the decomposition \eqref{eq:KLambdaKPullBack} more closely to ascertain the  symmetries of our new parameterization.

The decomposition \eqref{eq:KLambdaKPullBack} introduces a new redundancy in description.
We can always do a gauge transformation or diffeomorphism on the reference configurations and compensate for it, say, by 
changing $\{\varphi^a,c\}$, so that we end up getting the same physical hydrodynamic configuration. For example, consider the following gauge transformation for the charged fields:
\begin{equation}
\begin{split}
\Lref&\mapsto f^{-1}\, \Lref \,f-f^{-1}\,\Kbeta^\sigma \,\partial_\sigma f \\
c &\mapsto c\,f \,,
\end{split}
\end{equation}
where  $f$ is any flavour gauge transformation on $\Mref$.
It is then simple to see that this transformation leaves $\LambdaB$ unaffected.
Hence, these transformations should be thought of as a  gauge redundancy in our description which forces us to identify
\begin{equation}
\begin{split}
\{\Lref,\ c\}\sim \{f^{-1}\, \Lref \,f-f^{-1}\,\Kbeta^\sigma \partial_\sigma f, \ c\,f\}.
\end{split}
\end{equation}
as they give the same hydrodynamic field $\LambdaB.$

There is a similar redundancy in diffeomorphisms  on the reference manifold $\Mref$
given by
\begin{equation}
\begin{split}
\varphi^a &\;\mapsto\; f^a(\varphi) \\
\Lref[\varphi^a] &\;\mapsto\;  \Lref[f^a(\varphi)] \\
\Kref^a[\varphi^b]  &\;\mapsto\;  \frac{\partial f^a}{\partial \varphi^c}\; \Kref^c[f^b(\varphi)]  \\
\end{split}
\end{equation}
where $\varphi^a(x)$ are understood to be the coordinates on $\Mref$ obtained by pushing forward $x$-coordinates.
Hence, any two configurations which differ by such $\varphi$-diffeomorphisms should also be
thought of as the same fluid configuration as they lead to identical hydrodynamic fields 
$\{\Kbeta^\mu, \LambdaB\}$.

As with any gauge redundancy, it is convenient to pass to a formalism which is covariant with respect to this set of transformations.  Moreover, since the redundancy is the gauge and diffeomorphism properties of our fields, we can just introduce a metric and flavour gauge fields on $\Mref$ to properly account for it. Endowing $\Mref$  with these structures would allow us to covariantize all the transformations.  We will do this by first pushing forward the metric and flavour gauge fields on the actual spacetime ${\cal M}$ to $\Mref$ using $\{\varphi^a,c\}$ i.e., we define
\begin{equation}\label{eq:ghAh}
\begin{split}
\gref_{ab}[\varphi]&\equiv e_a^\mu\, e_b^\nu\, g_{\mu\nu}  \\
\Aref_a[\varphi] &\equiv e_a^\mu \,\brk{c^{-1}A_\mu c + c^{-1}\partial_\mu c } \,.
\end{split}
\end{equation}
Given the push-forward relation for the metric,  the Christoffel connection transforms in a related fashion:
\begin{equation}
\begin{split}
\Chref^a{}_{bc}[\varphi] &\equiv e_c^\lambda \; \partial_\mu \varphi^a \; \brk{
\Gamma^\mu{}_{\nu\lambda} \;e_b^\nu +
\partial_\lambda e_b^\mu } \,.
\end{split}
\end{equation}
We can use the connections $\Aref, \Chref$  to define covariant derivatives on $\Mref$ which can then be used to construct invariants of $\varphi$-diffeomorphisms and $c$ gauge transformations. One can also easily check that the connections on $\cal{M}$ and $\Mref$ are consistent with Lie transport. That is, for the pushforward of a general tensor field,\footnote{ We have written here the expression for an uncharged tensor field transforming solely under diffeomorphisms; including gauge transformations is straightforward which we leave as an exercise.}
\begin{equation}
  \mathbb{T}^{a_1 a_2 \cdots}{}_{b_1 b_2 \cdots}[\varphi] \equiv (\partial_{\mu_1} \varphi^{a_1}) (\partial_{\mu_2} \varphi^{a_2})\cdots   e^{\nu_1}_{b_1} e^{\nu_2}_{b_2}\cdots  \cdots T^{\mu_1\mu_2\cdots}{}_{\nu_1\nu_2\cdots} 	 \,,
\end{equation}
we have 
\begin{equation}
 (\partial_{\mu_1} \varphi^{a_1}) (\partial_{\mu_2} \varphi^{a_2})\cdots   e^{\nu_1}_{b_1} e^{\nu_2}_{b_2}\cdots 
    \left[\lieD_{\xi} T^{\mu_1\mu_2\cdots}{}_{\nu_1\nu_2\cdots}  \right]
 = \lieD_{(\xi.\partial)\varphi} \mathbb{T}^{a_1 a_2 \cdots}{}_{b_1 b_2 \cdots} \,.
\end{equation}

\subsection{Variational principle on reference manifold}
\label{sec:ReferenceVar}

We can now reformulate the variational principle that gives rise to the hydrodynamic
equations in terms of $\{\gref_{ab}, \Aref_a\}$. We begin by observing that invariance of
$S_\text{hydro}\brk{\hfields}$ under diffeomorphisms and gauge transformations means that
\begin{equation}
\begin{split}
S_\text{hydro}\brk{\hfields}
= \int_{\cal M} \sqrt{-g}\ \Lag\brk{\hfields}
= \int_{\Mref} \sqrt{-\gref}\ \Lag\brk{\gref, \Aref,\Kref,\Lref }
\equiv
\int_{\Mref} \sqrt{-\gref}\ \Lagref \brk{\hreffields}
\end{split}
\end{equation}
where we use the condensed notation $\hreffields$ to denote the collection of ``hydrodynamic'' fields on the reference manifold.

It follows then that we can get hydrodynamic equations by extremising
$\int_{\Mref} \sqrt{-\gref} \ \Lagref\brk{\hreffields}$ by varying $\{\varphi^a,c\}$
inside $\{\gref,\Aref\}$ keeping $\{\Kref,\Lref\}$ fixed. To see how
this works, we will begin by decomposing the variations of $\{\gref,\Aref\}$
into  (i) variation of reference sources and (ii) variation of the dynamical fields $\{\varphi^a,c\}$.
Using \eqref{eq:ghAh}, we obtain
\begin{equation}\label{eq:grefVar}
\begin{split}
\delta \gref_{ab}[\varphi] &= \delta (\gref_{ab}[\varphi])-\delta \varphi^c \;
\frac{\partial}{\partial \varphi^c} \gref_{ab}[\varphi]
= e_a^\mu \, e_b^\nu\; \delta g_{\mu\nu} -\delta_{\varphi} \gref_{ab} \\
\delta \Aref_{a}[\varphi] &= \delta(\Aref_{a}[\varphi])-\delta \varphi^c \frac{\partial}{\partial \varphi^c}
\Aref_{a}[\varphi]
= e_a^\mu \,c^{-1}(\delta A_\mu)\;c -\delta_{\varphi} \Aref_{a} \\
\end{split}
\end{equation}
where $\delta_\varphi$ is the Lie drag on $\Mref$  along $\{\delta \varphi^a,-c^{-1}\delta c\}$, viz.,
\begin{equation} \label{eq:varphiVar}
\begin{split}
\delta_{\varphi} \gref_{ab} &\equiv \Dref_{a} \delta \varphi_b+\Dref_{b} \delta \varphi_a\,,\\
\delta_{\varphi} \Aref_{a} &\equiv \Dref_{a}\prn{-c^{-1}\delta c+\Aref_b\; \delta \varphi^b}+\delta\varphi^b \;\Fref_{ba} \,.
\end{split}
\end{equation}
In the above expressions we have introduced the covariant derivatives $\Dref$ and field strength $\Fref$  which are defined with the reference connections $\Aref, \Chref$ respectively in the usual fashion.

With these definitions in place, let us then translate the variational calculus onto the reference manifold $\Mref$. 
First, when we vary $\int_{\Mref} \sqrt{-\gref}\ \Lagref\brk{\hreffields}$ with respect to $\{\gref_{ab},\Aref_a\}$, we get the currents:
\begin{equation} \label{eq:MrefVariation}
\begin{split}
\delta \int_{\Mref} \sqrt{-\gref}\ \Lagref\brk{\hreffields} 
  &= \int_{\Mref} \sqrt{-\gref} \; \prn{ \frac{1}{2}\, \Tref^{ab} \, \delta \gref_{ab} + \Jref^a \cdot \delta \Aref_a 
     + \Tref\, \aheatref_a \,\delta \Kref^a + \Tref \,\achargeref \cdot (\delta \Lref + \Aref_b\, \delta \Kref^b) } ,
\end{split}
\end{equation}
where all boldface currents are understood as the pushforwards of physical currents on ${\cal M}$ to the reference manifold. 
 If we further extremize  $\int_{\Mref} \sqrt{-\gref}\ \Lagref$ with respect to just the $\varphi$-part of $\{\gref_{ab},\Aref_a\}$ (holding fixed the functional form of $\{\Kref^a,\Lref\}$), we are led to conservation equations for the energy-momentum tensor and similarly for the charge current.  To obtain these we have to perform the constrained variation $\diffCons$ which amounts to Lie drags of the sources along $\{\delta \varphi^a,-c^{-1}\delta c\}$.  Using \eqref{eq:grefVar} we learn that one should employ the variations  
 \begin{equation}\label{eq:ConstrVar}
  \diffCons \gref_{ab} = -\delta_\varphi \gref_{ab} \,,\quad \diffCons  \Aref_a = -\delta_\varphi \Aref_a \,,\quad
  \diffCons \Kref^a = 0 \,, \quad \diffCons \Lref = 0 \,.
 \end{equation} 
 Applying this variation to \eqref{eq:MrefVariation} leads to
 \begin{equation}
 \begin{split}
  \diffCons \int_{\Mref} \sqrt{-\gref}\ \Lagref\brk{\hreffields}  &=  \int_{\Mref} \sqrt{-\gref}\, \bigg\{
    \delta \varphi^c \brk{ \Dref_a \Tref^a_c  -\Jref^a \cdot \Fref_{ca} }
   +(-c^{-1}\delta c + \Aref_a \delta \varphi^a) \cdot \Dref_c \Jref^c \bigg\} \,.   
 \end{split}
 \end{equation}
 We can now read off the hydrodynamic conservation equations on the reference manifold:
 \begin{equation}
 \Dref_a \Tref^a_c  -\Jref^a \cdot \Fref_{ca} = 0 \,, \qquad \Dref_c \Jref^c = 0 \,.
 \end{equation}
 This makes the variational principle on the reference manifold very simple in practice because it allows us to skip the computation of Bianchi identities\footnote{ While we have not explicitly indicated how to get the Bianchi identities, these follow in the same manner as on the physical spacetime ${\cal M}$. We simply vary $\int_{\Mref} \sqrt{-\gref}\ \Lagref$ with respect to 
 $\hreffields$ and follow the same set of arguments as in \S\ref{sec:LBianchi}. Modulo a font and index change, the algebra remains identical. } and obtain the correct hydrodynamic Ward identities directly. The covariance of our reference manifold formalism makes it very easy to translate these results back to equations of motion on the physical spacetime 
 ${\cal M}$: one just replaces Latin indices by Greek indices and switches bold-face letters back to normal font. This prescription is thus equivalent to the one given in \S\ref{sec:reffields}.

\subsection{Static gauge on the reference manifold \& hydrodynamic fields}
\label{sec:staticg}

Given a covariant form of an action with some redundancies it is
sometimes  convenient to pass to a gauge fixed version and focus on the physical degrees of freedom. To this end we can partially fix the gauge symmetries in the reference fields.
Ignoring any possible Gribov type topological ambiguities, let us use the gauge transformation and diffeomorphism  freedom on $\Mref$ to set
\begin{equation}
\begin{split}
\Lref=0 ,\qquad  \Kref^{a=0}=1\qquad \text{and} \qquad \Kref^{a={\scriptsize I}}=0\  \;\;\text{for}\
I\in\{1,\ldots,d-1\}.
\end{split}
\label{eq:staticref}
\end{equation}
In what follows, we will refer to this as the \emph{static reference gauge}. As is clear
from above, we will henceforth use uppercase Latin alphabets to denote
spatial indices on $\Mref$.

Let us now examine the residual gauge redundancy that is left unfixed
in the static reference gauge. The following set of $\varphi$-diffeomorphisms and
$c$ gauge transformations survive the static reference gauge fixing of \eqref{eq:staticref}
\begin{equation}
\begin{split}
\varphi^{_J} &\mapsto h^{_J}(\varphi^{_I})\,, \qquad  \text{det}\prn{\frac{\partial h^{_J}}{\partial \varphi^{_I}}} \neq 0 \,,\\
\varphi^0 &\mapsto \varphi^0 + g(\varphi^{_I})\,, \\
c &\mapsto c\, f(\varphi^{_I}) \,.
\end{split}
\end{equation}
We will refer to the transformations in first line as \emph{transverse diffeomorphisms} in
$\varphi$-space.  Further, we will call the shift symmetry in $\varphi^0$ in the second line
as the \emph{thermal shift} and  the (right) spatial gauge transformations of $c$ in
the third line as the \emph{chemical shift}.\footnote{ The rationale for this terminology, which is inspired from the formalism of non-dissipative fluids \cite{Dubovsky:2011sj}, will be come clear in Appendix \ref{sec:ndf}. We will demonstrate there how the class of adiabatic fluids encompasses the non-dissipative fluid framework explicitly.}

It is useful to compare the redundancies in the parameterization using the reference manifold variables  with those present in our discussion of the hydrostatic state in \S\ref{sec:hydrostatics}. One  can in fact check that the redundancies in the two cases are precisely the same.
For example, in static reference gauge, the metric and gauge field on $\Mref$
assume a `Kaluza-Klein' form familiar from hydrostatics:
\begin{equation}\label{eq:ghAhStatic}
\begin{split}
\gref_{ab} \, d\varphi^a d\varphi^b
&= -\frac{1}{\Tref^2}\; \left(d\varphi^0 + \akkref_{_I} \,d\varphi^{_I} \right)^2 + \pkkref_{_{IJ}} \;d\varphi^{_I} d\varphi^{_J} \\
\gref^{ab} \;\frac{\partial}{\partial \varphi^a}\otimes \frac{\partial}{\partial \varphi^b}
&= -\Tref^2\; \frac{\partial}{\partial \varphi^0}\otimes \frac{\partial}{\partial \varphi^0}
+ \pkkref^{_{IJ}}
\prn{\frac{\partial}{\partial \varphi^{_I}}- \akkref_{_I} \frac{\partial}{\partial \varphi^0}}
\otimes \prn{\frac{\partial}{\partial \varphi^{_J}}- \akkref_{_J} \frac{\partial}{\partial \varphi^0}}  \\
\Aref_a \,d\varphi^a
&= c^{-1} \left(\frac{\muref}{\Tref} \right) c \prn{d\varphi^0 + \akkref_{_I} d\varphi^{_I}}+ \Akkref_{_I}\, d\varphi^{_I}  \\
\ukkref^a \,\frac{\partial}{\partial \varphi^a}
&= \Tref \frac{\partial}{\partial \varphi^0}\ ,\quad
\ukkref_a \,d\varphi^a
= -\frac{1}{\Tref} \; \prn{d\varphi^0 + \akkref_{_I} \,d\varphi^{_I}} 
\end{split}
\end{equation}
where $\akkref_{_I}$ is a gauge field for the thermal shift, whereas $\Akkref_{_I}$ is the spatial component of the reference gauge field. As in hydrostatics,
we have the useful rule of thumb that the tensor components with only $0$'s in down
(covariant) indices and only $I$s in up (contravariant) indices are invariant under thermal shift.
This follows from the fact that the  thermal shift invariant basis on tangent and cotangent bundles of  $\Mref$ are given by
\begin{equation}
\begin{split}
\{ d\varphi^0 + \akkref_{_I} \; d\varphi^{_I}, d\varphi^{_I} \} \,, \qquad  \text{and} \qquad
\{ \frac{\partial}{\partial \varphi^0} , \frac{\partial}{\partial \varphi^{_I}}-
\akkref_{_I}\frac{\partial}{\partial \varphi^0} \}\,,
\end{split}
\end{equation}
respectively.
It is thus convenient to replace $\{ e^{\,\sigma}_{_I}\}$ which transforms under thermal shift
with a set of thermal shift invariant vectors $\{P^{\,\sigma}_{_I}\}$ instead. Let us define the hybrid hydrodynamic projectors
\begin{equation}\label{eq:PSigmaI}
\begin{split}
P^{\,\sigma}_{_I} \equiv
\frac{\partial x^\sigma}{\partial \varphi^{_I}}- \akkref_{_I} \;\frac{\partial x^\sigma}{\partial \varphi^0} \,.
\end{split}
\end{equation}
They satisfy
\begin{equation}
\begin{split}
P^{\,\alpha}_{_I} \; \partial_\beta \varphi^{_I} &=
P^{\,\alpha}_\beta = \delta^{\,\alpha}_\beta+  u^\alpha \, u_\beta\,,
\qquad
\partial_\alpha \varphi^{_I}\;  P^{\,\alpha}_{_J} = \delta^{\, {_I}}_{_J}\ ,\\
-T\, u_\sigma \, P^{\,\sigma}_{_I} &=
\left(\partial_\sigma \varphi^0+ \akkref_{_J} \;\partial_\sigma \varphi^{_J} \right)
P^{\,\sigma}_{_I}
=  P^{\,\sigma}_{_I}\; \partial_\sigma \varphi^0+\akkref_{_I} = 0\,,\\
g_{\alpha\beta}\; P^{\,\alpha}_{_I} \,P^{\,\beta}_{_J} &= \pkkref_{_{IJ}}\,,\qquad
\pkkref^{_{IJ}}\;P^{\,\alpha}_{_I}\,P^{\,\beta}_{_J} = P^{\alpha\beta}\,,\\
g_{\alpha\beta}\; P^{\,\beta}_{_I} &= \pkkref^{_{I}}{}_{_{J}} \,\partial_\alpha \varphi^{_J}\,,\qquad
\pkkref^{_{IJ}}\;P^{\,\alpha}_{_J} = g^{\alpha\beta}\,\partial_\beta \varphi^{_I}
\end{split}
\end{equation}
Using these thermal shift invariant vectors, we can write
\begin{equation}
\begin{split}
e^\sigma_a \,\delta \varphi^a = P^{\,\sigma}_{_I} \delta \varphi^{_I} + \Kbeta^\sigma \brk{\delta \varphi^0 +  \akkref_{_I} \delta \varphi^{_I} } \,.
\end{split}
\end{equation}
This thermal shift invariant form is often useful in explicit computations.
For example, when varying
$\sqrt{-\gref}\ \Lagref\brk{\hreffields}$, we get \eqref{eq:phiLagVar}
where  $\delta \varphi^a$ always occur in this combination.

The static reference gauge has the merit that once we adopt it, only dynamical fields show up in the description; all gauge redundancies are  eliminated.  Hydrodynamics in static reference gauge is  then completely described  by the following set of degrees of freedom:
\begin{itemize}
\item $(d-1)$ spatial $\varphi^{_I}$s which satisfy
$u^\sigma \, \partial_\sigma \varphi^{_I} =0$,
\item a  field $\varphi^0$ such that $T=u^\alpha\,\partial_\alpha \varphi^0$ and
\item a field $c$ such that $\mu=u^\alpha(\partial_\alpha c) c^{-1}+ u^\alpha \,A_\alpha$.
\end{itemize}
Further, we can solve for $u^\sigma$ itself directly in terms of these dynamical fields.
We get
\begin{equation}
\begin{split}
u^\sigma = \frac{1}{\sqrt{\text{det}_{d-1}\left( g^{\mu\nu}\,
\partial_\mu \varphi^{_I} \partial_\nu \varphi^{_J}\right) }}
\;
\varepsilon^{\sigma \alpha_1\ldots \alpha_{d-1}}\;
\partial_{\alpha_1}\varphi^1 \cdots \partial_{\alpha_{d-1}}\varphi^{d-1} \,.
\end{split}
\end{equation}
This follows from the fact that $u^\sigma$ is orthogonal to the
$(d-1)$ vector fields $\partial_\sigma \varphi^{_I}$
and is hence parallel to $\varepsilon^{\sigma \alpha_1\ldots \alpha_{d-1}} \;
\partial_{\alpha_1}\varphi^1 \cdots \partial_{\alpha_{d-1}}\varphi^{d-1}$. The square-root pre-factor
then ensures the correct normalization appropriate for a $d$-velocity.

A more elegant way of writing the above expressions is to introduce a spatial volume form on the space of $\varphi^{_I}$s using the spatial
part of the pushforward co-metric $\gref^{ab}$, i.e., we define
$\pkkref^{_{IJ}}\equiv g^{\mu\nu}\; \partial_\mu \varphi^{_I} \partial_\nu \varphi^{_J}$ whose inverse then defines a thermal shift invariant spatial metric
$\pkkref_{_{IJ}}$ introduced in  \eqref{eq:ghAhStatic}. Using this expression one  can work out the parameterization of hydrodynamic fields $\{\Kbeta^\mu, \LambdaB\}$ in terms of $\{\varphi^a,c\}$. For completeness let us record these expressions which read:
\begin{equation}\label{eq:uTmuStatic}
\begin{split}
u^\sigma &= \frac{1}{(d-1)!}\
\varepsilon^{\sigma \alpha_1\ldots \alpha_{d-1}} \;
\varepsilon^{(\pkkref)}_{{_I}_{_1}\ldots {_I}_{_{d-1}}} \; \prod_{i=1}^{d-1}
\partial_{\alpha_i}\varphi^{{_I}_{_i}} \,,\\
T &= \frac{1}{(d-1)!}\
\varepsilon^{\sigma \alpha_1\ldots \alpha_{d-1}} \;
\varepsilon^{(\pkkref)}_{{_I}_{_1}\ldots {_I}_{_{d-1}}}
\partial_{\sigma}\varphi^0\; \prod_{i=1}^{d-1} \partial_{\alpha_i}\varphi^{{_I}_{_i}} \\
&= \frac{1}{(d-1)!}\
\varepsilon^{\sigma \alpha_1\ldots \alpha_{d-1}}
\varepsilon^{(\pkkref)}_{{_I}_{_1}\ldots {_I}_{_{d-1}}}
\prn{\partial_{\sigma}\varphi^0+ \akkref_{_I} \; \partial_{\sigma}\varphi^{_I}} \; \prod_{i=1}^{d-1} \partial_{\alpha_i}\varphi^{{_I}_{_i}} \\
\mu &= \frac{1}{(d-1)!}\
\varepsilon^{\sigma \alpha_1\ldots \alpha_{d-1}}
\varepsilon^{(\pkkref)}_{{_I}_{_1}\ldots {_I}_{_{d-1}}}
\prn{(\partial_\sigma c) c^{-1}+  A_\sigma} \; \prod_{i=1}^{d-1} \partial_{\alpha_i} \varphi^{{_I}_{_i}}
\end{split}
\end{equation}
where $\varepsilon^{(\pkkref)}_{{_I}_{_1}\ldots {_I}_{_{d-1}}}$ is the spatial volume
form associated with $\pkkref_{_{IJ}}$. We can also give a similar expression for thermal shift invariant vectors   $\{P^{\,\sigma}_{_I}\}$ defined in \eqref{eq:PSigmaI}:
\begin{equation}
\begin{split}
P^{\,\sigma}_{_I} = \frac{1}{(d-2)!}\varepsilon^{\sigma\alpha\beta_1\ldots\beta_{d-2}}\;
\varepsilon^{(\pkkref)}_{{_I}{_J}_{_1}\ldots {_J}_{_{d-2}}}\;
u_\alpha \; \prod_{i=1}^{d-2} \partial_{\beta_i}\varphi^{{_J}_{_i}} 
\end{split}
\end{equation}
More generally one can express a $k$-fold anti-symmetric tensor product of the projection tensors $P^{\;\alpha}_{_I}$ in terms of the $\varphi^{_I}$ fields as
\begin{equation}
\begin{split}
(k+1)!\; u^{[\sigma} \; P^{\;\alpha_1}_{{_{[I}}_{_1}}\ldots P^{\;\alpha_k]}_{{_{{I}_{_k}]}}} &= \frac{1}{(d-1-k)!}
\varepsilon^{\sigma\alpha_1\ldots\alpha_k\beta_1\ldots\beta_{d-1-k}}\;
\varepsilon^{(\pkkref)}_{{_I}_{_1}\ldots {_I}_{_k} {_J}_{_1}\ldots {_J}_{_{d-1-k}}}
 \prod_{i=1}^{d-1-k} \partial_{\beta_i}\varphi^{{_J}_{_i}} \,.
\end{split}
\end{equation}

The expressions derived above,  express the hydrodynamic fields as gauge-invariant composite fields formed out of the basic dynamic fields $\{\varphi^0, \varphi^{_I},c\}$. While this is the most economical presentation in terms of the dynamical degrees of freedom, the gauge fixing introduces a degree of non-linearity in the mapping between the physical fields and the hydrodynamical variables.  Nevertheless, there is a certain simplicity to the parameterization: the hydrodynamic equations can then be obtained by writing
$\int \sqrt{-g}\ \Lag\brk{\hfields}$ or $\int_{\Mref} \sqrt{-\gref}\ \Lagref\brk{\hreffields}$ as a functional of  $\{g_{\mu\nu},A_\alpha,\varphi^0,\varphi^{_I},c\}$ and extremising with respect to
the dynamical fields $\{\varphi^0,\varphi^{_I},c\}$. Moreover, the hydrostatic limit in the static reference gauge can be obtained by setting
\begin{equation}
\begin{split}
c = \mathbb{1}\,,\qquad \partial_\sigma \varphi^0 = T_0\; \delta^{\,0}_\sigma\,,
\qquad \partial_\sigma \varphi^{_I} = \delta^{_I}_\sigma \,,
\end{split}
\end{equation}
i.e., we can just pull-back the reference configuration through what is essentially an
identity map between the spacetime and $\Mref$. This is to be expected given the close analogy between the two formalisms noted around \eqref{eq:ghAhStatic}. Then the formalism we just  described reduces to the hydrostatic formalism described in \cite{Banerjee:2012iz,Jensen:2012jh} as expected. One can easily derive explicit expressions for the partition function by recasting the results of \S\ref{sec:Lhstatic} in the static reference gauge.

The reader familiar with the discussion of non-dissipative fluids \cite{Dubovsky:2011sj,Bhattacharya:2012zx} will undoubtedly see some similarities with the variables used in that context. However, there are some subtle (but important) distinctions; we  are not yet within the remit of that framework.\footnote{ We draw the attention of the reader to the
fact that $T$ and $\mu$ are treated symmetrically in this description, with $e^{\varphi^0}$
playing an analogous role to $c$. This is similar to a  model considered in \cite{Dubovsky:2014ys}.} The connection between the formalism outlined herein and that used in the aforementioned references is explained in Appendix \ref{sec:ndf}, where we demonstrate that non-dissipative fluids (Class ND) are a subclass of adiabatic fluids.

\subsection{Field redefinitions in Class L}
\label{sec:fieldredef}

In this subsection, we will examine the  field redefinitions of the hydrodynamic fields
$\{u^\sigma,T,\mu\}$ which leaves the on-shell physics invariant. While we are allowed to do
a general redefinition of the hydrodynamic fields, this does not translate into a general
redefinition of the Lagrange multiplier fields $\{\Kbeta^\mu,\LambdaB\}$. We remind the 
reader that we have already used up a subset of field redefinitions  so as  to have a simple
relationship between the Lagrange multiplier fields $\{\Kbeta^\mu,\LambdaB\}$ and
the original hydrodynamic fields $\{u^\sigma,T,\mu\}$, see \S\ref{sec:amotive}. 
In this subsection, we will examine
the residual redefinitions which are allowed for the Lagrange multiplier fields $\{\Kbeta^\mu,\LambdaB\}$.

One of the advantages of shifting to   $\{\Kbeta^\mu,\LambdaB\}$ was that  the hydrostatic
configurations can simply be described by aligning $\{\Kbeta^\mu,\LambdaB\}$ to the Killing fields
$\{K^\mu,\Lambda_K\}$. An admissible field redefinition should preserve this feature. This then suggests 
that we consider redefinitions of the form
\begin{equation}\label{eq:KbetaRef}
\begin{split}
\Kbeta^\mu &\mapsto \Kbeta^\mu - \diffB V^\mu = \Kbeta^\mu +\lieD_V \Kbeta^\mu\ ,\\
\LambdaB &\mapsto \LambdaB - \diffB \Lambda_V =  \LambdaB + \lieD_V  \LambdaB + [\LambdaB,\Lambda_V] -\Kbeta^\sigma \partial_\sigma \Lambda_V
\end{split}
\end{equation}
for some general diffeomorphism and flavour parameter $\{V^\mu,\Lambda_V\}$. The presence of $\diffB$  ensures that the nice features of hydrostatics survive these redefinitions. This is the most general class of redefinitions that are admissible for $\{\Kbeta^\mu,\LambdaB\}$.

In Class L, there is a more concrete way of seeing why two fluids related by \eqref{eq:KbetaRef} should be considered on-shell equivalent. Using \eqref{eq:LagConstrainedVar} we can write down the change in Lagrangian induced by field redefinitions in  \eqref{eq:KbetaRef}:
\begin{equation}
\begin{split}
\Lag\brk{\hfields} \mapsto \Lag\brk{\hfields} &+ 
\bigg(\frac{1}{\sqrt{-g}}\diffB  \prn{\sqrt{-g}\ T \, \aheat_\sigma}
+ T\, \acharge \cdot \diffB  A_\sigma \bigg)\, V^\sigma
\\
& \qquad + \frac{1}{\sqrt{-g}}\diffB  \brk{\sqrt{-g}\ T\,\acharge} \cdot \prn{\Lambda_V+
A_\sigma \, V^\sigma}  +\nabla_\mu \prn{ \ldots } \,,
\end{split}
\end{equation}
viz., the Lagrangian is shifted by terms proportional to the equations of motion and a boundary term. In a field theory, this is the most general 
redefinition admissible in the Lagrangian density. What this means in practice is that we can effectively focus on the basis of on-shell independent 
scalars parameterizing $\Lag\brk{\hfields}$, which greatly simplifies the computation (see for example Appendix \ref{sec:neutral2d}). 

An alternate way to get at the same result is to shift to the description based on reference manifolds and pullback fields. So consider then 
replacing $\Lag\brk{\hfields}$ by the functional on the reference manifold $\Lagref\brk{\hreffields}$. Variation of this functional under arbitrary variation of
fields $\{\varphi^a, c \}$ leads to terms proportional to the equations of motion $\nabla_\nu T^{\mu\nu} = J_\nu \cdot F^{\mu\nu}$ and $D_\mu J^\mu =0$, which effectively means, using the notation introduced in \eqref{eq:consLvar},
\begin{equation}
\diffCons\Lag_k\brk{\hfields} = \prn{\nabla_\alpha T^{\alpha\mu}_{(k)}-J^{\alpha}_{(k)} \cdot F^{\alpha\mu} } \, e_{\mu a} \, \delta\varphi^a +
D_\alpha J^{\alpha}_{(k)}   \cdot \bigbr{-\delta c\ c^{-1}+e^\sigma_a \,\delta\varphi^a(\;
\partial_\sigma c\ c^{-1} + A_\sigma )}\ \,.
\label{eq:fredef}
\end{equation}
Here we are working order by order in the gradient expansion, which as explained earlier is completely kosher in the absence of anomalies. $\Lag_k$ denotes the scalar contribution at $k^{\rm th}$ order in gradients. Using \eqref{eq:deltaKLambdaK}, we can relate these field redefinitions  of the pull-back fields to the redefinitions of $\{\Kbeta^\mu,\LambdaB\}$ in 
\eqref{eq:KbetaRef}:
\begin{equation}
\begin{split}
V^\mu = e^\mu_a \, \delta\varphi^a\ ,\qquad \Lambda_V =  -\delta c\ c^{-1}+e^\sigma_a \,\delta\varphi^a \partial_\sigma c\ c^{-1} \,.
\end{split}
\end{equation}

In practice we can use this redefinition freedom as follows: say we are interested in contributions to the currents at $k^{\rm th}$ order in gradients. We can implement a shift of $\hreffields$ in all the terms up to the $(k-1)^{\rm st}$ order, so that we pick up a contribution to the Lagrangian proportional to the conservation equation to one lower order than we are interested in. In other words
\begin{equation}
\begin{split}
\sum_{l=1}^{k}\, \diffCons\Lag_l\brk{\hfields} &=  
\sum_{l=1}^{k-1}\, \prn{\nabla_\alpha T^{\alpha\mu}_{(l)}-J^{\alpha}_{(l)} F^{\alpha\mu} } \, e_{\mu a} \, \delta\varphi^a \\
&\qquad +
\sum_{l=1}^{k-1}\,\nabla_\alpha J^{\alpha}_{(l)}   \cdot \bigbr{-\delta c\ c^{-1}+e^\sigma_a \,\delta\varphi^a(\;
\partial_\sigma c\ c^{-1} + A_\sigma )}\ + \cdots
\end{split}
\end{equation}
where we have only retained terms up to $k^{\rm th}$ order in the gradients.  By a suitable choice of $\delta\varphi^a$ and $\delta c$ we can eliminate some of the terms in $\Lag_k\brk{\hfields}$.

The upshot of this discussion is that we can always choose to parameterize $\Lag\brk{\hfields}$ solely in terms of the on-shell independent scalars at a given order in the gradient expansion. This has a significant effect in simplifying the computations. An explicit verification of this statement at the level of neutral fluids at second order in gradients can be found in Appendix \ref{sec:neutral2d}.\footnote{ Field redefinitions affecting the entropy current have been described previously in \S\ref{sec:classDexamples}. }

\section{Applications of adiabatic fluids}
\label{sec:examples}

Having in the previous sections given a rather abstract discussion of the Class L adiabatic fluids, we now turn to some specific examples. We first describe how neutral fluids can be understood in this language and derive the constraints arising from demanding adiabaticity on such fluids up to the second order in hydrodynamic gradient expansion. We also comment briefly on the case of charged parity-odd fluids in $3$ dimensions working to first order in the gradient expansion.  We choose these specific examples for their simplicity and also because they have been previously analyzed in the framework of non-dissipative fluids (Class ND) in \cite{Bhattacharya:2012zx} and \cite{Haehl:2013kra} respectively. Later on in \S\ref{sec:8fold} we will also have occasion to describe charged fluids, when we illustrate the general classification scheme we develop.

\subsection{Neutral fluids up to second order in gradients}
\label{sec:neutral}

Consider a neutral non-anomalous fluid for which we wish to find the constraints imposed by adiabaticity.
Since there are no charges we ignore the field $\LambdaB$ and the corresponding gauge field source $A_\mu$; thus our Lagrangian is going to be a function only of the hydrodynamic field $\Kbeta^\mu$ and the background metric source $g_{\mu\nu}$. Our strategy will be to follow intuition from hydrodynamics and write down a Lagrangian density order by order in the gradients of these fields. So we have
\begin{equation}
\Lag\brk{\Kbeta^\mu, g_{\mu\nu} } = \Lag_0\brk{\Kbeta^\mu, g_{\mu\nu} }+\Lag_1\brk{\Kbeta^\mu, g_{\mu\nu} }+\Lag_2\brk{\Kbeta^\mu, g_{\mu\nu} }+ \cdots
\end{equation}
where $\Lag_k$ involves terms with exactly $k$ derivatives acting on the fields. We will now proceed to construct the first three terms in the above gradient expansion and derive the corresponding hydrodynamic constitutive relations.

\subsubsection{Zeroth order in gradients}

At leading order in the gradient expansion, we want a scalar function built out of $g_{\mu\nu}$ and $\Kbeta^\mu$. Clearly, there is only one such scalar which is the norm of $\Kbeta^\mu$, which we can trade for the temperature $T$ from \eqref{eq:Tumuinvert}. So we can write our leading Lagrangian as
\begin{equation}
\Lag_0\brk{\Kbeta^\mu, g_{\mu\nu} } = p(T) \,, \qquad \qquad  T =
\frac{1}{\sqrt{-g_{\mu\nu} \, \Kbeta^\mu\,\Kbeta^\nu}} \,.
\label{eq:lagn0}
\end{equation}

Now we can apply the variational calculus of \S\ref{sec:classL} to this Lagrangian and extract the currents. We already know the Bianchi identities and the dynamical equations they are supposed to satisfy on general grounds. Thus all we need is the analog of \eqref{eq:LagVar} for our specific choice of $\Lag_0$. A simple calculation using \eqref{eq:varrules} leads to\footnote{ Derivatives with respect to temperature
 are denoted by a prime, viz.,  $f'(T) = \frac{df}{dT}$.}
\begin{equation}
\frac{1}{\sqrt{-g}} \,\delta\prn{\sqrt{-g}\, \Lag_0} = \frac{1}{2}
 \brk{\left(T\,p'-p\right)u^\mu\,u^\nu +p\, P^{\mu\nu}} \delta g_{\mu\nu} +
T^3\, p'\, \Kbeta_\sigma \,\delta \Kbeta^\sigma
\end{equation}
There are no boundary terms, and the currents are just what we expect for an ideal fluid
\begin{equation}
\begin{split}
T_{(0)}^{\mu\nu}&= \left(T\,p' -p\right)u^\mu\,u^\nu +p\, P^{\mu\nu} \,, \qquad J^\mu_{S,(0)}= p'\, u^\mu \\
\aheat_{(0)}^\sigma &= p'(T)\, T^2\, \Kbeta^\sigma
\end{split}
\label{eq:n0d}
\end{equation}
where we identify $\epsilon(T)=  T\, p'(T) -p(T)$ with $p(T)$ being the pressure (or negative of the free energy). We have already verified that the ideal fluid satisfies the adiabaticity equation directly in \S\ref{sec:ideal}, but it of course now also follows from the variational calculus.

\subsubsection{First order in gradients}

 Moving to the next order in gradients, we find that there are two one derivative scalars that we can write down $\Kbeta^\sigma \nabla_\sigma T$ and $\nabla_\mu \Kbeta^\mu$, both of which can be multiplied by an arbitrary function of the scalar $T$. However, these two scalars are not independent as Lagrangian entries. They are equivalent up to a total derivative term owing to the identity: $f(T)  \,\nabla_\mu \Kbeta^\mu = \nabla_\mu \prn{f(T)\,\Kbeta^\mu} -  f'(T)\, \Kbeta^\mu\, \nabla_\mu T$. We will therefore only pick one of them to include in $\Lag_1$. Since it is simpler to vary the gradient of the temperature, we parameterize the first order Lagrangian as
\begin{align}
\Lag_1\brk{\Kbeta^\mu, g_{\mu\nu} } =  \Kbeta^\sigma \,\nabla_\sigma f_1(T)
\label{eq:lagn1}
\end{align}
whose variation again leads to
\begin{align}
\frac{1}{\sqrt{-g}} \,\delta\prn{\sqrt{-g}\, \Lag_1}
&= \frac{f_1'}{2}
\prn{ \Kbeta^\sigma\, \nabla_\sigma T\, g^{\mu\nu} -T\,  \nabla_\sigma\Kbeta^\sigma\, u^\mu\, u^\nu } \delta g_{\mu\nu}  + \nabla_\sigma\prn{f_1'\, \Kbeta^\sigma\, \delta T}
\nonumber \\
& \qquad +\;  f_1'\,\prn{\nabla_\mu T -  T^3\,(\nabla_\sigma \Kbeta^\sigma)\;
\Kbeta_\mu}\delta \Kbeta^\mu
\end{align}
The stress tensor arising from adding $\Lag_1$ is again of the perfect fluid form,  except that the definitions of the energy density and pressure are shifted by terms involving $\Kbeta^\sigma\nabla_\sigma T $ and
$\nabla_\sigma \Kbeta^\sigma$. The final expressions for the transport data are then:
\begin{equation}
\begin{split}
T_{(1)}^{\mu\nu} &= -f_1' \prn{ T\, \nabla_\sigma \Kbeta^\sigma + \Kbeta^\sigma \nabla_\sigma T}
u^\mu\, u^\nu + f_1'\, \Kbeta^\sigma\,\nabla_\sigma T\, P^{\mu\nu} \\
 J^\mu_{S,(1)} &= -\frac{1}{T} \, f_1' \prn{ u^\sigma\nabla_\sigma \log T + T\, \nabla_\sigma\Kbeta^\sigma}
\label{eq:n1dA}
\end{split}
\end{equation}
In addition we have the adiabatic heat  and  pre-symplectic currents given by
\begin{equation}
\begin{split}
\aheat_{(1)}^\mu &=   f_1'\,\prn{\nabla^\mu \log T -  T^2\,(\nabla_\sigma \Kbeta^\sigma)\;
\Kbeta^\mu} \\
(\PSymplPot{})^\mu & = f_1' \, \prn{T^2\, u_\alpha\, \delta \Kbeta^\alpha + \frac{1}{2}\, T\, u^\alpha\, u^\beta \, \delta g_{\alpha \beta} } \Kbeta^\mu
\label{eq:n1dAex}
\end{split}
\end{equation}

The stress tensor appears to be in the ideal fluid form, but given that there are various gradient terms lurking around, we would like to ascertain whether there are genuine viscous contributions. The hydrodynamic stress tensor at first order is expected to contain terms involving the shear and expansion of the fluid whose coefficients are the shear and bulk viscosities, cf., the Landau frame expression given in \eqref{eq:TLandau} (see also our discussion in \S\ref{sec:classD}). To compare with conventional expressions in hydrodynamics it is useful to write the answer for the stress tensor  \eqref{eq:n1dA} in a more familiar form.

Usually hydrodynamic stress tensors are given in terms of basis of  independent tensors which are identified by invoking on-shell relations at one lower order, cf., \cite{Bhattacharyya:2008jc} for a nice review of the procedure. For neutral fluids derivatives of the temperature are typically eliminated in favour of velocity gradients;  using the conservation of the ideal fluid we obtain
\begin{align}
\nabla_\mu T &\simeq \frac{\epsilon+ p}{\epsilon'(T)} \, \Theta \, u_\mu  -
\frac{\epsilon+ p}{p'(T)}\; \acc_\mu = T\, \vs \, \Theta\, u_\mu - T\, \acc_\mu \,.
\label{eq:gradT}
\end{align}
We have introduced the speed of sound
$\vs(T)$ to simplify future expressions:
\begin{align}
\vs \equiv \frac{dp}{d\epsilon} \,.
\end{align}
If we eliminate the temperature gradients using the above, we find for the gradients of $\Kbeta^\mu$ the following expressions:
\begin{equation}
\begin{split}
\nabla_\mu \Kbeta_\nu &\simeq\frac{1}{T} \prn{\sigma_{\mu\nu} + \omega_{\mu\nu}  + \frac{\Theta}{d-1} \, P_{\mu\nu} } -\frac{\vs}{T}\; \Theta\, u_\mu\,u_\nu \,, \\
\nabla_\sigma \Kbeta^\sigma &\simeq \frac{\Theta}{T}\, (1+ \vs)    \,.
\end{split}
\label{eq:gradKB}
\end{equation}

Armed with this information we can then rewrite the hydrodynamic currents in \eqref{eq:n1dA} as
\begin{equation}
\begin{split}
T_{(1)}^{\mu\nu} &
\;\;\ensuremath{\stackrel{\text{ideal}}{\simeq}} \;\;
 -f_1' \, \Theta\, \brk{ u^\mu\, u^\nu + \vs\,  P^{\mu\nu} } \,, \\
 J^\mu_{S,(1)} &
\;\;\ensuremath{\stackrel{\text{ideal}}{\simeq}} \;\;
-\frac{1}{T}\, f_1' \Theta\, u^\mu \,,
\end{split}
\label{eq:n1d}
\end{equation}
where we have made clear with the notation $\;\ensuremath{\stackrel{\text{ideal}}{\simeq}} \;$ that we are only taking the ideal part of the fluid on-shell. We also can check that the free energy current vanishes $\mathcal{G}_{(1)}^\sigma \simeq 0$ using  \eqref{eq:GDef}. Further, using that we have the pre-symplectic potential in \eqref{eq:n1dAex}, one can obtain the Komar charge using \eqref{eq:JSK}.  Noting as described in \S\ref{sec:NoetherL} that we only need the variation of the background metric  under Lie transport by $\Kbeta^\mu$ i.e.,  $\diffB g_{\mu\nu} = 2\, \nabla_{(\mu}\Kbeta_{\nu)}$  we have
\begin{equation}
 \Komar^{\mu\nu}[\Bfields]
\;\;\ensuremath{\stackrel{\text{ideal}}{\simeq}} \;\;
0 \,.
\end{equation}

We are now in a position to discuss some physical aspects of the first order Class L term \eqref{eq:lagn1}.
The first peculiar feature to note  is that  while the adiabaticity equation  \eqref{eq:naadiabatic} is clearly satisfied, the form of entropy current is counterintuitive. It is well known that a neutral fluid has no correction to the ideal fluid entropy current  at first order. In fact, by using the standard current algebra logic in hydrodynamics, one can show that an entropy current with non-negative divergence demands vanishing of the coefficient of $\Theta\, u^\mu$ at first order (similarly for the other a-priori allowed vector $\acc^\mu$)
 \cite{Bhattacharyya:2012nq}.

Clearly, in the present case what is happening is that the entropy production is being compensated for by the energy-momentum tensor off-shell (note that we have not imposed the first order conservation equations as yet). However, we can make a somewhat more clean statement by examining the stress tensor
itself.  Since we are free to make a certain amount of field redefinitions as discussed in \S\ref{sec:fieldredef}, which amount in hydrodynamics language to choice of fluid frame, in comparing the stress tensor we should account for this. At leading order in the gradient expansion the simplest way to proceed is to project the stress tensor \eqref{eq:n1d}
onto frame invariant (i.e., field redefinition independent) tensor structures. This is implemented by employing the tensor and scalar projectors $C_T^{\mu\nu}$ and $C_S$ respectively
\cite{Banerjee:2012iz}:
\begin{equation}
\begin{split}
C_T^{\mu\nu} &= P^{\mu\alpha}\, P^{\nu\beta}\, T^{(1)}_{\alpha \beta} -
\frac{1}{d-1}\, P^{\mu\nu}\, P^{\alpha\beta}\, T^{(1)}_{\alpha\beta} \,,
 \\
C_S &= \frac{1}{d-1} \, P^{\mu\nu} \, T^{(1)}_{\mu\nu} - \vs \, u^\mu\,u^\nu\,
T^{(1)}_{\mu\nu} \,.
\end{split}
\label{eq:frameproj}
\end{equation}
Acting with these operators on  \eqref{eq:lagn1}, we find the results to vanish identically. In other words, there is no frame independent on-shell information in the stress tensor. More importantly, the term in
\eqref{eq:lagn1} which resembles the bulk viscosity term $\Theta \, P^{\mu\nu}$ should not be interpreted as such; it is not a genuine contribution to the dissipative transport.

For the first order Lagrangian \eqref{eq:lagn1} we saw that the process of taking the ideal fluid part on-shell led to a stress tensor with no physical information.
We claim that the field redefinitions described in \S\ref{sec:fieldredef} can be used to remove \eqref{eq:lagn1} by setting $f_1 = {\rm constant}$, leading to the same conclusion as above.

Let us  see how this can be used at first order for the Lagrangian term \eqref{eq:lagn1}. We start with the ideal fluid contribution and write
\begin{align}
\diffCons\Lag_0 \brk{\Kbeta} + \Lag_1\brk{\Kbeta} &=  p'(T)\,
\bigg( \nabla_\mu \log T - \vs\, \Theta\, u_\mu + \acc_\mu \bigg)\;
e^\mu_a \, \delta\varphi^a  + f_1'(T)\, u^\sigma \nabla_\sigma \log T + \cdots
\end{align}
Writing $\vs\,p'\, \Theta = \prn{\vs\, p' - T\, \prn{\vs\,p'}' }\, u^\mu \nabla_\mu \log T $ up to a total derivative we see that picking
\begin{equation}
e^\mu_a \, \delta\varphi^a = - \frac{f_1'}{p' (1+\vs) - T\, \prn{\vs\,p'}' } \; u^\mu
\end{equation}
 we can eliminate the $\Lag_1$ completely as required.

\subsubsection{Second order in gradients}
\label{sec:2ndWeyl}

At the second order in gradient expansion we have many new scalar functions built from the background metric and hydrodynamic fields. One can use the standard fluid dynamical parameterization and write the terms as:\footnote{ The advantage of this parameterization is that it is easier to read off the energy-momentum tensor upon variation. It is straightforward to use \eqref{eq:hydrofields} to rewrite these in terms of $\Kbeta^\mu$ and its derivatives if necessary.}
\begin{equation}
\begin{split}
&\sigma^2 \equiv \sigma_{\mu\nu}\,\sigma^{\mu\nu}\,,\quad \omega^2 \equiv \omega_{\mu\nu}\,\omega^{\nu\mu}\,,\quad \acc^2 \equiv \acc_\mu\,\acc^\mu
\,, \quad\Theta^2 \,, \quad R \\
&  \nabla_\mu T\, \nabla^\mu T\,, \qquad
\Theta\, u^\mu\nabla_\mu T\,,\qquad \acc^\mu \,\nabla_\mu T\,,\qquad
u^\mu \nabla_\mu T \, u^\nu\nabla_\nu T\,, \\&
R_{00} \equiv R_{\mu\nu} \, u^\mu\,u^\nu \,,\quad u^\mu \nabla_\mu \Theta \,, \qquad
\nabla^2 T\,, \qquad u^\mu\,u^\nu \nabla_\mu\nabla_\nu T \,,
\end{split}
\label{eq:basisN2}
\end{equation}
where we have introduced the shear tensor $\sigma_{\mu\nu}$ and the vorticity tensor $\omega_{\mu\nu}$; cf., \eqref{eq:uder} and Table \ref{notation:tabFields} for their definition. A-priori we have thirteen independent functions of temperature multiplying these scalars and making for a rather formidable computation.\footnote{ A
classification of independent second order scalars was done in \cite{Bhattacharyya:2012nq}. Specifically, off-shell genuine second order scalars were shown to be five in total, which are the last four scalars in \eqref{eq:basisN2} along with the Ricci scalar $R$. The others are products of one-derivative objects. While \cite{Bhattacharyya:2012nq} took these to be the first four scalars of \eqref{eq:basisN2} after using
\eqref{eq:gradT},  we have a-priori included the terms in the second line since we choose to remain off-shell.}
However, there are some simplifications which we can exploit: 
\begin{enumerate}
\item[(i)] the four terms in the third line can be related to others up to total derivatives 
\item[(ii)] the four terms in the second line can be related to those in the first by a first order field redefinition (one chooses
$e^\mu_a\, \delta\varphi^a$ to be aligned along either $\acc^\mu$ or $\Theta \,u^\mu$).
\end{enumerate}
 All in all we have five independent terms to consider which still makes for a somewhat complex computation. The end result is that the adiabatic part of 15 independent transport coefficients for a neutral fluid at second order \cite{Bhattacharyya:2012nq} is determined in terms of five functions of temperature, pretty much along the lines of the non-dissipative effective action computation of \cite{Bhattacharya:2012zx}.  We postpone a full discussion of how this works in full detail (including explicit verification of our field redefinitions) to Appendix \ref{sec:neutral2d}.

For now we will restrict ourselves to Weyl invariant neutral fluids which are much easier to describe. For one there are only 5 independent transport coefficients \cite{Baier:2007ix,Bhattacharyya:2008jc}. Furthermore,  since a pre-requisite for Weyl invariance is that the Lagrangian must be invariant under Weyl rescalings of the background metric $g_{\mu\nu}$, we also have a reduction in the number of terms which enter the Lagrangian.  The Weyl transformation properties of various fields are well known. We further develop a Weyl covariant formulation of adiabatic hydrodynamics  to deal with fluids arising from conformal field theories, extending \cite{Loganayagam:2008is}, in Appendix \ref{sec:aweyl},  where the reader can find some of the necessary details for the computations below.  

The Weyl covariant second order scalars at our disposal are (each with Weyl weight\footnotemark\ $w=+2$)
\begin{align}
\sigma^2 \,,\qquad \omega^2 \,, \qquad (\RWeyl) \,,\qquad
g^{\mu\nu}\, (\DWeyl_\mu \log T)\, (\DWeyl_\nu \log T)\,.
\end{align}
\footnotetext{ A tensor is said to have Weyl weight $w$ if under Weyl rescalings of the background metric $g_{\mu\nu} \to e^{2\phi}\, {\tilde g}_{\mu\nu}$ it transforms homogeneously with a rescaling $e^{-w\phi}$. The metric itself has Weyl weight $w=-2$ with these conventions. For more details please see 
Appendix \ref{sec:aweyl}.}Using the identification \eqref{eq:HydroWeylW}, the Weyl covariant derivative is defined in \eqref{DWeyl:eq} and the associated  Ricci scalar in \eqref{eq:WeylcovR2}. 
Since the temperature $T$ has Weyl weight $w=+1$ it follows that the Lagrangian which is invariant under Weyl transformations has to take the form
\begin{equation}
\begin{split}
\Lag_2^\Wey &=
k_\sigma \,T^{d-2} \, \sigma^2+ k_\omega \,T^{d-2} \, \omega^2
 \\&
+k_R\, T^{d-2} \bigg[ R - (d-2)\, (d-1)\,\acc^2 +\frac{d-2}{d-1} \,\Theta^2 -2\,(d-2) (d-1) \, \acc^\alpha \nabla^\alpha \log T \\
  & \qquad\qquad\quad + 2\, (d-2)\,\Theta\, u^\alpha\nabla_\alpha \log T \bigg]
 \\ &
+ k_T\, T^{d-2}\, \prn{ (\nabla_\mu \log T)^2  + 2\, \acc^\mu \nabla_\mu \log T - \frac{2\,\Theta}{d-1} \, u^\mu
\, \nabla_\mu \log T + \acc^2 - \frac{\Theta^2}{(d-1)^2}
}
  \end{split}
\label{eq:weyl2d}
\end{equation}
where $k_\sigma$, $k_\omega$, $k_R$ and $k_T$ are constants. All the dependence on the thermal vector is implicit in \eqref{eq:weyl2d}; if necessary we can convert all the terms to appropriate combinations of $\Kbeta^\mu$ and its derivatives.  The field redefinition freedom discussed in
\S\ref{sec:fieldredef} allows us to set $k_T =0$ which we shall do forthwith (cf., Appendix \ref{sec:neutral2d} for further details).

The variation of the various terms in the Lagrangian can be computed using the rules given in \eqref{eq:varrules} in a straightforward (albeit tedious) manner. One of the advantages of using the standard parameterization in terms of the velocity and temperature instead of $\Kbeta^\mu$ is that simplifications at intermediate steps using hydrodynamic identities are transparent. We  present the variation of the full neutral fluid in Appendix \ref{sec:neutral2d}; see \eqref{eq:deltsigsq}-\eqref{eq:deltaR} from which the relevant details for the Weyl invariant fluid can be extracted. Let us therefore pass directly to a discussion of the stress tensor.

The Weyl covariant stress tensor for conformal fluid is expressed in a succinct manner in the following basis of five independent tensors \cite{Loganayagam:2008is} (see also \cite{Rangamani:2009xk}) as:\footnote{ When comparing with the expressions in these papers we warn the reader that there are some convention differences  (mostly involving factors of two and the sign in the definition of $\omega_{\mu\nu}$ and some index contractions). See footnote \ref{fn:conventions} for a mapping between conventions of various papers.}
\begin{equation}\label{eq:WeylT2}
\begin{split}
T_{(2),\Wey}^{\mu\nu} &=
 \tau\, u^\alpha \DWeyl_\alpha \, \sigma^{\mu\nu}
+  \kappa \, C^{\mu\alpha\nu\beta}\,u_\alpha \,u_\beta \\ &
\quad+\; \lambda_1\, \sigma^{\langle \mu \alpha}\, \sigma_{\alpha}{}^{\nu\rangle}+
\lambda_2\,  \sigma^{\langle \mu \alpha} \, \omega_{\alpha}{}^{\nu\rangle}+ \lambda_3\,
\omega^{\langle \mu \alpha}\, \omega_{\alpha}{}^{\nu\rangle} \,,
\end{split}
\end{equation}
where the longitudinal Weyl covariant derivative evaluates to 
\begin{equation}
 u^\alpha \DWeyl_\alpha \, \sigma^{\mu\nu} = P^\mu_\rho P^\nu_\sigma \, u^\alpha \, \nabla_\alpha \sigma^{\rho\sigma}+ \frac{\Theta}{d-1} \sigma^{\mu\nu}\,.
\end{equation}
This expression is written in the so called Landau frame where the corrections to the ideal fluid stress tensor are demanded to be perpendicular to the velocity field, i.e., $T^{\mu\nu} = T^{\mu\nu}_{(0)} +
\sum_{k\geq 1} \, T^{\mu\nu}_{(k)} $ with $u_\mu \,T^{\mu\nu}_{(k)} =0$.
We can equivalently write \eqref{eq:WeylT2} in a basis adapted to our classification scheme: 
\begin{equation}\label{eq:WeylT2eightfold}
\begin{split}
T_{(2),\Wey}^{\mu\nu} &= 
(\lambda_1 - \kappa)\, \sigma^{<\mu\alpha} \sigma_{\alpha}^{\nu>} + \left(\lambda_2 + 2\,\tau -2 \kappa\right)\, \sigma^{<\mu\alpha} \omega_{\alpha}^{\ \nu>} 
\\
& \quad
+ \tau \left( u^\alpha \DWeyl_\alpha \sigma^{\mu\nu} - 2\, \sigma^{<\mu\alpha} \omega_{\alpha}^{\ \nu>}\right)  
+ \lambda_3 \, \omega^{<\mu\alpha} \omega_{\alpha}^{\ \nu>}
\\
&\quad+ \kappa \,\left(C^{\mu\alpha\nu\beta}\,u_\alpha\,u_\beta + \sigma^{<\mu\alpha} \sigma_{\alpha}^{\nu>} + 2\, \sigma^{<\mu\alpha} \omega_{\alpha}^{\ \nu>} \right)\,.
\end{split}
\end{equation}
As we will see shortly, this adiabaticity adapted basis is more natural for it does not mix the different classes in the eightfold way; each term will turn out to be at home in a unique class. The first two terms will turn out to be forced to vanish in Class L, while the remaining three will be unconstrained. 

 The raw expressions obtained from the variation in Appendix \ref{sec:neutral2d} are somewhat unilluminating written as they are in a non-standard basis of tensors. As before we have to use the on-shell equations of motion for the ideal fluid \eqref{eq:gradT} to eliminate the thermal gradient terms. A somewhat more tricky proposition is the fact that the stress tensor which solves the adiabaticity equation is not necessarily in the Landau frame. Since the solution to the adiabaticity equation \eqref{eq:naadiabatic} in Class L for non-anomalous fluids has $J^\mu_S = s\, u^\mu$ one may in fact view the result as naturally being cast in the entropy frame (see also \cite{Bhattacharya:2012zx,Haehl:2013kra}). To compare the results with the Landau frame presentation, we first switch off the first order terms (since they carry no physical information). We then project the stress tensor computed by the variational principle  onto the  frame invariant tensor and scalar parts. This is a relatively trivial exercise and one can then read off the coefficients of the independent tensors used in \eqref{eq:TLandau}. The projectors in question are given explicitly in \eqref{eq:frameproj}. Carrying out the aforementioned computation we find the following set of transport coefficients for a Weyl invariant neutral fluid \cite{Bhattacharyya:2008mz}\footnote{ The first derivation of the second order transport coefficients was carried out explicitly  in  $d=4$ by \cite{Baier:2007ix,Bhattacharyya:2008jc}. }
\begin{equation}
\begin{split}
 \eta &= \zeta =0 \,, \\
 \tau &= - \prn{2\, (d-2)\, k_R +2\,k_\sigma}\, T^{d-2} \,, \\
 \kappa &= -2\, (d-2)\, k_R\, T^{d-2} \,,\\
 \lambda_1 &= -2\, (d-2)\, k_R\, T^{d-2} \,,\\
 \lambda_2 &=4\, k_\sigma T^{d-2} \,, \\
 \lambda_3 &= -2 \prn{ (d-2)\, k_R - 2\, k_\omega} T^{d-2}\,.
\end{split}
\label{eq:weyltcfs}
\end{equation}
The scaling with temperature can of course be determined on dimensional grounds. Equivalently, the Weyl covariant stress tensor in Class L is forced to take the form
\begin{equation} \label{eq:TWeylNeutral2ClassL}
\begin{split}
T_{(2),\Wey}^{\mu\nu} &=
\tau \left( u^\alpha \DWeyl_\alpha \sigma^{\mu\nu} - 2\, \sigma^{<\mu\alpha} \omega_{\alpha}^{\ \nu>}\right)  
+ \lambda_3 \, \omega^{<\mu\alpha} \omega_{\alpha}^{\ \nu>}
\\
&\quad + \kappa \,\left(C^{\mu\alpha\nu\beta}\,u_\alpha\,u_\beta + \sigma^{<\mu\alpha} \sigma_{\alpha}^{\nu>} + 2\, \sigma^{<\mu\alpha} \omega_{\alpha}^{\ \nu>} \right) 
\end{split}
\end{equation}
which is written in the basis of \eqref{eq:WeylT2eightfold} and can be derived from a two-derivative Lagrangian density
\begin{align}
\Lag_2^\Wey &=
\quarter \brk{ -\frac{2\,\kappa}{(d-2)} (\RWeyl) + 2\,(\kappa-\tau) \, \sigma^2+ 
(\lambda_3 -\kappa)  \, \omega^2}\,.
\label{eq:weyl2lambda}
\end{align}

What is interesting about the result \eqref{eq:weyltcfs} is the following: given that there are a-priori three parameters allowed in our Lagrangian, $\{k_\sigma, k_\omega, k_R\}$, after exploiting field redefinition freedom, we expect two relations between the five transport coefficients. These can be ascertained by inspection of \eqref{eq:TWeylNeutral2ClassL} to be the simple linear relations:
\begin{equation}
\lambda_1 = \kappa \,, \qquad  \lambda_2 = 2 \, (\kappa-\tau) \,.
\label{eq:weylrelns}
\end{equation}
These relations are actually quite fascinating;  we have an infinite class of hydrodynamic constitutive relations for which they hold thanks to the holographic fluid/gravity correspondence, cf., \cite{Hubeny:2011hd}.
We will return to a complete discussion of  holographic fluids and its relation to the adiabatic eightfold way
in \S\ref{sec:8fold}.

\subsection{Parity-odd fluids in 3 dimensions}
\label{sec:podd}

Our second example concerns the class of parity-odd charged fluids in 3 spacetime dimensions. This system has been described in the non-dissipative effective action framework originally \cite{Nicolis:2011ey} and was revisited more recently in \cite{Haehl:2013kra}. The investigations of the latter reference revealed that there is some tension in incorporating aspects of Hall transport in this framework (see also \cite{Bhattacharya:2012zx}). This has been addressed in an intriguing recent development \cite{Geracie:2014iva}, wherein the Hall viscosity coefficient was captured in terms of a non-local term in the effective action. For the present we will focus on local Lagrangians, but will comment on the non-local terms at the end. Given our discussion of the neutral fluid in \S\ref{sec:neutral} we will be a bit brief in the following, indicating just the salient results.

\subsubsection{Zeroth order in gradients}

Since we are dealing with charged fluids, we now have the full set of hydrodynamic fields $\hfields$ to consider.  At leading order in the gradient expansion, we need a scalar function of these fields. A moment's thought suffices to note that the only function of relevance is a scalar function of temperature and chemical potential (which will be the Gibbs free energy for the system). To wit, we have
\begin{equation}
\Lag_0\brk{\hfields} = p(T,\mu) \,.
\end{equation}

Applying the variational calculus of \S\ref{sec:classL} using \eqref{eq:varrules} leads to\footnote{ We will continue to denote temperature derivatives with a prime, while derivatives with respect to chemical potential are denoted with an over-dot. }
\begin{equation}
\begin{split}
\frac{1}{\sqrt{-g}} \,\delta\prn{\sqrt{-g}\, \Lag_0} &=
\frac{1}{2}  \prn{\left(T\,p' + \mu\, {\dot p} -p\right)u^\mu\,u^\nu +p\, P^{\mu\nu}} \delta g_{\mu\nu}
 + {\dot p}\, u^\sigma \delta A_\sigma \\
& \qquad
+\;\prn{T\, p'\,  + \mu \, \dot{p} }\, T^2\,  \Kbeta_\sigma \,\delta \Kbeta^\sigma
+ T\, {\dot p}\, (\delta \LambdaB  +A_\sigma\,\delta \Kbeta^\sigma)
\end{split}
\end{equation}
There are no boundary terms, and we have the currents for an ideal charged fluid
\begin{equation}
\begin{split}
T_{(0)}^{\mu\nu}&= \left(T\,p' + \mu\, {\dot p} - p\right)u^\mu\,u^\nu +p\, P^{\mu\nu} \,,
\qquad J^\mu_{(0)}= {\dot p}\, u^\mu \,,
\qquad J^\mu_{S,(0)}= p'\, u^\mu \\
\aheat_{(0)}^\sigma &=  \prn{ T\, p' + \mu\, {\dot p} }\, T\, \Kbeta^\sigma \,,\qquad
\acharge_{(0)} =   {\dot p}
\end{split}
\label{eq:p0d}
\end{equation}
In the present instance, $p$ is the pressure of the system and the charge density is given by the thermodynamics to be 
$q = \dot{p}$ and $\epsilon= \,p' + \mu\, {\dot p} - p$.

\subsubsection{First order in gradients}

 Moving to the next order in gradients, we find that there are no interesting parity-even one derivative scalars build from $\hfields$. The argument for this follows along similar lines as that presented in \S\ref{sec:neutral},
 so we will refrain from repeating it again here. Physically, of course, this is easily understood by noting that there are no non-trivial solutions to the adiabaticity equation at first order in gradients.

 However, if we have a system that is parity-odd, then in 3 spacetime dimensions we can write down two scalars
 which, following \cite{Haehl:2013kra}, we parameterize as
\begin{equation}
\Lag_1\brk{\hfields} =
\tilde{\mathfrak w}(T,\mu)\, \varepsilon^{\rho\sigma\lambda}\, u_\rho\, \nabla_\sigma u_\lambda
+ \tilde{\mathfrak b}(T,\mu)\,  \varepsilon^{\rho\sigma\lambda}\, u_\rho\, \nabla_\sigma A_\lambda \,.
\label{eq:lag1podd}
\end{equation}
In fact these are the two terms which are allowed in the hydrostatic partition function \cite{Jensen:2011xb}.
We have set to zero the parity-even first order terms such as the charged analog of the term discussed in \eqref{eq:lagn1}. They do not contribute to physical transport data. As a result, one can essentially view
$\Lag_1\brk{\hfields}$ as the off-shell extension of the equilibrium partition function.

Varying the Lagrangian density we find for the hydrodynamic currents
\begin{align}
T^{\alpha\beta}_{(1)}
&=
	2\,\varepsilon^{(\alpha \rho \sigma}\, u^{\beta)}
	\big(
	2\, \tilde{\mathfrak w} \, \nabla_\rho u_\sigma + \tilde{\mathfrak b}\, \nabla_\rho A_\sigma
	-u_\rho\, \brk{\tilde{\mathfrak w}' \nabla_\sigma T + \dot{\tilde{\mathfrak w}} \, \nabla_\sigma \mu }
	\big)
\nonumber \\
& \qquad
	+ \; u^\alpha \,u^\beta\,
	\prn{
	\brk{T\, \tilde{\mathfrak w}'  + \mu\, \dot{\tilde{\mathfrak w}} +2 \, \tilde{\mathfrak w}} \, \Omega +
	\brk{T\, \tilde{\mathfrak b}'  + \mu\, \dot{\tilde{\mathfrak b}} + \, \tilde{\mathfrak b}} \, \mathfrak{B} }
\nonumber \\
J^\alpha_{(1)}
&= 	\varepsilon^{\alpha \rho \sigma}\, \tilde{\mathfrak b}\, \nabla_\rho u_\sigma -
	\varepsilon^{\alpha \rho \sigma}\, u_\rho \prn{\tilde{\mathfrak b}' \,\nabla_\sigma T +
	\dot{\tilde{\mathfrak b}}\, \nabla_\sigma \mu }
	+ \brk{ \dot{\tilde{\mathfrak w}}\, \Omega + \dot{\tilde{\mathfrak b}}\, {\mathfrak B} }  u^\alpha
\label{eq:JTcl}
\end{align}
with the  pre-symplectic potential
\begin{align}
(\PSymplPot{})^\sigma_{(1)}
&= 	\tilde{\mathfrak w}(T,\mu)\, \varepsilon^{\rho\sigma\lambda}\, u_\rho\, \delta u_\lambda
+ \tilde{\mathfrak b}(T,\mu)\,  \varepsilon^{\rho\sigma\lambda}\, u_\rho\, \delta A_\lambda
\end{align}
and the adiabatic heat current and adiabatic charge density entering the Bianchi identities
\begin{align}
\aheat^\alpha_{(1)}
&= 	\prn{
	\brk{T\, \tilde{\mathfrak w}'  + \mu\, \dot{\tilde{\mathfrak w}} + 2 \,\tilde{\mathfrak w}} \, \Omega +
	\brk{T\, \tilde{\mathfrak b}'  + \mu\, \dot{\tilde{\mathfrak b}} +  \tilde{\mathfrak b}} \, \mathfrak{B} } \, u^\alpha
\nonumber \\
&\qquad + \varepsilon^{\alpha\rho\sigma} \left( 2\, \tilde{\mathfrak w}\, \nabla_\rho u_\sigma
           - \dot{\tilde{\mathfrak w}}\, u_\rho\, \nabla_\sigma \mu - \tilde{\mathfrak w}'\, u_\rho \,\nabla_\sigma T
           + \tilde{\mathfrak b}  \,\nabla_\rho\, A_\sigma \right)
\label{eq:POdd1}\\
\acharge_{(1)}
&= 	 \dot{\tilde{\mathfrak w}}  \, \Omega + \dot{\tilde{\mathfrak b}} \, \mathfrak{B}
\label{eq:POdd2}
\end{align}
where we defined the parity-odd scalars
\begin{align}
\Omega \equiv \varepsilon^{\rho\sigma\lambda}\, u_\rho\, \nabla_\sigma u_\lambda
\,, \qquad
\mathfrak{B} \equiv \frac{1}{2} \varepsilon^{\rho\sigma\lambda} \, u_\rho B_{\sigma\lambda} = \varepsilon^{\rho\sigma\lambda}\, u_\rho\, \nabla_\sigma A_\lambda \,.
\end{align}
The entropy current derived from (\ref{eq:POdd1}) and (\ref{eq:POdd2}) reads
\begin{equation}
  J_{S,(1)}^\mu = \left( \tilde{\mathfrak w}' \,\Omega + \tilde{\mathfrak b}' \, \mathfrak{B} \right)u^\mu \,.
\end{equation}
These expressions are reasonably similar to the ones derived in the non-dissipative effective action formalism by \cite{Haehl:2013kra}. One can pursue their algorithm to extract the transport coefficients as we summarize below.

Firstly, the most general parity-odd first order stress tensor and charge current are given by the following Landau frame expressions 
\cite{Jensen:2011xb}\footnote{ The analysis of  \cite{Jensen:2011xb} also a-priori allows the two further parity-even vectors contributions in the charge current,  viz., $ J^\alpha_{(1)} = -\chi_T \; T \,\acc^\alpha + \chi_E \; E^\alpha $.
 We have used the fact that the only parity-even contribution compatible with the second law is the conductivity term 
 exhibited in \eqref{eq:JEntr} and thus set $\chi_T = \chi_E =0$.} 
\begin{align}
T^{\alpha\beta}_{(1)} &= 
	\left(-\zeta \, \Theta +{\tilde \chi}_B \, {\mathfrak B} +{\tilde \chi}_\Omega \,  \Omega \right)\; P^{\alpha\beta} 
	- 2\,\eta \, \sigma^{\alpha\beta} - {\tilde \eta}_H \; \varepsilon^{\mu\nu(\alpha} \,u_\mu\,\sigma^{\beta)}_{\ \nu} \,,
\label{eq:TEntr} \\
J^\alpha_{(1)} &=  \sigma_{_{\text{Ohm}}} \; 
		\cv^\alpha + {\tilde \chi}_T \; \varepsilon^{\alpha\mu\nu}\,u_\mu\,\nabla_\nu T
		+ {\tilde \chi}_E\; \varepsilon^{\alpha\mu\nu}\,u_\mu\, E_\nu  
		+  {\tilde \sigma}_H \,   \varepsilon^{\alpha\mu\nu}\,u_\mu\, \cv_\nu  \,. 
\label{eq:JEntr}
\end{align}
where we use the parity-even vector introduced in \eqref{eq:cvdef}.  The coefficient ${\tilde \eta}_H$ is called the Hall viscosity and ${\tilde \sigma}_H$ is the (hydrodynamic) Hall conductivity. 
We will refer to ${\tilde \chi}_T$ as the odd Ohm conductivity and ${\tilde \chi}_T$ as the odd thermal conductivity. 

To compare this expression with \eqref{eq:JTcl} we have to project once again onto frame invariant combinations. 
Doing so we find the following relations (setting  $\mue \equiv \tfrac{\epsilon+p}{q} = \mu + \frac{sT}{q}$ 
and $\vs \equiv \left[\frac{\partial  p}{\partial \epsilon}\right]_q$ with $q ={\dot p}$ for brevity): 
\begin{equation}
\begin{split}
{\tilde \chi}_\Omega 
&= 
	\vs \left( 2\tilde{\mathfrak w} - T\,  \frac{\partial \tilde{\mathfrak w}}{\partial T}
       - \mu \frac{\partial \tilde{\mathfrak w}}{\partial \mu}\right) +\left[\frac{\partial  p}{\partial q}\right]_\epsilon \left( \tilde{\mathfrak b}- \frac{\partial \tilde{\mathfrak w}}{\partial \mu} \right) 
 \,,\\
{\tilde \chi}_B 
&=
	 \vs \left( \tilde{\mathfrak b}-T \frac{\partial \tilde{\mathfrak b}}{\partial T}
         - \mu \frac{\partial \tilde{\mathfrak b}}{\partial \mu}\right) -\left[\frac{\partial  p}{\partial q}\right]_\epsilon \,\frac{\partial \tilde{\mathfrak b}}{\partial \mu} 
\,, \\
T\, {\tilde \chi}_T 
&= 
	\left(\tilde{\mathfrak b} - T \frac{\partial \tilde{\mathfrak b}}{\partial T} - \mu \frac{\partial \tilde{\mathfrak b}}{\partial \mu} \right)
	- \frac{1}{\mue} \left( 2\tilde{\mathfrak w} - T \frac{\partial\tilde{\mathfrak w}}{\partial T} 
	- \mu \frac{\partial \tilde{\mathfrak w}}{\partial \mu} \right)  
\,, \\
{\tilde \chi}_E 
&= 
            -\frac{\partial \tilde{\mathfrak b}}{\partial \mu} +\frac{1}{\mue} \left( \frac{\partial \tilde{\mathfrak w}}{\partial\mu} -\tilde{\mathfrak b}\right) 
\,,\\
{\tilde \sigma}_H + {\tilde \chi}_E  
&= - \frac{2}{\mue} \left( \tilde{\mathfrak b} - \frac{1}{\mue} \, \tilde{\mathfrak w}\right)\, ,\\
 \sigma_{_{\text{Ohm}}} &= \eta =\zeta = {\tilde \eta}_H = 0 \, .
\end{split}
\label{eq:System2}
\end{equation}

We see from \eqref{eq:System2} that the stress tensor does not have any frame-invariant tensor data and thus the Hall viscosity
${\tilde \eta}_H$ predicted by \eqref{eq:lag1podd} vanishes, consistent with \cite{Bhattacharya:2012zx,Haehl:2013kra}. Examining the charge current, one finds that the Hall conductivity 
${\tilde \sigma}_H$ and ${\tilde \chi}_E$ can be taken to be the independent transport coefficients: 
\begin{equation}
\begin{split}
\tilde{\mathfrak w}  &= \frac{\mue^2}{2} \left[\frac{\mue\,\frac{\partial}{\partial\mu} \left({\tilde\chi}_E + {\tilde \sigma}_H\right)+2\tilde{\chi}_E}{\frac{\partial}{\partial \mu} \prn{\frac{sT}{q}}} \right] +\frac{\mue^2}{2}\, \left({\tilde \chi}_E+ {\tilde \sigma}_H\right)\,, \\
\tilde{\mathfrak b}  &= \frac{\mue}{2} \left[\frac{\mue\,\frac{\partial}{\partial\mu} \left({\tilde\chi}_E + {\tilde \sigma}_H\right)+2\tilde{\chi}_E}{\frac{\partial}{\partial \mu} \prn{\frac{sT}{q}}} \right]  \,.
\end{split}
\label{eq:wbsol}
\end{equation}
It is a-priori tempting, based on the linear relation involving the odd Ohm and Hall conductivities, to speculate that one can fix ${\tilde \sigma}_H$ in terms of the other transport coefficients. 

We expect in general three relations amongst the set 
$\{{\tilde \chi}_\Omega, {\tilde \chi}_B, {\tilde \chi}_T, {\tilde \chi}_E, {\tilde \sigma}_H\}$. One of these is the expression in the penultimate line of \eqref{eq:System2}. Another which can be ascertained by eliminating the Lagrangian scalars from the first four lines of \eqref{eq:System2} is
\begin{equation}
{\tilde \chi}_B - \frac{1}{\mue}\, {\tilde \chi}_\Omega - \vs \, T\, {\tilde \chi}_T = \left[\frac{\partial p}{\partial q} \right]_\epsilon
 {\tilde \chi}_E
\label{eq:poddPF1}
\end{equation}	
These results of course agree with those derived earlier in \cite{Jensen:2011xb} (see also \cite{Banerjee:2012iz}). The final relation can be written using \eqref{eq:wbsol}  eliminating the Lagrangian scalars $\tilde{\mathfrak w}, \tilde{\mathfrak b}$. We have not been able to derive a simple closed form answer, since we seem to need to employ thermodynamic identities in an involved fashion.

Passing to a simpler context of Weyl invariant fluids, reveals an unnecessary nuance. Now,   $\zeta = {\tilde \chi}_B = {\tilde \chi}_\Omega =0 $ by the tracelessness of the stress tensor and ${\tilde \chi}_T =0$ by virtue of $\nabla_\nu T$ not being homogeneous under Weyl rescaling (see Appendix \ref{sec:aweyl}). Then the only non-vanishing vector transport are the conductivities and the parity-odd ones have to be determined in terms of the Lagrangian functions.\footnote{ One can confirm that \eqref{eq:POdd1} trivializes in a Weyl invariant fluid.} One linear combination of ${\tilde \chi}_E$ and ${\tilde \sigma}_H$  gives a linear combination of the Lagrangian scalars, but this allows both of them to be independent.\footnote{ The one exception to this argument is the special case of an equation of state where we have a temperature independence, e.g., $p(T,\mu) = C\, \mu^3$ for some constant $C$ as one encounters for an extremal black hole (in the holographic context), since this implies that $\frac{\partial}{\partial \mu} \prn{\frac{sT}{q}}=0$.} 

Despite these complications we suspect that Class L does not allow for the most general form of the Hall conductivity ${\tilde \sigma}_H$. At the same time, a  curious fact of the Class L effective action is the vanishing  of the Hall viscosity as  has been noted earlier in  \cite{Haehl:2013kra}. In summary we see that in Class L
\begin{align}
{\tilde \eta}_H = 0  \,, \qquad {\tilde \sigma}_H  = \text{fixed} \,.
\label{eq:hallpred}
\end{align}

As we shall see later in  \S\ref{sec:classB}  these relations are akin to the relation between $\{\tau,\lambda_1, \lambda_2\}$ for the neutral fluid \eqref{eq:weylrelns}. Since ${\tilde \sigma}_H$ and ${\tilde \eta}_H$  are completely unconstrained by the second law \cite{Jensen:2011xb} one should find that any value for the Hall viscosity is acceptable in hydrodynamics.  We defer comments on non-local Lagrangians and the recent construction  of \cite{Geracie:2014iva} to \S\ref{sec:classB}, since understanding deviations from relations such as   \eqref{eq:weylrelns} and \eqref{eq:hallpred} is part of a more general endeavour of constructing actions for Berry-like transport. Once we understand that they are Berry-like terms (Class B), we will be able to find an effective action which allows arbitrary non-vanishing value of ${\tilde \eta}_H$ and ${\tilde \sigma}_H$ in our extended formalism (see \S\ref{sec:eightfoldLT}).

\section{Class B: Berry-like transport }
\label{sec:classB}

We have explored for the most part of our discussion, solutions  to the adiabaticity equation classified by a Lagrangian $\Lag\brk{\hfields}$, which we termed as Class L adiabatic fluids. This raises an interesting question: ``Are all solutions to \eqref{eq:Adiabaticity} obtained from a suitable Lagrangian?'' 
The answer surprisingly turns out to be no! There are several classes of adiabatic transport that do not appear to fit into a simple Lagrangian description. We have already hinted that anomalous transport (Class A) 
requires more structure. In particular, in \cite{Haehl:2013hoa} we argued that an anomalous hydrodynamic effective action necessarily involves a Schwinger-Keldysh doubling of fields in order to satisfy the correct Ward identities.  One might argue that such transport comprising typically of finite set of terms (those governed by the anomaly) is special.  

However, strangely enough, this doubling trick by itself does not appear to suffice in general. We find three additional classes of transport, of which two (Class B and Class $\GV$) are non-finite classes (the third Class C is finite). To complete our classification scheme and to understand the nature of adiabatic transport, we have to indeed analyze what such constitutive relations mean. Therefore, before trying to enlarge the set of Lagrangian Class L transport by incorporating Schwinger-Keldysh doubling and influence functionals (c.f., \S\ref{sec:skdouble}), let us now explicitly construct the parts of adiabatic transport which do not seem to fit into Class L. 

 We start with adiabatic transport that we call Class B (for Berry curvature inspired constitutive relations). These Berry terms actually arise in familiar contexts of hydrodynamic transport, and have indeed been encountered hitherto, without the general structure however being appreciated. The simplest examples of these arise in parity-odd fluids in 2+1 dimensions, where the Hall transport falls in this class, cf., \S\ref{sec:podd}.  We first introduce the basic  tensor structures and constitutive relations in \S\ref{sec:Bbasic} and then exhibit some examples to illustrate the construction.\footnote{ Using the differential operators introduced in \S\ref{sec:Ddiffops} one can in a single swoop construct all Class B constitutive relations. This discussion being somewhat abstract  is better appreciated once the basic story is laid out.}

\subsection{The Berry constitutive relations}
\label{sec:Bbasic}

We now describe a large set of solutions to the non-anomalous adiabaticity equation based purely on exploiting some tensor structures which conspire effectively. The logic is going to be similar to our discussion in \S\ref{sec:classD}. We will start with the grand canonical adiabaticity equation \eqref{eq:AdiabaticityG} reproduced here for convenience:
\begin{align}
-\nabla_\sigma\prn{\frac{\mathcal{G}^\sigma}{T}}&=
\half  T^{\mu\nu}\diffB  g_{\mu\nu} + J^\mu \cdot \diffB  A_\mu
\label{eq:AdGna}
\end{align}
We are going to use the fact that the r.h.s. involves an explicit contribution from the variation of the background fields along $\Bfields$ via the $\{\diffB g_{\mu\nu}, \diffB A_\mu\}$ insertions. Imagine picking an  ansatz for the conserved currents which also contains an explicit insertion of these operators, i.e., schematically consider $T^{\mu\nu}  \propto \diffB g_{\alpha\beta}$ and $J^\alpha \propto \diffB A_\beta$. If the intertwining tensors that complete the map above have the appropriate symmetries, then it is plausible that upon further contraction with $\diffB g_{\mu\nu}$ or $\diffB A_\mu$ we ensure that the divergence of the free energy current vanishes.  This means that we can solve the adiabaticity equation with the no free energy current; the conserved currents themselves conspire to ensure lack of dissipation. 

Inspired by the above argument, consider the following constitutive relations:\footnote{ The conflation of the notation with the tensor structures used for describing Class D constitutive relations in 
\S\ref{sec:classDexamples} is  intentional. It will allow us later to talk about a single tensor structure whose symmetric part contributes to Class D and anti-symmetric part to Class B.}
\begin{equation}\label{eq:TJBerry}
\begin{split}
(T^{\mu\nu})_\text{B} &\equiv
	 -\quarter \prn{ \BerryG^{(\mu\nu)(\alpha\beta)}-\BerryG^{(\alpha\beta)(\mu\nu)} }
	 \,  \diffB  g_{\alpha\beta} + \BerryGA^{(\mu \nu) \alpha} \cdot  \diffB  A_\alpha
\\
(J^\alpha)_\text{B} &\equiv
	- \half \BerryGA^{(\mu \nu) \alpha}  \diffB  g_{\mu\nu}
	- \BerryA^{[\alpha\beta]} \cdot \diffB  A_\beta
\end{split}
\end{equation}
where $\{\BerryG^{\mu\nu\alpha\beta},\BerryGA^{\mu \nu \alpha}, \BerryA^{\alpha\beta} \}$ are arbitrary tensors (modulo field redefinitions). Here $(\alpha\beta)$ and  $[\alpha\beta]$ indicates the usual projection to the symmetric and anti-symmetric parts respectively.

Substituting the above constitutive relations into  the adiabaticity equation in the grand canonical ensemble, we get
\begin{equation}
\begin{split}
\half  (T^{\mu\nu})_\text{B}&\,\diffB  g_{\mu\nu} + (J^\alpha)_\text{B} \cdot \diffB  A_\alpha
\\
&= -\frac{1}{8} \prn{ \BerryG ^{(\mu\nu)(\alpha\beta)}-\BerryG^{(\alpha\beta)(\mu\nu)} }
\diffB  g_{\mu\nu} \diffB  g_{\alpha\beta}\\
&\qquad  + \half \BerryGA^{(\mu \nu) \alpha} \cdot \prn{\diffB  g_{\mu\nu}\ \diffB  A_\alpha-\diffB  A_\alpha\ \diffB  g_{\mu\nu}} -  \diffB  A_\alpha \cdot \BerryA^{[\alpha\beta]} \cdot\diffB  A_\beta \\
&= 0
\end{split}
\end{equation}
we see that we solve the adiabaticity equation \eqref{eq:AdGna} if we simply take $(\mathcal{G}^\sigma)_\text{B} =0 $. 

In the micro-canonical ensemble, this is equivalent to taking the entropy current to have a purely canonical contribution, viz.,
\begin{equation}\label{eq:JSBerry}
\begin{split}
(J_S^\alpha)_\text{B} &\equiv  - \Kbeta_\beta    (T^{\alpha\beta})_\text{B} - \prn{\LambdaB + \Kbeta^\beta A_\beta} \cdot    (J^\alpha)_\text{B}  \\
&=   \frac{1}{T}\,
	\Bigg\{ \quarter   \prn{ \BerryG^{(\alpha\beta)(\mu\nu)}-\BerryG^{(\mu\nu)(\alpha\beta)} }
	 \, u_\beta + \half \,  \mu \cdot \BerryGA^{(\mu \nu) \alpha} \Bigg\} \diffB  g_{\mu\nu}
\\ &\qquad
	-\frac{1}{T}\, \Bigg\{ \BerryGA^{(\alpha\beta) \nu} \, u_\beta -  \mu\cdot\BerryA^{[\alpha\nu]} \Bigg\}
	\cdot \diffB  A_\nu
\end{split}
\end{equation}
Thus, equations \eqref{eq:TJBerry} and \eqref{eq:JSBerry} give a large set of adiabatic constitutive relations.  The set of constitutive relations parameterized by these expressions is what we term to be Class B.\footnote{ See, however, the slight generalization below.}

Before we proceed further with our analysis, let us pause to motivate our terminology. The tensors which multiply $\diffB g_{\mu\nu}$ and $\diffB A_\mu$ are anti-symmetric for the most part (the only symmetric tensor is the compensator $\BerryGA$ which mixes the two sources). Such anti-symmetric tensors may be viewed as curvatures in the phase space of hydrodynamic fields $\hfields$; indeed, they have the correct symmetries to be interpreted as such. Phase space connections and associated curvatures typically contribute to the Berry phase picked up by the system when it is made to traverse a closed loop in configuration space. While we have not quite justified why $\{\BerryG^{[(\mu\nu)|(\alpha\beta)]}, \BerryA^{[\mu\nu]}\}$ should be thought of as configuration space curvatures, supporting evidence for this interpretation can be advanced by examining the physics of Hall viscosity \cite{Read:2008rn}. As we will see below the Hall viscosity term is the simplest example of Class B transport, and the fact that it is associated with the dynamics of quantum states in the phase space makes it plausible to forward a rationale for our terminology.

All the Class B constitutive relations trivially satisfy hydrostatic principle because they vanish in hydrostatic equilibrium. They thus drop out of the hydrodynamic equations in the hydrostatic limit. These  are thus examples of  non-hydrostatic but non-dissipative constitutive relations.  In fact, some aspects of these as we shall see have been encountered in previous analyses but were not identified to belong to this general class.
For instance in the analysis of \cite{Jensen:2011xb} it was noticed that the Hall transport coefficients are unconstrained by any form of the second law, while \cite{Bhattacharya:2012zx} noticed a similar feature for a particular combination of second order transport coefficients for a neutral fluid. We will now show how these arise within the general construction above.

While a general Class B term can be characterized by the tensors $\{\BerryG, \BerryGA, \BerryA\}$ with the indicated symmetry properties, which can be classified by working in the gradient expansion, a slight generalization allows us to write a complete solution to Berry transport. Recall our discussion of  tensor valued derivative operators and the set of intertwining tensors used to describe dissipative Class D transport in \S\ref{sec:Ddiffops}. This construction can be exploited to give non-trivial Class B relations as well. This is not quite useful for the purposes of classification, but does provide an alternative perspective on the Berry-like transport whilst simultaneously indicating some degenerate situations where na\"ive Class D terms are actually adiabatic. Furthermore, it substantiates our earlier statement in \S\ref{sec:classD} regarding the adiabatic nature of non-symmetric intertwiners.

Firstly let us ask when $\Diss$ obtained in \eqref{eq:dissUp} vanishes. As the interwiners $\{\cdg,\cdA\}$  connect two identical representations, this will happen whenever  they transform in an anti-symmetric representation, i.e.,
\begin{equation}
\Diss = 0 \;\; \Longrightarrow \;\; 
\cdg \in \text{Asym}\left( \text{Tens}_\cdg \otimes \text{Tens}_\cdg \right) \,,\qquad 
\cdA \in \text{Asym}\left( \text{Tens}_\cdA \otimes \text{Tens}_\cdA \right) \,.
\label{eq:etaAntisymm}
\end{equation}	
So we clearly have an adiabatic constitutive relation.  Thus, equations \eqref{eq:TJClassV} and 
\eqref{eq:JSClassV} along with the conditions \eqref{eq:etaAntisymm} give a large set of Class B solutions. One can get even more by generalizing the intertwiner matrix in \eqref{eq:TJClassVMat} to contain off-diagonal mixed intertwiners $\cdgA\in \left( \text{Tens}_\cdg \otimes \text{Tens}_\cdA \right)$ as follows:
\begin{equation}\label{eq:TJClassVMatB}
\begin{split}
\prn{ \begin{array}{c} T^{\mu\nu} \\ J^\alpha \end{array} }_{\text{B}}
&=
- \prn{ \begin{array}{cc} \DVisc_{\cdg_g}^\dag & \DVisc_{\cdA_g}^\dag \\ \DVisc_{\cdg_A}^\dag & \DVisc_{\cdA_A}^\dag \end{array} }
\prn{ \begin{array}{rr} \cdg\ & \cdgA\ \\  -\cdgA\ & \cdA \end{array} }
\prn{ \begin{array}{cc} \DVisc_{\cdg_g} & \DVisc_{\cdg_A} \\ \DVisc_{\cdA_g} & \DVisc_{\cdA_A} \end{array} }
\prn{ \begin{array}{c} \half \diffB  g  \\ \diffB  A \end{array} } \,.
\end{split}
\end{equation}
The additional $\cdgA$-intertwiners with opposite sign drop out of the entropy production \eqref{eq:dissIntPart}, so they contribute to Class B. 

While the description in terms of the tensor valued differential operators  $\DVisc$ and intertwiners 
$\{\cdg, \cdgA, \cdA\}$ provides a complete description, there is one difference between this picture and the simpler parameterization introduced in \eqref{eq:TJBerry}, which should be borne in mind. This concerns the free energy flux. When the tensor operators  $\DVisc$ in \eqref{eq:TJClassV} are just tensors (i.e., they do not involve derivative operators), they reproduce the simple parametrization \eqref{eq:TJBerry}. For in  this case no integration by parts is necessary to obtain \eqref{eq:JSClassV}, so the Noether current is just zero (which is consistent with \eqref{eq:JSBerry}). However, in the more general case that the constitutive relations \eqref{eq:TJClassV} contain genuine derivative operators, one has to do an integration by parts in 
\eqref{eq:dissIntPart}, leaving behind some total derivative terms that give a non-canonical contribution to 
$(J^\mu_S)_\text{B}$.

Further, this construction makes it also clear that there are certain constitutive relations which we would want to think of as dissipative, which end up nevertheless in Class B. These are situations wherein the intertwiners $\{\cdg, \cdgA, \cdA\}$ themselves are built from higher order mixed symmetry representations. Consider for example, the following tensor representations
\begin{align}
\Yvcentermath1\yng(1) &\equiv \text{Sym}_2 \,, \nonumber \\
 {\cal N}_\text{B}^{[(\alpha\beta)(\mu\nu)]} =   
{\Yvcentermath1\yng(2,1)}^{\ (\alpha\beta)(\mu\nu)(\rho\lambda)} \diffB g_{\rho\lambda}
\,,\quad & \quad
{\cal N}_\text{D}^{((\alpha\beta)(\rho\lambda))} =   
{\Yvcentermath1\yng(2,1)}^{\ (\alpha\beta)(\mu\nu)(\rho\lambda)} \diffB g_{\mu\nu}
\end{align}
i.e., we obtain the intertwiners from the same underlying representation {\tiny \Yvcentermath1${\yng(2,1)}$} by contracting different sets of indices. However, when we compute $\Diss$ we find
\begin{equation}
\begin{split}
\Diss &=\diffB g_{\alpha\beta}\  {\cal N}_\text{D}^{((\alpha\beta)(\rho\lambda))} \ \diffB g_{\rho\lambda}  \\
&=\diffB g_{\alpha\beta}\  {\Yvcentermath1\yng(2,1)}^{\ (\alpha\beta)(\mu\nu)(\rho\lambda)} \; 
\diffB g_{\rho\lambda} \ \diffB g_{\mu\nu} \\
&= \diffB g_{\alpha\beta}\  {\cal N}_\text{B}^{[(\alpha\beta)(\mu\nu)]}  \ \diffB g_{\mu\nu}  \\
&=0 \,.
\end{split}
\end{equation}	
A similar analysis can be carried out for the flavour charges, by working with the representation $\text{Vect}$ instead. In fact, we can make the general statement by choosing to work with the representations $\text{Tens}_\cdg$ and $\text{Tens}_\cdA$ respectively. Let
\begin{align}
\Yvcentermath1\young(\cdg)  \equiv \text{Tens}_\cdg \,,& \qquad \Yvcentermath1\young(\cdA) \equiv \text{Tens}_\cdA \,,
\end{align}
and define
\begin{align}
\cdg_\text{D} \in \Yvcentermath1\young(\cdg\cdg,\cdg) \,,\qquad
\cdA_\text{D} \in \Yvcentermath1\young(\cdA\cdA,\cdA) 
\label{}
\end{align}
will provide examples of na\"ive Class D terms which  secretly belong to Class B, since they do not produce any entropy. 

Before proceeding with explicit examples of Class B terms, we should point out that the constitutive relations above \eqref{eq:TJBerry}  are subject to field redefinition ambiguities. For instance we can redefine the thermal vector and twist as in \eqref{eq:KbetaRef} which would then affect the intertwiners. We leave it as an exercise for the reader to work out what the changes induced are, noting that they involve rather messy algebra. We will however have a bit more to say about this as we develop the eightfold effective action  in \S\ref{sec:eightfoldLT}.

\subsection{Examples of Class B transport}
\label{sec:bex}

Let us now consider some examples of Berry transport. By construction, Class B constitutive relations have at least one derivative (since $\diffB g_{\mu\nu}$ and $\diffB A_\mu$ is linear in the gradients of
$\{\Kbeta^\mu,\LambdaB\}$). Thus, there are
no examples in zero derivative order.

\paragraph{Hall Transport in 3 dimensions:}
At one derivative order, in  3-dimensional parity violating fluids, there is an adiabatic constitutive relation
that can be obtained by setting
$\BerryG^{\mu\nu\alpha\beta}= - {\tilde \eta}_{_H} \,T\,u_\rho\,\varepsilon^{\rho\mu\alpha}\, P^{\nu\beta}$
along with  $\BerryGA^{\mu\nu\alpha} =0 $ and
$\BerryA^{\alpha\beta} =-{\tilde \sigma}_{_H} \,T\,u_\rho\,\varepsilon^{\rho\alpha\beta}$.
We obtain then for the currents
\begin{equation}
\begin{split}
(T^{\mu\nu})_\text{B}
&= -{\tilde \eta}_{_H} \, \varepsilon^{\alpha\beta(\mu} u_\alpha \sigma_\beta{}^{\nu)} \,,  \\
(J^\alpha)_\text{B} &= {\tilde \sigma}_{_H} \cdot \varepsilon^{\alpha\rho\sigma} u_\rho \cv_\sigma \,,\\
(J_S^\alpha)_\text{B} &= \frac{\mu}{T}\cdot {\tilde \sigma}_{_H} \cdot \varepsilon^{\alpha\rho\sigma} u_\rho \cv_\sigma  \,.
\end{split}
\label{eq:HallB}
\end{equation}
We recognize the transport coefficients  ${\tilde \sigma}_{_H} $ and  ${\tilde \eta}_{_H}$ as the Hall conductivity and Hall viscosity respectively, from our discussion in \S\ref{sec:podd}. As mentioned earlier the fact that the Hall transport terms on-shell lead to an exactly conserved entropy current (from the adiabaticity equation) was the reason that \cite{Jensen:2011xb} found in the current algebra approach no constraint on them from the second law. Since the tensor structures vanish in hydrostatics, so we have no information regarding these terms from the equilibrium partition function.

\paragraph{Berry terms in neutral fluids:} Our second example  for Class B constitutive relations is perhaps in the simplest hydrodynamic system imaginable, a neutral fluid! While there is no adiabatic transport at first order, we have seen that there are adiabatic parts to each of the 15 transport coefficients of a neutral fluid, cf., Appendix \ref{sec:neutral2d}.  Amongst these lurks a term of the form \eqref{eq:TJBerry}. Since
$\diffB g_{\mu\nu} = 2\,\nabla_{(\mu} \Kbeta_{\nu)}$ can be written using \eqref{eq:gradKB} in terms of the shear etc., and is clearly a first order term, we pick for the tensor $\BerryG^{\mu\nu \alpha \beta}$ another first order contribution. The symmetries  we require fix this tensor uniquely to be
\begin{equation}
\begin{split}
\BerryG^{\mu\nu\alpha\beta} = 2\,T\prn{\, \lambda_\sigma \, \sigma^{\mu\nu} \, P^{\alpha\beta} +
 \lambda_\omega \, \omega^{\mu\alpha}\, P^{\nu\beta}}
\end{split}
\end{equation}
Using the decomposition of the gradient of $\Kbeta^\mu$ we can express the stress tensor in a simple form:
\begin{equation}
(T^{\mu\nu})_\text{B}
=-\lambda_\sigma \prn{ \Theta\, \sigma^{\mu\nu}- \sigma^2 \, P^{\mu\nu}}  -
\lambda_\omega \prn{\omega^{\mu\alpha} \sigma_\alpha^\nu +\omega^{\nu\alpha} \sigma_\alpha^\mu}
\label{eq:Bn2}
\end{equation}
Let us compare this with the parametrization of the second order Landau frame stress tensor given in
\eqref{eq:TLandau}. Using two simple identities
\begin{equation}
\begin{split}
\lambda_0\, \Theta\, \sigma_{\mu\nu} + \xi_2\, P_{\mu\nu}\, \sigma^2 &= \frac{\lambda_0 +\xi_2}{2}\,
\prn{\Theta\, \sigma_{\mu\nu}+ P_{\mu\nu}\, \sigma^2} +  \frac{\lambda_0 -\xi_2}{2}\,
\prn{\Theta\, \sigma_{\mu\nu}-  P_{\mu\nu}\, \sigma^2} \\
\sigma_{\langle \mu}{}^{\alpha} \omega_{\alpha\nu\rangle} &= -\frac{1}{2} \prn{\omega^{\mu\alpha} \sigma_\alpha^\nu +\omega^{\nu\alpha} \sigma_\alpha^\mu }
\end{split}
\end{equation}
we identify the two coefficients $\lambda_\sigma$ and $\lambda_\omega$ as determining linear combinations of the transport coefficients, viz.,
\begin{equation}
\lambda_\sigma = \frac{\xi_2 - \lambda_0}{2} \,, \qquad \lambda_2 = 2\, \lambda_\omega \,.
\end{equation}
The fact that the two tensor structures  appearing in \eqref{eq:Bn2} are non-dissipative was in fact was noticed  in the analysis of \cite{Bhattacharya:2012zx}, but again it was not appreciated then that these were part of a larger set of adiabatic transport data in hydrodynamics.

\subsection{Embedding Class B in Class L?}
\label{sec:GSpodd}

Given a couple of examples at our disposal let us take stock of whether we can identify a way to embed Class B into Class L.  Each of our two examples has been explored in the non-dissipative effective action framework. So we can make some informed statements about whether or not this is possible. Since the details seem to be
a-priori distinct in the two cases we will address them in turn.

\paragraph{Hall transport:} The analysis of \cite{Haehl:2013kra}, building on earlier work of \cite{Nicolis:2011ey}
and \cite{Bhattacharya:2012zx}, argued that there is no local effective action that captures Hall viscosity. Furthermore, it was found in that construction that the Hall conductivity was not an independent transport coefficient, but rather a linear combination of it and the coefficient ${\tilde \chi}_E$ introduced in
\cite{Jensen:2011xb} was fixed by the effective action. More specifically, the tensor structures involved are
the ones displayed in \eqref{eq:HallB}.

We find a very similar relation in the Class L construction outlined in \S\ref{sec:podd}. In particular, in \eqref{eq:System2} we have derived the parity-odd transport coefficients in terms of the Lagrangian scalars $\{\tilde{\mathfrak b}, \tilde{\mathfrak w}\}$. As there are six parity-odd transport coefficients and only two scalar densities, we expect four relations amongst the transport. Two of these are hydrostatic relations which eliminate two combinations of of $\{{\tilde \chi}_\Omega, {\tilde \chi}_B, {\tilde \chi}_T\}$. One such relation is easy to find algebraically and  is given in \eqref{eq:poddPF1}; the other appears to be complicated to obtain in closed form. The third relation, which is not hydrodynamic, appears to fix Hall conductivity ${\tilde \sigma}_H$, which is invisible in hydrostatics, in terms of a hydrostatic response parameter ${\tilde \chi}_E$, \cite{Jensen:2011xb,Haehl:2013kra}. The final relation is the one that sets the Hall viscosity ${\tilde \eta}_H =0$.

A-priori, given that the Hall conductivity and viscosity are adiabatic, any value for these hydrodynamic transport coefficient is acceptable.  So it is in fact somewhat curious that the Class L theory fixes their value so specifically. The results obtained herein are consistent with the effective action analysis of  \cite{Haehl:2013kra}  (which involves a Legendre transformation -- see Appendix \ref{sec:ndf}). It should be noted that  recently \cite{Geracie:2014iva} have argued that a suitable non-local term allows one to at least obtain non-vanishing Hall viscosity. The construction involved constructing a Wess-Zumino term in the configuration space of fluids (using the Lagrangian scalar variables of the non-dissipative effective action formalism of \cite{Dubovsky:2011sj}).\footnote{ A general construction of Wess-Zumino terms for a wide class of physical systems with various choices of internal symmetries was described in \cite{Delacretaz:2014jka}.} While this construction does indeed give a non-vanishing Hall viscosity, it however constrains it to be of the specific functional form ${\tilde \eta}_H = s\, f(q/s)$, as opposed to a general function of $s$ and $q$ (or $T$ and $\mu$ after Legendre transformation). We believe this is significant and points to a different resolution of the puzzle of Class B Hall transport terms. Indeed, we will later exhibit a Lagrangian system in \S\ref{sec:classLT}  which will give us the most general form of Hall transport.

 \paragraph{Neutral fluids:} The situation in the neutral fluid case is similar. $\lambda_2$ is fixed in Class L, and is constrained to obeying the relations \eqref{eq:weylrelns} and  \eqref{eq:2drelnsA}, in situations with and without Weyl invariance respectively. Furthermore, these relations appear to be upheld in two extreme corners: for Weyl invariant strongly coupled holographic plasmas as well as in kinetic theory.\footnote{ In the holographic context the relations are only valid in two derivative Einstein-Hilbert theory. Higher derivative corrections appear to spoil the Class L relation fixing $\lambda_2$, see \cite{Shaverin:2012kv,Yarom:2014kx,Grozdanov:2014kva} and our discussion in \S\ref{sec:holofluids}. }  Once again we do not know of a simple modification to incorporate these terms in Class L, but we will make a case for an extended Lagrangian which allows arbitrary values for Class B transport coefficients in due course.

\section{Class C: Conserved entropy}
\label{sec:consJ}

In hydrodynamics, the conserved currents $\{T^{\mu\nu}, J^\mu\}$ are canonically defined, but the entropy current  $J^\mu_S$ is a more abstruse object. It has no microscopic counterpart, arising as it does due to coarse-graining inherent in the statistical description of the quantum system of interest. Per se one only requires an entropy current satisfying $\nabla_\mu J_S^\mu \geq 0$ or \eqref{eq:AdiabaticityD} to exist, with no implication of uniqueness. The ambiguities in entropy current have been well appreciated in various discussions, cf., \cite{Bhattacharyya:2008xc,Bhattacharyya:2012nq,Bhattacharyya:2014bha} for a sampling of  recent literature where this issue is clearly spelt out.

We will now argue that there is one more type  of adiabatic constitutive relation solving \eqref{eq:Adiabaticity} which relies potential ambiguities in the entropy current. Recall that  in  Class L
one can always add arbitrary Komar terms as in \eqref{eq:KomarDef} to any entropy current determined by the Noether construction (in the absence of interesting cohomology). Such ambiguities in the entropy current are physically uninteresting and we won't discuss them further. However, there may actually be other entropy current contributions which are cohomologically non-trivial but still identically conserved without producing energy-momentum or charge transport. These terms are not accounted for in our previous analysis because all adiabatic classes so far led to non-trivial energy-momentum or charge currents, even taking into account field redefinitions.

Let us therefore examine as our next class of non-anomalous adiabatic constitutive relation a very simple set of currents. At any order in the gradient expansion one can consider a family of exactly conserved vectors ${\sf J}^\mu$, $\nabla_\mu {\sf J}^\mu = 0$. Since we are only interested in solutions to \eqref{eq:Adiabaticity} we can simply set 
\begin{equation} \label{eq:TJC}
(J^\mu_S)_\text{C} = {\sf J}^\mu \,, \qquad (T^{\mu\nu})_\text{C} =0 \,, \qquad (J^\mu)_\text{C} = 0
\end{equation}	
and achieve this desired outcome! As long as we have conserved vector fields ${\sf J}^\mu\brk{\hfields}$ we have achieved a trivial adiabatic constitutive relation. 

 For reasons described earlier, not all conserved vector fields ${\sf J}^\mu$, or equivalently their dual current $(d-1)$-forms ${\mathbf  j}$, are physically interesting. 
A trivial class of conserved currents can be obtained by taking ${\sf J}^\mu = \nabla_\nu\, {\sf X}^{[\mu\nu]}$ for some anti-symmetric tensor ${\sf X}^{\mu\nu}$; in other words  the entropy current $(d-1)$ form is exact $\star {\mathbf j} = d(\star \,{\mathbf x})  \Longrightarrow d(\star\, {\mathbf j}) =0$. As in any physical application, we are interested in cohomologically non-trivial conserved currents. These are similar to the Komar terms encountered in Class L which are uninteresting as long as there are no boundaries. 
 We shall later see that in the extended Lagrangian theory these will correspond to total derivative boundary terms.  We will henceforth quotient the space of conserved currents by such exactly conserved currents and Class C will comprise of cohomologically non-trivial currents.

Since here we have no energy-momentum or charge transport, but solely entropy flux along the chosen vector field, one has a macroscopic manifestation of  entropy without any physical effect. While one might a-priori think that even non-trivial elements of the cohomology, i.e., non-exact $(d-1)$-current forms are uninteresting,  there are certain choices of ${\sf J}^\mu$ which are worth exploring closely.

To do so, let us consider some examples, starting as usual with parity-even charged fluids. For vectors built out of $\hfields$ and their gradients, it is clear that there is no conserved vector at first order in gradients; the three parity-even vectors 
$\acc^\mu$, $\Theta \, u^\mu$ and $\cv^\mu$ are generically non-conserved. At higher orders it is possible to find conserved vectors, but most of these are exact differentials of the form $\nabla_\nu\, {\sf X}^{[\mu\nu]}$. For instance, we have five such vectors at second order in gradients, since we have a plethora of  first order anti-symmetric tensors \cite{Bhattacharyya:2014bha},
\begin{align}
{\sf X}^{\mu\nu} \in \{ u^{[\mu}\, \acc^{\nu]} , \omega^{\mu\nu} , u^{[\mu}\, \cv^{\nu]}  
, u^{[\mu}\nabla^{\nu]}\left(\frac{\mu}{T}\right) , P^{\mu\alpha} P^{\nu \beta} \, F_{\alpha\beta} \} \,,
\label{eq:trivC}
\end{align}
which give an exactly conserved entropy current at second order. These we discard for being trivial cohomological elements. 

One however has a non-trivial conserved current in odd spacetime dimensions owing to topological considerations. The simplest 
example is in three-dimensional parity-even neutral fluids where, inspired by Wen-Zee shift current  \cite{Wen:1992ej} which appears in Hall transport, we have the following second order conserved vector:
\begin{equation}
\JWZ^\sigma = \frac{1}{2}c_{_{\text{Euler}}}\;\varepsilon^{\sigma\alpha\beta}
\; \varepsilon^{\mu\nu\lambda}\; u_\mu \prn{\nabla_\alpha u_\nu \nabla_\beta u_\lambda
- \frac{1}{2}R_{\nu\lambda\alpha\beta} } \,,
\label{eq:wz3}
\end{equation}	
where $R_{\alpha\beta\gamma\delta}$ is the Riemann tensor and $c_{_\text{Euler}}$ is an arbitrary constant. The nomenclature is motivated by the fact that the conserved topological charge associated with this current is the Euler characteristic of the codimension-one spatial slice normal to $u^\mu$ \cite{Golkar:2014paa}. It is easy to check conservation directly, though the analysis is greatly facilitated by writing the associated current $2$-form. We give a simple derivation of this fact and the generalization to arbitrary odd $d=2n+1$ dimensions in Appendix \ref{sec:WenZee}.\footnote{
These currents were recently revisited in the context of parity-odd Hall fluids in \cite{Golkar:2014wwa} and \cite{Golkar:2014paa}. The latter work independently generalized the construction to arbitrary odd dimensions.  Our discussion in 
Appendix \ref{sec:WenZee} provides a complementary perspective.
}
From there we find that in general 
\begin{align}
\JWZ^\sigma &= -\frac{1}{2^n}\; c_{_{\text{Euler}}} \;
\varepsilon^{\sigma\alpha_1\alpha_2\ldots \alpha_{2n-1}\alpha_{2n}}\;
u_\mu \; \varepsilon^{\mu\nu_1\nu_2 \ldots \nu_{2n-1}\nu_{2n}} 
\nonumber \\& \
\times \prn{\half R_{\nu_1\nu_2\alpha_1\alpha_2}-\nabla_{\alpha_1}u_{\nu_1}\nabla_{\alpha_2} u_{\nu_2} }
\ldots
\prn{\half R_{\nu_{2n-1}\nu_{2n}\alpha_{2n-1}\alpha_{2n}}-\nabla_{\alpha_{2n-1}}u_{\nu_{2n-1}}
\nabla_{\alpha_{2n}} u_{\nu_{2n}} } 
\label{eq:euler2}
\end{align}	
is a conserved current present at the $(2n)^{\rm th}$ derivative order and gives a Class C solution to \eqref{eq:Adiabaticity}. 

Let us understand  the physical consequence of the Euler current contribution to entropy current 
in $d=3$. The Euler current reduces to the Euler character of the spatial two manifold on which we place our fluid. Let us  for simplicity  take ${\cal M}_3 = {\mathbb R} \times \Sigma_2$  where $\Sigma_2$ is a compact two manifold. Then $\JWZ^\mu u_\mu$  is a measure of the topology of $\Sigma_2$, and in particular its integral gives the Euler character (and hence the genus) of this two-manifold. Since there is no a-priori reason to restrict attention to spherical or planar topology, we can consider fluids on negatively curved Riemann surfaces and extract a contribution from $\JWZ^\mu$. The topological contribution will compute a degeneracy in terms of the Euler character $s =  c_{_{\text{Euler}}} \, \chi$.

This situation can be realized holographically. A three-dimensional CFT such as the M2-brane worldvolume (ABJM) theory can be placed on ${\cal M}_3 = {\mathbb R} \times \Sigma_2$.\footnote{ These considerations extend to other three-dimensional QFTs; we simply choose the ABJM theory for illustrative purposes.} 
While the vacuum dynamics of this theory is ill-behaved owning to the conformal coupling of the massless scalars (transforming in the ${\bf 8}_v$ of 
$SO(8)_R$), it is plausible that the thermal corrections stabilize the theory. In the strong coupling limit the gravity dual is given by supergravity on $\text{AdS}_4 \times  {\bf S}^7$. The four-dimensional Gauss-Bonnet term is the leading correction to the two derivative Einstein-Hilbert dynamics. This term is however topological and integrates to a pure boundary term and thus does not affect dynamics. It does however change the  degeneracy of the thermal density matrix. In particular, in the presence of this Gauss-Bonnet term, a  black hole in $\text{AdS}_4$ picks up a contribution from the Wald functional \cite{Iyer:1994ys} proportional to the Euler character of the spatial two-manifold which is the bifurcation surface. For the CFT 
on $ {\mathbb R} \times \Sigma_2$ the black hole horizon is such that its spatial cross-sections and especially the  bifurcation surface have the same topology as $\Sigma_2$. Then the Wald entropy does get a contribution proportional to $\chi(\Sigma_2)$ which is indeed what we see purely from a field theory analysis. One can furthermore check that the pull-back of the Wald functional on the horizon onto the boundary, to define a boundary entropy current as in \cite{Bhattacharyya:2008xc} will indeed give a contribution of the form \eqref{eq:wz3}.  Similar remarks apply to higher (odd) dimensional CFTs on topologically non-trivial backgrounds. Indeed the entropy current \eqref{eq:euler2} is obtained from a particular Lovelock term in $\text{AdS}_{d+1}$.

The above discussion can be extended to charged fluids. For example we can consider the Chern current in odd spacetime dimensions 
\begin{equation}
\begin{split}
 \JChern^\sigma &= 
 	\frac{1}{2^n} \, c_{_{\text{Chern}}} \, \varepsilon^{\sigma \alpha_1\alpha_2 \cdots \alpha_{2n-1}\alpha_{2n}} 
 	F_{\alpha_1\alpha_2} \cdots F_{\alpha_{2n-1}\alpha_{2n}} \,,
\end{split}
\end{equation}
which despite being exact does contribute to the degeneracy of states and thus the entropy. One easy way to intuit this is to look at the a three-dimensional field theory again ($n = 1$). Now we have a background magnetic field in the spatial manifold which is well known to contribute to ground state degeneracy (e.g., classic Landau level physics). In higher dimensions we would be picking up contributions when the topology of the gauge bundle is non-trivial (e.g., instanton bundle in $d=5$). It would be interesting to investigate other such contributions from combinations of the background sources and realizations of such effects in physical fluids.

\section{The Vector Classes: Transverse free energy currents}
\label{sec:Cvector}

We now turn to another family of  solutions to \eqref{eq:Adiabaticity} which are not captured by Class L Lagrangians. The transport terms constructed in this section rely on the presence of a set of vector fields. These fields could be hydrostatic whence the transport will be characterized by the Class $\PV$ terms encountered in \S\ref{sec:hydrostatics}. It also transpires that we can have non-hydrostatic vector fields which give solutions to the adiabaticity equation; we will name the set of constitutive relations thus determined as belonging to Class $\GV$ (in analogy with our distinction in the scalar case).

\subsection{The hydrostatic Class $\PV$}
\label{sec:JLYtranscend}

In \S\ref{sec:hpfns} we have already mentioned the fact that hydrostatic partition functions are either classified by scalar densities or by conserved transverse vectors $P_V^\sigma$ which satisfy  
\begin{equation}
\KEq_\sigma \big(P_V^{\ \sigma}
\big)_\text{Hydrostatic}= \nabla_\sigma \big(P_V^{\ \sigma}\big)_\text{Hydrostatic}=
0 \,.
\end{equation}
Their contribution to the equilibrium partition function is as indicated in \eqref{eq:wpv}. We note that such terms 
have been studied in the context of  Cardy-like formulae in higher dimensions \cite{Landsteiner:2011cp,Loganayagam:2012zg,Jensen:2012kj,Jensen:2013rga}.
These terms first showed up as `integration constants' of the anomaly induced transport \cite{Neiman:2010zi,Loganayagam:2011mu}; that they contribute as vectors to the partition function was  first realized in \cite{Banerjee:2012iz}. 
These terms are sometimes termed transcendental anomaly induced transport terms in order to distinguish them from
the `rational' anomaly induced terms in Class A. These names emphasize the fact that Cardy-like formulae for 
Class $\PV$ transport always involve extra transcendental factors of $2\pi$ unlike the Class A transport.\footnote{ Such contributions are also well understood in the holographic context and we refer the reader to  \cite{Azeyanagi:2013xea}  and  
references therein for a detailed discussion.} The supersymmetric cousins of $\PV$ play a crucial role in the recent proposals
for Cardy-like formulae applicable to supersymmetric partition functions \cite{DiPietro:2014bca}.
We will now review  the structure of these terms as discussed in these references mainly to give a complete representation of all the classes of adiabatic transport.

Let us begin with a simple example in two-dimensional fluids ($d=2$). Consider the following adiabatic constitutive relations
\begin{equation}
\begin{split}
(T^{\alpha\beta})_{\PV} = -2\, \tilde{c}_g\, T^2\,  \varepsilon^{(\alpha\gamma} u_\gamma u^{\beta)}   
\,, \qquad    (J^\alpha)_{\PV} = 0 \,, \qquad
(J_S^\alpha)_{\PV} =-2\,\tilde{c}_g \,T \, \varepsilon^{\alpha\gamma}u_\gamma \,.
\end{split}
\end{equation} 
This corresponds to a transverse free energy current 
\begin{equation}
(\mathcal{G}^\alpha)_{\PV}= \tilde{c}_g \,T^2\, \varepsilon^{\sigma\gamma}u_\gamma\,.
\end{equation}	
These constitutive relations can be derived from a hydrostatic partition function
\begin{equation}
W_\text{Hydrostatic} = -\int_{\Sigma_E} \left(
\frac{1}{T} \, (\mathcal{G}^\sigma)_{\PV} \brk{\hfields_\Eqfields} \right)_\text{Hydrostatic}  \ d^{d-1}S_\sigma\,.
\end{equation}
It is easily checked that these constitutive relations solve adiabaticity equation \eqref{eq:AdiabaticityG} for an arbitrary number $\tilde{c}_g$. It was argued in \cite{Jensen:2012kj} that for a general field theory, we have  $\tilde{c}_g=2 (2\pi)^2  c_g$ where $c_g$ is the Lorentz anomaly of the underlying two-dimensional theory. For 2d CFTs, this is just the (parity-odd part of) Cardy formula. 
Thus, the parity-odd Cardy formula relates the coefficients that appear in the Class $\PV$ with the anomaly coefficients that  appear in Class A constitutive relations (which we will encounter in \S\ref{sec:anomalies}). 

The above construction (and the corresponding parity-odd Cardy-like formulae)  can be generalized to arbitrary even dimensions following \cite{Jensen:2013rga}. We will give a description of the construction below for completeness, but the reader may find the discussion below more comprehensible after reading through our Class A section, \S\ref{sec:anomalies}.

Given the close relation between the Class $\PV$ and Class A constitutive relations that Cardy formula implies, it is useful to set them in a common formalism. To this end, it is convenient to introduce a new gauge field 
$\AT_\mu$ and an associated chemical potential $\muT$. It turns out to be natural to treat the temperature as the chemical potential 
(by thinking of it as the twist in the thermal circle), so $\muT = T$. This is equivalent to introducing $\LambdaBT$ such that 
$\LambdaBT+\Kbeta^\sigma \AT_\sigma = 1 $.  We will take the field-strength $\FT_{\mu\nu}$ corresponding to this new gauge field to zero.\footnote{ In fact, a lot of this formalism will play an important role in the construction of an extended Lagrangian theory for adiabatic hydrodynamics in Part \ref{part:8classes}. We defer physical statements till 
\S\ref{sec:classLT}; for now the reader may simply take the introduction of $\AT$ and associated quantities a convenient way to encode Class $\PV$ constitutive relations as was done in \cite{Jensen:2013rga}. }

With this gauge field, Class $\PV$ constitutive relations take the same form as Class A constitutive relations (but now involving $\muT$ in addition) and the coefficients in Class $\PV$ correspond to pure and mixed  anomalies involving the new gauge field $\AT_\mu$. The higher  dimensional analogues of Cardy  formula can then be phrased as fixing the new anomaly coefficients  in terms of the usual anomaly coefficients. In $d=2n$ dimensions, Class $\PV$ constitutive relations are characterized by the  exact forms encoding these new anomalies. Thus, consider an anomaly polynomial relevant to Class $\PV$ of the form 
\begin{equation}
\begin{split}
\fP_{\PV} \equiv  \sum_{j=1}^{\lfloor\frac{n+1}{2}\rfloor} (\fFT\,)^{2j} \wedge \fP_{\PVj} [\fF,\fR] \,,
\end{split}
\end{equation}
where $\fP_{\PVj} [\fF,\fR]$ denotes an exact $2(n+1-2j)$-form made by wedging the flavour field strength 2-form $\fF$ and the Riemann curvature 2-form $\fR^\mu{}_\nu$. For a given set of $\lfloor\frac{n+1}{2}\rfloor$ exact 
forms $\{\fP_{\PVj} [\fF,\fR]\}_{j=1,2,\ldots,\lfloor\frac{n+1}{2}\rfloor}$, we can then construct a Class $\PV$ constitutive relation.\footnote{ The anomaly polynomial $\fP_{\PV}$ was called $\fP_{trans}$ in \cite{Jensen:2013rga}. We pretty much follow their notation for differential forms etc., and further notational conventions are as explained in \S\ref{sec:anomalies} (see also Appendix \ref{sec:Notation}).}  
We note that CPT invariance only allows $\fP_{\PV}$ which are even in $\FT$ \cite{Banerjee:2012cr,Jensen:2013rga}. 
This  also ensures that once $\FT$ is set to zero, all the new anomalies introduced via $\fP_{\PV}$ vanishes as they 
should.

As mentioned above, the detailed form of Class A constitutive relations and how they are derived starting from an anomaly polynomial are described in \S\ref{sec:anomalies} and Appendix \ref{sec:adcons}. It is a straightforward exercise to repeat the derivation with a new gauge field $\AT_\mu$ and the anomaly polynomial $\fP_{\PV}$ followed by  setting  $\FT_{\mu\nu}=0$ at the end. We will present here the result of this exercise and refer the reader to our sections on Class A for more details. We will 
need the definition of the spin chemical potential from equation \eqref{eq:DefOmega}
\begin{equation}
\Omega^\mu{}_\nu = \frac{1}{2}\,T \left( D_\nu \Kbeta^\mu - D^\mu \Kbeta_\nu\right) 
\end{equation}
along with the definition of ``hydrodynamical shadow'' gauge field and spin connection from \eqref{eq:hatAdef}
and \eqref{eq:hatGamdef} respectively. We have 
\begin{equation}
\begin{split}
\fAh &= \fA + \mu\, \fu\,,\\
\fGammah^\mu{}_\nu &=\fGamma^\mu{}_\nu+ \Omega^\mu{}_\nu \,\fu\,.
\end{split}
\end{equation}
In analogy with \eqref{eq:HallCurrentsDef}, let us also define the  bulk Hall currents for the sequence of anomaly 
polynomials $\{\fP_{\PVj} [\fF,\fR]\}_{j=1,2,\ldots,\lfloor\frac{n+1}{2}\rfloor}$ as 
\begin{equation}
\begin{split}
\hodgeB (\fJH)_{\PVj}  &= \frac{\partial \fP_{\PVj}}{ \partial \fF} \,, \qquad \hodgeB (\fSpH)_{\PVj}{}^b{}_a = 2\frac{\partial \fP_{\PVj}}{ \partial \fR^a{}_b} \,.
\end{split}
\end{equation}
We will denote by hats the corresponding objects evaluated for the shadow connections. 

We are now ready to present the general form of Class $\PV$ constitutive relations. Using \eqref{eq:InflowDef}
for the new anomaly polynomial and setting $\FT_{\mu\nu}=0$, we get for the  heat current $\fqPV$, the spin current 
$\fSPV{}$, the charge current $\fJPV$, and a contribution to the entropy current $(\fJpSPV$) the following expressions in differential form notation: 
\begin{equation}
\begin{split}
\star \fJPV 
&= - \sum_{j=1}^{\lfloor\frac{n+1}{2}\rfloor}  T^{2j} \,\fu \wedge (2\fomega)^{2j-1} \wedge (\fJHh)_{\PVj}
\,,\\
\star \fSPV{}^\beta{}_\alpha 
&= - \sum_{j=1}^{\lfloor\frac{n+1}{2}\rfloor}  T^{2j} \,\fu \wedge (2\fomega)^{2j-1} \wedge 
(\fSpHh)_{\PVj}{}^\beta{}_\alpha\,,\\
\star \fqPV 
&= - \sum_{j=1}^{\lfloor\frac{n+1}{2}\rfloor} (2j-1) \, T^{2j} \,\fu \wedge (2\fomega)^{2j-2} \wedge \fPh_{\PVj}
\,,\\
\star \fJpSPV 
&= - \sum_{j=1}^{\lfloor\frac{n+1}{2}\rfloor} 2\, j\, T^{2j-1} \,\fu \wedge (2\fomega)^{2j-1} \wedge (\fJHh)_{\PVj}
\,.\\
\end{split}
\end{equation}
This in turn gives (see Eqs.~\eqref{eq:bdyAnomCur} and \eqref{eq:ConstRelations})  
an energy momentum tensor and an entropy current of the form
\begin{equation}
\begin{split}
\TPV & = \qPV^\alpha u^\beta +  \qPV^\beta u^\alpha+ \frac{1}{2} \,D_\rho \prn{\SPV^{\alpha[\beta\rho]} + \SPV^{\beta[\alpha\rho]} -\SPV^{\rho(\alpha\beta)}}  \,, \\
(J_{S}^\alpha)_{\PV} &= \JpSPV^\alpha -\half \Kbeta_\sigma \, (\SpHh)_{\PV}^{\perp[\alpha\sigma]} \,.
\end{split}
\end{equation}
along with a charge current given by $\JPV$. Here, $(\SpHh)_{\PV}^{\perp[\alpha\sigma]}$ is defined via
\begin{equation}
\begin{split}
\star (\fSpHh)_{\PV}{}^\beta{}_\alpha 
&=  \sum_{j=1}^{\lfloor\frac{n+1}{2}\rfloor}  \brk{d(T\,\fu)}^{2j-1} \wedge 
(\fSpHh)_{\PVj}{}^\beta{}_\alpha\,.\\
\end{split}
\end{equation}
These constitutive relations then solve the adiabaticity equation \eqref{eq:Adiabaticity} without the anomalies (since the limit $\FT_{\mu\nu}=0$ sets all the new anomalies to zero).

The corresponding free energy current is given by $(\mathcal{G}^\sigma)_{\PV}=\GpPV +\half u_\alpha \, (\SpHh)_{\PV}^{\perp[\sigma\alpha]} $ with 
\begin{equation}
\begin{split}
\star \fGpPV 
&=  \sum_{j=1}^{\lfloor\frac{n+1}{2}\rfloor}  T^{2j} \,\fu \wedge (2\fomega)^{2j-2} \wedge \fPh_{\PVj}
\,.\\
\end{split}
\end{equation}
In the hydrostatic limit $u_\alpha\, (\SpHh)_{\PV}^{\perp[\sigma\alpha]} =0$ and only the $\GpPV$ part of the
free-energy current contributes to the hydrostatic partition function. Thus, 
\begin{equation}
\begin{split}
W_\text{Hydrostatic} &= -\int_{\Sigma_E} \left(
\frac{1}{T} \, (\mathcal{G}^{\sigma})_{\PV} \brk{\hfields_\Eqfields} \right)_\text{Hydrostatic}  \ d^{d-1}S_\sigma\,. \\
&=  -\int_{\Sigma_E} \sum_{j=1}^{\lfloor\frac{n+1}{2}\rfloor}\, T^{2j-1} \,\fu \wedge (2\fomega)^{2j-2} \wedge \fPh_{\PVj}\,.
\end{split}
\end{equation}
We note that this free-energy current is transverse, thus justifying our nomenclature in calling these class of terms
as Class $\PV$.

In any even space-time dimensions, given a set of exact forms $\{\fP_{\PVj} [\fF,\fR]\}_{j=1,2,\ldots,\lfloor\frac{n+1}{2}\rfloor}$ of appropriate degree, we can construct an adiabatic constitutive relation using the formulae above.
Additional physical considerations over and above second law such as Euclidean consistency, cf., 
\cite{Jensen:2012kj,Jensen:2013rga}, can be used to 
fix these exact forms in terms of the original anomaly polynomial of the theory. These `parity-odd Cardy formula'
or `replacement rule' thus fix $\fP_{\PV}$ that appears in the Class $\PV$ in terms of the 
anomaly polynomial $\fP[\fF,\fR]$  that controls Class A constitutive relations. Following \cite{Jensen:2012kj,Jensen:2013rga}, this relation takes the form
\begin{equation}\label{eq:ReplacementRule}
\begin{split}
\fP_{\PV}[\fF,\fR,\fFT\,] = \fP[\fF,{\rm tr} \fR^{2k} \mapsto {\rm tr} \fR^{2k} + 2(2\pi\, \fFT\,)^{2k} ] - \fP[\fF, {\rm tr} \fR^{2k} ]\,.
\end{split}
\end{equation}
For example, if the anomaly polynomial that controls Class A is taken to be $\fP= c_g ({\rm tr} \fR^2)^2 $, then
the  $\fP_{\PV}$ controlling Class $\PV$ is fixed by the Cardy formula to be 
$\fP_{\PV} =  4\,c_g\, (2\pi\, \fFT\,)^2\wedge {\rm tr} \fR^2+4\,c_g \,(2\pi\,\fFT\,)^4$.

This completes the discussion of hydrostatic transverse free energy currents which give rise to adiabatic constitutive relations.

\subsection{The non-hydrostatic Class $\GV$}
\label{sec:gibbsvec}

Let us now turn to another set of solutions to the adiabaticity equations involving vectorial degrees of freedom. We should be focusing now on vectorial contributions that vanish in hydrostatic equilibrium  (should they not do so, we would be able to include them in our discussion of Class $\PV$). 

\subsubsection{General construction of Class $\GV$}
\label{sec:GVconstruction}

 For the simplest way to motivate the construction, it is convenient to start with the non-anomalous adiabaticity equation in the grand canonical ensemble which we derived in \eqref{eq:AdiabaticityG}  and requoted in
 \eqref{eq:AdGna} during the Class B discussion of \S\ref{sec:classB}. The free energy current decomposes into  longitudinal  and transverse vectors  as in \eqref{eq:Gdecomp}. Longitudinal vectors can all be obtained directly from Class L; so we only need to focus now on transverse vectors to find the remaining solutions to  \eqref{eq:AdGna}.

The r.h.s.  of \eqref{eq:AdGna} involves at least one factor of $\diffB g_{\mu\nu}$ or $\diffB A_\mu$, which vanish in hydrostatics since $\diffEq g_{\mu\nu} = \diffEq A_\mu =0$. In other words, if we take the hydrostatic configurations off-shell by unlinking $\Bfields \neq \Eqfields $ then the Gibbs free energy flux is produced at ${\cal O}(\diffB)$. The statement of hydrostatic principle is simply at the this order we have compensating energy-momentum and charge flow to ensure adiabaticity.

However, now consider the situation where the Gibbs free energy flux is itself quadratic in departures from equilibrium, i.e.,
 $\mathcal{G}^\lambda \sim {\cal O}\left(\diffB^2\right)$. This would be invisible from an hydrostatic analysis. Taking divergence of such a term we should expect then that the r.h.s. of \eqref{eq:AdGna} would have contributions at ${\cal O}\left(\diffB^2\right)$  (when the derivative hits the tensor structure multiplying the $\diffB$ terms), as well as terms which behave as 
 $\diffB \, D_\mu\diffB$. Since the r.h.s. itself involves one $\diffB$ insertion, it follows that the terms of interest should have the currents containing combinations of $\diffB$ and $D \diffB$ terms. 

This simple reasoning then leads to the following ansatz for the energy-momentum and charge currents: 
\begin{equation}\label{eq:TJVec}
\begin{split}
(T^{\mu\nu})_{\GV} &\equiv
	  \half \brk{D_\rho{\mathfrak C}_{\BerryG}^{\rho (\mu\nu)(\alpha\beta)}
	  \, \diffB  g_{\alpha\beta} + 2\ {\mathfrak C}_{\BerryG}^{\rho (\mu\nu)(\alpha\beta)}
	  \, D_\rho \diffB  g_{\alpha\beta}} \\
&\qquad 	+ D_\rho{\mathfrak C}_{\BerryGA}^{\rho(\mu \nu) \alpha} \cdot \diffB  A_\alpha
	   + 2\ {\mathfrak C}_{\BerryGA}^{\rho(\mu \nu) \alpha} \cdot
  	  \, D_\rho \diffB  A_\alpha \\
(J^\alpha)_{\GV} &\equiv
	 \half \brk{ D_\rho{\mathfrak C}_{\BerryGA}^{\rho(\mu \nu) \alpha}  \diffB  g_{\mu\nu}
	   + 2\ {\mathfrak C}_{\BerryGA}^{\rho(\mu \nu) \alpha}
	  \, D_\rho \diffB  g_{\mu\nu} } \\
&\qquad 	+ D_\rho{\mathfrak C}_{\BerryA}^{\rho(\alpha\beta)} \cdot \diffB  A_\beta
	   + 2\ {\mathfrak C}_{\BerryA}^{\rho (\alpha\beta)} \cdot
  	  \, D_\rho \diffB  A_\beta
\end{split}
\end{equation}
where $ {\mathfrak C}_{\BerryG}^{\rho (\mu\nu)(\alpha\beta)}= {\mathfrak C}_{\BerryG}^{\rho (\alpha\beta)(\mu\nu)}$. These tensor fields are local functions of $\hfields$ and their gradients.

This solves adiabaticty equation with the free energy current
\begin{equation}\label{eq:GVec}
\begin{split}
(\N^\rho)_{\GV} &\equiv -\prn{\frac{\mathcal{G}^\rho}{T}}_{\GV} \\
&=	  \quarter \diffB  g_{\mu\nu}  {\mathfrak C}_{\BerryG}^{\rho (\mu\nu)(\alpha\beta)}
	  \, \diffB  g_{\alpha\beta}
	  +  \diffB  g_{\mu\nu} {\mathfrak C}_{\BerryGA}^{\rho(\mu \nu) \alpha} \cdot \diffB  A_\alpha
          +\diffB  A_\alpha \cdot {\mathfrak C}_{\BerryA}^{\rho(\alpha\beta)} \cdot \diffB  A_\beta
\end{split}
\end{equation}
As should be clear from the construction, the tensors $\{{\mathfrak C}_{\BerryG}, {\mathfrak C}_{\BerryGA},{\mathfrak C}_{\BerryA}\}$ are a-priori completely arbitrary with the indicated symmetry structure (modulo field redefinitions -- see below). Moreover, it is clear, that we will only obtain genuinely Class $\GV$ constitutive relations if we make sure that the free energy current will be transverse, i.e., we demand that the first index of the tensors be transverse: 
\begin{equation}
 {\mathfrak C}_{\BerryG}^{\rho(\mu\nu)(\alpha\beta)} \, u_\rho = {\mathfrak C}_{\BerryGA}^{\rho(\mu\nu)\alpha} \, u_\rho = {\mathfrak C}_{\BerryA}^{\rho(\alpha\beta)} \, u_\rho = 0 \,.
\end{equation}
The solution to \eqref{eq:AdGna} characterized by the constitutive relations \eqref{eq:TJVec} and the free energy current 
\eqref{eq:GVec} is the Class $\GV$ of Gibbsian vectors. 
The astute reader might wonder why we choose to call this Class $\GV$ as opposed to ${\text G}_V$ to indicate the Gibbsian structure employed in the construction. Our choice will be rationalized when we argue that these  non-hydrostatic vectors can be obtained from a generalized Lagrangian density (with enhanced symmetry) in \S\ref{sec:classLT}.

From the construction it is clear that $\GV$ terms contribute to the constitutive relations only from the second order in gradients; the presence of an explicit derivative and a single $\diffB$ in \eqref{eq:TJVec} ensures that we cannot get a first order contribution (even in the parity-odd case). In the context of proving the completeness of our classification (cf., Theorem \ref{thm:eight} in 
\S\ref{sec:8fold}), we will demonstrate that the parameterization \eqref{eq:TJVec}, \eqref{eq:GVec} is complete, i.e., every non-hydrostatic adiabatic constitutive relation with transverse free energy current is of this form.

\subsubsection{Example: second order charged fluid}
\label{sec:GVexample}

Let us look at an example to illustrate the construction.  Unfortunately the simplest setting where we encounter this class happens to be for a charged fluid at second order in gradients, owing to the observation above. We want a second order contribution to the free energy which implies that the tensors $\{{\mathfrak C}_{\BerryG}, {\mathfrak C}_{\BerryGA},{\mathfrak C}_{\BerryA}\}$ should be zero-derivative objects from \eqref{eq:TJVec}. A-priori we have the following inequivalent tensor structures at our disposal:
\begin{equation} \label{eq:HVbarTensors}
\begin{split}
{\mathfrak C}_\BerryG^{\rho(\mu\nu)(\alpha\beta)} &\in \{ P^{\rho(\mu} \Kbeta^{\nu)} P^{\alpha\beta},\, P^{\rho(\mu} \Kbeta^{\nu)} \Kbeta^\alpha \Kbeta^\beta ,\, P^{\rho(\mu} P^{\nu)(\alpha} \Kbeta^{\beta)} \} \,, \\
{\mathfrak C}_\BerryGA^{\rho(\mu\nu)\alpha} &\in 
\{ P^{\rho(\mu} \, P^{\nu) \alpha},\, P^{\mu\nu} \, P^{\rho\alpha} \} \,,\\
{\mathfrak C}_\BerryA^{\rho(\alpha\beta)} &\in \{ P^{\rho(\alpha}\Kbeta^{\beta)} \} \,,
\end{split}
\end{equation}	
and permutations thereof. These tensor structures have to be contracted with $\diffB g_{\mu\nu}$ and $\diffB  A_\alpha$ 
in the free energy current \eqref{eq:GVec}. These  linear variations can be expressed in terms of hydrodynamical objects using \eqref{eq:diffbga} which we reproduce here for convenience:
\begin{equation}
\begin{split}
\diffB g_{\mu\nu} &= 2\, \nabla_{(\mu} \Kbeta_{\nu)} 
 = \frac{2}{T}\, \brk{ \sigma_{\mu\nu}  + P_{\mu\nu}\, \frac{\Theta}{d-1} - \left(\acc_{(\mu} + \nabla_{(\mu} \log T \right) u_{\nu)} } \\
 \diffB A_\mu &
 = D_\mu(\LambdaB+\Kbeta^\nu A_\nu)+\Kbeta^\nu F_{\nu\mu}
 = u^\alpha\, D_\alpha\left(\frac{\mu}{T} \right)\, u_\mu - \frac{1}{T}\,\cv_\mu
\end{split}
\label{eq:diffbga2}
\end{equation}
with the vector $\cv^\mu$ defined in \eqref{eq:cvdef}. However, upon evaluating these variations on-shell as in 
\eqref{eq:diffbgaApp}, we find that the tensors in \eqref{eq:HVbarTensors} give only two different non-zero contributions to constitutive relations (others are linear combinations of these two).
Without loss of generality, we choose to parameterize the two non-trivial choices leading to inequivalent transverse vector contributions to the free energy current at this order as
\begin{align}
{\mathfrak C}_\BerryG^{\rho(\mu\nu)(\alpha\beta)} = {\mathfrak C}_\BerryA^{\rho(\alpha\beta)}  =0 \,, \quad 
{\mathfrak C}_\BerryGA^{\rho(\mu\nu)\alpha} = 
T\, C_1(T,\mu) \, P^{\rho<\mu} \, P^{\nu> \alpha} + T \, C_2 (T,\mu)\, P^{\mu\nu} \, P^{\rho\alpha} \,.
\end{align}
This gives the following transverse vector contributions to the free energy:
 \begin{equation}
 (\mathcal{G}^\rho)_{\GV} = 2\, C_1 \, \sigma^{\rho\alpha}\, \cv_\alpha
 + 2 \, C_2 \,  \Theta\, \cv^\rho \,.
 \end{equation}
The contributions to the energy-momentum tensor and charge current are rather cumbersome, so we give separately the terms that correspond to the $C_1(T,\mu)$ term and to the $C_2(T,\mu)$ term, respectively. From $C_1$ we get  
\begin{equation}
\begin{split}
 (T^{\mu\nu})_{\GV} &=  - C_1 \, \frac{dp}{dq} \, \Theta \sigma^{\mu\nu} - 2\, C_1 \, D^{<\mu} \cv^{\nu>} + \prn{\mathfrak{D}C_1-2\,C_1} \, \cv^{<\mu} \acc^{\nu>} - \dot{C}_{1} \, \cv^{<\mu} E^{\nu>} \\
 &\quad - \brk{\frac{q}{\varepsilon+p}(\mathfrak{D}C_1 - C_1)  - \dot{C}_1 } \cv^{<\mu} \cv^{\nu>}  
  + \frac{(3-d)C_1}{(d-1)}\, \brk{\sigma^{(\mu\rho} \cv_\rho \, u^{\nu)}  +\frac{1}{d-1} \, \Theta \, \cv^{(\mu} u^{\nu)}} \\
 &\quad - \frac{(d+1) C_1}{(d-1)}\omega^{(\mu\rho} \cv_\rho \, u^{\nu)}
 - C_1 \, \Theta \, \cv^{(\mu} \, u^{\nu)} 
 \,, \\
 (J^\alpha)_{\GV} &= \brk{-\, C_1 \, \sigma^2 } u^\alpha +  \brk{\frac{q}{\varepsilon+p}(\mathfrak{D}C_1 - C_1)  - \dot{C}_1 } \sigma^{\alpha\rho} \cv_\rho -  (\mathfrak{D}C_1) \, \sigma^{\alpha\rho} \acc_\rho \\
 &\quad + \dot{C}_1 \, \sigma^{\alpha\rho} E_\rho + 2 \, C_1 \, D_\rho \sigma^{\rho\alpha}+ \frac{(3-d)C_1}{2(d-1)}\,  \frac{q}{\varepsilon+p} \brk{ \sigma^{\rho\alpha} \cv_\rho + \frac{1}{d-1} \, \Theta \, \cv^\alpha}
 \\
 &\quad 
 - \frac{(d+1) C_1}{2(d-1)}\,  \frac{q}{\varepsilon+p} \,\omega^{\rho\alpha}\cv_\rho 
 - \frac{C_1}{2} \, \frac{q}{\varepsilon+p} \, \Theta \, \cv^\alpha \,,
\end{split}
\label{eq:HbarVex1}
\end{equation}
which is built from the following fluid frame invariant scalar, vector and tensor contributions:\footnote{ For the form of frame-invariant data we use the conventions of \cite{Banerjee:2012iz,
Haehl:2013kra} as described in \eqref{eq:frameproj}.}
\begin{equation}
\begin{split}
  C_S &= \frac{dp}{dq} \, C_1 \, \sigma^2 \,,\\
  C_V^\alpha &=   \brk{\frac{q}{\varepsilon+p}(\mathfrak{D}C_1 - C_1)  - \dot{C}_1 } \sigma^{\alpha\rho} \cv_\rho
  + \frac{(d+1)C_1}{(d-1)} \, \frac{q}{\varepsilon+p} \, \omega^{\alpha\rho}\cv_\rho -  (\mathfrak{D}C_1)\,  \sigma^{\alpha\rho} \acc_\rho \\
  &\quad + \dot{C}_1 \, \sigma^{\alpha\rho} E_\rho + 2 \, C_1 \, D_\rho \sigma^{\rho\alpha}\,, \\
  C_T^{\mu\nu} &=  - C_1 \, \frac{dp}{dq} \, \Theta \sigma^{\mu\nu} - 2\, C_1 \, D^{<\mu} \cv^{\nu>} + \prn{\mathfrak{D}C_1-2\,C_1} \cv^{<\mu} \acc^{\nu>} - \dot{C}_1 \, \cv^{<\mu} E^{\nu>} \\
   &\quad - \brk{\frac{q}{\varepsilon+p}(\mathfrak{D}C_1 - C_1)  - \dot{C}_1 } \cv^{<\mu} \cv^{\nu>}  \,.
\end{split}
\end{equation}
In the above expressions we use the abbreviation $\mathfrak{D}C_1 \equiv T \,C_1' + \mu\, {\dot C}_1$, where primes denote $T$-derivatives and over-dots denote $\mu$-derivatives.
The $C_2$ tensor structure on the other hand gives the following transport:
\begin{equation}\label{eq:HbarVex2}
\begin{split}
 (T^{\mu\nu})_{\GV} &=  \bigg\{ (\mathfrak{D}C_2) \, \acc^\alpha \cv_\alpha - \brk{\frac{q}{\varepsilon+p}(\mathfrak{D}C_2 - C_2)  - \dot{C}_2 } \cv^2 - \dot{C}_2 \, (\cv^\alpha \cdot E_\alpha)- 2 \,C_2 \, D_\alpha \cv^\alpha 
 \\
 &\quad - C_2 \, \frac{dp}{dq} \, \Theta^2\bigg\} P^{\mu\nu} -2\, C_2 \, \brk{ \sigma^{(\mu \rho} \, \cv_\rho \, u^{\nu)} - \omega^{(\mu\rho} \cv_\rho \, u^{\nu)} + \frac{1}{d-1}\, \Theta \, \cv^{(\mu} u^{\nu)}  }  
\,, \\
(J^\alpha)_{\GV} &= \brk{ C_2 \, \Theta^2  } u^\alpha   + \brk{\frac{q}{\varepsilon+p}(\mathfrak{D}C_2 - C_2)  - \dot{C}_2 } \Theta \, \cv^\alpha - \prn{ \mathfrak{D}C_2-2 \, C_2 } \Theta \, \acc^\alpha
\\
 &\quad  + \dot{C}_2\, \Theta E^\alpha  + 2\, C_2 \, P^{\alpha\rho} \nabla_\rho \Theta - \frac{q}{\varepsilon+p} \, C_2 \, \brk{\sigma^{\rho\alpha} \cv_\rho - \omega^{\rho\alpha}\cv_\rho + \frac{1}{d-1} \, \Theta \, \cv^\alpha} \,,
\end{split}
\end{equation}
with frame-invariant data as follows:
\begin{align}
  C_S &=  (\mathfrak{D}C_2) \, \acc^\alpha \cv_\alpha - \brk{\frac{q}{\varepsilon+p}(\mathfrak{D}C_2 - C_2)  - \dot{C}_2 } \cv^2- \dot{C}_2 \, (\cv^\alpha \cdot E_\alpha) - 2 \,C_2 \, D_\alpha \cv^\alpha \,, \notag \\
  C_V^\alpha &= \brk{\frac{q}{\varepsilon+p}(\mathfrak{D}C_2 - C_2)  - \dot{C}_2 } \Theta \, \cv^\alpha -\prn{ {\mathfrak D}C_2-2 \, C_2 } \Theta \, \acc^\alpha + \dot{C}_2\, \Theta E^\alpha  + 2\, C_2 \, P^{\alpha\rho} \nabla_\rho \Theta\,,  \notag \\
  C_T^{\mu\nu} &= 0 \,.
\end{align}
All the above expressions are written in the basis of independent scalars, vectors and tensors as listed in Table \ref{tab:ChargedTransport}. From that table and the list in Table \ref{tab:CountingEven}, it is evident that most of the above terms are fixed in terms of Class B and Class D parameters. The only terms that are not obvious to classify are the combinations $(C_V^\alpha,C_T^{\mu\nu}) = (D_\rho \sigma^{\rho\alpha},  D^{<\mu} \cv^{\nu>})$ and $(C_S,C_V^\alpha) = (D_\alpha \cv^\alpha, -P^{\alpha\rho}\nabla_\rho\Theta)$. It would be interesting to study the second order charged fluid in more detail and see if these combinations get fixed in Class L. 

From these calculations, we can now also see why there are no Class $\GV$ terms in neutral fluids at second order. In neutral fluids, only the tensor structure ${\mathfrak C}_\BerryG$ would be relevant. Hence we would take the tensors in the first line of \eqref{eq:HVbarTensors} and would compute the associated constitutive relations. To this end we would perform contractions with the on-shell expression of $\diffB g_{\mu\nu}$. However, as can be seen from \eqref{eq:diffbgaApp}, for uncharged fluids $\diffB g_{\mu\nu}$  only has pieces that are either completely transverse or completely longitudinal. A quick glance at the structure of the possible ${\mathfrak C}_\BerryG$ terms shows that in this way we could not get a transverse free energy current.

We find it  rather curious that Class $\GV$ constitutive relations are  sparse (at least at low orders in the gradient expansion). There is no reason for the number of transverse vectors to be limited at a given order in gradient expansion, but it appears that many such contributions are subsumed in other Classes. It would be useful to have a clear intuition for why this is the case. One hopes that by studying such constitutive relations in various hydrodynamic systems would help reveal some rationale for the paucity of Class $\GV$.

\section{Class A: Lagrangian solution to anomalous adiabaticity equation}
\label{sec:anomalies}

The framework of adiabatic fluids whilst sufficiently general to allow us to discuss anomalous hydrodynamics and being formulated as such to incorporate these effects, we have so far refrained from analyzing such systems explicitly.
The main novelty with the anomalous constitutive relations is that one has to account for contributions which account for a modification of the  equations of motion due to the presence of flavour and gravitational anomalies. Indeed it was in attempting to understand these constitutive relations that the adiabaticity equation was first proposed in \cite{Loganayagam:2011mu}.  

Thus far the only exposure to anomalies we have had has been in the context of hydrostatics. It has been well known for a while that Class H contains Class A, see for example the analyses of \cite{Banerjee:2012iz,Banerjee:2012cr,Jensen:2012kj,Jensen:2013kka,Jensen:2013rga}. However, we have established that Class $\PS$ $\subset$ Class L, which begs the question whether we can understand the anomalous transport in terms of an effective action. Indeed, one could take the philosophy that for the adiabatic fluid framework achieve its stated goal of enabling us understand how hydrodynamics can be derived from an action principle, we would need to demonstrate that the anomalous transport can be captured by a Lagrangian, thus establishing that Class A $\subset$ Class L.

There is reason for optimism on this front, since \cite{Haehl:2013hoa} have demonstrated that purely flavour anomalies can be captured by non-dissipative effective actions (Class ND); cf.,  Appendix \ref{sec:ndf}. So it would seem that by suitably reverse engineering the construction of \cite{Haehl:2013hoa} and implementing the Legendre transformation we should be able to solve for anomalous contributions to the adiabatic hydrodynamics. Indeed this is all that needs to be done in the case of flavour anomalies. The mixed flavour and gravitational anomaly story however turns out to be a bit more intricate. In fact as mentioned earlier, it provides us with a strong rationale to switch from the formalism of non-dissipative fluid effective actions to the framework proposed herein.

In this section we will show that a specific class of anomalous terms is a subset of Class L, i.e., they can be formulated in terms of a Lagrangian.  For the case of flavour anomalies, this is a simple modification of  \cite{Haehl:2013hoa} which we will use as a guiding template.  We will extend that analysis to the case of general mixed anomalies in what follows. We will focus on constructing particular solutions to the anomalous adiabaticity equation \eqref{eq:Adiabaticity}. This will be sufficient to capture all the flavour and mixed  contributions which are neither gauge nor diffeomorphism invariant, but will be insufficient to capture the gauge and diffeomorphism terms that are necessary for the consistency of the Euclidean partition function. 
The terms we are unable to include are the $\PV$ terms discussed in \S\ref{sec:Cvector}, which comprise of the  transcendental anomaly contributions \cite{Jensen:2013rga}.  In the present section  our main aim will be on finding solutions to the off-shell adiabaticity equation; only in \S\ref{sec:anward} will we worry about the on-shell conditions and the anomalous Ward identities for these require Schwinger-Keldysh doubling of the degrees of freedom. Subsequently in \S\ref{sec:classLT} we will give a prescription that does appear to capture all anomaly induced transport in an extended theory of adiabatic hydrodynamics.

\subsection{Flavour anomalies}
\label{sec:fanom}

Let us begin our discussion by recalling some salient facts from the analysis of \cite{Haehl:2013hoa} in the context of effective actions for flavour anomalies. In the framework of Class ND effective actions \cite{Haehl:2013hoa} showed that an effective action given as a transgression form provides a solution to \eqref{eq:Adiabaticity} with $\THall^{\mu\perp} =0$. More specifically, it was shown that for a hydrodynamic system in $d= 2n$ dimensions living on a spacetime manifold ${\cal M}$ one has a local effective action in one higher dimension.\footnote{ This follows from the fact that we can use the anomaly inflow mechanism \cite{Callan:1984sa} to construct a local effective action by coupling our anomalous quantum system to a topological theory in the higher dimension.} We have an effective action that can be succinctly written on an extended spacetime $\bulkM_{d+1}$ with
$\partial \bulkM_{d+1} = {\cal M}$ being the physical spacetime where the fluid propagates. The effective action takes the beguilingly simple form
\begin{equation}
S_{anom} = \int_{\bulkM_{d+1}} \sqrt{-g_{d+1}}\;\Lag_{anom}
=\int_{\bulkM_{d+1}} \VP[\fA, \fAh] = \int_{\bulkM_{d+1}} \frac{\fu}{2\fomega} \wedge \prn{\fP[\fF]-\fPh[\fFh]} .
\label{eq:flasanom}
\end{equation}
In the equation above, we have also provided  an explicit expression for the transgression form $\VP[\fA,\fAh]$ in terms of
in terms of the hydrodynamic velocity 1-form $\fu$, the vorticity 2-form $\fomega$ and the
anomaly polynomial $\fP[\fF]$ which is a $d+2= 2n+2$ form built from the gauge invariant field strengths.\footnote{ We will follow the notational conventions of \cite{Loganayagam:2011mu, Jensen:2012kj, Jensen:2013kka,Jensen:2013rga,Haehl:2013hoa} using bold-face letters to indicate differential forms etc.. Furthermore, to retain compact expressions we perform some formal manipulations with differential forms as in the aforementioned references. Divisions by a differential form implicitly indicates that the numerator when expanded out always has a factor which cancels the form we divide by; the expression in
\eqref{eq:flasanom} is a $2n+1$ form written as if it were a ratio of a $2n+3$ form and a $2$ form. We refer the reader to the above references where these concepts are explained in greater detail.} Note that $d\fu = 2\,\fomega - \fu \wedge \fa$ where $\fa$ is the acceleration 1-form.

The transgression form denoted herein as $\VP[\fA,\fAh]$ is a functional of two gauge connections $\fA$ and $\fAh$. The former is simply the background gauge field source in differential form notation, while the latter is what was called in \cite{Haehl:2013hoa} as the ``hydrodynamical shadow gauge field''. It is a linear combination of the background source and the hydrodynamic velocity field defined as
\begin{equation}
\fAh = \fA + \mu\, \fu\,,
\label{eq:hatAdef}
\end{equation}
or more directly in terms of the hydrodynamic fields $\Bfields$ we have for its components
\begin{equation}
\hat{A}_\mu = A_\mu + T^2 \, \Kbeta_\mu \, \left(\LambdaB + \Kbeta^\sigma\, A_\sigma\right) .
\label{eq:hatAkb}
\end{equation}
The symbol $\fPh$ denotes the anomaly polynomial evaluated over the shadow gauge field. This shadow field appears pretty much universally in all attempts to understand anomalous transport in hydrodynamics; it was first encountered during an attempt to solve the anomalous adiabaticity equation in \cite{Loganayagam:2011mu} and plays a significant role in the anomalous hydrostatic partition function (for reasons that will be transparent soon)
\cite{Banerjee:2012cr,Jensen:2012kj,Jensen:2013kka,Jensen:2013rga}.

As written the anomalous effective action is simply a functional of the background sources $\{g_{\mu\nu}, A_\mu\}$ and the hydrodynamic fields $\Bfields=\{\Kbeta^\mu,\LambdaB\}$. The gauge field dependence is manifest, while the velocity field
$u^\mu$ can be expressed in terms of $\Bfields$ using \eqref{eq:hydrofields}. What is perhaps less clear is the dependence on the background metric, but owing to the presence of the shadow field in the transgression form, one has a non-trivial metric dependence. To be sure we are extending our sources and hydrodynamic fields to live on ${\bulkM}_{d+1}$. We will use the same symbols to denote the {\em bulk} hydrodynamic fields only differentiating the components by the indices when necessary. Lowercase Latin indices from the later half of the alphabet will denote bulk indices, with $\perp$ being used to denote the direction normal to the physical spacetime ${\cal M}$. To wit,
\begin{equation}
\hfields_{d+1} = \{ g_{mn}, A_m, \Kbeta^m, \LambdaB\} \,,\qquad \Kbeta^\perp = 0 \,.
\end{equation}

Thus, despite its origins within the framework of non-dissipative effective actions in \cite{Haehl:2013hoa},  \eqref{eq:flasanom} should be viewed as a particular element of Class L for our purposes with $\Lag = \VP[\fA,\fAh]$. Strictly speaking we are now extending our definition of Class L to include local Lagrangians in one higher dimension, as we must, if we insist on dealing with anomalous symmetries.

Generically transgressions are defined on a space of interpolating connections. For instance, given two connections say $\fA_1$ and $\fA_2$ respectively,  the transgression form denoted more generally as $\form{{\cal T}}[\fA_1,\fA_2]$ can be viewed as a functional of a continuous set of connections $\fA_t$ with $t\in[0,1]$ interpolating between $\fA_{t=0} = \fA_1$ and $\fA_{t=1} = \fA_2$.  One can write this quite succinctly for gauge connections as
\begin{equation}
\form{{\cal T}}[\fA_1,\fA_2] = \int_0^1\, dt\, \left[\frac{d\fA_t}{dt} \cdot\left( \frac{\partial \fP}{\partial \fF}\right)_t\right] ,
\end{equation}
with
\begin{equation}
\fA_t = t\, \fA_{t=1} + (1-t)\, \fA_{t=0} \,.
\end{equation}
Having this explicit expression is useful for carrying out the variational calculus we need to do to check that the functional
$S_{anom}$ does indeed provide a solution to the anomalous adiabaticity equation \eqref{eq:Adiabaticity} with $\THall^{\mu\perp} = 0$.

For the particular choice of connections $\fA_{t=0}  = \fAh$ and $\fA_{t=1} = \fA$ we define
an interpolation from the hydrodynamic shadow field to the physical gauge field source via
$\fA_t= \fA+(1-t) \,\mu\, \fu$. The corresponding field-strengths are given by
\begin{equation}
\begin{split}
\fF &= d\fA + \fA^2  = \fB + \fu \wedge \fE \,,\\
\fFh &= d\fAh + \fAh^2 = \fBh + \fu \wedge \fEh
= \fB +2\fomega \mu + \fu \wedge \prn{\fE-D\mu-\fa \mu} .
\end{split}
\end{equation}
$\fa$ and $\fomega$ are the acceleration and vorticity defined after \eqref{eq:flasanom}, while 
$\fB$ and $\fE$ are the rest frame magnetic 2-form and electric 1-form respectively.
The interpolating field-strength is $ \fF_t = t \fF + (1-t) \fFh$ since $(\Delta \fA)^2 = 0$ .
One can, of course, check explicitly that
\begin{equation}\label{eq:HatUnhTransgr}
\begin{split}
\VP [\fA,\fAh] &\equiv \int_0^1 dt \brk{\frac{d \fA_t}{dt}  \cdot \prn{\frac{\partial \fP}{ \partial \fF} }_t
}
= \frac{\fu}{2\fomega} \wedge\int_0^1 dt \brk{\frac{d \fF_t}{dt}  \cdot \prn{\frac{\partial \fP}{ \partial \fF} }_t
} \\
&= \frac{\fu}{2\fomega} \wedge \prn{\fP-\fPh} ,\\
\end{split}
\end{equation}
as indicated above.

To compute the variation of these transgression forms, we need to evaluate $\delta \VP[\fA,\fAh]$. The explicit computation is described in Appendix D of \cite{Haehl:2013hoa} and we quote the final result:
\begin{equation}
\begin{split}
\delta \VP [\fA,\fAh]
&=\delta \fA \cdot \star_{2n+1} \fJH
- \delta \fAh \cdot \star_{2n+1} \fJHh
+d \bigbr{\delta \fA \cdot \star \fJP+ \delta \fu \wedge \star\fqP } \,.
\end{split}
\label{eq:fvpvar0}
\end{equation}
Here $\fJH$ is the Hall current defined directly in terms of the variation of the anomaly polynomial:
\begin{equation}
\star_{2n+1} \fJH = \frac{\partial \fP}{\partial \fF} \,,
\end{equation}
with a similar expression for the shadow Hall current $\fJHh$. The two other currents appearing in \eqref{eq:fvpvar0} are defined in terms of the boundary terms arising from the variation
\begin{equation}
\begin{split}
\int_0^1 dt \brk{\delta \fA_t   \cdot \prn{\frac{\partial^2 \fP}{ \partial \fF \partial \fF}}_t \cdot \frac{d \fA_t}{dt}
} &= \delta \fA \cdot \star \fJP+ \delta \fu \wedge \star\fqP\,,
\end{split}
\end{equation}
where we have used $\fu \wedge \frac{d \fA_t}{dt} =0$ and parameterized the terms involved in the variation in terms of gauge potential variation and the velocity field variation. These quantities $\fJP$ and $\fqP$ are determined directly from the variational calculus to be
\begin{equation}
\begin{split}
\star \fJP  &\equiv
\int_0^1 dt \brk{ \prn{\frac{\partial^2 \fP}{ \partial \fF \partial \fF}}_t \cdot \frac{d \fA_t}{dt} } \\
&= \frac{\fu}{2\fomega} \wedge \bigbr{\frac{\partial \fP}{ \partial \fF} -\frac{\partial \fPh}{ \partial \fFh} }\,,
\end{split}
\label{eq:fJPdef}
\end{equation}
and
\begin{equation}\label{eq:fqPdef}
\begin{split}
\star\fqP &= \int_0^1 ds \int_0^s dt\brk{\mu \cdot \prn{\frac{\partial^2 \fP}{ \partial \fF \partial \fF}}_t
\cdot \frac{d \fA_t}{dt}
} \\
&= -\frac{\fu}{(2\fomega)^2} \wedge
\bigbr{ \fP -  \fPh
- \prn{\fF-\fFh}\cdot  \frac{\partial \fPh}{ \partial \fFh }  }\,.
\end{split}
\end{equation}

So far the variational calculus did not call for any details of how we are parameterizing the hydrodynamic fields. While we have indeed written the final expression in terms of $\delta \fu$, the variation of the velocity field, it is easy to convert this to the hydrodynamic field variations using \eqref{eq:varrules}. Explicitly one can
evaluate variation of the action in terms of the sources and the hydrodynamic fields\footnote{ One can also convert this variation to one involving the reference fields introduced in \S\ref{sec:reffields}. These should also be viewed as living on the reference bulk spacetime since the hydrodynamic fields which they are a proxy for are defined in terms of maps from there to the physical spacetime. We will shortly come back to this viewpoint to facilitate some parts of the analysis.}
to obtain an explicit answer for the variation as
\begin{align}
\delta \int_{\bulkM_{d+1}} \VP[\fA, \fAh] &=\bulkint \,  \Biggl\{  \prn{ \JH^m -P^m_n \JHh^n } \cdot \delta A_m
- \mu \cdot \JHh^q \brk{P_q^{(m} u^{n)}\; \delta g_{mn} + \prn{P_{qm}+u_q\, u_m} T\;\delta \Kbeta^m}
\Biggr.   \nonumber \\
&\hspace{4cm}\Biggl.
- T\,u_q\,  \JHh^q \cdot  (\delta \LambdaB + A_m \delta \Kbeta^m)   \Biggr\}
\nonumber \\
&\quad + \int_{{\cal M}}  \sqrt{-g} \left[ \JP^\alpha \cdot \delta A_\alpha
+ \qP^{(\alpha} u^{\beta)}\delta g_{\alpha\beta}  \right] .
\label{eq:fvpvar}
\end{align}
In deriving the above we have used \eqref{eq:Ahatvar} to write the variation of $\fAh$ in terms of the
physical fields and their variations. For notational simplicity we have also abbreviated
\begin{equation}
\bulkint \;\; \equiv \;\;  \int_{{\bulkM}_{d+1}} \,\sqrt{-g_{d+1}}
\label{eq:intdef}
\end{equation}
so as to avoid cluttering up the equations.

As we see there are two types of contributions to the variation of our Lagrangian $\VP[\fA,\fAh]$. On the one hand, we have some bulk variations (the first two lines in the r.h.s of \eqref{eq:fvpvar}) which define the bulk currents
living on $\bulkM_{d+1}$. To wit, 
\begin{equation}
\begin{split}
\Tbulk^{mn} &= - \mu \cdot \JHh^q \prn{P_q^{m} u^{n} +P_q^{n} u^{m} } \,,\qquad \Jbulk^m =  \JH^m -P^m_n \;\JHh^n\,, \\
\aheatbulk_m &= - \mu \cdot \JHh^q \,\prn{P_{qm}+u_q\, u_m} T\,,\qquad\;\, \achargebulk =  T\,u_q\,  \JHh^q \,.
\end{split}
\end{equation}
These have to to satisfy the analog of the bulk adiabaticity equation. This can be shown directly by running our argument for the Bianchi identity in the bulk theory; cf., the discussion around \eqref{eq:DiffeoSanom3} for an explicit proof of this statement.

More interesting for us are the boundary terms in the last line of \eqref{eq:fvpvar} -- these are the contributions that arise from the inflow mechanism. In particular, they
capture the constitutive relations for  anomalous hydrodynamics. Since we have the terms explicitly in terms of the source variations we can directly read off from here using \eqref{eq:LagVar} the anomalous currents to be
\begin{equation}
\begin{split}
(T^{\alpha\beta})_\text{A} &= \qP^\alpha u^\beta +  \qP^\beta u^\alpha  \,, \qquad    (J^\alpha)_\text{A} = \JP^\alpha \,, \qquad
(J_{S}^\alpha)_\text{A}=0\,.
\end{split}
\label{eq:fanomcur}
\end{equation}

These currents satisfy the anomalous adiabaticity equation \eqref{eq:Adiabaticity} on the boundary manifold ${\cal M}$. This was first established in \cite{Loganayagam:2011mu} and follows immediately from the previous analysis of \cite{Haehl:2013hoa}.\footnote{ We will give a more detailed derivation for the mixed anomalies in \S\ref{sec:mixanom}. Setting the gravitational terms in that analysis to zero will demonstrate the claim herein explicitly.} As described there, by isolating the anomalous contributions and solving the adiabaticity equation to give the above particular solution \eqref{eq:fanomcur}, one has accounted for all flavour anomalies. One can then couple the anomalous Lagrangian $\Lag_{anom}$ to any non-anomalous adiabatic fluid Lagrangian system and continue to satisfy adiabaticity.

Finally, let us make a remark on the construction above which will be useful for generalizations. The anomalous Lagrangian density $\sqrt{g_{d+1}} \,\Lag_{anom}$ is a scalar density on the bulk spacetime manifold. Per se, in keeping with our general philosophy this is an off-shell quantity, since we have nowhere insisted in our construction above that the fields be on-shell.  However, restricting to hydrostatics by enforcing $\Bfields = \Eqfields$ one ends up with an on-shell construction which as we now appreciate is related to the hydrostatic partition function $W_\text{Hydrostatic}$ \eqref{eq:Lpfn}.

In general the relation between the hydrostatic partition function and the non-dissipative fluid formalism is  complicated 
by a non-linear Legendre transform (see \S\ref{sec:ndf}). However, for the flavour anomalies the fact that $\Lag_{anom}$ \eqref{eq:flasanom} is independent of the entropy density makes the Legendre transformation trivial.\footnote{ This was the reason why the direct comparison of the non-dissipative effective action with the hydrostatic partition function worked quite seamlessly in the consistency checks carried out in
\cite{Haehl:2013hoa}.} This also to some extent underscores the rationale for introduction of the shadow gauge field $\fAh$; the shadow field plays a crucial role in ensuring the correct properties of the hydrostatic partition function as has been described in earlier works.

\subsection{Mixed anomalies}
\label{sec:mixanom}

We would now like to generalize anomalous adiabatic fluids to the case where we have gravitational or mixed anomalies.
One of the motivations for reviewing in some detail the flavour case in the previous subsection, was that it provides a hint of how one should generalize the construction to incorporate gravitational effects. To a large extent a specific solution to the anomalous adiabaticity equation in the presence of mixed anomalies can be obtained by treating the gravitational field as a non-abelian flavour field. This is roughly the correct intuition, though as we will see in the course of a more thorough analysis below there are some subtleties we need to deal with. In particular, we will see that the entropy current is modified in the presence of gravitational effects, no longer vanishing as in \eqref{eq:fanomcur}.

We are going to start our discussion for the mixed anomaly by mimicking the discussion for the flavour case. Specifically, since there is a close connection between the off-shell Lagrangian and the on-shell hydrostatic partition function we are going to take inspiration from the latter which is by now well understood for mixed anomalies \cite{Jensen:2013kka}. Once again we imagine that the set of anomalies of our underlying quantum system is encoded in an anomaly polynomial
$\fP[\fA, \fGamma]$ with $\fA$ being the gauge connection and $\fGamma$ the spin connection for the background geometry.\footnote{ We will mostly work with the one-form spin connection since it is most conducive for the purposes of carrying out the formal manipulations. Translating this to the Christoffel connection is reasonably straightforward. We will in fact do so  when we write out some explicit components.} We are also unabashedly going to work in the bulk geometry $\bulkM_{d+1}$ with the physical spacetime ${\cal M}= \partial \bulkM_{d+1}$ as before. The general set of conditions we impose on the geometry is similar to that encountered in \S\ref{sec:fanom}, though we will have to add some new ingredients as we proceed.

In analogy with the flavour story, let us consider modifying Class L by adding to the Lagrangian a higher dimensional term of the form
\begin{equation}
\begin{split}
\int_{\bulkM_{d+1}} \sqrt{-g_{d+1}}\; \Lag_{anom} =
\int_{\bulkM_{d+1}} \VP[\fA, \fGamma,\fAh,\fGammah] =
\int_{\bulkM_{d+1}} \frac{\fu}{2\fomega}\wedge \prn{\fP[\fF,\fR]-\fPh[\fFh,\fRh]}
\end{split}
\label{eq:mixLanom}
\end{equation}
The specific choice of the Lagrangian is motivated by the fact that $\VP[\fA, \fGamma,\fAh,\fGammah] $ is the canonical form for the scalar part of the hydrostatic anomalous partition function (Class $\PS$).  The key difference from \eqref{eq:flasanom} is the dependence on the background geometry; the anomaly polynomial now is a functional both of the background field strength and the background curvature $\fR$.

Let us examine the dependence of the potential anomalous Lagrangian introduced above: apart from the backgound sources $\{g_{mn}, A_m\}$  we have also the shadow fields appearing in $\Lag_{anom}$. The shadow flavour gauge connection $\fAh$ is the same as before being given by \eqref{eq:hatAdef}. The shadow spin connection however is new and requires to be defined. A natural course of action is to follow the partition function analysis of \cite{Jensen:2013kka} and demand that this be given in terms of the background spin connection, the velocity field, and a {\em spin chemical potential} $\Omega^\mu{}_\nu$  as
\begin{equation}
\fGammah^\mu{}_\nu=\fGamma^\mu{}_\nu+ \Omega^\mu{}_\nu \,\fu\,.
\label{eq:hatGamdef}
\end{equation}
Modulo defining the spin chemical potential we are now equipped with a putative anomalous Lagrangian.

The spin chemical potential should couple to the background metric structure since its origins are in the Lorentzian structure of the local tangent space geometry on ${\cal M}$ (and hence by inflow on $\bulkM_{d+1}$). In hydrostatic equilibrium the analysis of \cite{Jensen:2013kka} shows that it is related to the gradient of the Killing vector $\KEq^\mu$
which extends trivially to a Killing field on $\bulkM_{d+1}$. The relation in the hydrodynamic variables living on the physical spacetime ${\cal M}$ is just
\begin{equation}
\prn{\Omega^\mu{}_\nu}_\text{Hydrostatic} = T\, D_\nu \KEq^\mu
\label{eq:spcpeq}
\end{equation}
with $T$ being the equilibrium temperature. We could take this expression off-equilibrium and off-shell by simply replacing 
$\KEq^\mu \mapsto \Kbeta^\mu$. However, we need to be careful with the symmetries: for a Killing vector field
$D_{(\mu} \KEq_{\nu)} =0$ by virtue of Killing's equation. Hence only the antis-symmetric part of the tensor
$D_\mu \KEq^\nu$ is non-zero. Away from equilibrium when we consider $D_\mu \Kbeta^\nu$ we are likely to encounter both the symmetric and anti-symmetric contributions. The na\"ive generalization $\KEq^\mu \mapsto \Kbeta^\mu$ would retain both, but we claim that the correct off-shell extension of \eqref{eq:spcpeq} should only keep the anti-symmetric part.\footnote{ We do not have an a-priori reason to motivate this particular choice; what we can see is a post-facto argument. Choosing the spin chemical potential to contain the symmetric part of the gradient results in a tension with the off-equilibrium adiabaticity equation; see the discussion around \eqref{eq:ConstRelations} for what goes wrong. We note that choosing the anti-symmetric part makes the shadow connection metric
compatible (but not torsion free). This property  however does not uniquely characterize our choice.
}  With this motivation we define:
\begin{equation}
\Omega^\mu{}_\nu = \frac{1}{2}\,T \left( D_\nu \Kbeta^\mu - D^\mu \Kbeta_\nu\right) \equiv T \, \fatQ^{\;\mu\alpha}_{\,\nu\beta}\; D_\alpha \Kbeta^\beta \,,
\label{eq:DefOmega}
\end{equation}
where we have introduced the antisymmetrizer $\fatQ^{\mu\alpha}_{\nu\beta} = \tfrac{1}{2} ( \delta^\mu_\beta\, \delta^\alpha_\nu - g^{\mu\alpha} \,g_{\nu\beta})$ for future convenience. By construction we ensure that in hydrostatic equilibrium we recover the spin chemical potential of  \cite{Jensen:2013kka}.

The main claim we wish to make is that the Lagrangian density $\VP[\fA,\fGamma,\fAh,\fGammah]$ provides a solution to the adiabaticity equation \eqref{eq:Adiabaticity}. Furthermore, the currents derived from this Lagrangian are consistent with those derived earlier in \cite{Jensen:2013kka} in hydrostatic equilibrium. We will establish this by a straightforward computation.

\subsection{Variational calculus for mixed anomalies}
\label{sec:varmix}

The anomalous Lagrangian density $\VP[\fA,\fGamma,\fAh,\fGammah]$ \eqref{eq:mixLanom} is once again a transgression form. Let us therefore focus on the transgression formula between the pairs of gauge and spin connections $\{\fA_{t=1}, \fGamma_{t=1}\} = \{ \fA, \fGamma\}$ and $\{\fA_{t=0}, \fGamma_{t=0}\} = \{ \fAh, \fGammah\}$ respectively. Since the gravitational connection $\fGamma$ behaves exactly like a non-abelian gauge connection, the calculation is a straightforward generalization of what we had to in the case of the flavour anomaly.

We begin by considering the first variation of the transgression form, which is given directly by the analog of \eqref{eq:fvpvar}, except that now we have terms coming from the spin connection. This can be written in a reasonably compact form by introducing bulk Hall currents and boundary anomalous currents as:
\begin{equation}
\begin{split}
\delta \VP \brk{\fA, \fGamma, \fAh,\fGammah}
  &= \delta \fA \wedge\cdot \hodgeB \fJH - \delta \fAh \wedge\cdot \hodgeB \fJHh \\
  &\qquad+\half \delta \fGamma^a{}_b \wedge \hodgeB \fSpH{}^b{}_a- \half \delta \fGammah^a{}_b\wedge \hodgeB \fSpHh{}^b{}_a  \\
  &\qquad +d \bigbr{ \delta \fA\wedge \cdot  \star \fJP + \half \delta \fGamma^\alpha{}_\beta \wedge  \star \fSP{}^\beta{}_\alpha + \delta \fu \wedge  \star \fqP
  } \,,
  \end{split}
  \label{eq:GravTransVar}
  \end{equation}

The bulk Hall currents are themselves given in terms of the derivatives of the anomaly polynomial with respect to the field strengths and are given by
\begin{equation}
\begin{split}
\hodgeB \fJH  &= \frac{\partial \fP}{ \partial \fF} \,, \qquad \hodgeB \fSpH{}^b{}_a = 2\frac{\partial \fP}{ \partial \fR^a{}_b} \,,
\end{split}
\label{eq:HallCurrentsDef}
\end{equation}
These currents will play a role as before in determining the amount of anomaly inflow into the boundary theory. The anomaly induced boundary currents can also be determined explicitly and the only change due to the  spin connection is a slight generalization of \eqref{eq:fqPdef} for the current $\fqP$  to include appropriate gravitational contributions. The flavour anomaly induced current $\fJP$  being blind to the spin connection remains unchanged as in \eqref{eq:fqPdef}. In addition we have a spin anomaly current $\fSP$. 
To write down the expressions for the currents, it is useful to introduce an electromagnetic splitting of the field strength and curvature:
\begin{equation}
\begin{split}
\fB &= \fF - \fu \wedge \fE\,, \qquad\qquad\qquad\qquad\; \fE = -\ic_u \fF \,, \\
(\fBR)^\alpha{}_\beta &= \fR^\alpha{}_\beta - \fu \wedge \fER{}^\alpha{}_\beta \,,\qquad\qquad \fER{}^\alpha{}_\beta = - \ic_u \fR^\alpha{}_\beta \,,
\end{split}
\end{equation}
where $\ic_u$ denotes contractions with $u^\mu$. With this information we can now write down all three currents, as we do below for completeness:\footnote{ For further details we refer the reader to \cite{Jensen:2013kka}.}
\begin{equation}
\begin{split}
\star \fJP &= \int_0^1 dt \brk{ \prn{\frac{\partial^2 \fP}{ \partial \fF \partial \fF}}_t \cdot \frac{d \fA_t}{dt}
+\left( \frac{\partial^2\fP}{\partial \fF \partial \fR^\alpha{}_\beta}\right) \frac{d(\fGamma^\alpha{}_\beta)_t}{dt}} \\
&= \frac{\fu}{2\fomega} \wedge \bigbr{\frac{\partial \fP}{ \partial \fF} -\frac{\partial \fPh}{ \partial \fFh} }
= \frac{\partial \VP}{\partial \fB}\,,\\
\star \fSP{}^\beta{}_\alpha &= 2\int_0^1 dt \brk{ \prn{\frac{\partial^2 \fP}{ \partial \fR^\alpha{}_\beta \partial \fR^\gamma{}_\delta}}_t \frac{d (\fGamma^\gamma{}_\delta)_t}{dt}
+ \left( \frac{\partial^2 \fP}{\partial \fR^\alpha{}_\beta \partial \fF} \right) \cdot \frac{d \fA_t}{dt}}\\
&= 2\frac{\fu}{2\fomega} \wedge \bigbr{\frac{\partial \fP}{ \partial \fR^\alpha{}_\beta} -\frac{\partial \fPh}{ \partial \fRh^\alpha{}_\beta} }
= 2\frac{\partial \VP}{\partial (\fBR)^\alpha{}_\beta}\,,\\
\star \fqP &= \int_0^1 dt \, (1-t)\bigg[ \mu \cdot  \prn{ \frac{\partial^2 \fP}{\partial \fF\partial \fF} \cdot \frac{d\fA_t}{dt} + \frac{\partial^2 \fP}{\partial \fF\partial \fR^\alpha{}_\beta}
\frac{d(\fGamma^\alpha{}_\beta)_t}{dt}} \\
&\qquad\qquad\qquad\;\;+ \Omega^\alpha{}_\beta \prn{ \frac{\partial^2 \fP}{\partial \fR^\alpha{}_\beta \partial \fR^\gamma{}_\delta} \frac{d(\fGamma^\gamma{}_\delta)_t}{dt} + \frac{\partial^2 \fP}{\partial \fR^\alpha{}_\beta \partial \fF} \cdot \frac{d\fA_t}{dt} }\bigg]\\
&  = -\frac{\fu}{(2\fomega)^2} \wedge \brk{ \fP - \fPh - (\fF - \fFh) \cdot \frac{\partial \fPh}{\partial \fFh}
- (\fR^\alpha{}_\beta - \fRh^\alpha{}_\beta) \frac{\partial \fPh}{\partial \fRh^\alpha{}_\beta}}
= \frac{\partial \VP}{\partial (2\fomega)} \,,
\end{split}
\label{eq:InflowDef}
\end{equation}

In order to obtain bulk and boundary Bianchi identities from the basic variation \eqref{eq:GravTransVar}, we need to follow the same logic as in the case of flavour anomalies. We perform the detailed computation in Appendix \ref{sec:AnomBianchi} and only quote the result for the Bianchi identities of the (physical) boundary theory:
\begin{equation}
\begin{split}
D_\beta & (T^{\alpha\beta})_\text{A}=  (J^\beta)_\text{A} \cdot F^{\alpha}{}_\beta + \frac{g^{\alpha\sigma}}{\sqrt{-g}} \diffB \brk{ \sqrt{-g} \, T (\qP)_\sigma } \\
& \hspace{2cm} +\half D_\gamma \prn{ \SpH^{\perp[\alpha\gamma]} - \SpHh^{\perp[\alpha\gamma]}}
- \prn{ \mu \cdot \JHh^\perp
+ \half \Omega^\nu{}_\mu \SpHh{}^{\perp \mu}{}_\nu } u^\alpha \,,
\end{split}
\label{eq:LorentzBdryBianchi1}
\end{equation}
and
\begin{equation}
D_\alpha (J^\alpha)_\text{A} = \JH^\perp - \JHh^\perp \,,
\label{eq:LorentzBdryBianchi2}
\end{equation}
which are satisfied by the anomalous currents 
\begin{equation}
\begin{split}
(T^{\alpha\beta})_\text{A} &= \frac{2}{\sqrt{-g}}\frac{\delta S_{anom}}{\delta g_{\alpha\beta}} \bigg{|}_{boundary} = \qP^\alpha u^\beta +  \qP^\beta u^\alpha+ \frac{1}{2} \,D_\rho \prn{\SP^{\alpha[\beta\rho]} + \SP^{\beta[\alpha\rho]} -\SP^{\rho(\alpha\beta)}} \,, \\
(J^\alpha)_\text{A}&= \frac{1}{\sqrt{-g}}\frac{\delta S_{anom}}{\delta A_{\alpha}} \bigg{|}_{boundary} = \JP^\alpha \,.
\end{split}
\label{eq:bdyAnomCur}
\end{equation}

We now want convert these Bianchi identities into an adiabaticity equation and check that \eqref{eq:Adiabaticity} is satisfied with an appropriate choice of currents. Since the energy-momentum and charge currents are defined by varying the Lagrangian with respect to the sources, these currents are already manifest in the above expressions. Plugging these in and demanding that the following adiabaticity equation be upheld
\begin{equation}
\begin{split}
D_\alpha (J_{S}^\alpha)_\text{A} &+ \Kbeta_\alpha \left[ D_\sigma (T^{\alpha\sigma})_\text{A} - 
(J^\sigma)_\text{A} \cdot F^\alpha{}_\sigma
- \half D_\gamma \SpH^{\perp[\alpha\gamma]} \right] \\
&+ (\LambdaB + \Kbeta^\alpha A_\alpha) \cdot \prn{ D_\sigma (J^\sigma)_\text{A} - \JH^\perp } = 0
\end{split}
\label{eq:AElorentz}
\end{equation}
results in a non-trivial solution for $(J_{S}^\alpha)_\text{A}$! More precisely, we find that in addition to \eqref{eq:bdyAnomCur}, we need to define the following entropy current in order to get a solution to \eqref{eq:AElorentz}:
\begin{equation}
\begin{split}
(J_{S}^\alpha)_\text{A} &= -\half \Kbeta_\sigma \, \SpHh^{\perp[\alpha\sigma]} \,.
\end{split}
\label{eq:ConstRelations}
\end{equation}

The expressions \eqref{eq:bdyAnomCur} and \eqref{eq:ConstRelations} define a required particular solution to \eqref{eq:Adiabaticity} that can be used to remove the anomaly terms (both flavour and Lorentz anomalies). What is curious in our construction is the fact that we have necessarily had to modify the entropy current in order to achieve this. Specifically, the entropy current $(J^\alpha_{S})_A$ does not quite satisfy \eqref{eq:sdef} anymore. We conclude that the solution to the anomalous adiabaticity equation requires modifying the entropy current apart from noting that given the variational principles, it is the only current that we are free to manipulate. 

While the reader might consider the above set of statements somewhat ad-hoc, we should add that the structure of the terms is rather tightly constrained. We have not been able to find any other Lagrangian solution to the anomalous adiabaticity equation. Moreover, small modifications such as allowing the spin chemical potential to be defined directly in terms of the gradient of the velocity field (i.e., without the anti-symmetrization introduced in \eqref{eq:DefOmega}) ends up destroying the structure. One intuition we can offer is the following: in contrast to the flavour anomaly discussion of \S\ref{sec:fanom} the new element we have to account for is the background metric variation. Since the diffeomorphism symmetry enters more universally any slight modification of the structures results in inconsistencies. We believe that this is indicative of some underlying structure that can be used to formulate our arguments more robustly -- we will pursue this line of thought in the future (see however \S\ref{sec:classLT} for some preliminary ideas on this front).

It is worth recording that in hydrostatic equilibrium $\SpHh^{\perp[\alpha\beta]}=0$ and we reproduce the result of \cite{Jensen:2013kka}. Furthermore, we also see that the anomalous stress-tensor and charge current in \eqref{eq:bdyAnomCur}  take precisely the form that is expected by na\"ively generalizing hydrostatic results. We take these to be important consistency checks of our construction. 
 Attempting to solve the adiabaticity equation directly to obtain off-shell currents, leads to somewhat different constitutive relations. While this will be discussed elsewhere \cite{Jensen:2014yg}, it is worth noting that this result uses a different spin chemical potential (in particular they take the gradient of the thermal vector eschewing the projection to the anti-symmetric part as in \eqref{eq:DefOmega}).

More generally, it is worth keeping in mind that the Class A constitutive relations are particular solutions to the inhomogeneous adiabaticity equation. As always these are ambiguous to shifts by homogeneous solutions. In terms of the current discussion, we have the freedom to add into Class A any other adiabatic constitutive relation, whilst maintaining adiabaticity. So two a-priori different looking solutions to the anomalous adiabaticity equation should be demonstrably related by adding in a linear combination of terms from the other six classes. More formally, Class A constitutive relations take values in the coset 
(Adiabatic constitutive relations)/$(\text{Class} \;\{\text{L}, \text{B},\text{C},\text{V}\})$ with $\text{V} = \PV\cup \GV$ denoting the vector classes.

\subsection{On-shell dynamics of anomalous adiabatic fluids}
\label{sec:mixos}

Given that we have off-shell adiabatic constitutive relations \eqref{eq:bdyAnomCur}, \eqref{eq:ConstRelations} we can ask whether our anomalous effective action \eqref{eq:mixLanom} satisfies the correct on-shell constraints. A-priori we expect based on our knowledge of the flavour anomaly discussion of \cite{Haehl:2013hoa}, that the on-shell Ward identities are not going to be obeyed by our Lagrangian system. We will show  in \S\ref{sec:anward} that a thermofield doubled construction can fix this problem. For now we  are simply going to use the construction of the previous subsections to show that the on-shell equations  we obtain from Class L anomalous hydrodynamics are incorrect.

To get started, let us assume that  $\int_{{\bulkM}_{d+1}} \VP[\fA,\fGamma,\fAh,\fGammah]$ provides for us a particular solution to the anomalous adiabaticity equation. The complete hydrodynamical system as we have discussed hitherto is then a combination of a {\em non-anomalous} part and the anomalous terms we have just taken care of. So we can write an effective action for our system as a sum of two contributions
\begin{equation}
S_{eff}\brk{\hfields} = \int_{\cal M} \sqrt{-g} \ \Lag_\text{n-a}\brk{\hfields} + \int_{{\bulkM}_{d+1}} \VP[\fA,\fGamma,\fAh,\fGammah]
\label{eq:SeffF}
\end{equation}
and treat the entire bulk + boundary dynamics as part of an extended Class L system.  In what follows we will  denote the contribution from the non-anomalous terms
in Class L arising from $\Lag_\text{n-a}$ by the subscript `{n-a}' so as to keep track of them explicitly.
These terms then are required to satisfy the non-anomalous Bianchi identities from \eqref{eq:LHydroEq}:
\begin{equation}
\begin{split}
D_\nu T^{\mu\nu}_\text{n-a}&= (J_\nu)_\text{n-a} \cdot F^{\mu\nu}
+\frac{g^{\mu\nu}}{\sqrt{-g}}\diffB\prn{\sqrt{-g}\ T\,(\aheat_\nu)_\text{n-a}}
+  g^{\mu\nu} T\,\acharge_\text{n-a} \cdot \diffB A_\nu \,,  \\
D_\sigma J^\sigma_\text{n-a} &= \frac{1}{\sqrt{-g}}\diffB\prn{\sqrt{-g}\ T\,\acharge_\text{n-a}}\,.
\end{split}
\label{eq:NonAnomConstrains}
\end{equation}
We note that we are not adding any bulk non-anomalous terms since the presumption is that the physical fluid lives on ${\cal M}$ with the bulk fields on $\bulkM_{d+1}$ simply providing us with an efficient way to keep track of the  inflow  and Hall currents.

Since we are interested in the on-shell dynamics, let us introduce the reference fields $\{\Kref^a,\Lref\}$
and their related pullback fields $\{\varphi^a,c\}$.\footnote{ For the discussion of anomalous fluids the reference fields and the reference manifold are taken to be ($d+1$)-dimensional. However, apart from using different indices as summarized in Table \ref{notation:tabVariations} we will refrain from introducing a new notation for the bulk reference quantities; hopefully it will be clear from the context whether we are discussing the bulk or the boundary reference data. }  The dynamical information of the theory is obtained by extremizing the  effective action $S_{eff}$ with respect to the pullback fields. 

Performing the required manipulations, we firstly find sensible equations of motion for bulk quantities which we quote in \S\ref{sec:OnShellAnom}. Similarly, for the boundary degrees of freedom we find that the extremization in the Lie orbit of the reference sources $\{\gref_{ab}, \Aref_a\}$ leads to
\begin{equation}
\begin{split}
\frac{g^{\mu\nu}}{\sqrt{-g}}&\diffB\prn{\sqrt{-g}\ T\,\brk{(\aheat_\nu)_\text{n-a}+(\qP)_\nu}}
+  g^{\mu\nu}\, T\,\acharge_\text{n-a} \cdot \diffB A_\nu \simeq 0\,,  \\
\frac{1}{\sqrt{-g}}&\diffB\prn{\sqrt{-g}\ T\,\acharge_\text{n-a}} \simeq 0\,.
\end{split}
\label{eq:AnomConstraints}
\end{equation}
Note that the anomalous part of the action only contributes a single term proportional $(\qP)_\mu$. This can be seen from \eqref{eq:DeltaSanom2} where all the boundary terms except the very last one give vanishing contribution when we restricted to the constrained variation in the Lie orbit of the reference sources.

Using then the Bianchi identities \eqref{eq:LorentzBdryBianchi1}, \eqref{eq:LorentzBdryBianchi2} together with the on-shell dynamical equations \eqref{eq:AnomConstraints} we find that the  on-shell fluid configurations on the boundary ${\cal M}$ obey
\begin{align}
&D_\beta\prn{T^{\alpha\beta}_\text{n-a}+ (T^{\alpha\beta})_\text{A}}  \nonumber \\
&\qquad\simeq (J^\sigma_\text{n-a}+(J^\sigma)_\text{A})\cdot F^\alpha{}_\sigma
+\half D_\gamma \prn{ \SpH^{\perp[\alpha\gamma]} - \SpHh^{\perp[\alpha\gamma]}}
- \prn{ \mu \cdot \JHh^\perp
+ \half \Omega^\nu{}_\mu \SpHh{}^{\perp \mu}{}_\nu } u^\alpha \,, \nonumber \\
& D_\sigma (J^\sigma_\text{n-a}+(J^\sigma)_\text{A}) \simeq \JH^\perp- \JHh^\perp \,.
\label{eq:FalseHydroEom}
\end{align}
We note that these are not quite the correct hydrodynamic equations. The anomalous Ward identities  \emph{should} rather be
\begin{equation}
\begin{split}
D_\beta\prn{T^{\alpha\beta}_\text{n-a}+ (T^{\alpha\beta})_\text{A}}
&\simeq (J^\sigma_\text{n-a}+(J^\sigma)_\text{A})\cdot F^\alpha{}_\sigma +\half D_\gamma \SpH^{\perp[\alpha\gamma]} \,,
\\
D_\sigma (J^\sigma_\text{n-a}+(J^\sigma)_\text{A}) &\simeq \JH^\perp \,.
\end{split}
\label{eq:TrueHydroEom}
\end{equation}
The troublesome terms are the shadow (hatted) currents on the r.h.s. of \eqref{eq:FalseHydroEom}.

Thus, we conclude that further modification is required in how one formulates the
field theory of the pullback fields for it to match with the usual hydrodynamics.
This will be the focus of \S\ref{sec:anward} where we will follow the analysis of \cite{Haehl:2013hoa}
to show that the correct Ward identities which require removing the shadow terms from the r.h.s.\ of
\eqref{eq:FalseHydroEom}, can be obtained by working with a doubled set of degrees of freedom. 
Before applying the Schwinger-Keldysh technology to Class A, let us now set up an appropriate general formalism.

\section{Schwinger-Keldysh formalism for Class L and application to Class A}
\label{sec:skdouble}

Thus far we have tried to formulate hydrodynamics in terms of response to a
single set of background fields $\{g_{\mu\nu},A_\mu\}$. However, since hydrodynamics
is ultimately a statistical system, we should allow for statistical fluctuations. By
the fluctuation-dissipation theorem these statistical fluctuations are closely tied to allowing statistical dissipation. The correct framework for dealing with this is the Schwinger-Keldysh formalism
\cite{Schwinger:1960qe,Keldysh:1964ud}  whereby the dynamical fields (and the background sources) are doubled.

Now the astute reader would wonder why we bring up the issue of doubling the fields since the basic premise of  the adiabatic fluid formalism is that it is conservative; on-shell the adiabaticity equation ensures that no entropy is produced thus allowing no dissipation. Nevertheless since we are interested in classifying hydrodynamic transport in general, it is worthwhile analyzing the situation in the adiabatic case which provides a useful context for the general discussion. Furthermore, our discussion of anomalous transport in \S\ref{sec:anomalies} confronts us with the issue of getting unwanted shadow contributions to the Ward identities. As we will show, this is an artifact of working with a single copy theory and doubling the degrees of freedom recovers for us the correct Ward identities, despite falling within the purview of adiabatic transport (Class A) as in \cite{Haehl:2013hoa}.

There are two issues we wish to highlight in the hydrodynamic Schwinger-Keldysh functionals that we will construct below. The first is the doubling of degrees of freedom and the associated symmetries. The second is the fact that such functionals a-priori allow interactions between the two sets of degrees of freedom. These terms are  are sometimes referred to as {\em influence functionals} following \cite{Feynman:1963fq}. In fact, our previous discussion of Lagrangian constructions of anomalous hydrodynamic effective actions  in \cite{Haehl:2013hoa} already exemplified the occurrence of such interaction terms.  We will however take this opportunity to rephrase the construction in a more canonical fashion. Along the way we will see some  advantages of the Class L reference field formalism for the Schwinger-Keldysh functionals. While this appears to hold useful clues in understanding how to apply the Schwinger-Keldysh construction in generic non-equilibrium dynamics, we will also find a certain tension with the adiabaticity equation. The following discussion should be viewed as a first step in setting up the general construction and we will in particular highlight some of the missing ingredients. A fuller exposition of these ideas will however be postponed to a future publication \cite{Haehl:2014kq}.

\subsection{Schwinger-Keldysh fields on the reference manifold}
\label{sec:skrefm}

With these facts in mind, let us now consider uplifting our construction from a single copy of fields $\hfields$ to the Schwinger-Keldysh doubled system. We start by doubling the fields to a pair of left and right fields indexed by sub/superscripts $\skLl$ and $\skRl$ respectively, $\{\hfields_\skL, \hfields_\skR\}$. In particular we not only double the background sources to $\{g_{\mu\nu}^\skL ,A_\mu^\skL\}$ and $\{g_{\mu\nu}^\skR,A_\mu^\skR\}$ but we also double the dynamical fields to
$\{\varphi_\skL,c_\skL\}$ and $\{\varphi_\skR,c_\skR\}$. We will want to couple both sets of these physical fields to the  background fields.

\begin{figure}
\setlength{\unitlength}{0.1\columnwidth}
\includegraphics[width=1.0\textwidth]{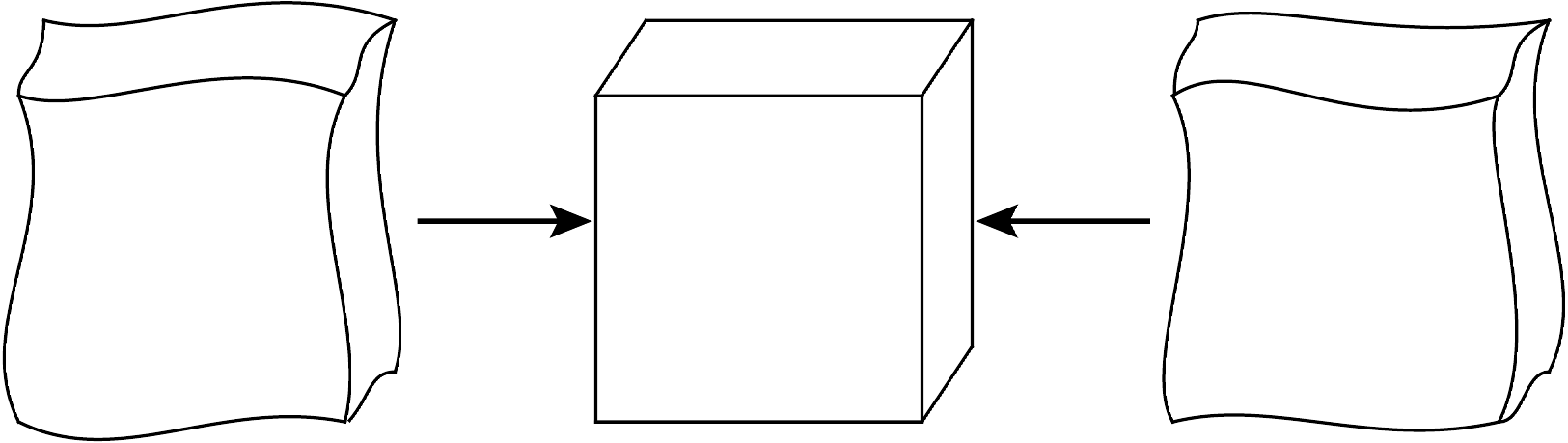}
\begin{picture}(0.3,0.4)(0,0)
\put(5,0.2){\makebox(0,0){$\Mref$}}
\put(1.3,0.2){\makebox(0,0){${\cal M}_{\skL}$}}
\put(8.7,0.2){\makebox(0,0){${\cal M}_{\skR}$}}
\put(1.1,2){\makebox(0,0){$\{g_{\mu\nu}^\skL, A_\mu^\skL\}$}}
\put(8.5,2){\makebox(0,0){$\{g_{\mu\nu}^\skR, A_\mu^\skR\}$}}
\put(4.84,2){\makebox(0,0){$\{\gref_{ab}^\skL, \Aref_a^\skL\}\,\{\gref_{ab}^\skR, \Aref_a^\skR\}$}}
\put(3.1,2.2){\makebox(0,0){$\{\varphi_\skL^a,c_\skL\}$}}
\put(6.9,2.2){\makebox(0,0){$\{\varphi_\skR^a,c_\skR\}$}}
\put(1.1,1.3){\makebox(0,0){$\{\Kbeta^\mu_\skL,\LambdaB^\skL\}$}}
\put(4.8,1.3){\makebox(0,0){$\{\Kref^a,\Lref\}$}}
\put(8.5,1.3){\makebox(0,0){$\{\Kbeta^\mu_\skR,\LambdaB^\skR\}$}}
\end{picture}
\caption{Illustration of the Schwinger-Keldysh setup. The physical spacetime manifold ${\cal M}$ has been doubled.
However, the two copies are not entirely independent as they are both related to the same reference configuration
on $\Mref$ via pull-backs using the dynamical fields $\{\varphi,c\}_{\skL,\skR}$. Despite the presence of two copies of
source fields on $\Mref$ there is only one diffeomorphism and gauge redundancy involved; invariance under this
symmetry implies Schwinger-Keldysh Bianchi identities.
}
\label{fig:SK}
\end{figure}
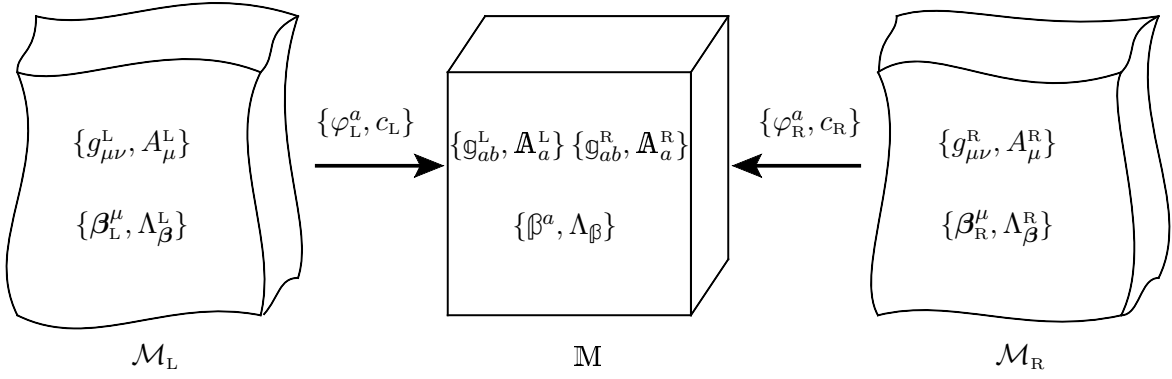

This has one important implication: since there are now two sets of pull-back fields which we can apply on the reference fields $\{\Kref^a,\Lref\}$ living on $\Mref$, we also have two sets of hydrodynamic fields $\{\Kbeta^\mu,\LambdaB\}_\skL$
and $\{\Kbeta^\mu,\LambdaB\}_\skR$ coupling to the corresponding backgrounds. Indeed this is to be expected since we literally wanted to double the numbers of degrees of freedom of our system.
Furthermore, it is clearly possible to derive both copies of the theory  on the physical spacetime ${\cal M}$ from the same reference configuration on $\Mref$. To wit, the $\skRl$-fields
are related to the reference fields via
\begin{equation}
\begin{split}
\Kbeta^\mu_\skR(x_\skR) &= (e_\skR)^\mu_a\;  \Kref^a[\varphi_\skR(x_\skR)]\,, \\
\Lambda_\Kbeta^\skR (x_\skR) &= c_\skR(x_\skR) \, \Lref[\varphi_\skR(x_\skR)] \, c^{-1}_\skR(x_\skR) + \Kbeta^\sigma(x_\skR)\; \partial_\sigma c_\skR(x_\skR) \, c_\skR^{-1} (x_\skR) \,,
\end{split}
\label{eq:pullbackR}
\end{equation}
and similarly for the $\skLl$-fields. On the other hand, if we push-forward the sources from the physical manifold ${\cal M}$ onto the reference manifold $\Mref$,  we get two copies of reference sources:
\begin{equation}
\begin{split}
g_{\mu\nu}^{\skR}(x_\skR) &= \partial_\mu \varphi_\skR^a\; \partial_\nu \varphi_\skR^b \;
\gref^{\skR}_{ab} [\varphi_\skR(x_\skR)]\,, \\
A_\mu^\skR(x_\skR) &= \partial_\mu \varphi_\skR^a\; \Big[ c_\skR(x_\skR) \; \Aref^{\skR}_a[\varphi_\skR(x_\skR)]\; c_\skR^{-1}(x_\skR)\Big] -
\partial_\mu c_\skR(x_\skR) \, c_\skR^{-1}(x_\skR) \,, \\
\end{split}
\label{eq:pullbackRgA}
\end{equation}
and similarly for the $\skLl$-sources. See Fig.\ \ref{fig:SK} for an illustration.

We now see a major advantage of introducing the reference configuration. In traditional treatments of the Schwinger-Keldysh formalism while one is fine with doubling the physical degrees of freedom, doubling of the background sources, especially the metric, poses an important puzzle. If we have a physical set of degrees of freedom living on a spacetime manifold ${\cal M}$ are we working with a pair of diffeomorphic manifolds ${\cal M}_\skR, {\cal M}_\skL$ in the doubled theory? If so, since points on two distinct  manifolds may only be related up to an overall diffeomorphism, how does one map degrees of freedom on the right to those on the left? The pay-off of introducing a reference manifold is that it provides a common ground for comparing the two different degrees of freedom. Since ${\cal M}_{\skL} \; \reflectbox{\ensuremath{\mapsto}}\; \;\Mref \;\mapsto {\cal M}_\skR $, we can bring all the physical fields onto the reference manifold using the pullback fields, perform all manipulations on the reference manifold and thence push forward to the physical spacetime. We believe this is the correct way to understand the symmetries of the Schwinger-Keldysh formalism. 

Having identified the relevant degrees of freedom we can write the Schwinger-Keldysh action generically as
\begin{equation}
\begin{split}
S_{SK}\brk{\hfields_\skR, \hfields_\skL} &= \int_{{\cal M}_\skR} \, \sqrt{-g_\skR} \, \Lag\brk{\hfields_\skR} -
\int_{{\cal M}_\skL} \, \sqrt{-g_\skL} \, \Lag\brk{\hfields_\skL}
+S_{_{IF}}\brk{\hfields_\skR, \hfields_\skL}\,, \\
&= \int_\Mref\, \bigg(\sqrt{-\gref_\skR} \; \Lagref\brk{\hreffields_\skR}
- \sqrt{-\gref_\skL} \; \Lagref\brk{\hreffields_\skR}+  \sqrt{-\gref_\skR} \;
\Lagref_{_{IF}}\brk{\hreffields_\skR,\hreffields_\skL} \bigg) \,.
\end{split}
\label{eq:Lsk}
\end{equation}
We have allowed here for the possibility of a non-vanishing influence functional that couples the two sets of degrees of freedom. 
In the process we have arbitrarily chosen to write $\Lagref_{_{IF}}$ as a scalar field with the metric on $\Mref$ taken to be $\gref_\skR$.\footnote{ Of course, the choice of the influence functional Lagrangian density being given in terms of the R metric is simply a matter of convenience. The choice matters in practical terms; when we have to define index contractions, covariant derivatives etc., which will be done with $g^\skR_{\mu\nu}$ with the above convention.}

For the situations we have discussed so far we do not need to include such terms, but anomalous (and also dissipative) parts of the constitutive relations will turn out to involve non-trivial influence functionals. At the moment we will not impose any constraints on $\Lag_{IF}$  (apart from the obvious symmetry requirements).

We now want to describe the variational principle on the reference manifold for Schwinger-Keldysh actions of the form \eqref{eq:Lsk}. Varying $S_{SK}$ with respect to the various fields and sources yields a-priori two sets of currents associated with the two sets of degrees of freedom:
\begin{align}\label{eq:SKrefVar}
 \delta S_{SK}\brk{\hfields_\skR, \hfields_\skL} &=  \int_{\Mref} \sqrt{-\gref_\skR}\, \bigg\{ \frac{1}{2} \Tref^{ab}_\skR \delta \gref_{ab}^\skR - \frac{1}{2} \,\gratio\, \Tref^{ab}_{\skL} \delta \gref^\skL_{ab}
    + \Jref_\skR^a\cdot \delta \Aref_a^\skR - \gratio\,\Jref_\skL^a\cdot \delta \Aref_a^\skL  \\
    & + \prn{ \Tref_\skR\,\aheatref^\skR_a -\gratio \, \Tref_\skL\, \aheatref^\skL_a} \delta \Kref^a
    + \Tref_\skR\,\achargeref_\skR \cdot(\delta \Lref +\Aref_a^\skR \delta\Kref^a  ) 
    - \Tref_\skL\,\gratio \, \achargeref_\skL \cdot(\delta \Lref +  \Aref_a^\skL\delta \Kref^a) \bigg\} 
    \notag
\end{align}
where we abbreviated $\gratio = \sqrt{-\gref_\skL}/\sqrt{-\gref_\skR}$.\footnote{ Note that temperature and chemical potential come as $\skR$ and $\skL$ versions since they depend on the metric and gauge field, respectively: 
$$\Tref_{\skR/\skL} = \frac{1}{\sqrt{- \gref_{ab}^{\skR/\skL} \Kref^a \Kref^b}} \,, \qquad \frac{\muref_{\skR/\skL}}{\Tref_{\skR/\skL}} = \Lref+\Kref^c \Aref_c^{\skR / \skL}$$} While this defines the general Schwinger-Keldysh constitutive relations, the equations of motion are obtained by extremizing $S_{SK}$ with respect to $\{\varphi^a,c\}_{\skR,\skL}$ inside $\{\gref_{ab}[\varphi], \Aref_a[\varphi]\}_{\skR,\skL}$, holding fixed $\{\Kref^a,\Lref\}$. Since there are now two copies of pullback fields $\{\varphi^a,c\}_{\skR,\skL}$, there are a-priori two copies of diffeomorphism and gauge symmetries. For example, we can obtain equations of motion for $\{\Tref_\skR^{ab},\Jref_\skR^a\}$ by infinitesimally varying the right pullback fields. To wit, consider a variation $\{ \delta \varphi_\skR^a , -c_\skR^{-1}\delta c_\skR\}$ inside the R-fields:
\begin{equation}\label{eq:SKconstrainedVar}
\begin{split}
\diffCons \gref^{\skR}_{ab} = -\delta_\varphi \,\gref^{\skR}_{ab} \,,\quad &\diffCons \Aref^{\skR}_{a} = -\delta_\varphi \,\Aref^{\skR}_{a} \,,\\
 \diffCons \gref^{\skL}_{ab}=0 \,,\quad & \diffCons \Aref^{\skL}_{a} = 0 \,, \\
 \diffCons \Kref^a = 0 \,,\quad &\diffCons \Lref = 0 \,.
\end{split}
\end{equation}
Applying this variation to \eqref{eq:SKrefVar}, we obtain
\begin{equation}
\begin{split}
 \diffCons S_{SK}\brk{\hfields_\skR, \hfields_\skL} &=  \int_{\Mref} \sqrt{-\gref_\skR}\, \bigg\{
   \delta \varphi^c_\skR \brk{ \Dref_a (\Tref_\skR)^a_c  -\Jref_\skR^a \cdot \Fref_{ca}^\skR }
  +(-c_\skR^{-1}\delta c_\skR + \Aref^\skR_a \delta \varphi^a_\skR) \cdot \Dref_c \Jref_\skR^c \bigg\} \,.   
\end{split}
\end{equation}
From this we can read off the usual equations of motion for $\{\Tref^{ab}_\skR, \Jref^a_\skR\}$. Equations of motion for the left copy can be derived in an analogous fashion. 

However, note that it is not obvious how the two copies should be coupled. In order to make this deficiency of a na\"ive Schwinger-Keldysh formalism more apparent, let us briefly demonstrate how it fails to give a sensible adiabaticity equation. To this end, let us consider Bianchi identities on the reference manifold as obtained from a common diffeomorphism and gauge transformation along $\{\xiref^a,\Lambdaref\}$:
\begin{equation}
\begin{split}
\diffF \gref^{\skR/\skL}_{ab} &= \lieD_\xiref \,\gref^{\skR/\skL}_{ab} \,, \quad
\diffF \Kref^a =\lieD_\xiref \, \Kref^a \,, \\
 \diffF \Aref^{\skR/\skL}_{a} &= \lieD_\xiref \,\Aref^{\skR/\skL}_{a} + [\Aref^{\skR/\skL}_a, \Lambdaref] + \partial_a \Lambdaref \,,\\
 \diffF\Lref + \Aref_a^{\skR/\skL} \diffF \Kref^a &= \xiref^a \Dref_a [ \Lref+\Kref^b \Aref^{\skR/\skL}_b] - \Kref^a \Dref_a [\Lambdaref+ \xiref^b \Aref^{\skR/\skL}_b] \\
 &\qquad  - \xiref^a \Kref^b \, \Fref^{\skR/\skL}_{ab} + [\Lref + \Kref^a \Aref_a^{\skR/\skL}, \Lambdaref+ \xiref^b \Aref_b^{\skR/\skL}] \,.
\end{split}
\end{equation}
Inserting this in \eqref{eq:SKrefVar} we can do exactly the same integration by parts as in \S\ref{sec:LBianchi} to obtain:
\begin{align}
 &\diffF S_{SK}\brk{\hfields_\skR, \hfields_\skL} 
 \notag \\
 &\quad= \int_{\Mref}\sqrt{-\gref_\skR}\, \xiref_a \bigg[ - \Dref_b \prn{\Tref^{ab}_\skR - \gratio \Tref^{ab}_\skL} + \prn{\Jref^\skR_b \cdot \Fref^{ab}_\skR - \gratio\Jref^\skL_b \cdot \Fref^{ab}_\skL} 
\notag \\
 &\quad\qquad + \frac{\gref_\skR^{ab}}{\sqrt{-\gref_\skR}} \diffBref \prn{ \sqrt{-\gref_\skR} \, [\Tref_\skR\,\aheatref^\skR_b -\Tref_\skL\, \gratio \,\aheatref^\skL_b]}
 + \prn{\gref_\skR^{ab} \, \Tref_\skR \, \achargeref_\skR \cdot \diffBref \Aref_b^\skR 
   - \gref_\skR^{ab} \, \Tref_\skL \,\gratio\, \achargeref_\skL \cdot \diffBref  \Aref_b^\skL }\bigg] 
   \notag \\
 &\qquad+ \int_{\Mref}\sqrt{-\gref_\skR}\, (\Lambdaref+\xiref^c \Aref_c^\skR) \cdot \left[ -\Dref_c \Jref^c_\skR + \frac{1}{\sqrt{-\gref_\skR}} \diffBref \prn{\sqrt{-\gref_R} \, \Tref_\skR\, \achargeref_\skR} \right] 
 \notag \\
 &\qquad- \int_{\Mref}\sqrt{-\gref_\skR}\, (\Lambdaref+\xiref^c \Aref_c^\skL) \cdot \left[ -\Dref_c (\gratio  \Jref^c_\skL) + \frac{1}{\sqrt{-\gref_\skR}} \diffBref \prn{\sqrt{-\gref_R} \, \Tref_\skL\, \gratio \, \achargeref_\skL} \right] \,.
  \label{eq:SKrefBianchiVar}
\end{align}
where $\diffBref$ denotes Lie transport along $\Breffields=\{\Kref^a,\Lref\}$.
We can directly read off Bianchi identities from this variation. We define the reference manifold entropy current in the same way as on physical spacetime ${\cal M}$, i.e.\ $\Jref_S^a = (\Tref_\skR \, \sref_\skR -  \gratio \,\Tref_\skL \, \sref_\skL) \, \Kref^a$ with $\sref_{\skR/\skL}$ defined as the functional derivative of $S_{SK}$ with respect to $\Tref_{\skR/\skL}$ exactly as in \eqref{eq:sdef}. Using the Bianchi identities from 
\eqref{eq:SKrefBianchiVar} it is then straightforward to derive the following adiabaticity equation in complete analogy to \eqref{eq:Laddemo}:
\begin{equation}
\begin{split}
 \Dref_a \Jref_S^a &+ \Kref_a \brk{ \Dref_b \prn{\Tref_\skR^{ab}-\gratio\,\Tref^{ab}_\skL} - \prn{ \Jref_b^\skR \cdot \Fref_\skR^{ab} - \gratio\, \Jref^\skL_b\cdot \Fref_\skL^{ab}}} 
 \notag \\
 &\qquad+\brk{ (\Lref+\Kref^c \Aref_c^\skR) \cdot \Dref_a \Jref_\skR^a 
   - (\Lref+\Kref^c \Aref_c^\skL) \cdot \Dref_a (\gratio \, \Jref_\skL^a) } \\
 & = \frac{1}{\sqrt{-\gref_\skR}} \diffBref \bigg( \sqrt{-\gref_\skR} \brk{ (\Tref_\skR \, \sref_\skR - \gratio \,\Tref_\skL \,  \sref_\skL) + \uref^a (\aheatref^\skR_a - \gratio\, \aheatref^\skL_a) + (\Tref_\skR \,\muref_\skR \cdot \achargeref_\skR -\gratio \,\Tref_\skL \,\muref_\skL\cdot  \achargeref_\skL)  }\bigg) \\
 &= 0 \,.
\end{split}
\end{equation}
Clearly this form of a Schwinger-Keldysh adiabaticity equation is not satisfactory: in the hydrodynamic limit where right and left quantities coincide, the equation collapses to something trivial. This indicates that a proper Schwinger-Keldysh formalism must contain a rule to connect the right and left contours.  Note that in the absence of influence functionals we can separately derive the Bianchi identities on the left and right and check that they are upheld. While this in itself is not ideal, at least in the absence of influence functionals adiabaticity continues to hold in the doubled theory. The issues alluded to above, start arising only when the two sides start talking to each other.

In \S\ref{sec:classLT} we will construct a doubled formalism where an additional symmetry ensures a sensible adiabaticity equation. For the moment, we leave it with the observation that a na\"ive Schwinger-Keldysh formalism as developed in the present section suffers from various deficiencies. 

\subsection{Hydrodynamic currents in common/difference basis}
\label{sec:SKcommondifference}

Regardless of the problems pointed out in the previous subsection, from working on the reference manifold we nevertheless gained a distinct advantage: we have a way of defining currents in a basis of common and difference fields. Let us now briefly outline how this allows to compute the hydrodynamic currents of interest in a rather simple way. By taking linear combinations of the sources we can consider the average source and the difference source which will be useful in our discussion. Generalizing the notion to also include the hydrodynamic fields, we define the following average and difference fields on the reference manifold:
\begin{equation}
\dRLhfref \equiv \hreffields_\skR - \hreffields_\skL \,, \qquad
\aRLhfref  = \frac{1}{2}\, \left(\hreffields_\skR +\hreffields_\skL \right) \,.
\end{equation}
For completeness we record the explicit definition of the difference hydrodynamic fields which will play an important role in what follows:
\begin{equation}
\begin{split}
\dRLgref_{ab}(\xref) &= \gref^\skR_{ab}(\xref) - \gref_{ab}^\skL(\xref) \,, \\
\dRLAref_{a}(\xref) &= \Aref_a^\skR(\xref) - \Aref_a^\skL(\xref)  \,,
\end{split}
\label{eq:DiffFields}
\end{equation}
where $\xref^a$ are some coordinates on $\Mref$; for instance, one could consider the above difference fields as functionals of the diffeomorphism fields by setting $\xref^a = \varphi^a_\skR(x)$.

 In any system out of equilibrium, we are interested in analyzing the causal response of sources. This amounts to considering linear combinations of correlation functions with operator insertions in both the R and L copies of the theory. The causal correlation functions involve a single variation with respect to the difference source \cite{Chou:1984es}. It is therefore useful to remember that the Schwinger-Keldysh construction \eqref{eq:Lsk} couples the difference source to the average current and the average source to the difference current:
 \begin{align}
  \delta S_{SK}\brk{\hfields_\skR, \hfields_\skL} &=  \int_{\Mref} \sqrt{-\breve\gref}\, \bigg\{ \frac{1}{2} \prn{ \gratio_\skR\,\Tref^{ab}_\skR- \gratio_\skL\, \Tref^{ab}_{\skL}} \delta \breve{\gref}_{ab} + \frac{1}{4} \prn{ \gratio_\skR\,\Tref^{ab}_\skR+ \gratio_\skL\, \Tref^{ab}_{\skL}} \delta \bar{\gref}_{ab} 
  \notag \\
   &\qquad\qquad
     + \prn{ \gratio_\skR\,\Jref^{a}_\skR- \gratio_\skL\, \Jref^{a}_{\skL}} \cdot \delta \breve{\Aref}_{a} 
     + \frac{1}{2}\prn{ \gratio_\skR\,\Jref^{a}_\skR+ \gratio_\skL\, \Jref^{a}_{\skL}} \cdot \delta \bar{\Aref}_{a} 
\label{eq:SKrefVar2} \\
     & \qquad\qquad+ \prn{ \Tref_\skR\, \gratio_\skR\,\aheatref^\skR_a - 
     \Tref_\skL\, \gratio_\skL\, \aheatref^\skL_a} \, \delta \Kref^a
     \notag \\
     & \qquad\qquad+  \prn{\Tref_\skR\,\gratio_\skR\achargeref_\skR- \Tref_\skL\,\gratio_\skL \, \achargeref_\skL} \cdot(\delta \Lref + \breve{\Aref}_a \delta \Kref^a) 
   \notag\\  
      &\qquad \qquad  + \frac{1}{2}\prn{\Tref_\skR\,\gratio_\skR\achargeref_\skR+\Tref_\skL\,\gratio_\skL  \achargeref_\skL} \cdot (  \bar{\Aref}_a\delta\Kref^a)  \bigg\} 
       \notag
 \end{align}
where $\gratio_{\skR,\skL} \equiv \sqrt{-\gref_{\skR,\skL}} / \sqrt{-\breve \gref}$. 

While the above discussion remains valid for generic non-equilibrium dynamics, our interest is in the hydrodynamic limit where we only allow  small (long-wavelength) departures from thermodynamic equilibrium. One useful consequence is that we can obtain the hydrodynamic currents by working to linear order in deviations from the equilibrium configuration wherein the left and right degrees of freedom are identified. Said differently, we consider linear deviations of the diffeomorphism and gauge transformation fields about a common equilibrium configuration: $\varphi^a_\skR(x) = \varphi^a_\skL(x) \equiv \varphi^a(x)$ and  $c_\skR(x) = c_\skL(x) \equiv c(x)$. From the action (\ref{eq:Lsk}) one obtains the hydrodynamic constitutive relations by varying with respect to the reference sources $\{\gref, \Aref\}_{\skR,\skL}$ and taking the hydrodynamic limit:
\begin{equation}\label{eq:hydroCur}
\begin{split}
\Tref_{hydro}^{ab} &= \frac{2}{\sqrt{-\gref_\skR}} \left( \frac{\delta S_{SK}}{\delta \gref_{ab}^\skR[\varphi_\skR]} -
\frac{\delta S_{SK}}{\delta \gref_{ab}^\skL[\varphi_\skL]} \right)
\bigg{|}_{\tiny\begin{aligned}&\varphi_\skR^a(x)=\varphi_\skL^a(x) \equiv \varphi^a(x)\\&
c_\skR(x) = c_\skL(x) \equiv c(x)\end{aligned}} \\
\Jref_{hydro}^{a} &= \frac{1}{\sqrt{-\gref_\skR}} \left( \frac{\delta S_{SK}}{\delta \Aref_a^\skR[\varphi_\skR]}
- \frac{\delta S_{SK}}{\delta \Aref_a^\skL[\varphi_\skL]} \right)
\bigg{|}_{\tiny\begin{aligned}&\varphi_\skR^a(x)=\varphi_\skL^a(x)\equiv \varphi^a(x)\\&c_\skR(x) = c_\skL(x) \equiv c(x) \end{aligned}}
\end{split}
\end{equation}
where $\varphi^a= \frac{1}{2} (\varphi_\skR^a+\varphi_\skL^a)$ is the common part of $\varphi_\skR$ and $\varphi_\skL$ which coincides with $\varphi_\skR$, $\varphi_\skL$ in the hydrodynamic limit. From \eqref{eq:SKrefVar2} it transpires that the hydrodynamic currents \eqref{eq:hydroCur} are the common currents which can equivalently be obtained by varying $S_{SK}$ with respect to $\{\bar{\gref}_{ab},\bar{\Aref}_a\}$ and then taking the hydrodynamic coincidence limit. Note however, that such a variation yields the same expressions as the right hand side of \eqref{eq:hydroCur} only in the hydrodynamic limit (i.e., to linear order in difference fields). Beyond linear order, the natural Schwinger-Keldysh currents are defined by \eqref{eq:SKrefVar}.

Similarly, the desired equations of motion in the hydrodynamic limit can be obtained making use of this simplified variational principle: in analogy to the discussion of \S\ref{sec:ReferenceVar} we can do a variation of $\{\varphi^a,c\}$ inside the difference sources, holding the reference configuration of $\{\Kref^a,\Lref\}$ fixed:
\begin{equation}\label{eq:ConsVarRefSK}
\begin{split}
  \diffF \bar{\gref}_{ab} = - \delta_\varphi \bar{\gref}_{ab}\,,&\quad \diffF\bar{\Aref}_a = -\delta_\varphi \bar{\Aref}_a \,, \\
  \diffF \Kref^a = 0 \,, & \quad \diffF \Lref = 0 \,.
\end{split}
\end{equation} 
This variational principle applied to \eqref{eq:SKrefVar2} directly yields equations of motion for the hydrodynamic common currents. Its relation to the variational principle of the previous subsection should be seen as being consistent to linear order in difference fields (which is good enough if we take the hydrodynamic limit at the end of the day). 

While it is clear that the above Schwinger-Keldysh formalism can achieve some things, it is certainly not entirely satisfying. By postulating a reference manifold configuration underlying both copies of the physical theory together with a way the symmetries act there, we are able to obtain equations of motion for the hydrodynamic currents. However, we have no obvious way of defining an entropy current on the reference manifold from first principles. In order to obtain an adiabaticity equation for the currents obtained in this subsection, we need to construct the reference manifold entropy current by hand such that adiabaticity is satisfied. This is the strategy that we will follow in the construction of Class A constitutive relations (see  \S\ref{sec:anward} below).

\subsection{Anomalous Ward identities in the Schwinger-Keldysh formalism}
\label{sec:anward}

The analysis of \S\ref{sec:anomalies} underscores the fact that, while the anomalous adiabaticity equation can be solved within the framework of Class L adiabatic fluids, one fails to recover the desired on-shell Ward identities. The reason for this failure can be traced to the fact that the amount of anomaly inflow into a single copy theory, respecting the adiabatic principle, is a bit too much \cite{Haehl:2013hoa}. As one can see from the transgression form characterization of the anomalous Lagrangian \eqref{eq:mixLanom}, the inflow to the boundary manifold ${\cal M}$ from the bulk topological theory comprises not only of the anomaly in the physical background sources  $\{\fA,\fGamma\}$ but also involves an extra bit of inflow from the shadow fields $\{\fAh, \fGammah\}$. The latter have to be removed from the system in order to obtain the correct physical Ward identities.\footnote{ We recall that the extra terms in \eqref{eq:FalseHydroEom} all involve the shadow fields in the r.h.s.\ of the physical conservation equations.}

Fortunately we now know a cure for this problem; as discussed in \cite{Haehl:2013hoa} and
reviewed in the previous subsections the Schwinger-Keldysh formalism provides a natural framework to understand the Ward identities. In particular, we will find it quite useful in the following discussion to be able to have non-trivial influence functional contributions. The symmetries of the Schwinger-Keldysh construction pick out a unique influence functional which in turn implies the desired anomalous Ward identities
\eqref{eq:TrueHydroEom}.

Let us hark back to the discussion of \S\ref{sec:mixos} where we took our anomalous effective action in the single copy theory to be $S_{eff}\brk{\hfields}$, cf., \eqref{eq:SeffF}. From that discussion, it is clear that we need to add to the total action $
S_{eff}$ another term which fixes the dynamics by ensuring that we have the correct amount of inflow.
In the double-field context, we are thus looking for a total action of the form
\begin{equation}
\begin{split}
S_{tot} &\equiv S_{tot}[\hfields_\skL,\hfields_\skR]
\\
&= S_{eff}\brk{\hfields_\skR}- S_{eff}\brk{\hfields_\skL}     + S_{_{IF}}[\hfields_\skR, \hfields_\skL] \,,
\end{split}
\end{equation}
with $S_{eff}\brk{\hfields}$ being given in \eqref{eq:SeffF} and  $S_{_{IF}}$ is a cross-contour term that involves fields from both copies of the theory.  

It proves convenient for reasons mentioned above to write the action $S_{tot}$ on the reference 
manifold directly. This can always be achieved using \eqref{eq:pullbackR}.
As explained in \S\ref{sec:skrefm} one then has a single background geometry where all the currents live. 
The resulting action will depend on $\hreffields_{\skR} = \{ \gref_{ab}^\skR[\varphi_\skR], \Aref_a^\skR[\varphi_\skR],
\Kref^a[\varphi_\skR], \Lref[\varphi_\skR] \}$ and similarly for
$\hreffields_{\skL}$, i.e., we effectively view it as an effective action for two copies of sources while keeping one copy of hydrodynamic degrees of freedom $\{\Kref^a,\Lref\}$. We thus write
\begin{equation}
\begin{split}
S_{tot} \equiv S_{tot}[\varphi_\skR, c_\skR,\varphi_\skL,c_\skL] &= \left( S_\text{n-a}[\hreffields_{\skR}]+ \int_{\Mref_{d+1}} \VP[\hreffields_{\skR}]  \right) \\
&\quad  - \left( S_\text{n-a}[\hreffields_{\skL}]+ \int_{\Mref_{d+1}} \VP[\hreffields_{\skL}]  \right)
+ S_{_{IF}}[\hreffields_{\skR}, \hreffields_{\skL}]
\end{split}
\label{eq:SK_Stot}
\end{equation}
As remarked earlier, by working in terms of the reference manifold $\Mref$ (and its bulk extension
which we denote as $\Mref_{d+1}$),  we circumvent potential confusions about the presence of two copies of the spacetime manifold with two metrics and two
gauge and diffeomorphism symmetries.  Despite the fact that there are still two copies of source fields living on
$\Mref$, there is only one physical gauge and diffeomorphism invariance involved in \eqref{eq:SK_Stot}. Equations of motion in the hydrodynamic limit can then be obtained from a simple variational principle as described in \S\ref{sec:SKcommondifference}.

Since the contributions from $S_{eff}\brk{\hreffields}$ have already been computed in \S\ref{sec:mixos}, we now turn to an explicit description of the contributions that come from $S_{_{IF}}$. Using the same arguments as \cite{Haehl:2013hoa}, we can infer what the form of $S_{_{IF}}\brk{\hreffields_\skR, \hreffields_\skL} $ ought to be. Its form is pretty much dictated by ensuring that we have the correct amount of inflow: it needs to be a transgression form between the two sets of shadow fields. We therefore claim that the precise term to add as our anomalous influence functional is the transgression from hatted fields on the right towards hatted fields on the left contour, i.e.,
\begin{equation}
S_{_{IF}} = \int_{{\Mref}_{d+1}} \VP[ {\hat \Aref}_\skR, {\hat \Chref}{}_\skR; {\hat \Aref}_\skL, {\hat \Chref}{}_\skL]
\equiv \int_{\Mref_{d+1}} \VP\left(\fAh[ \hreffields_\skR] , \fGammah[\hreffields_\skR]; \fAh[ \hreffields_\skL], \fGammah[\hreffields_\skL]\right) \,,
\label{eq:ifanom}
\end{equation}
The main thing we need to check is that the above cross-term influence functional provides the right correction terms necessary to fix the anomalous hydrodynamic Ward identities \eqref{eq:FalseHydroEom} without influencing the physical currents \eqref{eq:ConstRelations}.  We can use the same kind of manipulations as in \S\ref{sec:AnomBianchi} to verify this. As we pointed out in the general discussion of Class L Schwinger-Keldysh formalism, \S\ref{sec:skrefm}, a non-linear treatment requires carefully separating R and L degrees of freedom. However, we are eventually interested in the hydrodynamic limit of the currents, i.e., the coincidence limit where difference fields are set to zero. Therefore we can employ the simpler formalism where we vary the Schwinger-Keldysh action directly with respect to difference sources, disregarding any current contributions that contain hydrodynamically vanishing quantities, c.f., \S\ref{sec:SKcommondifference}.
After explicit computation (see \S\ref{sec:SKvariation}), we obtain for the variation of the entire anomaly part of the action in the hydrodynamic coincidence limit\footnote{ We adhere to our index conventions stated earlier: 
\begin{itemize}
 \item Boundary physical manifold ${\cal M}$: indices from the Greek alphabet.
 \item Bulk physical manifold $\bulkM_{d+1}$: indices from the second half of the lowercase Latin alphabet. 
 \item Boundary reference manifold $\Mref$: indices from the first half of the lowercase Latin alphabet. 
 \item Bulk reference manifold $\Mref_{d+1}$: indices from the second half of the uppercase Latin alphabet.
\end{itemize}
}
\begin{align}
\label{eq:DeltaSanomFinal}
\delta S_{anom}\bigg{|}_{hydro} &\equiv \delta \left( S_{_{IF}} +
\int_{\Mref_{d+1}}\left( \VP[\hreffields_\skR] - \VP[\hreffields_\skL]  \right) \right)\bigg{|}_{hydro} \notag \\
& = \bulkintref \left[  \frac{1}{2} \, \Dref_\pb \left( \SpHref^{\mb[\nb\pb]}+\SpHref^{\nb[\mb\pb]}-\SpHref^{\pb(\mb\nb)}\right)
\half \delta \dRLgref_{\mb\nb}
+ \JHref^\mb \cdot \delta \dRLAref_{\mb} \right] \\
&\quad + \int_\Mref \sqrt{-\breve{\gref}} \; \Bigg\{
\brk{ \frac{1}{2} \,\Dref_c \prn{\SPref^{a[bc]} + \SPref^{b[ac]} -\SPref^{c(ab)}}
+ 2\,\qPref^{(a} \uref^{b)} } \, \half \delta \dRLgref_{ab}
+\;   \JPref^a \cdot \delta \dRLAref_{a}  \Bigg\} . \notag
\end{align}
where $hydro$ denotes the limit where all the expressions that are not variations of difference fields are
evaluated at $\varphi_\skR^a(x)=\varphi_\skL^a(x)\equiv \varphi^a(x)$ and $c_\skR(x) = c_\skL(x) \equiv c(x)$. Further, $\{\bar{\gref}_{ab},\bar{\Aref}_a\}$ denote the difference sources as introduced in \S\ref{sec:SKcommondifference}.
Equations of motion are now obtained by varying $\{\varphi^a,c\}$ inside the difference sources on $\Mref$, i.e.\ \eqref{eq:ConsVarRefSK}. Following that logic, we immediately obtain the bulk on-shell equation of motion, which we now write for the reference fields
\begin{equation}
\half  \Dref_{\nb} \Dref_{\pb} \left( \SpHref^{\mb[\nb\pb]}+\SpHref^{\nb[\mb\pb]}-\SpHref^{\pb(\nb\mb)}\right) \simeq 
\Fref_{\mb\pb} \cdot\JHref^\pb  \,,
\qquad \Dref_{\pb} \JHref^\pb \simeq 0 \,.
\end{equation}
The boundary equations of motion from $S_{anom}$ then take the form
\begin{equation}\label{eq:DeltaSanom}
 \Dref_b \brk{ \half \,\Dref_{c} \prn{\SPref^{a[bc]} + \SPref^{b[ac]} -\SPref^{c(ab)}}
+ 2 \, \qPref^{(a} \uref^{b)} } \simeq  \Fref^a{}_{c} \cdot\JPref^c  
+\half \Dref_c  \SpHref^{\perp[ac]} \,, \qquad
 \Dref_{a} \JPref^a \simeq \JHref^\perp \,,
 \end{equation}
where all fields are in the hydrodynamic limit, i.e., the $\skRl$- and $\skLl$-fields have been identified.
After combining the $\skRl$- and $\skLl$-pieces, the additional non-anomalous contributions to the boundary
equations of motion are the same as in \eqref{eq:NonAnomConstrains} with \eqref{eq:AnomConstraints}. Putting all of this together we get the hydrodynamic equations of motion for the action \eqref{eq:SK_Stot}:
\begin{equation}\label{eq:SKeom}
\begin{split}
&\Dref_{b} \left[ \Tref^{ab}_\text{n-a} +\half \Dref_{c} \prn{\SPref^{a[bc]} + \SPref^{b[ac]} -\SPref^{c(ab)}}
+ 2 \, \qPref^{(a} u^{b)}\right]  \\
&\qquad\quad \simeq \Fref^a{}_c \cdot \left( \Jref^c_\text{n-a} + \JPref^c \right)+\half \Dref_c  \SpHref^{\perp[ac]}  \,,\\
& \Dref_{a} \brk{ \Jref^a_\text{n-a} + \JPref^a } \simeq \JHref^\perp \,.
\end{split}
\end{equation}
When written in terms of quantities on the physical spacetime ${\cal M}$, these are the usual equations of motion (\ref{eq:TrueHydroEom}) for hydrodynamics with mixed flavour and gravitational anomaly.\footnote{ The translation between the physical and reference manifolds is simple: we replace $\hreffields \mapsto \hfields$ and change indices back to Greek. } 

This completes for us the derivation of the equations of motion for the full Schwinger-Keldysh action. We conclude that a Schwinger-Keldysh formulation with suitable Feynman-Vernon term is capable of imposing the correct dynamics on our theory. Working in a formalism with doubled set of degrees of freedom is inevitable if we want to circumvent having various Ward identities contaminated by unwanted hatted anomaly contributions.

\subsection{Effective actions for Class D?}
\label{sec:skcritical}

In \S\ref{sec:classD} we have seen that dissipative transport is well under control. The positivity of the leading order transport coefficients and lack of further constraints is reminiscent of general structures of effective field theories, wherein unitarity imposes positivity controls on kinetic terms but leaves typically higher order interactions arbitrary.\footnote{ This analogy comes with caveats. In relativistic quantum field theories there are sign-definiteness constraints on the leading corrections to the positive definite quadratic kinetic terms arising from causality as discussed for instance in \cite{Adams:2006sv}.}

Ideally, it would be great if we can give a complete picture for dissipative transport by constructing an effective action; once this is understood we would be able to make a clear analysis from a microscopic perspective.  It has long been understood that such an effective action has to be described using the Schwinger-Keldysh construction. We would like to offer some critical thoughts at this juncture, why this construction requires further bolstering, paving the way for a general picture in \S\ref{sec:classLT}. 

To incorporate dissipation in the Schwinger-Keldysh doubled formalism, it is necessary to incorporate interactions between the left and right fields via the Feynman-Vernon influence functionals \cite{Feynman:1963fq}. Then, upon integrating out one set of degrees of freedom that we take to be the difference fields ($\hfields_\skL-\hfields_\skR$), one obtains an effective action for the common fields $\hfields_\skL  + \hfields_\skR$ with dissipative interactions. What we are after then is a constraint on the influence functionals ensuring that such interactions are compatible with the second law. We should emphasize that this is a very physical requirement since  the second law of thermodynamics is a macroscopic manifestation of microscopic unitarity.

If we na\"ively construct influence functionals without any constraint, then we have seen that the adiabaticity equation fails. One should then worry about terms that violate the second law. In particular, using generic influence functionals one can construct effective actions which allow non-vanishing Class $\PF$ terms, which as we have seen are forbidden by adiabaticity (in fact hydrostatics). One potential issue is that unconstrained influence functions violate the fluctuation-dissipation relations which typically are encoded by the Kubo-Martin-Schwinger (KMS) condition in non-equilibrium dynamics.  While there is some understanding of how these relations are to be imposed in the Schwinger-Keldysh formalism we are as yet unaware of a complete treatment in the hydrodynamic context (see for instance \cite{Kovtun:2014hpa,Grozdanov:2013dba,Grozdanov:2015nea} for some progress in this direction).

Ideally, we would like some element in the Schwinger-Keldysh construction which forbids influence functionals that lead to disallowed constitutive relations (such as Class $\PF$). Given our previous discussions and in particular \S\ref{sec:anward}, let us take a step back and see what we can learn from the adiabatic effective actions. For Class L constitutive relations, the Schwinger-Keldysh construction is rather straightforward. We simply have \eqref{eq:Lsk} with $\Lagref_{_{IF}} = 0$ since there is no need for any interaction terms in either Class $\PS$  or Class $\LS$.
So this does not provide us with much guidance on how to proceed. 

Anomalous transport of Class A is however more interesting, as we need a non-trivial influence functional \eqref{eq:ifanom}, required in order to satisfy the Ward identities. While our construction was guided by the symmetries and the rigidity of anomalies, it should be borne in mind that the final answer in \eqref{eq:SK_Stot} is not derived from first principles. While we have presumably fixed the non-covariant part of the influence functional correctly by demanding the Ward identities, it is plausible that there are additional pieces which ensure that the term  we propose satisfies the KMS condition. 

In any event the story about influence functionals for adiabatic transport is incomplete -- we have not been able thus far to incorporate the vector classes $\PV$, $\GV$ and C into an effective action framework, nor the Berry-like transport terms of Class B. It appears intuitive that writing an effective action for such transport does require some form of Schwinger-Keldysh doubling; indeed we will see glimpses of such a structure in \S\ref{sec:classLT}-\S\ref{sec:eightfoldLT}.  We will argue there for a new symmetry principle which circumvents all these problems. The set of influence functionals will be constrained in precisely the right way by the presence of a new symmetry to ensure adiabaticity. Furthermore, we will see a more natural variational principle deriving the Ward identities of hydrodynamics.  We however forewarn the reader that whilst this structure is tantalizing, we postpone a complete discussion of the implications vis a vis the Schwinger-Keldysh constructions to a future publication \cite{Haehl:2014kq}.

\newpage
\part{The Eightfold Way to Dissipation and its Lagrangian Unification}
\label{part:8classes}
\hspace{1cm}

\section{The Eightfold Way}
\label{sec:8fold}

We are now in a position to outline the complete classification of hydrodynamic transport at arbitrary orders in the gradient expansion building on the results derived in \S\ref{sec:hydrostatics}-\S\ref{sec:skdouble}. We will give the algorithm for the intrepid hydrodynamicist to implement the construction at any desired order.

\subsection{The route to classification}
\label{sec:rclassify}

We will work off-shell in the most general fluid frame. We first compile a list of all tensor structures that can appear in constitutive relations $\hcur\brk{\hfields}$. We will sequentially eliminate elements of this collection by assigning them to distinct classes suggested by the eightfold way. The algorithm for understanding the transport classification can be implemented in the following sequence:
\begin{itemize}
\item The first step of our analysis is to remove the particular solutions of Class A by picking the correct particular solution leading to the anomalous constitutive relations, i.e.,  $\{({\cal G}^\sigma)_\text{A},(T^{\mu\nu})_\text{A},(J^\mu)_\text{A}\}$ given in terms of the anomaly polynomial $\fP[\fF,\fR]$. These anomalous currents take the form 
\begin{equation}\label{eq:AnomConstRep}
\begin{split}
(J_S^\alpha)_\text{A} &=  -\half \Kbeta_\sigma \, \SpHh^{\perp[\alpha\sigma]} \,, \\
(T^{\alpha\beta})_\text{A}  & = \qP^\alpha u^\beta +  \qP^\beta u^\alpha+ \frac{1}{2} \,D_\rho \prn{\SP^{\alpha[\beta\rho]} + \SP^{\beta[\alpha\rho]} -\SP^{\rho(\alpha\beta)}} \,, \\
(J^\alpha)_\text{A} &= \JP^\alpha \,,
\end{split}
\end{equation}
where the various pieces in these currents are given in terms of the transgression form $\VP \equiv \frac{\fu}{2\fomega}\wedge \big(\fP[\fF,\fR]-\fPh[\fFh,\fRh]\big)$ as 
\begin{equation}
\begin{split}
 \star \fqP &= \frac{\partial \VP}{\partial (2\fomega)} \,,\qquad\;\;\;
 \star \fSP{}^\beta{}_\alpha = 2\frac{\partial \VP}{\partial (\fBR)^\alpha{}_\beta} \,,\\
 \star \fJP &= \frac{\partial \VP}{\partial \fB} \,,\qquad
 \hodgeB \fSpH{}^b{}_a = 2\frac{\partial \fP}{ \partial \fR^a{}_b} \,.
\end{split}
\end{equation}

\item We then remove the terms that are forbidden by the hydrostatic analysis. This involves discarding  Class $\PF$ terms $\{({\cal G}^\sigma)_{\PF},(T^{\mu\nu})_{\PF},(J^\mu)_{\PF}\}$ from the constitutive relations. These terms are tensor structures allowed by symmetry, but forbidden by the second law.
\item There are combinations which can never be removed irrespective of choice of
entropy current. These irreducibly dissipative combinations belong to Class D .
All other combinations solve non-anomalous adiabaticity equation which is homogeneous
in derivative order. We will henceforth proceed derivative order by derivative order
with no mixing of adiabatic constitutive relations at different  orders. The dissipative constitutive relations take the form \eqref{eq:TJClassV}:
\begin{equation}\label{eq:TJClassVrep}
\begin{split}
(T^{\mu\nu})_{\text{D}} &\equiv \; 
	  -\half \; \brk{\DVisc_{\cdg_g}^\dag\ \cdg\  \DVisc_{\cdg_g}
	  + \DVisc_{\cdA_g}^\dag\ \cdA\  \DVisc_{\cdA_g}}^{(\mu\nu)(\alpha\beta)}
	  \, \diffB  g_{\alpha\beta} 
\\&	  \qquad \qquad 
	   -\; \brk{\DVisc_{\cdg_g}^\dag\ \cdg\  \DVisc_{\cdg_A}
	  + \DVisc_{\cdA_g}^\dag\ \cdA\  \DVisc_{\cdA_A}}^{(\mu \nu) \alpha} \cdot 
	  \diffB  A_\alpha
\\
(J^\alpha)_{\text{D}} &\equiv \; 
	-\half \; \brk{\DVisc_{\cdg_A}^\dag\ \cdg\  \DVisc_{\cdg_g}
	  + \DVisc_{\cdA_A}^\dag\ \cdA\  \DVisc_{\cdA_g}}^{\alpha(\mu \nu)}
	  \diffB  g_{\mu\nu} 	  
\\ & 	\qquad \qquad	  
	  -\; \brk{\DVisc_{\cdg_A}^\dag\ \cdg\  \DVisc_{\cdg_A}
	  + \DVisc_{\cdA_A}^\dag\ \cdA\  \DVisc_{\cdA_A}}^{\alpha\beta} \cdot \,
	  \diffB  A_\beta \,.
\end{split}
\end{equation}
where the dissipative Noether current $(\N^\sigma)_{\text{D}}$ is determined by integration by parts as in 
\eqref{eq:dissIntPart}.  As shown in \S\ref{sec:classDexamples}, a large subset of Class D constitutive relations up to the second order in derivative expansion can be obtained instead by the simpler task of classifying transverse tensor structures $\{\BerryG^{\mu\nu\rho\sigma},\BerryGA^{\mu\nu\alpha},\BerryA^{\alpha\beta}\}$ and plugging them into
\begin{align}\label{eq:TJDissSimplerep}
(T^{\mu\nu})_{\text{D},2\partial} &\equiv \prn{-2\, \eta\,\sigma^{\mu\nu} - \zeta\,\Theta\,P^{\mu\nu} }
	 -\quarter \prn{ \BerryG^{(\mu\nu)(\alpha\beta)}+\BerryG^{(\alpha\beta)(\mu\nu)} }
	 \,  \diffB  g_{\alpha\beta} + \BerryGA^{(\mu \nu) \alpha} \cdot \diffB  A_\alpha\,,
\nonumber \\(
J^\alpha)_{\text{D},2\partial} &\equiv \prn{ \sigma_{_{\text{Ohm}}} \, \cv^\alpha}+
	\half \BerryGA^{(\mu \nu) \alpha}  \diffB  g_{\mu\nu}
	- \BerryA^{(\alpha\beta)} \cdot\diffB  A_\beta \,,
\nonumber \\(
{\cal G}^\sigma)_{\text{D},2\partial} & = 0 \,.
\end{align}

\item We then remove the Class C constitutive relations by eliminating the non-trivial conserved vectors that can serve to provide contributions to the entropy current:
\begin{equation}
(T^{\mu\nu})_\text{C} =0 \,, \qquad (J^\mu)_\text{C} = 0 \,, \qquad ({\cal G}^\sigma)_\text{C} = -T\, {\sf J}^\sigma  \,,
\end{equation}	
where ${\sf J}^\sigma$ are identically conserved topological currents. 
\item At the next step, we will remove Class B by looking at all 
combinations in that derivative order that solve adiabaticity equation with zero free energy current. 
In order to adhere to the derivative counting one should simple classify the intertwining tensors $\{\BerryG,\BerryGA,\BerryA\}$
without the derivative operators $\DVisc$ as the latter mixes derivative orders:\footnote{ Note that this argument seems to indicate that  $\DVisc$ construction while useful in constructing solutions for Class B, serves little purpose in the classification program, where all we care about is the set of solutions at a given derivative order. }
\begin{equation}\label{eq:TJBerryrep}
\begin{split}
(T^{\mu\nu})_\text{B} &\equiv
	 -\quarter \prn{ \BerryG^{(\mu\nu)(\alpha\beta)}-\BerryG^{(\alpha\beta)(\mu\nu)} }
	 \, \diffB  g_{\alpha\beta} + \BerryGA^{(\mu \nu) \alpha} \cdot  \diffB  A_\alpha \,,
\\
(J^\alpha)_\text{B} &\equiv
	- \half \BerryGA^{(\mu \nu) \alpha} \diffB  g_{\mu\nu}
	- \BerryA^{[\alpha\beta]} \cdot \diffB  A_\beta \,,
\\
({\cal G}^\sigma)_\text{B} & = 0 \,.
\end{split}
\end{equation}
\end{itemize} 

At a given derivative order, say $k^{\rm th}$, let there be $N_{tot}$ functions which parameterize 
constitutive relations solving the non-anomalous adiabaticity equation and have non-trivial free energy  current. We will examine the grand canonical adiabaticity equation \eqref{eq:AdiabaticityG} and focus on the expression
for the most general adiabatic free  energy current. This is in fact easier to deal with than the stress-tensor and charge currents owing to the fact that we only have to classify vectors and not symmetric tensors. As we have already taken the effort to remove Class B terms in the preceding steps, the most general adiabatic free  energy current
will then be written in terms of just these $N_{tot}$ functions we enumerate.

Let us decompose the adiabatic free energy current into a longitudinal scalar and vector part by the ansatz:
\begin{align}
-\frac{\mathcal{G}^\mu}{T} = \Lag \; \Kbeta^\mu -P^\mu _{\ \sigma}\; \frac{\mathcal{G}^\sigma}{T} 
\label{eq:feq1}
\end{align}
using the hydrodynamic field and the transverse spatial projector $P_{\mu\nu}$.  Further, let $N_\Lag$ be the number of functional combinations that appear in the scalar part $\Lag$. Without loss of generality, let us assume our parametrization is such that we can then divide the $N_{tot}$ number of  functions in $\mathcal{G}^\sigma$
into $N_\Lag$ functions that  appear in $\Lag$ and  the reminder (that do not appear in $\Lag$).
Now, set the $N_{tot}-N_\Lag$ number of  functions that do not appear in $\Lag$ to zero. 
We are then left with a
$N_\Lag$ functions worth of solution of adiabaticity equation which we will denote by 
$\{\mathcal{G}_{Sc}^\sigma,\; T_{Sc}^{\mu\nu},\; J_{Sc}^\mu\}$. After subtracting this solution, we have $N_{tot}-N_\Lag$ solutions with purely transverse free energy current  denoted by $\{\mathcal{G}_V^\sigma,\; T_V^{\mu\nu},\; J_V^\mu\}$.

Let us now focus on $\{\mathcal{G}_{Sc}^\sigma,\; T_{Sc}^{\mu\nu}, \; J_{Sc}^\mu\}$. 
The decomposition for these solutions \eqref{eq:feq1} then reduces to
\begin{align}
-\frac{\mathcal{G}_{Sc}^\mu}{T} =\Lag\; \Kbeta^\mu -P^\mu_{\ \sigma} \,\frac{\mathcal{G}_{Sc}^\sigma}{T}  
\end{align}
where the data is now parameterized by scalar functions.

In the next step, let us use $\Lag$ as the Lagrangian and then construct $N_\Lag$ functions worth
of Class L constitutive relations which are of the form
\begin{equation}
\begin{split} 
 {\cal G}^\sigma_\Lag &= -T \bigg({\Kbeta^\sigma \Lag-(\PSymplPot{\Bfields})^\sigma +\nabla_\nu \Komar^{\sigma\nu}[\Bfields]}\bigg)\,,\\
 T^{\mu\nu}_\Lag &= \frac{2}{\sqrt{-g}} \frac{\delta (\sqrt{-g}\,\Lag)}{\delta g_{\mu\nu}} \,,\qquad 
 J^\mu_\Lag = \frac{1}{\sqrt{-g}} \frac{\delta (\sqrt{-g}\,\Lag)}{\delta A_\mu}\,.
\end{split}
\end{equation}
Since this accounts for all the $N_\Lag$  solutions, the difference $\{\mathcal{G}_{Sc}^\sigma-\mathcal{G}_\Lag^\sigma,\; 
T_{Sc}^{\mu\nu}-T_\Lag^{\mu\nu},\; J_{Sc}^\mu-J_\Lag^\mu\}$ can only contain repetitions, trivial solutions or a mix of other
classes, such as Class B or even $\GV$ type solutions.\footnote{ It is useful to know that even though we are generating solutions from a scalar Lagrangian density we can indeed get some transverse vector components in the free energy current. } 
All that matters for our discussion is that the currents are derivable from some form of a generating function (as in, e.g., Class L).
In that case we can set the controlling functions  to be some
functionals of the hydrostatic and hydrodynamic parameters in $\Lag$. We can discard them without loss of generality, and choose a basis of solutions
such that
$$ \{\mathcal{G}_{Sc}^\sigma,T_{Sc}^{\mu\nu},
J_{Sc}^\mu\} = \{\mathcal{G}_\Lag^\sigma,T_\Lag^{\mu\nu},
J_\Lag^\mu\} $$

\begin{itemize}
\item  The above discussion takes care of all Class L (which divides further into $\PS$ and $\LS$) constitutive relations. At this point, we have accounted for five adiabatic classes $\{\PS,\LS,{\rm A},{\rm B}, {\rm C}\}$ in addition to the Class D terms (and we have eliminated $\PF$ terms).
\item We are now left with the remaining constitutive relations with transverse free energies, 
$\{\mathcal{G}_V^\sigma,\; T_V^{\mu\nu},\; J_V^\mu\} =  \{(\mathcal{G}^\sigma)_{\PV}+(\mathcal{G}^\sigma)_{\GV},
\; (T^{\mu\nu})_{\PV}+(T^{\mu\nu})_{\GV},\; (J^\mu)_{\PV}+(J^\mu)_{\GV}\}$.
\item The Class $\PV$ terms can be eliminated by invoking the replacement rule arising from Euclidean consistency and thus employing a similar trick as in the Class A discussion earlier. We define a modified anomaly polynomial $\fP$ via 
\begin{equation}\label{eq:ReplacementRulerep}
\begin{split}
\fP_{\PV}[\fF,\fR,\fFT\,] = \fP[\fF,{\rm tr} \fR^{2k} \mapsto {\rm tr} \fR^{2k} + 2(2\pi\, \fFT\,)^{2k} ] - \fP[\fF, {\rm tr} \fR^{2k} ]\,.
\end{split}
\end{equation}
and find the constitutive relations 
\begin{equation}
\begin{split}
 (J_{S}^\alpha)_{\PV} &= \JpSPV^\alpha -\half \Kbeta_\sigma \, (\SpHh)_{\PV}^{\perp[\alpha\sigma]} \,, \\
\TPV & = \qPV^\alpha u^\beta +  \qPV^\beta u^\alpha+ \frac{1}{2} \,D_\rho \prn{\SPV^{\alpha[\beta\rho]} + \SPV^{\beta[\alpha\rho]} -\SPV^{\rho(\alpha\beta)}} \,,  \\
\JPV &= J_{_{\PV}}^\alpha
\end{split}
\end{equation}
where
\begin{equation}
\begin{split}
 \star \fqPV &= \frac{\partial \VPPV}{\partial (2\fomega)} \,,\qquad\;\;\;
 \star \fSPV{}^\beta{}_\alpha = 2\frac{\partial \VPPV}{\partial (\fBR)^\alpha{}_\beta} \,,\\
 \star \fJPV &= \frac{\partial \VPPV}{\partial \fB} \,,\qquad
  \hodgeB (\fSpH)_{\PV}{}^b{}_a = 2\frac{\partial \fP_{\PV}}{ \partial \fR^a{}_b} \,.
\end{split}
\end{equation}
One main difference from the Class A constitutive relations (apart from the presence of $\AT$ in $\fP_{\PV}$) is the fact that the entropy current (and thus the free energy current) gets a non-trivial contribution in this class. The additional contribution to the  entropy current $\JpSPV^\alpha$ is defined in terms of the $\PV$ transgression form: 
\begin{equation}
\star \fJpSPV =   \frac{\partial (\VPPV)}{\partial \fBT} \,.
\end{equation}
where $\fBT$ is the two-form magnetic field for $\AT$, $\fBT =\fFT -\fu \wedge i_{\fu} \, \fFT$.

\item Having dealt with the other classes  the rest will go into Class $\GV$ and takes the general form
\begin{align}\label{eq:TJVecrep}
(\mathcal{G}^\rho)_{\GV} 
&= -T \left[  \quarter \diffB  g_{\mu\nu}  {\mathfrak C}_{\BerryG}^{\rho ((\mu\nu)|(\alpha\beta))}
	  \, \diffB  g_{\alpha\beta}
	  +  \diffB  g_{\mu\nu} {\mathfrak C}_{\BerryGA}^{\rho(\mu \nu) \alpha} \cdot \diffB  A_\alpha
          +\diffB  A_\alpha \cdot {\mathfrak C}_{\BerryA}^{\rho(\alpha\beta)} \cdot \diffB  A_\beta \right]
\notag \\
(T^{\mu\nu})_{\GV} &=
	  \half \brk{D_\rho{\mathfrak C}_{\BerryG}^{\rho ((\mu\nu)|(\alpha\beta))}
	  \, \diffB  g_{\alpha\beta} + 2\ {\mathfrak C}_{\BerryG}^{\rho ((\mu\nu)|(\alpha\beta))}
	  \, D_\rho \diffB  g_{\alpha\beta}} 
\notag \\
&\qquad 	+ D_\rho{\mathfrak C}_{\BerryGA}^{\rho(\mu \nu) \alpha} \cdot \diffB  A_\alpha
	   + 2\ {\mathfrak C}_{\BerryGA}^{\rho(\mu \nu) \alpha} \cdot
  	  \, D_\rho \diffB  A_\alpha 
\notag \\
(J^\alpha)_{\GV} &=
	 \half \brk{ D_\rho{\mathfrak C}_{\BerryGA}^{\rho(\mu \nu) \alpha}  \diffB  g_{\mu\nu}
	   + 2\ {\mathfrak C}_{\BerryGA}^{\rho(\mu \nu) \alpha}
	  \, D_\rho \diffB  g_{\mu\nu} } 
\notag \\
&\qquad 	+ D_\rho{\mathfrak C}_{\BerryA}^{\rho(\alpha\beta)} \cdot \diffB  A_\beta
	   + 2\ {\mathfrak C}_{\BerryA}^{\rho (\alpha\beta)} \cdot
  	  \, D_\rho \diffB  A_\beta \,.
\end{align}
\item Once we have accounted for these terms we have exhausted all possible hydrodynamic constitutive relations; the eightfold path is complete and the most general constitutive relations allowed by symmetries at a given order in derivatives can be written as
\begin{equation} \label{eq:GeneralConstRel}
\begin{split}
 {\cal G}^\sigma &= ({\cal G}^\sigma)_\text{A} + ({\cal G}^\sigma)_{\PF} + ({\cal G}^\sigma)_\text{D} + ({\cal G}^\sigma)_\text{C} + ({\cal G}^\sigma)_\text{B} + {\cal G}^\sigma_\Lag + {\cal G}^\sigma_{V} \,, \\
 T^{\mu\nu} &= (T^{\mu\nu})_\text{A} + (T^{\mu\nu})_{\PF} + (T^{\mu\nu})_\text{D} + (T^{\mu\nu})_\text{C} + (T^{\mu\nu})_\text{B} + T^{\mu\nu}_\Lag + T^{\mu\nu}_{V} \,, \\
 J^\mu &= (J^\mu)_\text{A} + (J^\mu)_{\PF} + (J^\mu)_\text{D} + (J^\mu)_\text{C} + (J^\mu)_\text{B} + J^\mu_\Lag + J^\mu_{V} \,.
\end{split}
\end{equation}
\end{itemize}

\begin{theorem}
\label{thm:eight} 
All hydrodynamic transport is exhaustively classified by one of the aforementioned seven adiabatic classes, viz.,  $\{\PS,\LS,{\rm A},{\rm B}, {\rm C}, \PV, \GV\}$ and the forbidden constitutive relations of Class $\PF$, 
in addition to the dissipative Class {\rm D} $=\Dv \cup \Ds$.
\end{theorem}

The constructive algorithm described above outlines the general structure of the proof we would like to present. However, in order to complete the proof, we need a precise argument stating that our constructions exhaust the terms in the non-Lagrangian classes $\{\GV,{\rm B},{\rm D}\}$ completely. We will give such an argument in the following subsection, \S\ref{sec:completeness}. However, before we give an independent proof, let us anticipate the result of \S\ref{sec:classLT}-\S\ref{sec:eightfoldLT} which makes it clear that there exists a master effective field theory which will encompass precisely the adiabatic classes and thus provides a much more direct proof of the completeness of our classification: 

\begin{theorem}
\label{thm:classLT}
 The sevenfold classes of adiabatic hydrodynamic transport can be obtained from a scalar Lagrangian density $\LagT\brk{\Kbeta^\mu, \LambdaB, g_{\mu\nu}, A_{\mu}, 
\tildeg_{\mu\nu},\tildeA_{\mu}, \AT_\mu}$:
\begin{align}
\LagT &= \frac{1}{2}\, T^{\mu\nu}\, \tildeg_{\mu\nu} + J^\mu\cdot \tildeA_{\mu} 
+\left(J_S^\sigma+\Kbeta_\nu T^{\nu\sigma}+(\LambdaB+\Kbeta^\nu A_\nu)\cdot J^\sigma\right)\AT_\sigma
\end{align}
 As indicated the Lagrangian density depends not only on the hydrodynamic fields and the background sources, but also the `Schwinger-Keldysh' partners of the sources $\{\tildeg_{\mu\nu}, \tildeA_\mu\}$ and a new KMS-gauge field $\AT_\mu$. This Lagrangian is invariant under diffeomorphisms and gauge transformations\footnote{ Anomalies if present are dealt with using the inflow mechanism \cite{Callan:1984sa}. $\LagT$  then includes a topological theory  in $d+1$ dimensions coupled to the physical $d$-dimensional QFT (at the boundary/edge).} and under $\UT$ which acts only on the sources as a thermal diffeomorphism or gauge transformation along $\Bfields$. The $\UT$ gauge invariance implies a Bianchi identity, which is nothing but the adiabaticity equation \eqref{eq:Adiabaticity}. Furthermore, a constrained variational principle for the fields $\{\Kbeta^\mu,\LambdaB\}$ ensures that the dynamics of the theory is simply given by conservation.
\end{theorem}

Given the Lagrangian $\LagT$ we are essentially done, since all we need to do is to  show that by picking appropriate scalar densities in the extended space of fields gives rise to a solution in one of the aforementioned eight classes.  This is relatively straightforward as we shall see in \S\ref{sec:eightfoldLT}. What is less apparent at first sight is the rationale for the existence of the extended set of degrees of freedom and the extra $\UT$ symmetry. The reader might take these as part of our construction for the present, though we believe that the Class $\LT$ story we are about to present hints at some fundamental truisms that ought to be valid in non-equilibrium dynamics of QFTs (and potentially fixing some of the problems described in \S\ref{sec:skdouble}). 

Before presenting the detailed construction of $\LagT$ in \S\ref{sec:classLT}-\S\ref{sec:eightfoldLT}, we shall now give an independent proof of the completeness claimed in Theorem \ref{thm:eight} and then illustrate our eightfold classification for various fluid systems.

\subsection{Completeness of the adiabatic taxonomy}
\label{sec:completeness}

In the lead up to the statement of Theorem \ref{thm:eight} we have already covered a reasonable amount of ground vis a vis a proof of the statement. We will in the following complete some of the open issues which that discussion left out and argue that out eightfold classification is complete. The proof per se will be phrased in a physical language; it can be made mathematically rigorous as necessary but we prefer to illustrate the basic statements in a fashion that makes them more intuitive.

The key component of the proof is to realize that one needs to control the free energy current ${\mathcal G}^\sigma$ in order to ascertain the behaviour of transport. In much of our discussion, including \S\ref{sec:rclassify}, we have emphasized the fact that the free energy current is a spacetime vector simplifies the classification scheme. Accounting for all possible vectors that can appear in 
$\mathcal{G}^\sigma$ would suffice for our purposes of demonstrating completeness. 

Let us first examine adiabatic constitutive relations. We invoke the decomposition \eqref{eq:feq1} of the free energy current. Using the argument following this equation in  \S\ref{sec:rclassify}, it is clear that the entire contribution to the longitudinal part of the free energy current is captured by the Class L ($= \PS \cup \LS$) by picking a set of generating scalars which are either hydrostatic ($\PS$)  of hydrodynamic ($\LS$). Furthermore, anomalies are dealt with using the particular solutions of the adiabaticity equation, leaving us then with situations of traverse vector free energy current and some situations where the free energy is vanishing (Class B) or identically conserved (Class C).  In addition we have the dissipative constitutive relations.  Of these Class C terms are easy to handle and like with the anomalies one quickly can exhaust the space of cohomologically non-trivial conserved currents. 

Thus for a full proof of Theorem \ref{thm:eight} we need to ascertain that the parameterizations we gave for Classes $\{ \PV, \GV,{\rm B},{\rm D}\}$ are exhaustive.  Let us make a couple of remarks:
\begin{itemize}
\item For Class D, the argument is clear since we can always recast the most general positive definite form 
using a set of tensor valued differential operators  $\DVisc$ and suitably chosen intertwiners $\{\cdg, \cdA\}$ as 
discussed in \S\ref{sec:Ddiffops}.
\item For Class B (without $\DVisc$ operators), one can argue that in the vector space of terms, all the combinations orthogonal to $\{\diffB g_{\mu\nu}, \diffB A_\mu \}$ necessarily take the form given in  Eqs.~\eqref{eq:TJBerry} and \eqref{eq:JSBerry}.
\item The hydrostatic transverse vector free energy currents are likewise easy to tackle by focusing on a limited set of terms that survive equilibrium.
\item The only unresolved problem is how to argue that our construction for $\GV$ does not miss any terms. The issue here
is that there are too few of them in our explicit examples to see how they work in general. We will present an argument in favour of the completeness of our classification below. 
\end{itemize}

To prove the completeness of our parametrization \eqref{eq:TJVecrep} of Class $\GV$, we need to consider the set of all possible transverse free energy currents $({\cal G}^{\sigma})_{\GV}$. Since we are only interested in non-hydrostatic currents (otherwise we could describe them in Class $\PV$), they need to contain at least one factor of $\diffB g_{\mu\nu}$.\footnote{ In this subsection, we will content ourselves with the discussion of neutral fluids. In charged fluids, one could also use $\diffB A_\mu$ instead of $\diffB g_{\mu\nu}$ to describe deviations from equilibrium, which leads to a completely analogous discussion.} Let us parameterize such currents as
\begin{equation}\label{eq:HbarVansatz}
 (\N^\sigma)_{\GV} \equiv -\frac{1}{T}\,({\cal G}^\sigma)_{\GV} = {\mathfrak C}^{\sigma\mu\nu} \, \diffB g_{\mu\nu}  \qquad \text{with} \qquad u_\sigma {\mathfrak C}^{\sigma\mu\nu} = 0 \,.
\end{equation}
We first consider the case where ${\mathfrak C}^{\sigma\mu\nu}$ is some tensor (not a derivative operator). Now consider the divergence of this current:
\begin{equation}\label{eq:Nbardiv}
 \nabla_\sigma (\N^\sigma)_{\GV} = \prn{\nabla_\sigma {\mathfrak C}^{\sigma\mu\nu}} \, \diffB g_{\mu\nu} + {\mathfrak C}^{\sigma\mu\nu}\, \nabla_\sigma \prn{\diffB g_{\mu\nu}} \,.
\end{equation}
From the structure of the adiabaticity equation \eqref{eq:AGdiss}, it is clear that this free energy current is only adiabatic if the divergence \eqref{eq:Nbardiv} is proportional to $\diffB g_{\mu\nu}$. This is clear by examining the adiabaticity equation \eqref{eq:AdiabaticityG}.  In \eqref{eq:Nbardiv} first term in the r.h.s. has an explicit $\diffB g_{\mu\nu}$ which would serve. The second term however contains descendant operators (using the terminology from \S\ref{sec:classD}). These are independent tensors and do not contain a free $\diffB g_{\mu\nu}$.  Alternately, one simply notes that  there are fluid configurations which are hydrostatic locally (i.e., $\diffB g_{\mu\nu} = 0$ at some point on ${\cal M}$)  but such that $\nabla_\sigma \prn{\diffB g_{\mu\nu}} \neq 0$, the above divergence can only be consistent with adiabaticity if one of the following two scenarios occurs:\footnote{ The authors thank Akash Jain for pointing out to us the existence of the second scenario and its importance for the justification of Class B with differential operators.}
\begin{itemize}
\item Either: every term in \eqref{eq:Nbardiv} contains at least one undifferentiated factor of $\diffB g_{\mu\nu}$. Then we require ${\mathfrak C}^{\sigma\mu\nu}$ to contain a factor of $\diffB g_{\mu\nu}$ in general. However, this requirement implies that $({\cal G}^\sigma)_{\GV}$ contains two factors of $\diffB g_{\mu\nu}$ and hence is precisely captured by our Class $\GV$ parametrization \eqref{eq:TJVecrep}.  
\item Or: the second term in \eqref{eq:Nbardiv} is zero and hence does not contribute. This happens if we have 
\begin{equation}
  {\mathfrak C}^{\sigma\mu\nu} = \bar{\mathfrak{C}}^{[\sigma\mu\nu|\alpha\beta\gamma]} \nabla_\alpha (\diffB g_{\beta\gamma})
\end{equation}
for some tensor $\bar{\mathfrak{C}}$. 
Then we would simply find 
\begin{equation}\label{eq:Nbardiv55}
 \nabla_\sigma (\N^\sigma)_{\GV} = \prn{\nabla_\sigma\bar{\mathfrak{C}}^{[\sigma\mu\nu|\alpha\beta\gamma]} \nabla_\alpha (\diffB g_{\beta\gamma}) + \bar{\mathfrak{C}}^{[\sigma\mu\nu|\alpha\beta\gamma]} \nabla_\sigma \nabla_\alpha (\diffB g_{\beta\gamma})
} \, \diffB g_{\mu\nu}  \,.
\end{equation}
This case does not provide any new transport since it is captured by Class B with non-trivial $\Upsilon$-operators. The easiest way to see this is by recognizing that the divergence \eqref{eq:Nbardiv55} is of the form of a Class B divergence \eqref{eq:dissIntPart} with 
\begin{equation}
\begin{split}
  \prn{\DVisc_{\cdg_g}}_\lambda = 2\,\nabla_\lambda \,, \quad \cdg^{\sigma\mu\nu\alpha\beta\gamma} = - \bar{\mathfrak{C}}^{[\sigma(\mu\nu)|\alpha(\beta\gamma)]}\,.
\end{split}
\end{equation}
This case justifies the existence of Class B with non-trivial $\Upsilon$-operators, for without them we would not achieve completeness here.
\end{itemize}

Let us now consider the slightly more general situation where the object ${\mathfrak C}$ in \eqref{eq:HbarVansatz} is not a tensor, but a tensor-valued derivative operator. W.l.o.g.\ we can parameterize this situation as
\begin{equation}\label{eq:HbarVansatz2}
 (\N^\sigma)_{\GV} = {\mathfrak C}^{\sigma\lambda(\mu\nu)} \, \nabla_\lambda (\diffB g_{\mu\nu} )\qquad \text{with} \qquad u_\sigma {\mathfrak C}^{\sigma\lambda\mu\nu} = 0 \,,
\end{equation}
where ${\mathfrak C}^{\sigma\lambda\mu\nu}$ is some tensor. Let us consider separately the cases where ${\mathfrak C}^{\sigma\lambda(\mu\nu)}$ is symmetric and anti-symmetric in its first two indices, respectively:
\begin{itemize}
\item Anti-symmetric case (${\mathfrak C}^{\sigma\lambda(\mu\nu)} = {\mathfrak C}^{[\sigma\lambda](\mu\nu)}$): 
One can easily see that the anti-symmetric component does not provide anything new as it can always be canceled by a combination of terms of other classes and an uninteresting Komar piece; to see this, observe that $(\N^\sigma)_{\GV} $ can be decomposed as follows:
\begin{align}\label{eq:HbarVansatz3}
 (\N^\sigma)_{\GV} &= {\mathfrak C}^{[\sigma\lambda](\mu\nu)} \, \nabla_\lambda (\diffB g_{\mu\nu} ) 
 \notag \\
 &= \nabla_\lambda \prn{{\mathfrak C}^{[\sigma\lambda](\mu\nu)}\diffB g_{\mu\nu}} - \prn{\nabla_\lambda {\mathfrak C}^{[\sigma\lambda](\mu\nu)}}\diffB g_{\mu\nu} \\
 &= \nabla_\lambda \prn{{\mathfrak C}^{[\sigma\lambda](\mu\nu)}\diffB g_{\mu\nu}} - P^\sigma_\rho \prn{\nabla_\lambda {\mathfrak C}^{[\rho\lambda](\mu\nu)}}\diffB g_{\mu\nu} + u^\sigma \prn{ u_\rho \nabla_\lambda {\mathfrak C}^{[\rho\lambda](\mu\nu)}}\diffB g_{\mu\nu} \,,
 \notag
\end{align}
In the last line the first term is Komar (hence uninteresting), the second is captured by $\GV$ (it's of the form \eqref{eq:HbarVansatz}) and the last one is longitudinal (and thus in Class L). We conclude that this does not lead to any transport not captured by either $\GV$ or in one of the other Classes we have already accounted for.

\item Symmetric case (${\mathfrak C}^{\sigma\lambda(\mu\nu)} = {\mathfrak C}^{(\sigma\lambda)(\mu\nu)}$): In the symmetric case, consider the divergence of the free energy current \eqref{eq:HbarVansatz2}: 
\begin{equation} \label{eq:HbarVcalc}
\begin{split}
 \nabla_\sigma (\N^\sigma)_{\GV} 
 &= \prn{ \nabla_\sigma {\mathfrak C}^{(\sigma\lambda)(\mu\nu)} } \nabla_\lambda (\diffB g_{\mu\nu}) + {\mathfrak C}^{(\sigma\lambda)(\mu\nu)} \nabla_\sigma \nabla_\lambda (\diffB g_{\mu\nu}) \,.
 \end{split}
\end{equation}

W.l.o.g.\ we can assume that ${\mathfrak C}^{\sigma\lambda\mu\nu}$ does not contain any undifferentiated $\diffB g_{\mu\nu}$ (if it did, we would be back in the case parameterized by \eqref{eq:HbarVansatz}). A-priori the second term in \eqref{eq:HbarVcalc} does then not contain any undifferentiated $\diffB g_{\mu\nu}$. Hence the only way that this situation can be compatible with adiabaticity, is that a cancellation between the two terms happens such that their sum is proportional to $\diffB g_{\mu\nu}$. 

Let us see under what conditions it could happen that the second term is canceled by the first. Note that $\nabla_{(\sigma} \nabla_{\lambda)} (\diffB g_{\mu\nu})$ is a genuine $3^\text{rd}$ order descendant object which is not a product of lower order tensors. Since the first term in \eqref{eq:HbarVcalc} contains a $2^\text{nd}$ order descendant factor, a cancellation between terms can only happen if ${\mathfrak C}^{\sigma\lambda\mu\nu}$ itself contains a factor of $\nabla_\kappa (\diffB g_{\alpha\beta})$. If the two terms were to cancel, we would hence have to require that
\begin{equation}
 {\mathfrak C}^{(\sigma\lambda)(\mu\nu)} = \bar{{\mathfrak C}}^{(\sigma\lambda\kappa)[(\mu\nu)|(\alpha\beta)]} \nabla_\kappa(\diffB g_{\alpha\beta})  \,.
\end{equation}
But then we find that the Noether current is just zero and hence does not provide new data:
\begin{equation}\label{eq:HVfinal0}
  (\N^\sigma)_{\GV} = \bar{{\mathfrak C}}^{(\sigma\lambda\kappa)[(\mu\nu)|(\alpha\beta)]}  \nabla_\lambda(\diffB g_{\mu\nu})\nabla_\kappa(\diffB g_{\alpha\beta}) = 0 \,.
\end{equation}
%
\end{itemize}
These considerations show that an ansatz of the form \eqref{eq:HbarVansatz2} does not lead to any constitutive relations that are not captured by the ones we already have. 

 Similarly, one can proceed and consider ans\"atze where $({\cal G}^\sigma)_{\GV}$ contains higher derivatives of $\diffB g_{\mu\nu}$, but analogous arguments as the one presented above would show that this is only consistent with adiabaticity if $({\cal G}^\sigma)_{\GV}$ is secretly a combination of Class $\GV$, B and Komar terms as we parameterized them in \S\ref{sec:rclassify}.
 Together with our earlier comments, this completes our proof that the parametrization \eqref{eq:TJVec}, \eqref{eq:GVec} exhausts all possible non-hydrostatic, adiabatic transverse free energy currents.

\subsection{Example I: Charged parity-even fluids}
\label{sec:counting}

To exemplify our general story we turn to an example that has been discussed in some detail in \cite{Bhattacharyya:2014bha}, viz., a charged parity-even fluid. Neutral fluids are clearly a subset obtained by setting the chemical potential and charge density to zero. We will describe first outline the classification in general and then indicate in \S\ref{sec:weylsum} how to specialize to Weyl invariant case (which has the advantage of being able to be tested holographically).

We begin by counting the total number of transport coefficients: there is one frame invariant scalar (for definiteness,
let us take it to be in $P^{\mu\nu}$ part of  the energy momentum tensor), one frame invariant transverse vector
(for definiteness, let us take it to be in  the charge current) and one frame invariant transverse traceless tensor
 occurring as a part of the energy momentum tensor. The final result of this counting is summarized in Tables
 \ref{tab:ChargedResponse} and \ref{tab:ChargedTransport}.

The first order constitutive relations are a-priori parameterized by one scalar $\Theta$, three transverse vectors $\{\acc^\mu, E^\mu, \cv^\mu\}$, and one transverse traceless tensor $\sigma^{\mu\nu}$. The constitutive relations
\begin{equation}
\begin{split}
T^{\mu\nu} &=  \epsilon\, u^\mu\,u^\nu + p\, P^{\mu\nu} - 2\,\eta\, \sigma^{\mu\nu} - \zeta\, \Theta \, P^{\mu\nu}  \\
J^\mu & = q\, u^\mu + \sigma_{_{\text{Ohm}}}\, \cv^\mu +\chi_E\, E^\mu - \chi_T\, T\, \acc^\mu \\
J_S^\mu &= s\, u^\mu + \alpha_1\, \acc^\mu + \alpha_2\, \Theta\, u^\mu + \alpha_3\, \cv^\mu + \alpha_4\, E^\mu
\end{split}
\end{equation}	
satisfy the second law of thermodynamics provided the following relations hold:
\begin{itemize}
\item The coefficient of sign-indefinite terms in $\Diss$ vanish. These are the $\PF$ constraints and at this order they can be shown to imply:
\begin{align}
\epsilon + p   = T\,s + q\,\mu \,,\qquad d\epsilon = T\, ds + \mu\, dq \,.
\end{align}
If these conditions which are familiar from thermodynamics (as the Euler relation and first law) are not satisfied then there is an obstruction to the existence of a hydrostatic partition function.
\item In addition one finds that (see \cite{Bhattacharyya:2014bha} or earlier works such as \cite{Jensen:2011xb} for a derivation)
\begin{align}
\alpha_1 = \alpha_2 = \alpha_3 = \alpha_4 =0 \,, \qquad \chi_E = \chi_T =0 \,.
\end{align}
\item The coefficient of sign-definite terms contributing to $\Diss$ must be correct. Evaluating the relevant terms we obtain
\eqref{eq:visent} as expected and learn that viscosities and conductivities are positive definite. In \S\ref{sec:Ddiffops} we have already given the result that these terms can be obtained from the Class D parametrization \eqref{eq:TJClassVrep} by choosing $\DVisc_{\cdg_g}=\DVisc_{\cdA_A} = \text{Id}$ and $\DVisc_{\cdA_g}=\DVisc_{\cdg_A}=0$ along with the Class $\Dv$ tensor structures
\begin{equation}
\begin{split}
 \cdg^{\mu\nu\alpha\beta} = T\, \zeta \, P^{\mu\nu}P^{\alpha\beta} + 2\,T\,\eta \,P^{\alpha<\mu} P^{\nu>\beta}\,,\qquad
 \cdA^{\alpha\beta} = T\, \sigma_{_{\text{Ohm}}} \, P^{\alpha\beta} \,.
\end{split}
\end{equation}
\end{itemize}

\begin{table}[h]
\centering
\begin{tabular}{|| c | c | c || }
\hline\hline
\multicolumn{3}{||c||}{{\shadeB $2^{\rm nd}$ order charged hydrostatic response}}\\
\hline
 Scalars & Vectors & Tensors \\
 \hline
$\omega^2$   &  $\omega^{\mu\nu}\acc_\nu$  & $\omega^{\alpha<\mu}\omega^{\nu>}{}_\alpha$ \\
$ \omega_{\alpha\beta}B^{\alpha\beta} $ & $\omega^{\mu\nu}E_\nu$ & $\omega^{\alpha<\mu}B^{\nu>}{}_\alpha$ \\
$B^2$ &   $B^{\mu\nu}E_\nu$ & $B^{\alpha<\mu}B^{\nu>}{}_\alpha$\\
$\acc_\alpha E^\alpha$ &   $B^{\mu\nu}\acc_\nu$  &  $\acc^{<\mu}E^{\nu>}$ \\
$E^2$                            &     & $E^{<\mu}E^{\nu>}$ \\
$\acc^2$                        &     &  $\acc^{<\mu}\acc^{\nu>}$ \\
$R$ & & $R^{<\mu\nu>}$\\
\hline
 $R_{\alpha\beta} u^\alpha u^\beta$  & $P^\mu_\nu D_\lambda \omega^{\nu\lambda}$ &  $F_R^{<\mu\nu>}\equiv u^\alpha u^\beta R^{<\mu}{}_\alpha{}^{\nu>}{}_\beta$ \\
 $D_\mu E^\mu$ & $P^\mu_\nu D_\lambda B^{\nu\lambda}$ &  $D^{<\mu}E^{\nu>}$ \\
\hline
\multicolumn{3}{||c||}{\shadeR{$9S+6V+9T={\bf 24}=17\,\PF+7\, \PS+0\, \PV+0 \,\text{A}$}}\\
\hline\hline
\end{tabular}
\caption{The $24$ hydrostatic response terms for  parity-even charged  fluids at $2^{\rm nd}$ order in derivative expansion. Among them, $\PF=17$ combinations are forbidden by hydrostatic principle whereas the remaining $\PS=7$ combinations are generated by using the first $7$
scalars in the Lagrangian.}
\label{tab:ChargedResponse}
\end{table}

\begin{table}[h]
\centering
\begin{tabular}{|| c | c | c || }
\hline\hline
\multicolumn{3}{||c||}{\shadeB{$2^{\rm nd}$ order charged non-hydrostatic transport}} \\
\hline
 Scalars & Vectors & Tensors \\
\hline
$\Theta^2$                 &   & $\sigma^{\alpha<\mu}\sigma^{\nu>}{}_\alpha$\\
$\sigma^2$                 &       & $\Theta \sigma^{\mu\nu}$ \\
$\cv_\alpha \acc^\alpha$  &    $\Theta \acc^\mu$ & \\
$\cv_\alpha E^\alpha$  &     $\Theta E^\mu$  & \\
$\cv^2$ &    $\Theta \cv^\mu$  & \\
  & $\sigma^{\mu\nu}\cv_\nu$ & $\cv^{<\mu}\cv^{\nu>}$ \\
 & $\sigma^{\mu\nu}\acc_\nu$ & $\cv^{<\mu}\acc^{\nu>}$ \\
                                    & $\sigma^{\mu\nu}E_\nu$ & $\cv^{<\mu}E^{\nu>}$ \\
                                    &  $\omega^{\mu\nu}\cv_\nu$   & $\sigma^{\alpha<\mu}\omega^{\nu>}{}_\alpha$ \\
                                   & $B^{\mu\nu}\cv_\nu$ & $\sigma^{\alpha<\mu}B^{\nu>}{}_\alpha$ \\
 \hline
$(u^\alpha D_\alpha) \Theta$  & $P^\mu_\nu (u^\alpha D_\alpha)E^\nu$ & $(u^\alpha D_\alpha) \sigma^{\mu\nu}$  \\
 $D_\mu \cv^\mu$ & $P^{\mu\nu}D_\nu \Theta$ &  $D^{<\mu}\cv^{\nu>}$ \\
                & $P^\mu_\nu D_\lambda \sigma^{\nu\lambda}$  & \\
\hline
\multicolumn{3}{||c||}{\shadeR {$7S+11V+9T={\bf 27}=5\,\LS+2\,\GV+11\, \text{B}+9\, \text{D}$}}\\
\hline\hline
\end{tabular}
\caption{The $27$ non-hydrostatic transport terms for  parity-even charged  fluid at $2^{\rm nd}$ order in derivative expansion. Among these, $\LS=5$ combinations are generated by inserting the first $5$ non-hydrostatic scalars into the Lagrangian. Two combinations are generated by inserting  terms proportional to the non-hydrostatic vectors $\{\sigma^{\mu\nu}\cv_\nu, \Theta \cv^\mu \}$ in the free energy current, thus $\GV=2$. Among the rest, there are $11$ combinations in Class B and
$9$ combinations in Class D. Explicit expressions for these $20$ combinations are given in Table \ref{tab:CountingEven}.}
\label{tab:ChargedTransport}
\end{table}

Having dispensed with the leading order, let us move to the second order constitutive relations. 
There are in total $51$ parity-even $2$-derivative terms that can appear in the charged fluid constitutive relations 
\cite{Banerjee:2012iz,Bhattacharyya:2014bha} among which $9$ scalars, $6$ transverse vectors and $9$ transverse traceless tensors (i.e., a total of $24$ terms as in Table  \ref{tab:ChargedResponse}) enter the hydrostatic description. The existence of hydrostatics imposes a series of constraints
 on these $24$ terms. The most straightforward way to derive these constraints is to write down the most general
 hydrostatic partition function using the first  $7$ hydrostatic scalars multiplied by arbitrary functions of ${T,\mu}$.
Note that we have  discarded the last $2$ hydrostatic scalars which can be removed by an integration by parts in the partition
function. Hence, we conclude $\PS=7$. For this system, there are no hydrostatically conserved
 vectors ($\PV=0$) and no possible anomalies ($\text{A}=0$).

By varying this partition function, we then get the most general hydrostatic energy momentum tensor
consistent with hydrostatic principle. This procedure then fixes the $24$ response coefficients in terms of $7$ functions that
appear in the partition function.  Eliminating these $7$ functions, we get $24-7=17$ relations thus giving the
number of hydrostatically forbidden combinations as $\PF=17$.

We now turn to the non-hydrostatic transport parameters listed in Table \ref{tab:ChargedTransport}. There are
$7$ scalars, $11$ transverse vectors and $9$ transverse traceless tensors which add up to $27$ non-hydrostatic
transport coefficients. Of these, $5$ combinations can be obtained from including the first $5$ non-hydrostatic
scalars into a Lagrangian (the last $2$ scalars can be discarded via integration by parts). These $5$ combinations
give $\LS=5$.

The remainder of the analysis involves figuring out which of the residual $27-5 = 22$ terms can be obtained in the three non-Lagrangian classes $\{\GV,\text{B}, \text{D}\}$. These are classified by the set of admissible tensor structures which were described in the preceding. We have already shown in \S\ref{sec:gibbsvec} that $\GV =2$ by explicit construction; this involved considering all zeroth order tensor structures $\{{\mathfrak C}_{\BerryG}^{\rho ((\mu\nu)|(\alpha\beta))}, {\mathfrak C}_{\BerryGA}^{\rho(\mu\nu)\alpha} , {\mathfrak C}_{\BerryA}^{\rho(\alpha\beta)} \}$ and plugging them into \eqref{eq:TJVecrep}. Independent data was obtained from the two combinations
\begin{equation}\label{eq:HVBconstex}
 \begin{split}
  {\mathfrak C}_{\BerryGA}^{\rho(\mu\nu)\alpha} &= T\, C_1(T,\mu)\, P^{\rho<\mu} P^{\nu>\alpha} + T\, C_2(T,\mu)\, P^{\mu\nu} P^{\rho\alpha} \\
  \Rightarrow \quad {\cal G}^\rho &= 2\,C_1\, \sigma^{\rho\alpha}\cv_\alpha + 2 \, C_2\, \Theta \, \cv^\rho \,.
 \end{split}
\end{equation}

Let us now turn to Classes B and D. All independent Class B and D constitutive relations are shown in 
Table \ref{tab:CountingEven}. The $11+9$ combinations presented there can be constructed by classifying all possible first order transverse tensor structures $\{\BerryG^{\mu\nu\alpha\beta}, \BerryGA^{\mu\nu\alpha},\BerryA^{\alpha\beta}\}$ in the constitutive relations \eqref{eq:TJBerryrep} and \eqref{eq:TJDissSimplerep}, respectively. All inequivalent Class B terms can be obtained (up to numerical factors) by plugging the following $11$ tensor structures into \eqref{eq:TJBerryrep}:\footnote{ Note that each of the tensor structures appears with an arbitrary function of $T,\mu$ multiplying it. The transport coefficients themselves are determined by suitable (differential) linear combinations of these functions. }
\begin{equation}\label{eq:Bconstex}
\begin{split}
  \BerryG^{\mu\nu\alpha\beta} &\in T\{ \sigma^{\mu\nu} P^{\alpha\beta} \,, \, \omega^{\mu\alpha} P^{\nu\beta}\,,\, B^{\mu\alpha} P^{\nu\beta} \} \,,\\
  \BerryGA^{\mu\nu\alpha} &\in T\{ P^{\mu\nu} \cv^\alpha \,,\, P^{\mu\nu} E^\alpha \,,\, P^{\mu\nu} \acc^\alpha \,,\, \cv^{<\mu} P^{\nu>\alpha} \,,\,
  E^{<\mu} P^{\nu>\alpha}\,,\, \acc^{<\mu} P^{\nu>\alpha} \} \,,\\
  \BerryA^{\alpha\beta} &\in T\{ \omega^{\alpha\beta} \,,\, B^{\alpha\beta} \}\,.
\end{split}
\end{equation}
The $9$ Class D terms listed in Table \ref{tab:CountingEven} can be obtained by plugging the following tensor structures into \eqref{eq:TJDissSimplerep}:
\begin{equation}\label{eq:Dconstex}
\begin{split}
   \BerryG^{\mu\nu\alpha\beta} &\in T\{ \sigma^{\mu\nu} P^{\alpha\beta} \,, \, \sigma^{\mu\alpha}P^{\nu\beta} \,,\, \Theta \, P^{\mu\nu} P^{\alpha\beta}\}  \,,\\
  \BerryGA^{\mu\nu\alpha} &\in T\{ P^{\mu\nu} \cv^\alpha \,,\, P^{\mu\nu} E^\alpha \,,\, P^{\mu\nu} \acc^\alpha \,,\, \cv^{<\mu} P^{\nu>\alpha} \,,\,
  E^{<\mu} P^{\nu>\alpha}\,,\, \acc^{<\mu} P^{\nu>\alpha} \} \,,\\
  S^{\alpha\beta} &= 0 \,.
\end{split}
\end{equation}
These lists for Classes B and D are exhaustive in the sense that any other tensor structure leads to Class B and D constitutive relations with frame invariant data being the same as of those already obtained (or linear combinations thereof).\footnote{ For instance, an obvious structure one might want to add to Class D parameterization is $\BerryA^{\alpha\beta} = \sigma^{\alpha\beta}$. However, this gives constitutive relations which are a linear combination of the Class B terms and the Class D terms originating from $\BerryGA^{\mu\nu\alpha} = P^{\mu\nu} \cv^\alpha$.} Note that we are not guaranteed that the parametrization \eqref{eq:TJDissSimplerep} is exhaustive -- we only verify this a-posteriori by observing that we have found a total of $51$ inequivalent terms in all the classes of transport which matches the total number of inequivalent tensor structures that are possible for the second order charged fluids.

We can generalize the above  discussion to parity-odd transport using the results of \cite{Bhattacharyya:2013ida}. There are 
2 additional parity-odd pseudo-vectors at first order in $d=4$ (the magnetic field vector $B^\mu$ and the vorticity vector 
$\ell^\mu = \varepsilon^{\mu\alpha\rho\sigma}\, u_\alpha\, \nabla_\rho\, u_\sigma$).
27 parity-odd terms at second order (27 = $6\,S + 9\, V + 12\, T$). Out of these second order terms $12 = 4\, S+2\, V+6 \,T$ are hydrostatic and can be obtained from an equilibrium partition function parameterized by two scalars: hence  $\PS =2$ and $\PF =10$ (which includes A). Additionally one can see from their table 2 that $\LS =2$. This leaves us with 13 further terms ($7V+6T$) which should belong to the other classes. We leave it as an exercise for the reader to complete the classification for this case.

\subsection{Example II: Weyl invariant fluid dynamics}
\label{sec:weylsum}

While the second order charged fluid allows us to illustrate the complete set of constitutive relations within the eightfold way, it is useful to record some examples which can be tested at least theoretically using the fluid/gravity correspondence \cite{Bhattacharyya:2008jc,Hubeny:2011hd}. We therefore give a summary of how various terms for Weyl invariant fluids (both neutral and charged) fit into our classification scheme. In \S\ref{sec:holofluids} we will provide explicit evidence of the eightfold classification in holography.

Let us now consider the case of second order fluid dynamics with Weyl invariance. The case of uncharged fluids is summarized in Tables \ref{tab:NeutralResponseW} and \ref{tab:NeutralTransportW}. 

\begin{table}[h]
\centering
\begin{tabular}{|| c | c | c || }
\hline\hline
\multicolumn{3}{||c||}{{\shadeB $2^{\rm nd}$ order Weyl invariant neutral fluids: hydrostatic response}}\\
\hline
 Scalars & Vectors & Tensors \\
 \hline
$\omega^2$   &    & $\omega^{\alpha<\mu}\omega^{\nu>}{}_\alpha$ \\
$\RWeyl$ & & $C^{<\mu\alpha \nu>\beta} u_\alpha u_\beta $\\
\hline
& $P^\mu_\nu \DWeyl_\lambda \omega^{\nu\lambda}$ &  \\
\hline
\multicolumn{3}{||c||}{{\shadeR $2T={\bf 2}=0\, \PF+2\, \PS+0\,\PV+0\, \text{A}$}} \\
\hline\hline
\end{tabular}
\caption{The $2$ hydrostatic response terms for Weyl invariant parity-even neutral fluid at $2^{\rm nd}$ order in derivative expansion. We have listed
the scalars and vectors though they do not contribute to frame-invariant transport data. Both the relevant symmetric tensors can be obtained by using the  $2$ scalars in the Lagrangian, see \S\ref{sec:neutral} for details.}
\label{tab:NeutralResponseW}
\end{table}

\begin{table}[h]
\centering
\begin{tabular}{|| c | c | c || }
\hline\hline
\multicolumn{3}{||c||}{{\shadeB $2^{\rm nd}$ order Weyl invariant neutral fluids: non-hydrostatic transport}} \\
\hline
 Scalars & Vectors & Tensors \\
\hline
 $\sigma^2$            &    & $\sigma^{\alpha<\mu}\sigma^{\nu>}{}_\alpha$\\
 &     & $\sigma^{\alpha<\mu}\omega^{\nu>}{}_\alpha$ \\
 \hline
& $P^\mu_\nu \DWeyl_\lambda \sigma^{\nu\lambda}$  & $(u^\alpha \DWeyl_\alpha) \sigma^{\mu\nu}$  \\
 \hline
\multicolumn{3}{||c||}{{\shadeR $3T={\bf 3}=1\,\LS+0\,\GV+1\, \text{B}+1\, \text{D}$}} \\
\hline\hline
\end{tabular}
\caption{The $3$ non-hydrostatic transport terms for  Weyl invariant parity-even neutral  fluid at $2^{\rm nd}$ order in derivative expansion.
We have listed  the scalars and vectors though they do not contribute to frame-invariant transport data.  Among the symmetric tensors,
$\LS=1$ combination is generated by inserting $\sigma^2$ into the Lagrangian. We have  $\GV=0$
since $\sigma^{\alpha<\mu}\omega^{\nu>}{}_\alpha$ is  in Class B and $\sigma^{\alpha<\mu}\sigma^{\nu>}{}_\alpha$ is in Class D.}
\label{tab:NeutralTransportW}
\end{table}

There is a total of $5$ Weyl invariant second order symmetric tensors that can enter the most general symmetry allowed constitutive relations, c.f.\ \eqref{eq:WeylT2}. Our Lagrangian analysis in \S\ref{sec:2ndWeyl} showed explicitly that the $2$ combinations $\omega^{\alpha<\mu}\omega^{\nu>}{}_\alpha$ and $C^{<\mu\alpha \nu>\beta} u_\alpha u_\beta $ are in the hydrostatic Class $\PS$ and a third term $(u^\alpha \DWeyl_\alpha) \sigma^{\mu\nu}$ is Class $\LS$. These $3$ terms can be obtained from a Lagrangian constructed out of the $3$ Weyl invariant scalars. $2$ more tensor structures in Table \ref{tab:NeutralTransportW}, $\sigma^{\alpha<\mu}\omega^{\nu>}{}_\alpha$ and $\sigma^{\alpha<\mu}\sigma^{\nu>}{}_\alpha$, are non-Lagrangian terms in Class B and Class D, respectively. Both these
combinations are surprisingly absent in fluids dual to Einstein gravity -- we only generate the particular linear combination of the $3$ non-hydrostatic terms that is in $\LS$, cf., \S\ref{sec:neutral} for a discussion.

We now turn to the case of second order charged fluids with Weyl invariance. All symmetry allowed tensor structures are summarized in Tables \ref{tab:ChargedResponseW} and \ref{tab:ChargedTransportW}.

\begin{table}[h]
\centering
\begin{tabular}{|| c | c | c || }
\hline\hline
\multicolumn{3}{||c||}{{\shadeB $2^{\rm nd}$ order Weyl invariant charged fluids: hydrostatic response}}\\
\hline
 Scalars & Vectors & Tensors \\
 \hline
$\omega^2$   &    & $\omega^{\alpha<\mu}\omega^{\nu>}{}_\alpha$ \\
$ \omega_{\alpha\beta}B^{\alpha\beta} $ & $\omega^{\mu\nu}E_\nu$ & $\omega^{\alpha<\mu}B^{\nu>}{}_\alpha$ \\
$B^2$ &   $B^{\mu\nu}E_\nu$ & $B^{\alpha<\mu}B^{\nu>}{}_\alpha$\\
$E^2$                            &     & $E^{<\mu}E^{\nu>}$ \\
$\RWeyl$ & & $C^{\mu\alpha\nu\beta}u_\alpha u_\beta $\\
\hline
 $\DWeyl_\mu E^\mu$ & $P^\mu_\nu \DWeyl_\lambda B^{\nu\lambda}$ &  $\DWeyl_{<\mu}E_{\nu>}$ \\
  & $P^\mu_\nu \DWeyl_\lambda \omega^{\nu\lambda}$ &   \\
\hline
\multicolumn{3}{||c||}{{\shadeR $4V+6T={\bf 10}=5\,\PF+5\, \PS+0\,\PV+0\, \text{A}$}} \\
\hline\hline
\end{tabular}
\caption{The $10$ hydrostatic response terms for  Weyl invariant, parity-even, charged  fluid at $2^{\rm nd}$ order in derivative expansion.
We have also enumerated the scalars  despite the fact that they do not  appear in the constitutive relations of a Weyl invariant fluid. Among the constitutive relations generated by the vectors and symmetric tensors,
$\PF=5$ combinations are forbidden by hydrostatic principle whereas the remaining $\PS=5$ combinations are generated by using the first $5$ scalars in the Lagrangian.}
\label{tab:ChargedResponseW}
\end{table}

\begin{table}[h]
\centering
\begin{tabular}{|| c | c | c || }
\hline\hline
\multicolumn{3}{||c||}{{\shadeB $2^{\rm nd}$ order Weyl invariant charged fluids: non-hydrostatic transport}} \\
\hline
 Scalars & Vectors & Tensors \\
\hline
$\sigma^2$                 &       & $\sigma^{\alpha<\mu}\sigma^{\nu>}{}_\alpha$\\
$\cv_\alpha E^\alpha$  &       & \\
$\cv^2$ &     & \\
  & $\sigma^{\mu\nu}\cv_\nu$ & $\cv^{<\mu}\cv^{\nu>}$ \\
                                    & $\sigma^{\mu\nu}E_\nu$ & $\cv^{<\mu}E^{\nu>}$ \\
                                    &  $\omega^{\mu\nu}\cv_\nu$   & $\sigma^{\alpha<\mu}\omega^{\nu>}{}_\alpha$ \\
                                   & $B^{\mu\nu}\cv_\nu$ & $\sigma^{\alpha<\mu}B^{\nu>}{}_\alpha$ \\
 \hline
  & $P^\mu_\nu (u^\alpha \DWeyl_\alpha)E^\nu$ & $(u^\alpha \DWeyl_\alpha) \sigma^{\mu\nu}$  \\
 $\DWeyl_\mu \cv^\mu$ & $P^\mu_\nu \DWeyl_\lambda \sigma^{\nu\lambda}$   &  $\DWeyl_{<\mu}\cv_{\nu>}$ \\
\hline
\multicolumn{3}{||c||}{{\shadeR $6V+7T={\bf 13}=3\,\LS+1\,\GV+6\,\text{B}+3\, \text{D}$}}\\
\hline\hline
\end{tabular}
\caption{The $13$ non-hydrostatic transport terms for  Weyl invariant, parity-even charged  fluid at $2^{\rm nd}$ order in derivative expansion. Among these,
$\LS=3$ combinations are generated by inserting the first $3$ non-hydrostatic scalars into the Lagrangian. One combination is generated by inserting a term proportional
to the non-hydrostatic vector $\sigma^{\mu\nu}\cv_\nu$ in the free energy current, thus $\GV=1$. Explicit expressions for the $6$ combinations in Class B and $3$ combinations in Class D can be found in Table \ref{tab:CountingEven}.}
\label{tab:ChargedTransportW}
\end{table}

The hydrostatic case is easy to intuit. We have $10$ hydrostatic vectors and symmetric tensors which generate constitutive relations. $\PS=5$ of these terms can be obtained from a Lagrangian that contains the $5$ hydrostatic Weyl invariant scalars. The remaining $\PF=5$ are forbidden by the second law constraint. Beyond hydrostatics, there are another $13$ vectors and tensors, $\LS = 3$ of which can be generated by Lagrangians. The remaining $10$ non-Lagrangian terms split into $\{\GV,\text{B},\text{D}\} = \{1,6,3\}$. Explicit expressions for the $10$ combinations are given in Table \ref{tab:CountingEven}. Their corresponding constitutive relations as listed in Table \ref{tab:CountingEven} are precisely those that are obtained in the general constructions \eqref{eq:HVBconstex}, \eqref{eq:Bconstex} and \eqref{eq:Dconstex}, but restricting to Weyl invariant tensor structures. Practically, this means deleting all those constitutive relations that were obtained in the non-Lagrangian classes before, which contain non-Weyl invariant objects such as $\Theta$ and $\acc^\mu$.

\subsection{Adiabatic fluids in holography and kinetic theory}
\label{sec:holofluids}

Given the relations obtained for adiabatic Weyl invariant fluids in Class L up to second order in gradients, we have now some structural understanding of the class of hydrodynamic systems we are dealing with. We can for instance ask if there are physical theories which are aware of the adiabatic eightfold classification. The answer turns out to be a resounding confirmation of using adiabaticity to classify hydrodynamic transport. We will explain this statement, by examining to distinct hydrodynamic systems that are physically motivated and for which we have data to compare transport coefficients. First, we look at holographic fluids arising in the fluid/gravity correspondence \cite{Bhattacharyya:2008jc,Hubeny:2011hd} which arise from strongly coupled gauge theory plasmas. Subsequently we examine the fluids arising from weakly coupled quantum field theories which have been understood using kinetic theory. We will focus exclusively on second order transport since first order transport is entirely in Class $\Dv$.

\subsubsection{Holographic fluids and adiabaticity}
\label{sec:hoload}

We begin by examining the transport properties of holographic fluids using the AdS/CFT correspondence.
As is well known the class of fluids dual to two derivative Einstein-Hilbert gravity saturate the KSS bound \cite{Kovtun:2004de} for shear viscosity $\eta/s = \frac{1}{4\pi}$  at first order. Since $\eta\,\sigma^{\mu\nu}$ is a Class $\Dv$ term, we have no information to gain from an adiabatic analysis. It is nevertheless interesting to note that entropy production encoded in $\Diss = 2\,\eta\, \sigma_{\mu\nu}\, \sigma^{\mu\nu}$ is minimized (assuming the KSS bound  $\eta/s \geq \frac{1}{4\pi}$).

At second order in the gradient expansion we have more to say, since there are definitely adiabatic parts to transport as we have discovered above.  Let us start by understanding the relations 
\eqref{eq:weylrelns} in the eightfold way. The transport coefficient $(\lambda_1 - \kappa)$ is dissipative (Class $\Ds$); indeed, this term contributes to entropy production as $\nabla_\mu J_S^\mu \sim (\lambda_1-\kappa) \,\sigma_{\alpha\nu} \sigma^{\nu\beta} \sigma^\alpha{}_\beta$ which is sub-dominant to the leading order $\eta \,\sigma^{\mu\nu} \sigma_{\mu\nu}$ entropy production. A-priori this ought not to be visible in an effective action (either Class L or $\LT$). On the other hand, $(\lambda_2+2\tau-2\kappa)$ is a Class B term, for which we ought to be writing a Class $\LT$ effective action. The three remaining terms in \eqref{eq:WeylT2eightfold}  encode the adiabatic part of the second order transport and  are in one-to-one correspondence with the three free functions in the Lagrangian \eqref{eq:weyl2lambda} parameterizing the 
Class L Landau-Ginzburg free energy. Correspondingly they are unconstrained by any hydrodynamic  analysis. This is the general expectation from our classification scheme. Let us now turn to seeing what has been computed in the literature so far.

Firstly, we find a relation between the transport coefficients $\lambda_1$ and $\kappa$: 
\begin{equation}
\lambda_1 -\kappa =0\,.
\label{eq:lam1kap}
\end{equation}	
This constraint follows directly in all dimensions for any large central charge quantum field theory whose holographic  dual is given by Einstein-Hilbert gravity. This can be ascertained from the analysis of
\cite{Bhattacharyya:2008mz}, see their Eq.\ (4.6). It also holds in the case of charged fluids in $d=4$ (we believe it probably holds in all dimensions) as can be verified from Eqs.~(4.14) and (5.2) of \cite{Plewa:2012vt}. Its validity in more general theories has not been checked as far as we are aware; it would be interesting to examine this relation in more general theories of gravity. As demonstrated in \cite{Bhattacharya:2012zx} it is a necessary consequence of non-dissipation; one can show that the on-shell entropy current is conserved iff $\kappa = \lambda_1$ as we expect from the adiabaticity perspective. The relation we note is not visible in hydrostatics since $\sigma_{\mu\nu} $ vanishes in equilibrium and one therefore is unable to fix the value of $\lambda_1$; $\kappa$ on the other hand is part of thermodynamic response,
 cf., \cite{Banerjee:2012iz}. On the other hand we have seen that our Class L Landau-Ginzburg free energy lands us precisely on this subspace of allowed constitutive relations, cf., \eqref{eq:weylrelns}.

Curiously, we also find a second relation which fixes $\lambda_2$ despite it being a Class B term. In Class L we find this captured by the second relation  \eqref{eq:weylrelns}. 
In fact, using \eqref{eq:lam1kap} we can express this after eliminating $\kappa$ 
as a relation between  $\tau, \lambda_1, \lambda_2$ when it  is even more fascinating and easily recognized in holography: 
\begin{equation}
\tau = \lambda_1 - \frac{1}{2}\, \lambda_2\,.
\label{eq:hyrel}
\end{equation}	
This is precisely the universal relation between second order transport shown to hold in a very broad class of theories by \cite{Haack:2008xx}. They derived the relation in two derivative theories of gravity coupled to arbitrary matter fields (scalars and gauge fields).\footnote{ The relation given in Eq.~(9) of \cite{Haack:2008xx} uses different conventions and also has a small typo. The authors define the shear tensor with an extra factor of 2 relative to our definition. Furthermore, the sign in the definition of $\omega_{\mu\nu}$ should be flipped; this affects the sign of $\lambda_2$. We thank Michael Haack and Amos Yarom for double checking these results carefully and for discussions on implications and generalizations of this statement.}
This relation also holds naturally in the non-dissipative effective action approach, but is not demanded per se from entropy conservation \cite{Bhattacharya:2012zx}; the latter analysis leaves $\lambda_2$ unconstrained. From our modern perspective of adiabaticity, this makes sense as the term is part of Class B transport. It is rather surprising that not only the Class L theories fix the value of $\lambda_2$ but they also do so in a manner consistent with holography!\footnote{ The relation holds without imposing Weyl invariance in Class L as can be seen explicitly in Appendix \ref{sec:neutral2d}. }

Going beyond the two derivative gravity theories, we can ask if the relations \eqref{eq:weylrelns} or 
\eqref{eq:hyrel} hold once we include higher order corrections to the gravitational Lagrangian. This would correspond to the finite coupling corrections to the strong coupling limit of the holographic plasma. So far it has been checked that \eqref{eq:hyrel} holds   perturbatively in Gauss-Bonnet theories to leading order in the higher-derivative coupling \cite{Shaverin:2012kv}, though not to next to leading order \cite{Grozdanov:2014kva,Yarom:2014kx}. Curiously enough, higher derivative corrections that arise in string theory (from Type IIB flux compactification on ${\bf S}^5$)  uphold this relation to one additional order 
\cite{Grozdanov:2014kva} (to ${\cal O}(\lambda^{-3/2})$ in the strong coupling perturbation expansion for the ${\cal N}= 4$ SYM plasma). However, the original relations as stated in \eqref{eq:weylrelns} are satisfied only to leading order in the higher derivative correction to gravity. From the adiabatic fluid perspective, 
\eqref{eq:weylrelns} is a bit more fundamental since $\kappa-\lambda_1$ provides a measure for entropy production. 

Viewing these relations as fixing a Class D and Class B term respectively is itself an interesting statement, independent of the precise values. While any physical fluid would of course have specific values of transport coefficients, one generically expects that the second order Weyl transport is a point in the five-dimensional space of parameters. Having extra constraints fixing two parameters in terms of the others is an interesting statement which deserves to be understood better. Moreover, the value chosen for $\lambda_1$
is such that no entropy is produced. This is rather remarkable hinting that holographic fluids are even more perfect than hitherto believed to be.

Finally, for completeness let us record the values of $\{k_\sigma, k_\omega, k_R\}$ that are suggested by holography. Translating the results of \cite{Bhattacharyya:2008mz} we have\footnote{ We use $c_\text{eff}$ to denote the effective central charge of the field theory; $c_\text{eff} = \frac{\ell_\text{AdS}^{d-1}}{16\pi\, G_N}$. For ${\cal N}=4$ SYM  in $d=4$ with gauge group $SU(N)$ this is $\frac{1}{8\pi^2} \, N^2$.
}
\begin{equation}
\begin{split}
k_R &= -\frac{c_\text{eff}}{d-2} \prn{\frac{4\pi}{d}}^{d-2} \,, \\
k_\omega &= \frac{d-2}{2}\, k_R \,, \\
k_\sigma &= \frac{c_\text{eff}}{2\,d}\, \prn{\frac{4\pi}{d}}^{d-2}  \, \text{Harmonic}\prn{\frac{2}{d}-1} \,,
\end{split}
\end{equation}
where $\text{Harmonic}(x) = \gamma_e + \frac{\Gamma'(x)}{\Gamma(x)}$ is the Harmonic number function
($\gamma_e$ is Euler's constant). Thus, the fluid-gravity result for second order neutral fluid transport can be determined explicitly from a Lagrangian density
\begin{align}
\Lag^\Wey &=
c_\text{eff} \prn{\frac{4\pi T}{d}}^{d}-c_\text{eff} \prn{\frac{4\pi T}{d}}^{d-2}\brk{ \frac{\RWeyl}{(d-2)}  +\half  \, \omega^2 +\frac{1}{\,d}\,  \text{Harmonic}\prn{\frac{2}{d}-1}\, \sigma^2}
\label{eq:weyl2lambdaFG}
\end{align}
where we have included also the zero derivative pressure term. 

It is really amazing that the simple effective action \eqref{eq:weyl2lambdaFG} captures all the non-trivial results about the thermodynamics of a strongly coupled plasma along with the non-linear part of transport. Only the value of the first order Class D term, shear viscosity, is undetermined and indeed modulo this contribution (which is of course important), holographic plasmas are effectively adiabatic. Coupled with the low value of shear viscosity \cite{Kovtun:2004de}, it follows that flows of these plasmas tend to minimize the amount of dissipation. The nearly perfect fluid picture, persists even more strongly perhaps at second order in gradients. We argue in \S\ref{sec:discuss} that this suggests a minimum entropy production conjecture, which would be fascinating to understand in greater detail than explored herein.

It is an interesting challenge in fluid-gravity correspondence to give a gravity prescription to directly derive this expression. We advocate this as  a sharp test for  the proposals  on how to think about AdS/CFT effective actions in
the presence of horizons (see, for example \cite{Nickel:2010pr}). 

\subsubsection{Kinetic theory and adiabatic fluids}
\label{sec:kinectic}
We will now turn to examine the existing results in weakly coupled field theories in the light of our eightfold classification. Computations in weak coupling are surprisingly more difficult than the 
AdS/CFT computations in the previous subsection.

The transport coefficients in the hydrostatic Classes $\PS$ and $\PV$  are computable  via straightforward 
Euclidean methods without any subtle issues regarding analytic continuation. At weak coupling, the leading contributions to these coefficients are generically given by free-theory results which  can be then systematically corrected via perturbative expansion.  These coefficients are also  more amenable to numerical methods in lattice.  The most common example is the computation of pressure as a function of temperature and chemical potential   in various weakly coupled field theories.

The above state is to be  contrasted with the non-hydrostatic classes which require real-time (Schwinger-Keldysh) techniques for their computation. Further, the leading contributions to  these non-hydrostatic coefficients behave generically as inverse powers of coupling (and inverse powers of the logarithm of couplings) and often require careful resummations to   deal with infrared issues. A common simplification in this context is to pass to an effective kinetic description. A paradigmatic example in this regard is the computation of viscosities and conductivities  for weakly coupled theories \cite{Arnold:2000dr,Arnold:2003zc}.

Computations of higher order transport coefficients are less common. Let us specialize to the case of ($3+1$)-dimensional parity-even, Weyl invariant neutral fluid transport coefficients at second order in derivative expansion. The hydrostatic transport coefficients in this case were computed by Moore and Sohrabi \cite{Moore:2012tc} whose result can be stated in our notations as
\begin{equation}\label{eq:MSTherm}
\begin{split}
-\lambda_3 = 3\kappa = \frac{T^2}{48}\prn{-4 N_S+2N_{WF}+ 16 N_V } + O(g) \,,
\end{split}
 \end{equation}
where $N_S$ is the number of real scalars, $N_{WF}$ is the number of Weyl fermions and $N_V$ is the number of massless vectors in the theory. Here,  $O(g)$ represents sub-leading corrections due to interactions.\footnote{ We will refer the reader to \cite{Manes:2012hf,Manes:2013kka,Megias:2014mba} for a generalization 
of these results to parity odd transport coefficients. There is an unresolved discrepancy in the value of $\lambda_3$
between the results of \cite{Moore:2012tc} and \cite{Megias:2014mba}. } 

The non-hydrostatic transport coefficients are more difficult to compute. For the ($3+1$)-dimensional parity-even, 
Weyl invariant neutral fluid, the leading order answers have been computed using kinetic theory in \cite{York:2008rr} (see also
\cite{Baier:2007ix}). These leading answers are proportional to inverse powers of coupling (and inverse powers of the logarithm of couplings) as expected. Thus,  in contrast to the hydrostatic coefficients in \eqref{eq:MSTherm}  which are known up to zeroth order in coupling, the non-hydrostatic coefficients  are known less precisely. The results of \cite{York:2008rr} translated into our notation takes the form\footnote{\label{fn:conventions} Unfortunately the literature is littered with a multitude of conventions for various hydrodynamic tensors which affects the numerical values of transport coefficients. As far as we have been able to ascertain the following is a useful dictionary to aid the translation between the references cited:
\begin{equation}
\begin{split}
 & \omega =  -\omega_\text{\tiny{\cite{Bhattacharyya:2008jc}}}  =  \omega_\text{\tiny{\cite{Bhattacharyya:2008mz}}} =  
 - \omega_\text{\tiny{\cite{Haack:2008xx}}} = -\omega_\text{\tiny{\cite{Rangamani:2009xk}}} 
= \omega_\text{\tiny{\cite{Baier:2007ix}}}=\omega_\text{\tiny{\cite{York:2008rr}}} \,, \\
& \sigma = \sigma_\text{\tiny{\cite{Bhattacharyya:2008jc}}}  =  \sigma_\text{\tiny{\cite{Bhattacharyya:2008mz}}} =  
\frac{1}{2}\, \sigma_\text{\tiny{\cite{Haack:2008xx}}} =
\sigma_\text{\tiny{\cite{Rangamani:2009xk}}}  =
 \frac{1}{2}\, \sigma_\text{\tiny{\cite{Baier:2007ix}}}= \frac{1}{2}\, \sigma_\text{\tiny{\cite{York:2008rr}}} \,, \\
& C = C_\text{\tiny{\cite{Bhattacharyya:2008mz}}} =  C_\text{\tiny{\cite{Rangamani:2009xk}}} =
 -\frac{1}{d-2}\, C_\text{\tiny{\cite{Baier:2007ix}}}= -\frac{1}{d-2}\, C_\text{\tiny{\cite{York:2008rr}}}
\,,
\end{split}
\label{eq:conventions1}
\end{equation}
with the un-subscripted symbols corresponding to the ones used in this text and $C$ is the tensor structure governing the  curvature coupling of the fluid (given by Weyl tensor term $C^{\mu\rho\nu\sigma}\,u_\rho\,u_\sigma$ in our conventions).
We have also taken the liberty to correct the sign in the definition of \cite{Haack:2008xx} based on a private correspondence with the authors (their Eq.~(7) should have an extra sign in front).
In addition to further complicate signs, we have $\lambda_2$ multiplying different contracting of $\sigma^{\mu\nu}$ and $\omega^{\mu\nu}$ giving an additional sign to keep track of:
\begin{equation}
\begin{split}
\text{coefficient} \; \lambda_2: \qquad &(\sigma_{\alpha}^{\ <\mu}  \,\omega^{\alpha\nu>}) \\
&
 (\sigma_{\alpha}^{\ <\mu}  \,\omega^{\nu>\alpha})_\text{\tiny{\cite{Bhattacharyya:2008jc}}} 
 \,,\qquad (\sigma_{\alpha}^{\ <\mu}  \,\omega^{\nu>\alpha})_\text{\tiny{\cite{Bhattacharyya:2008mz}}} 
\,,\qquad (\sigma_{\alpha}^{\ <\mu} \,\omega^{\alpha\nu>})_\text{\tiny{\cite{Haack:2008xx}}}  \\
&
(\sigma_{\alpha}^{\ <\mu}  \,\omega^{\nu>\alpha})_\text{\tiny{\cite{Baier:2007ix}}} 
\,,\qquad  (\sigma_{\alpha}^{\ <\mu}  \,\omega^{\nu>\alpha})_\text{\tiny{\cite{York:2008rr}}} 
\,,\qquad  (\sigma_{\alpha}^{\ <\mu}  \,\omega^{\nu>\alpha})_\text{\tiny{\cite{Rangamani:2009xk}}}  
\end{split}
\label{eq:conventions2}
\end{equation}	
}
\begin{equation}\label{eq:MSAdiab}
\begin{split}
\tau_1 &= \frac{2\eta^2}{\epsilon+p} \times 5.9\ \text{to}\ 5.0 \;\text{ (varies with coupling)}\ :\ 6.10517 \text{ in $\phi^4$ theory}\,,\\
\lambda_1 &= \frac{4\eta^2}{\epsilon+p} \times  5.2\ \text{to}\ 4.1 \;\text{ (varies with coupling)}\ :\ 6.13264 \text{ in $\phi^4$ theory}\,,\\
\kappa &=0 \,, \qquad  \lambda_2 = 2\,\tau\,.
\end{split}
\end{equation}
These results are valid for Debye screening lengths of the order of temperature (see \cite{York:2008rr} for more detailed plots
of these transport coefficients as  functions of coupling and the approximations involved). 

We will draw the readers attention to a relation obtained by combining the last two transport results in \eqref{eq:MSAdiab}
\begin{equation}
\lambda_2 \,\textcolor{red}{-}\, 2(\tau-\kappa) = 0\,.
\label{eq:kinl2}
\end{equation}	
The relation $\lambda_2 \textcolor{red}{-} 2\,\tau = 0$ is a universal prediction of kinetic theory which follows naturally from the Boltzmann equation \cite{York:2008rr} and is consistent with earlier derivations  \cite{Baier:2007ix}. The fact that \cite{Baier:2007ix}  could ascertain $\lambda_2$ without knowing the collision kernel in the Boltzmann equation tells us that this is indeed a non-dissipative part of transport. Along with the  fact that $\kappa=0$ at this order in coupling leads to the relation we have quoted above. We have however chosen to highlight the color of the sign, since it differs from the holographic result in \eqref{eq:weylrelns}.
Indeed had the sign been consistent with the holographic result, we would have concluded that even in kinetic theory $\lambda_2$ would have had a value determined by a Class L  Landau-Ginzburg free energy. It might be useful to cross-check this result independently (despite two independent confirmations above) to demonstrate  that $\lambda_2$ is not obtainable from an effective action in kinetic theory.\footnote{ We thank Andrei Starinets for useful discussions on this point.}

We see that this combination chosen by kinetic theory is exactly  the combination that we have identified as the Class B transport coefficient (and it is also one of the combinations which are zero in the two derivative gravity). We can thus state this universal result from kinetic theory as the statement that the Class B term is absent in the constitutive relation
derived from kinetic theory. We will take this as an evidence that our eightfold classification gives a natural framework from which the 
kinetic theory results could be understood. It would be an interesting exercise to try to see whether one could simplify these kinetic theory computations using various techniques introduced in this paper.

\subsection{Eightfold classification for various fluid systems}
\label{sec:8classvar}

In the preceding subsections we have seen evidence for the eightfold classification of transport in various physical systems and we have also outlined how to transcribe the eightfold path in certain examples. We now give a comprehensive summary of the results in a tabular form for future reference.  Tables \ref{tab:CountingEven} and \ref{tab:CountingOdd} provide a classification of the total number of vector and tensor structures that give constitutive relations allowed by symmetry in various fluid systems up to second order in the derivative expansion.
 
The terms listed for Classes $\PS$ and $\LS$ are scalars that can be used in a Lagrangian to generate independent constitutive relations. Similarly, Classes $\PV$ and $\GV$ provide partition function vectors, i.e., free energy currents.  Classes B and D are non-Lagrangian, so we give directly the expressions for stress tensors (and charge currents) that can be generated using the rules given in \S\ref{sec:classB} and \S\ref{sec:classD}, respectively. Similarly, Class A terms are given directly as constitutive relations of stress tensor and charge current. Further taking into account a number $\PF$ of terms that are disallowed by the existence of an equilibrium configuration (or equivalently by the second law of thermodynamics), the terms listed in each row of the tables exhaust the number of independent transport data.\footnote{ We refrain from listing Class C constitutive relations in the Tables as they are dimension specific and depend on the topology of the background ${\cal M}$ on which the fluid lives.}

\afterpage{\clearpage
\begin{sidewaystable}[H]
\small
\centering
\begin{tabular}{||c  || c | c | c | c |H  H  c | c ||}
\hline\hline
	{\shadeB Case}      & {\shadeR $\PF$} & {\shadeR $\LS$} & {\shadeR $\PS=L/\LS$} &   {\shadeR $\GV$} & {\shadeR $\PV$} & {\shadeR A} & {\shadeR B}   & {\shadeR D}   \\
\hline
{\shadeB Charged}         & ${\bf 2}$  &          &      &       &  &  &  & ${\bf 3} \prn{{\bf 2}\ \text{in}\ d=2}$  \\
 {\shadeB Even  parity}                   &           &     &   &      &  &  &
                                             &  $ \prn{\sigma^{\mu\nu},0} \text{in}\ d\geq3$, \\
{\shadeB $1^{\rm st}$ order}        &             &      &  &      &  &         & & $\prn{\Theta P^{\mu\nu},0}, \prn{0,\cv_\mu}$ \\
\hline
{\shadeB Charged}         & ${\bf 17}$   & ${\bf 5}$         & ${\bf 7}$ & ${\bf 2}$       &       &  & ${\bf 11}$  & ${\bf 9}$  \\
 {\shadeB Even  parity}                    &           & $\sigma^2,\Theta^2,$    & $\omega^2,R,\acc^2$ & $\sigma^{\mu\nu}\cv_\nu,$ &      &  & $\prn{\Theta \sigma^{\mu\nu}-\sigma^2 P^{\mu\nu},0},\prn{\sigma^{\alpha<\mu}\omega^{\nu>}{}_\alpha,0},$
                                             &  $ (\Theta \sigma^{\mu\nu}+\sigma^2 P^{\mu\nu},0),$ \\
{\shadeB $2^{\rm nd}$ order}        &             &  $\cv_\alpha E^\alpha,$    & $ \omega_{\alpha\beta}B^{\alpha\beta}, $  & $\Theta \cv^\mu$ &      &  & $\prn{\sigma^{\alpha<\mu}B^{\nu>}{}_\alpha,0},\prn{0,\omega^{\mu\nu}\cv_\nu},
\prn{0,B^{\mu\nu}\cv_\nu},$
                                             &  $(\sigma^{\alpha<\mu}\sigma^{\nu>}{}_\alpha,0), (\Theta^2 P^{\mu\nu},0) ,$ \\
    {\shadeB }    &             &  $\cv_\alpha \acc^\alpha,$   & $E^2,B^2,$ &   &   &  & $\prn{\cv^2 P^{\mu\nu},-\Theta \cv^\mu},\prn{\cv_\alpha E^\alpha P^{\mu\nu},-\Theta E^\mu}$ &
                                              $\prn{\cv^2 P^{\mu\nu},\Theta \cv^\mu},\prn{\cv_\alpha E^\alpha P^{\mu\nu},\Theta E^\mu}$  \\
  {\shadeB }       &             &  $\cv^2$   &  $\acc_\alpha E^\alpha$ &  &    &  & $\prn{\cv_\alpha \acc^\alpha P^{\mu\nu},-\Theta \acc^\mu},\prn{\cv^{<\mu}\cv^{\nu>},-\sigma^{\mu\nu}\cv_\nu},$ &
                                              $\prn{\cv_\alpha \acc^\alpha P^{\mu\nu},\Theta \acc^\mu},\prn{\cv^{<\mu}\cv^{\nu>},\sigma^{\mu\nu}\cv_\nu},$  \\
   {\shadeB }      &             &     &  &    &  &  & $\prn{\cv^{<\mu}\acc^{\nu>},-\sigma^{\mu\nu}\acc_\nu},\prn{\cv^{<\mu}E^{\nu>},-\sigma^{\mu\nu}E_\nu}$ &
                                              $\prn{\cv^{<\mu}\acc^{\nu>},\sigma^{\mu\nu}\acc_\nu},\prn{\cv^{<\mu}E^{\nu>},\sigma^{\mu\nu}E_\nu}$  \\
\hline
 {\shadeB Ditto with}        & ${\bf 5}$  & ${\bf 3}$         & ${\bf 5}$  & ${\bf 1}$    &      &  & ${\bf 6}$  & ${\bf 3}$  \\
 {\shadeB Weyl inv.}                     &           & $\sigma^2,\cv^2,$    & $\omega^2,\RWeyl,$ &
                                  $\sigma^{\mu\nu}\cv_\nu$   &      &  & $\prn{\sigma^{\alpha<\mu}\omega^{\nu>}{}_\alpha,0}, \prn{\sigma^{\alpha<\mu}B^{\nu>}{}_\alpha,0},$ & $\prn{\sigma^{\alpha<\mu}\sigma^{\nu>}{}_\alpha,0},   $\\
    {\shadeB }    &             & $\cv_\alpha E^\alpha$     & $  \omega_{\alpha\beta}B^{\alpha\beta},$  &  &     &  & $\prn{0,\omega^{\mu\nu}\cv_\nu},
\prn{0,B^{\mu\nu}\cv_\nu},$
                                             &  $\prn{\cv^{<\mu}\cv^{\nu>},\sigma^{\mu\nu}\cv_\nu},$ \\
      {\shadeB }   &             &     & $E^2,B^2$ &  &    &  & $\prn{\cv^{<\mu}\cv^{\nu>},-\sigma^{\mu\nu}\cv_\nu},\prn{\cv^{<\mu}E^{\nu>},-\sigma^{\mu\nu}E_\nu}$
                                             & $\prn{\cv^{<\mu}E^{\nu>},\sigma^{\mu\nu}E_\nu}$   \\
\hline
{\shadeB Neutral  }           & ${\bf 5}$     & ${\bf 2}$         & ${\bf 3}$ &      &      &  & ${\bf 2}$     & ${\bf 3}$  \\
 {\shadeB Even parity }                    &           &$\sigma^2,\Theta^2$    & $\omega^2,R,\acc^2$  &  &      &  & $(\Theta \sigma^{\mu\nu}-\sigma^2 P^{\mu\nu},0),$ &
                                              $ (\Theta \sigma^{\mu\nu}+\sigma^2 P^{\mu\nu},0), $ \\
{\shadeB $2^{\rm nd}$ order}         &           &     &  &   &   &  & $(\sigma^{\alpha<\mu}\omega^{\nu>}{}_\alpha,0)$ &
                                               $(\sigma^{\alpha<\mu}\sigma^{\nu>}{}_\alpha,0), (\Theta^2 P^{\mu\nu},0)$ \\
\hline
 {\shadeB Ditto with}    &    & ${\bf 1}$          & ${\bf 2}$   &   &       &  & ${\bf 1}$   & ${\bf 1}$  \\
 {\shadeB Weyl inv.}                                             &           & $\sigma^2$    & $\omega^2,\RWeyl$  &  &    &  & $(\sigma^{\alpha<\mu}\omega^{\nu>}{}_\alpha,0)$ &
                                             $(\sigma^{\alpha<\mu}\sigma^{\nu>}{}_\alpha,0)$ \\
 \hline\hline
\end{tabular}
\caption{{Eightfold way of non-linear transport in parity-even fluid dynamics. The tensor structures in the columns $\LS$, $\PS$, $\GV$ are partition function scalars and vectors, respectively (i.e., they contribute as scalar and vector parts of free energy currents to constitutive relations). The objects in the columns for Classes B and D are constitutive relations $(T^{\mu\nu},J^\mu)$.} }
\label{tab:CountingEven}
\end{sidewaystable}

\begin{sidewaystable}[H]
\small
\centering
\begin{tabular}{||c  || c | c | c | c | c | c H H ||}
\hline\hline
{\shadeB Case}                      & {\shadeR $\PF$} & {\shadeR $\LS$} & {\shadeR $\PS$} & {\shadeR $\PV$} & {\shadeR $A$} & {\shadeR B}  &   {\shadeR $\GV$} & {\shadeR D}   \\
\hline
{\shadeB Charged}  &           &          &       & ${\bf 1}$         &     ${\bf 1}$    &    &  &   \\
 {\shadeB Odd   parity}  &  &                &     & $T^2\, \varepsilon^{\mu\nu}u_\nu $     & $\big(-4\mu^2\, \varepsilon^{(\mu\rho}u_\rho u^{\nu)},\, -2\mu\, \varepsilon^{\mu\rho} u_\rho\big) $           &       &  &   \\
 {\shadeB $d=2$, $0^{\rm th}$ order}  &  &                &            &            &          &           &  &   \\
\hline
{\shadeB Charged}  &      ${\bf 2}$     &         & ${\bf 2}$       &         &         &  ${\bf 2}$   &  &   \\
{\shadeB  Odd   parity}  &  &                &   $\varepsilon^{\alpha\mu\nu}u_\alpha \omega_{\mu\nu}, $         &            &          &  $\prn{\sigma^{<\mu}{}_\alpha\varepsilon^{\alpha\beta \nu>} u_\beta,0}, \prn{0,\varepsilon^{\mu\alpha\beta} u_\alpha \cv_\beta}$        &  &   \\
 {\shadeB $d=3$, $1^{\rm st}$ order}  &  &                & $\varepsilon^{\alpha\mu\nu}u_\alpha B_{\mu\nu}$            &            &          &           &  &   \\
 \hline
{\shadeB Charged}  &        &        &        &      ${\bf 1}$    & ${\bf 1}$        &     &  &   \\
 {\shadeB Odd   parity}  &  &     &            &   $\mu T^2(\varepsilon^{\mu\lambda\alpha\beta} u_\lambda\omega_{\alpha\beta})$         &     $\big(\varepsilon^{(\mu \lambda\rho\sigma} u_\lambda u^{\nu)} [-4\mu^3\omega_{\rho\sigma}- 3\mu^2 B_{\rho\sigma}], $       &      &  &   \\
 {\shadeB $d=4$, $1^{\rm st}$ order}  &  &                 &             &            &    $ \varepsilon^{\mu \lambda\rho\sigma} u_\lambda [-3\mu^3\omega_{\rho\sigma}- 3\mu^2 B_{\rho\sigma}] \big) $ &           &  &   \\ 
 \hline\hline
\end{tabular}
\caption{Eightfold way of non-linear transport in parity-odd fluid dynamics. The tensor structures in the columns $\LS$, $\PS$, $\GV$ are partition function scalars and vectors, respectively (i.e., they contribute as free energy currents to constitutive relations). The objects in the column for Classes A and B are constitutive relations $(T^{\mu\nu},J^\mu)$.}
\label{tab:CountingOdd}
\end{sidewaystable}
\clearpage }

\newpage
\section{Class $\LT$: Eightfold Lagrangian}
\label{sec:classLT}

Our discussion thus far has focused on generating adiabatic constitutive relations and demonstrating how these help us classify hydrodynamic transport at arbitrary orders in the gradient expansion. This discussion is encapsulated in the statement of Theorem \ref{thm:eight}. We have argued in that context for the completeness of our classification based on the structure of the adiabaticity equation.

 We would now like to justify the statement more directly and in the process explain the basic rationale for considering adiabatic transport which has played a starring role.  As a result we now introduce a novel ingredient in our analysis, which involves constructing a master Lagrangian. The adiabaticity equation is not derived from this master equation as a Bianchi identity of the underlying diffeomorphism and gauge symmetries as in the Class L discussion of \S\ref{sec:classL}, but rather follows as the statement of invariance under a new  abelian gauge symmetry. The corresponding gauge field $\AT_\mu$ and the associated gauge group $\UT$ will be motivated below and argued to ensure that the associated Gauss Law  translates directly into  the statement of adiabaticity. The main upshot of this construction is then to provide a constructive proof of  
 Theorem \ref{thm:classLT}.  

 The framework which we christen Class $\LT$ (since it extends Class L to include the non-Lagrangian solutions to adiabaticity equation) involves not only a new symmetry, but also introduces some additional background fields. These are analogous to the Schwinger-Keldysh counterparts of the metric and gauge field sources. The strategy we follow is to guess at a set of fields and invariances that are suggestive from our adiabatic analysis of  Part \ref{part:adiabatic}. This allows us to postulate a master Lagrangian which generates \emph{precisely} the constitutive relations consistent with adiabaticity equation, deriving for us the eightfold classification in the process. Furthermore, we will argue for the existence of an  appropriate variational principle  which yields exactly the hydrodynamic equations of motion.  While satisfactory in terms of helping us complete our taxonomy, we don't provide here a full-fledged argument for why certain sources are doubled, nor do we give the actual relation between this construction and that of the Schwinger-Keldysh analysis of \S\ref{sec:skdouble}.\footnote{ We will however note in the course of the discussion that the construction in Class $\LT$ appears to be consistent with the Schwinger-Keldysh doubling required for the anomalous hydrodynamic transport discussed in \S\ref{sec:skdouble}.}  We will just note here that from our preliminary analysis, the Class $\LT$ framework appears to capture some of the basic features necessary in the analysis of non-equilibrium transport. The lessons that we can glean from fleshing out this statement and the connections with the Schwinger-Keldysh construction will be deferred to a separate future publication \cite{Haehl:2014kq}. 	

\subsection{Introducing $\UT$ invariance}
\label{sec:U1T}

We initially motivated the adiabaticity equation to capture the part of transport where off-shell entropy production was compensated by a flow of energy-momentum and charge. The latter currents are of course conserved as a consequence of diffeomorphism and gauge invariance of the underlying microscopic theory. In Class L we noticed further that working with a thermal density matrix parameterized by $\{\Kbeta^\mu,\LambdaB\}$  allowed us to extract a set of Bianchi identities which together imply the adiabaticity equation as a natural corollary, cf., \S\ref{sec:classL}. We have however seen that Class L Lagrangians are not exhaustive in describing the space of adiabatic transport coefficients.

On the one hand this is to be expected because a genuine treatment of non-equilibrium field theory should require a 
Schwinger-Keldysh doubling of the degrees of freedom. From this viewpoint it is rather surprising that Class L Lagrangians already capture many aspects of adiabatic transport. On the other hand as we argued in \S\ref{sec:skcritical} there is something missing in a simple-minded construction of Schwinger-Keldysh doubled Lagrangian effective field theories. The problem is not so much in simply doubling the degrees of freedom and writing down independent Lagrangians for the left and right degrees of freedom, but rather in constraining the interactions between the two sets. The challenge is to keep the doubled degrees of freedom under control after coupling the two copies via Feynman-Vernon influence functionals. A-priori there would be two independent copies of diffeomorphism and flavour gauge symmetries that act on the system independently on the left and right. One might imagine breaking these down to the corresponding diagonal symmetries upon introduction of the influence functionals. This however does not suffice to forbid terms that allow for violation of the second law of thermodynamics. 

The principle of adiabaticity introduced in \S\ref{sec:adiabat} allows us to focus on the marginal case of zero entropy production.
Based on this we would like to argue that we should understand first a basic principle that guarantees \eqref{eq:Adiabaticity} as a statement of invariance. Since it asserts effectively that the entropy current is conserved on-shell (up to anomalous contributions), it is tempting to posit a gauge invariance whose associated current conservation leads directly to the adiabaticity equation. This we claim that will suffice to impose sufficient conditions on the  Feynman-Vernon terms to ensure consistency with the second law of thermodynamics. More precisely, it ensures that such terms are consistent with the microscopic KMS condition.

Let us then record the ingredients we deem necessary to construct an effective Lagrangian for all the adiabatic constitutive relations. Firstly we have the  low energy fluid degrees of freedom $\{\Kbeta^\mu, \LambdaB\}$ and the background sources 
$\{g_{\mu\nu}, A_\mu\}$. These can be viewed as functionals of the maps from some reference configuration, whence we can directly deal with the Goldstone bosons as described in \S\ref{sec:skdouble}. We will in addition postulate the existence of a second set of sources, which we call  $\{\tildeg_{\mu\nu},\tildeA_\mu\}$. 

The abelian symmetry which we will henceforth refer to as $\UT$ can be viewed as a KMS-gauge symmetry.\footnote{ The nomenclature is suggestive, but as we have explained in the text we will not justify this at present.} This symmetry acts as 
a thermal diffeomorphism or flavour gauge transformation on the sources. As indicated above it corresponds to difference diffeomorphisms/gauge transformations which are aligned with the hydrodynamic fields $\{\Kbeta^\mu,\LambdaB\}$. The KMS-gauge field corresponding to the $\UT$ symmetry will be denoted as $\AT_\mu$.  We have in addition an associated  holonomy field $\LambdaBT$ and a $\UT$ chemical potential $\LambdaBT+\Kbeta^\mu \AT_\mu$. 

The diffeomorphism and flavour transformations on the fields in an obvious manner. On the contrary $\UT$ acts nonlinearly and mixes with flavour and diffeomorphism transformations:
 \begin{itemize}
\item On  all fields, $\UT$ acts as  a longitudinal  diffeomorphism and flavour  gauge transformation along $\{\Kbeta^\mu, \LambdaB\}$.
\item In addition, on $\{\tildeg_{\mu\nu},\tildeA_\mu\}$, there  is a further shift  by $\{\diffB g_{\mu\nu}, \diffB A_\mu\}$.
\item The field $\AT_\mu$ transforms as a connection for $\UT$ and  $\LambdaBT$ acts like a gauge transformation parameter,
viz., $\LambdaBT+\Kbeta^\sigma \AT_\sigma$ is invariant. 
\end{itemize}

It is worth noting that from  a Schwinger-Keldysh point of view, these transformation rules are not the most natural ones.
It would have been more natural to retain the abelian part of the non-diagonal diffeomorphism and flavour gauge symmetries  along $\Bfields$. We anticipate that the difference is due to the fact that the natural basis of sources chosen here is not the canonical Schwinger-Keldysh choice. In fact it seems plausible to conjecture that  
\begin{equation}
\begin{split}
g^\skR_{\mu\nu} & = g_{\mu\nu} \,, \\ 
A^\skR_\mu &= A_\mu \\
g^\skL_{\mu\nu} &=    g_{\mu\nu} -\tildeg_{\mu\nu} - \Kbeta_{\mu}\, \AT_{\nu} -\Kbeta_\nu\, \AT_\mu \,,  \\
A^\skL_\mu &=  A_\mu-\tildeA_\mu - \left(\LambdaB+\Kbeta^\alpha\, A_\alpha\right)\, \AT_\mu
\end{split}
\label{eq:skLTdef}
\end{equation}
as the appropriate identifications for the right (R) and left (L) sources, respectively. 
We will however not flesh this out in great detail, since it (a) appears much cleaner in the formalism we introduce to 
write down $\UT$ invariant Lagrangians and (b) the connections with the Schwinger-Keldysh construction are being deferred to a separate publication \cite{Haehl:2014kq} anyway.  For the present the reader may therefore take our prescription merely as a technical tool to prove the completeness of our eightfold classification without worrying about the profound physical consequences.

\subsection{The fields and their transformation properties}
\label{sec:fieldsLT}

Let us start by writing down the extended set of fields and transformation properties based on the above discussion. We have  the following fields  which form the building blocks for the master Lagrangian: 
\begin{enumerate}
\item the sources $\{g_{\mu\nu},A_\mu\}$,
\item the fluid fields $\{\Kbeta^\mu, \LambdaB\}$,
\item partners for the sources $\{\tildeg_{\mu\nu},\tildeA_\mu\}$ which are a symmetric tensor and a vector transforming in the adjoint representation of the flavour symmetry,
\item an additional $\UT$ gauge field $\AT_\mu$ and its holonomy field $\LambdaBT$. 
\end{enumerate}
When necessary we will collectively refer to these fields as $\hfieldsT$.  The symmetries that any effective Lagrangian needs to preserve are diagonal diffeomorphisms/flavour gauge transformations (acting equally on sources and their partners) and in addition the abelian $\UT$ thermal shift symmetry (which we claim enforces consistency of Feynman-Vernon terms).

Let us now record the transformation rules for the fields $\hfieldsT$. We denote the transformation parameters of diffeomorphism, flavour, and $\UT$ transformations by $\{\xi,\Lambda,\LambdaT\}$ respectively. In terms of these independent parameters, $\UT$ has a twisted action on the various fields. 
This is because fields transform non-linearly under it and part of the $\UT$ transformation involves diffeomorphisms and flavour gauge transformations. 
We will deal with the non-trivial mixing between diffeomorphism and flavour transformations on the one hand and $\UT$ on the other hand using the following trick: instead of using the original transformation parameters, we will move to a new basis of transformation parameters $\{\txi^\mu, \tLam, \LambdaTb \}$ which generate combinations of the original transformations which do not mix with each other. The original transformation parameters are related to these via
\begin{subequations}
\begin{align}
\xi^\mu &\equiv \txi^\mu -(\LambdaTb+\txi^\sigma \,\AT_\sigma ) \, \Kbeta^\mu\,,
& \quad  
\txi^\mu  &\equiv \xi^\mu  +(\LambdaT+\xi^\sigma\,\AT_\sigma ) \, \Kbeta^\mu 
 \,,\\
\Lambda  &\equiv  \tLam - (\LambdaTb+\txi^\sigma \,\AT_\sigma ) \,\LambdaB \,,
&\quad
\tLam  &\equiv  \Lambda  + (\LambdaT+\xi^\sigma\,\AT_\sigma )\,\LambdaB\,,
\\
\LambdaT &\equiv  \LambdaTb + (\LambdaTb+\txi^\sigma \,\AT_\sigma ) \, \Kbeta^\nu\,\AT_\nu \,,
&\quad 
\LambdaTb &\equiv\LambdaT - (\LambdaT+\xi^\sigma \,\AT_\sigma )\, \Kbeta^\nu \,\AT_\nu\,.
\end{align}
\end{subequations}
We have given the translation between the two sets of gauge transformation parameters 
$\{\xi^\mu,\Lambda, \LambdaT\}$ and $\{\txi^\mu,\tLam, \LambdaTb\}$
in both forward and reverse directions to facilitate translation between them in the future.
A useful relation in converting between these parameters is
\[ \LambdaT+\xi^\sigma \,\AT_\sigma = \LambdaTb+\txi^\sigma\, \AT_\sigma \,. \]

\paragraph{The transformation rules:}
Armed with this we are now in a position to write down the explicit transformations of various fields which takes a simple form in terms of the untwisted transformation parameters $\{\txi^\mu, \tLam, \LambdaTb \}$:\footnote{ We denote the derivative operator which covariantly transforms under diffeomorphisms, flavour gauge, and $\UT$ transformations by $D_\mu$ in what follows. It is defined by appropriately extending \eqref{eq:CovDer} to incorporate $\UT$ transformations as well.}
\begin{equation}\label{eq:TactgA}
\begin{split}
\diffF  g_{\mu\nu} &\equiv 
	\lieD_{\txi} g_{\mu\nu} 
	= D_\mu \txi_\nu + D_\nu  \txi_\mu  \,,\\
\diffF  A_\mu &\equiv 
	\lieD_{\txi} A_\mu + [A_\mu,\tLam ]+\partial_\mu \tLam
 	=D_\mu\prn{\tLam +\txi^\nu  A_\nu  }+\txi^\nu F_{\nu\mu}\,,\\
\diffF  \Kbeta^\mu &\equiv 
	\lieD_{\txi} \Kbeta^\mu
	=\txi^\nu D_\nu \Kbeta^\mu  - \Kbeta^\nu D_\nu\txi^\mu \,,\\
\diffF  \LambdaB + A_\nu \,\diffF  \Kbeta^\nu
&\equiv
	 \txi^\mu  \,\diffB A_\mu-\Kbeta^\mu \,D_\mu\prn{\tLam + \txi^\nu \, A_\nu}
	+ [\LambdaB+\Kbeta^\lambda A_\lambda,\tLam + \txi^\nu  A_\nu]\,.
\end{split}
\end{equation}

In terms of the original transformation parameters $\{\xi^\mu, \Lambda, \LambdaT \}$, these transformations would mix diffeomorphism and flavour transformations with $\UT$. The advantage gained from working with $\{\txi,\tLam,\LambdaTb\}$ is an untwisting of $\UT$ such that $\{g_{\mu\nu},A_\mu,\Kbeta^\mu,\LambdaB\}$ are blind to it. The partner sources $\{\tildeg_{\mu\nu}, \tildeA_\mu\}$ transform similarly, but in addition pick up an inhomogeneous piece which contains a source Lie-dragged along $\Bfields$ under the $\UT$ action: 
\begin{equation}
\begin{split}
\label{eq:TactgAbar}
\diffF  \tildeg_{\mu\nu} &
	\equiv \lieD_{\txi} \,\tildeg_{\mu\nu} +\LambdaTb\,  \diffB g_{\mu\nu}  
 \\ &
	= 2\,\tildeg_{\sigma(\mu}D_{\nu)} \txi^\sigma  
	+ \txi^\sigma \prn{ D_\sigma \tildeg_{\mu\nu} - \AT_\sigma \, \diffB g_{\mu\nu} } 
	+  \prn{\LambdaTb+\txi^\sigma \,\AT_\sigma } \diffB g_{\mu\nu} 
 \\
\diffF  \tildeA_\mu &
	\equiv  \lieD_{\txi} \,\tildeA_\mu + [\tildeA_\mu,\tLam]  + \LambdaTb \,\diffB A_\mu   
\end{split}
\end{equation}
Finally, the transformation of the $\UT$ connection and its holonomy are given by
\begin{equation}\label{eq:TactAT}
\begin{split}
\diffF \AT_\mu &\equiv \lieD_{\txi} \,\AT_\mu +\partial_\mu \LambdaTb
 =D_\mu\prn{\LambdaTb +\txi^\nu \,\AT _\nu  }+\txi^\nu\, \FT_{\nu\mu} \,,\\
 \diffF  \LambdaBT + \AT_\nu \, \diffF  \Kbeta^\nu
&\equiv \txi^\mu  \,\diffB \AT_\mu-\Kbeta^\mu \,D_\mu\prn{\LambdaTb + \txi^\nu  \,\AT_\nu} . 
\end{split}
\end{equation}
The first line is just the usual transformation rule for the gauge field of an abelian symmetry; the second line is such that 
$\LambdaBT+\Kbeta^\nu \,\AT_\nu$ is invariant. In fact, as we will see the present formalism is a natural extension of the Class $\PV$ formalism of \S\ref{sec:JLYtranscend}. To obtain consistency between the two formalisms (see \S\ref{sec:LTanom}), we are led to a natural choice for fixing the above invariant combination. We will choose
\begin{equation}
\LambdaBT+\Kbeta^\sigma \,\AT_\sigma = 1 \,.
\end{equation}

Given that the transformations rules have thus far been ``pulled out of a hat'', we demonstrate that they are consistent in
Appendix \ref{sec:appendixT}. In particular, we  will will check that they form an algebra such that the usual Wess-Zumino consistency conditions are satisfied. 
This allows us to proceed with confidence about these transformations.

\paragraph{Difference source combinations:}
While this completes the basic transformation rules from which all the subsequent expressions can be derived, it is convenient to 
consider a linear combination of the sources $\{g_{\mu\nu}, A_\mu\}$ and their partners $\{\tildeg_{\mu\nu}, \tildeA_\mu\}$
which is simplifies the expressions somewhat. 

To appreciate this let us define the shifted partner sources
\begin{equation}
\begin{split}
 \tgdiff_{\mu\nu} &\equiv  g_{\mu\nu} -\tildeg_{\mu\nu}  \\
\tAdiff_\mu &\equiv A_\mu-\tildeA_\mu 
\end{split}
\label{eq:diffs}
\end{equation}	
as well as the associated covariant derivative $\tDdiff$ and field strength $\tFdiff$ respectively.\footnote{ Some useful identities for converting between covariant
derivatives of two different metrics are the following: 
\begin{equation}
\begin{split}
D'_\mu ( g'_{\nu\sigma} V^\sigma ) &= D_\mu ( g'_{\nu\sigma} V^\sigma ) + \half V^\sigma \prn{D_\sigma g'_{\mu\nu}- D_\mu g'_{\nu\sigma}- D_\nu g'_{\mu\sigma} } \\
D'_\nu ( g'_{\sigma\mu} T^{\mu\nu} ) &= \frac{\sqrt{-g}}{\sqrt{-g'}} D_\nu ( \frac{\sqrt{-g'}}{\sqrt{-g}} g'_{\sigma\mu} T^{\mu\nu} ) - \half T^{\mu\nu} D_\sigma g'_{\mu\nu}
\end{split}
\end{equation}
where $ V^\sigma $ and $T^{\mu\nu}$ are some general vector and a symmetric tensor respectively.}
We define them as 
\begin{equation}
\begin{split}
\tDdiff_\alpha X^{\mu\cdots\nu}{}_{\rho\cdots\sigma} &= \nabla'_\alpha X^{\mu\cdots\nu}{}_{\rho\cdots\sigma} + 
[\tAdiff_\alpha, \, X^{\mu\cdots\nu}{}_{\rho\cdots\sigma}] \,, \\
\tFdiff_{\mu\nu} &= \nabla'_\mu \tAdiff_\nu -  \nabla'_\mu \tAdiff_\nu + [\tAdiff_\mu,\tAdiff_\nu]
\end{split}
\label{eq:CovDerdiff}
\end{equation}
The primed covariant derivative acts on tensors as in \eqref{eq:CovDer2}. 

Then it a simple exercise  to see that we can rewrite \eqref{eq:TactgAbar} as 
\begin{equation}
\begin{split}
\label{eq:TactgAbard}
 \diffF  \tildeg_{\mu\nu} &  
	= \diffF  g_{\mu\nu}  +  (\LambdaTb+\txi^\sigma \,\AT_\sigma )\ \diffB g_{\mu\nu} 
	-\bigbr{ 2\, \tDdiff_{(\mu} \prn{ \tgdiff_{\nu) \rho} \,\txi^\rho }
	+ \txi^\sigma \AT_\sigma  \diffB g_{\mu\nu}   }\,,
 \\
 \diffF  \tildeA_\mu &
	= \diffF A_\mu +  (\LambdaTb+\txi^\sigma \,\AT_\sigma )\ \diffB A_\mu  
	-  \bigbr{ \tDdiff_\mu \prn{\tLam+\txi^\sigma \, \tAdiff_\sigma }
	+ \txi^\sigma \, \tFdiff_{\sigma\mu} +   \txi^\sigma\, \AT_\sigma  \;\diffB A_\mu  }\,.
\end{split}
\end{equation}
In fact, we can more conveniently merge \eqref{eq:TactgA} and \eqref{eq:TactgAbard} into a transformation rule for the 
partner fields themselves. To wit, 
\begin{equation}\label{eq:Tactgmgbar}
\begin{split}
\diffF  \tgdiff_{\mu\nu}
&=
	\tDdiff_\mu \prn{\tgdiff_{\nu\rho} \, \txi^\rho}+ \tDdiff_\nu \prn{\tgdiff_{\mu\rho}\,\txi^\rho}
	+ \txi^\sigma \AT_\sigma  \diffB g_{\mu\nu}  
	-  (\LambdaTb+\txi^\sigma \AT_\sigma )\ \diffB g_{\mu\nu} \,,\\
\diffF  \tAdiff_\mu 
&= 
	\tDdiff_\mu  \prn{\tLam+\txi^\sigma \,\tAdiff_\sigma }
	+ \txi^\sigma \,\tFdiff_{\sigma\mu} +   \txi^\sigma\, \AT_\sigma  \,\diffB A_\mu 
	 -  (\LambdaTb+\txi^\sigma \, \AT_\sigma )\ \diffB A_\mu  \,.\\
\end{split}
\end{equation}

\paragraph{Schwinger-Keldysh inspired combinations:}
Above we have chosen to take the partner sources $\{\tgdiff_{\mu\nu},\tAdiff_\mu\}$ without any potential contamination from $\AT_\mu$.  However, attempts to reconcile the construction here with the Schwinger-Keldysh picture developed for Class A suggests that the combination that may be relevant is instead given as in  \eqref{eq:skLTdef}. Taking this seriously let us consider the twisted partner sources \eqref{eq:skLTdef} with suggestive names inspired by Schwinger-Keldysh construction. 
We can then rewrite \eqref{eq:TactgAbar} as 
\begin{equation}
\begin{split}
\label{eq:TactgAbardSK}
 \diffF  \tildeg_{\mu\nu} &  
	= \diffF  g_{\mu\nu}  +  (\LambdaTb+\txi^\sigma \,\AT_\sigma )\ \diffB g_{\mu\nu} 
	-\;\bigbr{ 2\, \bDdiff_{(\mu} \prn{ \bgdiff_{\nu) \rho} \,\txi^\rho }
	+ \txi^\sigma \AT_\sigma  \,\diffB g_{\mu\nu} 
   }\,,
 \\
 \diffF  \tildeA_\mu &  
	= \diffF A_\mu +  (\LambdaTb+\txi^\sigma \,\AT_\sigma )\ \diffB A_\mu  
	-  \biggl\{ \bDdiff_\mu \prn{\tLam+\txi^\sigma \, \bAdiff_\sigma }
	+ \txi^\sigma \, \bFdiff_{\sigma\mu} +   \txi^\sigma\, \AT_\sigma  \;\diffB A_\mu 
	\biggr\} \,.
\end{split}
\end{equation}
and merge \eqref{eq:TactgA} and \eqref{eq:TactgAbardSK} into 
\begin{align}\label{eq:TactgmgbarSK}
\diffF  \bgdiff_{\mu\nu}
&=
	2\,\bDdiff_{(\mu }\prn{\bgdiff_{\nu)\rho} \, \txi^\rho}
	+ \txi^\sigma \AT_\sigma\,  \diffB g_{\mu\nu}^\skR  -  2\, \lieD_{\txi} \prn{\Kbeta_{(\mu} \,\AT_{\nu)} } 
	-  (\LambdaTb+\txi^\sigma \AT_\sigma )\ \diffB g_{\mu\nu}^\skR \,, \notag \\
\diffF  \bAdiff_\mu 
&= 	
	\bDdiff_\mu \prn{\tLam+\txi^\sigma \, \bAdiff_\sigma }
	+ \txi^\sigma \, \bFdiff_{\sigma\mu} +   \txi^\sigma\, \AT_\sigma  \;\diffB A_\mu^\skR 
	-\lieD_{\txi} \prn{(\LambdaB + \Kbeta^\sigma \, A_\sigma^\skR)\, \AT_\mu} 
 \notag \\& \qquad	
	- [\LambdaB + \Kbeta^\sigma  A_\sigma^\skR, \tLam] \,\AT_\mu
	 -  (\LambdaTb+\txi^\sigma \, \AT_\sigma )\ \diffB A_\mu^\skR  \,.
\end{align}

In much of our discussion we will only use the difference sources $\{\tgdiff_{\mu\nu}, \tAdiff_\mu\}$ and only briefly in the discussion involving anomalous hydrodynamics revert to the Schwinger-Keldysh inspired $\{\bgdiff_{\mu\nu}, \bAdiff_\mu\}$.
The translation between the two sets of languages being straightforward (the basic formulae are all given above), it should be simple to translate statements between the two if necessary.

\subsection{Bianchi identities in Class $\LT$}
\label{sec:LTBianchi}

We can now use the various fields introduced in the previous subsection to construct Lagrangians 
$\LagT\brk{\hfieldsT}$ invariant under diffeomorphism, flavour, and $\UT$ transformations. This invariance yields Bianchi identities which we will now show imply the adiabaticity equation in the hydrodynamic limit (i.e., to linear order in the Schwinger-Keldysh difference fields). Let us parameterize the variation of $\LagT$ by 
\begin{equation}\label{eq:LagTVar}
\begin{split}
\frac{1}{\sqrt{-g}}&\delta\prn{\sqrt{-g}\ \LagT} -\nabla_\mu (\PSymplPot{}^\smallT)^\mu
\\ &= 
	\half \; \TL^{\mu\nu}\;\delta g_{\mu\nu} + \JL^\mu \cdot \delta A_\mu 
	+ T \,\aheat_\sigma \;\delta \Kbeta^\sigma
	+ T\, \acharge \cdot \prn{\delta\LambdaB+ A_\sigma \,\delta \Kbeta^\sigma} 
\\ &  \quad
	+\;
	\half \; \TLc^{\mu\nu}\;\delta \tildeg_{\mu\nu} + \JLc^\mu \cdot \delta \tildeA_\mu 
	+ \JT^\sigma \;\delta\AT_\sigma+ T\, \,\achargeT  \prn{\delta\LambdaBT
	+ \AT_\sigma \,\delta \Kbeta^\sigma} \,.\\
\end{split}
\end{equation}
The subscript $\text{L}$ is supposed to indicate that these constitutive relations can be obtained from Class L as discussed in  \S\ref{sec:classL}. Similarly, the subscript $\text{L}^c$ suggests that these will be all remaining adiabatic constitutive relations not obtainable from Class L. 

The variation \eqref{eq:LagTVar} defines the constitutive relations obtained from $\LagT$. We can now explicitly perform the diffeomorphism, flavour gauge, and $\UT$ transformations using the explicit variations given in \S\ref{sec:fieldsLT}.  We  simply replace $\delta$ in \eqref{eq:LagTVar} by $\diffF$ as defined by \eqref{eq:TactgA}-\eqref{eq:TactAT} and perform some necessary integration by parts to isolate the coefficients of the transformation parameters $\{\txi^\mu, \tLam, \LambdaTb\}$. As this exercise is a straightforward generalization of the analysis in Class L we simply quote the final answers for the Bianchi identities (some useful intermediate steps are given in Appendix \ref{sec:appendixT}). 

\begin{itemize}
\item The diffeomorphism Bianchi identity is
\begin{equation}\label{eq:TBianchiLT}
\begin{split}
&D_\mu (\TLLc)^\mu_{\ \sigma}
 -\JLLc^\nu \cdot F_{\sigma\nu}
- \JT^\nu \cdot \FT_{\sigma\nu}  
\\ & \qquad =
	\frac{1}{\sqrt{-g}} \diffB \prn{\sqrt{-g}\ T\ \aheat_\sigma} + T\,\acharge\cdot \diffB A_\sigma 
	+ T\,\achargeT\cdot \diffB \AT_\sigma 
\\
&\qquad  \qquad 
	+\; D_\nu \prn{\tgdiff_{\sigma\mu}\, \TLc^{\mu\nu}} 
	 - \half \, \TLc^{\mu\nu} D_\sigma \tgdiff_{\mu\nu}-  \JLc^\nu\cdot \tFdiff_{\sigma\nu}  
\\&\qquad   \qquad  
-\AT_\sigma \prn{   \half \TLc^{\mu\nu}\, \diffB g_{\mu\nu}  +\JLc^\mu\cdot  \diffB A_\mu }
-\tildeA_\sigma \cdot  \bigg(D_\mu \JLc^\mu  - [\tildeA_\mu, \JLc^\mu] \bigg) \,.
\end{split}
\end{equation}
\item 
The flavour Bianchi identity is given by
\begin{equation}\label{eq:JBianchiLT}
\begin{split}
D_\mu  \JLLc^\mu &= 
	 \frac{1}{\sqrt{-g}} \diffB \prn{\sqrt{-g}\ T\ \acharge}  +  
	\bigg(D_\mu \JLc^\mu  - [\tildeA_\mu, \JLc^\mu] \bigg) \,.
\end{split}
\end{equation}
\item Finally, the $\UT$ Bianchi identity reads (after setting $\LambdaBT+\Kbeta^\sigma \,\AT_\sigma = 1$):
\begin{align}\label{eq:U1TBianchi}
D_\mu \NT^\mu&=\half \TL^{\mu\nu} \; \diffB g_{\mu\nu}  +\JL^\mu\cdot  \diffB A_\mu
+ \half \, \TLc^{\mu\nu} \; \diffB \tildeg_{\mu\nu}  +\JLc^\mu\cdot  \diffB \tildeA_\mu 
+ \JT^\mu \; \diffB \AT_\mu 
\,.
\end{align}
%
\end{itemize}

We have skipped several steps in the derivation of \eqref{eq:U1TBianchi}, which unlike the diffeomorphism and flavour Bianchi identities does require isolating the $\UT$ transformation explicitly. The intermediate steps can be found in 
Appendix \ref{sec:appendixT}.  The essential steps involve reverting back to  the 
 original  (twisted) transformation parameters  $\{\xi^\mu, \Lambda, \LambdaT \}$. It should hopefully be clear that this can be
 achieved without modifying the diffeomorphism and flavour Bianchi identities. 
 After performing the required shift and defining 
\begin{align}
\NT^\mu \equiv - \frac{\GT^\mu}{T} 
&\equiv
	 \JT^\mu +\Kbeta_\nu \TLLc^{\mu\nu}  + \prn{\LambdaB +\Kbeta^\nu  A_\nu } \cdot \JLLc^\mu  
	 - \bigbr{\aheat_\sigma\, \Kbeta^\sigma+  \acharge \cdot 
	 \prn{\LambdaB + \Kbeta^\nu  A_\nu} + \,\achargeT} \, u^\mu 
 \notag\\ &\qquad   \qquad 
	 - \Kbeta^\rho \; \tgdiff_{\nu\rho}  \; \TLc^{\mu\nu}
	 - (\LambdaB +\Kbeta^\nu \,\tAdiff_\nu )  \cdot  \JLc^\mu\ \,.
\label{eq:utNTdef}
\end{align}
we arrive at \eqref{eq:U1TBianchi}. 

We note in passing that we can further simplify some expressions, by noting a particularly interesting combination of the Bianchi identities that follows straightforwardly in the derivation (see Appendix \ref{sec:appendixT}). One finds that the grand canonical adiabaticity equation for $\{\TLc^{\mu\nu}, \JLc^\mu \}$ holds identically, i.e., 
\begin{align}
& D_\mu \JT^\mu  =\half \, \TLc^{\mu\nu}\; \diffB g_{\mu\nu} + \JLc^\mu \cdot \diffB A_\mu +  
\frac{1}{\sqrt{-g}} \diffB \prn{\sqrt{-g}\ T\ \,\achargeT}\,,
\label{eq:JTBianchiLT} \\
&\qquad \Longrightarrow \qquad
D_\mu (\JT^\mu-\,\achargeT u^\mu)  =\half \, \TLc^{\mu\nu} \; \diffB g_{\mu\nu} 
+\JLc^\mu \cdot \diffB A_\mu\,.
\label{eq:U1TBianchiBar}
\end{align}
This identity can be useful when we have anomalous terms, as it helps at various stages of the derivation to keep track of the origins of various contributions.

\subsection{The adiabatic Lagrangian $\LagT$}
\label{sec:LagTdefine}

Given the Bianchi identities for the various symmetries in Class $\LT$ we can now examine the implications for the adiabaticity equation. In particular, we shall prove that a given constitutive relation furnishes a solution to the adiabaticity equation \eqref{eq:Adiabaticity} if and only if it can be derived from a master Lagrangian
 $\LagT$ that preserves diffeomorphism, flavour, and $\UT$ invariance. 

Firstly, we should make contact with the adiabaticity equation. Unlike in Class L where we gave a prescription for the entropy density, we have not yet done so for $\LagT\brk{\hfieldsT}$. Intuitively, we want to identify the $\UT$ invariance as being responsible for adiabaticity. However, it is clear that $\JT^\mu$ should play some role in its definition since $\UT$ symmetry was introduced to ensure adiabaticity. In fact,  the redefined counterpart $\NT^\mu$ is closely related to the grand canonical free energy current.  

In fact we can make this a bit more precise by viewing the Bianchi identities inspired by our discussion about the Schwinger-Keldysh construction in \S\ref{sec:skdouble}. Consider using the reference configuration in the Schwinger-Keldysh construction to pull-back the L fields onto the R-manifold. Then we have all the 
background sources living on the right manifold. Furthermore, working with the common hydrodynamic fields, we can attempt to derive a set of Bianchi-identities under the right diffeomorphism and flavour symmetries. This leads to an analog of \eqref{eq:SKrefBianchiVar}, with the sole difference being that all the fields are on the R-manifold and we are working with the physical degrees of freedom. This set of identities in fact can be shown to be closely related to Eqs.~\eqref{eq:TBianchiLT}-\eqref{eq:JBianchiLT}. To see this, note that the defining variation \eqref{eq:LagTVar} can be rewritten as
\begin{equation}\label{eq:LagTVar2}
\begin{split}
\frac{1}{\sqrt{-g}}&\delta\prn{\sqrt{-g}\ \LagT} -\nabla_\mu (\PSymplPot{}^\smallT)^\mu
\\ &= 
	\half \; \TLLc^{\mu\nu}\;\delta g_{\mu\nu} -\;
	\half \; \TLc^{\mu\nu}\;\delta \tgdiff_{\mu\nu}
	+ \JLLc^\mu \cdot \delta A_\mu 
	- \JLc^\mu \cdot \delta \tAdiff_\mu  \\ &  \quad
	+ T \,\aheat_\sigma \;\delta \Kbeta^\sigma
	+ T\, \acharge \cdot \prn{\delta\LambdaB+ A_\sigma \,\delta \Kbeta^\sigma} 
	+ \JT^\sigma \;\delta\AT_\sigma+ T\, \,\achargeT  \prn{\delta\LambdaBT
	+ \AT_\sigma \,\delta \Kbeta^\sigma} \,.\\
\end{split}
\end{equation}
Apart from the last two terms (which are tied to the presence of the $\UT$ symmetry principle), this expression has the same structure as the Schwinger-Keldysh variation \eqref{eq:SKrefVar} (after pulling it to the R-manifold). Aided by this observation, we tentatively forward the following hypothesis: in the absence of anomalies the Schwinger-Keldysh sources can be identified with $\{g^{\skR}_{\mu\nu}, A^{\skR}_\mu\} \mapsto \{ g_{\mu\nu}, A_\mu\} $ and 
$\{g^{\skL}_{\mu\nu}, A^{\skL}_\mu\} \mapsto \{ \tgdiff_{\mu\nu}, \tAdiff_\mu\} $ respectively.\footnote{ In the present context we view this identification as heuristic. Studying anomalies as in \S\ref{sec:LTanom} shows that the identification of Schwinger-Keldysh fields should actually be twisted to involve the field $\AT_\mu$ as in \eqref{eq:skLTdef}.} Accordingly, \eqref{eq:LagTVar2} suggests that the currents map as $\{T_\skR^{\mu\nu},J_\skR^\mu\} \mapsto \{\TLLc^{\mu\nu},\JLLc^\mu\}$ and $\{T_\skL^{\mu\nu},J_\skL^\mu\} \mapsto \{\TLc^{\mu\nu},\JLc^\mu\}$. The constrained (Schwinger-Keldysh) variational principle \eqref{eq:SKconstrainedVar}, would then inspire us to put forward a constrained variational principle in Class $\LT$ which takes the following form: under diffeomorphisms and flavour gauge transformations only $\{\Kbeta^\mu,\LambdaB,\tgdiff_{\mu\nu},\tAdiff_\mu\}$ transform while the R-sources $\{g_{\mu\nu},A_\mu\}$ are held fixed. Modulo a sensible extension to the new fields $\{\AT_\mu,\LambdaBT\}$ and to $\UT$ transformations, this is indeed what we will find in \S\ref{sec:refLT}.

This close analogy to the Schwinger-Keldysh doubled formalism then suggests that in the hydrodynamic limit, where we want to consider only the fully retarded correlators, we should be working to leading order in the difference fields which are now $\{\tildeg_{\mu\nu}, \tildeA_\mu\}$
for reasons explained in \S\ref{sec:skdouble} (see also \cite{Haehl:2013hoa}). For the present analysis it implies that once we are done with the various variations we should set
$\tildeg_{\mu\nu} \to 0$ and $\tildeA_\mu \to 0$.  In addition, as far as the new $\UT$ symmetry is concerned, the hydrodynamic limit corresponds to setting $\AT_\mu = 0$ and $\LambdaBT = 1$.

Once we set all the auxiliary fields  $\{\tildeg_{\mu\nu},\tildeA_\mu,\AT_\mu\}$ to zero and $\LambdaBT=1$, the $\UT$ Bianchi identity \eqref{eq:U1TBianchi} takes the form:\footnote{ Note that in this limit, we have $\diffB \tildeg_{\mu\nu} = \diffB g_{\mu\nu} \neq 0$.}
\begin{equation}\label{eq:U1TAdiab}
\begin{split}
& D_\mu  \NT^\mu    \Big{|}_{\haux}
=\half \TLLc^{\mu\nu}\Big{|}_{\haux} \; 
\diffB g_{\mu\nu}  +\JLLc^\mu\Big{|}_{\haux} \cdot  \diffB A_\mu \;\;
\\ 
& \text{where} \quad
\haux =\{\tildeg_{\mu\nu} = \tildeA_\mu = \AT_\mu = 0 \,, \;\; \LambdaBT=1, \;\& \; \hfields = \text{arbitrary} \} \,.
\end{split}
\end{equation}
Thus $\{ \NT^\mu , \TLLc^{\mu\nu},\JLLc^\mu\}\big|_{\haux} $ is a constitutive relation that solves adiabaticity equation. 

This shows that the constitutive relations derived from any diffeomorphism, flavour, and $\UT$ invariant Lagrangian $\LagT$ are always guaranteed to be solutions of the adiabaticity equation. We will now argue that the converse is also true: every adiabatic constitutive relation can be obtained by this method. 

Consider an arbitrary set of covariant constitutive relations $\{ \N^\mu = -{\mathcal G}^\mu/T, T^{\mu\nu},J^\mu\}$ such that they solve the grand canonical adiabaticity equation \eqref{eq:AdiabaticityG}, i.e., they identically satisfy
\begin{equation}
\begin{split}
D_\mu  \N^\mu
&=\half T^{\mu\nu} \,\diffB g_{\mu\nu}  +J^\mu \cdot  \diffB A_\mu \,.
\end{split}
\end{equation}
This implies that the combination
 $ \N^\mu  \AT_\mu + \half T^{\mu\nu} \tildeg_{\mu\nu}  +J^\mu \cdot  \tildeA_\mu$
is invariant under flavour, diffeomorphism and $\UT$ transformations up to boundary terms. Hence, the following master Lagrangian provides an allowed effective description: 
\begin{equation}\label{eq:MasterLag}
\begin{split}
 \LagT =  \N^\mu  \AT_\mu + \half T^{\mu\nu} \tildeg_{\mu\nu}  +J^\mu \cdot  \tildeA_\mu \,.
 \end{split}
\end{equation}

It is now easy to see that our procedure for generating  adiabatic constitutive relation from  $\LagT\brk{\hfieldsT}$ exactly reproduces the original
constitutive relations we started with, i.e.,
\begin{equation}
\{ \NT^\mu , \TLLc^{\mu\nu},\JLLc^\mu\}\big{|}_{\haux} 
= \{ \N^\mu , T^{\mu\nu},J^\mu\} 
\end{equation}
Thus, we have shown any adiabatic constitutive relation can be obtained from some $\LagT\brk{\hfieldsT}$. This establishes $\LagT$ to be the generating function for all adiabatic constitutive relations and completes the proof of Theorem \ref{thm:classLT}. In \S\ref{sec:eightfoldLT} we will give a more detailed discussion of this fact and indeed show explicitly how the eightfold way is implemented in Class $\LT$. Before doing so let us however show how hydrodynamic equations of motion and entropy current conservation are obtained in Class $\LT$.

\section{Hydrodynamic Ward identities and the Second Law in Class $\LT$}
\label{sec:LTvariational}
Having constructed the basic formalism for the construction of $\UT$ invariant Lagrangians in Class $\LT$, we now turn to demonstrating that the implied equations of motion are the usual conservation equations of hydrodynamics. We do this by giving a constrained variational principle that is completely analogous to the procedure in Class L, c.f., \S\ref{sec:Leoms}. The new features are, of course, that Class $\LT$ captures all of adiabatic transport and the presence of the additional $\UT$ symmetry whose associated Ward identity will turn out to be entropy current conservation.

\subsection{The Class $\LT$ variational principle}
\label{sec:LTvar}

Having derived a master Lagrangian $\LagT$ that determines precisely those constitutive relations that satisfy adiabaticity equation, it remains to show that  currents involved satisfy the  correct hydrodynamical equations of motion. Our next goal is thus to define a constrained variational principle for 
$\LagT\brk{\hfieldsT}$ which leads to the desired on-shell Ward identities.

Before we get into the technicalities, it should be clear that any variational principle must  reduce to the Class L variational principle \eqref{eq:consLvar} when the auxiliary fields are absent. This means that the constrained variational principle described there must be extended in an $\UT$ invariant manner to the auxiliary fields $\{\tildeg_{\mu\nu}, \tildeA_\mu, \AT_\mu, \LambdaBT\}$.

Consider a constrained variation $\diffCons$ of the following form: it acts on the single copy fields exactly as in Class L, i.e.,  
\begin{align}
\diffCons: \quad \diffCons \Kbeta^\mu =\diffF \Kbeta^\mu \,,\qquad
\diffCons  \LambdaB=\diffF\LambdaB\,,\qquad
\diffCons g_{\mu\nu} = \diffCons  A_\mu =0 \,,
\label{eq:consLTvar0}
\end{align}
and it acts on the copy sources and $\UT$ fields in a similar way:
\begin{equation}\label{eq:consLTvar}
\begin{split}
 \diffCons: 
    \quad &\diffCons \tildeg_{\mu\nu} = -\diffF\tgdiff_{\mu\nu} 
	\; \Longrightarrow \;
	\diffCons \tgdiff_{\mu\nu} = \diffF \tgdiff_{\mu\nu} 
	\\
	&\diffCons  \tildeA_\mu =-\diffF\tAdiff_\mu 
	\;\; \Longrightarrow \;
	\diffCons  \tAdiff_\mu \,=\diffF \tAdiff_\mu\,,
    \\
    &
	\diffCons  \AT_\mu =\diffF \AT_\mu \,, \quad \;\;\;\,
	\diffCons \LambdaBT = \diffF \LambdaBT\,.
\end{split}
\end{equation}
The choices for the variations of the 
auxiliary fields made in \eqref{eq:consLTvar} involves treating them like the physical hydrodynamic fields. In other words the constrained variation 
$\diffCons$ consists of varying the fields $\{\Kbeta^\mu, \LambdaB,\tgdiff_{\mu\nu} ,
\tAdiff_\mu,\AT_\mu,\LambdaBT\}$ along a Lie-orbit
while keeping the sources $\{g_{\mu\nu} ,A_\mu\}$ fixed. As alluded to in \S\ref{sec:LagTdefine}, these transformation rules are very canonical if we consider our Class $\LT$ Lagrangian as a natural extension of the Schwinger-Keldysh formalism developed \S\ref{sec:skrefm}.

Given this constrained variational principle, we can plug the explicit variations into \eqref{eq:LagTVar}, integrate by parts where necessary and end up with the basic statement:\footnote{ The quickest derivation of this expression involves starting from 
\eqref{eq:LTvarfinal} which gives the unconstrained variation and setting the currents which arise from the variation of the physical sources, viz.,  $\TLLc^{\mu\nu}$ and $\JLLc^\mu$ to zero. The latter currents arise from the variations of the physical sources which are forbidden in \eqref{eq:consLTvar}.}
\begin{equation}
\begin{split}
\frac{1}{\sqrt{-g}}&\diffCons \prn{\sqrt{-g}\ \LagT} - \text{Boundary terms}
\\& 
	= \prn{\Lambda +\xi^\nu  A_\nu } \cdot \Biggl\{   \frac{1}{\sqrt{-g}} \diffB \prn{\sqrt{-g}\ T\ \acharge} 
	+\bigg(D_\mu \JLc^\mu  - [\tildeA_\mu, \JLc^\mu] \bigg) \Biggr\}
 \\ & 
 	+\; \xi^\sigma 
	\Biggl\{
		\frac{1}{\sqrt{-g}} \diffB \prn{\sqrt{-g}\ T\ \aheat_\sigma} + T\,\acharge\cdot \diffB A_\sigma 
		+ T\,\achargeT\cdot \diffB \AT_\sigma +\JT^\nu \cdot \FT_{\sigma\nu} 
	\Biggr.	
 \\ &\qquad  
	 \Biggl.\qquad
	+\; D_\nu \prn{\tgdiff_{\sigma\mu}\, \TLc^{\mu\nu}} 
	 - \half \, \TLc^{\mu\nu} D_\sigma \tgdiff_{\mu\nu}-  \JLc^\nu\cdot \tFdiff_{\sigma\nu} 
	 \Biggr.
\\ &\qquad  
	\Biggl.\qquad 
		 -\;A^{(T)}_\sigma \prn{   \half \,\TLc^{\mu\nu} \, \diffB g_{\mu\nu}  
		 +\JLc^\mu\cdot  \diffB A_\mu } 
		 -\tildeA_\sigma \cdot  \bigg(D_\mu \JLc^\mu  - [\tildeA_\mu, \JLc^\mu] \bigg)
	\Biggr\} \\
& 	+(\LambdaT + \xi^\sigma \AT_\sigma ) 
	\Biggl\{ 
	-D_\mu  \JST^\mu
	 + \;\half \, \TLc^{\mu\nu} \diffB \tildeg_{\mu\nu}  +\JLc^\mu\cdot  \diffB \tildeA_\mu +\JT^\mu  \diffB \AT_\mu
	 \Biggr.
\\ &\qquad
	\Biggl.\qquad
	+\; 
	\prn{D_\mu \JT^\mu - \half\TLc^{\mu\nu} \diffB g_{\mu\nu}  -\JLc^\mu\cdot  \diffB A_\mu 
	- \frac{1}{\sqrt{-g}} \diffB \prn{\sqrt{-g}\ T\ \,\achargeT} } \Biggr\}
 \end{split}
 \label{eq:LagTCvarfinal}
\end{equation}
In deriving this expression we have to bear in mind the fact that the physical sources do not vary. As a result the contribution to the current  $\NT^\mu$ from  $\TLLc^{\mu\nu}$ and $\JLLc^\mu$ is missing. We  have chosen to indicate this by defining the 
entropy current in Class $\LT$ directly via 
\begin{equation}
\begin{split}
\JST^\mu = \NT^\mu -\Kbeta_\nu \, \TLLc^{\mu\nu} - (\LambdaB+\Kbeta^\nu\,A_\nu) \cdot \JLLc^\mu  \,.
\end{split}
\label{eq:JSTdef}
\end{equation}

Thus extremizing $\LagT$ for  an arbitrary constrained  variation  \eqref{eq:consLTvar} parameterized by  
 $\{\xi^\sigma,\Lambda\}$)  yields the equations:
\begin{equation}\label{eq:HydroLT1}
\begin{split}
0&\simeq 
\frac{1}{\sqrt{-g}} \diffB \prn{\sqrt{-g}\ T\ \aheat_\sigma} + T\,\acharge\cdot \diffB A_\sigma 
		+ T\,\achargeT\cdot \diffB \AT_\sigma +\JT^\nu \cdot \FT_{\sigma\nu} 	
 \\ &\qquad \qquad 
\qquad
	+\; D_\nu \prn{\tgdiff_{\sigma\mu}\, \TLc^{\mu\nu}} 
	 - \half \, \TLc^{\mu\nu} D_\sigma \tgdiff_{\mu\nu}-  \JLc^\nu\cdot \tFdiff_{\sigma\nu} 
\\ &\qquad \qquad \qquad 
		 -\;\AT_\sigma \prn{   \half \,\TLc^{\mu\nu} \, \diffB g_{\mu\nu}  
		 +\JLc^\mu\cdot  \diffB A_\mu } 
		 -\tildeA_\sigma \cdot  \bigg(D_\mu \JLc^\mu  - [\tildeA_\mu, \JLc^\mu] \bigg) \\
0&\simeq \frac{1}{\sqrt{-g}} \diffB \prn{\sqrt{-g}\ T\ \acharge} +\bigg(D_\mu \JLc^\mu  - [\tildeA_\mu, \JLc^\mu] \bigg)\ .
 \end{split}
\end{equation}
Comparing this against the Bianchi identities \eqref{eq:TBianchiLT} and   \eqref{eq:JBianchiLT}, we get the hydrodynamic equations we expect, viz.,
the usual hydrodynamic equations of motion
\begin{equation}\label{eq:HydroLT2}
\begin{split}
D_\mu (\TLLc)_\sigma^\mu &\simeq \JLLc^\nu \cdot F_{\sigma\nu}\,, \\
D_\mu  \JLLc^\mu &\simeq 0\,.
\end{split}
\end{equation}

Since we have an additional symmetry $\UT$ we should also examine the coefficient of 
$(\LambdaT + \xi^\sigma\, \AT_\sigma)$ to see what further constraints are imposed on dynamics. We find now
\begin{equation}\label{eq:HydroLT2b}
\begin{split}
 D_\mu \JST^\mu &\simeq \half \, \TLc^{\mu\nu} \diffB \tildeg_{\mu\nu}  
 +\JLc^\mu\cdot  \diffB \tildeA_\mu +\JT^\mu  \diffB \AT_\mu
  \\ &\qquad
	+
	\prn{D_\mu \JT^\mu - \half\TLc^{\mu\nu} \diffB g_{\mu\nu}  -\JLc^\mu\cdot  \diffB A_\mu 
	- \frac{1}{\sqrt{-g}} \diffB \prn{\sqrt{-g}\ T\ \,\achargeT} } \,.
\end{split}
\end{equation}
Comparing with the $\UT$ Bianchi identity \eqref{eq:U1TBianchi} we can simplify the expression  above to read:
\begin{equation}
D_\mu \big[\NT^\mu - \JST^\mu\big] \simeq \frac{1}{2}\,\TLLc^{\mu\nu}\, \diffB g_{\mu\nu} + \JLLc \cdot \diffB A_\mu \,.
\label{eq:utdynA}
\end{equation}	
This is an additional {\em dynamical} equation of motion that should be satisfied by systems described by $\LagT$. A-priori we seem to have too much, for in \eqref{eq:HydroLT2} we have all the equations of motion we actually want.  
When we however examine the hydrodynamic limit by setting the auxiliary fields to zero, 
$\hfieldsT =\haux$, as described  around \eqref{eq:U1TAdiab}, we can further eliminate terms from \eqref{eq:utdynA} to arrive at 
\begin{equation}
D_\mu \, \JST^\mu \simeq 0 \,,
\label{eq:HydroLT3}
\end{equation}	
which is simply the statement of on-shell entropy conservation in Class $\LT$. As promised the $\UT$ transformations ensure adiabaticity off-shell, which in turn implies that the entropy current is conserved on-shell.

\subsection{Reference fields for Class $\LT$}
\label{sec:refLT}

Let us now try to introduce new fields to convert this into an unconstrained variational problem.
To do this, we will pass to a description whereby we factorise the dynamical fields into a rigid
reference configuration and a pullback diffeomorphism and flavour transformation. The hydrodynamic
equations will then be generated by the variations of the pull back fields keeping the reference
fields fixed. 
Note that here we will refrain from giving a prescription for the precise form of $\UT$ action on the reference manifold.
While it is clear that such a formalism must exist (in fact it is not very hard to write the analogue of \eqref{eq:ConstrVar} including the $\UT$ action),
its detailed workings may depend on other issues which we defer to future work \cite{Haehl:2014kq}. For the moment we just illustrate that the reference manifold variational 
principle of \S\ref{sec:reffields} can be extended to the new field content, thus obtaining the hydrodynamic equations of motion.

We begin by imagining a copy of hydrodynamic fields   $ \{\Kref^a,\Lref\}$  living in a
reference manifold  $\Mref$. In addition, we will posit  a metric, a gauge field and a copy of $\{\AT_\sigma,\LambdaBT\}$,
viz.,  $\{\tgdiffref_{ab},\tAdiffref_a,\ArefT,\LrefT\}$ on $\Mref$. The actual fields  are obtained by introducing
a diffeomorphism field  $\varphi^a(x)$ and a gauge transformation field  $c(x)$ from physical spacetime ${\cal M}$
to $\Mref$ and then using them to  pull-back $\hreffieldsT \equiv \{\Kref^a,\Lref,\gref_{ab},\Aref_a,\tgdiffref_{ab},\tAdiffref_a,\ArefT_a,\LrefT\}$.
Now consider constrained variations as defined by \eqref{eq:consLTvar0}, \eqref{eq:consLTvar}. Their diffeomorphism and flavour parts (disregarding the $\UT$ action for now) are implemented on the reference manifold as follows:
\begin{equation}\label{eq:diffConstrainedRef}
\begin{split}
\diffCons \gref_{ab} &= - \delta_\varphi \gref_{ab} \,, \qquad
\diffCons \Aref_a =- \delta_\varphi \Aref_a \,, \\
\diffCons \Kref^a &= 0 \,, \qquad \diffCons \tgdiffref_{ab} = 0  
\,, \qquad \diffCons \ArefT_a = 0 \\
\diffCons \Lref &= 0 \,, \qquad
\diffCons \tAdiffref_a = 0 \,, \qquad\,
\diffCons \LrefT = 0\,,
\end{split}
\end{equation}
where $\delta_\varphi$ denotes Lie drag on $\Mref$ along $\{e^\mu_a \delta \varphi^a,\, -(\delta c) c^{-1} \}$.
Let us see how the variational principle works on $\Mref$: 
\begin{equation}
\begin{split}
&\frac{1}{\sqrt{-\gref}} \, \diffCons \prn{\sqrt{-\gref} \;\Lagref_\smallT} 
\\
&\qquad = \frac{1}{2} \TrefLLc^{ab} \, \diffCons \gref_{ab}
- \frac{1}{2} \TrefLc^{ab} \, \diffCons \tgdiffref_{ab}
+ \JrefLLc^a \cdot \diffCons \Aref_a - \JrefLc^a \cdot \diffCons \tAdiffref_a 
\\
&\qquad\quad+ \Tref\, \aheatref_a \, \diffCons \Kref^a + \Tref\, \achargeref \cdot (\diffCons \Lref+\Aref_c \, \diffCons \Kref^c) + \JTref^a \, \diffCons \ArefT_a + \Tref\, \achargeTref \prn{ \diffCons \LrefT + \ArefT_c \, \diffCons \Kref^c } 
\\
&\qquad = -\frac{1}{2} \TrefLLc^{ab} \prn{\Dref_a \delta\varphi_b + \Dref_b \delta\varphi_a  }  
 - \JrefLLc^a \cdot\brk{ \Dref_a \prn{ -c^{-1}\delta c + \Aref_b \, \delta\varphi^b } + \delta\varphi^b \Fref_{ba} }\,.
\end{split}
\end{equation}
As usual, we perform an integration by parts and obtain
\begin{equation}\label{eq:LTrefVar}
\begin{split}
&\frac{1}{\sqrt{-\gref}} \, \diffCons \prn{\sqrt{-\gref} \;\Lagref_\smallT} + \nabla_a ( \,\cdots )
\\
&\qquad = \delta \varphi_a \brk{ \Dref_b\TrefLLc^{ab} - \JrefLLc^b \cdot \Fref^a{}_b} + \prn{-c^{-1} \delta c + \Aref_b \, \delta \varphi^b }\cdot \Dref_a \, \JrefLLc^a 
 \,.
\end{split}
\end{equation}
Demanding invariance under any such variation, we can read off the reference manifold version of the conservation equations \eqref{eq:HydroLT2} from \eqref{eq:LTrefVar}.

\section{Eightfold adiabatic transport in Class $\LT$}
\label{sec:eightfoldLT}

We have now all the ingredients in Class $\LT$ to demonstrate how all classes of adiabatic transport can be realized by a Lagrangian description. For this purpose, we will distinguish constitutive relations that are Lagrangian ($\Lag = \PS \cup \LS$), non-Lagrangian ($\Lag^c = \text{B} \cup \text{C} \cup \GV$),  and anomaly induced ($\text{A} \cup \PV$). The most general adiabatic constitutive relations thus split into\footnote{ For the purposes of giving simple expression, we have chosen here to group the adiabatic classes somewhat differently; in particular, we choose to group Class $\PV$ with Class A as this is quite natural in the $\UT$ invariant formalism.} 
\begin{equation}
\begin{split}
 {\cal G}^\sigma &=  {\cal G}^\sigma_\Lag + ({\cal G}^\sigma)_{\Lag^c}+({\cal G}^\sigma)_{\text{A},\PV}\,, \\
 T^{\mu\nu} &=    T^{\mu\nu}_\Lag +(T^{\mu\nu})_{\Lag^c}  +(T^{\mu\nu})_{\text{A},\PV}\,, \\
 J^\mu &=   J^\mu_\Lag  + (J^\mu)_{\Lag^c} +(J^\mu)_{\text{A},\PV}\,,
\end{split}
\end{equation}
where ${\cal G}^\sigma \equiv -T \, \N^\sigma$.
Comparing with \eqref{eq:GeneralConstRel}, we see that Classes D and $\PF$ have been discarded from the most general currents that would be allowed just by classifying off-shell inequivalent tensor structures. Since such terms are either dissipative or forbidden by the second law, we don't expect to see them in Class $\LT$ (which we claim captures precisely the second law-allowed adiabatic transport).

The fact that Class D  and Class $\PF$ cannot be realized in Class $\LT$ is easy to see: these constitutive relations explicitly break the $\UT$ invariance by  virtue of not satisfying the adiabaticity equation.  This is only to be expected, of course, for the Class $\LT$ formalism to achieve the desired goal of providing a description precisely of adiabatic transport. 

In turn, we will now show how all the adiabatic constitutive relations which are allowed by second law can be obtained from a Class $\LT$ effective action.  

\subsection{$\LagT$ for Class L constitutive relations}
\label{sec:LTclassL}

We start our discussion of the eightfold way in Class $\LT$ by reproducing Class L constitutive relations. As we will demonstrate momentarily, Class $\LT$ provides a natural extension of Class L, for it contains Class L as a rather trivial special case. 

By definition Class L currents $\{{\cal G}^\sigma_\Lag, \, T^{\mu\nu}_\Lag,\, J^\mu_\Lag\}$ are precisely those which can be obtained from a scalar Lagrangian $\Lag[ \hfields ] \equiv \Lag [g_{\mu\nu},A_\mu, \Kbeta^\mu, \LambdaB]$. In this case there is thus no need to build a master Lagrangian that involves the enhanced field content of Class $\LT$ as in \eqref{eq:MasterLag}; instead we can simply take
\begin{equation}\label{eq:ClassLTL}
\LagT[\hfieldsT] = \Lag[\hfields] \,.
\end{equation}
This simplified prescription for Class L terms has the advantage that it is obvious how $\int \sqrt{-g} \, \LagT$ reduces in hydrostatics to the equilibrium partition function. Indeed, precisely the same argument as for Class L (see \S\ref{sec:Lhstatic}) proves that this consistency condition is met. Equivalently one can consider starting with $\LagT$ and gauge fix the auxiliary fields 
$\{\tildeg_{\mu\nu}, \tildeA_\mu, \AT_\mu\}$ differently from $\haux$ in \eqref{eq:U1TAdiab}. 
For instance given that the Noether current is roughly determined by the Class L Lagrangian, 
$ \N^\mu = \Kbeta^\mu \,\Lag  -(\PSymplPot{\Bfields})^\sigma +\nabla_\nu \Komar^{\sigma\nu}[\Bfields]$, suggests that one might choose the following set of $\{\AT_\mu = -T^2\, \Kbeta_\mu, \LambdaB =0, \tildeg_{\mu\nu} =0, \tildeA_\mu =0\}$ to recover the Lagrangian density $\Lag$.\footnote{ This statement should account for the contribution from the pre-symplectic current which is not always transverse. One might however be able to field redefine this contribution away, though we have not checked this statement in detail.}

The constitutive relations obtained from the Class $\LT$ variational principle can be read off from \eqref{eq:LagTVar} and \eqref{eq:utNTdef}:
\begin{equation}
\begin{split}
 \TL^{\mu\nu} &= T_\Lag^{\mu\nu} \,,\quad \JL^\mu = J^\mu_\Lag \,,\quad
 \GT^\sigma = {\cal G}_\Lag^\sigma \,, \\
 \TLc^{\mu\nu} &= 0 \,, \qquad  \JLc^\mu = 0 \,,\quad\;\;\, \JT^\mu = 0 \,.
\end{split}
\end{equation}
From this it is clear that also the dynamics implied by the variational principle in \S\ref{sec:LTvariational} reproduces everything we developed in Class L. Hence the choice \eqref{eq:ClassLTL} identically reproduces Class L with the nice additional feature that the adiabaticity equation (and entropy conservation as an equation of motion) is manifestly satisfied a-priori because the Lagrangian is trivially $\UT$ invariant.

\subsection{$\LagT$ for non-Lagrangian constitutive relations (Classes B, C and $\GV$)}
\label{sec:LTclassLc}

Having reproduced Class L in Class $\LT$ in a rather trivial way, let us now turn to constitutive relations which were not captured by na\"ive Lagrangians of Class L. To wit, consider adiabatic constitutive relations $\{({\cal G}^\sigma)_{\Lag^c},\, (T^{\mu\nu})_{\Lag^c},\, (J^\mu)_{\Lag^c} \}$ which subsume terms of Classes B, C and $\GV$ as constructed in \eqref{eq:TJClassVMatB}, \eqref{eq:TJC} and \eqref{eq:TJVec}. While a Class L Lagrangian giving such terms does not exist, we can use our new machinery to construct the associated effective master Lagrangian as in \eqref{eq:MasterLag}:
\begin{equation}\label{eq:MasterLagLc}
\begin{split}
 \LagT[\hfieldsT] =  -\frac{({\cal G}^\mu)_{\Lag^c}}{T}\,  \AT_\mu + \half (T^{\mu\nu})_{\Lag^c} \,\tildeg_{\mu\nu}  +(J^\mu)_{\Lag^c} \cdot  \tildeA_\mu \,.
 \end{split}
\end{equation}
Via \eqref{eq:LagTVar} this defines some constitutive relations $\{\GT^\sigma,\TLLc^{\mu\nu},\JLLc^\mu\}$ which mix the full field content $\{\Kbeta^\mu,\LambdaB,g_{\mu\nu},A_\mu,\tildeg_{\mu\nu},\tildeA_\mu,\AT_\mu\}$. The dynamical equations implied by the variational principle of section \S\ref{sec:LTvar} are precisely the hydrodynamic conservation equations and the conservation of the entropy current 
\begin{equation}
\JST^\mu = -\frac{\GT^\mu}{T} -\Kbeta_\nu \, \TLLc^{\mu\nu} - (\LambdaB+\Kbeta^\nu\,A_\nu) \cdot \JLLc^\mu \,.
\end{equation}

As demonstrated in \S\ref{sec:LagTdefine}, after setting the auxiliary fields to zero, these currents reduce to the desired ones:
\begin{equation}
 \{\GT^\sigma,\TLLc^{\mu\nu},\JLLc^\mu\} \big{|}_{\haux}  = \{({\cal G}^\sigma)_{\Lag^c},\, (T^{\mu\nu})_{\Lag^c},\, (J^\mu)_{\Lag^c} \}
\end{equation}
and their Ward identities are still the standard hydrodynamic equations of motion together with conservation of the entropy current 
\begin{equation}
\JST^\mu \big{|}_{\haux}  = -\frac{({\cal G}^\mu)_{\Lag^c}}{T} - \Kbeta_\nu (T^{\mu\nu})_{\Lag^c} - (\LambdaB+\Kbeta^\nu\,A_\nu) \cdot (J^\mu)_{\Lag^c} = (J_S^\mu)_{\Lag^c} \,.
\end{equation}
%

\subsection{$\LagT$ for anomalies (Classes A and $\PV$)}
\label{sec:LTanom}

The construction of the master Lagrangian $\LagT$ heavily relies on a doubling of the field content and demanding invariance under $\UT$. We already gave indications that this can be linked back to the Schwinger-Keldysh doubling that we employed previously in \S\ref{sec:anward} to describe anomalies. We would thus like to show that the master Lagrangian $\LagT$ is capable of describing anomalies in an analogous (but not quite identical) fashion. In this section we will discuss Class A terms, but it is clear from \S\ref{sec:JLYtranscend} that this immediately incorporates Class $\PV$, as well. The reason for this is that, once terms associated to some anomaly polynomial $\fP[\fF,\fR]$ are dealt with, the presence of $\AT_\mu$ and $\LambdaBT$ allows us to perform the Class $\PV$ generalization of $\fP[\fF,\fR]$ directly by using the replacement rule \eqref{eq:ReplacementRule}. The fact that we are free to set $\LambdaBT+\Kbeta^\mu \AT_\mu = 1$ in the Class $\LT$ formalism ensures consistency with the Class $\PV$ discussion of \S\ref{sec:JLYtranscend}. 

Consider a particular constitutive relation $\{ ({\cal G}^\mu)_\text{A} \equiv -T(\N^\mu)_\text{A}, (T^{\mu\nu})_\text{A},(J^\mu)_\text{A}\}$ that satisfies the  adiabaticity equation with covariant anomalies
\begin{equation}
\begin{split}
D_\mu  (\N^\mu)_\text{A}
&=\half (T^{\mu\nu})_\text{A} \diffB g_{\mu\nu}  +(J^\mu)_\text{A} \cdot  \diffB A_\mu + \N_H^\perp
\end{split}
\end{equation}
with $\N_H^\perp = \Kbeta_\sigma \THall^{\perp\sigma}+(\LambdaB+\Kbeta^\nu A_\nu) \cdot \JH^\perp$.
Then, the combination 
\begin{equation}
 \LagT[\hfieldsT] = (\N^\mu)_\text{A}  \,\AT_\mu + \half (T^{\mu\nu})_\text{A} \,\tildeg_{\mu\nu}  
 +(J^\mu)_\text{A} \cdot  \tildeA_\mu  
 \end{equation}
is no more $\UT$ invariant. In fact, we get
\begin{equation}
\begin{split}
 \diffF \int \sqrt{-g}\;\LagT &= -\int \sqrt{-g}\;\LambdaTb \N_H^\perp
 = -\int \sqrt{-g}\,\brk{ \LambdaTb \Kbeta_\sigma \THall^{\perp\sigma}+\LambdaTb (\LambdaB+\Kbeta^\nu A_\nu) \cdot \JH^\perp }\,.
 \end{split}
 \label{eq:U1TanomLag}
\end{equation}

In order to account for this anomaly, we extend our natural construction of master Lagrangians $\LagT$ to the bulk theory, using Hall currents as the bulk constitutive relations. We claim that the following Class $\LT$ action gives the correct anomalous boundary constitutive relations and adiabaticity equation as a result of bulk inflow:
\begin{equation}
 S_{\,\smallT} = \int_{\cal M} \sqrt{-g} \; \LagT+ \bulkint \prn{ \N_H^m\, \AT_m + \frac{1}{2} \THall^{mn} \tildeg_{mn} + \JH^m \cdot \tildeA_m } \,,
 \label{eq:LagTanom}
\end{equation}
where $\N_H^m = s u^m + ( \Kbeta_s \THall^{ms} + (\LambdaB+\Kbeta^n A_n) \JH^m )$ is the non-canonical part of bulk entropy current. It is easy to see that the combination \eqref{eq:LagTanom} is $\UT$ invariant: 
\begin{align}
\diffF S_{\,\smallT} &= - \int_{\cal M} \sqrt{-g} \; \LambdaTb \N^\perp_H + \bulkint \brk{\N_H^m \, D_m \LambdaTb +  \LambdaTb \prn{\THall^{mn} \diffB g_{mn} + \JH^m \diffB A_m }} 
\notag \\
&= - \int_{\cal M} \sqrt{-g} \; \LambdaTb \N^\perp_H + \bulkint D_m \brk{ \LambdaTb \prn{ su^m + \Kbeta_n \THall^{mn}+ (\LambdaB+\Kbeta^p A_p) \cdot \JH^m }} 
\notag \\
& \qquad - \bulkint  \LambdaTb \brk{ D_m (s u^m) + \Kbeta^m \prn{ D_n (\THall)^n_m - \JH^n \cdot F_{mn} } + (\LambdaB + \Kbeta^p A_p) \cdot D_m \JH^m } 
\notag \\
&= 0 \,,
\end{align}
where in the last step we evaluated the integral over the total derivative to cancel the boundary anomaly (using $u^\perp = 0$). Furthermore, we used a bulk adiabaticity equation to kill the second bulk integral (equivalently we could have computed the bulk Bianchi identities explicitly to see that such a bulk adiabaticity equation is satisfied).

While the Lagrangian \eqref{eq:LagTanom} is very natural and simple from the Class $\LT$ point of view, it is less clear how this can be made consistent with the Class A Lagrangian that we gave before in \eqref{eq:SK_Stot}. In order to get a step closer towards the Class A Schwinger-Keldysh Lagrangian, let us rotate to another basis of sources: it turns out that, while $\{g_{\mu\nu},A_\mu,\tildeg_{\mu\nu}, \tildeA_\mu\}$ are convenient to construct the master Lagrangian, anomalies are easier to describe in a Schwinger-Keldysh inspired basis $\{g_{\mu\nu}^\skR,A_\mu^\skR,g_{\mu\nu}^\skL,A_\mu^\skL\}$ as we defined it in \eqref{eq:skLTdef}. 
The twisted difference sources $\{ \dRLg_{\mu\nu},\dRLA_\mu\}$ then have a nice $\UT$ transformation:
\begin{equation}
\begin{split}
 \diffF \, \dRLg_{\mu\nu} = \diffLB \, g_{\mu\nu}  \,, \qquad
 \diffF \, \dRLA_\mu = \diffLB \, A_\mu \,,
\end{split}
\end{equation}
where $\diffLB$ denotes Lie drag along $ \{\LambdaTb \Kbeta^\mu, \LambdaTb \LambdaB\}$.

We can account for the anomaly by viewing it as originating from inflow due to a bulk Chern-Simons theory. To this end, we need to switch from a description of covariant currents to one in terms of consistent currents, c.f., Appendix \ref{sec:adcons}. The Bardeen-Zumino currents by which the two descriptions differ will be seen as part of the bulk inflow. To wit, the consistent Lagrangian is related to $\LagT$ as follows: 
\begin{equation}
\begin{split}
\LagT^{\text{cons}} &\equiv (\LagT)_\text{A} -  (\LagT)_{BZ} \\
&= \big( (\N^\mu)_\text{A}-\N_{BZ}^\mu\big) \AT_\mu + \frac{1}{2} \big( (T^{\mu\nu})_\text{A}-T_{BZ}^{\mu\nu}) \big)\,\tildeg_{\mu\nu} + 
\big( (J^\mu)_\text{A}-J_{BZ}^\mu\big) \cdot \tildeA_\mu \\
&= \big( (J_S^\mu)_\text{A}-J_{S,BZ}^\mu\big)  \AT_\mu + \frac{1}{2} \big( (T^{\mu\nu})_\text{A}-T_{BZ}^{\mu\nu} \big)\, \dRLg_{\mu\nu} + 
\big( (J^\mu)_\text{A}-J_{BZ}^\mu\big)  \cdot \dRLA_\mu \,,
\end{split}
\end{equation}
where $J_{S,BZ}^\mu = - \Kbeta^\nu \mathcal{T}_\nu{}^\mu$ as in Appendix \ref{sec:adcons}.  
In order to show that the philosophy of \S\ref{sec:anward} is being upheld by the present construction, we want to demonstrate that the Bardeen-Zumino currents can be interpreted as an anomaly inflow due to the difference of two Chern-Simons terms. The bulk contribution of these Chern-Simons terms should give rise to the Hall currents. 
Indeed, we observe that the Bardeen-Zumino and Hall part of the action \eqref{eq:LagTanom} can be written as 
\begin{equation}\label{eq:LTanomBZ}
\begin{split}
  \int_{\cal M} \sqrt{-g}&\; (\LagT)_{BZ} + \bulkint \prn{ \N_H^m\, \AT_m + \frac{1}{2} \THall^{mn} \tildeg_{mn} + \JH^m \cdot \tildeA_m } \\
  &= \int_{\cal M}\sqrt{-g} \; \prn{J_{S,BZ}^\mu\, \AT_\mu + \frac{1}{2}\,T_{BZ}^{\mu\nu}\, \dRLg_{\mu\nu} + J_{BZ}^\mu \cdot \dRLA_\mu } \\
  &\quad + \bulkint \prn{ J_{S,H}^m\, \AT_m + \frac{1}{2} \THall^{mn} \dRLg_{mn} + \JH^m \cdot \dRLA_m } \\
  &= \bulkint \Big( \ICS[\fA_{\skR},\fGamma_\skR] -\ICS[\fA_{\skL},\fGamma_\skL] \Big)  \\
  &\quad +\int_{\cal M}\sqrt{-g} \; \prn{J_{S,BZ}^\mu\, \AT_\mu} + \bulkint \Big( J_{S,H}^m\, \AT_m \Big) + {\cal O}\big((\bar{\hreffields}_\smallT)^2\big)\,,
\end{split}
\end{equation}
where we used in the last step that the non-entropic currents in the second and third line are the linear terms in when we expand a difference of Chern-Simons terms. Since our
answers are insensitive to terms that are quadratic in difference sources, we can perform the last step. Terms that are discarded since they are quadratic in difference fields are denoted as ${\cal O}\big((\bar{\hreffields}_\smallT)^2\big)$. 

As it stands \eqref{eq:LTanomBZ} demonstrates consistency with the methods used in \S\ref{sec:anward} (Schwinger-Keldysh for Class A). What we see in the final expression of \eqref{eq:LTanomBZ} is precisely the difference of Chern-Simons terms that could be converted into an appropriate transgression form. Since we are already in the doubled construction, the hatted Chern-Simons terms which would have been present appear to have cancelled out against the influence functional. However, we are forced in our $\UT$-invariant formalism to introduce some new pieces of data -- the terms in the last line of \eqref{eq:LTanomBZ} involving the Bardeen-Zumino current are not present in the Schwinger-Keldysh formalism. These terms determine the entropy current both in bulk and physical manifold uniquely in terms of the anomaly polynomial. Of course, these terms depend on the presence of $\AT_m$ and could therefore not be seen in our previous na\"ive Schwinger-Keldysh treatment of Class A. Note also that in the Schwinger-Keldysh construction with the influence functionals one is unable to determine the precise form of the entropy current owing to our failure to derive the adiabaticity equation. As a result what we can compare naturally in the two formalisms is the constitutive relations and these agree as they must for consistency. 
We leave it as an open and interesting problem to finish delineating the connections between the two formalisms, at the very least to check how $\UT$ invariance helps restore adiabaticity explicitly. For now we simply take comfort in the fact that we have a Class $\LT$ effective action that gives us a consistent adiabatic solution for the anomalous constitutive relations.

\subsection{Field redefinitions}
\label{sec:LTredef}

An important consistency requirement on our eightfold classification is, of course, that it is (on-shell) invariant under field redefinitions. While we commented at various occasions on the most general allowed field redefinitions, the unified framework of Class $\LT$ simplifies the discussion considerably.\footnote{ For example, an isolated proof for the field redefinition-invariance of Class B would be notoriously complicated.} Consider the most general field redefinitions that preserve the property of $\{\Kbeta^\mu,\LambdaB\}$ being in equilibrium aligned with the symmetry generators $\{K^\mu,\Lambda_K\}$:
\begin{equation}\label{eq:KbetaRefLT}
\begin{split}
\Kbeta^\mu &\mapsto \Kbeta^\mu - \diffB {\bar V}^\mu = \Kbeta^\mu +\lieD_{\bar V} \Kbeta^\mu\,,\\
\LambdaB &\mapsto \LambdaB - \diffB \Lambda_{\bar V} =  \LambdaB + \lieD_{\bar V}  \LambdaB + [\LambdaB,{\bar \Lambda}_V] -\Kbeta^\sigma \partial_\sigma{\bar \Lambda}_V\,,\\
\tgdiff_{\mu\nu} & \mapsto \tgdiff_{\mu\nu}  + \lieD_{\bar V} \tgdiff_{\mu\nu} - \bar{\Lambda}^\smallTbrk_V \, \diffB\tildeg_{\mu\nu}  \,,\\
\tAdiff_\mu & \mapsto \tAdiff_\mu + \lieD_{\bar V} \tAdiff_\mu + [\tAdiff_\mu , {\bar \Lambda_V}] - \bar{\Lambda}^\smallTbrk_V \, \diffB \tildeA_\mu \,,\\
\AT_\mu & \mapsto \AT_\mu + \lieD_{\bar V} \AT_\mu + \partial_\mu {\bar \Lambda}^\smallTbrk_V \,,\\
\LambdaBT &\mapsto \LambdaBT + \lieD_{\bar V}  \LambdaBT  -\Kbeta^\sigma \partial_\sigma{\bar \Lambda}^\smallTbrk_V\,,\\
g_{\mu\nu} & \mapsto g_{\mu\nu} \,, \qquad A_\mu  \mapsto A_\mu \,.
\end{split}
\end{equation}
for some general diffeomorphism, flavour and $\UT$ parameters $\{V^\mu,\Lambda_V,\Lambda^\smallTbrk_V\}$ and their twisted counterparts $\{\bar{V}^\mu,{\bar \Lambda}_V, {\bar \Lambda}_V^\smallTbrk\}$. An argument completely analogous to the one presented for Class L (see \S\ref{sec:fieldredef}) shows that under these transformations the Lagrangian $\LagT$ only changes by terms proportional to the equations of motion \eqref{eq:HydroLT1}, \eqref{eq:HydroLT2b}. This can easily be seen from the fact that the above field redefinitions take the form of a constrained variation \eqref{eq:consLTvar0}, \eqref{eq:consLTvar}. 

Since the eightfold Lagrangian $\LagT$ encompasses all seven classes of adiabatic transport and disallows Class $\PF$, we see that the field redefinitions \eqref{eq:KbetaRefLT} preserve this structure. This provides a very general argument for our eightfold classification being on-shell field redefinition-invariant.

\newpage
\part{Future Prospects}
\label{part:discuss}
\hspace{1cm}

\section{Discussion}
\label{sec:discuss}

We have described a framework for non-linear hydrodynamics that gives a complete classification of hydrodynamic transport, arguing in the process for an eightfold way of hydrodynamic dissipation. The key idea was to exploit the natural decomposition of transport in two primary categories: adiabatic and dissipative. While the notion of adiabaticity we introduce reduces  on-shell to a more intuitive notion of non-dissipative (i.e., entropy conserving) dynamics, it more generally involves playing off entropy production against energy momentum and charge transport (off-shell). One of the remarkable outcomes of our analysis was demonstrating that much of transport beyond leading order is in fact adiabatic. 

We have established that adiabatic transport is cleanly organized into seven distinct classes of constitutive relations. Some of these are easily understood such as the hydrostatic scalar Class $\PS$ and anomaly induced Class A, which have been explored extensively in recent literature, whilst certain others are perhaps a bit more exotic and unfamiliar. Curiously, all but one of these classes (Class $\GV$) have been encountered in earlier investigations, though neither the structural aspects nor their import has been completely appreciated  hitherto. Our aim in the sections above has been to clearly bring out these aspects for every one of these adiabatic classes.

Having figured out how to tackle adiabatic transport, we further argued that dissipative hydrodynamics is under much better control than one might have a-priori anticipated. The primary result in this context has already been derived in \cite{Bhattacharyya:2013lha,Bhattacharyya:2014bha}. The adiabatic analysis allows for an alternative perspective which complements and bolsters the central point: {\em dissipative transport is constrained by the second law of thermodynamics to obey sign-definiteness constraints only at the leading order in the gradient expansion}.  

From the point of view of the hydrodynamic effective field theory, the combination of adiabaticity and the fact that dissipative parts of higher order transport are unconstrained, restores a sense of democracy to hydrodynamics. Specifically, the standard current algebra description of hydrodynamic constitutive relations could have been a simple exercise in representation theory, were it not for the constraint imposed by the second law. As has been argued in the text and in the aforementioned references, the constraints at arbitrary orders in the gradient expansion are completely captured by demanding the existence of hydrostatic equilibrium. This class of forbidden constitutive relations, Class $\PF$, is easily ascertained by writing down all symmetry allowed tensor structures for the basic currents (which survive the hydrostatic limit) and eliminating ones that do not arise from an effective action (the hydrostatic partition function). Thus, modulo the Class $\PF$ constitutive relations, one finds the task of a hydrodynamicist is rather simple for every other constitutive relation is physically admissible and satisfies the second law (beyond leading order).

While the major part of our construction of adiabatic constitutive relations was  carried out straightforwardly using conventional techniques familiar in hydrodynamic analyses, we have also uncovered some new structures in our quest for constructing an effective action for hydrodynamics. We have argued for a new symmetry principle, an Abelian gauge symmetry $\UT$ arising as a consequence of the eightfold way. We have primarily used this symmetry  to write down a  generating functional for the adiabatic classes. Let us therefore take stock of the physical implications of this construction. The discussion below is meant to be heuristic and we hope to provide concrete evidence for it in the future \cite{Haehl:2014kq}.

One naively might imagine that a hydrodynamic effective action in the absence of dissipation should be rather simple, following the usual rules of effective field theory. This is indeed the case for the Class L constitutive relations where we have a straightforward construction of a Landau-Ginzburg functional (Lagrangian) in terms of the effective infra-red degrees of freedom (the thermal vector and twist). However, as we have explicitly seen this is inadequate to capture all of adiabatic transport. Not only are the transverse vector based Classes $\PV$ and $\GV$ outside the remit of such a Lagrangian density, but we have furthermore evidence that Berry-like (Class B) terms are unaccounted for. Moreover, an analysis of anomalous constitutive relations in Class A both herein and in our earlier work \cite{Haehl:2013hoa} has revealed the impossibility of satisfying both the Ward identities and the second law simultaneously, unless one is willing to enlarge the set of degrees of freedom. 

This perspective is natural when one considers the passage to a complete theory of hydrodynamics including dissipation, or more generally views hydrodynamics as  governing infra-red fluctuations of a density matrix. One naturally then anticipates invoking a real-time non-equilibrium formalism such as the Schwinger-Keldysh  construction \cite{Schwinger:1960qe,Keldysh:1964ud}, thereby motivating the exploration of implications of doubling the degrees of freedom inherent in this formalism. This doubling in the fluid dynamical context is often called Martin-Siggia-Rose or Janssen-deDominicis-Politi formalism (see 
\cite{Kamenev:2011aa}).

In the context of anomalies we have shown that this intuition can indeed be made to work, in part since the structure of allowed terms is tightly constrained by the underlying symmetry (flavour gauge and diffeomorphism). However, once we double the degrees of freedom  and consider a pure, entangled state in the doubled theory as the starting point of our discussion, we are led naturally to inquire after the most general set of interactions allowed in the effective field theory. The key issue concerns the constraints on the Feynman-Vernon influence functionals \cite{Feynman:1963fq} describing the interactions between the two sets of degrees of freedom. Obtaining the correct anomalous Ward identities without modifying the single copy constitutive relation forces upon us  a particular set of anomaly induced influence functionals. 

If we were to allow arbitrary influence functionals however, then it is easy to generate Class $\PF$ constitutive relations at will, thus explicitly violating  the  second law. This is equivalent to stating that, given the usual symmetries, the emergence of entropy, dissipation and second law at long distances is not `natural' in the 't Hooft-Wilson sense. Thus one needs an additional principle to forbid arbitrary influence functionals and solve this naturalness  problem of second law. Our claim is that in the hydrodynamic effective field theory the $\UT$ symmetry by ensuring adiabaticity guarantees that the second law is upheld.

Let us then sketch a scenario how this $\UT$ symmetry works in ensuring the second law in the hydrodynamic limit.  We first recall that the hydrodynamic gradient expansion requires that the fluctuations about the underlying Gibbs density matrix are sufficiently long wavelength. Operationally this almost always implies that we are in the high temperature regime. In this limit Schwinger-Keldysh construction for the equilibrium density matrix is such that the common or average fields are macroscopic, but the difference fields are retained only to linear order (this is required to ensure that we extract the correct retarded correlators) \cite{Chou:1984es}. Pictorially we can then imagine that while the two sets of degrees of freedom $\hfields_\skL$ and $\hfields_\skR$ live on different background spacetimes (or parts of the Schwinger-Keldysh contour), the effective spacing between them is vanishingly small in the Euclidean time direction.  Alternately, the difference fields encode the fluctuations or noise in the system and they are Avagadro-suppressed in a fluid.

We would like to view the fluid phase as the {\em Higgs phase} for $\UT$ with the difference fields acting as the Higgs fields which transform non-linearly under $\UT$ .  This ensures that both the $\UT$ gauge field and the difference fields are invisible at long distances.  A crucial result in this paper is to show that, ignoring effects of  $\UT$ ghost fields, one can posit nonlinear transformations for difference fields (see Eq.~\eqref{eq:TactgAbar}) that ensure the removal of Class $\PF$ and reproduce
the seven adiabatic classes. It would be enormously satisfying to show that this success somehow extends to dissipative class when appropriate ghost fields are added and their
effects are taken into account. Indeed we have noted in our Class D description that the intertwiners  $\BerryG$ and $\BerryA$ are required to be positive definite (symmetric) quadratic forms in some tensor representation. It is worthwhile noting that these intertwiners typically take on the role of the kinetic terms of certain ghosts which naturally occur
in the  Martin-Siggia-Rose dissipative effective action  \cite{Kovtun:2014hpa}. Such a structure makes it natural to demand positivity for these quadratic forms which along with  removal of  Class $\PF$ would then ensure second  law. It would be interesting whether these set of statements can be made precise in order to provide a Wilsonian  explanation for second law of thermodynamics.

We see that the well-behavedness of the hydrodynamic effective field theory appears from this perspective to be tied to the underlying presence of the $\UT$ symmetry. In some sense, this Abelian symmetry is the macroscopic manifestation of the  Kubo-Martin-Schwinger (KMS) condition which encodes the near-thermal correlations between two copies of 
Schwinger-Keldysh construction. Of course, it would be fascinating to develop these heuristic ideas and derive the consequences of the $\UT$ symmetry not only in the hydrodynamic limit but more generally in non-equilibrium QFT.  Based on this intuitive picture it is tempting to christen the symmetry as {\em KMS-gauge invariance}, but we presently refrain from doing so, hoping to  provide more concrete evidence in its favour in the near future.

Having described some of the conceptual implications of our construction of the eightfold Lagrangian, we now turn to  other aspects of our analysis. One useful perspective provided by the hydrodynamic effective actions in Classes L and $\LT$ relates to the entropy current. It is usually said that  the notion of the entropy current is mysterious from a microscopic perspective. However, from our  effective action viewpoint, the entropy current (or equivalently the free-energy current) is quite canonically derived as a Noether charge. This is a simple consequence of diffeomorphism invariance. It is well appreciated that in a dynamical theory of gravity where diffeomorphisms are gauged, there is a notion of entropy for spacetimes with horizons. This horizon entropy itself is constructed as a Noether charge \cite{Iyer:1994ys}. 
More generally one may ascribe a gravitational entropy to any diffeomorphism invariant dynamics. This point was  appreciated by the
authors of \cite{Iyer:1994ys} and was amplified in the context of ideal fluids by \cite{Iyer:1996ky}. This then begs a related question:
``Can we use the hydrodynamic entropy current to provide a Noether current for black holes out of equilibrium?'' The answer to this question is clearly in the affirmative based on the direct isomorphism between the hydrodynamic entropy current and black hole entropy in the fluid/gravity context \cite{Bhattacharyya:2008xc}. This then provides a possible avenue for ascertaining the non-equilibrium gravitational entropy current in time-dependent situations.\footnote{ We would like to thank Sayantani Bhattacharyya and Shiraz Minwalla for extensive discussions on this issue.}  

We noted in \S\ref{sec:intro} that often in addition to the second law of thermodynamics, one requires hydrodynamic transport to satisfy the Onsager relations  \cite{Onsager:1931fk,Onsager:1931uq}. This requirement is based for the most part on the empirical observation that most physical fluids satisfy these relations. One can phrase the Onsager relations in our language as follows: ``In systems with time-reversal symmetry there are no anti-symmetric contributions to transport.'' Paraphrasing this statement into our adiabatic classification, we learn that the Onsager relations forbid the presence of Class B terms in hydrodynamic transport. However, one must pay attention to the underlying assumptions about the dynamics; for example, the derivations in \cite{Onsager:1931fk,Onsager:1931uq}  rely  on either assuming that physical systems produce currents out of equilibrium only to extremize dissipation or (equivalently by the fluctuation-dissipation theorem) assume a Gaussian spectrum of fluctuations. 

We are for the most part agnostic about these relations, though our framework is broad enough to allow  exploration of such transport. For example, the relations we have found constraining Class B terms when deriving adiabatic constitutive relations from a Class L scalar Lagrangian density, such as the vanishing of Hall viscosity or the relation \eqref{eq:hyrel} in neutral fluids, can be viewed as higher-order versions of the Onsager relations. Indeed  one might argue that demanding that all non-dissipative transport outside equilibrium be derivable from a Landau-Ginzburg free energy functional as in Class L (without Schwinger-Keldysh doubling) provides an alternate route to ensuring the Onsager relations are satisfied, even in time-reversal violating systems. 

Perhaps more curiously, there appears to be an interesting constraint on  Class B terms in holographic fluids. Firstly all known examples of such transport obey a membrane paradigm like formula. By this we mean that the precise value of the Class B transport coefficient is given in terms of some geometric quantity evaluated on the horizon of an asymptotically AdS black hole. For instance, 
Eq.\ (32) of \cite{Saremi:2011ab} expresses Hall viscosity as such, while Eq.~(6.24) of \cite{Jensen:2011xb} provides an analogous expression for Hall conductivity. For parity-even fluids, the derivation of the universal Haack-Yarom relation between second order transport coefficients \eqref{eq:hyrel} also expresses the appropriate combination this way, cf., Eq.~(47) of \cite{Haack:2008xx}.
In all these cases the relevant membrane-paradigm quantity then vanishes at the horizon in the two-derivative theory because of some general feature of horizon geometry, forcing thence the vanishing of a Class B term. This seems rather generic in two derivative theories of gravity; inclusion of higher derivative interactions appears to allow non-vanishing Class B terms.  For example, vis a vis the Haack-Yarom relation, one finds that it is upheld to linear order in Gauss-Bonnet corrections to gravity \cite{Shaverin:2012kv}, but not beyond \cite{Yarom:2014kx,Grozdanov:2014kva}. Recent analysis by \cite{Grozdanov:2014kva} in the physically more interesting situation of string theory induced derivative corrections to Type IIB supergravity appear to uphold this to one higher order.  It would be interesting to analyze what feature of black hole horizons and gravitational dynamics plays a role in determining these aspects of transport.

Holographic fluids via the fluid/gravity correspondence \cite{Bhattacharyya:2008jc,Hubeny:2011hd} provide an ideal environment to test various statements about fluid dynamics in general by allowing one to be able to compute explicit constitutive relations. We find it reassuring to see evidence for the eightfold way in the examples studied hitherto. We outline in \S\ref{sec:homework} a set of questions that should give us even better insight into how holography implements the classification. What is perhaps curious however, is certain specificity in these fluids. While there is no a-priori reason that every aspect of transport allowed by the second law should find a holographic manifestation, it is nevertheless amazing to find further evidence for the near-idealness of these systems. It has been known for a very long time, starting from the seminal work of \cite{Policastro:2001yc}, that holographic fluids tend to want to minimize the shear viscosity and saturate the famous bound $\frac{\eta}{s} \geq \frac{1}{4\pi}$ \cite{Kovtun:2004de}.\footnote{ We refer the reader to  \cite{Buchel:2008vz,Cremonini:2011iq} for critical discussions of this bound and survey of attempts to violate it in various systems.} The low value of shear viscosity implies that these fluids minimize their entropy production for arbitrary flows. Fascinatingly, this statement also appears to be upheld at the next order in gradients. 

For Weyl invariant holographic fluids we find that the subleading contribution (third order) to entropy production is forced to vanish if the Class D transport coefficient $\lambda_1-\kappa$ vanishes.  This relation appears to hold in all two-derivative theories of gravity explored so far (but is violated upon inclusion of higher derivative interactions \cite{Grozdanov:2014kva}). This then suggests that holographic fluids obtained in the long-wavelength limit of strongly interacting quantum dynamics, obey a {\em principle of minimum dissipation}. One might suspect that all such transport be derivable from a  (Class L) Landau-Ginzburg free energy functional, which in turn should be obtainable from the bulk gravitational dynamics. The simplest test would be to derive \eqref{eq:weyl2lambdaFG} directly from the Einstein-Hilbert dynamics for gravity in asymptotically AdS spacetimes. It would be fascinating to develop this line of thought, for it should provide us with a geometric underpinning for deriving effective actions for generic non-equilibrium quantum dynamics. More generally the interplay of gravity with the extended framework of $\UT$ KMS-gauge invariant effective field theories could potentially provide important insights into formulating Wilsonian low energy dynamics for QFTs in mixed states.
\section{Homework problems}
\label{sec:homework}

We have at various stages of our discussion exemplified adiabatic fluids and the efficacy of the eightfold way using simple examples of hydrodynamic systems. While the structural aspects of our discussion are completely clear in the abstract, it is remarkable and reassuring that physical fluid systems are aware of the eightfold classification. We gave some arguments in favour of this using the analysis of holographic fluids (particularly the neutral Weyl invariant fluid) as well as in kinetic theory in \S\ref{sec:holofluids}.  

Ideally, of course, we would have liked to give many more examples and furthermore argue that the classification scheme can be used as an efficient organizational principle vis a vis actual computations. While we do believe this to be true, making further progress requires data in other hydrodynamic systems, which we do have at hand at present. This provides us with an opportunity to outline a set of problems, which we think are solvable, some of which perhaps more straightforwardly than others. We therefore leave our readers who have made it this far with a few problems that we feel are worth exploring.

One natural venue for exploration is simply to obtain explicit results for constitutive relations using the fluid/gravity correspondence for holographic fluids. Weyl invariant neutral fluids are already covered in our analysis. An obvious next step would be to examine Weyl invariant charged fluid dynamics by explicitly studying the Einstein-Maxwell (for parity-even) or Einstein-Maxwell-Chern-Simons (for anomalous parity-odd) examples. We note that \cite{Erdmenger:2008rm,Banerjee:2008th,Plewa:2012vt,Megias:2013joa} have studied specific aspects of transport in this particular  set-up up to second order in gradients, but the full set of constitutive relations require turning on all possible background sources  has yet to be done. Another direction to consider is neutral fluids without Weyl invariance, but since the simplest set-ups realizable holographically secretly enjoy higher dimensional conformal invariance \cite{Kanitscheider:2009as} (see also \cite{David:2009np,David:2010qc}) it is easy to intuit that these analyses won't provide more detailed 
insight (see however \cite{Bigazzi:2010ku} for explicit conformal symmetry breaking by sources). One could perhaps also make use of higher derivative gravitational theories, though in that case care must be exercised to ensure that one is dealing with a unitary QFT dual.  With this preamble let us state a few questions for each of our classes and some specific issues that can be understood.

\begin{itemize}
\item \textbf{Class $\PF$:} The hydrostatic analysis for charged fluids implies 17 hydrostatic forbidden constitutive relations at second order in gradients \cite{Banerjee:2012iz}. Find explicit expressions for these  constraints and further specialize to the case with Weyl invariance. Use the set-up described above to demonstrate that fluid/gravity automatically incorporates the hydrostatic constraints. Furthermore,  demonstrate that holographic fluids always respect the $\PF$ constraints at any order in the gradient expansion.
\item \textbf{Class L:} Write down the $\PS \cup \LS$ terms describing a fluid dual to Einstein-Maxwell system using the aforementioned holographic analysis, i.e., the analog of  \eqref{eq:weyl2lambda}. More generally give a bulk prescription to compute the effective action for this system from the gravitational dynamics (perhaps building on the ideas in \cite{Nickel:2010pr}).
\item \textbf{Class $\PV$:} In the holographic context demonstrate how $\PV$ terms arise from the extension of the Noether procedure of \cite{Iyer:1994ys} by \cite{Tachikawa:2006sz} to account for the presence of Chern-Simons interactions, i.e., extend the  leading order argument in \cite{Azeyanagi:2013xea,Azeyanagi:TobePublished} to arbitrary derivative order.
\item  \textbf{Class $\GV$:} This is the most interesting class of adiabatic constitutive relations. Thus far we have no data on such terms in any realistic hydrodynamic context. Do these terms arise in fluid-gravity correspondence at all? Or are they constrained to vanish as with certain Class B terms we have described in the text? Does there exist a simple membrane paradigmesque formula for them in terms of data on the horizon \cite{Iqbal:2008by}?
\item \textbf{Class B:}  Formulate a general membrane paradigm formula to compute all these terms in terms of horizon
data, extending the results of \cite{Saremi:2011ab}. Is there a role for an attractor like mechanism (for non-extremal black holes) which requires that these coefficients are in certain sense robust? 
\item \textbf{Class C:} Explore the presence of exactly conserved topological currents in fluid/gravity. For example show how the Lovelock terms in the gravitational description of even-dimensional AdS spacetimes map to Euler currents in the boundary fluid entropy current using \cite{Bhattacharyya:2008xc}.
\item \textbf{Class D:}  Derive a membrane paradigm formula for general Class D transport coefficients. Is there a role for the tensor valued differential operators $\Upsilon$ to show up in the bulk? Do holographic fluids always attempt to minimize the entropy production for a given fluid flow. Give evidence or disprove the {\em minimum entropy production conjecture} for two derivative theories of gravity. What is the corresponding statement in higher derivative theories of gravitational dynamics? 
\item \textbf{Class A:} Derive the Feynman-Vernon terms for Class A described in \cite{Haehl:2013hoa} and \S\ref{sec:anomalies}  from the fluid/gravity correspondence.
\item \textbf{Eightfold way:} Derive the eightfold way of transport from a  general Schwinger-Keldysh analysis \cite{Haehl:2014kq}. Extend this eightfold classification to systems with spontaneously broken symmetries and other Goldstone modes such as
superfluids (cf., \cite{Bhattacharya:2011tra} for a comprehensive treatment of superfluid dynamics and \cite{Sonner:2010yx,Herzog:2011ec,Bhattacharya:2011eea} for discussions about holographic superfluids). Is there a relation between the eightfold way and Hohenberg-Halperin classification \cite{Hohenberg:1977ym} of dynamic critical phenomena?

\end{itemize}

\newpage
\acknowledgments

It is a pleasure to thank Koushik Balasubramanian, Jyotirmoy Bhattacharya, Sayantani Bhattacharyya, Veronika Hubeny, Kristan Jensen, Hong Liu, Juan Maldacena, Shiraz Minwalla, Guy Moore,  Dam Son,  Andrei Starinets, and  Amos Yarom 
for enjoyable discussions  on various aspects of hydrodynamics.
FH  and MR would like to thank the Institute for Advanced Study, Princeton for hospitality during the course of this project. MR would in addition like to thank Yukawa Institute for Theoretical Physics, Kyoto, Univ. of Amsterdam, Aspen Center for Theoretical Physics and Tata Institute for Fundamental Research, Mumbai, for hospitality during the course of this project.

FH is supported by a Durham Doctoral Fellowship. RL gratefully acknowledges support from Institute for Advanced Study, Princeton. MR was supported in part by the Ambrose Monell foundation,  by the STFC Consolidated Grants ST/J000426/1 and ST/L000407/1, by the NSF grant under Grant No. PHY-1066293, and by the European Research Council under the European Union's Seventh Framework Programme (FP7/2007-2013), ERC Consolidator Grant Agreement ERC-2013-CoG-615443: SPiN (Symmetry Principles in Nature).

\appendix
\newpage

\part{Interconnections \& Generalizations}
\label{part:igeneral}
\hspace{1cm}

\section{Adiabaticity equation for consistent currents}
\label{sec:adcons}

We have elected to work with covariant currents and covariant anomalies in the bulk of the paper.
But, for some applications, it is useful to work with consistent currents and consistent  anomalies instead.
In this appendix we describe various statements about how adiabaticity equation changes if one works with consistent currents. Some salient results (such as those necessary for hydrostatics) have already been quoted in the main text; the derivation of these results will be found herein.

\subsection{Bardeen-Zumino currents}

To begin with, we need to pass to one of the consistent anomalies from the covariant
anomalies we have been working with earlier. To do this, we choose
a Chern-Simons form $\ICS$ such that $d\ICS=\fP$ where $\fP$ is the anomaly
polynomial of the system under consideration. The consistent anomalies are then given by
\begin{equation}
\mathcal{J}\  \hodge 1 \equiv  \frac{\partial \ICS}{\partial \fA}\ ,\quad
\mathcal{T}_\mu{}^\nu\  \hodge 1 \equiv  \frac{\partial \ICS}{\partial \fGamma^\mu{}_\nu}\ .
\end{equation}
It is also useful to work with Bardeen-Zumino current and energy momentum tensor $\{\JBZ^\mu, \TBZ^{\mu\nu} \}$ which are local functions of sources $\{A_\mu, g_{\mu\nu} \}$ given by
\begin{equation}
\begin{split}
\JBZ^\mu\  \hodge dx_\mu &\equiv  \frac{\partial \ICS}{\partial \fF}\ ,\quad
(\SpBZ)^{\sigma\nu}{}_\mu\  \hodge dx_\sigma \equiv  2\frac{\partial \ICS}{\partial \fR^\mu{}_\nu}\ , \\
\TBZ^{\mu\nu} &\equiv \half \nabla_{\sigma} \brk{(\SpBZ)^{\mu[\nu\sigma]} + (\SpBZ)^{\nu[\mu\sigma]} - (\SpBZ)^{\sigma(\nu\mu)} } .
\end{split}
\end{equation}

We will not need the detailed forms above for our analysis, though given any quantum system we can determine the anomaly polynomial and thence the currents if necessary. All we need for the moment is the following fact: the
Bardeen-Zumino current and energy momentum tensor obey conservation type equations with a right hand side that has the difference between  the covariant anomaly and the consistent anomaly, viz.,
\begin{equation}
\begin{split}
D_\mu\JBZ^\mu & =  \JH^\perp-\mathcal{J} \,, \\
\nabla_\nu \TBZ^{\mu\nu} -(\JBZ)_\nu \,F^{\mu\nu} &=\THall^{\mu\perp} +A^\mu \mathcal{J} +\frac{g^{\mu\alpha}}{\sqrt{-g}}\partial_\nu\prn{\sqrt{-g}\ \mathcal{T}_\alpha{}^\nu }\,.
\end{split}
\end{equation}

We now want to derive an adiabaticity type equation for  Bardeen-Zumino currents using these
identities. To this end, consider the combination
\begin{equation}
\begin{split}
\Kbeta_\mu &\prn{\nabla_\nu \TBZ^{\mu\nu} -(\JBZ)_\nu \,F^{\mu\nu} - \THall^{\mu\perp} }
+ (\LambdaB +  \Kbeta^\sigma A_\sigma) \cdot \prn{ D_\mu\JBZ^\mu -  \JH^\perp } \\
& = \frac{ \Kbeta^\mu }{\sqrt{-g}}\partial_\nu\prn{\sqrt{-g}\ \mathcal{T}_\mu{}^\nu }
- \LambdaB  \cdot \mathcal{J} \\
& =\nabla_\sigma\prn{ \Kbeta^\nu \mathcal{T}_\nu{}^\sigma } - (\partial_\nu \Kbeta^\mu)  \,\mathcal{T}_\mu{}^\nu
- \LambdaB  \cdot \mathcal{J} \\
\end{split}
\end{equation}

If we define the  Bardeen-Zumino entropy current as
\begin{equation}\label{eq:JSBZ}
J_{S,BZ}^\sigma \equiv  - \Kbeta^\nu \,\mathcal{T}_\nu{}^\sigma
\end{equation}
we can then write the adiabaticity equation satisfied by Bardeen-Zumino currents as
\begin{equation}\label{eq:AdiabBZ}
\begin{split}
\nabla_\sigma J_{S,BZ}^\sigma +  \Kbeta_\mu &\prn{\nabla_\nu \TBZ^{\mu\nu} -(\JBZ)_\nu F^{\mu\nu} -
\THall^{\mu\perp} }
+ (\LambdaB +  \Kbeta^\sigma A_\sigma) \cdot \prn{ D_\mu\JBZ^\mu -  \JH^\perp } \\
&=  - (\partial_\nu \Kbeta^\mu) \, \mathcal{T}_\mu{}^\nu
- \LambdaB  \cdot \mathcal{J} \\
\end{split}
\end{equation}

The \textbf{consistent} currents are defined as the difference of the covariant currents and the Bardeen-Zumino currents, viz.,
\begin{equation}
\begin{split}
J_{S,cons}^\mu \equiv  J_S^\mu -  J_{S,BZ}^\mu\ ,\qquad
T_{cons}^{\mu\nu} \equiv  T^{\mu\nu} - \TBZ^{\mu\nu} \ ,\qquad
J_{cons}^\mu \equiv  J^\mu -  \JBZ^\mu\ .
\end{split}
\end{equation}
By subtracting \eqref{eq:AdiabBZ} from \eqref{eq:Adiabaticity}, we obtain the adiabaticity equation written
in terms of consistent currents and consistent anomalies:
\begin{equation}\label{eq:AdiabCons}
\begin{split}
\nabla_\mu J_{S,cons}^\mu &+ \Kbeta_\mu\prn{\nabla_\nu T_{cons}^{\mu\nu}-(J_{cons})_\nu \cdot F^{\mu\nu} }\\
&+ (\LambdaB+\Kbeta^\lambda A_\lambda) \cdot D_\nu J^\nu =
(\partial_\nu \Kbeta^\mu)  \mathcal{T}_\mu{}^\nu
+ \LambdaB  \cdot \mathcal{J}
\end{split}
\end{equation}
This then is the analog of \eqref{eq:Adiabaticity} in terms of the consistent currents. This expression is useful when we directly want to work with effective actions (without introducing Chern-Simons terms).

\subsection{The Consistent Gibbs current}

 If we wish to work directly with the Gibbs current as in \S\ref{sec:afree}, then one can introduce the Bardeen-Zumino free energy current given \eqref{eq:JSBZ} via:
\begin{equation}
\begin{split}
\mathcal{G}_{BZ}^\mu &\equiv -T\brk{ J_{S,BZ}^\mu+\Kbeta_\nu \,\TBZ^{\mu\nu}+
(\LambdaB + \Kbeta^\sigma A_\sigma) \cdot \JBZ^\mu } \\
&=u_\nu \mathcal{T}^{\mu\nu} -
\brk{ u_\nu \,\TBZ^{\mu\nu} + \mu \cdot \JBZ^\mu } .
\end{split}
\end{equation}
This Bardeen-Zumino free energy current satisfies the grand canonical counterpart of \eqref{eq:AdiabBZ}
\begin{equation}\label{eq:AdiabBZG}
\begin{split}
-\brk{\nabla_\sigma\prn{\frac{\mathcal{G}_{BZ}^\sigma}{T}}-\frac{\mathcal{G}_{_H}^\perp}{T}}&=
\half  \TBZ^{\mu\nu}\diffB  g_{\mu\nu} + \JBZ^\mu \cdot \diffB  A_\mu
- (\partial_\nu \Kbeta^\mu)  \mathcal{T}_\mu{}^\nu
- \LambdaB  \cdot \mathcal{J} \,.
\end{split}
\end{equation}
Subtracting this equation from \eqref{eq:AdiabaticityG}, we get
\begin{equation}\label{eq:AdiabConsG}
\begin{split}
-\nabla_\sigma\prn{\frac{\mathcal{G}_{cons}^\sigma}{T}}&=
\half  T_{cons}^{\mu\nu}\;\diffB  g_{\mu\nu} + J_{cons}^\mu \cdot \diffB  A_\mu
+  (\partial_\nu \Kbeta^\mu)  \mathcal{T}_\mu{}^\nu
+\LambdaB  \cdot \mathcal{J} \,,
\end{split}
\end{equation}
where we have defined the consistent free energy current
\begin{equation}
\mathcal{G}_{cons}^\sigma\equiv
\mathcal{G}^\sigma - \mathcal{G}_{BZ}^\sigma \,.
\end{equation}
Equation \eqref{eq:AdiabConsG} plays a useful role in constructing the equilibrium partition function for anomalous hydrodynamics.  The current $\mathcal{G}_{cons}^\sigma/T$ is conserved in the hydrostatic limit
(unlike $\mathcal{G}^\sigma/T$ itself) and hence can be integrated over the base space to give a generating function of hydrostatic correlation functions.

\subsection{Noether construction for free-energy current }

In this subsection, we will show that the free-energy currents for Class A can be  obtained by a Noether type construction. To do this, we begin by considering the variation of the   transgression form introduced in \cite{Jensen:2013kka}, see \S\ref{sec:anomalies}
\[ \VP \equiv \frac{\fu}{2\fomega}\wedge \big(\fP[\fF,\fR]-\fPh[\fFh,\fRh]\big).\]
Using the identities
\begin{equation}
\begin{split}
\delta (2\fomega) &= \frac{1}{T} d(T\delta \fu) + T\delta \fu \wedge \diffB \fu + \fu \wedge(\ldots)\\
\delta \fB  &= D(\delta \fA + \mu \delta \fu) -\frac{\mu}{T} d(T\delta \fu) + T\delta \fu \wedge \diffB \fA + \fu \wedge(\ldots)\\
\delta (\fBR)^\mu{}_\nu &= D(\delta \fGamma^\mu{}_\nu + T D_\nu\Kbeta^\mu \delta \fu) -D_\nu\Kbeta^\mu d(T\delta \fu) + T\delta \fu \wedge \diffB \fGamma^\mu{}_\nu + \fu \wedge(\ldots)\\ 
\end{split}
\end{equation}
we can write the variation of $\VP$ in the following form:
\begin{equation}
\begin{split}
\delta \VP &= T\delta \fu \wedge 
\Bigl\{ -\ic_\Kbeta \VP 
+  \diffB \fu \wedge  \frac{\partial \VP}{\partial (2\fomega)} +   \diffB \fA \wedge  \frac{\partial \VP}{\partial \fB} 
+   \diffB \fGamma^\mu{}_\nu  \wedge  \frac{\partial \VP}{\partial (\fBR)^\mu{}_\nu} \Bigr.\\
&\qquad
+   \half g^{\mu\sigma} \diffB g_{\nu\sigma}\  D\prn{ \frac{\partial \VP}{\partial (\fBR)^\mu{}_\nu} }
\\
&\qquad
+ \frac{\mu}{T} \cdot \brk{ D\prn{ \frac{\partial \VP}{\partial \fB} } + \frac{\partial \fPh}{\partial \fFh} }
+ \frac{1}{T} \Omega^\mu{}_\nu  \brk{ D\prn{ \frac{\partial \VP}{\partial (\fBR)^\mu{}_\nu} } + \frac{\partial \fPh}{\partial \fRh^\mu{}_\nu} }
\\
&\qquad \Bigl. 
+ d\brk{\frac{1}{T} \frac{\partial \VP}{\partial (2\fomega)} 
-\frac{\mu}{T} \cdot \frac{\partial \VP}{\partial \fB} 
- \frac{1}{T} \Omega^\mu{}_\nu  \frac{\partial \VP}{\partial (\fBR)^\mu{}_\nu}
- \half g^{\mu\sigma} \diffB g_{\nu\sigma}  \ \frac{\partial \VP}{\partial (\fBR)^\mu{}_\nu} 
} 
\Bigr\} \\
&\qquad 
+ \delta \fA \cdot D\prn{ \frac{\partial \VP}{\partial \fB} }
+   \delta \fGamma^\mu{}_\nu  \wedge  D\prn{ \frac{\partial \VP}{\partial (\fBR)^\mu{}_\nu} }\\
&\qquad
\qquad- \delta(\mu\fu) \frac{\partial \fPh}{\partial \fFh} -  \delta(\Omega^\mu{}_\nu \fu) \frac{\partial \fPh}{\partial \fRh^\mu{}_\nu}\\
&\qquad
+ d \bigbr{ \delta \fu \wedge \frac{\partial \VP}{\partial (2\fomega)} + \delta \fA \wedge  \frac{\partial \VP}{\partial \fB} 
+   \delta \fGamma^\mu{}_\nu  \wedge  \frac{\partial \VP}{\partial (\fBR)^\mu{}_\nu} } \,.
\end{split}
\end{equation}

Following the authors of \cite{Jensen:2013kka}, we also introduce a transgression form 
\[ \WCS \equiv \frac{\fu}{2\fomega}\wedge \big(\ICS[\fF,\fR]-\ICS[\fFh,\fRh]\big)\,, \]
which is related to $\VP$ via the relation $\VP+d\WCS=\ICS[\fF,\fR]-\ICS[\fFh,\fRh]$. Similar
manipulations for variation of $\WCS$ yields
\begin{equation}
\begin{split}
\delta \WCS &= T\delta \fu \wedge 
\Bigl\{  -\ic_\Kbeta \WCS -\LambdaB \frac{\partial \WCS}{\partial \fA} - \partial_\nu \Kbeta^\mu \frac{\partial \WCS}{\partial \fGamma^\mu{}_\nu} 
\Bigr.\\
&\qquad 
+  \diffB \fu \wedge  \frac{\partial \WCS}{\partial (2\fomega)} +   \diffB \fA \wedge  \frac{\partial \WCS}{\partial \fB} 
+   \diffB \fGamma^\mu{}_\nu  \wedge  \frac{\partial \WCS}{\partial (\fBR)^\mu{}_\nu} 
+   \half g^{\mu\sigma} \diffB g_{\nu\sigma}\  D\prn{ \frac{\partial \WCS}{\partial (\fBR)^\mu{}_\nu} }
\\
&\qquad + \frac{\mu}{T} \cdot \brk{\frac{\partial \WCS}{\partial \fA} + D\prn{ \frac{\partial \WCS}{\partial \fB} } + \frac{\partial \ICS[\fAh,\fGammah]}{\partial \fFh} } \\
&\qquad
+ \frac{1}{T} \Omega^\mu{}_\nu  \brk{ \frac{\partial \WCS}{\partial \fGamma^\mu{}_\nu}+ D\prn{ \frac{\partial \WCS}{\partial (\fBR)^\mu{}_\nu} } + \frac{\partial \ICS[\fAh,\fGammah]}{\partial \fRh^\mu{}_\nu} }
\\
&\qquad \Bigl.+ d\brk{\frac{1}{T} \frac{\partial \WCS}{\partial (2\fomega)} 
-\frac{\mu}{T} \cdot 
\frac{\partial \WCS}{\partial \fB}
- \frac{1}{T} \Omega^\mu{}_\nu  \frac{\partial \WCS}{\partial (\fBR)^\mu{}_\nu}
- \half g^{\mu\sigma} \diffB g_{\nu\sigma}  \ \frac{\partial \WCS}{\partial (\fBR)^\mu{}_\nu} 
} 
\Bigr\}\\
&\qquad 
+ \delta \fA \cdot \bigbr{ \frac{\partial \WCS}{\partial \fA} + D\prn{ \frac{\partial \WCS}{\partial \fB} } } 
+   \delta \fGamma^\mu{}_\nu  \wedge  D\prn{ \frac{\partial \WCS}{\partial (\fBR)^\mu{}_\nu} }\\
&\qquad
\qquad- \delta(\mu\fu) \frac{\partial \ICS[\fAh,\fGammah]}{\partial \fFh} -  \delta(\Omega^\mu{}_\nu \fu) \frac{\partial \ICS[\fAh,\fGammah]}{\partial \fRh^\mu{}_\nu}\\
&\qquad
+ d \bigbr{ \delta \fu \wedge \frac{\partial \WCS}{\partial (2\fomega)} + \delta \fA \wedge \frac{\partial \WCS}{\partial \fB} 
+   \delta \fGamma^\mu{}_\nu  \wedge  \frac{\partial \WCS}{\partial (\fBR)^\mu{}_\nu}}
\end{split}
\end{equation}

We can then use   $\delta\VP+d\delta \WCS=\delta\ICS[\fF,\fR]-\delta\ICS[\fFh,\fRh]$ to get the following identities:
\begin{equation}
\begin{split}
\star \fJP &\equiv \frac{\partial \VP}{\partial \fB} = 
-\bigbr{ \frac{\partial \WCS}{\partial \fA} + D\prn{ \frac{\partial \WCS}{\partial \fB} } } +\frac{\partial \ICS}{\partial \fF} - \frac{\partial \ICS[\fAh,\fGammah]}{\partial \fFh}  \\
D\star \fJP &= \frac{\partial \fP}{\partial \fF} - \frac{\partial \fPh}{\partial \fFh} \\
(\star \fSP)^\nu{}_\mu &\equiv 2\frac{\partial \VP}{\partial (\fBR)^\mu{}_\nu}= 
-2\bigbr{ \frac{\partial \WCS}{\partial \fGamma^\mu{}_\nu} + D\prn{ \frac{\partial \WCS}{\partial (\fBR)^\mu{}_\nu} } } +2\frac{\partial \ICS}{\partial \fR^\mu{}_\nu} -2 \frac{\partial \ICS[\fAh,\fGammah]}{\partial \fRh^\mu{}_\nu}  \\
(D\star \fSP)^\nu{}_\mu  &= 2\frac{\partial \fP}{\partial \fR^\mu{}_\nu} - 2\frac{\partial \fPh}{\partial \fRh^\mu{}_\nu} \,.
\end{split}
\end{equation}
Further, we get an identity of the form 
\begin{equation}
\begin{split}
\hodgeB \form{N}_{_H} &\equiv \frac{\mu}{T} \cdot \frac{\partial \fP}{\partial \fF} 
+ \frac{1}{T} \Omega^\mu{}_\nu   \frac{\partial \fP}{\partial \fR^\mu{}_\nu} 
\\
&  = \ic_\Kbeta \VP 
- \brk{  \diffB \fu  \wedge \frac{\partial \VP}{\partial (2\fomega)} +   \diffB \fA \wedge \frac{\partial \VP}{\partial \fB} +  \diffB \fGamma^\mu{}_\nu  \wedge  \frac{\partial \VP}{\partial (\fBR)^\mu{}_\nu}} \\
&\qquad - d\brk{\frac{1}{T} \frac{\partial \VP}{\partial (2\fomega)} 
-\frac{\mu}{T} \cdot \frac{\partial \VP}{\partial \fB} 
- \frac{1}{T} \Omega^\mu{}_\nu  \frac{\partial \VP}{\partial (\fBR)^\mu{}_\nu}
- \half g^{\mu\sigma} \diffB g_{\nu\sigma}  \ \frac{\partial \VP}{\partial (\fBR)^\mu{}_\nu} 
} \,. \\
\end{split}
\end{equation}
This is the statement that the bulk free energy current is same as the bulk Noether current corresponding to a bulk Lagrangian $\VP$
up to Komar terms. Taking the normal component and restricting it to the boundary leads to the statement that the covariant free energy current  in the boundary, defined via
\begin{equation}
\begin{split}
\star \form{N}_{_{\fP,cov}}&\equiv -\frac{1}{T} \frac{\partial \VP}{\partial (2\fomega)} 
+\frac{\mu}{T} \cdot \frac{\partial \VP}{\partial \fB} 
+ \frac{1}{T} \Omega^\mu{}_\nu  \frac{\partial \VP}{\partial (\fBR)^\mu{}_\nu}
+\half g^{\mu\sigma} \diffB g_{\nu\sigma}  \ \frac{\partial \VP}{\partial (\fBR)^\mu{}_\nu} \,,
\end{split}
\end{equation}
satisfies the adiabaticity equation with covariant anomaly.
 
The last identity we obtain takes the form 
\begin{equation}
\begin{split}
\star \form{N}_{_{\fP,cons}} &\equiv
 -\frac{1}{T} \frac{\partial \VP}{\partial (2\fomega)} + \frac{\mu}{T} \cdot 
\brk{ \frac{\partial \VP}{\partial \fB} - \frac{\partial \ICS}{\partial \fF} } 
+ \frac{1}{T} \Omega^\mu{}_\nu  \brk{\frac{\partial \VP}{\partial (\fBR)^\mu{}_\nu} - \frac{\partial \ICS}{\partial \fR^\mu{}_\nu} }
\\
&=
 -\ic_\Kbeta \WCS -\LambdaB \frac{\partial \WCS}{\partial \fA} - \partial_\nu \Kbeta^\mu \frac{\partial \WCS}{\partial \fGamma^\mu{}_\nu}  \\
&+  \diffB \fu \wedge \frac{\partial \WCS}{\partial (2\fomega)} +   \diffB \fA  \wedge \frac{\partial \WCS}{\partial \fB} 
+   \diffB \fGamma^\mu{}_\nu  \wedge  \frac{\partial \WCS}{\partial (\fBR)^\mu{}_\nu} 
+   \half g^{\mu\sigma} \diffB g_{\nu\sigma}\  D\prn{ \frac{\partial \WCS}{\partial (\fBR)^\mu{}_\nu} }
\\
&+ d\brk{\frac{1}{T} \frac{\partial \WCS}{\partial (2\fomega)} -\frac{\mu}{T} \cdot 
\frac{\partial \VP}{\partial \WCS} } 
\end{split}
\end{equation}
This is the statement that the consistent free energy current  in the boundary  is same as the boundary Noether current  corresponding to a boundary Lagrangian density $\WCS$ up to  Komar terms.

\section{Class ND: From adiabatic to non-dissipative fluids}
\label{sec:ndf}

As presaged at the end of  \S\ref{sec:Leoms}, there is a striking similarity between the Class L Lagrangian solutions to the adiabaticity equation and the effective action formalism for non-dissipative fluids which was developed earlier in \cite{Dubovsky:2011sj} and explored in \cite{Bhattacharya:2012zx,Haehl:2013kra}.\footnote{ There are also results in the non-dissipative effective action framework for anomalous transport \cite{Dubovsky:2011sk, Haehl:2013hoa} which we will defer till \S\ref{sec:anomalies}. As a result the discussion in this section will be restricted to non-anomalous, non-dissipative fluids.} Intuitively, it is clear that the family of non-dissipative fluids where every on-shell solution to the dynamical equations of motion is constrained to not produce entropy, should be a special case of adiabatic fluids, which are engineered to uplift the statement of entropy conservation off-shell in a suitable manner.
The similarities are also striking in the explicit examples discussed in \S\ref{sec:examples}.

However, at a basic level there is a crucial distinction between the framework of non-dissipative fluids and the story we have developed thus far for adiabatic fluids. To wit, the physical fields in the adiabatic fluid formalism are the fluid velocity and the intensive (local) thermodynamic parameters characterizing the fluctuating Gibbs density matrix. On the other hand the non-dissipative fluids use entropy density as a primary variable instead of the temperature. One can nevertheless pass between the two constructions by realizing that temperature and entropy being conjugate variables one can exchange the two by the simple expedient of a Legendre transform.

Aided by this observation, we now consider the Legendre transform of $\Lag\brk{\hfields}$  trading the temperature $T$ for its conjugate variable $s$. Part of our basic motivation is of course to   make contact with the  existing literature on effective actions for hydrodynamics. We will argue upon carrying out this  Legendre transform and thence passing to a suitable gauge, we recover the  effective action formalism. 

In effect we will establish that the family of non-dissipative fluids which we call Class ND are encapsulated within Class L family of adiabatic fluids as presaged in \S\ref{sec:intro}. This map will in particular make transparent the physical origin of the symmetries employed in the effective action formalism.

\subsection{Legendre transformation to an entropic description}
\label{sec:legent}

Let us begin by shifting  from a description in terms of $\{\Kbeta^\sigma,\LambdaB\}$
to a description in terms of a $(d-1)$ form and a $d$-form which we denote as
$\{\Snd_{\alpha_1\ldots\alpha_{d-1}}, (\Lnd)_{\alpha_1\ldots\alpha_d}\}$ respectively. They are defined  via their Hodge duals:
\begin{equation}
\begin{split}
\epform{\Snd}^\sigma &\equiv
\frac{1}{(d-1)!}\, \varepsilon^{\sigma\alpha_1\ldots\alpha_{d-1}} \;
{\bf S}_{\alpha_1\ldots\alpha_{d-1}}
= T \,s\, \Kbeta^\sigma = s \,u^\sigma \\
\epform{\!\Lnd} & \equiv
\frac{1}{d!}\; \varepsilon^{\alpha_1\ldots\alpha_d} \;
(\Lnd)_{\alpha_1\ldots\alpha_d}
= T \,s \,\LambdaB = s\; (\mu-u^\alpha\, A_\alpha)\,.
\end{split}
\label{eq:legvar}
\end{equation}
These forms $\{\Snd_{\alpha_1\ldots\alpha_{d-1}}, (\Lnd)_{\alpha_1\ldots\alpha_d}\}$ will play
a role analogous to $\{\Kbeta^\sigma,\LambdaB\}$ in the Legendre transformed description.
To avoid unnecessary clutter, we have adopted above a concise notation for
the contraction with $\varepsilon$ tensor which we will use from now on. For any
$p$-form ${\bf V}_{\alpha_1\alpha_2\ldots \alpha_p}$, we define
\begin{equation}
\begin{split}
\epform{\bf V}^{\sigma_1\sigma_2\ldots \sigma_{d-p}} &\equiv
\frac{1}{p!}\;
\varepsilon^{\sigma_1\sigma_2\ldots \sigma_{d-p}\alpha_1\alpha_2\ldots \alpha_p}
{\bf V}_{\alpha_1\alpha_2\ldots \alpha_p} \\
\end{split}
\end{equation}

Let us first attempt to make contact with the variational principle for Lagrangian theories
\eqref{eq:LagVar} by re-expressing them in term of our new variables. To do so we begin
with the following identities obtained by varying \eqref{eq:legvar}:
\begin{align}
T\; \delta \Kbeta^\sigma &=\frac{1}{s}\; \epform{\!\delta \Snd}^\sigma
-\frac{\Kbeta^\sigma}{s\,\sqrt{-g}}\;\delta\prn{\sqrt{-g}\ T\,s} \,,
\nonumber \\
T\;\delta \LambdaB &=\frac{1}{s}\; \epform{\!\delta \Lnd}
-\frac{\LambdaB }{s\,\sqrt{-g}}\;\delta\prn{\sqrt{-g}\ T\, s} \,.
\end{align}
From these expressions one can easily check that
\begin{equation}
\begin{split}
T \,\aheat_\sigma \,\delta \Kbeta^\sigma &
+ T\, \acharge \cdot \prn{\delta\LambdaB+ A_\sigma \delta \Kbeta^\sigma}-
\frac{1}{\sqrt{-g}}\delta\prn{\sqrt{-g}\ T\,s}\\
&= \frac{\aheat_\sigma}{s} \; \epform{\Snd}^\sigma
+\frac{\acharge}{s}\cdot \bigg( \epform{\Lnd} +  A_\sigma \, \epform{\Snd}^\sigma  \bigg) .
\end{split}
\end{equation}
In deriving the above we have used  the identity
$s+\aheat_\sigma \,\Kbeta^\sigma+\acharge\cdot(\LambdaB+A_\sigma\, \Kbeta^\sigma)=0$ from
\eqref{eq:sVZeta}. We can then rewrite \eqref{eq:LagVar} as
\begin{equation}\label{eq:LagVarLegendre}
\begin{split}
\frac{1}{\sqrt{-g}}\delta\bigg(\sqrt{-g}\ (\Lag-Ts)\bigg) &= \half T^{\mu\nu}\delta g_{\mu\nu} + J^\mu \cdot \delta A_\mu +\nabla_\mu
(\PSymplPot{})^\mu\\
&\qquad +
\frac{\aheat_\sigma}{s} \; \epform{\Snd}^\sigma
+\frac{\acharge}{s}\cdot \bigg( \epform{\Lnd} +  A_\sigma \, \epform{\Snd}^\sigma  \bigg)
\end{split}
\end{equation}

The rewritten variational equation can be interpreted as follows. We think of
$\Lag-T\,s$ as a functional of the variables
$\{g_{\mu\nu},A_\mu,{\bf S}_{\alpha_1\ldots\alpha_{d-1}}, {\bf \Lambda_S}_{\alpha_1\ldots\alpha_d}\}$ which we can collectively call $\hfields_S$ with the subscript denoting passing to the entropic description. To wit, we simply write
\begin{equation}
\int \sqrt{-g}\ \Lag_S\brk{\hfields_S}  \equiv \int \sqrt{-g}\ \prn{\Lag\brk{\hfields}-T\,s }
\bigg|_{\{\Kbeta,\LambdaB\} \mapsto \{{\bf S},{\bf \Lambda_S} \}}
\end{equation}

Then \eqref{eq:LagVarLegendre} is the defining variational formula for us and gives us the functionals $\{\aheat_\sigma,\acharge\}$ in this Legendre transformed description. Since entropy density $s$ has taken a primary role, it follows that the temperature $T$ is then a derived  quantity which can then be computed from
\begin{equation}
T=-\frac{\delta}{\delta s}\int\sqrt{-g}\ \Lag_S\brk{\hfields_S}
= -\frac{1}{s^2}\; \bigg(\aheat_\sigma \; \epform{\Snd}^\sigma+\acharge\cdot \prn{
\epform{\Lnd}+A_\sigma\, \epform{\Snd}^\sigma} \bigg)
= -\frac{1}{s}\prn{\mu\cdot\acharge+u^\sigma \,\aheat_\sigma}
\end{equation}

There is no loss of information in the translation between the two descriptions as long as we restrict to
non-anomalous fluids. More specifically, much of our analysis in previous sections can be repeated in this Legendre transformed
variables. For instance,  the Bianchi identities \eqref{eq:LHydroEq}  can be translated into our new language as 
\begin{equation}\label{eq:LTsHydroEq}
\begin{split}
\nabla_\nu T^{\mu\nu}&= J_\nu \cdot F^{\mu\nu}
+\frac{g^{\mu\nu}}{\sqrt{-g}}\; \delta_{\epform{\Snd}}\prn{\sqrt{-g}\ \frac{1}{s}\aheat_\nu}
+  g^{\mu\nu} \;\frac{1}{s}\acharge \cdot \delta_{\epform{\Snd}} A_\nu +\nabla^\mu(T\,s)\\
D_\sigma J^\sigma &= \frac{1}{\sqrt{-g}}\delta_{\epform{\Snd}}\prn{\sqrt{-g}\ \frac{1}{s}\,\acharge}\ .
\end{split}
\end{equation}
where $\delta_{\epform{\Snd}}$ denotes Lie derivative along $\{\epform{\Snd}^\sigma,\epform{\Lnd}\}$. Moreover, the Noether current can be shown in this description to be given by
\begin{equation}\label{eq:NchiS}
\begin{split}
{}^{\bf S}\N^\mu\brk{\Xfields} &\equiv \xi_\nu T^{\mu\nu}+ (\Lambda+\xi^\lambda A_\lambda) \cdot J^\mu\\
&\quad +\frac{\aheat_\nu}{s}
\prn{ \xi^\mu \epform{\Snd}^\nu- \epform{\Snd}^\mu\, \xi^\nu }
+\frac{\acharge}{s}\cdot \prn{\xi^\mu ( \epform{\Lnd}+\epform{\Snd}^\lambda A_\lambda)-
\epform{\Snd}^\mu \,(\Lambda+\xi^\lambda A_\lambda) }
\end{split}
\end{equation}
providing the translation of \eqref{eq:Nchi}.
As before, \eqref{eq:LTsHydroEq} then imply adiabaticity equation as a corollary.

It is also easy to see that the hydrodynamic equations can be obtained by
extremizing $\int \sqrt{-g}\ \Lag_S\brk{\hfields_S}$ with respect to arbitrary Lie drags of
the forms $\{\Snd_{\nu_1\ldots\nu_{d-1}}, (\Lnd)_{\nu_1\ldots\nu_d}\}$.
However, written in this form, it is not evident that these equations obey the hydrostatic
principle. In general, this Legendre transformed description obscures
the physics of the hydrostatic state which is better dealt with using
$\Lag$ instead. This makes intuitive physical sense, since the hydrostatic configurations when thought about in terms of the Euclidean background geometry give the temperature a geometric interpretation in terms of the size of the thermal circle, but do not accord a similar preferred status to the entropy density.\footnote{ The development of the adiabatic fluid formalism was in fact precipitated by attempting to recover the hydrostatic partition function from Class ND.}

\subsection{Gauge fixing and the non-dissipative effective action}
\label{sec:gfnd}

Now, we would like to pass to a description in terms of a set of physical dynamical
fields $\{\phi^a, \cnd\}$ such that extremizing $\int \sqrt{-g}\ \Lag_S\brk{\hfields_S}$ with respect to them 
gives the correct hydrodynamic equations. Without further ado, based on our preceding analysis of \S\ref{sec:reffields} let us introduce a reference manifold $\Mref$ endowed
with reference forms $\{\Sndref_{ a_1\ldots a_{d-1}}[\phi], (\Lndref)_{a_1\ldots a_d}[\phi]\}$ living on it.
We then pull-back these reference forms to real spacetime using maps $\{\phi^a,\cnd\}$.\footnote{ Since we are working with the Legendre transformed system the diffeomorphism and gauge transformation fields $\{\phi^a,\cnd\} $  are related in a complicated manner to the fields $\{\varphi, \cnd\}$ we introduced earlier in \S\ref{sec:refgauge}. We prefer to therefore use a different notation (at least stylistically) to emphasize the distinction. As we will see soon the fields introduced in this section are the ones that naturally arise in the non-dissipative fluid effective actions studied in
\cite{Bhattacharya:2012zx,Haehl:2013kra,Haehl:2013hoa}.} We have
\begin{equation}
\begin{split}
\Snd_{\nu_1\ldots\nu_{d-1}} &= \Sndref_{ a_1\ldots a_{d-1}} \prod_{i=1}^{d-1}\partial_{\nu_i}\phi^{a_i}
\\
(\Lnd)_{\nu_1\ldots\nu_d} &= \cnd\; (\Lndref)_{a_1\ldots a_d} \;\cnd^{-1}\  \prod_{i=1}^{d}\partial_{\nu_i}\phi^{a_i} + d\; (\partial_{[\nu_1} \cnd)\ \cnd^{-1}\;
\Snd_{\nu_2\ldots\nu_{d}]}
\end{split}
\label{eq:RefND}
\end{equation}
or equivalently more compactly as just
\begin{equation}
\begin{split}
\epform{\Snd}^\sigma &= e^\sigma_a\  \epformref{\Sndref}^a\\
\epform{\Lnd} &= \cnd\; \epformref{\Sndref} \;\cnd^{-1} +
e^\sigma_a\  \epformref{\Sndref}^a \; (\partial_{\sigma} \cnd)\ \cnd^{-1}
\end{split}
\end{equation}
where we have defined also an $\varepsilon$-symbol  (denoted ${\mathbb e}$) associated with the push forward metric $\gref$:
\begin{equation}
{\mathbb e}^{\,a_1\ldots a_d} \equiv \varepsilon^{\nu_1\ldots \nu_d} \; \prod_{i=1}^{d}
\partial_{\nu_i}\phi^{a_i}
\end{equation}

To compute the equations of motion, etc., we obtain first by varying this expression,
\begin{equation}
\begin{split}
\epform{\!\delta \Snd}^\sigma &= e^\sigma_a \;\epformref{\! \delta\Sndref}
^a+ \epform{\!\delta_{e\delta\phi} \Snd}^\sigma\\
&= e^\sigma_a \;\epformref{\! \delta\Sndref}-\delta_{\epform{\Snd}}(e^\sigma_a\;\delta \phi^a)+ \Snd^\sigma \, \nabla_\lambda(e^\lambda_a\,\delta \phi^a)\\
\epform{\Lnd} &= \cnd\;(\epformref{\! \delta \Lndref})\; \cnd^{-1} + e^\sigma_a \;\epformref{\!\delta\Sndref}^a (\partial_{\sigma} \cnd)\ \cnd^{-1}\\
&\quad +\delta_{\epform{\Snd}}\brk{(\delta \cnd) \cnd^{-1}-e^\sigma_a\,(\partial_\sigma \cnd) \,\cnd^{-1}}+ \epform{\Lnd} \; \nabla_\lambda(e^\lambda_a\;\delta \phi^a)
\end{split}
\end{equation}

The derivation follow a very similar pattern as  \eqref{eq:deltaKLambdaK}. Substituting the above expression into  \eqref{eq:LagVarLegendre}, we finally get
\begin{equation}\label{eq:phiLagVarLegendre}
\begin{split}
\frac{1}{\sqrt{-g}}\delta\prn{\sqrt{-g}\ \Lag_S\brk{\hfields_S} }
&= \half T^{\mu\nu}\delta g_{\mu\nu} + J^\mu \cdot \delta A_\mu -\nabla_\mu B^\mu
\\
& \quad
+e^\sigma_a\;\brk{\aheat_\sigma+\acharge \cdot(\partial_\sigma \cnd\ \cnd^{-1} + A_\sigma) }\frac{1}{s}
\epformref{\!\delta \Sndref}^a
+\cnd^{-1}\,\acharge \,\cnd\cdot \frac{1}{s}\; \epformref{\!\delta \Lndref} \\
&\quad -
\frac{1}{\sqrt{-g}}\; \delta_{\epform{\Snd}}
\prn{\sqrt{-g}\ \frac{\acharge}{s}}  \cdot \bigbr{(\delta \cnd)\,\cnd^{-1}-e^\sigma_a \;\delta\phi^a\;
(\partial_\sigma \cnd\ \cnd^{-1} + A_\sigma )}\\
&\quad
+e^\sigma_a \; \delta\phi^a \brk{  \frac{1}{\sqrt{-g}}\delta_{\epform{\Snd}}\;
\prn{\sqrt{-g}\ \frac{ \aheat_\sigma}{s} } + \frac{\acharge}{s} \cdot\delta_{\epform{\Snd}} A_\sigma + \nabla_\sigma(T\,s)
}
\end{split}
\end{equation}
with the boundary term
\begin{equation}
B^\mu=
(\PSymplPot{})^\mu - T \,s\  e^\mu_a\, \delta\phi^a
+ u^\mu\, \acharge \cdot (\delta \cnd)\cnd^{-1}  -u^\mu \,e^\sigma_a \,\delta\phi^a \brk{\aheat_\sigma+\acharge \cdot(\partial_\sigma \cnd\ \cnd^{-1} + A_\sigma)}
\end{equation}

This then is the analogue of \eqref{eq:phiLagVar} in the Legendre transformed description. Thus, extremizing $\int \sqrt{-g}\ \Lag_S\brk{\hfields_S}$ with  respect to $\{\phi^a,\cnd\}$ variations gives the correct energy-momentum and charge conservation equations i.e.,
\eqref{eq:LTsHydroEq} as required.   Furthermore as discussed before we can equivalently work with the variational principle in terms of push forward metric and gauge fields
$\{\gref_{ab},\Aref_a\}$.

\subsection{Static gauge: symmetries of non-dissipative effective actions}
\label{sec:staticnd}

As a final piece of our discussion we now turn to the symmetries in the Legendre transformed variables. We move to the static gauge by using the redundancies in the reference configuration, to facilitate the analysis.

The redundancies to start with, are
same as before -- we have the diffeomorphisms and gauge transformations on $\Mref$.
Let us explicitly gauge fix these by picking a particular frame for the tensor fields
$\{\Sndref, \Lndref\}$. A convenient choice happens to be
\begin{equation}
\begin{split}
\Sndref_{123\cdots(d-1)} = 1\ ,\qquad   \Sndref_{0 I_1I_2\cdots I_{(d-2)}} = 0 \ ,\qquad
(\Lndref)_{0123\ldots (d-1)} = 0\ .
\end{split}
\end{equation}
where as before $I_k\in \{1,2,\ldots,(d-1)\}$. Let us now enumerate the residual gauge redundancies which are left unfixed by the conditions above:
\begin{equation}
\begin{split}
\phi^{_J} &\mapsto h^{_J}(\phi^{_I})\ ,  \quad  \text{det}\prn{\frac{\partial h^{_J}}{\partial \phi^{_I}}} =1 \\
\phi^0 &\mapsto g (\phi^{_I},\phi^0)\ ,  \quad  \frac{\partial g}{\partial \phi^0}\neq 0 \\
\cnd &\mapsto \cnd\, f(\phi^{_I})
\end{split}
\end{equation}
Thus, the spatial $\phi$-diffeomorphisms get reduced to the subset of volume preserving diffeomorphisms whereas the analogue of thermal shift gets enhanced to a $\phi^0$-dependent shift. In fact, $\phi^0$ completely drops out of all hydrodynamic fields altogether in this gauge.\footnote{ The reader might worry at this point that
our gauge fixing has lost us an equation of motion. This is indeed true and in this sense static gauge is a bit like temporal gauge in electromagnetism where Gauss constraint is lost after gauge fixing. But as we will see in a moment, unlike electromagnetism, adiabaticity equation comes to our rescue and restores the equation that is lost.} The chemical shift remains unchanged. The first and the third expressions which transform $\{ \phi^{_I}, \cnd \}$ have been described in \cite{Haehl:2013kra,Haehl:2013hoa} as $\widetilde{\text{Sdiff}}_{{\cal M}_{\phi,\cnd}}$, the extended volume preserving diffeomorphism symmetry.

Much of the structure we had in static gauge before Legendre transform survives with small modifications. Since the hydrodynamic fields $u^\sigma$ and $\mu$ are unaffected
by Legendre transform, they retain their definitions given in \eqref{eq:uTmuStatic} with the simple replacement
$\{\varphi,c\} \mapsto \{\phi,\cnd\}$. We have
$(d-1)$ spatial fields $\phi^{_I}$ such that $u^\sigma\partial_\sigma \phi^{_I}=0$
and a flavour field $\cnd$. Using these we define in analogy with the discussion of
\S\ref{sec:staticg} a distance measure on the spatial geometry of the reference manifold in terms of the matrix
$\pndref^{_{IJ}}\equiv g^{\mu\nu} \partial_\mu \phi^{_I} \partial_\nu \phi^{_J}$. Its inverse then defines a spatial metric
$\pndref_{_{IJ}}$ as in \S\ref{sec:staticg}. We can then derive the expressions for the hydrodynamic fields as
\begin{equation}\label{eq:umuS}
\begin{split}
s &= \frac{1}{\sqrt{\pndref}} \\
u^\sigma &= \frac{1}{(d-1)!}\
\varepsilon^{\sigma \alpha_1\cdots \alpha_{d-1}} \;
\varepsilon^{(\pndref)}_{{_I}_{_1} \ldots {_I}_{_{d-1}}}
\prod_{i=1}^{d-1}  \partial_{\alpha_i}\phi^{{_I}_{_i}}
\\
\mu &= u^\sigma \brk{(\partial_\sigma \cnd) \cnd^{-1}+  A_\sigma}
\end{split}
\end{equation}
where, as before, $\varepsilon^{(\pndref)}_{{_I}_{_1}\ldots {_I}_{_{d-1}}}$ is the spatial volume
form associated with $\pndref_{_{IJ}}$ and $\pndref$ denotes the determinant of $\pndref_{_{IJ}}$.

As advertised, the field $\phi^0$ completely drops out of all hydrodynamic fields altogether in this gauge. Since there is no $\delta \phi^0$ in the variation, thermal shift invariance dictates that all the  $e^\sigma_a \, \delta \phi^a$ factors in \eqref{eq:phiLagVarLegendre} reduce to $P^{\,\sigma}_{_I} \delta \phi^{_I}$. In turn, this means that the absence of $\phi^0$ leads to  loss of the energy-momentum conservation in the longitudinal direction (along
$u^\mu$). This however is not as problematic as it seems because of a particular feature of this description: the entropy conservation in this description is automatic. We have
\begin{equation}\label{eq:umuS2}
\begin{split}
\nabla_\sigma J_S^\sigma  &= \frac{1}{\sqrt{-g}} \partial_\sigma\prn{ \sqrt{-g}\ s\, u^\sigma}
= \frac{1}{\sqrt{-g}} \partial_\sigma\prn{ \frac{\sqrt{-g}}{\sqrt{\pndref}}\  u^\sigma}  =0
\end{split}
\end{equation}
This equation then via adiabaticity equation leads to the energy-momentum conservation along $u^\mu$. This then is an additional equation which compensates for the loss of $\phi^0$ equation. Effectively we have used the conservation of entropy of the non-dissipative fluids
as a dynamical equation of motion.

With these manipulations we have made explicit that the adiabatic fluid formalism along with an appropriate Legendre transform reduces to the standard effective action formalism for hydrodynamics used  in \cite{Dubovsky:2011sj,Bhattacharya:2012zx,Haehl:2013hoa} and other references mentioned earlier. In particular, it emphasizes the importance of the symmetries of the latter formalism. The $\widetilde{\text{Sdiff}}_{{\cal M}_{\phi,c}}$ symmetry as indicated in \cite{Haehl:2013hoa} is an important part of the framework arising as it does by gauge fixing a covariant formalism involving the background sources and the hydrodynamic fields $\{\Snd, \Lnd\}$. While we could have perfectly well started with these variables and eschewed the introduction of the $\{\Kbeta, \LambdaB\}$ description used in the preceding sections.  However, this formalism for reasons outlined in the preamble to this section suffers when we attempt to describe hydrostatics. 
Likewise the fact that the entropy current is held rigid in the non-dissipative fluids of Class ND is a substantial obstacle to describing in full generality a Lagrangian solution to adiabatic anomalous hydrodynamics, cf., \S\ref{sec:mixanom}.

\section{Topological  currents in odd spacetime dimensions}
\label{sec:WenZee}

On general grounds, one would expect every conserved topological charge on codimension-one spatial slices to be associated with an identically conserved entropy current. In this appendix we provide a general Euler current (with associated charge being the Euler characteristic) and a flavour-charged topological Chern current.

\subsection{Generalized Euler current}
\label{sec:Euler}

Let $\wzD$ denote the covariant exterior derivative acting on tensor-valued forms and $u_\mu$ be the fluid velocity vector. Then,
$u^\mu u_\mu =-1$ implies that $\wzD u_\mu$ is a transverse vector valued 1-form. Further, we will also need the following  transverse anti-symmetric tensor-valued 2-form
\begin{equation}
\begin{split}
P_\mu^\alpha P_\nu^\beta \fR_{\alpha\beta} = \fR_{\mu\nu} + \fR_{\mu\alpha} u^\alpha u_\nu + u_\mu u^\alpha \fR_{\alpha\nu} =  \fR_{\mu\nu} + (\wzD^2 u_\mu) u_\nu - u_\mu \wzD^2 u_\nu
\end{split}
\end{equation}
where $\fR^\mu{}_\nu$ is the curvature 2-form.

Say our fluid is living in $d=3$ spacetime dimensions. Then, we have the following identity \cite{Golkar:2014wwa} due to the transversality properties mentioned above:
\begin{equation}
\begin{split}
\wzD
&\brk{ \varepsilon^{\mu\nu\lambda} u_\mu \prn{\wzD u_\nu \wedge \wzD u_\lambda -\fR_{\nu\lambda} } } \\
&= \varepsilon^{\mu\nu\lambda} \wzD u_\mu \wedge \wzD u_\nu \wedge \wzD u_\lambda - \varepsilon^{\mu\nu\lambda} \wzD u_\mu \wedge \prn{ \fR_{\nu\lambda}
+ (\wzD^2 u_\nu) u_\lambda - u_\nu \wzD^2 u_\lambda } \\
&= 0
\end{split}
\end{equation}
This implies that  we  can define an identically conserved current (which we will term the \emph{Euler current}\footnote{ Sometimes this current is called as the \emph{Wen-Zee current}, following \cite{Wen:1992ej} who realized that 3d Hall currents are often shifted by such a term. The coefficient of this term is hence sometimes called the `shift'.}) via
\begin{equation}
\begin{split}
\hodge \JWZ &\equiv \frac{1}{2}c_{_{\text{Euler}}} \; \varepsilon^{\mu\nu\lambda} \; u_\mu \prn{\wzD
u_\nu \wedge \wzD u_\lambda - \fR_{\nu\lambda} }
\nonumber \\
\JWZ^\sigma &\equiv \frac{1}{2}c_{_{\text{Euler}}}\;\varepsilon^{\sigma\alpha\beta}
\; \varepsilon^{\mu\nu\lambda}\; u_\mu \prn{\nabla_\alpha u_\nu \nabla_\beta u_\lambda
- \frac{1}{2}R_{\nu\lambda\alpha\beta} }
\end{split}
\end{equation}
which satisfies $\nabla_\mu \JWZ^\mu= 0 $. Note that this is a parity-even current appearing in second order in derivative expansion.

Let us now generalize this construction to arbitrary odd spacetime dimensions.\footnote{ During the course of preparation of this paper \cite{Golkar:2014paa} also constructed the generalization of the Euler current to arbitrary odd dimensions. Our presentation in terms of vector valued one-forms is complementary and equivalent to their explicit construction.}  Again we let $\wzD$ denote the  covariant exterior derivative acting on tensor-valued forms. It is uniquely defined by zero-torsion condition  $\wzD (dx^\mu)=0$ and metric compatibility  condition $\wzD g_{\mu\nu} = 0$.  Let   $u_\mu$ be the fluid velocity vector.  Then,  $u^\mu u_\mu =-1$ implies that $\wzD u_\mu$ is a transverse vector
valued 1-form.

For what follows, it is useful to define a new covariant exterior derivative $\Dp$ using the connection 1-form
\begin{equation}\label{eq:GammapDef}
\begin{split}
\fGammap^\mu{}_\nu \equiv \fGamma^\mu{}_\nu+(\wzD u^\mu)u_\nu-u^\mu  \wzD u_\nu
\end{split}
\end{equation}
This connection has a torsion $\Dp (dx^\mu)=(\wzD u^\mu)\fu-u^\mu \wzD\fu$ but it is still metric compatible $\Dp g_{\mu\nu} = 0$.
In addition, it is also velocity compatible $\Dp u^\mu = 0$, though it is not unique in being  velocity compatible. We can work out the curvature 2-form for this connection as
\begin{equation}\label{eq:RpDef}
\begin{split}
\fRp^\mu{}_\nu = \fR^\mu{}_\nu -\wzD u^\mu\wedge \wzD u_\nu + \wzD^2 u^\mu u_\nu -  u^\mu \wzD^2 u_\nu
\end{split}
\end{equation}
This satisfies $\Dp\prn{\fRp^\mu{}_\nu}=0$ and further $\fRp^\mu{}_\nu$ is completely transverse.
It then follows that the following form is $\Dp$-closed (and is hence $\wzD$-closed):
\begin{equation}
\begin{split}
\hodge \JWZ \equiv  -\frac{1}{n!}c_{_{\text{Euler}}}\; u_\mu
\; \varepsilon^{\mu\nu_1\nu_2 \nu_3
\nu_4\ldots \nu_{2n-1}\nu_{2n}}
\;  \fRp_{\nu_1\nu_2} \fRp_{\nu_3\nu_4} \ldots \fRp_{\nu_{2n-1}\nu_{2n}}
\end{split}
\end{equation}
Here $c_{_{\text{WZ}}}$ is some arbitrary numerical constant and we are working in
spacetime dimensions $d=2n+1$. To see how this generalizes  the $d=3$ Euler current, we rewrite the above as
\begin{equation}
\begin{split}
\hodge \JWZ &\equiv
-\frac{1}{2^n}c_{_{\text{Euler}}} \; u_\mu \; \varepsilon^{\mu\nu_1\nu_2 \ldots \nu_{2n-1}\nu_{2n}} \\
&\qquad \times \prn{\fR_{\nu_1\nu_2}-\wzD u_{\nu_1}\wedge \wzD u_{\nu_2} }
\ldots \prn{\fR_{\nu_{2n-1}\nu_{2n}}-\wzD u_{\nu_{2n-1}}\wedge \wzD u_{\nu_{2n}} }
 \\
\JWZ^\sigma &\equiv -\frac{1}{2^n}\; c_{_{\text{Euler}}} \;
\varepsilon^{\sigma\alpha_1\alpha_2\ldots \alpha_{2n-1}\alpha_{2n}}\;
u_\mu \; \varepsilon^{\mu\nu_1\nu_2 \ldots \nu_{2n-1}\nu_{2n}} \\
&\
\times \prn{\half R_{\nu_1\nu_2\alpha_1\alpha_2}-\nabla_{\alpha_1}u_{\nu_1}\nabla_{\alpha_2} u_{\nu_2} }
\ldots
\prn{\half R_{\nu_{2n-1}\nu_{2n}\alpha_{2n-1}\alpha_{2n}}-\nabla_{\alpha_{2n-1}}u_{\nu_{2n-1}}
\nabla_{\alpha_{2n}} u_{\nu_{2n}} } \\
\end{split}
\end{equation}
This is  then a parity-even, identically conserved current with $(d-1)$ derivatives that generalizes the $d=3$
construction of \cite{Golkar:2014wwa} to an arbitrary odd $d$ (see also \cite{Golkar:2014paa}).

One of the reasons for our interest in the Euler current $\JWZ^\sigma$ is that it provides a simple homogeneous solution to the adiabaticity equation. One can simply take the entropy current term to be $\JWZ^\sigma$ and  set all other contributions to zero. Thus, given any solution to the adiabaticity equation, we have the freedom to shift the entropy current in odd dimensions by the Euler current (with an arbitrary constant).

\subsection{Chern current}
\label{sec:Chern}

Another identically conserved current for flavour-charged fluids in $d=2n+1$ dimensions is easily constructed:
\begin{equation}
\begin{split}
 \JChern^\sigma &= 
 	\frac{1}{2^n} \, c_{_{\text{Chern}}} \, \varepsilon^{\sigma \alpha_1\alpha_2 \cdots \alpha_{2n-1}\alpha_{2n}} 
 	F_{\alpha_1\alpha_2} \cdots F_{\alpha_{2n-1}\alpha_{2n}} \,.
\end{split}
\end{equation}
Despite being  an exact form (it is the gradient of the Chern-Simon form) as we argue in the main text, it does contribute to the topological degeneracy of states.

It would be interesting to examine if there are other conserved currents that like the Euler current constructed above, include contributions from the background gauge field. We have not been able to find any, but haven't quite proved a no-go theorem either.
\section{Adiabatic hydrodynamics with Weyl invariance}
\label{sec:aweyl}

The hydrodynamics of a CFT has an additional Weyl symmetry over and above the Poincar\'e and the
flavour symmetries. This can be dealt with by treating Weyl symmetry like any other flavour symmetry
except for a few additional complications that stem from the fact that Weyl symmetry is a spacetime
symmetry. In this appendix we give a self-contained construction of the Weyl invariant structures building on earlier work in 
\cite{Loganayagam:2008is}. 

\subsection{Weyl transformation}
\label{sec:weyltrans}
Let us begin by adding in a Weyl transformation parameter $\LambdaW$ to flavour and diffeomorphism $\Xfields\equiv\{\xi^\mu,\Lambda\}$.
We will denote by $\diffFW$ the combined flavour, Weyl and diffeomorphisms generated by $\Xfields^\Wey\equiv\{\xi^\mu,\Lambda,\LambdaW\}$.
Thus, we can write
\begin{equation}\label{eq:AgGamWeyl}
\begin{split}
\diffFW  A_\mu &= \diffF A_\mu  \\
\diffFW  g_{\mu\nu} &= \diffF g_{\mu\nu} + 2 \LambdaW\ g_{\mu\nu} \\
\diffFW \Gamma^\nu{}_{\lambda\mu}&=\diffF \Gamma^\nu{}_{\lambda\mu} + \delta^{\nu}_{\lambda}\partial_{\mu}\LambdaW\ + \delta^{\nu}_{\mu}\partial_{\lambda}\LambdaW\
- g_{\lambda\mu}g^{\nu\sigma}\partial_{\sigma}\LambdaW \,.
\end{split}
\end{equation}
A \emph{Weyl covariant} tensor $Q^{\mu\ldots}_{\nu\ldots}$  of weight $w$ is a tensor whose Weyl variation is given by
\begin{equation}\label{eq:genWeyl}
\begin{split}
\diffFW  Q^{\mu\ldots}_{\nu\ldots} &= \diffF Q^{\mu\ldots}_{\nu\ldots}- w\ \LambdaW\ Q^{\mu\ldots}_{\nu\ldots} \\
&=[Q^{\mu\ldots}_{\nu\ldots},\Lambda] - w\ \LambdaW\ Q^{\mu\ldots}_{\nu\ldots}  + \xi^\alpha \partial_\alpha Q^{\mu\ldots}_{\nu\ldots}
- (\partial_\alpha \xi^\mu)  Q^{\alpha\ldots}_{\nu\ldots} +\ldots \\
&\qquad + (\partial_\nu \xi^\alpha)  Q^{\mu\ldots}_{\alpha\ldots} +\ldots
\end{split}
\end{equation}
The flavour gauge field $A_\mu$ has a Weyl weight $w=0$ whereas metric $ g_{\mu\nu}$ has a Weyl weight
$w=-2$. The hydrodynamic fields $\{\Kbeta^\mu,\LambdaB\}$ are invariant under Weyl transformation.\footnote{ In  hydrodynamics, as elsewhere,  a useful thumb rule to determine the Weyl weights is
\begin{equation*}
 \text{Weyl weight} = \text{mass dimension }+ \text{No. of upper indices} - \text{No. of lower indices}\ .
\end{equation*}
}
It follows that the velocity $u^\mu$, the temperature $T$ and the flavour chemical potential $\mu$ all have
$w=1$.

\subsection{Weyl connection}
\label{sec:weylcon}
To mimic our construction for flavour symmetries, we extend the sources by adding in  a Weyl connection
(or a gauge field) $\AWeyl_\mu$ which transforms as
\begin{equation}\label{eq:WeylCon}
\begin{split}
\diffFW  \AWeyl_\mu &= \diffF \AWeyl_\mu  + \partial_\mu \LambdaW\ .
\end{split}
\end{equation}

We can then construct  Weyl-invariant Christoffel symbols
\begin{equation}\label{eq:GamWeyl}
\begin{split}
(\GamWeyl)^\mu{}_{\nu\lambda} &\equiv \Gamma^\mu{}_{\nu\lambda} + g_{\nu\lambda}\AWeyl^{\mu} - \delta^{\mu}_{\nu}\AWeyl_\lambda- \delta^{\mu}_{\lambda}\AWeyl_{\nu}
\end{split}
\end{equation}
such that
\begin{equation}\label{eq:GamWeylTransf}
\begin{split}
\diffFW (\GamWeyl)^\mu{}_{\lambda\alpha}  = \diffF (\GamWeyl)^\mu{}_{\lambda\alpha} \,.
\end{split}
\end{equation}

In turn, this can be used to define  a Weyl-covariant derivative \cite{Loganayagam:2008is}
\begin{equation}\label{DWeyl:eq}
\begin{split}
\DWeyl_\lambda\ Q^{\mu\ldots}_{\nu\ldots} &\equiv D_\lambda\ Q^{\mu\ldots}_{\nu\ldots} + w\  \AWeyl_{\lambda} Q^{\mu\ldots}_{\nu\ldots} \\ &+\brk{{g}_{\lambda\alpha}\AWeyl^{\mu} - \delta^{\mu}_{\lambda}\AWeyl_\alpha  - \delta^{\mu}_{\alpha}\AWeyl_{\lambda}} Q^{\alpha\ldots}_{\nu\ldots} + \ldots\\
&-\brk{{g}_{\lambda\nu}\AWeyl^{\alpha} - \delta^{\alpha}_{\lambda}\AWeyl_\nu  - \delta^{\alpha}_{\nu}\AWeyl_{\lambda}} Q^{\mu\ldots}_{\alpha\ldots} - \ldots
\end{split}
\end{equation}
where in analogy with flavour covariant derivative,   we have  added a $w\  \AWeyl_{\lambda}$ term. Further,
the additional terms in the definition occur due to the fact that Weyl symmetry in a spacetime symmetry under
which Christoffel symbols transform inhomogenously. It is easily checked that these terms serve to correct the Christoffel
symbols in $D_\lambda$ into Weyl invariant Christoffel symbols. An often useful spacial case is the action on covariant and contravariant vectors
\begin{equation}\label{DWeylV:eq}
\begin{split}
\DWeyl_\lambda\ V^\sigma &\equiv D_\lambda\ V^\sigma + (w-1)\  \AWeyl_{\lambda} V^\sigma +   \AWeyl^\sigma V_{\lambda}
 - \delta^{\sigma }_{\lambda}(\AWeyl. V) \\
 \DWeyl_\lambda\ V_\sigma &\equiv D_\lambda\ V_\sigma + (w+1)\  \AWeyl_{\lambda} V_\sigma +   \AWeyl_\sigma V_{\lambda}
 - g_{\lambda\sigma}(\AWeyl. V) \\
\end{split}
\end{equation}
and
\begin{equation}\label{DWeylJ:eq}
\begin{split}
\DWeyl_\sigma\ J^\sigma &\equiv D_\sigma\ J^\sigma + (w-d)\  \AWeyl_{\sigma} J^\sigma  \\
 \DWeyl_\nu\ T^{\mu\nu} &\equiv D_\nu\ T^{\mu\nu} +  (w-d-2)\ \AWeyl_\nu\ T^{\mu\nu} + \AWeyl^\mu\  T^\sigma{}_\sigma +  \AWeyl_\nu \prn{ T^{\mu\nu} -T^{\nu\mu} } 
\end{split}
\end{equation}

The above Weyl-covariant derivative is metric compatible ($\DWeyl_{\lambda}g_{\mu\nu} =0 $) and is torsionless when acting on
Weyl-invariant scalar fields. The familiar variational formula for Christoffel symbols
\begin{equation}
\begin{split}
\delta \Gamma^\mu{}_{\nu\lambda} = \half g^{\mu\alpha} \prn{ \nabla_\nu \delta g_{\lambda\alpha} + \nabla_\lambda \delta g_{\nu\alpha} - \nabla_\alpha \delta g_{\nu\lambda} }
\end{split}
\end{equation}
has a Weyl-covariant counterpart
\begin{equation}\label{eq:delGamdelgW}
\begin{split}
\delta \Gamma^\mu{}_{\nu\lambda} + \delta\brk{ g_{\nu\lambda} g^{\mu\alpha} } \AWeyl_{\alpha} = \half g^{\mu\alpha} \prn{ \DWeyl_\nu \delta g_{\lambda\alpha} + \DWeyl_\lambda \delta g_{\nu\alpha} - \DWeyl_\alpha \delta g_{\nu\lambda} }
\end{split}
\end{equation}
This in particular implies that the combination $\delta \Gamma^\mu{}_{\nu\lambda} + \delta\brk{ g_{\nu\lambda} g^{\mu\alpha} } \AWeyl_{\alpha}$ is
Weyl-invariant\footnote{ A quick proof for the Weyl invariance of this combination follows from noting that
$$ \delta(\GamWeyl)^\mu{}_{\nu\lambda} = \delta \Gamma^\mu{}_{\nu\lambda} + \delta\brk{ g_{\nu\lambda} g^{\mu\alpha} } \AWeyl_{\alpha}
+   g_{\nu\lambda} g^{\mu\alpha}  \delta\AWeyl_{\alpha} - \delta^{\mu}_{\nu}\delta\AWeyl_\lambda- \delta^{\mu}_{\lambda}\delta\AWeyl_{\nu} $$
and using the statement that $\delta\AWeyl_{\alpha}$  being the \emph{difference} of two Weyl connections, is Weyl-invariant.}
which would prove useful later on.

The  curvatures  associated with the Weyl-covariant derivative can be defined by the usual procedure of evaluating the commutator between two covariant derivatives on more general fields.  For a covariant vector field $V_\mu$ of weight $w$, we get
\begin{equation}\label{eq:Weylcovcomm}
\begin{split}
[\DWeyl_\mu,\DWeyl_\nu]V_\lambda &= [F_{\mu\nu},V_\lambda]+ w\ (\FWeyl)_{\mu\nu}\ V_\lambda - (\RWeyl)^\alpha{}_{\lambda\mu\nu}  V_\alpha\quad \quad \text{with}\\
F_{\mu\nu} &\equiv \nabla_\mu A_\nu - \nabla_\nu A_\mu + [A_\mu,A_\nu] \\
(\FWeyl)_{\mu\nu} &\equiv \partial_\mu \AWeyl_\nu - \partial_\nu \AWeyl_\mu \\
(\RWeyl)^\alpha{}_{\lambda\mu\nu} &\equiv \partial_\mu (\GamWeyl)^\alpha{}_{\lambda\nu} - \partial_\nu (\GamWeyl)^\alpha{}_{\lambda\mu}
+ (\GamWeyl)^\alpha{}_{\beta\mu} (\GamWeyl)^\beta{}_{\lambda\nu} - (\GamWeyl)^\alpha{}_{\beta\nu} (\GamWeyl)^\beta{}_{\lambda\mu}
\end{split}
\end{equation}
As is evident from their definitions, all these field strengths are Weyl-invariant. A more convenient expression
for the Weyl covariant Riemann tensor is given by the formula\footnote{ Note that our Riemann
tensor notion is slightly different from those defined in \cite{Loganayagam:2008is} which is responsible for
different signs appearing in our expression.}
\begin{equation}\label{eq:WeylcovR1}
\begin{split}
(\RWeyl)_{\mu\nu\lambda\sigma}+g_{\mu\nu} (\FWeyl)_{\lambda\sigma}&= R_{\mu\nu\lambda\sigma}   - 
4\, \delta^\alpha_{[\mu}g_{\nu][\lambda}\delta^\beta_{\sigma]}\left(\nabla_\alpha \AWeyl_\beta + \AWeyl_\alpha \AWeyl_\beta - \frac{\AWeyl^2}{2} g_{\alpha\beta} \right) .
\end{split}
\end{equation}
These curvatures obey Bianchi identities of the form
\begin{equation}\label{eq:WeylBianchi}
\begin{split}
(\RWeyl)^\mu{}_{\nu\lambda\sigma} + (\RWeyl)^\mu{}_{\lambda\sigma\nu} + (\RWeyl)^\mu{}_{\sigma\nu\lambda} &= 0 \\
\DWeyl_{\nu} (\RWeyl)^{\alpha\beta}{}_{\lambda\sigma} +\DWeyl_{\lambda} (\RWeyl)^{\alpha\beta}{}_{\sigma\nu} + \DWeyl_{\sigma} (\RWeyl)^{\alpha\beta}{}_{\nu\lambda} &= 0 \\
\DWeyl_{\nu} (\FWeyl)_{\lambda\sigma} +\DWeyl_{\lambda} (\FWeyl)_{\sigma\nu} + \DWeyl_{\sigma} (\FWeyl)_{\nu\lambda} &= 0 \\
\end{split}
\end{equation}
and
\begin{equation}\label{eq:WeylRiemSymm}
\begin{split}
(\RWeyl)_{\mu\nu\lambda\sigma} + (\RWeyl)_{\nu\mu\lambda\sigma} & =  - 2 (\FWeyl)_{\lambda\sigma} g_{\mu\nu}  \\
(\RWeyl)_{\mu\nu\lambda\sigma} - (\RWeyl)_{\lambda\sigma\mu\nu} & =(\FWeyl)_{\mu\nu} g_{\lambda\sigma}   -(\FWeyl)_{\lambda\sigma} g_{\mu\nu}
-4\,  (\FWeyl)_{\alpha\beta}  \delta^\alpha_{[\mu}g_{\nu][\lambda}\delta^\beta_{\sigma]}
\end{split}
\end{equation}
We can use the Weyl covariant Riemann tensor to define the Weyl covariant Ricci tensor, Ricci scalar and Schouten tensor\footnote{ The Schouten tensor
in defined as
$$ S_{\mu\nu} \equiv  \frac{1}{d-2} \brk{ R_{\mu\nu} - \frac{1}{2(d-1)}R\ g_{\mu\nu} } $$
It is often used in defining the Weyl curvature part of Riemann curvature via $C_{\mu\nu\lambda\sigma} \equiv R_{\mu\nu\lambda\sigma}+\delta^\alpha_{[\mu}g_{\nu][\lambda}\delta^\beta_{\sigma]}\ S_{\alpha\beta}$. Its significance in Weyl-invariant theories arises from the fact that it acts like a connection for the special conformal
transformations.}
 via
\begin{equation}\label{eq:WeylcovR2}
\begin{split}
(\RWeyl)_{\mu\nu}+(\FWeyl)_{\mu\nu} &= R_{\mu\nu} +\prn{(d-2)\delta^\alpha_\mu \delta^\beta_\nu+g_{\mu\nu} g^{\alpha\beta}}  \prn{ \nabla_\alpha \AWeyl_\beta + \AWeyl_\alpha \AWeyl_\beta - \frac{\AWeyl^2}{2} g_{\alpha\beta} } \\
\RWeyl &= R +2(d-1) g^{\alpha\beta}  \prn{ \nabla_\alpha \AWeyl_\beta + \AWeyl_\alpha \AWeyl_\beta - \frac{\AWeyl^2}{2} g_{\alpha\beta} }\\
(\SWeyl)_{\mu\nu}+\frac{1}{d-2}(\FWeyl)_{\mu\nu}
&= S_{\mu\nu} + \nabla_\mu \AWeyl_\nu + \AWeyl_\mu \AWeyl_\nu - \frac{\AWeyl^2}{2} g_{\mu\nu}\,.
\end{split}
\end{equation}

\subsection{Covariant form of Weyl transformations}
\label{sec:weylcov}

As in the flavour case, we can  rewrite the flavour and diffeomorphism Weyl variations  \eqref{eq:WeylCon}  and \eqref{eq:GamWeylTransf} in terms of the Weyl-covariant derivative
\begin{equation}\label{eq:Weyldiff}
\begin{split}
\diffFW  \AWeyl_\mu &= \DWeyl_\mu \prn{\LambdaW +\xi^\alpha \AWeyl_\alpha } + \xi^\alpha\ (\FWeyl)_{\alpha\mu} \\
\diffFW (\GamWeyl)^\mu{}_{\nu\lambda}
&= \DWeyl_\lambda (\DWeyl_\nu \xi^\mu ) + \xi^\sigma (\RWeyl)^\mu{}_{\nu\sigma\lambda}
\end{split}
\end{equation}
where we have assumed $\xi^\alpha$ to be a Weyl-invariant vector. The equation for and \eqref{eq:genWeyl} for a Weyl-covariant tensor becomes
\begin{equation}\label{eq:WeyldiffGen}
\begin{split}
\diffFW  Q^{\mu\ldots}_{\nu\ldots} &=[Q^{\mu\ldots}_{\nu\ldots},\Lambda + \xi^\alpha A_\alpha] - w\ \prn{\LambdaW+\xi^\alpha \AWeyl_\alpha}\ Q^{\mu\ldots}_{\nu\ldots}  + \xi^\alpha \DWeyl_\alpha Q^{\mu\ldots}_{\nu\ldots}\\
&\qquad - (\DWeyl_\alpha \xi^\mu)  Q^{\alpha\ldots}_{\nu\ldots} +\ldots
+ (\DWeyl_\nu \xi^\alpha)  Q^{\mu\ldots}_{\alpha\ldots} +\ldots
\end{split}
\end{equation}
In particular, for the metric $g_{\mu\nu}$ we have
\begin{equation}\label{eq:Weyldiffg}
\begin{split}
\diffFW  g_{\mu\nu}&=\DWeyl_\mu  \xi_\nu + \DWeyl_\nu  \xi_\mu +2\prn{\LambdaW+\xi^\alpha \AWeyl_\alpha} g_{\mu\nu}
\end{split}
\end{equation}
We can then write
\begin{equation}
\begin{split}
\diffFW  \Gamma^\mu{}_{\nu\lambda} + \diffFW \brk{ g_{\nu\lambda} g^{\mu\alpha} } \AWeyl_{\alpha} = \half g^{\mu\alpha} \brk{ \DWeyl_\nu (\diffFW  g_{\lambda\alpha}) + \DWeyl_\lambda (\diffFW  g_{\nu\alpha}) - \DWeyl_\alpha (\diffFW  g_{\nu\lambda}) }\,.
\end{split}
\end{equation}
%

\subsection{Weyl covariance and conservation equations}
\label{sec:weylcons}

In this subsection, we will study the conservation equations in a zero temperature field theory with Weyl invariance. Consider the path integral
of this field theory with the background metric and Weyl connection turned on.\footnote{ For simplicity, we will consider the case with no
flavour symmetries - the expressions in this subsection can be trivially generalized to account for flavour symmetries if present.}
We can  write the variation of the logarithm of this path integral (up to boundary terms) as
\begin{equation}
\begin{split}
-i\ \delta \ln Z = \int d^dx \sqrt{-g} \brk{ \half t^{\alpha\beta} \delta g_{\alpha\beta}  +j^\mu_{\Wey}\ \delta \AWeyl_\mu +\half  (\Sp_\Wey)^{\sigma\mu}{}_\nu\  \delta (\GamWeyl)^\nu{}_{\mu\sigma} }
\end{split}
\label{eq:weyvar1}
\end{equation}
where we  treat $\{g_{\alpha\beta},\AWeyl_\mu,(\GamWeyl)^\nu{}_{\mu\sigma}\}$ as independent sources for later convenience.
We will call  $(\Sp_\Wey)^{\sigma\mu}{}_\nu$ as the {\em Weyl spin current}.  
The tensors $\{t^{\alpha\beta},j^\mu_{\Wey}\}$ are related to the orbital  energy-momentum tensor and the orbital virial current in a way we will make precise below.

Let us now eliminate $(\GamWeyl)^\nu{}_{\mu\sigma}$ variations in favor of variations of the basic sources $\{g_{\alpha\beta},\AWeyl_\mu\}$. First, we  use the identity
\begin{equation}
\begin{split}
\half  (\Sp_\Wey)^{\sigma\mu}{}_\nu\  \delta (\GamWeyl)^\nu{}_{\mu\sigma}
&= \half  (\Sp_\Wey)^{\sigma\mu}{}_\nu\  \brk{\delta \Gamma^\nu{}_{\mu\sigma} + \delta (g_{\mu\sigma} g^{\nu\alpha}) \AWeyl_\alpha } \\
&\qquad + \half g_{\alpha\beta} \prn{ \Sp_\Wey^{\alpha [\beta\sigma]} + \Sp_\Wey^{\beta [\alpha\sigma]} -  \Sp_\Wey^{\sigma (\alpha\beta)} } \delta \AWeyl_\sigma \,,
\end{split}
\end{equation}
to  write \eqref{eq:weyvar1} in an equivalent form
\begin{equation}
\begin{split}
-i\ \delta \ln Z = \int d^dx \sqrt{-g} \bigbr{ \half t^{\alpha\beta} \delta g_{\alpha\beta}  +J^\mu_{\Wey}\ \delta \AWeyl_\mu
+\half  (\Sp_\Wey)^{\sigma\mu}{}_\nu\  \brk{\delta \Gamma^\nu{}_{\mu\sigma} + \delta (g_{\mu\sigma} g^{\nu\alpha}) \AWeyl_\alpha }  } ,
\end{split}
\end{equation}
where we have defined the total virial current $J^\mu_{\Wey}$ as the sum
\begin{equation}
\begin{split}
J^\sigma_{\Wey} \equiv
j^\sigma_{\Wey}
+ \half g_{\alpha\beta} \prn{ \Sp_\Wey^{\alpha [\beta\sigma]} + \Sp_\Wey^{\beta [\alpha\sigma]} -  \Sp_\Wey^{\sigma (\alpha\beta)} } .
\end{split}
\end{equation}
Next we  integrate by parts using \eqref{eq:delGamdelgW} and discard the boundary terms to get
\begin{equation}
\begin{split}
-i\ \delta \ln Z = \int d^dx \sqrt{-g} \bigbr{ \half T_{\Wey}^{\alpha\beta} \delta g_{\alpha\beta}  +J^\mu_{\Wey}\ \delta \AWeyl_\mu  }
\end{split}
\end{equation}
where the Weyl energy-momentum tensor $T_{\Wey}^{\alpha\beta}$ can be defined by more familiar looking expressions involving the orbital energy-momentum current and Weyl spin current as:
\begin{equation}
\begin{split}
T_{\Wey}^{\alpha\beta}\equiv
t^{\alpha\beta}
+ \half  \DWeyl_\sigma \prn{ \Sp_\Wey^{\alpha [\beta\sigma]} + \Sp_\Wey^{\beta [\alpha\sigma]} -  \Sp_\Wey^{\sigma (\alpha\beta)} } .
\end{split}
\end{equation}

We are now ready to derive the conservation equations in the fields theory that follow from Weyl and diffeomorphism symmetries.
Assuming there are no Weyl or diffeomorphism anomalies, the path-integral is then invariant under the Weyl and diffeomorphism
symmetries of the theory. This gives
\begin{equation}
\begin{split}
0 &= \int d^dx \sqrt{-g} \bigbr{ \half T_{\Wey}^{\alpha\beta} (\diffFW g_{\alpha\beta})  +J^\mu_{\Wey}\ (\diffFW \AWeyl_\mu)  } \\
&= - \int d^dx \sqrt{-g}\ \xi_\alpha \bigbr{ \DWeyl_\beta T_{\Wey}^{\alpha\beta} - (J_{\Wey})_\beta (\FWeyl)^{\alpha\beta} } \\
&\qquad - \int d^dx \sqrt{-g}\ (\LambdaW+ \xi^\sigma\AWeyl_\sigma) \bigbr{ \DWeyl_\beta J^\beta_{\Wey} - g_{\alpha\beta} T_{\Wey}^{\alpha\beta} }
\end{split}
\end{equation}
where in the second step, we  have integrated by parts and discarded the boundary terms. thus, for a general Weyl-invariant field theory
we obtain the conservation equations
\begin{equation}
\begin{split}
 \DWeyl_\beta T_{\Wey}^{\alpha\beta}  &= (J_{\Wey})_\beta (\FWeyl)^{\alpha\beta} \\
\DWeyl_\beta J^\beta_{\Wey}  &= g_{\alpha\beta} T_{\Wey}^{\alpha\beta}
\end{split}
\end{equation}
with
\begin{equation}
\begin{split}
T_{\Wey}^{\alpha\beta} &\equiv
t^{\alpha\beta}
+ \half  \DWeyl_\sigma \prn{ \Sp_\Wey^{\alpha [\beta\sigma]} + \Sp_\Wey^{\beta [\alpha\sigma]} -  \Sp_\Wey^{\sigma (\alpha\beta)} } \\
J^\sigma_{\Wey} &\equiv
j^\sigma_{\Wey}
+ \half g_{\alpha\beta} \prn{ \Sp_\Wey^{\alpha [\beta\sigma]} + \Sp_\Wey^{\beta [\alpha\sigma]} -  \Sp_\Wey^{\sigma (\alpha\beta)} }
\end{split}
\end{equation}
The energy-momentum conservation equation is reasonably familiar with all the currents now taking their Weyl invariant form. For the Weyl current conservation we see that the stress tensor trace contributes a source term.

\subsection{Velocity compatible Weyl connection}
\label{sec:vweyl}

While the above discussion  was quite general we now would like to specialize to hydrodynamics, where there is a unique Weyl connection  which satisfies the velocity compatibility conditions \cite{Loganayagam:2008is}
\begin{equation}
u^\sigma \DWeyl_\sigma u_\mu = 0 \ ,\quad \DWeyl_\sigma u^\sigma = 0\ ,
\end{equation}
which can equivalently be stated in terms of $\Kbeta^\mu$ as
\begin{equation}
\Kbeta^\sigma \DWeyl_\sigma \Kbeta^\mu = \Kbeta^\mu \DWeyl_\sigma \Kbeta^\sigma \ .
\end{equation}
Imposing these velocity compatibility conditions gives
\begin{equation} \label{eq:HydroWeylW}
\AWeyl_\sigma \equiv u^\alpha\nabla_\alpha u_\sigma -\frac{\nabla_\alpha u^\alpha}{d-1}  \, u_\sigma= 
\acc_\sigma - \frac{\Theta}{d-1}\,  u_\sigma 
\end{equation}

We can then compute the variation of $\AWeyl_\mu$ as follows: varying the velocity compatibility conditions, we get
\begin{equation}
\begin{split}
0= \delta \prn{u^\sigma \DWeyl_\sigma u_\mu}
&=
\delta u^\sigma \DWeyl_\sigma u_\mu
 +u^\sigma \prn{ \DWeyl_\sigma \delta u_\mu
 - u_\lambda \delta \Gamma^\lambda{}_{\mu\sigma}
 +\delta \AWeyl_\mu u_\sigma  - u_\alpha \delta ( g_{\mu\sigma} \AWeyl^\alpha)
 }
 \\
 &=
 \delta u^\sigma\ \DWeyl_\sigma u_\mu
 + u^\sigma\DWeyl_\sigma \delta u_\mu
  - u_\lambda u^\sigma \prn{\delta \Gamma^\lambda{}_{\mu\sigma}
+\delta\brk{g_{\mu\sigma}g^{\alpha\lambda}} \AWeyl_\alpha } -P_\mu^\nu \delta \AWeyl_\nu
 \\
0= \delta \prn{ u_\mu \DWeyl_\sigma u^\sigma} &=
u_\mu \DWeyl_\sigma \delta u^\sigma + u_\mu u^\sigma \prn{\delta \Gamma^\lambda{}_{\lambda\sigma}+\delta\brk{g_{\lambda\sigma}g^{\alpha\lambda}} \AWeyl_\alpha }
-(d-1) u_\mu u^\nu\ \delta\AWeyl_\nu \\
\end{split}
\end{equation}
We can use these equations solve for $\delta \AWeyl_\mu $ to get
\begin{equation}
\begin{split}
\delta \AWeyl_\mu
 &=
 \delta u^\sigma\ \DWeyl_\sigma u_\mu
 + u^\sigma\DWeyl_\sigma \delta u_\mu
 - \frac{1}{d-1} u_\mu \DWeyl_\sigma \delta u^\sigma \\
&\qquad  \qquad- u^\sigma \prn{ u_\lambda \delta_\mu^\beta + \frac{1}{d-1} u_\mu \delta_\lambda^\beta }\prn{\delta \Gamma^\lambda{}_{\beta\sigma}
+\delta\brk{g_{\beta\sigma}g^{\alpha\lambda}} \AWeyl_\alpha }
\end{split}
\end{equation}

We can then use $ \delta u^\sigma\ \DWeyl_\sigma u_\mu  =
 T\delta \Kbeta^\sigma\ \DWeyl_\sigma u_\mu $ and
\begin{equation}
\begin{split}
u^\sigma& \DWeyl_\sigma \delta u_\mu
 - \frac{1}{d-1} u_\mu \DWeyl_\sigma \delta u^\sigma\\
&=  \prn{u^\lambda P_{\mu\sigma} - \frac{1}{d-1} u_\mu P^\lambda_\sigma}\DWeyl_\lambda \prn{ T\delta \Kbeta^\sigma }
  +  \half   \prn{P_\mu^\alpha \, u^\beta+P_\mu^\beta \,u^\alpha - \frac{d}{d-1} u_\mu \,u^\alpha  \,u^\beta} (u.\DWeyl)  \delta g_{\alpha\beta}
 \end{split}
\end{equation}
to write
\begin{equation}
\begin{split}
\delta \AWeyl_\mu
 &=
 \half   \prn{P_\mu^\alpha \, u^\beta+P_\mu^\beta \,u^\alpha - \frac{d}{d-1} u_\mu \,u^\alpha  \,u^\beta} (u.\DWeyl)  \delta g_{\alpha\beta} \\
 &\qquad  \qquad- u^\sigma \prn{ u_\lambda \delta_\mu^\beta + \frac{1}{d-1} u_\mu \delta_\lambda^\beta }\prn{\delta \Gamma^\lambda{}_{\beta\sigma}
+\delta\brk{g_{\beta\sigma}g^{\alpha\lambda}} \AWeyl_\alpha } \\
&\qquad \qquad +T\delta \Kbeta^\sigma\ \DWeyl_\sigma u_\mu
+ \prn{u^\lambda P_{\mu\sigma} - \frac{1}{d-1} u_\mu P^\lambda_\sigma}\DWeyl_\lambda \prn{ T\delta \Kbeta^\sigma }
\end{split}
\end{equation}
A useful corollary of this result is
\begin{equation}\label{eq:WeylCorollary}
\begin{split}
 J^\mu_{\Wey}\delta \AWeyl_\mu
&= -  \half  \delta g_{\alpha\beta}  \prn{P_\mu^\alpha \, u^\beta+P_\mu^\beta \,u^\alpha - \frac{d}{d-1} u_\mu \,u^\alpha  \,u^\beta}   (u.\DWeyl) J^\mu_{\Wey}
\\
&\qquad  \qquad- u^\sigma  \prn{  J_{\Wey}^\mu u_\lambda + \frac{1}{d-1} (u. J_{\Wey} ) \delta_\lambda^\mu }\prn{\delta \Gamma^\lambda{}_{\mu\sigma}
+\delta\brk{g_{\mu\sigma}g^{\alpha\lambda}} \AWeyl_\alpha }\\
&\qquad  \qquad + T\delta \Kbeta^\lambda  \bigbr{\ J^\mu_{\Wey}\DWeyl_\lambda u_\mu
 - \DWeyl_\sigma\brk{ J^\mu_{\Wey}  \prn{ u^\sigma P_{\mu\lambda} - \frac{1}{d-1}  u_\mu P^\sigma_\lambda } }\   } + \DWeyl_\sigma (\ldots)\\
\end{split}
\end{equation}
%

\subsection{Class L for Weyl covariant fluids}
\label{sec:Lweyl}

In this subsection, we will study the Class L for Weyl covariant fluids. In analogy with our previous discussion for a zero temperature
field theory, we can write the variation of Lagrangian density (up to boundary terms) as
\begin{equation}
\begin{split}
 \int d^dx \sqrt{-g} \brk{ \half t^{\alpha\beta} \delta g_{\alpha\beta} +J^\mu_{\Wey}\delta \AWeyl_\mu +\half  (\Sp_\Wey)^{\sigma\mu}{}_\lambda \prn{\delta \Gamma^\lambda{}_{\mu\sigma}
+\delta\brk{g_{\mu\sigma}g^{\alpha\lambda}} \AWeyl_\alpha }
 + T  \aheat^\Wey_\lambda \delta \Kbeta^\lambda
 }
\end{split}
\label{eq:weyvarHyd}
\end{equation}
We can then integrate by parts using \eqref{eq:WeylCorollary} to get
\begin{equation}
\begin{split}
\half  t^{\alpha\beta} & \delta g_{\alpha\beta}  +J^\mu_{\Wey}\delta \AWeyl_\mu +\half  (\Sp_\Wey)^{\sigma\mu}{}_\lambda \prn{\delta \Gamma^\lambda{}_{\mu\sigma}
+\delta\brk{g_{\mu\sigma}g^{\alpha\lambda}} \AWeyl_\alpha } +  T  \aheat^\Wey_\lambda \delta \Kbeta^\lambda \\
&=
T\delta \Kbeta^\lambda  \bigbr{\   \aheat^\Wey_\lambda+ J^\mu_{\Wey}\DWeyl_\lambda u_\mu
 - \DWeyl_\sigma\brk{ J^\mu_{\Wey}  \prn{ u^\sigma P_{\mu\lambda} - \frac{1}{d-1}  u_\mu P^\sigma_\lambda } }\   } \\
&\qquad    + \half  \delta g_{\alpha\beta} \bigbr{ t^{\alpha\beta} -\prn{P_\mu^\alpha \, u^\beta+P_\mu^\beta \,u^\alpha - \frac{d}{d-1} u_\mu \,u^\alpha  \,u^\beta}   (u.\DWeyl) J^\mu_{\Wey} }
\\
&\qquad  \qquad+\half \brk{ \Sp_\Wey^{\sigma\mu}{}_\lambda - 2u^\sigma  \prn{  J_{\Wey}^\mu u_\lambda + \frac{1}{d-1} (u. J_{\Wey} ) \delta_\lambda^\mu } }\prn{\delta \Gamma^\lambda{}_{\mu\sigma}
+\delta\brk{g_{\mu\sigma}g^{\alpha\lambda}} \AWeyl_\alpha }\\
&\qquad  \qquad + \DWeyl_\sigma \brk{\ldots} \\
\end{split}
\end{equation}

Another integration by parts gives us the fluid energy-momentum tensor $T_{\Wey}^{\alpha\beta}$ 
\begin{equation}
\begin{split}
T_{\Wey}^{\alpha\beta}&=
t^{\alpha\beta} -\prn{P_\mu^\alpha \, u^\beta+P_\mu^\beta \,u^\alpha - \frac{d}{d-1} u_\mu \,u^\alpha  \,u^\beta}   (u.\DWeyl) J^\mu_{\Wey} \\
&\quad + \half  \DWeyl_\sigma \prn{ \Sp^{\alpha [\beta\sigma]} + \Sp^{\beta [\alpha\sigma]} -  \Sp^{\sigma (\alpha\beta)} } .
\end{split}
\end{equation}
with 
\[ \Sp^{\sigma\mu}{}_\lambda \equiv \Sp_\Wey^{\sigma\mu}{}_\lambda - 2u^\sigma  \prn{  J_{\Wey}^\mu u_\lambda + \frac{1}{d-1} (u. J_{\Wey} ) \delta_\lambda^\mu } \]
along with the adiabatic heat current 
\begin{equation}
\begin{split}
  \aheat_\lambda &= \aheat^\Wey_\lambda+ J^\mu_{\Wey}\DWeyl_\lambda u_\mu
 - \DWeyl_\sigma\brk{ J^\mu_{\Wey}  \prn{ u^\sigma P_{\mu\lambda} - \frac{1}{d-1}  u_\mu P^\sigma_\lambda } }
 \end{split}
\end{equation}
Thus, with these virial corrections the Class L for Weyl-covariant fluids reduces to the  Class L for the usual fluids with the above energy momentum tensor
and adiabatic heat current.

\newpage
\part{Computational details}
\label{part:technical}
\hspace{1cm}

\section{Useful variational formulae}
\label{sec:varform}
In this Appendix we collect various useful variational formulae and some derivations filling in the gaps for various results used in the main text.

\subsection{Mapping variations of hydrodynamic fields}
\label{sec:HydroVariations}

If we denote the hydrodynamic projector by
$P_{\alpha\beta}\equiv g_{\alpha\beta}+u_\alpha u_\beta$, we can derive from
\eqref{eq:hydrofields} the following translation between variations of
$\{\Kbeta^\mu, \LambdaB\}$  and those of the  more traditional fields
$\{u^\mu, T, \mu\}$:
\begin{equation}
\begin{split}
\delta u^\alpha &= T \,P^\alpha_\beta \, \delta \Kbeta^\beta
+  \half \,u^\alpha \,u^\beta \, u^\rho \,\delta g_{\beta\rho} \\
\delta u_\alpha &= T \,P_{\alpha\lambda} \, \delta \Kbeta^\lambda
+ \half \prn{P_\alpha^\beta \, u^\rho+P_\alpha^\rho \,u^\beta -u_\alpha \,u^\beta  \,u^\rho}
\delta g_{\beta\rho} \\
\delta T &= T^2 \,u_\alpha \,\delta \Kbeta^\alpha
+ \half \,T\, u^\alpha\, u^\beta \, \delta g_{\alpha\beta} \\
\delta \mu &= T \prn{ \mu\, u_\sigma \,\delta \Kbeta^\sigma +\delta\LambdaB +A_\sigma
\,\delta \Kbeta^\sigma  } + u^\sigma \,\delta A_\sigma
+ \half\, \mu\, u^\alpha \,u^\beta \, \delta g_{\alpha\beta} \,.
\label{eq:varrules}
\end{split}
\end{equation}
These equations follow from the basic definition of the hydrodynamic fields. Having the explicit expressions at hand comes in handy while deriving various results in the text. 

For reference, we also note the variation of the sources along a configuration $\Bfields = \{\Kbeta^\mu,\LambdaB\}$ and evaluate them on-shell:
\begin{align}
\begin{split}
\diffB g_{\mu\nu} &= 2\, \nabla_{(\mu} \Kbeta_{\nu)} 
 = \frac{2}{T}\, \brk{ \sigma_{\mu\nu}  + P_{\mu\nu}\, \frac{\Theta}{d-1} - \left(\acc_{(\mu} + \nabla_{(\mu} \log T \right) u_{\nu)} }
 \\
 &\simeq \frac{2}{T} \brk{\sigma_{\mu\nu} + P_{\mu\nu} \, \frac{\Theta}{d-1} - \vs \,\Theta \, u_{\mu} u_{\nu} - \frac{q}{\varepsilon+p} \, \cv_{(\mu} u_{\nu)} }+ 2^{\rm nd}\, \text{order}
 \\
 \diffB A_\mu &
 = D_\mu(\LambdaB+\Kbeta^\nu A_\nu)+\Kbeta^\nu F_{\nu\mu}
 = u^\alpha\, D_\alpha\left(\frac{\mu}{T} \right)\, u_\mu - \frac{1}{T}\,\cv_\mu 
 \\
 &\simeq -\frac{1}{T}\frac{dp}{dq} \, \Theta \, u_\mu - \frac{1}{T} \cv_\mu + 2^{\rm nd}\, \text{order} \,.
 \end{split}
\label{eq:diffbgaApp}
\end{align}
%

\subsection{Relating variations of hydrodynamic fields to reference parameterization}
\label{sec:refpar}

We now turn to the important task of relating the variations of the physical hydrodynamic fields $\{\Kbeta^\mu, \LambdaB\}$ to the parameterization in terms of the reference fields (rigid) and physical fluctuating fields introduced in \S\ref{sec:reffields}. Our goal is to start with the definition \eqref{eq:KLambdaKPullBack} and derive \eqref{eq:deltaKLambdaK}.

Let us begin by varying $\Kbeta^\mu$ starting from its definition in the first line of \eqref{eq:KLambdaKPullBack}. An explicit computation gives:
 \begin{equation}
 \begin{split}
 \delta \Kbeta^\mu
 &=\Kref^a[\varphi]  \; \delta e^\mu_a   + e^\mu_b \ \delta \varphi^a
 \; \frac{\partial}{\partial \varphi^a} \Kref^b[\varphi] +  e^\mu_a \; \delta \Kref^a[\varphi] \\
 &= -e^\mu_a \Kbeta^\nu\partial_\nu \delta\varphi^a  + e^\mu_b \delta \varphi^a
 \frac{\partial \Kref^b}{\partial \varphi^a} +  e^\mu_a \delta \Kref^a \\
 &= -\diffB \prn{e^\mu_a \delta\varphi^a}
 +\brk{\diffB  e^\mu_a + e^\mu_b \frac{\partial \Kref^b}{\partial \varphi^a}
 }\delta\varphi^a +  e^\mu_a \delta \Kref^a
 \end{split}
 \end{equation}
To further simplify the expression, consider first the middle term in the above expression, which we argue vanishes.
\begin{equation}
\begin{split}
\diffB  e^\mu_a + e^\mu_b \frac{\partial \Kref^b}{\partial \varphi^a}
&= \Kbeta^\nu \partial_\nu e^\mu_a - e^\nu_a  \partial_\nu \Kbeta^\mu + e^\mu_b \frac{\partial \Kref^b}{\partial \varphi^a}\\
&= \Kref^b e^\nu_b \partial_\nu e^\mu_a - \Kref^b e^\nu_a  \partial_\nu e^\mu_b
- e^\nu_a e^\mu_b  \partial_\nu \varphi^c \frac{\partial \Kref^b}{\partial \varphi^c}+ e^\mu_b \frac{\partial \Kref^b}{\partial \varphi^a}\\
&= \Kref^b \prn{e^\nu_b \partial_\nu e^\mu_a - e^\nu_a  \partial_\nu e^\mu_b} =0
\end{split}
\end{equation}
where in the last step we have used the fact that the Lie commutator between two
coordinate basis vectors is zero. Thus, we finally obtain
\begin{equation}
\begin{split}
\delta \Kbeta^\mu
&=   e^\mu_a \delta \Kref^a -\diffB \prn{e^\mu_a \delta\varphi^a}
\end{split}
\label{eq:delKrefvar}
\end{equation}
As advertised, the variation of $\varphi^a$ enters only as a change along Lie orbit.

We now turn to variation of $\LambdaB$:
\begin{equation}
\begin{split}
\delta\LambdaB
&= [\delta c\ c^{-1},c\ \Lref\ c^{-1}] + c\ \delta\Lref\ c^{-1}+ \delta \varphi^a c\
\frac{\partial \Lref}{\partial \varphi^a}\ c^{-1}+ \delta \Kbeta^\sigma (\partial_\sigma c)c^{-1} \\
&\qquad+\Kbeta^\sigma\partial_\sigma\prn{\delta c\ c^{-1}}+[\delta c\ c^{-1},\Kbeta^\sigma (\partial_\sigma c)c^{-1}]\\
&= c\ \delta\Lref\ c^{-1}+ \delta \varphi^a c\
\frac{\partial \Lref}{\partial \varphi^a}\ c^{-1}+ e_a^\sigma\delta \Kref^a  (\partial_\sigma c)c^{-1}
-\diffB (e^\sigma_a \delta \varphi^a)\ \partial_\sigma c\ c^{-1}\\
&\qquad+\Kbeta^\sigma\partial_\sigma\prn{\delta c\ c^{-1}}+[\delta c\ c^{-1},\LambdaB]\\
&=  \delta \varphi^a \brk{e^\sigma_a \diffB \prn{ \partial_\sigma c\ c^{-1}}
+ c\
\frac{\partial \Lref}{\partial \varphi^a}\ c^{-1}
}+
c\ \delta\Lref\ c^{-1}+ e_a^\sigma\delta \Kref^a  (\partial_\sigma c)c^{-1} \\
&\qquad +\diffB  \brk{\delta c\ c^{-1}- e^\sigma_a \delta \varphi^a \ \partial_\sigma c\ c^{-1}}
\end{split}
\end{equation}
We now focus on the first term which will end up vanishing after a suitable amount of massaging:
\begin{equation}
\begin{split}
\diffB \prn{ \partial_\sigma c\ c^{-1}}
&= -\partial_\sigma \LambdaB + \Kbeta^\alpha\partial_\alpha(\partial_\sigma c\ c^{-1})
+\partial_\sigma \Kbeta^\alpha (\partial_\alpha c\ c^{-1}) + [\partial_\sigma c\ c^{-1},\LambdaB]\\
&= -\partial_\sigma (\LambdaB-\Kbeta^\alpha\partial_\alpha  c\ c^{-1})
+  [\partial_\sigma c\ c^{-1},\LambdaB-\Kbeta^\alpha\partial_\alpha  c\ c^{-1}]\\
&= -\partial_\sigma (c\Lref c^{-1})
+  [\partial_\sigma c\ c^{-1},c\Lref c^{-1}]\\
&= -c\ \partial_\sigma \Lref\ c^{-1} = -\partial_\sigma \varphi^a  c\ \frac{\partial \Lref}{\partial \varphi^a}\ c^{-1}\,.
\end{split}
\end{equation}
Using the above result we can write
\begin{equation}
\begin{split}
e^\sigma_a \diffB \prn{ \partial_\sigma c\ c^{-1}}
+ c\
\frac{\partial \Lref}{\partial \varphi^a}\ c^{-1} = 0\,,
\end{split}
\end{equation}
so that in the end
\begin{equation}
\delta\LambdaB =
c\ \delta\Lref\ c^{-1}+ e_a^\sigma\delta \Kref^a  (\partial_\sigma c)c^{-1}
+\diffB  \brk{\delta c\ c^{-1}- e^\sigma_a \delta \varphi^a \ \partial_\sigma c\ c^{-1}}
\label{eq:delLrefvar}
\end{equation}
Equations \eqref{eq:delKrefvar} and \eqref{eq:delLrefvar} form the basic map between the variation of the physical fields and those of the reference fields and they are quoted in the text as \eqref{eq:deltaKLambdaK}.

\subsection{Variational rules for anomalous hydrodynamics}
\label{sec:varanomhat}
We collect in this appendix various useful formulae for checking the results in \S\ref{sec:fanom} and
\S\ref{sec:varmix}. For most of the fields we already know the variations in terms of the hydrodynamic fields and the sources. The new objects whose variations we need in the anomaly discussion are the shadow fields
$\fAh$  and $\fGammah$. These are however conveniently defined for us in terms of the hydrodynamic fields so it is quite simple to see how to write down their variations in terms of our preferred set of fields.

Consider first the flavour shadow connection defined in \eqref{eq:hatAdef}; from the basic variations (\ref{eq:varrules}) it follows immediately that
\begin{equation}
\begin{split}
\delta \Ah_\lambda &=
\mu \,P_\lambda^{(\alpha}\, u^{\beta)} \; \delta g_{\alpha\beta}
+P_\lambda^\alpha \,\delta A_\alpha  \\
&\quad
+ \mu \prn{P_{\lambda\alpha}+u_\lambda u_\alpha} \,T\, \delta \Kbeta^\alpha
+ u_\lambda \,T\,(\delta \LambdaB + A_\alpha \delta \Kbeta^\alpha) \,.
\end{split}
\label{eq:Ahatvar}
\end{equation}

For the mixed anomaly discussion we also need the variation of the spin connection shadow fields. In order to get the variation $\delta \Gammah^\rho{}_{\sigma\lambda}$, we first start with the spin chemical potential defined in \eqref{eq:DefOmega} and observe by explicit variation that
\begin{equation}
\begin{split}
\delta \Omega^\rho{}_\sigma &= \frac{\delta T}{T} \,\Omega^\rho{}_\sigma 
+ T \,\fatQ^{\rho\mu}_{\sigma\kappa} \, D_\mu\delta \Kbeta^\kappa
+ T \,\Kbeta^\nu \,\fatQ^{\rho\mu}_{\sigma\kappa} \,\delta \Gamma^\kappa{}_{\mu\nu}
+ T \,(D_\mu \Kbeta^\kappa ) \,\delta \fatQ^{\rho\mu}_{\sigma\kappa}\\
& =  T \,\Omega^\rho{}_\sigma \, u_\alpha \delta \Kbeta^\alpha + \frac{1}{2} \, \Omega^\rho{}_\sigma u^\mu u^\nu \delta g_{\mu\nu}
+ T \,\fatQ^{\rho\mu}_{\sigma\nu} \, D_\mu\delta \Kbeta^\nu
+ T\, \Kbeta^\nu\, \fatQ^{\rho\mu}_{\sigma\kappa} \,\delta \Gamma^\kappa{}_{\mu\nu} \\
&\quad   - \frac{1}{2} \, T \left( \delta^\mu_\sigma D^\rho \Kbeta^\nu - g^{\rho\mu} \nabla^\nu \Kbeta_\sigma \right)\delta g_{\mu\nu} \,.
\end{split}
\label{eq:Omegavar}
\end{equation}
From this we immediately infer
\begin{equation}
\begin{split}
\delta \Gammah^\rho{}_{\sigma\lambda} &= \left[\Omega^\rho{}_\sigma P_{\lambda}^{(\mu} u^{\nu)} - \half \, T \,u_\lambda \left(
\delta^\mu_\sigma \nabla^\rho \Kbeta^\nu - g^{\rho\mu} \nabla^\nu \Kbeta_\sigma \right)\right] \delta g_{\mu\nu}
+ \fatP^{\rho\mu\nu}_{\sigma\kappa\lambda} \,\delta \Gamma^\kappa{}_{\mu\nu} \\
&\quad
+ \Omega^\rho{}_\sigma \prn{ P_{\lambda\alpha}+u_\lambda u_\alpha} T \,\delta \Kbeta^\alpha
+ u_\lambda T\, \fatQ^{\rho\mu}_{\sigma\kappa} \,D_\mu \delta \Kbeta^\kappa \,,
\end{split}
\label{eq:Ghatvar}
\end{equation}
where we use the abbreviation $\fatP^{\rho\mu\nu}_{\sigma\kappa\lambda} = \delta^\rho_\kappa \delta^\mu_\sigma \delta^\nu_\lambda
+ \fatQ^{\rho\mu}_{\sigma\kappa} \, u^\nu u_\lambda$.

\section{Details of the neutral fluid computation at second order}
\label{sec:neutral2d}

In this appendix we work out the Class L theory describing neutral fluids at second order in the gradient expansion.  In \S\ref{sec:neutral} we have described the basic set-up for this problem. The task at hand is to take the $13$ scalar terms given in \eqref{eq:basisN2} and work out their variations. Once we do that we will be in a position to work out the stress tensor and read off the physical quantities.

\subsection{Variational calculus for the second order scalars }
\label{sec:}

Let us parameterize the general second order Lagrangian using the basis \eqref{eq:basisN2} as follows:
\begin{align}
\Lag_2 & =
K_\sigma(T)\, \sigma^2 + K_\omega(T)\,  \omega^2 + K_a(T)\,  \acc^2 + K_\Theta(T)\, \Theta^2
+ K_R(T)\,  R
\nonumber \\
&\qquad +\;
K_t(T)\,  \nabla_\mu T\, \nabla^\mu T   +   K_u(T)\,  \Theta\, u^\mu \nabla_\mu T +
K_x(T) \, \acc^\mu \,\nabla_\mu T +  K_y(T)\, (u^\mu \nabla_\mu T)^2
\nonumber \\
& \qquad
+ f_a(T)\,  R_{00}  + f_b(T) \, u^\mu \nabla_\mu \Theta + f_c(T)\, \nabla^2 T + f_d (T)\, u^\mu\, u^\nu \nabla_\mu\, \nabla_\nu T
\label{eq:LbasN2}
\end{align}

We have singled out the last four terms since by a suitable integration by parts they can be eliminated in favour of the nine  terms in the first two lines. More specifically we have
 \begin{subequations}
 \begin{align}
\,f_a(T)\, R_{00} &=
	-\,\brk{ f_a'(T) \,\acc^\mu \nabla_\mu T
	+f_a(T)\prn{\sigma^2 + \omega^2 + u^\mu \nabla_\mu \Theta + \frac{\Theta^2}{d-1} } }
\nonumber \\ &
\qquad \qquad
	+\;\, \nabla_\mu \prn{f_a(T)\, \acc^\mu} \,,
\\
\,f_b(T)\, u^\mu\nabla_\mu \Theta &=
	-\, \brk{f_b(T)\, \Theta^2 + f_b'(T)\,\Theta\, u^\mu\nabla_\mu T} +
	\nabla_\mu \prn{f_b(T)\, \Theta\, u^\mu }
\end{align}
\begin{align}
f_c(T)\, \nabla^2 T &=
	-\,\brk{ f_c'(T) \,\nabla^\mu T\, \nabla_\mu T }
	+  \nabla_\mu \prn{f_c(T)\, \nabla^\mu T} \,.
\\
f_d(T)\, u^\mu\, u^\nu\, \nabla_\mu\nabla_\nu T &=
	-\,\brk{ f_d'(T) \, \prn{ u^\mu  \nabla_\mu T }^2
	+f_d(T)\, \Theta\, u^\mu \,\nabla_\mu T + f_d(T)\, \acc^\mu\, \nabla_\mu T}
\nonumber \\ &
 \qquad \qquad
	+\;\, \nabla_\mu \prn{f_d(T)\, u^\mu\, u^\nu \nabla_\nu T} \,.
\end{align}
\end{subequations}
where we dropped the integral signs to keep the expressions compact.

Furthermore, we have at our disposal the field redefinition freedom described in \S\ref{sec:fieldredef}. Since we set the first order gradient terms to vanish $\Lag_1 =0$, the freedom we have is field redefinition of the ideal fluid Lagrangian \eqref{eq:lagn0}. We claim that by suitable choice of $\delta \varphi^a$ we can eliminate all the four terms in the second line of \eqref{eq:LbasN2}. Explicitly, under a field redefinition we find as before
\begin{align}
\diffCons\Lag_0 \brk{\Kbeta} + \Lag_2\brk{\Kbeta} &=  p'(T)\,
\bigg( \nabla_\mu \log T - \vs\, \Theta\, u_\mu + \acc_\mu \bigg)\;
e^\mu_a \, \delta\varphi^a  +\Lag_2\brk{\Kbeta}+ \cdots \,.
\end{align}
Now what we want to do is to eliminate all terms involving the gradients of temperature. That this is possible is manifest from the equation above, since by choosing appropriate values of $e^\mu_a\,\delta\varphi^a$ we can set to zero the coefficient functions $\{K_t(T),K_u(T), K_x(T), K_y(T)\}$. However, in doing so we will shift the coefficient functions involving at least one factor of $\Theta$ or $\acc^\mu$. It is then easy to see that the combinations
\begin{align}
{\widetilde K}_a = K_a +T^2\, K_t - T\, K_x \,, \qquad
{\widetilde K}_\Theta = K_\Theta -T^2\, \vssq \, K_t-T\, \vs\, K_u +T^2\, \vssq  \, K_y \,.
\label{eq:Kfredef}
\end{align}
are field redefinition invariant. So even if we failed to implement the field redefinitions any transport coefficient not involving these particular combinations would point to an error in the computation.

This means that we only have to compute the variation of the $5$ terms in the first  line of
\eqref{eq:LbasN2}. We can do so in a straightforward manner using the basic variational formulae quoted in
\eqref{eq:varrules}. However, in order to demonstrate the efficacy of our field redefintions together with the potential cross-check it offers on the result, we actually vary all the $9$ terms in the first two lines of
\eqref{eq:LbasN2}.

In what follows we will write the answer eschewing the integrals and factors of $\sqrt{-g}$ though for completeness we will indicate the total derivative pieces we encounter in the process of integrating by parts to facilitate reading off the pre-symplectic potential. In addition to fit the expressions compactly we introduce a shorthand for terms on the l.h.s. viz.,
 $\delta (K_X X) \equiv \frac{1}{\sqrt{-g}} \,\delta\prn{\sqrt{-g}\, K_X\, X} $. For ease of visualization we have also indicated the total derivative terms in a different color; these will be  useful later in the computation of the free energy current.
 \begin{subequations}
\begin{align}
\delta\prn{ K_\sigma\, \sigma^2}   & =
	\bigg( K_\sigma\, \sigma^2\, g^{\mu\nu} + T\, K_\sigma' \, \sigma^2\, u^\mu\,u^\nu -4\, K_\sigma\,
	\sigma^{\mu\alpha}\,\sigma_\alpha^{\;\nu} - 4\, K_\sigma\, \frac{\Theta}{d-1}\, \sigma^{\mu\nu}
\nonumber \\&
	 -\;4\, K_\sigma\, u_\alpha \prn{u^\rho\,\nabla_\rho\, \sigma^{\alpha(\mu}}\, u^{\nu)}
	 -4\, u^{(\nu}\,\nabla_\alpha\prn{K_\sigma \,\sigma^{\mu)\alpha}}
	 +2\, K_\sigma\, \sigma^2\, u^\mu\,u^\nu \bigg) \frac{1}{2}\, \delta g_{\mu\nu} 
\nonumber \\
& \quad
	-\bigg( 2\, K_\sigma\, \sigma^{\alpha\nu} \, u_\mu\bigg)
	\delta \Gamma_{\nu\alpha}^\mu
	+ \MR{\nabla_\mu\prn{2\,K_\sigma\, \sigma^{\mu\nu} \delta u_\nu}}
\nonumber \\
& \quad
	+ \bigg( T\, K_\sigma'\, \sigma^2 \, u_\alpha
	-2\,  P_{\nu\alpha}\,\nabla_\mu\prn{K_\sigma\, \sigma^{\mu\nu}}
	+ 2\, K_\sigma\, a^\mu\,\sigma_{\mu\alpha}\bigg) \,T\,\delta\Kbeta^\alpha \,,
\label{eq:deltsigsq}
\\
\delta\prn{ K_\omega\, \omega^2}   & =
	\bigg( K_\omega\, \omega^2\, g^{\mu\nu} + \prn{2\, K_\omega +
	 T\, K_\omega'} \, \omega^2\, u^\mu\,u^\nu
	 - 4\, K_\omega\, \omega^{\mu\alpha}\,\omega_\alpha^{\;\nu}
	 + 4\, K_\omega \omega^{(\mu\alpha}\,u^{\nu)}\, \acc_{\alpha}
 \nonumber \\
& \qquad
 	+\; 4\, \nabla_\alpha\prn{K_\omega\, \omega^{\alpha(\mu} }\,u^{\nu)} \bigg)
 	\frac{1}{2}\, \delta g_{\mu\nu}
 	-\MR{\nabla_\mu\prn{2\,K_\omega\, \omega^{\mu\nu} \delta u_\nu}}
\nonumber \\
& \quad
	+ \bigg( T\, K_\omega'\, \omega^2 \, u_\alpha
	 +2\, T\, P_{\nu\alpha}\, \nabla_\mu\prn{K_\omega\, \omega^{\mu\nu}}
	 +2\, T\, K_\omega\, \omega_\alpha^{\ \mu}\, \acc_\mu
	 \bigg) \,T\,\delta\Kbeta^\alpha \,,
\label{eq:deltaomsq}
\\
\delta\prn{ K_a\, \acc^2}   & =
	\bigg( K_a\, \acc^2\, g^{\mu\nu} + K_a' \, T\, \acc^2\, u^\mu\,u^\nu
	+ 2\, K_a\, \acc^\mu\,\acc^\nu + 4\, K_a\, \acc^2\, u^\mu \, u^\nu \bigg)
	\frac{1}{2}\, \delta g_{\mu\nu}
\nonumber \\
& \quad
	+ \bigg( 2\, K_a\, \acc_\mu \, u^\nu\, u^\alpha\bigg)
	\delta \Gamma_{\nu\alpha}^\mu
	+ \MR{\nabla_\mu\prn{2\,K_a \, \acc_\nu  u^\mu \delta u^\nu}}
\nonumber \\
& \quad
	+ \bigg(2\, K_a\, \acc_\mu \, P^\nu_\alpha\, \nabla_\nu\, u^\mu - 2\, \nabla_\nu
	\prn{K_a\, \acc_\mu\, u^\nu} \, P_\alpha^\mu + T\, K_a'\, \acc^2 \, u_\alpha
	\bigg) \,T\, \delta\Kbeta^\alpha \,,
\label{eq:deltaasq}
\end{align}
\begin{align}
\delta\prn{ K_\Theta \, \Theta^2}   & =
	\bigg( K_\Theta\, \Theta^2\, g^{\mu\nu} + \prn{T\, K_\Theta' \, \Theta^2
	+ 2\,  T\,\vs\;K_\Theta' \Theta^2 \,
	-2\, K_\Theta\, u^\alpha\, \nabla_\alpha \Theta} u^\mu\,u^\nu
	\bigg) \frac{1}{2}\, \delta g_{\mu\nu}
\nonumber \\
& \quad
	+ \bigg( 2\, K_\Theta\, \Theta \, \delta^\nu_\mu\, u^\alpha\bigg)
	\delta \Gamma_{\nu\alpha}^\mu
	+ \MR{\nabla_\mu\prn{2\,K_\Theta\, \Theta\, \delta u^\mu}}
\nonumber \\
& \quad
	+ \bigg( T\, K_\Theta'\, \Theta^2 \, u_\alpha
	- 2\,  P^\mu_\alpha\, \nabla_\mu\prn{\Theta\, K_\Theta}
	\bigg)\, T\,\delta\Kbeta^\alpha \,,
\label{eq:deltathsq}
\\
\delta\prn{ K_R\, R}  & =
	\bigg( K_R \, R\, g^{\mu\nu}+ K_R'\,T\, R\, u^\mu\,u^\nu
	- 2\, K_R\,R^{\mu\nu} +2\, \nabla^\mu \nabla^\nu K_R
	- \;  2\, g^{\mu\nu} \nabla^2\, K_R \bigg) \, \half\,\delta g_{\mu\nu}
\nonumber \\
&\qquad
+ \MR{
 \nabla_\mu \bigbr{ 2\ \delta g_{\alpha\beta} \brk{g^{\alpha\beta}\, \nabla^\mu -g^{\mu\beta}\nabla^{\alpha}}K_R\;
    - \brk{g^{\alpha\beta}\, \nabla^\mu -g^{\mu\beta}\nabla^{\alpha}}\prn{ K_R \ \delta g_{\alpha\beta} }  }	}
\nonumber \\
&\qquad
	+ T^2\, K_R'\, R\, u_\alpha\, \delta \Kbeta^\alpha \,,
\label{eq:deltaR}
\\
\delta\prn{ K_t\, \nabla_\alpha T \nabla^\alpha T}   & =
	\bigg( K_t\, (\nabla T)^2\, g^{\mu\nu} -2\, K_t\, \nabla^\mu T\, \nabla^\nu T
\nonumber \\
& \qquad  \quad
	- T\, \prn{K_t' \, (\nabla T)^2 + 2\, K_t\, \nabla^2 T}  u^\mu\,u^\nu
	\bigg) \frac{1}{2}\, \delta g_{\mu\nu}
		+ \MR{\nabla_\mu\prn{2\,K_t \, \nabla^\mu T  \;\delta T}}
\nonumber \\
& \quad
	 - \bigg(K_t' (\nabla T)^2 + 2\, K_t\, \nabla^2 T \bigg)\,
	T^2\, u_\alpha \delta\Kbeta^\alpha \,,
\label{eq:deltadelTsq}
\\
\delta\prn{ K_u\, \Theta\, u^\mu\,\nabla_\mu T}&   =
	\bigg( K_u \, \Theta \, u^\alpha \nabla_\alpha T\, g^{\mu\nu}
	+\bigg[ K_u\,\Theta\, u^\alpha\,\nabla_\alpha T
	-  T\, K_u\, \prn{\Theta^2 + u^\alpha \nabla_\alpha \Theta}
\nonumber \\
& \hspace{4.2cm}
	- u^\alpha \nabla_\alpha\prn{K_u\, u^\beta\, \nabla_\beta T}
	\bigg] u^\mu\,u^\nu	\bigg) \frac{1}{2}\, \delta g_{\mu\nu}
\nonumber \\
& \quad
	+\bigg(  P_\alpha^\mu \big( K_u\Theta\,\nabla_\mu  T
	-  \nabla_\mu \prn{ K_u\, u^\mu \nabla_\mu T} \big)
	- T\, K_u\, \big( \Theta^2 + u^\mu \nabla_\mu \Theta\big)\; u_\alpha
	\bigg)\, T\, \delta \Kbeta^\alpha
\nonumber \\
& \quad
	+ K_u\, u^\beta \nabla_\beta T\, u^\nu\, \delta^\alpha_\mu \, \delta \Gamma^{\mu}_{\alpha\nu}
	+ \MR{\nabla_\mu\prn{K_u \Theta \,u^\mu \, \delta T + K_u\, u^\alpha \nabla_\alpha T \delta u^\mu}} \,,
\label{eq:deltaudelT}
\\
\delta\prn{ K_x\,\acc^\mu \nabla_\mu T}   & =
	\bigg( K_x\, \acc^\alpha\nabla_\alpha T\, g^{\mu\nu}
	+ \bigg[ K_x \, \acc^\alpha \nabla_\alpha T -T\, K_x\, \nabla_\alpha \acc^\alpha
\nonumber \\
& \hspace{4.2cm}
	- u^\alpha \nabla_\beta  \prn{K_x u^\beta\,\nabla_\alpha T}	\bigg]  u^\mu\,u^\nu
	\bigg) \frac{1}{2}\, \delta g_{\mu\nu}
\nonumber \\
& \quad
	+\bigg(P_\alpha^\mu \big( K_x\, \nabla_\mu u^\nu \,\nabla_\nu T
	- \nabla_\nu \prn{ K_x\, \nabla_\mu T\, u^\nu} \big)
	- T\, K_x\, \nabla_\mu \acc^\mu \; u_\alpha
	\bigg)\, T\, \delta \Kbeta^\alpha
\nonumber \\
& \quad
	+ K_x\, \nabla_\mu T\, u^\alpha\, u^\nu \, \delta \Gamma^{\mu}_{\alpha\nu}
	+ \MR{\nabla_\mu\prn{K_x \,u^\mu \, \nabla_\nu T  \;\delta u^\nu + K_x\, \acc^\mu \,
	\delta T}} \,,
\label{eq:deltaadelT}
\\
\delta\prn{ K_y\, \prn{ u^\mu\,\nabla_\mu T}^2} &   =
	\bigg( K_y \, (u^\alpha \nabla_\alpha T)^2\,  g^{\mu\nu}
	+\bigg[ (T\, K_y'  +2\, K_y)\, (u^\alpha \nabla_\alpha T)^2
\nonumber \\
& \hspace{4.2cm}
	- 2\, T\, \nabla_\alpha \prn{K_y\, u^\alpha\, u^\beta\, \nabla_\beta T}
	\bigg] u^\mu\,u^\nu	\bigg) \frac{1}{2}\, \delta g_{\mu\nu}
\nonumber \\
& \quad
	+\bigg( 2\, K_y\, u^\beta\, \nabla_\beta T \; P_\alpha^\mu  \,\nabla_\mu T
	-2\, T\, \nabla_\mu\prn{K_y\, u^\beta\nabla_\beta T\, u^\mu } u_\alpha
\nonumber \\
& \quad
	+ \;T\, K_y' \prn{u^\beta\nabla^\beta T} u_\alpha
	\bigg)\, T\, \delta \Kbeta^\alpha
	+ \MR{\nabla_\mu\prn{2\, K_y \, u^\alpha \nabla_\alpha T\, u^\mu\, \delta T}} \,.
\label{eq:deltaudelTsq}
\end{align}
\end{subequations}

In the course of the derivation, we have used for the variation of the curvature term the standard identity
\begin{align}
g^{\alpha\beta}\delta R_{\alpha \beta} = \nabla^\mu\brk{ \nabla^\nu \delta g_{\mu\nu}
- g^{\alpha\beta} \nabla_\mu \delta g_{\alpha \beta} } \,.
\end{align}
The stress tensor can be read off from the above expressions as the coefficient of $\frac{1}{2}\, \delta g_{\mu\nu}$. However, to do so we need to convert $\delta \Gamma^\mu_{\nu\alpha}$ variations in some of the terms into $\delta g_{\mu\nu}$ variations. This can be done easily using the identity
\begin{equation}
\begin{split}
X^{\alpha \sigma}{}_\rho \; \delta \Gamma^{\rho}_{\alpha \sigma} &= \nabla_\rho\brk{ X^{\alpha[\beta\rho]} +
X^{\beta[\alpha\rho]} - X^{\rho(\alpha\beta)}}\, \delta g_{\alpha\beta} \\
& \quad  -\nabla_\rho\bigbr{\brk{ X^{\alpha[\beta\rho]} +
X^{\beta[\alpha\rho]} - X^{\rho(\alpha\beta)}}\, \delta g_{\alpha\beta} }
\end{split}
\end{equation}
Using this we compute the combined contribution from \eqref{eq:deltsigsq}-\eqref{eq:deltaudelT} separately to be
\begin{align}
2\bigg( &
 	K_\Theta\, \Theta \,\delta^\nu_\mu \, u^\alpha + K_a \, \acc_\mu \, u^\nu\,u^\alpha - K_\sigma \, \sigma^{\alpha\nu}\, u_\mu
 \bigg)  \delta \Gamma^\mu_{\nu\alpha}=
 \nonumber \\
&
	2\, \nabla_\rho
	\bigg(
		-K_\Theta\, \Theta\, u^\rho\, g^{\mu\nu} + K_a\prn{ u^\mu\,u^\nu \,\acc^\rho
		- 2\,\acc^{(\mu} u^{\nu)}\, u^\rho}
		-K_\sigma\prn{\sigma^{\mu\nu}\, u^\rho - 2\, \sigma^{\rho(\mu} \,u^{\nu)}}
	\bigg ) \frac{1}{2}\,\delta g_{\mu\nu}
\nonumber \\
& +  \MR{ \nabla_\mu \brk{ K_a\prn{ -u^\alpha\,u^\beta \,\acc^\mu
		+2\,\acc^{(\alpha} u^{\beta)}\, u^\mu} \delta g_{\alpha\beta} } }
\nonumber \\
 &+  \MR{ \nabla_\mu \brk{ K_\Theta\, \Theta\, u^\mu\, g^{\alpha\beta}\delta g_{\alpha\beta}
 +K_\sigma\prn{\sigma^{\alpha\beta}\, u^\mu - 2\, \sigma^{\mu(\alpha} \,u^{\beta)}} \delta g_{\alpha\beta} } }
\nonumber \\
\bigg( &
  	K_x \, \nabla_\mu T \, u^\alpha\,u^\nu +  K_u \, u^\beta \nabla_\beta T\, \delta^\alpha_\mu\, u^\nu
\bigg)  \delta \Gamma^\mu_{\nu\alpha}=
 \nonumber \\
&
	   \nabla_\rho
	\bigg(
		K_x\prn{ u^\mu\,u^\nu \nabla^\rho T - 2\, u^\rho \, u^{(\mu} \,\nabla^{\nu)}T}
		- K_u\, u^\beta\, \nabla_\beta T\; u^\rho \, g^{\mu\nu}
	\bigg ) \frac{1}{2}\,\delta g_{\mu\nu} 
 \nonumber \\
& + \MR{  \nabla_\mu\bigbr{\brk{-K_x\prn{ u^\alpha\,u^\beta \nabla^\mu T - 2\, u^\mu \, u^{(\alpha} \,\nabla^{\beta)}T}
		+ K_u\, u^\rho \nabla_\rho T\; u^\mu \, g^{\alpha\beta}
	} \frac{1}{2}\,\delta g_{\alpha\beta}} } \,.
\label{eq:christvar}
\end{align}
All in all we find the currents for the second order neutral fluid to be:
\begin{align}
T^{\mu\nu}_{(2)}
& =
	K_\sigma\, \sigma^2\, g^{\mu\nu} + T\, K_\sigma' \, \sigma^2\, u^\mu\,u^\nu -4\, K_\sigma\,
	\sigma^{\mu\alpha}\,\sigma_\alpha^{\;\nu} - 4\, K_\sigma\, \frac{\Theta}{d-1}\, \sigma^{\mu\nu}
\nonumber \\
&\quad
	 -\;4\, K_\sigma\, u_\alpha \prn{u^\rho\,\nabla_\rho\, \sigma^{\alpha(\mu}}\, u^{\nu)}
	 -4\, u^{(\nu}\,\nabla_\alpha\prn{K_\sigma \,\sigma^{\mu)\alpha}}
	 +2\, K_\sigma\, \sigma^2\, u^\mu\,u^\nu
\nonumber \\
&\quad
	 + 2\,\nabla_\rho \prn{K_\sigma \prn{\sigma^{\mu\nu}\, u^\rho - 2\, \sigma^{\rho(\mu} \,u^{\nu)}} }
	+ K_\omega\, \omega^2\, g^{\mu\nu} +
	\prn{2\, K_\omega +  T \, K_\omega'} \, \omega^2\, u^\mu\,u^\nu
\nonumber \\
& \quad
	 - 4\, K_\omega\, \omega^{\mu\alpha}\,\omega_\alpha^{\;\nu}
	 + 4\, K_\omega \,\omega^{(\mu\alpha}\,u^{\nu)}\, \acc_{\alpha}
 	+\; 4\, \nabla_\alpha\prn{K_\omega\, \omega^{\alpha(\mu} }\,u^{\nu)}
 	+ {\widetilde K}_a\, \acc^2\, g^{\mu\nu}
\nonumber \\
& \quad
	+ {\widetilde K}_a' \, T\, \acc^2\, u^\mu\,u^\nu
	+ 2\, {\widetilde K}_a\, \acc^\mu\,\acc^\nu + 4\, {\widetilde K}_a\, \acc^2\, u^\mu \, u^\nu
	 + 2\,\nabla_\rho \prn{{\widetilde K}_a\prn{ u^\mu\,u^\nu \,\acc^\rho
		- 2\,\acc^{(\mu} u^{\nu)}\, u^\rho}}
\nonumber \\
& \quad
	+ {\widetilde K}_\Theta\, \Theta^2\, g^{\mu\nu}  + \prn{T\, {\widetilde K}_\Theta' \, \Theta^2
	+ 2\,  T\,\vs\;{\widetilde K}_\Theta' \Theta^2 \,
	-2\, {\widetilde K}_\Theta\, u^\alpha\, \nabla_\alpha \Theta} u^\mu\,u^\nu
	-2 \,\nabla_\rho\prn{{\widetilde K}_\Theta\, \Theta\, u^\rho\, g^{\mu\nu}}
\nonumber \\
& \quad
	+ K_R \, R\, g^{\mu\nu}+ K_R'\,T\, R\, u^\mu\,u^\nu
	- 2\, K_R\,R^{\mu\nu} +2\, \nabla^\mu \nabla^\nu K_R
	- \;  2\, g^{\mu\nu} \nabla^2\, K_R \,,
\label{eq:n2traw}
\end{align}
while
\begin{align}
\aheat_{(2)}^\alpha &= T\, K_\sigma'\, \sigma^2 \, u^\alpha
	-2\,  P_{\nu}^\alpha\,\nabla_\mu\prn{K_\sigma\, \sigma^{\mu\nu}}
	+ 2\, K_\sigma\, a^\mu\,\sigma_\mu{}^\alpha
	+ T\, K_\omega'\, \omega^2 \, u^\alpha
	 +2\, T\, P_\nu^\alpha\, \nabla_\mu\prn{K_\omega\, \omega^{\mu\nu}}
\nonumber \\
&\quad
	 +2\, T\, K_\omega\, \omega^{\alpha\mu}\, \acc_\mu
	 +2\, {\widetilde K}_a\, \acc_\mu \, P^{\nu\alpha}\, \nabla_\nu\, u^\mu - 2\, \nabla_\nu
	\prn{{\widetilde K}_a\, \acc_\mu\, u^\nu} \, P^{\mu\alpha} + T\, {\widetilde K}_a'\, \acc^2 \, u^\alpha
\nonumber \\
&\quad
	+ T\, {\widetilde K}_\Theta'\, \Theta^2 \, u^\alpha
	- 2\,  P^{\alpha\mu}\, \nabla_\mu\prn{\Theta\, {\widetilde K}_\Theta}
	+ T\, K_R'\, R\, u^\alpha \,,
\label{eq:n2vraw}
\end{align}
where we have written the answer in the field redefinition invariant combination. The pre-symplectic potential for this system
can be read off from the boundary terms as 
\begin{equation}
\begin{split}
 (\PSymplPot{})^\mu_{(2)} &=
 	 \prn{ 2\, K_\sigma \, \sigma^{\mu\nu}-2\,   K_\omega \, \omega^{\mu\nu}  }\, \delta u_\nu
 	+  2\, K_a \,u^\mu \; \acc_\nu \delta u^\nu
 	+  2\, K_\Theta \Theta \, \delta u^\mu  \\
&\quad 
	+ 2\ \delta g_{\alpha\beta} \brk{g^{\alpha\beta}\, \nabla^\mu -g^{\mu\beta}\nabla^{\alpha}}K_R\;
   	 - \brk{g^{\alpha\beta}\, \nabla^\mu -g^{\mu\beta}\nabla^{\alpha}}\prn{ K_R \ \delta g_{\alpha\beta} }  \\
&\quad 
	+ \; \prn{2\,K_t \, \nabla^\mu T   + K_u \,\Theta \,u^\mu + K_x\, \acc^\mu + 2\, K_y \,  u^\mu u^\rho \nabla_\rho T } 
	\delta T \\
&\quad  
	+\;  u^\rho \nabla_\rho T\  \prn{ K_u\, \delta u^\mu  }
	+ K_x \,u^\mu \, \nabla_\nu T  \;\delta u^\nu  \\
& \quad	
	+  K_a\prn{ -u^\alpha\,u^\beta \,\acc^\mu+2\,\acc^{(\alpha} u^{\beta)}\, u^\mu} \delta g_{\alpha\beta}
 \\
 &\quad 
 	+  K_\Theta\, \Theta\, u^\mu\; g^{\alpha\beta}\delta g_{\alpha\beta} 
 	+K_\sigma\prn{\sigma^{\alpha\beta}\, u^\mu - 2\, \sigma^{\mu(\alpha} \,u^{\beta)}} \delta g_{\alpha\beta}
 \\
 & \quad
 	+\brk{ -K_x\prn{ u^\alpha\,u^\beta \nabla^\mu T - 2\, u^\mu \, u^{(\alpha} \,\nabla^{\beta)}T}
		+ K_u\, u^\rho \nabla_\rho T\; u^\mu \, g^{\alpha\beta}} \frac{1}{2}\,\delta g_{\alpha\beta}	\\
 \end{split}
 \label{eq:neutralpresymp2}
\end{equation}
where the first four lines are obtained from the explicit boundary terms in the variations 
\eqref{eq:deltsigsq}-\eqref{eq:deltaudelTsq} and the last three lines are obtained from the integration by parts on the variation of the Christoffel symbols \eqref{eq:christvar}.

\subsection{Transport coefficients for neutral fluids}
\label{sec:n2transport}

 The expression for the stress tensor as written in \eqref{eq:n2traw} is rather unilluminating (not to mention formidable), so we need to massage it further to extract some physical information and compare with results in the literature. It is traditional to present the result for the energy momentum tensor in the Landau frame, where the corrections (dissipative or otherwise) to the ideal fluid stress-tensor in the gradient expansion are orthogonal to the fluid velocity field. One can then express the result up to second order   \cite{Bhattacharyya:2012nq}\footnote{ Note that we have not isolated an explicit factor of $T$ in front of the second order transport coefficients as in \cite{Bhattacharyya:2012nq}. Also some authors (cf., \cite{Baier:2007ix}) prefer to make explicit that some second order transport is inherited from first order viscous terms, e.g., it is common to find $\tau = \eta\, \tau_\pi$. We refrain from making such choices to keep expressions manageable.}
 \begin{equation}\label{eq:TLandau}
\begin{split}
T_{\mu\nu} =~& \epsilon\, u_\mu\,u_\nu +  p\;P_{\mu\nu}
-2\,\eta\, \sigma_{\mu\nu} - \zeta \, P_{\mu\nu} \Theta\\
~&+\;\bigg[ \tau \, u^\alpha \nabla_\alpha \sigma_{\langle\mu\nu\rangle} + \kappa_1 \, R_{\langle \mu\nu\rangle} + \kappa_2 \, (F_R)_{\langle \mu\nu\rangle} +\lambda_0\,  \Theta\, \sigma_{\mu\nu}\\
&\qquad +\lambda_1\, {\sigma_{\langle \mu}}^\alpha\, \sigma_{\alpha\nu\rangle}+ \lambda_2\,  {\sigma_{\langle \mu}}^\alpha \, \omega_{\alpha\nu\rangle}+ \lambda_3\, {\omega_{\langle \mu}}^\alpha\, \omega_{\alpha\nu\rangle} + \lambda_4\, \acc_{\langle\mu}\acc_{\nu\rangle}\bigg]\\
&+\, P_{\mu\nu}\bigg[\zeta_1\, u^\alpha\nabla_\alpha \Theta + \zeta_2 \, R + \zeta_3\, R_{00}
 + \xi_1 \, \Theta^2 + \xi_2\,  \sigma^2+ \xi_3 \, \omega^2
+\xi_4 \, \acc^2 \bigg]\,.
\end{split}
\end{equation}
Most of the fluid dynamical tensors which we are using in the above are given in Table \ref{notation:tabFields} and the angle bracket notation indicates projection to the symmetric part transverse to the velocity.
\begin{equation}
A_{\langle\mu\nu\rangle} \equiv P_\mu^\alpha P_\nu^\beta\left(\frac{A_{\alpha\beta} + A_{\beta\alpha}}{2} - \left[\frac{A_{\rho\sigma}P^{\rho\sigma}}{d-1}\right]g_{\alpha\beta}\right) \,.
\label{eq:ttproj}
\end{equation}
In addition we have a few combinations of the curvatures which are defined as
\begin{equation}\label{eq:RFdef}
\begin{split}
&F_R^{\mu\nu} = R^{\mu \alpha \nu \beta}\, u_\alpha u_\beta, \qquad R^{\mu\nu}
= R^{\alpha\mu \beta\nu}g_{\alpha \beta}\,, \qquad R_{00} = R^{\mu\nu} \,u_\mu\,u_\nu \\
\end{split}
\end{equation}
with $R_{\alpha\beta\gamma\delta}$ being the Riemann tensor of the background geometry.

\begin{table}[h]
\centering
\begin{tabular}{|| c | c | c || }
\hline\hline
\multicolumn{3}{||c||}{\shadeB{$2^{\rm nd}$ order neutral fluids: hydrostatic response}}\\
\hline
 Scalars & Vectors & Tensors \\
 \hline
$\omega^2$   &  $\omega^{\mu\nu}\acc_\nu$  & $\omega^{\alpha<\mu}\omega^{\nu>}{}_\alpha$ \\
$\acc^2$                        &     &  $\acc^{<\mu}\acc^{\nu>}$ \\
$R$ & & $R^{<\mu\nu>}$\\
\hline
 $R_{\alpha\beta} u^\alpha u^\beta$ & $P^\mu_\nu D_\lambda \omega^{\nu\lambda}$ &  $F_R^{<\mu\nu>}\equiv u^\alpha u^\beta R^{<\mu}{}_\alpha{}^{\nu>}{}_\beta$ \\
\hline
\multicolumn{3}{||c||}{{\shadeR $4S+4T={\bf 8}=5\,\PF+3 \,\PS+0\,\PV+0\,\text{A}$}}\\
\hline\hline
\end{tabular}
\caption{The $8$ hydrostatic response terms for  parity-even neutral fluid at $2^{\rm nd}$ order in derivative expansion. We have listed
the vectors though they do not contribute to frame-invariant transport data. Among the $8$ terms,
$\PF=5$ combinations are forbidden by hydrostatic principle whereas the remaining $\PS=3$ combinations are generated by using the first $3$ scalars in the Lagrangian.}
\label{tab:NeutralResponse}
\end{table}

\begin{table}[h]
\centering
\begin{tabular}{|| c | c | c || }
\hline\hline
\multicolumn{3}{||c||}{{\shadeB $2^{\rm nd}$ order neutral fluid: non-hydrostatic transport}} \\
\hline
 Scalars & Vectors & Tensors \\
\hline
$\Theta^2$                 & $\Theta \acc^\mu$   & $\sigma^{\alpha<\mu}\sigma^{\nu>}{}_\alpha$\\
$\sigma^2$                 &       & $\Theta \sigma^{\mu\nu}$ \\
 & $\sigma^{\mu\nu}\acc_\nu$     & $\sigma^{\alpha<\mu}\omega^{\nu>}{}_\alpha$ \\
 \hline
$(u^\alpha D_\alpha) \Theta$  & $P^{\mu\nu}D_\nu \Theta$  & $(u^\alpha D_\alpha) \sigma^{\mu\nu}$  \\
                & $P^\mu_\nu D_\lambda \sigma^{\nu\lambda}$  & \\
\hline
\multicolumn{3}{||c||}{{\shadeR $3S+4T={\bf 7}=2\,\LS+0\,\GV+2\,\text{B}+3\,\text{D}$}}\\
\hline\hline
\end{tabular}
\caption{The $7$ non-hydrostatic transport terms for  parity-even neutral  fluid at $2^{\rm nd}$ order in derivative expansion.
We have listed the vectors though they do not contribute to frame-invariant transport data.  Among the $7$ terms,
$\LS=2$ combinations are generated by inserting the first $2$ non-hydrostatic scalars into the Lagrangian. We have  $\GV=0$
and there are $2$ combinations in Class B and $3$ combinations in Class D, which are given in Table \ref{tab:CountingEven}.}
\label{tab:NeutralTransport}
\end{table}

The comparison with the expression for the stress tensor \eqref{eq:TLandau} is however tricky as written for two reasons:
\begin{itemize}
\item[(a)]  The result \eqref{eq:TLandau} is given in a basis of  independent tensors wherein derivatives of the temperature have been eliminated in favour of those involving the velocity gradients.
\item[(b)] It is also presented in the Landau frame, while the stress tensor we compute will naturally be in a frame where the entropy current is $J^\mu_S = s\, u^\mu $. The latter has been called {\em entropy frame} in \cite{Bhattacharya:2012zx}.
\end{itemize}

Using the conservation of the ideal fluid \eqref{eq:gradT} we can eliminate temperature gradients and obtain the result for the stress tensor in terms of velocity gradients alone on-shell.  Then one can attempt to convert the answer to the Landau frame by an explicit field redefinition. Fortunately, if we set the first order gradient terms to vanish, we can employ a shortcut as discussed in \cite{Bhattacharya:2012zx,Haehl:2013kra} which we used in \S\ref{sec:neutral}. We simply project our result for $T^{\mu\nu}$ onto the invariant tensor and scalar parts using \eqref{eq:frameproj} and read off the coefficients of the independent tensors used in \eqref{eq:TLandau}. We implement this for each term and use \eqref{eq:gradT} at intermediate stages to simplify the computation. When the dust settles we find:
\begin{subequations}
\label{fconst1}
\begin{align}
&
	\eta = \zeta =0 \,,
\\ &
	\tau =- 2\,T\,K_R' - 2\,K_\sigma \qquad
\label{eq:tausol}
\\&
	\kappa_1  = -2\,K_R \qquad
	\kappa_2= -2\,T\, K_R' \,,
\label{eq:kappasol}
\\  &
	 \lambda_0=  \prn{2\,\vs -\frac{4}{d-1} } T\, K_R'-2\,K_\sigma
 	+2\,\vs\,T \, K_\sigma'
 \label{eq:lam0sol}
 \\ &
	\lambda_1 = -2\,T\, K_R' \,,\qquad
	\lambda_2 = 4\,K_\sigma\,, \qquad
	\lambda_3 = -2\,T\, K_R' + 4\, K_\omega\,,
\label{eq:lam13sol}
\\ &
	\lambda_4= -2 \, {\widetilde K}_a
	+2\, T\,\prn{T\,K_R}''  \,,
\nonumber \\
&
	\zeta_1 = -2\,\vs \, {\widetilde K}_a
	+2\,\frac{d-2}{d-1}\, T\, K_R' -2\, {\widetilde K}_\Theta \,,
\nonumber \\  &
	\zeta_2 = \prn{\frac{d-3}{d-1} +\vs}\, K_R - \vs\, T\, K_R'
\nonumber \\ &
	\zeta_3 = -2\,\vs \, {\widetilde K}_a+2\,\prn{\frac{d-2}{d-1} +\vs} T\, K_R'
	+ 2\,\prn{\vs - \frac{1}{d-1}}\, K_R  	\,,
\label{eq:zeta123sol}
	\\
&	\xi_1 =  -\frac{2}{d-1}\,\vs\, {\widetilde K}_a  +2\, \vs\,T\, \prn{\vs\,T\, K_R'}'
	+ 2\,\prn{\frac{d-2}{d-1}  +\vs}\, \prn{\frac{1}{d-1}-\vs}\,T\, K_R'
\nonumber \\
&	\hspace{2cm}
	 -\; \brk{ (1+ \vs) \, {\widetilde K}_\Theta
	- \vs\, T\,{\widetilde K}_\Theta' } ,
\nonumber \\&
	\xi_2 =  -2\,\vs\, {\widetilde K}_a
	+2\, \prn{\frac{d-2}{d-1}+\vs}\, T\, K_R' + \prn{1-\vs} K_\sigma
	-\vs\, T\,K_\sigma'
\nonumber \\ &
	\xi_3 =	-2\,\vs \, {\widetilde K}_a +2\, \prn{\frac{d-2}{d-1}+\vs}\, T\,K_R' +
	\prn{\frac{d-5}{d-1}+3\,\vs} K_\omega
	-\vs\, T\,K_\omega'
\nonumber \\ &
	\xi_4 =  \brk{ \prn{\frac{d-3}{d-1}+ \vs}  \,{\widetilde K}_a + \vs \, T\, {\widetilde K}_a' }
	 -2\, \prn{\frac{d-2}{d-1}+\vs}\,  T\,\prn{T\,K_R'}'	.
\end{align}
\end{subequations}

These are the physically interesting results for the transport coefficients written in terms of the field redefinition invariant combinations of the coefficient functions
$K_\sigma(T)$, $K_\omega(T)$, ${\widetilde K}_a(T)$, ${\widetilde K}_\Theta(T)$ and $K_R(T)$.
There are several interesting relations that these hydrodynamic data obey: for one because $15$
a-priori independent data are expressed in terms of $5$ functions, we expect to see $10$ relations between them (which can be obtained by eliminating the $K_i$).
More explicitly, we obtain the first 5 relations by substituting
\begin{equation}
\begin{split}
K_R &= -\half \, \kappa_1 \\
K_\omega &= \quarter \brk{\lambda_3- T \,\kappa_1' }\\
{\widetilde K}_a &=  -\half  \brk{\lambda_4 + T(T\kappa_1)'' }
\end{split}
\end{equation}
into the expressions for $\{\kappa_2,\zeta_2,\zeta_3,\xi_3,\xi_4\}$. These 5 relations are the ones that appear in the hydrostatic partition
function analysis \cite{Banerjee:2012iz}. 
In addition, if we substitute
\begin{equation}
\begin{split}
K_\sigma &=\frac{1}{2}\brk{T\, \kappa_1'-\tau}
 \\
{\widetilde K}_\Theta &= - \frac{1}{2} \brk{\zeta_1+ \frac{d-2}{d-1}\, T\, \kappa_1' - \vs \prn{\lambda_4 + T(T\kappa_1)'' } }
\end{split}
\end{equation}
into the expressions for $\{\lambda_0,\lambda_1,\lambda_2,\xi_1,\xi_2\}$ we get 5 additional relations which cannot be obtained from
hydrostatic  analysis. Among these 5 relations into the 2 relations for $\{\lambda_0-\xi_2,\lambda_2\}$ which remove
the two Class B transport coefficients whereas the 3 relations for  $\{\lambda_0+\xi_2,\lambda_1,\xi_1\}$ remove the three Class D transport coefficients.
Further, we can clearly see that the two relations we highlighted for the Weyl invariant fluid  in (\ref{eq:weylrelns}) continue to hold even for the general neutral fluid:
\begin{equation}
\tau  = \lambda_1 -\frac{1}{2}\,\lambda_2 \,, \qquad \lambda_1 = \kappa_2\,.
\label{eq:2drelnsA}
\end{equation}
These relations serve to project out one Class B transport coefficient and one Class D transport coefficient respectively.

\subsection{Entropy current for the neutral fluid}
\label{sec:jsn2}

The entropy current can be easily computed by variation of the Lagrangian with respect to the temperature. The quickest way is to use the identity \eqref{eq:sVZeta} in terms of the adiabatic heat and charge currents. Using the result obtained in 
\eqref{eq:n2vraw} for the adiabatic heat current it is trivial to compute the second order corrections to the entropy current.  We find a simple expression:
\begin{equation}
\label{eq:2ndEntopyCur}
  J^\mu_{S,(2)} = \left( K_\sigma'\, \sigma^2 + K_\omega'\,\omega^2+{\widetilde K}_a'\, \acc^2+{\widetilde K}_\Theta'\, \Theta^2+K_R' R^2 \right) u^\mu \,.
\end{equation}
which of course fits with the explicit $T$ dependence of $\Lag_2$. 

It is instructive to examine the Noether current in Class L for this system, which encodes the non-canonical part of the entropy current (upto Komar terms). To achieve this we need the pre-symplectic potential which  has been computed in 
\eqref{eq:neutralpresymp2}. One then computes:
\begin{align}
\N^\mu_{(2)}\brk{\Bfields} &=  \brk{ \Kbeta^\mu \Lag- (\PSymplPot{\Bfields})^\mu }_{(2)}
\nonumber \\ &=  
	 \frac{K_\sigma}{T} \prn{\sigma^2 \, u^\mu - 2 \,\sigma^{\mu\nu} \,T\, \diffB u_\nu
	- T \prn{\sigma^{\alpha\beta}\, u^\mu - 2\, \sigma^{\mu(\alpha} \,u^{\beta)}} \diffB g_{\alpha\beta} } 
\nonumber \\ & \quad	 + \;
	\frac{K_\omega}{T} \prn{\omega^2 \,u^\mu+ 2 \,\omega^{\mu\nu}\, T\, \diffB u_\nu } 
\nonumber \\ & \quad 	+\;  
	\frac{K_a}{T} \, \brk{ u^\mu \prn{\acc^2 - 2 \acc_\nu \, T\, \diffB u^\nu } 
	+ T\prn{ u^\alpha\,u^\beta \,\acc^\mu-2\,\acc^{(\alpha} u^{\beta)}\, u^\mu} \diffB g_{\alpha\beta}
	}
\nonumber \\ & \quad 	+\;  
 	 \frac{K_\Theta}{T}  \prn{\Theta^2 u^\mu - 2 \,\Theta\, T\diffB u^\mu 
 	-  T\, \Theta\, u^\mu\; g^{\alpha\beta}\diffB g_{\alpha\beta} } 
\nonumber \\ & \quad 	+\;  
	\frac{K_R}{T}\, R\, u^\mu- 2\, \diffB g_{\alpha\beta} \brk{g^{\alpha\beta}\, \nabla^\mu -g^{\mu\beta}\nabla^{\alpha}}K_R\;
   	 + \brk{g^{\alpha\beta}\, \nabla^\mu -g^{\mu\beta}\nabla^{\alpha}}\prn{ K_R \ \diffB g_{\alpha\beta} } 
\nonumber \\ & \quad 	+\;  
	 \frac{K_t}{T} \prn{\nabla_\alpha T\, \nabla^\alpha T\, u^\mu -2\, T\, \nabla^\mu T\, \diffB T}
\nonumber \\ & \quad 	+\;  
	 \frac{K_u}{T} \prn{ \Theta\, u^\alpha \nabla_\alpha T\, u^\mu - T\, \Theta\, u^\mu \diffB T 
	 - T\, u^\alpha\, \nabla_\alpha T\; \diffB u^\mu 
	 - \frac{1}{2} \, T\, u^\rho\, \nabla_\rho T\, u^\mu\, g^{\alpha\beta} \diffB g_{\alpha\beta} }
\nonumber \\ & \quad 	+\;  
	 \frac{K_x}{T} \prn{ \acc^\alpha\, \nabla_\alpha T \, u^\mu- T\,\acc^\mu \, \diffB T - T\, u^\mu \nabla_\nu T\, \diffB u^\nu 
	 + T  \prn{ \frac{1}{2}u^{\alpha}u^{\beta} \nabla^\mu T -  u^\mu \,u^{(\alpha}\, \nabla^{\beta)}T }  \diffB g_{\alpha\beta} }
\nonumber \\ & \quad 	+\;  
	\frac{K_y}{T}\prn{(u^\alpha \nabla_\alpha T)^2 \, u^\mu  -  2\, T \,  u^\mu \, u^\rho \nabla_\rho T \, \diffB T}	 
\end{align}

Using the on-shell expressions:
\begin{align}
&\diffB u^\mu \simeq-\Theta\, \vs\, \Kbeta^\mu \,, 
\qquad 
\diffB u_\mu \simeq  \Theta\, \vs\, \Kbeta_\mu 
\,, \qquad
\diffB T \simeq -\Theta\, \vs \,,
\nonumber \\
& \diffB g_{\alpha\beta} = \frac{2}{T}
\prn{\sigma_{\alpha\beta} + P_{\alpha\beta}\, \frac{\Theta}{d-1} - \Theta\, \vs \, u_\alpha\, u_\beta} 
\end{align}	
we can simplify the above to
\begin{align}
\N^\mu_{(2)}\brk{\Bfields} &=   -\frac{K_\sigma}{T} \,\sigma^2 \, u^\mu +\frac{K_\omega}{T} \,\omega^2 \,u^\mu 
	+ \frac{K_a}{T}\, \prn{\acc^2\, u^\mu - 2\, \vs\, \Theta\, \acc^\mu}
	-\frac{K_\Theta}{T}\, \Theta^2\, u^\mu 
\nonumber \\ & \quad 	+\;  
	\frac{K_R}{T}\, R\, u^\mu- 2\, \diffB g_{\alpha\beta} \brk{g^{\alpha\beta}\, \nabla^\mu -g^{\mu\beta}\nabla^{\alpha}}K_R\;
   	 + \brk{g^{\alpha\beta}\, \nabla^\mu -g^{\mu\beta}\nabla^{\alpha}}\prn{ K_R \ \diffB g_{\alpha\beta} } 
\nonumber \\ & \quad 	+\;  
	T\, K_t \prn{ \acc^2\, u^\mu + \Theta^2\, \vssq \, u^\mu -2\,\Theta\, \vs\, \acc^\mu} + K_u\, \, \Theta^2\, \vs\, u^\mu
\nonumber \\ & \quad 	+\;
	  K_x\, \prn{ -\acc^2\, u^\mu + 2\, \Theta\, \vs \, \acc^\mu} -T\, K_y\, \Theta^2\, \vssq  \,u^\mu  	
\nonumber \\ &	
	=   -\frac{K_\sigma}{T} \,\sigma^2 \, u^\mu +\frac{K_\omega}{T} \,\omega^2 \,u^\mu 
	+ \frac{{\widetilde K}_a}{T}\, \prn{\acc^2\, u^\mu - 2\, \vs\, \Theta\, \acc^\mu}
	-\frac{{\widetilde K}_\Theta}{T}\, \Theta^2\, u^\mu 
\nonumber \\ & \quad 	+\;  
	\frac{K_R}{T}\, \prn{R\, u^\mu 
	+2\, \nabla^\mu\prn{\Theta\, (1+\vs)} -2\, \nabla_\alpha\prn{\sigma^{\alpha \mu}+ P^{\alpha\mu}\,\frac{\Theta}{d-1}
	 - \Theta\, \vs\, u^\alpha\, u^\mu } }
\nonumber \\ & \quad 	-\;  
	2\, \frac{(T\, K_R)'}{T} \prn{\Theta^2\, \vs\, u^\mu - 
	\Theta\, \acc^\mu \, \prn{\frac{d-2}{d-1}+ \vs}  + \sigma^{\mu\alpha}\, \acc_\alpha }	 
\label{eq:necurf}
\end{align}
It is useful to note that the free energy current is field redefinition invariant as can be explicitly seen from the fact that the terms combine into the combinations suggested in \eqref{eq:Kfredef}.

\section{The hydrostatic entropy current}
\label{sec:sayantani}

In \S\ref{sec:hydrostatics} and \S\ref{sec:classL} we constructed an entropy current starting from a basic variational principle, which we used, for example, in deriving \eqref{eq:necurf}.  
As  we have mentioned in the course of our discussions, a very impressive analysis of the entropy current arising from hydrostatics was described in \cite{Bhattacharyya:2013lha,Bhattacharyya:2014bha}. 
We revisit that argument in our language providing a simple translation to the considerations of these papers to the current discussion.  

\subsection{The entropy analysis of Bhattacharyya}
\label{sec:sbcompare}

Consider a system in hydrostatic equilibrium as described in \S\ref{sec:hydrostatics}, for which we can write down an equilibrium partition function. 
To understand the structure of the entropy current, we will need to depart from hydrostatics, which we shall do in the gradient expansion, by introducing time dependence
as described in \cite{Bhattacharyya:2013lha,Bhattacharyya:2014bha}.  A useful proxy for the time dependence is the operator $\diffB $, since in equilibrium $\diffB \mid_\text{Hydrostatics} = \diffEq$ 
annihilates the background sources.  Thus introducing linear time dependence is tantamount to working at linear order in variations captured by $\diffB$. 

With this understanding, let us state the various results obtained in \cite{Bhattacharyya:2013lha} in order:
\begin{itemize}
\item  The second law of thermodynamics implies that every equilibrium configurations is associated  with a partition function.
\item The leading ${\cal O}(\diffB)$ terms in the entropy current are determined by this partition function. 
\item  Demanding on-shell conservation of this entropy current to linear order in $\diffB$ is same as demanding that energy-momentum and charge currents
be derived by varying the associated partition function with respect to background sources. This gives all the equality type constraints.
\item The divergence of entropy current at quadratic order in $\diffB$ expansion leads  to  inequality type constraints. 
\item Terms in the entropy current at higher order in the $\diffB$ expansion can then be arranged as to  give a non-negative definite quadratic form for the total entropy production $\Diss$. They do not
produce any new constraints.
\end{itemize}
The crucial step here is, of course, the  construction of an entropy current from the hydrostatic partition function.  Paraphrasing our discussion in  \S\ref{sec:hscurrents}, we can describe this construction 
as follows: 
\begin{enumerate}
\item We begin with the thermodynamic formula for entropy 
\begin{equation}
\begin{split}
\text{Total entropy}
 &= \frac{\partial}{\partial T}\int_{\Sigma_E}
\brk{ \Kbeta^\sigma \Lag }_\text{Hydrostatic}
\ d^{d-1}S_\sigma\,.  \\
\end{split}
\end{equation}
which can then be rewritten up to boundary terms as
\begin{equation}
\begin{split}
\text{Total entropy}
 &= \int_{\Sigma_E}
\brk{- \Kbeta_\lambda \,T^{\sigma\lambda}-
(\LambdaB+\Kbeta^\lambda A_\lambda) \cdot J^\sigma +\Kbeta^\sigma \Lag }_\text{Hydrostatic}
\ d^{d-1}S_\sigma\,.  \\
\end{split}
\label{eq:SBtent}
\end{equation}
\item Next we compute the time derivative of this  entropy. Whenever the hydrostatic equality type constraints are satisfied, the time derivative of the first two terms in \eqref{eq:SBtent} gives a boundary term which is just the total free energy current. 
In particular, we can read off the spatial component of the free energy current from this boundary term.
\item  In turn, this gives an expression for the hydrostatic contribution to the entropy current $(J_S^\mu)_\text{Hydrostatic}$.
\end{enumerate}
Instead of working in the microcanonial ensemble,   we can directly move to canonical ensemble in the first step, and focus on free energy current instead of entropy current.  
With this change in ensemble, the above set of arguments  is then  equivalent to the algorithm we described in \S\ref{sec:hscurrents}.

\subsection{Comparison with the charged fluid analysis of Bhattacharyya}
\label{sec:sbcharged}

Armed with this understanding we can compare the analysis of \cite{Bhattacharyya:2014bha} for  parity-even charged fluids with the arguments presented in \S\ref{sec:counting}. 
\begin{enumerate}
\item    One begins by counting the total number of  transport coefficients in frame invariant language. This gives
16 Scalars + 17 Vectors + 18 Tensors = 51 a-priori different transport coefficients.
\item  Class A:  Remove Class A using the particular combination from anomaly-induced transport theory.
In this example, there is nothing to remove, which gives $\text{A}=0$.
\item  Classes $\{\PS , \PV , \PF \}$ (or the 3-fold fate of non-anomaly induced
hydrostatic transport coefficients): Focus on the remaining terms that survive hydrostatics; these amount to
9 Scalars+ 6 vectors + 9 Tensors = 24 a-priori hydrostatic transport coefficients.
Thus, we are leaving out 7 Scalars + 11 vectors + 9 Tensors = 27 transport coefficients that do not
survive hydrostatic limit. By looking at the partition function we see that, of these 24, 7 come from $\PS$ terms. This gives $\PS=7, \PV=0$ as obtained in \cite{Bhattacharyya:2014bha}. Thus we have $\PF = 24- (7+0) = 17$.

The first half of \cite{Bhattacharyya:2014bha}  (and  Appendix A therein) is devoted to showing
that one can complete these to 7 solutions of adiabaticity equation. This involves constructing the entropy current etc.. 
By this point, we understand how to do this very well covariantly, so we can just skip ahead and construct a covariant entropy current as outlined in the main text.  Henceforth, we have to only worry about the 27 non-hydrostatic terms.
\item  Class B: Next we examine  $\half T^{\mu\nu} \diffB g_{\mu\nu} + J^\mu \cdot\diffB A_\mu$ for these 
non-hydrostatic terms. This is contained in  Eqs.~(5.6) and (5.7) and Appendix B of \cite{Bhattacharyya:2014bha}.
Before we proceed, let us note that the last two terms of $\text{ (5.7)}_{\text{\tiny{\cite{Bhattacharyya:2014bha}}}}$ with 
$T_8$ and $T_9$ respectively are identically zero -- the tensor term summation should stop with $T_7$.
With this small amendment we have\\
2 Scalars + 3 Vectors + 2 Tensors = 7 terms in Eq.~$\text{(5.6)}_{\text{\tiny \cite{Bhattacharyya:2014bha}}}$\\
5 Scalars + 0 vectors + 4 tensors = 9 terms in Eq.~$\text{(5.7)}_{\text{\tiny \cite{Bhattacharyya:2014bha}}}$

Comparing this against the total count of 7 Scalars + 11 Vectors + 9 Tensors for
non-hydrostatic terms, we conclude that 0 Scalars + 8 Vectors + 3 Tensors = 11 terms go away at this step.
This gives us 11 terms in Class B in agreement with our counting. We remove these and thence focus on the 
16 terms that are left.

\item  Class  $\LS$: At this point, \cite{Bhattacharyya:2014bha} argues that 5 out of the 16 terms that survived this far, can  be absorbed as total derivatives into the entropy current. We know independently that this is the correct counting based on $\LS = 5$, following from the procedure described below equation \eqref{eq:feq1}. At this point we are left with 11 terms. 

\item Class D: All of the 11 terms are now dissipative. Moreover, their contribution to $\Diss$ can be explicitly assembled schematically into the combinations
\begin{equation}
\begin{split}
T\, \Diss = &2\,\eta\, (\sigma+ \diffB^2{\cal O}_1)^2 + \zeta\, (\Theta + \diffB^2 {\cal O}_2)^2 + \sigma_{_{\text{Ohm}}}\, (\cv +\diffB^2\, {\cal O}_3)^2 + {\cal O}(\partial^4) 
\end{split}
\end{equation}
for some operators ${\cal O}_i$, which can be obtained from the  explicit construction in \cite{Bhattacharyya:2014bha} if necessary.  We see that this works directly by using the differential operators at our disposal. 
\end{enumerate}
This completes then a cross-check of our results with the analysis of \cite{Bhattacharyya:2014bha}.

\section{Bianchi identities for anomalous hydrodynamics}
\label{sec:AnomBianchi}

In this appendix we derive Bianchi identities and on-shell constraints for anomalous hydrodynamics. This  fills in the details and complements the discussion in \S\ref{sec:anomalies}.

\subsection{Bianchi identities from anomalous part of effective action}

Our goal here is to evaluate Eq.\ (\ref{eq:GravTransVar}) which we reproduce for convenience: 
\begin{equation}
\begin{split}
\delta \VP \brk{\fA, \fGamma, \fAh,\fGammah}
  &= \delta \fA \wedge\cdot \hodgeB \fJH - \delta \fAh \wedge\cdot \hodgeB \fJHh \\
  &\qquad+\half \delta \fGamma^a{}_b \wedge \hodgeB \fSpH{}^b{}_a- \half \delta \fGammah^a{}_b\wedge \hodgeB \fSpHh{}^b{}_a  \\
  &\qquad +d \bigbr{ \delta \fA\wedge \cdot  \star \fJP + \half \delta \fGamma^\alpha{}_\beta \wedge  \star \fSP{}^\beta{}_\alpha + \delta \fu \wedge  \star \fqP
  } \,,
  \end{split}
  \label{eq:GravTransVarApp}
  \end{equation}
In order to further evaluate this expression, we need variations of various objects. We begin
by first using the relations derived in Appendix \ref{sec:varanomhat}. In particular substituting \eqref{eq:Ahatvar} and \eqref{eq:Ghatvar} into \eqref{eq:GravTransVar} we find that the variation of the transgression form
takes the following form:
\begin{align}
\delta &\int_{{\bulkM}_{d+1}} \VP[\fA,\fGamma; \fAh , \fGammah] \nonumber \\
&=\bulkint \,
\brk{ \prn{ \JH^m -P^m_n \,\JHh^n } \cdot \delta A_m
+ \half \prn{ \SpH{}^{ms}{}_r -  \fatP^{qsm}_{prn} \;\SpHh{}^{np}{}_q } \delta \Gamma^r{}_{sm} }
\nonumber \\
&\quad-\bulkint \,\brk{ \prn{ \mu \cdot \JHh^q
+ \half \Omega^r{}_s \SpHh{}^{qs}{}_r } P_q^{(m}\,u^{n)} -\frac{1}{4} \, T \,u_q \, \SpHh{}^{qs}{}_r \left( \delta^m_s
D^r \Kbeta^n - g^{rm} D^n \Kbeta_s\right) }\;\delta g_{mn}
\nonumber \\
&\quad
-\bulkint \;  \prn{ \mu \cdot \JHh^q + \half \Omega^r{}_s \SpHh{}^{qs}{}_r } \left(P_{qm}+u_q\, u_m\right) T\,\delta \Kbeta^m
\nonumber \\
&\quad
-\bulkint \; T\,u_m \brk{  \JHh^m \cdot  (\delta \LambdaB + A_n \delta \Kbeta^n)
+ \half \, \fatQ^{qs}_{pr} \; \SpHh{}^{mp}{}_q \,  D_s \delta \Kbeta^r }
\nonumber \\
&\quad + \int_{\cal M} \sqrt{-g} \left[ \JP^\alpha \cdot \delta A_\alpha + \half \SP{}^{\alpha\sigma}{}_\rho \, \delta \Gamma^\rho{}_{\sigma\alpha}
+ \qP^{(\alpha} u^{\beta)}\delta g_{\alpha\beta}
+ (\qP)_\sigma \, T \delta \Kbeta^\sigma \right]
\label{eq:DeltaSano}
\end{align}
where we have introduced a new projector $\fatP^{\rho\mu\nu}_{\sigma\kappa\lambda} = \delta^\rho_\kappa \, \delta^\mu_\sigma\,  \delta^\nu_\lambda  + \fatQ^{\rho\mu}_{\sigma\kappa} \, u^\nu u_\lambda$ to keep the expression compact.

The variational formula needs to be massaged further to bring it into an amenable form from which we can read off the bulk and boundary currents. For one the variation of the Christoffel symbols need to be converted to metric variations. For another the bulk term involving $D_s \delta \Kbeta^r$ should be integrated by parts and will thus contribute some boundary terms. Both of these features arise from the gravitational contribution. Indeed setting the spin connection terms to zero we see that \eqref{eq:DeltaSano} reduces to \eqref{eq:fvpvar}.

Firstly, the variation of the Christoffel symbols can be converted into a variation of the metric by observing the identity
\begin{equation}
\begin{split}
\int_{\cal M}\sqrt{-g} \; \half \SP{}^{\alpha\sigma}{}_\rho \, \delta \Gamma^\rho{}_{\sigma\alpha}
&= \int_{\cal M}\sqrt{-g} \; \frac{1}{2} D_\rho \brk{\SP^{\alpha[\beta\rho]} + \SP^{\beta[\alpha\rho]} -\SP^{\rho(\alpha\beta)}} \, \half\delta g_{\alpha\beta} \,,
\end{split}
\end{equation}
We use this expression on the boundary ${\cal M}$ to simplify the term in the last line of \eqref{eq:DeltaSano}. In the intermediate step we have discarded a total derivative term using the fact that $\partial {\cal M}=0$.
For the bulk term however we have to do a bit more work since now the boundary contributions from total derivative terms cannot be ignored. These can however be accounted for by recalling that our coordinatization of the bulk spacetime $\bulkM_{d+1}$ was such that the normal direction to ${\cal M}$ was denoted as $\perp$. Putting this together we find from the bulk term involving the Christoffel symbol variation
\begin{align}
\bulkint\;&\half \prn{\SpH{}^{ms}{}_r - \fatP^{qsm}_{prn} \; \SpHh{}^{np}{}_q } \delta \Gamma^r{}_{sm}
\nonumber \\&  = \bulkint\;
\Biggl\{\frac{1}{2} \,D_k \brk{\SpH^{m[nk]}+\SpH^{n[mk]}-\SpH^{k(mn)}} \half \delta g_{mn}
\Biggl.    \nonumber \\
&\hspace{3cm} \Biggr.
-\frac{1}{2} \,D_k \brk{
\prn{\fatP^{q[nm}_{prs} g^{k]r}+ \fatP^{q[mn}_{prs} g^{k]r} -
\fatP^{q(mk}_{prs} g^{n)r} }\; g_{ql}\; \SpHh^{spl}}\half \delta g_{mn} \biggr\}
\nonumber \\
&\qquad \quad- \int_{\cal M} \sqrt{-g} \;
\half\prn{ \fatP^{\eta(\alpha \perp}_{\rho \gamma \lambda} g^{\beta) \gamma} g_{\eta \kappa}
\SpHh^{\lambda\rho\kappa}}
\half\delta g_{\alpha\beta}
\nonumber \\
&= \bulkint \;\frac{1}{2} \,D_p \brk{\left( \SpH^{m[np]}+\SpH^{n[mp]}-\SpH^{p(mn)}\right)
-\left( P^m_q  \SpHh^{q[np]}+ P^n_q\SpHh^{q[mp]}-   \SpHh^{p(mn)}\right) } \half\delta g_{mn}
\nonumber \\
&\qquad  \quad- \int_{d}\sqrt{-g} \;  \half \SpHh^{\perp(\mu\nu)}  \half\delta g_{\mu\nu}  \,,
\end{align}
where extrinsic boundary terms of the form $\SpH^{\alpha\perp\beta}+\SpH^{\beta\perp\alpha}$ have been set to zero and in the second step we used the following identities to implement some simplifications:
\begin{equation}
\begin{split}
\fatP^{q[nm}_{prs} g^{k]r} g_{ql} &= \delta_p^{[n} \delta_l^{k]} \delta_s^m +
\half\left(\delta_p^{[n} \delta_l^{k]}  - \delta_l^{[n} \delta_p^{k]} \right) u^m u_s
= \delta_p^{[n} \delta_l^{k]} P_s^m \,, \\
\fatP^{q(mk}_{prs} g^{n)r} g_{ql} &= \delta_p^{(m} \delta_l^{n)} \delta_s^lk +
\half \left( \delta_p^{(m} \delta_l^{n)}
- \delta_l^{(m} \delta_p^{n)} \right) u^k \,u_s = \delta_p^{(m} \delta_l^{n)} \delta_s^k \,.
\end{split}
\end{equation}

Finally, performing an integration by parts also in the third line of Eq.\ (\ref{eq:DeltaSano}), we obtain the final simplified variational formula of interest
\begin{equation}
\begin{split}
\delta &\int_{{\bulkM}_{d+1}} \VP[\fA,\fGamma;\hat{\fA},\hat{\fGamma}] \\
&=\bulkint \, \prn{ \Jbulk^m \cdot \delta A_m
+  \half\delta g_{mn} \; \Tbulk^{mn}
+ \,T \; \aheatbulk_m \delta \Kbeta^m
+\,T \; \achargebulk \cdot (\delta \LambdaB + A_a \delta \Kbeta^a)
}  \\
&\quad + \int_{\cal M} \sqrt{-g} \; \left( (J^\alpha)_\text{A}\cdot \delta A_\alpha + \frac{1}{2} \delta g_{\alpha\beta} \,  (T^{\alpha\beta})_\text{A}
+ (\qP)_\sigma \, T \delta \Kbeta^\sigma \right) \\
\end{split}
\label{eq:DeltaSanom2}
\end{equation}
where the bulk currents now take the form
\begin{align}
\Tbulk^{mn} &= \frac{1}{2} \, T \,u_p \; \SpHh{}^{ps}{}_r \left( \delta^m_s D^r \Kbeta^n - g^{rm} D^n
\Kbeta_s\right)  -\prn{ \mu \cdot \JHh^p
+ \half \Omega^r{}_s \SpHh{}^{ps}{}_r } \prn{P_p^m u^n + P_p^n u^m}  
\nonumber\\
&\qquad   +\;\frac{1}{2} \,D_p \brk{\left( \SpH^{m[np]}+\SpH^{n[mp]}-\SpH^{p(mn)}\right)
-\left( P^m_q  \SpHh^{q[np]}+ P^n_q\,\SpHh^{q[mp]}- \SpHh^{p(mn)} } \right)\  
\nonumber\\
\Jbulk^m &=  \JH^m -P^m_n\, \JHh^n  
\nonumber\\
\aheatbulk_m &= -\prn{ \mu \cdot \JHh^p
+ \half \Omega^r{}_s \, \SpHh{}^{ps}{}_r }  \left(P_{pm}+u_p \,u_m\right)
+    \frac{1}{2\,T} D_s\prn{ T u_r \, \fatQ^{qs}_{pm} \,\SpHh{}^{rp}{}_q }  
\nonumber\\
\achargebulk &= - u_m \, \JHh^m
\label{eq:bulkAnomCur}
\end{align}
and boundary currents turn out to be
\begin{equation}
\begin{split}
 (T^{\alpha\beta})_\text{A} &= \qP^\alpha u^\beta +  \qP^\beta u^\alpha+ \frac{1}{2} \,D_\rho \prn{\SP^{\alpha[\beta\rho]} + \SP^{\beta[\alpha\rho]} -\SP^{\rho(\alpha\beta)}}
- \half \SpHh^{\perp(\alpha\beta)} \,, \\
(J^\alpha)_\text{A}&= \JP^\alpha \,.
\end{split}
\label{eq:bdyAnomCurApp}
\end{equation}
We have written this expression, allowing a-priori for a contribution to the stress tensor involving the shadow Hall current  
$\half \SpHh^{\perp(\alpha\beta)}$. In the main text, \eqref{eq:bdyAnomCur}, the stress tensor is quoted without this term. This is due to the fact that $\SpHh^{\perp(\alpha\beta)} = 0$ always holds for our choice of  spin chemical potential. To prove this, observe that the connection $\fGammah^\mu{}_\nu$ is metric compatible due to our spin chemical potential being anti-symmetric: 
\begin{equation}
\widehat{\nabla}_\sigma g_{\mu\nu} = - u_\sigma (\Omega_{\mu\nu}+ \Omega_{\nu\mu}) = 0 \,.
\end{equation}
From metric compatibility it follows immediately that the associated curvature tensor $\fRh^\nu{}_\mu$ is anti-symmetric. From the definition \eqref{eq:HallCurrentsDef} one can see that $(\fSpHh)^\mu{}_\nu$ inherits this antisymmetry. In what follows we will therefore often set 
\begin{equation}
 \SpHh^{\perp(\alpha\beta)} = 0 \,.
\end{equation}

Equation \eqref{eq:DeltaSanom2} is our master equation for the Lagrangian variation. Our main interest is not in a generic variation, but rather in the variations engendered by diffeomorphisms and gauge transformations of the fields on $\bulkM_{d+1}$, which is what is needed to derive the Bianchi identities.
Using the general formula (\ref{eq:NoetherHydroPre}) for the bulk integral, we find for the particular case where $\delta = \diffF$ is a gauge transformation and diffeomorphism:\footnote{ As in all of our discussion of anomalies, the fields $\Xfields =\{\xi^m, \Lambda\}$ are taken to live on $\bulkM_{d+1}$.}
\begin{equation}
\begin{split}
\diffF &\int_{{\bulkM}_{d+1}} \VP[\fA,\fGamma\hat{\fA},\hat{\fGamma}] \\
&=
\bulkint \xi_m \Biggl\{ -D_n \prn{\; \Tbulk^{mn}} + F^m{}_n \cdot \prn{\JH^n-P^n_p \,\JHh^p}
\Biggl.\\
&\hspace{3cm}  \Biggr.
+ g^{mn} \, T\;\achargebulk \cdot \diffB A_n +\frac{g^{mn}}{\sqrt{-g_{d+1}}}
\diffB \brk{\sqrt{-g_{d+1}}\ T\;\aheatbulk_m} \Biggr\}
\\
&\quad-\bulkint\ \prn{\Lambda+\xi^n\, A_n}\cdot \Biggl\{
D_m\prn{\; \Jbulk^m} -  \frac{1}{\sqrt{-g_{d+1}}} \diffB \brk{\sqrt{-g_{d+1}}\;T\;\achargebulk  }
\Biggr\}   \\
&\quad
+\int_{\cal M} \sqrt{-g}\;\xi_\alpha\,\Biggl( \half D_\gamma \prn{ \SpH^{\perp[\alpha\gamma]}
- \SpHh^{\perp[\alpha\gamma]}}
- \prn{ \mu \cdot \JHh^\perp
+ \half \Omega^\nu{}_\mu \SpHh{}^{\perp \mu}{}_\nu } u^\alpha \Biggr)  \\
&\quad
+\int_{\cal M} \sqrt{-g}\, \prn{\Lambda+\xi^\alpha A_\alpha}
\cdot\prn{  \JH^\perp- \JHh^\perp} \\
&\quad
+ \int_{\cal M} \sqrt{-g} \;
\prn{ (J^\alpha)_\text{A}\cdot \diffF A_\alpha + (T^{\alpha\beta})_\text{A} \, \half\diffF g_{\alpha\beta}
+ (\qP)_\sigma \, T \diffF \Kbeta^\sigma }\\
\end{split}
\label{eq:DiffeoSanom}
\end{equation}
In writing the above expression we have done some integration by parts mostly to remove the derivatives of the diffeomorphism field $\xi^m$.

We can now directly read off the Bianchi identities for the bulk theory: these are simply given by the first two lines of \eqref{eq:DiffeoSanom}. They satisfy the expected form of the equations \eqref{eq:LHydroEq} as derived earlier from general considerations. Indeed in so far as the bulk theory is concerned, we have a gapped topological system which obeys bulk diffeomorphism and gauge invariance and so  we should have a-priori expected to see this work out as stated. Note that upon setting the spin currents to zero we recover the flavour Bianachi identities as indicated in \S\ref{sec:fanom}.

Once the bulk Bianchi identities are satisfied for arbitrary bulk $\Xfields$ we see that the variation of the anomalous Lagrangian is purely a boundary term. This has both the physics of the hydrodynamic system of interest as well as the anomaly inflow term that enable us to write down the expressions for the covariant currents. We have one final manipulation to do to bring this into a canonical form. Expressing the variations $\diffF g_{\alpha\beta} $ and $\diffF A_\alpha $ in terms of the gauge transformation fields
$\Xfields$ and performing yet another integration by parts we finally convert \eqref{eq:DiffeoSanom} into
\begin{equation}
\begin{split}
\diffF &\int_{{\bulkM}_{d+1}} \VP[\fA,\fGamma;\hat{\fA},\hat{\fGamma}] \\
&=
\int_{\cal M} \sqrt{-g}\;\xi_\alpha\;\Biggl(
\half D_\gamma \prn{ \SpH^{\perp[\alpha\gamma]} - \SpHh^{\perp[\alpha\gamma]}}
- \prn{ \mu \cdot \JHh^\perp
+ \half \Omega^\nu{}_\mu \SpHh{}^{\perp \mu}{}_\nu } u^\alpha \Biggl.  \\
&\hspace{2cm} \Biggr.
- D_\beta  (T^{\alpha\beta})_\text{A}
+(J^\beta)_\text{A} \cdot F^\alpha{}_\beta
+ \frac{g^{\alpha\sigma}}{\sqrt{-g}} \diffB \brk{ \sqrt{-g} \, T (\qP)_\sigma } \Biggr) \\
&\quad + \int_{\cal M}\sqrt{-g} \; \prn{\Lambda+\xi^\alpha A_\alpha}\cdot\prn{
-D_\alpha (J^\alpha)_\text{A}+ \JH^\perp- \JHh^\perp }
\end{split}
\label{eq:DiffeoSanom3}
\end{equation}

We are now in a position to read off the boundary Bianchi identities which are obeyed by our anomalous fluid. We find that these take the form (picking out coefficients of the arbitrary $\xi_\alpha$ and $(\lambda + \xi^\alpha A_\alpha)$ from the above expression)
\begin{equation}
\begin{split}
D_\beta & (T^{\alpha\beta})_\text{A}=  (J^\beta)_\text{A} \cdot F^{\alpha}{}_\beta + \frac{g^{\alpha\sigma}}{\sqrt{-g}} \diffB \brk{ \sqrt{-g} \, T (\qP)_\sigma } \\
& \hspace{2cm} +\half D_\gamma \prn{ \SpH^{\perp[\alpha\gamma]} - \SpHh^{\perp[\alpha\gamma]}}
- \prn{ \mu \cdot \JHh^\perp
+ \half \Omega^\nu{}_\mu \SpHh{}^{\perp \mu}{}_\nu } u^\alpha \,,
\end{split}
\label{eq:LorentzBdryBianchi1App}
\end{equation}
and
\begin{equation}
D_\alpha (J^\alpha)_\text{A}= \JH^\perp - \JHh^\perp \,.
\label{eq:LorentzBdryBianchi2App}
\end{equation}
The terms on the r.h.s. of the expressions  of \eqref{eq:LorentzBdryBianchi1App} and
\eqref{eq:LorentzBdryBianchi2App} with the one $\perp$ component of the Hall currents are due to bulk inflow.

\subsection{On-shell constraints from the full Lagrangian}
\label{sec:OnShellAnom}

For reference, we quote both the bulk and the boundary on-shell constraints that are obtained in \S\ref{sec:mixos} by extremizing the full effective Lagrangian 
$\Lag_{eff}\brk{\hfields} = d \Lag_\text{n-a} \brk{\hfields} + \VP[\fA,\fGamma,\fAh,\fGammah]$ with respect to the pullback fields. For the bulk theory, this yields the following on-shell constraints:
\begin{align}
&  g^{mn} \, T\;\achargebulk \cdot \diffB A_n +\frac{g^{mn}}{\sqrt{-g_{d+1}}}
\diffB \brk{\sqrt{-g_{d+1}}\ \aheatbulk_m}
\simeq 0 \,,\nonumber \\
&\frac{1}{\sqrt{-g_{d+1}}} \diffB \brk{\sqrt{-g_{d+1}}\;T\; \achargebulk  }  \simeq 0\,.
\label{eq:BulkAnomConstraints}
\end{align}
The boundary on-shell constraints, on the other hand, are given by 
\begin{align}
\frac{g^{\mu\nu}}{\sqrt{-g}}&\diffB\prn{\sqrt{-g}\ T\brk{(\aheat_\nu)_\text{n-a}+(\qP)_\nu}}
+  g^{\mu\nu}\, T\,\acharge_\text{n-a} \cdot \diffB A_\nu \simeq 0\,, \nonumber \\
\frac{1}{\sqrt{-g}}&\diffB\prn{\sqrt{-g}\ T\,\acharge_\text{n-a}} \simeq 0\,.
\label{eq:AnomConstraintsBdyApp}
\end{align}

Combining these on-shell constraints with the anomalous Bianchi identities we obtain the equations of motion. In the bulk, we find from \eqref{eq:DiffeoSanom},
\begin{equation}\label{eq:EOManom1}
D_n \prn{\; \Tbulk^{mn}} = F^m{}_n \cdot \prn{\JH^n-P^n_p \,\JHh^p} \,, 
\qquad D_m\prn{\; \Jbulk^m} = 0 \,.
\end{equation}
Similarly, in the boundary theory, we obtain 
\begin{align}\label{eq:EOManom2}
&D_\beta\prn{T^{\alpha\beta}_\text{n-a}+ (T^{\alpha\beta})_\text{A}}  \nonumber \\
&\qquad\simeq (J^\sigma_\text{n-a}+(J^\sigma)_\text{A})\cdot F^\alpha{}_\sigma
+\half D_\gamma \prn{ \SpH^{\perp[\alpha\gamma]} - \SpHh^{\perp[\alpha\gamma]}}
- \prn{ \mu \cdot \JHh^\perp
+ \half \Omega^\nu{}_\mu \SpHh{}^{\perp \mu}{}_\nu } u^\alpha \,, \nonumber \\
& D_\sigma (J^\sigma_\text{n-a}+(J^\sigma)_\text{A}) \simeq \JH^\perp- \JHh^\perp \,.
\end{align}

We note that we could have arrived at these results on a slightly easier path. As discussed in general in \S\ref{sec:ReferenceVar}, equations of motion are easier to obtain on the reference manifold where the relevant part of the constrained variation is already built in. Concretely, we could have started from \eqref{eq:DeltaSanom2} written on the reference manifold $\Mref$ by replacing all currents by boldface currents and changing indices from Greek to Latin. Then directly performing the variation \eqref{eq:ConstrVar} would give 
\begin{equation}
\begin{split}
-\,\diffCons &\int_{{\Mref}_{d+1}} \VP[\Aref_\pb,\Chref^\pb_{\mb\nb};\hat{\Aref}_\pb,\hat{\Chref}^\pb_{\mb\nb}] \\
&\quad=\bulkintref \, \prn{ \Jbulkref^\mb \cdot \diffCons \Aref_\mb +  \half\diffCons \gref_{\mb\nb} \; \Tbulkref^{\mb\nb} } 
 + \int_{\Mref} \sqrt{-\gref} \; \prn{ (\Jref^{a})_\text{A} \cdot \diffCons \Aref_a + \frac{1}{2} \delta \gref_{ab} \, (\Tref^{ab})_\text{A} }\\
&\quad=
\bulkintref \delta \varphi_\mb \brk{ -\Dref_\nb \prn{\; \Tbulkref^{\mb\nb} + \Fref^\mb{}_\nb \cdot \prn{\JHref^\nb-P^\nb_\pb \,\JHhref^\pb}}} \\
&\qquad -\bulkintref\ \prn{-c^{-1}\delta c + \Aref_\mb \delta \varphi^\mb}\cdot \Dref_\mb\prn{\; \Jbulkref^\mb}\\
&\qquad
+\int_{\Mref} \sqrt{-\gref}\;\delta \varphi_a\,\brk{\half \Dref_c \prn{ \SpHref^{\perp[ac]}
- \SpHhref^{\perp[ac]}}
- \prn{ \muref \cdot \JHhref^\perp
+ \half \Omegaref^b{}_c\, \SpHhref{}^{\perp c}{}_b } \uref^a}  \\
&\qquad
+\int_{\Mref} \sqrt{-\gref}\, \prn{-c^{-1}\delta c + \Aref_b \delta \varphi^c}
\cdot\prn{  \JHref^\perp- \JHhref^\perp} \\
&\qquad
+ \int_{\Mref} \sqrt{-\gref} \;
\brk{ \delta\varphi_a \prn{ - \Dref_b (\Tref^{ab})_\text{A} + (\Jref^{b})_\text{A} \cdot \Fref^a{}_b } - \prn{-c^{-1}\delta c + \Aref_b\, \delta \varphi^c} \Dref_a (\Jref^{a})_\text{A}} \,.
\end{split}
\label{eq:DeltaSanomRef}
\end{equation}
From this we can immediately read off the anomalous part of the equations of motion \eqref{eq:EOManom1} and \eqref{eq:EOManom2}. While working on the reference manifold is thus manifestly easier, we still gain computational insight from doing the full analysis from a point of view of physical $\cal{M}$.

\subsection{Bianchi identities of anomalous Schwinger-Keldysh action}
\label{sec:SKvariation}

We will now give some details of the derivation of anomalous Schwinger-Keldysh currents as outlined in \S\ref{sec:anward}. Since we double also the sources, we want to avoid complications that arise from an analysis on two different manifolds and work instead on the unambiguously defined reference manifold as described in \S\ref{sec:skdouble}. For simplicity we can work in the hydrodynamic limit from the beginning, i.e., we only keep track of terms linear in difference fields. 
Let us start by computing the variation of the influence functional adapting the results form \cite{Haehl:2013hoa}. We have
\begin{equation}
\begin{split}
\delta  S_{_{IF}}  &= \int_{\Mref_{d+1}} \Biggl\{\delta \fAh[ \hreffields_\skR] \wedge  \hodgeB \fJHh[ \hreffields_\skR] - \delta \fAh[\hreffields_\skL] \wedge  \hodgeB \fJHh[\hreffields_\skL] \Biggl. \\
&\hspace{2.5cm} \Biggl.
+\; \delta \fGammah[\hreffields_\skR]^\mb{}_\nb\wedge \hodgeB \fSpHh[\hreffields_\skR]^\nb{}_\mb
- \delta \fGammah[\hreffields_\skL]^\mb{}_\nb\wedge \hodgeB \fSpHh[\hreffields_\skL]^\nb{}_\mb \Biggr\} \\
&\quad +\int_{\Mref}  \Big\{ \text{Boundary terms}\brk{\hreffields_\skR,\hreffields_\skL}  \approx {\cal O}\left((\hreffields_\skR-\hreffields_\skL)^2\right)\Big\}\,.
\end{split}
\label{eq:DeltaScross}
\end{equation}

Isolating the variations of difference fields makes the linearization in difference fields manifest and
we can for ease of notation immediately take the hydrodynamic limit of all the remaining quantities involved.
Using  the variation rules \eqref{eq:Ahatvar}, \eqref{eq:Ghatvar} and integration by parts, the
variation of the influence functional $S_{_{IF}}$, \eqref{eq:DeltaScross}, with respect to the difference fields takes the form\footnote{ We adhere to our conventions stated earlier: boundary reference manifold indices are from the earlier part of the lowercase Latin alphabet, while the bulk reference manifold is denoted as $\Mref_{d+1}$ and indexed by letters from the second half of the uppercase Latin alphabet.}
\begin{equation}
\begin{split}
\delta  S_{_{IF}} \bigg{|}_{hydro}
&=\bulkintref \prn{(\Tbulkref^{\mb\nb})_{_{IF}}\, \half \, \delta \dRLgref_{\mb\nb}  +  \;
(\Jbulkref^\mb)_{_{IF}}  \cdot \delta \dRLAref_\mb } 
\,,
\end{split}
\label{eq:AnomVar1}
\end{equation}
where $hydro$ denotes the limit where all the expressions that are not variations of difference fields are
evaluated at $\varphi_\skR^a(x)=\varphi_\skL^a(x)\equiv \varphi^a(x)$ and $c_\skR(x) = c_\skL(x) \equiv c(x)$
and we find the following bulk currents generated by the influence functional:
\begin{align}
(\Tbulkref^{\mb\nb})_{_{IF}} &=
2 \left( \muref \cdot \JHhref^\pb + \half \,\Omegaref^\rb{}_\sbb \,\SpHhref{}^{\pb\sbb}{}_\rb \right)  P_\pb^{(\mb}\,\uref^{\nb)}
+\frac{1}{2} \, \Dref_\pb \left( P^\mb_\qb  \;\SpHhref^{\qb[\nb\pb]}+ P^\nb_\qb\;\SpHhref^{\qb[\mb\pb]}- \SpHhref^{\pb(\mb\nb)}\right)
\nonumber \\
&\qquad  -\; \half \, \Tref\, \uref_\qb \, \SpHhref{}^{\qb\sbb}{}_{\rb} \left(
\delta_\sbb^\mb \Dref^\rb \Kref^\nb - g^{\rb\mb} \Dref^\nb \Kref_\sbb \right)
\nonumber \\
\left(\Jbulkref^\mb\right)_{_{IF}} &= P^\mb_\nb \JHhref^\nb
\end{align}

We now consider the variation of the left and right anomalous terms in $S_{tot}$. These are completely analogous to
our earlier discussion in \S\ref{sec:varmix} except that we have two sets of contributions to keep track of. Sticking to the hydrodynamic limit, we find that one copy each of the $\skRl$ and $\skLl$ anomalous variation
\eqref{eq:DeltaSanom2} combine in the hydrodynamic limit to give\footnote{ The currents written in \eqref{eq:AnomVar2} strictly speaking live on the reference manifold. However, we have refrained from introducing further notation to distinguish them from their physical counterparts; they are distinguish by their indices which indicate their origins. They are of course pushed-forward along the diffeomorphism and gauge transformation fields like any other tensor field defined on $\Mref$.}
\begin{align}
\delta &\left(\int_{\Mref_{d+1}} \VP[\hreffields_\skR] - \VP[\hreffields_\skL] \right) \bigg{|}_{hydro} \notag\\
&\quad = \bulkintref \, \prn{ \; \Tbulkref^{\mb\nb}\, \delta\dRLgref_{\mb\nb}+\; \Jbulkref^\mb\, \delta\dRLAref_\mb} 
 + \int_\Mref \sqrt{-\breve{\gref}}\; \prn{
(\Tref^{ab})_\text{A} \, \half\delta \dRLgref_{ab}
+ (\Jref^{a})_\text{A} \cdot \delta \dRLAref_{a}} \,.
\label{eq:AnomVar2}
\end{align}
The currents showing up in the above equation are the ones we have already recorded in \eqref{eq:bulkAnomCur} and \eqref{eq:bdyAnomCur} (just on the reference manifold).
As expected, to linear order in the variations as required for the hydrodynamic limit the contributions simply combine to give the common current times the difference of the $\skRl$ and $\skLl$ sources.

We can now put everything together to find the bulk and boundary Bianchi identities and the dynamical equations of motion.  Firstly, combining Eqs. \eqref{eq:AnomVar1} and \eqref{eq:AnomVar2}, we obtain for the variation of the entire anomaly part of the action
\begin{equation}
\begin{split}
\delta S_{anom}\bigg{|}_{hydro} &\equiv \delta \left( S_{_{IF}} +
\int_{\bulkM_{d+1}}\left( \VP[\hreffields_\skR] - \VP[\hreffields_\skL]  \right) \right)\bigg{|}_{hydro} \\
& = \bulkintref \left[  \frac{1}{2} \, \Dref_\pb \left( \SpHref^{\mb[\nb\pb]}+\SpHref^{\nb[\mb\pb]}-\SpHref^{\pb(\mb\nb)}\right)
\half \delta \dRLgref_{\mb\nb}
+ \JHref^\mb \cdot \delta \dRLAref_{\mb} \right] \\
&\quad + \int_\Mref \sqrt{-\breve{\gref}} \; \Bigg\{
\brk{ \frac{1}{2} \,\Dref_c \prn{\SPref^{a[bc]} + \SPref^{b[ac]} -\SPref^{c(ab)}}
+ 2\,\qPref^{(a} \uref^{b)} } \, \half \delta \dRLgref_{ab}
+\;   \JPref^a \cdot \delta \dRLAref_{a}  \Bigg\} .
\end{split}
\label{eq:DeltaSanomFinalApp}
\end{equation}
From this one can readily compute the equations of motion (see \S\ref{sec:anward}).

\section{Class $\LT$ details}
\label{sec:appendixT}

This appendix collects some intermediate steps of the computations  relevant for the Class $\LT$ discussions of 
\S\ref{sec:classLT}. In \S\ref{sec:utwz} we check that our  diffeomorphism, flavour gauge, and $\UT$ transformations form an algebra. \S\ref{sec:utbder} fills in some intermediate steps involved in the derivation of the $\UT$  Bianchi identity.

\subsection{Consistency of $\UT$ transformations}
\label{sec:utwz}

We would like to check that the diffeomorphism, flavour gauge and $\UT$ transformations given in \eqref{eq:TactgA}, \eqref{eq:TactgAbar} and \eqref{eq:TactAT} form a Lie algebra. In particular, we would like to ensure that the Wess-Zumino consistency conditions are satisfied. This in particular requires that the commutator of  two transformations parameterized by 
$\Xfields_1  = \{\txi^\mu_1,\tLam_1, \LambdaTb_1\}$ and $\Xfields_2  = \{\txi^\mu_2,\tLam_2, \LambdaTb_2\}$,
is itself given by a diffeomorphism, flavour and $\UT$ transformation with parameters 
 $\Xfields_3  = \{\txi^\mu_3,\tLam_3, \LambdaTb_3\}$.  Note that we work directly with in the untwisted formalism; the conclusions will of course be unchanged should we switch to the twisted transformation variables parameterizing individual elements of the Class $\LT$ symmetries.

 We will proceed systematically analyzing the commutator of two successive transformations. A-priori the transformation of the basic hydrodynamic fields   $\hfields$ will fix the dependence of $ \{\txi^\mu_3,\tLam_3\}$ on $\Xfields_1$ and 
 $\Xfields_2 $. However, since the $\UT$ transformation mixes with diffeomorphisms, it also follows that the parameter $\LambdaTb_3$ is already constrained.  Consistency of our transformations requires that  the partner sources and $\AT$ transform by the now determined values of $\Xfields_3$. Ensuring that this is upheld will form the main check of the analysis below.

 Let us start with the transformation for the background metric $g_{\mu\nu}$. From \eqref{eq:TactgA} we find
\begin{align}
[\delta_{\Xfields_1}, \delta_{\Xfields_2}] g_{\mu\nu} = [\lieD_{\txi_1}, \lieD_{\txi_2}] g_{\mu\nu} \equiv \lieD_{\txi_3} g_{\mu\nu}
\label{}
\end{align}
with 
\begin{equation}\label{eq:X3a}
\begin{split}
\txi_3^\sigma &= [\txi_1,\txi_2]^\sigma \\
\xi_3^\sigma  &= 
	[\xi_1,\xi_2]^\sigma + \prn{ \lT_2 \, \lieD_{\xi_1}\, \Kbeta^\sigma - \lT_1 \, \lieD_{\xi_2}\, \Kbeta^\sigma } + x \, \Kbeta^\sigma \\	
\lT_3 & = 
	\lieD_{\xi_1} \lT_2- \lieD_{\xi_2} \lT_1 
	+ \lT_1 \lieD_{\Kbeta}\lT_2 - \lT_2 \lieD_{\Kbeta} \lT_1 - x \,,
\end{split}
\end{equation}
where we define 
\begin{align}
\lT_k \equiv \LambdaTb_k + \txi_k^\sigma\, \AT_\sigma = \LambdaT_k + \xi_k^\sigma \, \AT_\sigma 
\label{eq:lTdef}
\end{align}
to keep the expressions compact. Further, $x$ is an arbitrary scalar parameterizing a family of relations that are all consistent with $\txi_3^\sigma = \xi_3^\sigma + \lT_3 \, \Kbeta^\sigma$. The value of $x$ will have to be determined later on to ensure consistency of the formalism. 

The parameter $\txi_3$ encodes the effective diffeomorphism in the untwisted variables. Once we have ascertained this it is clear that the $\Kbeta$ transformation in \eqref{eq:TactgA} is consistent with \eqref{eq:X3a} since $\Kbeta^\mu$ itself is Lie transported along  $\Xfields$. 

Let us then turn to the flavour gauge transformations, taking  the commutator of the gauge transformations we have
using \eqref{eq:X3a} 
\begin{align}
[\delta_{\Xfields_1}, \delta_{\Xfields_2}] A_\mu 
&= \lieD_{\txi_1} \prn{ \lieD_{\txi_2} A_\mu + [A_\mu,\tLam_2] + \partial_\mu \tLam_2} + [ \lieD_{\txi_2} A_\mu + [A_\mu,\tLam_2] + \partial_\mu \tLam_2 \,,\, \tLam_1 ] \\
&\quad -\lieD_{\txi_2} \prn{ \lieD_{\txi_1} A_\mu + [A_\mu,\tLam_1] + \partial_\mu \tLam_1} -[ \lieD_{\txi_1} A_\mu + [A_\mu,\tLam_1] + \partial_\mu \tLam_1 \,,\, \tLam_2 ] \\
&= \lieD_{\txi_3} A_\mu + \lieD_{\txi_1} D_\mu \tLam_2 - \lieD_{\txi_2} D_\mu \tLam_1 + [A_\mu,[\tLam_2,\tLam_1]] \notag\\
&\quad + [\lieD_{\txi_2} A_\mu + \partial_\mu \tLam_2, \, \tLam_1] -[\lieD_{\txi_1} A_\mu + \partial_\mu \tLam_1, \, \tLam_2] \notag \\
&= \lieD_{\txi_3} A_\mu + D_\mu \tLam_3
\label{}
\end{align}
where we defined $\tLam_3 \equiv \Lambda_3 + \lT_3 \LambdaB$ with 
\begin{align}
\tLam_3 & = \lieD_{\txi_1} \tLam_2 - \lieD_{\txi_2} \tLam_1 - [\tLam_1,\tLam_2] \,,\\
\Lambda_3 &=  \lT_2 \prn{\lieD_{\xi_1} \LambdaB + [\LambdaB,\Lambda_1]} - \lT_1 \prn{\lieD_{\xi_2} \LambdaB + [\LambdaB,\Lambda_2]}  \\
  &\quad + \lieD_{\txi_1} \Lambda_2 - \lieD_{\txi_2} \Lambda_1 -[\Lambda_1,\Lambda_2] - \lT_1 \lT_2 \, [\LambdaB,\LambdaB] + x \LambdaB \,,
\end{align}
Having ensured that the flavour gauge field transformation works, it is clear that the transformation of $\LambdaB$ will also follow along similar lines. 

We can now fix the free parameter $x$ by demanding consistency of the $\UT$ gauge field transformation. To wit, $\AT_\mu$ transforms like an abelian gauge field, so it works in a way similar to the flavour field $A_\mu$:
\begin{align}
[\delta_{\Xfields_1}, \delta_{\Xfields_2}] \AT_\mu & = \lieD_{\txi_3} \AT_\mu + \lieD_{\txi_1} \partial_\mu \LambdaTb_2 - \lieD_{\txi_2} \partial_\mu \LambdaTb_1   \notag\\
&= \lieD_{\txi_3} \AT_\mu + \partial_\mu \LambdaTb_3
\label{}
\end{align}
with transformation parameters
\begin{equation}\label{eq:LambdaDef}
\begin{split}
\LambdaTb_3 & = \lieD_{\txi_1} \LambdaTb_2 - \lieD_{\txi_2} \LambdaTb_1 \,, \\
\LambdaT_3 &=  \lT_2 \lieD_{\xi_1} \LambdaBT  - \lT_1 \lieD_{\xi_2} \LambdaBT  
  + \lieD_{\txi_1} \LambdaT_2 - \lieD_{\txi_2} \LambdaT_1 - x \, (\Kbeta^\nu \AT_\nu)\,,
\end{split}
\end{equation}
such that consistently $\LambdaTb_3 = \LambdaT_3 - \lT_3 (\Kbeta^\nu \AT_\nu)$. From demanding the relation \eqref{eq:lTdef} to hold, we can fix the parameter $x$. Indeed, using the definitions \eqref{eq:X3a} and \eqref{eq:LambdaDef}, the relation $\lT_3 = \LambdaTb_3 + \txi_3^\sigma\, \AT_\sigma$ only holds if
\begin{equation}
x = \txi^\mu_1 \txi^\nu_2 \, \FT_{\mu\nu}\,.
\end{equation}

Let us now consider the sources $\{\tildeg_{\mu\nu},\tildeA_\mu\}$. For the partner metric source $\tildeg_{\mu\nu}$ we find:
\begin{equation}
\begin{split}
[\delta_{\Xfields_1}, \delta_{\Xfields_2}] \tildeg_{\mu\nu} & = 
\lieD_{\txi_1} \prn{ \lieD_{\txi_2} \tildeg_{\mu\nu} + \LambdaTb_2 \, \diffB g_{\mu\nu} } + \LambdaTb_1\, \delta_{\Xfields_2} (\diffB g_{\mu\nu}) \\
&\quad - \lieD_{\txi_2} \prn{ \lieD_{\txi_1} \tildeg_{\mu\nu} + \LambdaTb_1 \, \diffB g_{\mu\nu} } - \LambdaTb_2 \,\delta_{\Xfields_1} (\diffB g_{\mu\nu}) \\
&= \lieD_{\txi_3} \tildeg_{\mu\nu} + \LambdaTb_3\, \diffB g_{\mu\nu}  
\end{split}
\label{eq:LTgtildeTrf}
\end{equation}
where we used 
\begin{equation}
\delta_{\Xfields_1} (\diffB g_{\mu\nu}) 
= \delta_{(\delta_{\Xfields_1} \Bfields)} \, g_{\mu\nu} + \diffB (\delta_{\Xfields_1} g_{\mu\nu})
= \lieD_{[\txi_1,\, \Kbeta]} g_{\mu\nu} + \lieD_{\Kbeta} (\lieD_{\txi_1} g_{\mu\nu})
= \lieD_{\txi_1} (\diffB g_{\mu\nu})
\end{equation}
and similarly for $\delta_{\Xfields_2} (\diffB g_{\mu\nu})$. Note that the successive application of transformations in \eqref{eq:LTgtildeTrf} introduces an inhomogeneous term (the last terms in the first and second line) which must be evaluated on the new configuration of fields and sources. 

The commutator of two transformations of $\tildeA_\mu$ works in a similar fashion:
\begin{equation}
\begin{split}
[\delta_{\Xfields_1}, \delta_{\Xfields_2}] \tildeA_\mu & = 
\lieD_{\txi_1} \prn{ \lieD_{\txi_2} \tildeA_\mu + \partial_\mu \tLam_2 + [A_\mu,\tLam_2]+ \LambdaTb_2 \, \diffB A_\mu } \\
&\quad\;\;\;\, + \brk{\lieD_{\txi_2} \tildeA_\mu + \partial_\mu \tLam_2 + [A_\mu,\tLam_2] + \LambdaTb_2 \, \diffB A_\mu \,, \, \tLam_1}  + \LambdaTb_1 \,\delta_{\Xfields_2} (\diffB A_\mu)  \\
&\quad- \lieD_{\txi_2} \prn{ \lieD_{\txi_1} \tildeA_\mu + \partial_\mu \tLam_1 +[A_\mu,\tLam_1]+ \LambdaTb_1 \, \diffB A_\mu } \\
&\quad\;\;\;\, - \brk{\lieD_{\txi_1} \tildeA_\mu + \partial_\mu \tLam_1 +[A_\mu,\tLam_1] + \LambdaTb_1 \, \diffB A_\mu \,, \, \tLam_2} - \LambdaTb_2 \,\delta_{\Xfields_1} (\diffB A_\mu)  \\
&= \lieD_{\txi_3} \tildeA_\mu + D_\mu \tLam_3 + \LambdaTb_3\, \diffB A_\mu \,,
\end{split}
\end{equation}
using $\delta_{\Xfields_1} (\diffB A_\mu) = \lieD_{\txi_1} (\diffB A_\mu) + [\diffB A_\mu , \, \tLam_1]$ and similarly for $\delta_{\Xfields_2} (\diffB A_\mu)$.

To summarize, the above computations show that the Class $\LT$ transformation rules \eqref{eq:TactgA}, \eqref{eq:TactgAbar} and \eqref{eq:TactAT} form a Lie algebra where the commutator of two transformations $\Xfields_1  = \{\txi^\mu_1,\tLam_1, \LambdaTb_1\}$ and $\Xfields_2  = \{\txi^\mu_2,\tLam_2, \LambdaTb_2\}$ is another diffeomorphism/gauge/$\UT$ transformation with parameters
\begin{equation}
\begin{split}
\txi_3^\sigma &= [\txi_1,\txi_2]^\sigma \,,\qquad
\tLam_3 = \lieD_{\txi_1} \tLam_2 - \lieD_{\txi_2} \tLam_1 - [\tLam_1,\tLam_2] \,,\qquad
\LambdaTb_3  = \lieD_{\txi_1} \LambdaTb_2 - \lieD_{\txi_2} \LambdaTb_1 \,.
\end{split}
\end{equation}
%

\subsection{Deriving Class $\LT$ Bianchi identities}
\label{sec:utbder}

We now outline some of the computations relevant to obtain Bianchi identities 
Eqs.~\eqref{eq:TBianchiLT}-\eqref{eq:JTBianchiLT} from the master Lagrangian $\LagT\brk{\hfieldsT}$.
The steps are straightforward, but rendered somewhat complex by the sheer number of fields in $\hfieldsT$. 
We start with the computation of a full diffeomorphism, flavour, and $\UT$ transformation on $\LagT$
using the untwisted transformation parameters introduced in \S\ref{sec:fieldsLT}:\footnote{ We thank A.~Jain for pointing out to us some computational mistakes in an earlier version of this appendix.}
\begin{align}\label{eq:U1TbianchiVar}
&\frac{1}{\sqrt{-g}}\diffF \prn{\sqrt{-g}\ \LagT} -\nabla_\mu (\PSymplPot{\Xfields}^\smallT)^\mu
\notag\\ &= 
	\half \; \TL^{\mu\nu}\;\diffF g_{\mu\nu} + \JL^\mu \cdot \diffF A_\mu +
	T \,\aheat_\sigma \;\diffF \Kbeta^\sigma
	+ T\, \acharge \cdot \prn{\diffF\LambdaB+ A_\sigma \,\diffF \Kbeta^\sigma} 
\notag\\ &\qquad 
	+\; \half \; \TLc^{\mu\nu}\;\diffF \tildeg_{\mu\nu} + \JLc^\mu \cdot \diffF \tildeA_\mu 
	+\JT^\sigma \;\diffF \AT_\sigma + T\, \,\achargeT  \prn{\diffF\LambdaBT+ \AT_\sigma \,\diffF \Kbeta^\sigma} 
\notag\\ & =
	 \txi^\sigma 
	\Biggl\{
	-  D_\mu (\TLLc)_\sigma^\mu +\JLLc^\nu \cdot F_{\sigma\nu}+ \JT^\nu \, \FT_{\sigma\nu}  
	\Biggr.
\notag\\ &\qquad 
	\Biggl. \qquad \quad
	+\;\frac{1}{\sqrt{-g}} \diffB \prn{\sqrt{-g}\ T\ \aheat_\sigma} + T\,\acharge\cdot \diffB A_\sigma 
	+ T\,\achargeT\cdot \diffB \AT_\sigma  +\; D_\nu \prn{\tgdiff_{\sigma\mu}\, \TLc^{\mu\nu}} 
	\Biggr.
\notag\\ &  \qquad 
	\Biggl. 
	\qquad\quad
	-\; \half \TLc^{\mu\nu} \, D_\sigma \tgdiff_{\mu\nu} 
	-  \JLc^\nu\cdot \,\tFdiff_{\sigma\nu} 
	-\;\AT_\sigma \prn{  \half \; \TLc^{\mu\nu} \diffB g_{\mu\nu}  +\JLc^\mu\cdot  \diffB A_\mu } 
	\Biggr\} 
\notag\\  & \qquad 
	 -\prn{\tLam+\txi^\nu \, A_\nu } \cdot 
	 \Biggl\{ D_\mu  \JLLc^\mu  
	 -\frac{1}{\sqrt{-g}} \diffB \prn{\sqrt{-g}\ T\ \acharge}  \Biggr\}
\notag\\ &\qquad 
	+ \prn{\tLam+\txi^\nu \,\tAdiff_\nu } \cdot  
	\bigg[D_\mu \JLc^\mu  - [\tildeA_\mu, \JLc^\mu] \bigg]
\notag\\&\qquad  
	- (\LambdaTb+\txi^\sigma \AT_\sigma ) 
	\Biggl\{D_\mu \JT^\mu -\half \TLc^{\mu\nu} \diffB g_{\mu\nu}  - \JLc^\mu \cdot \diffB A_\mu   
	- \frac{1}{\sqrt{-g}} \diffB \prn{\sqrt{-g}\ T\ \,\achargeT} \Biggr\}
\notag\\&\qquad
	+\; D_\mu \Biggl(\ \txi_\nu\, \TLLc^{\mu\nu}  + \prn{\tLam+\txi^\nu  A_\nu } \cdot \JLLc^\mu 
	+ (\LambdaTb + \txi^\nu  \AT_\nu) \JT^\mu  \Biggr.
\notag\\ &\qquad
	 \Biggl.
	 \qquad \qquad
	 -\; \Kbeta^\mu \bigbr{ T\, \aheat_\sigma \,\txi^\sigma+  T\,\acharge \cdot \prn{\tLam
	 + \txi^\nu  A_\nu} + T\,\achargeT\ (\LambdaTb + \txi^\nu  \AT_\nu) }  
	 \Biggr.
\notag\\ &\qquad 
	\Biggl. 
	\qquad\qquad  
	-\;\txi^\sigma \, \tgdiff_{\alpha\sigma}\,  \TLc^{\alpha\mu} 
	- \prn{\tLam+\txi^\nu  \tAdiff_\nu} \cdot  \JLc^\mu\
	\Biggr) \,.
\end{align}
In deriving this we have used the transformation rules given in Eqs.~\eqref{eq:TactgA}, \eqref{eq:TactgAbar} and 
\eqref{eq:TactAT} and integrated  terms by parts where necessary. Furthermore, we used $\diffB\AT_\mu = \Kbeta^\nu\FT_{\nu\mu}$.

From this one can read off the diffeomorphism and flavour Bianchi identities: 
\begin{itemize}
\item The diffeomorphism identity \eqref{eq:TBianchiLT} can be read off as  the coefficient of $\txi^\sigma$. This is clear since switching off $\LambdaT$ implies that $\txi^\sigma \to \xi^\sigma$ and indeed the coefficient of the latter is the desired term. The answer then if the set of terms in the curly braces of the first three lines and in addition, the contribution from the fifth line. This is because we have an isolated $\txi^\sigma\, \tildeA_\sigma$ term, which is not the flavour invariant combination and should be accounted for in the diffeomorphism transformations.\footnote{ This term and contributions involving $\tildeg_{\mu\nu}$ or $\tgdiff_{\mu\nu}$ as we have written owe their origin to the tensorial properties of $\{\tildeg_{\mu\nu}, 
\tildeA_\mu\}$ which matters when we perform diffeomorphisms.}
\item The flavour Bianchi identity \eqref{eq:JBianchiLT} is  given by the coefficient of   $(\bar{\Lambda}+\txi^\nu A_\nu)$. This gets contributions from the fourth and fifth lines respectively.
\end{itemize}

We can read off another interesting identity  from the coefficient of $(\LambdaTb+\txi^\sigma \AT_\sigma )$ in  \eqref{eq:U1TbianchiVar}:
\begin{equation}
\begin{split}
D_\mu \JT^\mu =\half \TLc^{\mu\nu} \diffB g_{\mu\nu} + \JLc^\mu \cdot \diffB A_\mu + 
 \frac{1}{\sqrt{-g}} \diffB \prn{\sqrt{-g}\ T\ \,\achargeT}
\end{split}
\label{eq:utbalt}
\end{equation}
This is just the grand canonical adiabaticity equation for $\{\TLc^{\mu\nu},\JLc^\mu\}$, where the free energy current is identified with $\JT^\mu - T\,\achargeT\, \Kbeta^\mu$. One might at this point think that we are done with deriving the three Bianchi identities following from the diffeomorphism, flavour, and $\UT$ transformations. However, owing to  the twisted character of the $\UT$ symmetry, \eqref{eq:utbalt} is not the actual $\UT$ Bianchi identity, but rather a combination of it with the other Bianchi identities. To extract the actual $\UT$ Bianchi identity, we need to shift back to the original transformation parameters  $\{\xi^\mu, \Lambda, \LambdaT \}$. We now turn to this exercise.

As mentioned in the discussion above, upon shifting to the original transformation parameters  $\{\xi^\mu, \Lambda, \LambdaT \}$ the diffeomorphism and flavour Bianchi identities do not change. However, to get the actual $\UT$ Bianchi identity, we have to do this shift and then read off the coefficient of $(\LambdaT+\xi^\sigma \AT_\sigma )$. To do this, it is convenient to define
the $\UT$ Noether current as in \eqref{eq:utNTdef}.

We can then use this expression to simplify  \eqref{eq:U1TbianchiVar}. Let us temporarily group together the terms in  the r.h.s. of this expression into the total derivative and non-derivative terms
\begin{equation}
\frac{1}{\sqrt{-g}}\diffF \prn{\sqrt{-g}\ \LagT} -\nabla_\mu (\PSymplPot{\Xfields}^\smallT)^\mu 
\equiv {\mathfrak O}\brk{\txi^\mu, \tLam,\LambdaTb} + D_\mu \prn{{\mathfrak T}^\mu	\brk{\txi^\mu, \tLam,\LambdaTb}}	
\end{equation}	
We can now use \eqref{eq:utNTdef} to simplify the two contributions separately. Firstly we find that  
a straightforward substitution leads to a simplification of ${\mathfrak T}^\mu\brk{\txi^\mu, \tLam,\LambdaBT}$ to
 %
\begin{align}
{\mathfrak T}^\mu \brk{\bar\xi^\mu, \bar\Lambda,\LambdaTb} &= 
	 \xi_\nu \,\TLLc^{\mu\nu}  + \prn{\Lambda +\xi^\nu  A_\nu } \cdot \JLLc^\mu + 
	 (\LambdaT + \xi^\nu  \AT_\nu) \, \NT^\mu [\Kbeta]
 \notag \\ &  
 	 -\;  \Kbeta^\mu \bigbr{ T\ \aheat_\sigma \,\xi^\sigma+  T\, \acharge \cdot \prn{\Lambda + \xi^\nu  A_\nu} } 
 	 - \xi^\sigma \, \tgdiff_{\sigma\alpha}\, \TLc^{\mu\alpha} 
	- \prn{\Lambda+\xi^\nu  \tAdiff_\nu} \cdot  \JLc^\mu  
\end{align}
upon defining the Noether current
\begin{align}
\NT^\mu[\bar\xi\,]
&\equiv
	 \JT^\mu +\bar{\xi}_\nu \TLLc^{\mu\nu}  + \prn{\Lambda +\bar{\xi}^\nu  A_\nu } \cdot \JLLc^\mu  
	 - \bigbr{\aheat_\sigma\, \bar{\xi}^\sigma+  \acharge \cdot 
	 \prn{\Lambda + \bar{\xi}^\nu  A_\nu} + (\LambdaTb+\bar{\xi}^\sigma \AT_\sigma)\,\achargeT} \, u^\mu 
 \nonumber \\ &\qquad   \qquad 
	 - \bar{\xi}^\sigma \; \tgdiff_{\sigma\nu} \;  \TLc^{\mu\nu}
	 - (\Lambda +\bar{\xi}^\nu \,\tAdiff_\nu )  \cdot  \JLc^\mu\ \,.
\label{eq:utNTdefApp}
\end{align}
and evaluating it on $\bar \xi^\mu = \Kbeta^\mu$.

Further, we can simplify the non-total derivative pieces combined into ${\mathfrak O}$ by using the easily verified identity
\begin{equation}
\begin{split}
&{\mathfrak O}\brk{\bar\xi^\mu,\bar \Lambda,\LambdaTb}
=  
	 -D_\mu \prn{\NT^\mu[\bar{\xi}]  + (\LambdaTb+\bar{\xi}^\sigma \AT_\sigma-1)\JT^\mu   } 
	+ \JT^\mu  \diffF \AT_\mu 
\\ &\qquad\qquad 
	+ \; \half \TL^{\mu\nu} \, \diffF g_{\mu\nu}  
	+ \JL^\mu\cdot  \diffF A_\mu  
	+ \half\TLc^{\mu\nu} \diffF \tildeg_{\mu\nu} 
	+\JLc^\mu\cdot  \diffF \tildeA_\mu 
\\ &\qquad\qquad
+ T \, \aheat_\sigma \, \diffF \Kbeta^\sigma 
	+ T\, \,\acharge  \prn{\diffF\LambdaB+ A_\sigma \,\diffF \Kbeta^\sigma}
	+ T\, \,\achargeT  \prn{\diffF\LambdaBT+ \AT_\sigma \,\diffF \Kbeta^\sigma}\,,
\end{split}
\end{equation}
%
Evaluated on $\{\Kbeta^\mu,\LambdaB,\LambdaBT\}$, this becomes
\begin{equation}
\begin{split}
&{\mathfrak O}\brk{\Kbeta^\mu,\LambdaB,\LambdaBT}
=  
	 -D_\mu \prn{\NT^\mu[\Kbeta]  + (\LambdaBT+\Kbeta^\sigma \AT_\sigma-1)\JT^\mu   } 
	+ \JT^\mu  \diffB \AT_\mu
\\ &\qquad\qquad 
	+ \; \half \TL^{\mu\nu} \, \diffB g_{\mu\nu}  
	+ \JL^\mu\cdot  \diffB A_\mu  
	+ \half\TLc^{\mu\nu} \diffB \tildeg_{\mu\nu} 
	+\JLc^\mu\cdot  \diffB \tildeA_\mu \,,
\end{split}
\end{equation}
%
Plugging this into the full variation \eqref{eq:U1TbianchiVar} we find
\begin{align}
\frac{1}{\sqrt{-g}}&\diffF \prn{\sqrt{-g}\ \LagT} -\nabla_\mu (\PSymplPot{\Xfields}^\smallT)^\mu
\notag \\
 &= 
	\xi^\sigma \Biggl\{
	-  D_\mu (\TLLc)_\sigma^\mu +\JLLc^\nu \cdot F_{\sigma\nu}+ \JT^\nu \, \FT_{\sigma\nu}  
	\Biggr.
\notag\\ &\qquad 
	\Biggl. \qquad \quad
	+\;\frac{1}{\sqrt{-g}} \diffB \prn{\sqrt{-g}\ T\ \aheat_\sigma} + T\,\acharge\cdot \diffB A_\sigma 
	+ T\,\achargeT\cdot \diffB \AT_\sigma  +\; D_\nu \prn{\tgdiff_{\sigma\mu}\, \TLc^{\mu\nu}} 
	\Biggr.
\notag\\ &  \qquad 
	\Biggl. 
	\qquad\quad
	-\; \half \TLc^{\mu\nu} \, D_\sigma \tgdiff_{\mu\nu} 
	-  \JLc^\nu\cdot \,\tFdiff_{\sigma\nu}   
	-\;\AT_\sigma \prn{  \half \; \TLc^{\mu\nu} \diffB g_{\mu\nu}  +\JLc^\mu\cdot  \diffB A_\mu } 
	\Biggr\} 
\notag\\  & \qquad 
	 -\prn{\Lambda+\xi^\nu \, A_\nu } \cdot 
	 \Biggl\{ D_\mu  \JLLc^\mu  
	 -\frac{1}{\sqrt{-g}} \diffB \prn{\sqrt{-g}\ T\ \acharge}  \Biggr\}
\notag\\ &\qquad 
	+ \prn{\Lambda+\xi^\nu \,\tAdiff_\nu } \cdot  
	\bigg[D_\mu \JLc^\mu  - [\tildeA_\mu, \JLc^\mu] \bigg]
\notag\\&\qquad  
	- \; (\LambdaT+\xi^\sigma \AT_\sigma ) 
	\Biggl\{ - {\mathfrak O}\brk{\Kbeta^\mu,\LambdaB,\LambdaBT} 
\notag\\&\qquad
	-
	(\LambdaBT+\Kbeta^\sigma \AT_\sigma -1) 
	\Big[ D_\mu \JT^\mu -\half \TLc^{\mu\nu} \diffB g_{\mu\nu}  - \JLc^\mu \cdot \diffB A_\mu   
	- \frac{1}{\sqrt{-g}} \diffB \prn{\sqrt{-g}\ T\ \,\achargeT} \Big]
\Biggr\}
\notag \\ & \qquad
	+\; D_\mu\Big({\mathfrak T}^\mu\brk{\xi, \Lambda, \LambdaT} 
	\Big)
\label{eq:LTvarfinal}
\end{align}
The $\UT$ Bianchi identity can now be read off as the coefficient of $(\LambdaT+\xi^\sigma \AT_\sigma )$ in the three lines preceding the total derivative term. We have chosen to leave the latter in terms of the intermediate quantities defined earlier, to avoid unnecessary repetition. After setting $(\LambdaT+\xi^\sigma \AT_\sigma ) =1$, the $\UT$ Bianchi identity takes the form \eqref{eq:U1TBianchi}: $ {\mathfrak O}\brk{\Kbeta^\mu,\LambdaB,\LambdaBT}=0$.

\newpage

\section{Notation and conventions}
\label{sec:Notation}

\begin{table}[h!]
\centerline{
\begin{tabular}{||r|l||r|l||}
\hline
   \bf{Symbol} & \bf{Definition} & \bf{Symbol} & \bf{Definition} \\
   \hline \hline
   \multicolumn{4}{||c||}{{\shadeB{\it Basic hydrodynamical variables $\hfields$ on physical spacetime ${\cal M}$}}} \\
   \hline
   $g_{\mu\nu}$ & Background metric &
   $A_\mu$ & Background gauge field \\
   $\Kbeta^\mu$ & Thermal vector $=\frac{1}{T}\, u^\mu$ &
   $\LambdaB$& Thermal twist $=\frac{\mu}{T} - \Kbeta^\nu A_\nu$\\
   \hline
   \multicolumn{4}{||c||}{{\shadeR \it Hydrostatic variables $\Eqfields$}} \\
   \hline
   $\KEq^\mu$ & Hydrostatic thermal vector &
   $\LambdaEq$ & Hydrostatic thermal twist \\
   \hline
   \multicolumn{4}{||c||}{{\shadeB \it Hydrodynamic tensors at first order in gradients}} \\
   \hline
   $\sigma^{\mu\nu}$ &  Shear tensor
    $= P^{\mu\alpha}P^{\nu\beta} \left(\nabla_{(\alpha}u_{\beta)} - \frac{\Theta}{d-1}\,
   P_{\alpha\beta}\right)$&
   $\Theta$ & Expansion $ = \nabla_\mu u^\mu$\\
    $\omega^{\mu\nu}$ &     Vorticity tensor
    $= P^{\mu\alpha}P^{\nu\beta}\, \nabla_{[\alpha}u_{\beta]}$&
   $\acc^\mu$&  Acceleration $= u^\nu \nabla_\nu u^\mu$\\
  $\cv^\mu$ & Potential gradient =  $E^\mu - TP^{\mu\nu}\, \nabla_\nu \left(\frac{\mu}{T}\right) $ & 
$E^\mu$ & Electric field = $F^{\mu\nu} u_\nu$\\
  $B^{\mu\nu} $ & Magnetic field $= P^{\mu\alpha}P^{\nu\beta}F_{\alpha\beta} $ &  & \\
      \hline 
     \multicolumn{4}{||c||}{{\shadeR \it Currents $\hcur[\hfields]$}} \\
      \hline
      $T^{\mu\nu}$ & Energy-momentum tensor &
      $J^\mu$ & Charge current \\
      $J_S^\mu$ & Fluid entropy current &
      $\mathcal{G}^\mu$& Gibbs free energy current (\ref{eq:GDef})\\
      \hline
      \multicolumn{4}{||c||}{{\shadeB \it Further physical currents}} \\
      \hline
      $\aheat_\sigma$ & Adiabatic heat current (\ref{eq:LagVar}) &
      $\acharge$ & Adiabatic charge density (\ref{eq:LagVar})\\
      $(\PSymplPot{})^\mu$ & Presymplectic potential (\ref{eq:LagVar}) &
      $\Komar^{\mu\nu}$ & Komar charge (\ref{eq:KomarDef}) \\
      $\N^\mu$ & Noether current (\ref{eq:Nchi}) &
      & \\ 
      \hline
\end{tabular}
}
\caption{Basic fields, sources and hydrodynamic quantities on physical spacetime manifold ${\cal M}$.}
\label{notation:tabFields}
\end{table}

\begin{table}[h!]
\centerline{
\begin{tabular}{||r|l||r|l||}
\hline
   \bf{Symbol} & \bf{Definition} & \bf{Symbol} & \bf{Definition} \\
   \hline \hline
   \multicolumn{4}{||c||}{{\shadeB \it Basic hydrodynamical variables $\hreffields$ on reference spacetime $\Mref$}} \\
   \hline
   $\gref_{ab}$ & Background metric \eqref{eq:ghAh} &
   $\Aref_a$ & Background gauge field \eqref{eq:ghAh}\\
   $\Kref^a$ & Thermal vector $=\frac{1}{\Tref}\, \uref^a$ &
   $\Lref$& Thermal twist $=\frac{\muref}{\Tref} - \Kref^a \Aref_a$ \\
   \hline
   \multicolumn{4}{||c||}{{\shadeR \it Derived hydrodynamical variables on reference spacetime $\Mref$}} \\
   \hline
   $\Tref$ & Temperature &
   $\muref$ & Chemical potential\\
   $\uref^a$ & Velocity &
   & \\
   \hline
   \multicolumn{4}{||c||}{{\shadeB \it Transition functions from ${\cal M}$ to $\Mref$}} \\
   \hline
   $\varphi^a$ & Diffeomorphism field (\ref{eq:KLambdaKPullBack}) &
   $c$ & Gauge transformation  (\ref{eq:KLambdaKPullBack})\\
   \hline
   \multicolumn{4}{||c||}{{\shadeR \it Currents \eqref{eq:MrefVariation}}} \\
   \hline
   $\Tref^{ab}$ & Energy-momentum tensor &
   $\Jref^a$ & Charge current \\
   $\Jref^a_S$ & Fluid entropy current & 
   $\aheatref_a$ & Adiabatic heat current \\
   $\achargeref$ & Adiabatic charge density &
    & \\
   \hline    
\end{tabular}
}
\caption{Basic sources, fields and currents on reference manifold $\Mref$.}
\label{notation:tabFieldsRef}
\end{table}

\begin{table}[h!]
\centerline{
\begin{tabular}{||r|l||r|l||}
\hline
   \bf{Symbol} & \bf{Definition} & \bf{Symbol} & \bf{Definition} \\
   \hline \hline
   \multicolumn{4}{||c||}{{\shadeR \it Variational symbols and derivatives}} \\
   \hline
   $\delta$ & Unconstrained variation &
   $\diffF$ & Generic diffeo/gauge transf.\ (\ref{eq:GenericDiffeo})\\ 
   $\diffB$ & Diffeo/gauge trf.\ w.r.t.\ $\Bfields = \{\Kbeta^\mu,\LambdaB\}$ (\ref{eq:delBdef})  &
   $\diffBref$ & Diffeo/gauge trf.\ on $\Mref$ w.r.t.\ $\Breffields = \{\Kref^a,\Lref\}$  \\
   $\delta_\varphi$ & Lie drag on $\Mref$ along $\{\delta \varphi^a, -c^{-1}\delta c\}$ \eqref{eq:varphiVar} & 
   $\diffCons$ & Constrained variation along Lie orbits (\ref{eq:consLvar})  \\
   $D_\alpha$ &  Gauge covariant derivative (\ref{eq:CovDer}) &
   $\lieD_\xi$ & Lie derivative along $\xi^\mu$ \\
   $\DWeyl_\lambda$ & Weyl covariant derivative (\ref{DWeyl:eq}) &
   $\diffFW$ & $\diffF+$ Weyl transformation (\ref{eq:AgGamWeyl})  \\
   \hline
   \multicolumn{4}{||c||}{{\shadeB \it Indices}} \\
   \hline  
   \hspace{-.15cm}\scriptsize{$\alpha,\beta,\mu,\nu,\ldots$} & Physical manifold ${\cal M}$ &
   \scriptsize{$a,b,c,\ldots$} & Reference manifold $\Mref$ \\
   \scriptsize{$m,n,p,\ldots$} & Physical bulk manifold $\bulkM_{d+1}$ & 
   \scriptsize{$\mb,\nb,\pb,\ldots$} & Reference bulk manifold $\Mref_{d+1}$ \\
   \hline
\end{tabular}
}
\caption{Variational symbols, derivatives, index conventions.}
\label{notation:tabVariations}
\end{table}

\begin{table}[h!]
\centerline{
\begin{tabular}{||r|l||r|l||}
\hline
   \bf{Symbol} & \bf{Definition} & \bf{Symbol} & \bf{Definition} \\
   \hline \hline
   \multicolumn{4}{||c||}{{\shadeB \it Shadow connections}} \\
   \hline
   $\hat{\Gamma}^\mu{}_{\nu\rho}$ & Shadow spin connection $=\Gamma^\mu{}_{\nu\rho} + \Omega^\mu{}_\nu \,u_\rho$ &
   $\hat{A}^\mu$ & Shadow gauge field $=A^\mu+\mu\,u^\mu$ \\
   $\Omega^\mu{}_\nu$ & Spin potential $=\tfrac{1}{2} T(D_\nu \Kbeta^\mu - D^\mu \Kbeta_\nu)$ & & \\
   \hline
   \multicolumn{4}{||c||}{{\shadeR \it Anomaly induced currents}} \\
   \hline
   $\SpH{}^{ab}{}_c$ & Bulk Hall spin current (\ref{eq:HallCurrentsDef}) &
   $\JH^a $ & Bulk Hall charge current (\ref{eq:HallCurrentsDef}) \\
   $\THall^{\mu\perp} $ & Covariant Lorentz anomaly $=\frac{1}{2} D_\nu \SpH^{\perp[\mu\nu]}$ &
   $\JH^\perp$ & Covariant flavour anomaly \\
   $\JP^\mu$ & Anomalous flavour current (\ref{eq:InflowDef}) &
   $\SP$ & Anomalous spin current (\ref{eq:InflowDef}) \\
   $\qP^\mu$ & Mixed anomalous current (\ref{eq:InflowDef}) &
   $(J^\alpha_{S})_\text{A}$ & Anomalous entropy current (\ref{eq:ConstRelations}) \\
   $(T^{\alpha\beta})_\text{A}$ & Anomalous boundary stress tensor (\ref{eq:bdyAnomCur}) &
   $(J^\alpha)_\text{A}$ & Anomalous boundary current (\ref{eq:bdyAnomCur}) \\
   $\Tbulk^{mn}$ & Anomalous bulk stress tensor (\ref{eq:DeltaSanom2})&
   $\Jbulk^m$ & Anomalous bulk current (\ref{eq:DeltaSanom2})\\
   $\aheatbulk_m$ & Anomalous bulk heat current (\ref{eq:DeltaSanom2})&
   $\achargebulk$ & Anomalous bulk charge density (\ref{eq:DeltaSanom2})\\
   \hline
   \multicolumn{4}{||c||}{{\shadeB \it Basic fields as differential forms}} \\
   \hline   
   $\fu$ & Velocity 1-form $=u_\mu \, dx^\mu$ & 
   $\iota_{\fu} \form{X}^{\text{\tiny{(2)}}}$ & Velocity contraction $=u^\mu X^{\text{\tiny{(2)}}}_{\mu\nu} dx^\nu$ \\
   $\fa$ & Acceleration 1-form $=\acc_\mu \, dx^\mu$ &   
   $2\fomega$ & Vorticity 2-form $=d\fu + \fu \wedge \fa$ \\
   $\fA$ & Gauge field 1-form $=A_\mu \, dx^\mu$ &
   $\fGamma^\mu{}_\nu$ & Connection 1-form $=\Gamma^\mu{}_{\nu\rho}\, dx^\rho$ \\
   $\fAh$ & Shadow gauge field $=\fA + \mu \, \fu$ &
   $\fGammah^\mu{}_\nu$ & Shadow connection $ =\fGamma^\mu{}_\nu + \Omega^\mu{}_\nu\, \fu$ \\
   $\fF$ & Field strength  $= d\fA + \fA \wedge \fA$ &
   $\fR^\mu{}_\nu$ & Curvature  $=d\fGamma^\mu{}_\nu + \fGamma^\mu{}_\rho \wedge \fGamma^\rho{}_\nu$ \\
   $\fE$ & Electric field $=-\iota_{\fu} \fF$ &
   $(\fER)^\mu{}_\nu$ &  Electric curvature $=-\iota_{\fu} \fR^\mu{}_\nu$ \\
   $\fB$ & Magnetic field $=\fF-\fu\wedge \fE$ &
   $(\fBR)^\mu{}_\nu$ & Magnetic curvature $=\fR^\mu{}_\nu - \fu\wedge (\fER)^\mu{}_\nu$ \\
   $\AT$ & $\UT$ gauge field 1-form &
   $\fFT$ & $\UT$ field strength 2-form $= d\AT$ \\
   \hline
   \multicolumn{4}{||c||}{{\shadeB \it Abbreviations}} \\
   \hline
   $\fatQ^{\mu\alpha}_{\nu\beta}$ & Antisymmetrizer $= \tfrac{1}{2} ( \delta^\mu_\beta\, \delta^\alpha_\nu - g^{\mu\alpha} \,g_{\nu\beta})$ &
   $\fatP^{\rho\mu\nu}_{\sigma\kappa\lambda}$ & Projector $= \delta^\rho_\kappa \, \delta^\mu_\sigma\,  \delta^\nu_\lambda  + \fatQ^{\rho\mu}_{\sigma\kappa} \, u^\nu u_\lambda$ \\
   $\bulkint$ & Bulk integral $=\int_{{\bulkM}_{d+1}} \,\sqrt{-g_{d+1}}$ &
   $\skR,\skL$ & Shortcuts for right and left SK copies \\
   \hline
\end{tabular}
}
\caption{Class A and Schwinger-Keldysh (SK) quantities on physical spacetime ${\cal M}$.}
\label{notation:tabAnom}
\end{table}

\begin{table}[h!]
\centerline{
\begin{tabular}{||r|l||r|l||}
\hline
   \bf{Symbol} & \bf{Definition} & \bf{Symbol} & \bf{Definition} \\
   \hline \hline
   \multicolumn{4}{||c||}{{\shadeB \it Shadow connections}} \\
   \hline
   $\hat{\Chref}^a{}_{bc}$ & Shadow spin connection $=\Chref^a{}_{bc} + \Omegaref^a{}_b \,\uref_c$ &
   $\hat{\Aref}^a$ & Shadow gauge field $=\Aref^a+\muref\,\uref^a$ \\
   $\Omegaref^a{}_b$ & Spin potential $=\tfrac{1}{2} \Tref(\Dref_b \Kref^a - \Dref^a \Kref_b)$ & & \\
   \hline
   \multicolumn{4}{||c||}{{\shadeR \it Anomaly induced currents}} \\
   \hline
   $\SpHref{}^{\mb\nb}{}_\pb$ & Bulk Hall spin current &
   $\JHref^\mb $ & Bulk Hall charge current  \\
   $\JPref^a$ & Anomalous flavour current (\ref{eq:DeltaSanomFinal}) &
   $\SPref$ & Anomalous spin current (\ref{eq:DeltaSanomFinal}) \\
   $\qPref^a$ & Mixed anomalous current current (\ref{eq:DeltaSanomFinal}) &
   $(\Jref^a_{S})_\text{A}$ & Anomalous entropy current \\
   $(\Tref^{ab})_\text{A}$ & Anomalous boundary stress tensor  &
   $(\Jref^a)_\text{A}$ & Anomalous boundary current \\
   $\Tbulkref^{\mb\nb}$ & Anomalous bulk stress tensor &
   $\Jbulkref^\mb$ & Anomalous bulk current \\
   $\Tref_{hydro}^{ab}$ & Hydrodynamic SK stress tensor \eqref{eq:hydroCur} &
   $\Jref_{hydro}^a$ & Hydrodynamic SK charge current \eqref{eq:hydroCur} \\
   $(\Tbulkref^{\mb\nb})_{_{IF}}$ & Cross contour stress tensor \eqref{eq:AnomVar1} &
   $(\Jbulkref^{\mb})_{_{IF}}$ & Cross contour charge current \eqref{eq:AnomVar1} \\
   \hline
   \multicolumn{4}{||c||}{{\shadeB \it Abbreviations}} \\
   \hline
   $\gratio$ & Ratio of measures $=\sqrt{-\gref_\skL}/\sqrt{-\gref_\skR}$ &
   $\gratio_{\skR,\skL}$ & Ratio of measures $=\sqrt{-\gref_{\skR,\skL}}/\sqrt{-\breve\gref}$ \\
   $\dRLhfref$ & Difference fields $=\hreffields_\skR - \hreffields_\skL$ &
   $\aRLhfref$ & Common fields $=\tfrac{1}{2} (\hreffields_\skR+\hreffields_\skL)$ \\
   $\bulkintref$ & Bulk integral $=\int_{{\Mref}_{d+1}} \,\sqrt{-\gref_{d+1}}$ &
   & \\
   \hline
\end{tabular}
}
\caption{Class A and Schwinger-Keldysh (SK) quantities on reference manifold $\Mref$.}
\label{notation:tabAnomRef}
\end{table}

\begin{table}[h!]
\centerline{
\begin{tabular}{||r|l||r|l||}
\hline
   \bf{Symbol} & \bf{Definition} & \bf{Symbol} & \bf{Definition} \\
   \hline \hline
   \multicolumn{4}{||c||}{{\shadeB \it Basic fields}} \\
   \hline
    $\tildeg_{\mu\nu}$ & Partner metric & $\tildeA_\mu$ & Partner gauge field \\
    $\tgdiff_{\mu\nu}$ & Shifted partner metric $=g_{\mu\nu}-\tildeg_{\mu\nu}$ & 
    $\tAdiff_\mu$ & Shifted partner gauge field $=A_\mu-\tildeA_\mu$ \\
    $\bgdiff_{\mu\nu}$ & SK difference $=\tgdiff_{\mu\nu} 
 - \Kbeta_{\mu}\, \AT_{\nu} -\Kbeta_\nu\, \AT_\mu$ &
    $\bAdiff_\mu$ & SK difference $=\tAdiff_\mu - (\LambdaB + \Kbeta^\alpha A_\alpha) \AT_\mu$ \\ 
    $\AT_\mu$ & $\UT$ KMS gauge field & $\LambdaBT$ & $\UT$ holonomy field \\
    $\hfieldsT$ & $=\{\Kbeta^\mu,\LambdaB, g_{\mu\nu},A_\mu,\tildeg_{\mu\nu},\tildeA_\mu,\AT_\mu,\LambdaBT\}$ & & \\
   \hline
   \multicolumn{4}{||c||}{{\shadeR \it Currents}} \\
   \hline
   $\TL^{\mu\nu}$ & Stress tensor associated to $g_{\mu\nu}$ &
   $\JL^\mu$ & Charge current associated to $A_\mu$ \\
   $\TLc^{\mu\nu}$ & Stress tensor associated to $\tildeg_{\mu\nu}$ & 
   $\JLc^\mu$ & Charge current associated to $\tildeA_\mu$ \\
   $\JT^\mu$ & $\UT$ KMS current & $\achargeT$ & Adiabatic $\UT$ charge density \\
   $\NT^\sigma$ & Class $\LT$ Noether current \eqref{eq:utNTdef} &
   $\GT^\sigma$ & Class $\LT$ free energy current $ = -T \, \NT^\sigma$ \\ 
   $\JST^\mu$ & Class $\LT$ entropy current  \eqref{eq:JSTdef} & & \\ 
   \hline
\end{tabular}
}
\caption{Class $\LT$ quantities.}
\label{notation:tabLT}
\end{table}

\begin{table}[h!]
\centerline{
\begin{tabular}{||r|l||r|l||}
\hline
   \bf{Symbol} & \bf{Definition} & \bf{Symbol} & \bf{Definition} \\
   \hline \hline
   \multicolumn{4}{||c||}{{\shadeR \it Hydrodynamical fields for Legendre transform to entropic description}} \\
   \hline
   $\Snd_{\alpha_1\cdots\alpha_{d-1}}$ & $(d-1)$-form entropy current dual to $T\,s\,\Kbeta^\mu$&
   $\epformref{\Sndref}^a$ & Dual of $\Snd_{\alpha_1\cdots\alpha_{d-1}}$ on $\Mref$ (\ref{eq:RefND}) \\
   $(\Lnd)_{\alpha_1\ldots\alpha_d}$ & $d$-form gauge parameter dual to $T\,s\,\LambdaB$ &
   $\epformref{\Lndref}$ & Dual of $(\Lnd)_{\alpha_1\ldots\alpha_d}$ on $\Mref$ (\ref{eq:RefND}) \\
   \hline
\end{tabular}
}
\caption{Class ND fields.}
\label{notation:tabND}
\end{table}

\clearpage

\providecommand{\href}[2]{#2}\begingroup\raggedright\endgroup

\end{document}